\documentclass[12pt,a4paper,twoside,openany]{book}
\usepackage[top=2.55cm, bottom=3.0cm, left=3.7cm, right=2.55cm]{geometry}
\usepackage{graphicx}
\usepackage{epsfig}
\usepackage{amssymb,amsmath}
\usepackage[nottoc]{tocbibind}
\usepackage[compact]{titlesec}  
\usepackage{setspace,cite} 

\usepackage[usenames,dvips]{color}
\usepackage{longtable,lscape}
\usepackage{hvfloat}
\usepackage{url}
\usepackage{epsfig}
\usepackage{subfigure}
\usepackage{booktabs}
\usepackage{multirow}
\usepackage{rotating}

\usepackage[T1]{fontenc}
\usepackage{fourier}
\usepackage{bm}

\usepackage{array}

\def\baselinestretch{1.5}

\numberwithin{equation}{section}

\graphicspath{{figures/},{figs_ch1/},{figs_ch2/},{figs_ch3/},{figs_ch4/},{figs_ch5/},{figs_ch6/},{figs_ch7/},{figs_ch8/},{figs_ch9/}}

\usepackage[small,normal,bf,up]{caption2}

\usepackage{fancyhdr}			
\def\lazz{\mathrel{\mathchoice {\vcenter{\offinterlineskip\halign{\hfil
$\displaystyle##$\hfil\cr<\cr\sim\cr}}}
{\vcenter{\offinterlineskip\halign{\hfil$\textstyle##$\hfil\cr<\cr\sim\cr}}}
{\vcenter{\offinterlineskip\halign{
\hfil$\scriptstyle##$\hfil\cr<\cr\sim\cr}}}
{\vcenter{\offinterlineskip\halign{\hfil$\scriptscriptstyle##
$\hfil\cr<\cr\sim\cr}}}}}
\def\gazz{\mathrel{\mathchoice {\vcenter{\offinterlineskip\halign{\hfil
$\displaystyle##$\hfil\cr>\cr\sim\cr}}}
{\vcenter{\offinterlineskip\halign{\hfil$\textstyle##$\hfil\cr>\cr\sim\cr}}}
{\vcenter{\offinterlineskip\halign{
\hfil$\scriptstyle##$\hfil\cr>\cr\sim\cr}}}
{\vcenter{\offinterlineskip\halign{\hfil$\scriptscriptstyle##
$\hfil\cr>\cr\sim\cr}}}}}
\def\pr{\prime}
\def\be{\begin{equation}}
\def\lan{\left\langle}
\def\ran{\right\rangle}
\def\ee{\end{equation}}
\def\barr{\begin{array}}
\def\earr{\end{array}}

\def\nn8{\\}
\def\l{\left}
\def\r{\right}
\def\dis{\displaystyle}
\def\ed{\end{document}}

\def\dg{\dagger}

\def\bv{{\mbox{\boldmath $V$}}}

\def\bF{{\mbox{\boldmath $F$}}}
\def\bee{{\mbox{\boldmath $E$}}}
\def\coo{\co^\dagger \co}
\def\bd{{\widetilde {\cal D}}}

\def\co{{\cal O}}
\def\cg{{\cal G}}
\def\ce{{\cal E}}
\def\cx{{\cal X}}
\def\ch{{\cal H}}
\def\cn{{\cal N}}
\def\cac{{\cal C}}

\def\cd{{\cal D}}
\def\cads{{\mbox{\boldmath $d$}}}
\def\cx{{\cal X}}

\def\cav{{\cal V}}

\def\cq{{\cal Q}}
\def\cii{{\cal I}}
\def\caa{{\cal A}}
\def\caf{{\cal F}}
\def\cas{{\cal S}}
\def\cm{{\bf m}}
\def\cs{{\bf s}}

\def\bl{{\bf l}}

\def\cS{{\bf S}}
\def\ct{{\bf t}}
\def\cf{{\cal F}}
\def\spin{\frac{1}{2}}
\def\v{\mbox{v}}

\def\wx{{\widehat {x}}}
\def\whm{{\widehat {M}}}
\def\wj{{\widehat {J}}}
\def\we{{\widehat {E}}}
\def\wy{{\widehat {y}}}
\def\wh{{\widehat {H}}}
\def\whh{{\widehat {h}}}
\def\wv{{\widehat {V}}}

\def\wD{{\widetilde {D}}}
\def\wmp{{\widetilde {m}_1}}
\def\wmn{{\widetilde {m}_2}}
\def\wtM{{\widetilde {M}}}
\def\wm{{\widetilde {m}}}

\def\Gm{\Gamma}

\def\crh{{\textcolor{red}{H(k_H)}}}
\def\cbh{{\textcolor{blue}{H(k_H)}}}
\def\crd{{\textcolor{red}{D}}}
\def\cbd{{\textcolor{blue}{D}}}
\def\crdt{{\textcolor{red}{\wD}}}
\def\cbdt{{\textcolor{blue}{\wD}}}

\begin{document}

\frontmatter
\begin{titlepage}
\begin{center}
    {\bf\Large{SOME STUDIES ON TWO-BODY RANDOM MATRIX \\
    ENSEMBLES}}
    \vskip0.7in
    {\large Thesis submitted to}
    \vskip0.5cm
    \includegraphics[width=2.25cm,height=2.25cm]{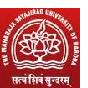}
    \vskip0.25cm
    {\large The Maharaja Sayajirao University of Baroda \\
    Vadodara - 390 002, India}
    \vskip0.5in
    for the degree of
    \vskip0.3in
    {\bf \Large Doctor of Philosophy in Physics}
    \vskip0.3in
    \large{by}
    \vskip0.3in
    {\bf\Large{MANAN VYAS}}
    \vskip0.75cm
    \includegraphics[width=2cm,height=2cm]{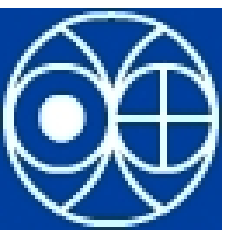}
    \vskip0.3cm
    {Theoretical Physics Division \\
    Physical Research Laboratory \\
    Ahmedabad - 380 009, India}
    \vskip0.5in
    {\Large May, 2011}
\end{center}
\end{titlepage}

\chapter*{}

\thispagestyle{empty}

\thispagestyle{empty}
\null\vspace{\stretch{1}}
\begin{center}
{\Huge To my family}
\end{center}
\vspace{\stretch{2}}\null

\chapter*{}

\thispagestyle{empty}

\chapter*{Acknowledgments}

This dissertation would not have been possible without the meticulous guidance
of Prof. V. K. B. Kota. First of all, I am grateful to him for giving me an
opportunity to work with him. His motivation, enthusiasm, immense knowledge, and
attention to details has helped me a lot in all the time of research and writing
of this thesis. I sincerely thank him for his patience and efforts to educate me
with the basics necessary for the work in this thesis.   

During this work I have collaborated with Dr. K. B. K. Mayya,  Dr. N. D. Chavda,
and Dr. P. C. Srivastava and I extend my warmest thanks to them. I am thankful
to Dr. N. D. Chavda  for his help in the initial stages of my thesis work. I am
grateful to Prof. V. N. Potbhare and Dr. N. D. Chavda for their help during my
visits to M. S. University of Baroda. I  thank Prof. A. C. Sharma for his
interest in the thesis work. I am grateful to him for his guidance, 
encouragement and help. I wish to express my warm and sincere thanks to Prof.
Steven Tomsovic for some useful discussions. I sincerely thank Dr. Dilip Angom
for encouragement and discussions. I also offer my sincere thanks to Prof.
Panigrahi for guiding me in my initial stages of research. I thank Prof. A. K.
Dutta and Prof. K. P. Maheshwari for their excellent teaching of basic physics
in my master degree course.

This acknowledgment will be incomplete if I do not express my gratitude to the 
various facilities availed in PRL and the people associated with them. To
mention a few, this includes Dr. P. Sharma, Mr. G. Dholakia, Mrs. Nishtha
AnilKumar, Mrs. Pauline Joseph, Mrs. Parul Parikh, Miss Jayshree, Mr. Shashi, 
Mr. Gangadharia, Mr. B. M. Joshi, Mr. Ghanshyam Patel, Mr. Ranganathan, Dr.
Sheetal Patel, Miss Pragya,  Mr. N. P. M. Nair, Miss Miral Patel, and Mrs.
Sujata. I have extensively used POWER5 machine and HPC facility for my thesis
work with the help of computer center staff. I also extend my thanks to the
staff of M. S. University of Baroda for their kind co-operation and prompt
services. 

I express my special gratitude to Prof. V. K. B. Kota and  Mrs. Vijayalakshmi
Kota for hospitality during my academics related visits to their place. I am
deeply touched by Mrs. Kota's motherly care during these visits. Thanks are also
due to all my friends in PRL for their affectionate company and best wishes. Dr.
Rajneesh Atre has been a great senior and has stood with me in thick and thin.
Eventually, he has become a part of my family. I thank him for everything he has
done for me. I express my gratitude to my parents and my brother Manthan for
their loving support and encouragement. My sincere thanks are due to my
parents-in-law for their untiring efforts and boundless love. Words fail me to
express my appreciation for my husband Harinder whose patience and encouragement
have never let my spirits down. I thank him for his unconditional support. I
wish to thank my entire extended family for their patience and moral support.  

Finally, I feel great pleasure to express my gratitude to all others who have
directly or indirectly contributed to this thesis. 

\chapter*{Abstract}
\addcontentsline{toc}{chapter}{Abstract}

Random matrix theory (RMT) has been established to be one of the central
themes in quantum physics during the end of 20th century. This theory has
emerged as a powerful statistical approach leading to paradigmatic models
describing generic properties of complex systems. On the other hand, with
scientific developments it was clear by mid 20th century that deterministic
ideas are not valid for microscopic systems and this led to the development
of a new field of research called `quantum chaos'. One-body chaos is well
understood by 90's with RMT playing a key role. More specifically, the
spectral statistics predicted by RMT is a characteristic of quantum systems
whose classical analogue is chaotic. However, most of the real systems are
many-body in character. The classical Gaussian orthogonal (GOE), unitary
(GUE) and symplectic (GSE) ensembles, introduced by Wigner and Dyson, are
ensembles of multi-body interactions. In various quantum many-body systems
such as nuclei, atoms, mesoscopic systems like quantum dots and small
metallic grains, interacting spin systems modeling quantum computing core
and BEC, the interparticle interactions are essentially two-body in nature.
This together with nuclear shell-model examples led to the introduction of
random matrix ensembles generated by two-body interactions  in 1970-1971.
These two-body ensembles are defined by representing the two-particle
Hamiltonian by one  of the classical ensembles and then the $m$ ($m>2$) 
particle $H$-matrix is generated by the Hilbert space geometry. Thus the
random matrix ensemble in the two-particle spaces is embedded in the
$m$-particle $H$-matrix and therefore these ensembles are generically called
embedded ensembles (EEs). Simplest of these ensembles is the embedded
Gaussian orthogonal ensemble of random matrices generated by two-body
interactions for spinless fermion [boson] systems, denoted by EGOE(2)
[BEGOE(2); here `B' stands for bosons]. In addition to the complexity
generating two-body interaction, Hamiltonians for realistic systems consist
of a mean-field one-body part.  Then the appropriate random matrix ensembles
are EE(1+2). The spinless fermion/boson EGEs (orthogonal and unitary
versions) have been explored in detail from 70's with a major revival from
1994.  It is now well understood that EGEs generate paradigmatic models for
many-body chaos or stochasticity exhibited by isolated finite interacting
quantum systems. Besides the mean-field and the two-body character,
realistic Hamiltonians also carry a variety of symmetries. In many
applications of EGEs, generic properties of EGE for spinless fermions are
`assumed'  to extend to symmetry subspaces. More importantly,  there are
several  properties of real systems that require explicit inclusion of
symmetries and they are defined by a variety of Lie algebras. The aim of
the present thesis is to identify and systematically analyze many different
physically relevant EGEs with symmetries by considering a variety of
quantities and measures that are important for finite interacting quantum
systems mentioned above. The embedded ensembles investigated to this end and 
the corresponding results are as follows.

The thesis contains nine chapters. Chapter \ref{ch1} gives an introduction
to the subject of two-body random matrix ensembles. Also, the known results
for spinless fermion/boson EGEs are described briefly for completeness
and for easy reference in the following chapters.

Finite interacting Fermi systems with a mean-field and a chaos generating
two-body interaction are modeled, more realistically, by one plus two-body
embedded Gaussian  orthogonal ensemble of random matrices with spin degree
of freedom [called EGOE(1+2)-$\cs$]. Numerical calculations are used to
demonstrate that, as $\lambda$, the strength of the interaction (measured in
the  units of the average spacing of the single particle levels defining the
mean-field), increases, generically there is Poisson to GOE transition  in
level fluctuations, Breit-Wigner to Gaussian transition in strength
functions (also called local density of states) and also a duality  region
where information entropy will be the same in both the mean-field and
interaction defined basis. Spin dependence of the transition points
$\lambda_c$, $\lambda_F$ and  $\lambda_d$, respectively, is described using
the propagator for the spectral variances and the analytical formula for the
propagator is derived. We further  establish that the duality region
corresponds to a region of thermalization. For this purpose we have compared the
single particle entropy defined by the occupancies of the single particle
orbitals with thermodynamic entropy and information entropy for various
$\lambda$ values  and they are very close to each other at $\lambda =
\lambda_d$. All these results are presented in Chapter \ref{ch2}.

EGOE(1+2)-$\cs$  also provides a model for understanding general structures
generated by pairing correlations.  In the space defined by EGOE(1+2)-$\cs$
ensemble for fermions, pairing defined by the algebra $U(2\Omega) \supset
Sp(2\Omega) \supset SO(\Omega) \otimes  SU_S(2)$ is identified and some of
its properties are derived. Using numerical calculations it is shown that in
the strong coupling limit, partial densities defined over pairing subspaces
are close to Gaussian form and propagation formulas for their centroids and
variances are derived. As a part of understanding pairing correlations in
finite Fermi systems, we have shown that pair transfer strength sums (used
in nuclear structure) as a function of excitation energy (for fixed $S$), a
statistic for onset of chaos, follows, for low spins, the form derived for 
spinless fermion systems, i.e., it is close to a ratio of Gaussians. Going
further, we have considered a quantity in terms of ground state energies,
giving conductance peak spacings in mesoscopic systems at low temperatures,
and studied its distribution over EGOE(1+2)-$\cs$ by including both  pairing
and exchange interactions. This model is shown to generate  bimodal to
unimodal  transition in the distribution of conductance peak spacings. All
these results are presented in Chapter \ref{ch3}.

For $m$ fermions in $\Omega$ number of single particle orbitals, each 
four-fold degenerate, we have introduced and analyzed in detail embedded
Gaussian unitary ensemble of random matrices generated by random two-body
interactions that are $SU(4)$ scalar [EGUE(2)-$SU(4)$]. Here, the $SU(4)$
algebra corresponds to the Wigner's  supermultiplet  $SU(4)$ symmetry in
nuclei. Embedding algebra for the EGUE(2)-$SU(4)$ ensemble is $U(4\Omega)
\supset U(\Omega) \otimes SU(4)$. Exploiting the Wigner-Racah algebra of the
embedding algebra, analytical expression for the ensemble average of the
product of any two $m$-particle Hamiltonian  matrix  elements is derived.  
Using this, formulas for a special class of $U(\Omega)$ irreducible
representations (irreps) $\{4^r,p\}$, $p=0$, $1$, $2$, $3$ are derived for
the ensemble  averaged spectral  variances and also for the covariances in
energy centroids and spectral variances. On the other hand, simplifying the
tabulations available for $SU(\Omega)$ Racah coefficients, numerical
calculations are carried out for general $U(\Omega)$ irreps. Spectral
variances clearly show, by applying the so-called Jacquod and Stone
prescription, that the EGUE(2)-$SU(4)$ ensemble generates ground state
structure just as the quadratic Casimir invariant $(C_2)$ of $SU(4)$. This
is further corroborated by the calculation of the expectation values of
$C_2[SU(4)]$ and the four periodicity in the ground state energies.
Secondly, it is found that the covariances in energy centroids and spectral
variances increase in magnitude considerably as we go from EGUE(2) for
spinless fermions to EGUE(2) for fermions with spin to  EGUE(2)-$SU(4)$
implying that the differences in ensemble and spectral averages grow with
increasing symmetry. Also for EGUE(2)-$SU(4)$ there are, unlike for GUE,
non-zero cross-correlations in energy centroids and spectral variances
defined over spaces with different particle numbers and/or  $U(\Omega)$
[equivalently $SU(4)$] irreps. In the dilute limit defined by $\Omega \to
\infty$, $r >> 1$ and $r/\Omega \to 0$, for the $\{4^r,p\}$ irreps,  we have
derived analytical results for these correlations. All correlations are
non-zero for finite $\Omega$ and they tend to zero as $\Omega \to \infty$.
All these results are presented in Chapter \ref{ch4}.

One plus two-body embedded Gaussian orthogonal ensemble of random matrices with
parity [EGOE(1+2)-$\pi$] generated by a random two-body interaction (modeled by
GOE in two particle spaces) in the presence of a mean-field, for spinless
identical  fermion systems, is defined in terms of two mixing parameters and a
gap between the positive $(\pi=+)$ and negative $(\pi=-)$ parity single particle
states.  Numerical calculations are used to demonstrate, using realistic values
of the mixing parameters, that  this ensemble generates Gaussian form (with
corrections) for fixed parity state densities. The random matrix model also
generates many features in parity ratios of state densities that are similar to
those predicted by a method based on the Fermi-gas model for nuclei. We have
also obtained a simple formula for the spectral variances defined over
fixed-$(m_1,m_2)$ spaces where $m_1$ is the number of fermions in the $+$ve
parity single particle states and $m_2$ is the number of fermions in the $-$ve
parity single particle states. The smoothed densities generated by the sum of
fixed-$(m_1,m_2)$ Gaussians with lowest two shape corrections describe the
numerical results in many situations.  The model  also generates preponderance
of $+$ve  parity ground states for small  values of the mixing parameters and
this is a feature seen in nuclear shell-model results.  All these results are
presented in Chapter \ref{ch5}.

For $m$ number of bosons, carrying spin ($\cs=\spin$) degree of freedom, in
$\Omega$ number of single particle orbitals, each doubly degenerate, we have
introduced and analyzed embedded Gaussian orthogonal ensemble of random matrices
generated by random two-body interactions that are spin ($S$) scalar
[BEGOE(2)-$\cs$]. The ensemble BEGOE(2)-$\cs$ is  intermediate to the BEGOE(2)
for spinless bosons and for bosons with spin $\cs=1$ which is relevant for
spinor BEC. Embedding algebra for the BEGOE(2)-$\cs$ ensemble and also for
BEGOE(1+2)-$\cs$ that includes the mean-field one-body part  is $U(2\Omega)
\supset U(\Omega) \otimes SU(2)$ with $SU(2)$ generating spin.  A method for
constructing the ensembles in fixed-($m,S$) spaces has been developed. Numerical
calculations show that the  fixed-$(m,S)$ density of states
is close to Gaussian and generically there is Poisson to GOE transition in level
fluctuations as the interaction strength (measured in the units of the average
spacing of the single particle levels defining the mean-field) is  increased.
The interaction strength needed for the onset of the transition  is found to
decrease with increasing $S$. Propagation formulas for the fixed-$(m,S)$ space
energy centroids and ensemble averaged spectral variances
are derived.
Using these, covariances in energy centroids and spectral variances are
analyzed.  Variance propagator clearly shows that the BEGOE(2)-$\cs$ ensemble
generates ground states with spin $S=S_{max}$. This is further corroborated by
analyzing the structure of the ground states in the  presence of the exchange
interaction $\hat{S}^2$ in BEGOE(1+2)-$\cs$.  Natural spin ordering ($S_{max}$,
$S_{max}-1$, $S_{max}-2$, $\ldots$, $0$ or $\spin$) is also observed with random
interactions. Going beyond these,  we have also introduced pairing symmetry in
the space defined by BEGOE(2)-$\cs$. Expectation values of the pairing
Hamiltonian show that random interactions exhibit pairing correlations in the
ground state region. All these results are presented in Chapter \ref{ch6}.

Parameters defining many of the important spectral distributions (valid in the
chaotic region), generated by EGEs, involve traces of product of four two-body
operators. For example, these higher order traces are required for calculating
nuclear structure matrix elements for $\beta\beta$ decay and also for
establishing Gaussian density of states generated by various embedded ensembles.
Extending the binary correlation approximation method for two different
operators and for traces over two-orbit configurations, we have derived
formulas, valid in the dilute limit, for the skewness and excess parameters for
EGOE(1+2)-$\pi$ ensemble. In addition, we have derived a formula for the traces
defining the correlation coefficient of the bivariate transition strength
distribution generated by the two-body transition operator  appropriate for
calculating 0$\nu$-$\beta \beta$ decay nuclear transition matrix elements and
also for other higher order traces required for justifying the bivariate
Gaussian form for the strength distribution. With applications in the subject of
regular structures generated by random interactions, we have also derived
expressions for the coefficients in the expansions to order $[J(J+1)]^2$ for the
energy centroids $E_c(m,J)$ and spectral variances $\sigma^2(m,J)$ generated by
EGOE(2)-$J$ ensemble members for the single-$j$ situation. These expansion coefficients
also involve traces of four two-body operators. All these results are presented 
in Chapter \ref{ch7}. 

In Chapter \ref{ch8}, to establish random matrix structure of nuclear shell
model Hamiltonian matrices, we have presented a comprehensive analysis of the
structure of  Hamiltonian matrices based on visualization of the matrices in
three dimensions as well as in terms of measures for GOE, banded and embedded
random matrix ensembles. We have considered two nuclear shell-model examples,
$^{22}$Na with $J^\pi T = 2^+0$ and $^{24}$Mg with $J^\pi T = 0^+0$  and, for
comparison we have also considered  SmI atomic example with $J^\pi = 4^+$. It is
clearly established that the matrices are  neither GOE nor banded. For the EGOE
[strictly speaking, EGOE(2)-$JT$ or EGOE(2)-$J$] structure we have examined the
correlations between diagonal elements and eigenvalues, fluctuations in the
basis states variances and structure of the two-body part of the Hamiltonian in
the eigenvalue basis. Unlike the atomic example, nuclear examples show that the
nuclear shell-model Hamiltonians can be well represented by EGOE. 

Finally, Chapter \ref{ch9} of the thesis gives conclusions and future outlook. 
To summarize, we have obtained large number of new results for embedded
ensembles and in particular for EGOE(1+2)-$\cs$, EGUE(2)-$SU(4)$,
EGOE(1+2)-$\pi$ and BEGOE(1+2)-$\cs$, with EGUE(2)-$SU(4)$  introduced for the
first time in this thesis. Moreover, some results are presented for EGOE(2)-$J$
and for the first time BEGOE(1+2)-$\cs$  has been explored in detail in this
thesis. In addition,  formulas are derived, by extending the binary correlation
approximation method, for higher order traces for embedded ensembles with  $U(N)
\supset U(N_1) \oplus U(N_2)$ embedding and some of these are needed for new
applications of statistical nuclear spectroscopy. Results of the present thesis
establish that embedded Gaussian ensembles can be used gainfully to study a
variety of problems in many-body quantum physics and this includes quantum
information science and the thermodynamics of isolated finite interacting
quantum systems.

\tableofcontents
\chapter*{}

\thispagestyle{empty}

\mainmatter
\chapter{Introduction}
\label{ch1}

\section{Random Matrix Theory, Quantum Chaos and Finite Quantum Systems}

\begin{figure}[tp]
\centering
\includegraphics[width=3in,height=5in,angle=-90]{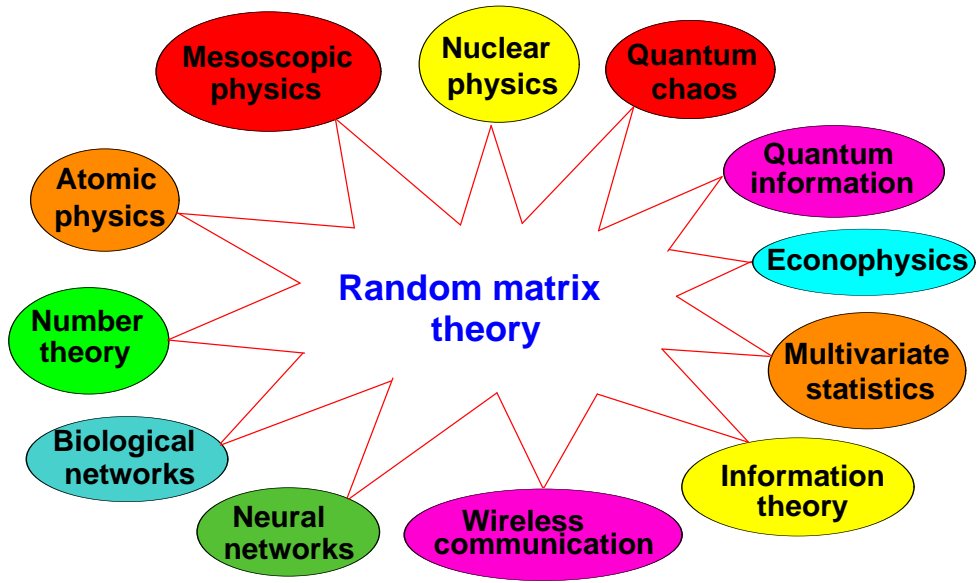}
\caption{Figure showing wide ranging applications of random matrix 
theory.}
\label{appl}
\end{figure}

Random matrix theory (RMT), starting with Wigner and Dyson's Gaussian random
ensembles introduced to describe neutron resonance data, has emerged as a
powerful statistical approach leading to paradigmatic models describing generic
properties of complex systems. Importance of RMT has been recognized almost
since its inception in physics by Wigner in 1955 \cite{Wi-55} to explain the
compound nucleus resonance data. Though random matrices were encountered much
earlier in 1928 by Wishart \cite{Wis-28} in the context of multivariate
statistics and later in 1967 by Pastur \cite{Ma-67}, their extensive study began
with the pioneering work of Wigner \cite{Wi-67,Po-65}. Though not referred
explicitly, Bohr's idea of compound nucleus \cite{Bo-36} almost certainly
motivated Wigner to introduce random matrices. Porter's book \cite{Po-65} gives
a good introduction to classical random matrix ensembles with a detailed
derivation of the joint probability distribution for these ensembles along with
an impressive and instructive collection of papers on the subject till 1965.
Mathematical foundations of RMT  are well described by Mehta \cite{Me-04} (the
first edition of Mehta's book was published in 1967). The classical random
matrix ensembles are developed and applied during 1955-1972 by Dyson, Mehta,
Porter and others \cite{Po-65,Br-81}. In the last three decades, RMT has been
successfully used in diverse areas, as shown in Fig. \ref{appl}, 
with wide ranging
applicability to various mathematical, physical and engineering branches
\cite{Po-65, Me-04, Br-81,  Gu-98, Tu-04, St-06, Br-06, Ul-08, We-09a, Fo-10,
Wr-10, An-10, Ba-10, Ha-10, Mi-10}.

RMT helps to analyze statistical properties of physical systems whose exact
Hamiltonian is too complex to be studied directly. The exact Hamiltonian of the
system under consideration is represented by an ensemble of random matrices that
incorporate generic symmetry properties of the system. As stated by Wigner
\cite{Wi-79}: {\it The assumption is that the Hamiltonian which governs the
behavior of a complicated system is a random symmetric matrix, with no special
properties except for its symmetric nature}. More importantly, as emphasized by
French \cite{Fr-65}: {\it with one short step beyond this, specifically
replacing ``complicated'' by ``non-integrable'', this paper would have led to
the foundations of quantum chaos. Perhaps it should be so regarded even as it
stands}. Depending on the global symmetry properties, namely rotational and
time-reversal, Dyson classified the classical random matrix ensembles into three
classes- Gaussian orthogonal (GOE), unitary (GUE) and symplectic (GSE) ensembles
\cite{Dy-62}. As the names suggest,  these ensembles will be invariant under
orthogonal, unitary, and symplectic transformations, respectively. The
corresponding matrices will be real symmetric, complex hermitian and real
quaternion matrices. In order to study symmetry breaking effects on level and
strength fluctuations and order-chaos transitions, it is necessary to consider
interpolating or deformed random matrix ensembles \cite{Pa-81,FKPT}. Earliest
examples include banded random matrix ensembles, the Porter-Rosenzweig model and
2$\times$2 GOE due to Dyson \cite{Po-65}.

\begin{figure}
    \centering
    \subfigure[]
    {
        \includegraphics[width=2.5in,height=1.5in]{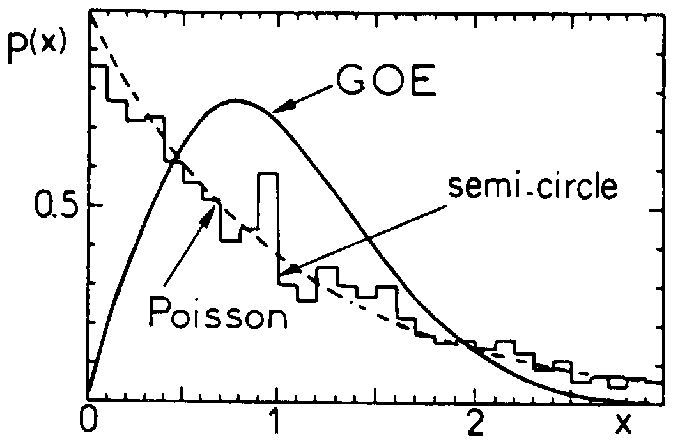}
        \label{semi-circ}
    }
    \subfigure[]
    {
        \includegraphics[width=2.5in,height=1.5in]{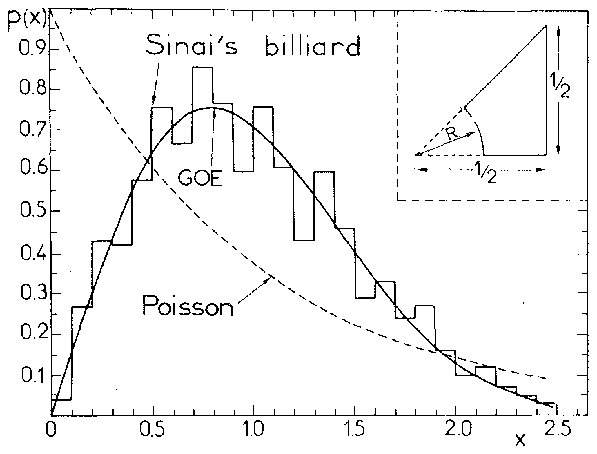}
        \label{sinai-nnsd}
    }
    \\
    \subfigure[]
    {
        \includegraphics[width=2.5in,height=1.5in]{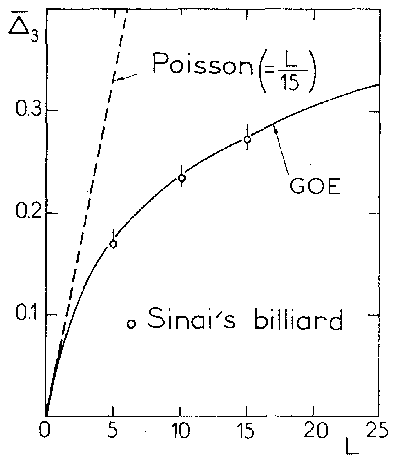}
        \label{sinai-del3}
    }
    \subfigure[]
    {
        \includegraphics[width=2.5in,height=1.5in]{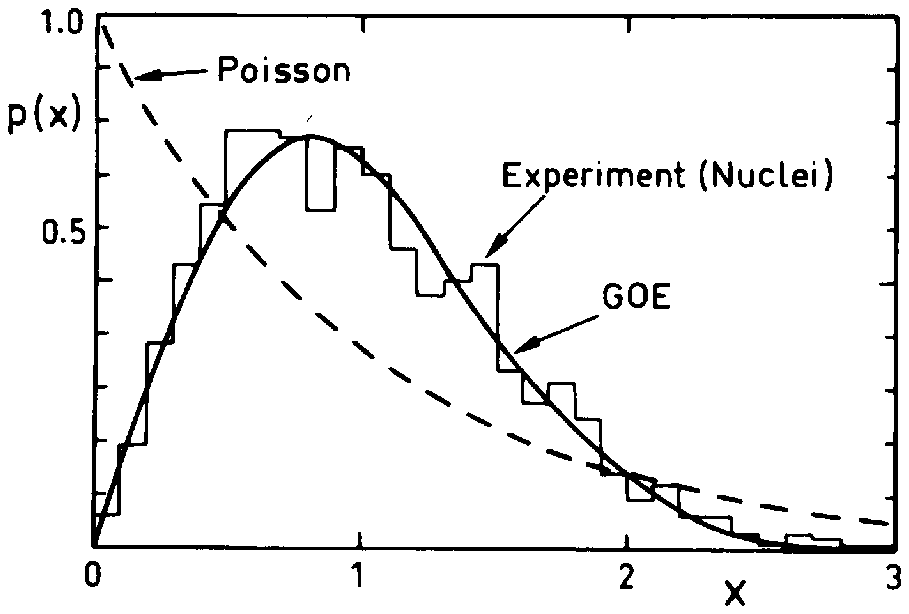}
        \label{nde}
    }
    \caption{Figure illustrating the connection between RMT and quantum
    chaos. (a) NNSD for the regular (integrable)
    semi-circular billiard follows
    Poisson distribution $P(X) dX = e^{-X} dX$ 
    (taken from \cite{Bo-84a}); (b) NNSD for the
    chaotic Sinai billiard follows GOE (taken from \cite{Bo-84b}) and the GOE
    Wigner form is $P(X) dX = \frac{\pi}{2} X \exp{-\frac{\pi X^2}{4}} dX$; (c)
    Dyson-Mehta statistic $\Delta_3(L)$ for Sinai billiard also follows
    GOE (taken from \cite{Bo-84b}) and for GOE, 
    $\Delta_3(L) \sim \ln L$; and (d)  NNSD for the nuclear data
    ensemble (NDE) follows GOE \cite{Bo-83}. Though we haven't shown, the
    $\Delta_3(L)$ for $L \leq 20$ also follows GOE for the NDE
    \cite{Ha-82}. Note that $X$ is the level 
    spacing normalized to the average level spacing and $L$ is the
    length of the energy interval over which $\Delta_3$ is calculated.
    It is clearly seen that, unlike regular billiard, chaotic billiard
    follows RMT and more importantly, the NDE follows RMT 
    establishing that the neutron resonance region is a region of chaos.}
    \label{ch1-nnsd}
\end{figure}

RMT has been established to be one of the central themes in quantum physics with
the recognition that quantum systems whose classical analogues are chaotic,
follow RMT. The   BGS conjecture \cite{Bo-84b} is the corner stone for this and
the earlier work on this is due to  McDonald and Kaufman \cite{Mc-79}, Casati et
al \cite{Ca-80} and Berry \cite{Be-81}. The BGS conjecture is: {\it Spectra of
time-reversal-invariant systems whose classical analogues are K systems show the
same fluctuation properties as predicted by GOE}. Also as stated by BGS: {\it if
this conjecture happens to be true, it will then have established the
`universality of the laws of level fluctuations' in quantal spectra already
found in nuclei and to a lesser extent in atoms. Then, they should also be found
in other quantal systems, such as molecules, hadrons etc.} The details of the
developments establishing the connection between RMT and the  spectral
fluctuation properties of quantum systems whose classical analogues are chaotic
are summarized in \cite{Ha-10,St-06}. Recently, Haake et al gave a proof for the
BGS conjecture using semi-classical methods \cite{He-07}. Combining BGS work
with that of Berry on integrable systems \cite{Be-77},  as summarized by
Altshuler in the abstract of the colloquium he gave in memory of J.B. French at
the university of Rochester in 2004: {\it Classical dynamical systems can be
separated into two classes - integrable and chaotic. For quantum systems this
distinction manifests itself, e.g. in spectral statistics. Roughly speaking
integrability leads to Poisson distribution for the energies while chaos implies
Wigner-Dyson statistics of levels, which are characteristic for the ensemble of
random matrices. The onset of chaotic behavior for a rather broad class of
systems can be understood as a delocalization in the space of quantum numbers
that characterize the original integrable system $\ldots$} Following this, as
stated by Papenbrock and Weidenm\"{u}ller \cite{Pa-07}: {\it We speak of chaos
in quantum systems if the statistical properties of the eigenvalue spectrum
coincide with predictions of random-matrix theory.} For example,  the nearest
neighbor spacing distribution (NNSD) showing von-Neumann Wigner \cite{Wi-29} 
`level repulsion' and the Dyson-Mehta \cite{Dy-63} $\Delta_3$ statistic showing
`spectral rigidity'  are exhibited by quantum systems whose classical analogues
are chaotic; see Fig. \ref{ch1-nnsd}.  It is now well recognized that chaos is a
typical feature of atomic nuclei and other self bound Fermi systems. 

\begin{figure}[htp]
\centering
\includegraphics[width=4in,height=3.5in]{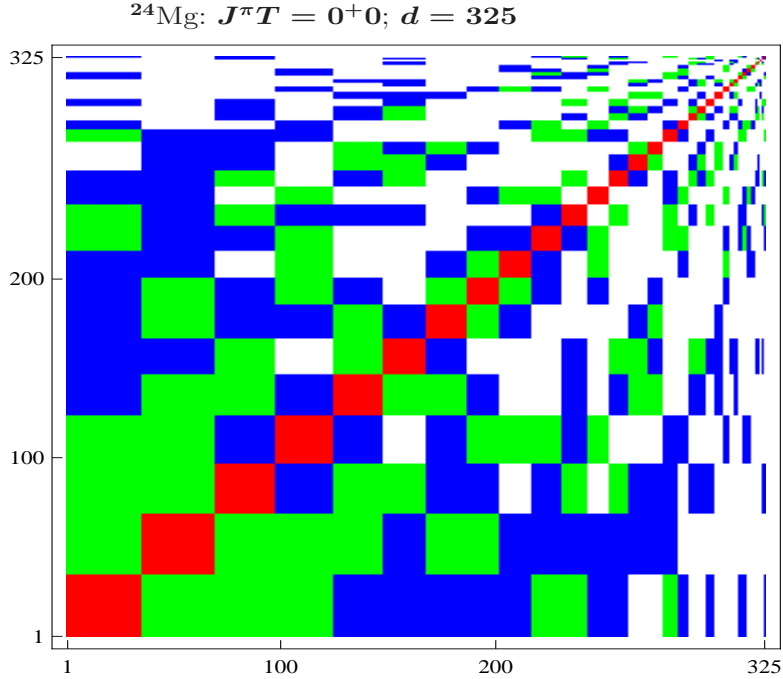} 
\caption{Block matrix structure of the $H$ matrix of
$^{24}$Mg displaying two-body selection rules. Total number of blocks
are $33$, each labeled by the spherical configurations $(m_1,m_2,m_3)$.
The diagonal blocks are shown in red and within these blocks there will
be no change in the  occupancy of the nucleons in the three $sd$ orbits.
Green corresponds to the region (in the matrix) connected by the
two-body interaction that involve change of occupancy of one nucleon.
Similarly, blue corresponds to change of occupancy of two nucleons.
Finally, white  correspond to the region forbidden by the two-body
selection rules. This figure was first reported by us in \cite{Ma-10b} 
and a similar figure was given earlier in \cite{Pa-05} for
$^{28}$Si with $(J^\pi T)=(0^+0)$. Section \ref{c8s2} gives 
further discussion.}
\label{m8-block}
\end{figure}

Finite quantum systems such as nuclei, atoms, quantum dots, small metallic
grains, interacting spin systems modeling quantum computing core and BEC,
share one common property - their constituents (predominantly) interact
via  two-particle interactions. As pointed out by French \cite{Fr-80}: 
{\it The GOE, now almost universally regarded as a model for a
corresponding chaotic system is an ensemble of multi-body, not two-body
interactions. This difference shows up both in one-point (density of
states) and two-point (fluctuations) functions generated by nuclear shell
model.} Therefore, it is more appropriate to represent an isolated  finite
interacting quantum system by random matrix models generated by random
{\it two-body} interactions (in general, by $k$-body interactions with $k
<$ particle number $m$). Matrix ensembles generated by random two-body
interactions, called two-body random ensembles (TBRE),  model what one may
call many-body chaos or stochasticity or complexity exhibited by these
systems. These ensembles are defined by representing the two-particle
Hamiltonian by one of the classical ensembles (GOE or GUE or GSE) and then
the $m > 2$ particle $H$ matrix is  generated by the $m$-particle Hilbert
space geometry \cite{Fr-70,Bo-71,MF-75} and with GOE(GUE) embedding, they
will be EGOE(EGUE). The key element here is  the recognition that there is
a Lie algebra that transports the information in the two-particle spaces
to many-particle spaces \cite{MF-75,Ko-05,Be-01}. Thus the random matrix
ensemble in the two-particle spaces is embedded  in the $m$-particle $H$
matrix and therefore these ensembles are  more generically called embedded
ensembles (EE) \cite{MF-75,Br-81}. Due to the two-body selection rules,
many of the $m$-particle matrix elements will be zero.  Figure
\ref{m8-block} gives an example of a $H$-matrix displaying the structure
due to two-body selection rules which form the basis for the EE
description. At this stage, it is appropriate to recall the purpose, as
stated by the organizers Altshuler, Bohigas and Weidenm\"{u}ller, of a
workshop (held at ECT*, Trento in February 1997) on chaotic dynamics of
many-body systems: {\it The study of quantum manifestations of classical 
chaos has known important developments, particularly for systems with few
degrees of freedom. Now, we understand much better how the universal and
system-specific properties of `simple chaotic systems' are connected with
the underlying classical dynamics. The time has come to extend, from this
perspective, our understanding to objects with many degrees of freedom,
such as interacting many-body systems. Problems of nuclear, atomic, and
molecular theory as well as the theory of mesoscopic systems will be
discussed at the workshop.} Note that, chaos implies RMT and the new
emphasis is on many-body chaos. Recent thinking  is that  EE generate
paradigmatic models for many-body chaos \cite{Ko-01,Go-11}  (one-body
chaos is well understood using classical ensembles). The present thesis is
devoted to developing and analyzing a variety of EE so that one can
quantify and apply the results of many-body chaos.

Simplest of EE is the embedded Gaussian orthogonal ensemble of random
matrices for spinless fermion/boson systems generated by random two-body
interactions. However, unlike for fermion systems, there are only a few EE
investigations for finite interacting boson systems \cite{PDPK, Ag-01,
Ag-02, Ch-03, Ch-04}; the corresponding EE are called BEE (B stands for
bosons). The spinless fermion/boson EGEs (orthogonal and unitary versions)
have been explored in detail from 70's \cite{Br-81,Ko-01,Go-11} 
with a major revival from mid 90's 
\cite{Fl-94, Fl-96a, Fl-97, Ho-95, Fr-96, Ja-97, Ko-98}. Before proceeding
further, we briefly describe the known results for spinless fermion/boson
EGEs for completeness and for easy reference in the following chapters.

\section{Embedded Ensembles for Spinless Fermion Systems}
\label{sec1p2}

The embedding algebra for EGOE$(k)$ and EGUE$(k)$ [also BEGOE$(k)$ and
BEGUE$(k)$] for a system of $m$ spinless particles (fermions or bosons) in
$N$ single particle  (sp) states with $k$-body interactions $(k < m)$ is
$SU(N)$. These ensembles are defined by the three parameters $(N,m,k)$. A
large number of asymptotic results are derived for EGOE$(k)$ and EGUE$(k)$
using Wigner's binary correlation approximation \cite{MF-75, Br-81, FKPT}
and, more importantly,  also some exact analytical results are derived
using $SU(N)$ Wigner-Racah algebra  \cite{Ko-05, Be-01, Pl-02}. For
bosons, the dense limit studies are interesting as this limit does not
exist for fermion systems \cite{KP-80, PDPK, Ag-01, Ag-02, Ch-03, Ch-04}.
Now we will briefly discuss the definition, construction and the known
results for these ensembles.  

\subsection{EGOE(2) and EGOE$(k)$ ensembles}
\label{sec1p2p1}

\begin{figure}[htp]
\centering
\includegraphics[width=2.5in,height=2.5in]{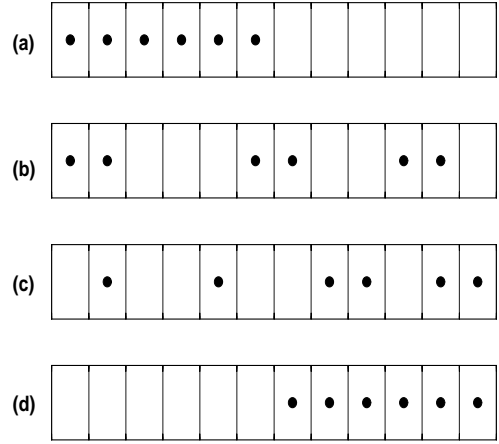}
\caption{Figure showing some configurations for the distribution of $m=6$
spinless fermions in $N=12$ single particle states. Distributing the $m$
fermions in all possible ways in $N$ single particle states generates
the $m$-particle configurations or basis states. This is similar to
distributing $m$ particles in $N$ boxes with the conditions that the
occupancy of each box can be either zero or one and the total number of
occupied boxes equals $m$.  In the figure, (a) corresponds to the basis
state $\l| \nu_1 \nu_2 \nu_3 \nu_4 \nu_5  \nu_6 \ran$ , (b) corresponds
to the basis state $\l| \nu_1 \nu_2 \nu_6 \nu_7 \nu_{10} \nu_{11} \ran$,
(c) corresponds to the basis state  $\l| \nu_2 \nu_5 \nu_8 \nu_9
\nu_{11} \nu_{12} \ran$ and  (d) corresponds to the basis state $\l|
\nu_7 \nu_8 \nu_9 \nu_{10}  \nu_{11} \nu_{12} \ran$.}
\label{sp-fermion}
\end{figure}

The EGOE(2) ensemble for spinless fermion systems is generated by defining
the two-body Hamiltonian $H$ to be GOE in two-particle spaces and then
propagating it to many-particle spaces by using the geometry of the many-particle 
spaces [this is in general valid for $k$-body Hamiltonians, with
$k < m$, generating EGOE$(k)$]. Let us consider a system of $m$ spinless
fermions occupying $N$ sp states. Each possible distribution of fermions
in the sp states generates a configuration or a basis state; see Fig.
\ref{sp-fermion}. Given the sp states $\l| \l. \nu_i \ran \r.$, $i=1, 2,
\ldots, N$,  the EGOE(2) is defined by the Hamiltonian operator, 
\be
\wh = \dis\sum_{\nu_i < \nu_j,\;\nu_k < \nu_\ell}
\lan \nu_k\;\nu_\ell
\mid \wh \mid \nu_i\;\nu_j \ran \, a^\dg_{\nu_\ell}\,a^\dg_{\nu_k}\,
a_{\nu_i}\,a_{\nu_j}\;.
\label{eq.bpp1a}
\ee
The action of the Hamiltonian operator defined by Eq. (\ref{eq.bpp1a}) on
the basis states $\l| \nu_1 \nu_2 \cdots \nu_m \ran$  (see Fig.
\ref{sp-fermion} for examples) generates the EGOE(2) ensemble in $m$-particle
spaces. The
symmetries for the antisymmetrized two-body matrix elements  $\lan
\nu_k\;\nu_\ell \mid \wh \mid \nu_i \;\nu_j \ran$ are
\be
\barr{c}
\lan \nu_k\;\nu_\ell \mid \wh \mid \nu_j\;\nu_i\ran = -\lan
\nu_k\;\nu_\ell \mid \wh \mid \nu_i\;\nu_j\ran \;,\\
\lan \nu_k\;\nu_\ell \mid \wh \mid \nu_i\;\nu_j\ran = 
\lan \nu_i\;\nu_j \mid \wh \mid \nu_k\;\nu_\ell\ran \;.\\
\earr \label{eq.bpp1}
\ee
Note that $a_{\nu_i}$ and $a^\dg_{\nu_i}$ in Eq. (\ref{eq.bpp1a})
annihilate and create a fermion in the sp state $\l| \l. \nu_i \ran \r.$,
respectively.  The Hamiltonian matrix $H(m)$ in $m$-particle spaces
contains three different types of non-zero matrix elements (all other
matrix elements are zero due to two-body selection rules) and explicit
formulas for these are \cite{Ko-01},
\be
\barr{rcl}
\lan  \nu_1 \nu_2 \cdots \nu_m \mid \wh \mid 
\nu_1 \nu_2 \cdots \nu_m \ran & = & 
\dis\sum_{\nu_i < \nu_j \leq \nu_m }\; 
\lan \nu_i \nu_j \mid \wh \mid \nu_i \nu_j \ran \;,\\
\\
\lan \nu_p \nu_2 \nu_3 \cdots \nu_m \mid \wh \mid 
\nu_1 \nu_2 \cdots \nu_m \ran & = & 
\dis\sum_{\nu_i=\nu_2}^{\nu_m}\; 
\lan \nu_p \nu_i \mid \wh \mid
\nu_{1} \nu_i \ran \;,\\
\\
\lan \nu_p \nu_q \nu_3 \cdots \nu_m \mid \wh \mid 
\nu_1 \nu_2 \nu_3 \cdots \nu_m \ran & = & 
\lan \nu_p \nu_q \mid \wh \mid \nu_1 \nu_2 \ran \;.
\earr \label{eq.bpp2}
\ee
Note that, in Eq. (\ref{eq.bpp2}), the notation  $\l| \nu_1 \nu_2 \cdots
\nu_m \ran$ denotes the orbits occupied by the $m$ spinless fermions.  The
EGOE(2) is defined by Eqs. (\ref{eq.bpp1}) and (\ref{eq.bpp2}) with GOE
representation for $\wh$ in the two-particle spaces, i.e.,
\be
\barr{l}
\lan \nu_k\;\nu_\ell \mid \wh \mid \nu_i\;\nu_j \ran \;\;{\mbox{are
independent Gaussian random variables}} \\ \\
\overline{\lan \nu_k\;\nu_\ell \mid \wh \mid \nu_i\;\nu_j \ran} = 
0\;, \\ \\
\overline{\l| \lan \nu_k\;\nu_\ell
\mid \wh \mid \nu_i\;\nu_j \ran \r|^2} = v^2 \, \l( 1+
\delta_{(ij),(k\ell)}\r) \;.
\earr \label{eq.bpp3}
\ee
In Eq. (\ref{eq.bpp3}), `overline' indicates ensemble average and $v$ is a
constant. Now the $m$-fermion EGOE(2) Hamiltonian matrix ensemble is denoted by 
$\{H(m)\}$ where $\{\ldots\}$ denotes ensemble, with $\{H(2)\}$ being GOE. Note
that, the $m$-particle $H$-matrix dimension is $d_f(N,m)=\binom{N}{m}$ and the
number of independent matrix elements is  $d_f(N,2)[d_f(N,2)+1]/2$; 
the subscript `$f$' in $d_f(N,m)$ stands for `fermions'.  A computer code for
constructing EGOE(2) ensemble is available in our group \cite{Ko-01}; many other
groups in the world have also developed codes for EGOE(2). 

Just as the EGOE(2) ensemble, it is possible to define $k$-body ($k < m$)
ensembles EGOE($k$) (these are also called $2$-BRE, $3$-BRE,  $\cdots$ in
\cite{Vo-08}). Some of the generic results,  derived numerically and
analytically, for EGOE($k$) are as follows: (i) state densities approach
Gaussian  form for large $m$ and they exhibit, as $m$ increases from $k$,
semi-circle to Gaussian transition  with $m=2k$ being the  transition
point \cite{Br-81, Be-01}; (ii) level and strength fluctuations follow GOE
(as far as one can infer from  numerical examples) \cite{Br-81}; (iii)
there is average fluctuation separation with increasing $m$ and the
averages are determined  by a few long wavelength modes in the normal mode
decomposition of the density of states \cite{MF-75, Br-81, Le-08}; (iv)
smoothed (ensemble  averaged) transition strength densities take bivariate
Gaussian form and as a consequence transition strength sums originating
from a given eigenstate will  be close to a ratio of two Gaussians
\cite{FKPT}; (v) cross-correlations between spectra with different
particle numbers will be non-zero \cite{Pa-06, Ko-06a}. For reviews on
EGOE, see \cite{Br-81, Be-03, Ko-01}. 

\subsection{EGOE(1+2) ensemble}

Besides the two-body interaction, Hamiltonians for realistic systems  also
contain a mean field one-body part (generating shell structure) and  therefore a
more appropriate  random matrix ensemble for finite quantum systems is
EGOE(1+2)\footnote{At this point it is also useful to mention that EGOE(1+2)'s
[and EGOE(2)'s] are also called TBRE in literature; Sec. 5.7 in \cite{Go-11}
gives clarifications on this nomenclature. As Brody et al state \cite{Br-81}:
{\it The most severe mathematical difficulties with TBRE are due to angular
momentum constraints $\ldots$ Another type of ensemble, $\ldots$ much closer to
being mathematical tractable abandons the $J$ restrictions entirely $\ldots$ an
embedded GOE, or EGOE for short.}},  the embedded GOE of one plus two-body
interactions  \cite{Fl-96a,Fl-96b,Fl-97,Ko-01}. Given the mean-field Hamiltonian
$\whh(1) = \sum_i \epsilon_i \hat{n}_i$, where $\hat{n}_i$ are number operators
and  $\epsilon_i$, $i=1,2,\ldots,N$ are the sp energies, and the two-body
interaction $\wv(2)$ (this is same as $\wh(2)$ defined in Sec. \ref{sec1p2p1}), 
EGOE(1+2) is defined by the operator
\be
\{\wh\} = \whh(1) + \lambda \, \{\wv(2)\} \;.
\label{eq.bpp4}
\ee
The $\{\wv(2)\}$ ensemble in two-particle spaces is represented by GOE(1)  and
$\lambda$ is the strength of $\wv(2)$.  Note that GOE($v^2$) means GOE with
variance $v^2$ for the off-diagonal matrix elements and $2 v^2$ for the diagonal
matrix elements; see Eq. (\ref{eq.bpp3}). The mean-field one-body Hamiltonian
$\whh(1)$ in Eq. (\ref{eq.bpp4}), in our studies, 
is a  fixed one-body operator defined by the sp
energies $\epsilon_i$ with average spacing $\Delta$. It is important to note
that  the operators $h(1)$ and $V(2)$ are independent. Without loss  of
generality, we put $\Delta=1$ so that $\lambda$, the strength of the
interaction,  is in the units of $\Delta$. Thus, EGOE(1+2) is defined by the
three parameters $(N, m, \lambda)$. It is possible to draw the $\epsilon_i$'s
from the eigenvalues of a random  ensemble and then the  corresponding EGOE(1+2)
is called two-body random interaction model (TBRIM) \cite{Fl-97} or from the
center of a GOE and then the corresponding EGOE(1+2) is called random
interaction matrix model (RIMM) \cite{Al-00,Al-01}. Construction of the
EGOE(1+2) ensemble in $m$-fermion spaces follows easily from the results in Sec.
\ref{sec1p2p1}. The notation used in Eq. (\ref{eq.bpp4}) implies that the action
of the operator $\{\wh\}$ on the $m$-particle basis space generates EGOE(1+2)
Hamiltonian matrix ensemble in $m$-particle spaces. It should be noted that the 
embedding for EGOE(1+2) is also generated by the $SU(N)$ group and   the
propagation formulas for the energy centroids and variances of the eigenvalue
densities follow from the unitary decomposition of $H$ with respect to the
$U(N)$ algebra; see Appendix \ref{c2a2}.

The most significant aspect of EGOE(1+2) is that the ensemble admits three
chaos markers as $\lambda$ is increased form zero. Firstly,  eigenvalue
(state) densities approach Gaussian form for large $m$ for all values of
$\lambda$. As the value of $\lambda$ increases from zero, level
fluctuations exhibit transition from Poisson  to GOE at
$\lambda=\lambda_c$  \cite{Ja-97}. With further increase in the $\lambda$
value, strength functions (also called local density of states) make a 
transition from Breit-Wigner (BW) to Gaussian form at $\lambda=\lambda_F
>> \lambda_c$ \cite{Ge-97, KS-01, Ja-02}. Beyond this  point, there is a
region around $\lambda=\lambda_d$ where entropies and other statistics
become same in the eigenbasis of the mean-field Hamiltonian and the pure
two-body Hamiltonian \cite{Ho-95, Ko-02}  (though not yet proved, this
result  perhaps extends to any basis \cite{Ko-03}). Equivalently, all
different definitions  for thermodynamic  properties like entropy,
temperature etc. coincide at $\lambda=\lambda_d$. It should be stressed
that the  chaos markers form the basis \cite{Ko-03} for statistical
spectroscopy \cite{Ko-01, Fr-82, Kar-94, Fl-99, Fr-06, KH-10}. The parametric
dependence of $\lambda_c$, $\lambda_F$ and $\lambda_d$ is also known and
this will be discussed in detail in Chapter \ref{ch2}. Besides these,
generic properties of EGOE(2) are valid for EGOE(1+2) in the strong
coupling limit; i.e., for $\lambda >> \lambda_F$.  Detailed 
discussion of the three chaos markers $(\lambda_c, \lambda_F, \lambda_d)$
generated by EGOE(1+2) and also applications of the ensemble are given in
\cite{Ko-01, Ko-03, An-04, Br-08, Go-11} and references therein. Now, we will
turn to embedded Gaussian unitary ensembles for spinless fermions.

\subsection{EGUE(2) and EGUE$(k)$ ensembles}
\label{egue1}

For a system of $m$ fermions occupying $N$ sp states, all the $N_m =
d_f(N,m) = \binom{N}{m}$ antisymmetric states transform as the  irreducible
representation (irrep)
$f_m=\{1^m\}$, in Young tableaux notation, with respect to the $U(N)$
algebra. With only two-body interactions among the fermions, the
Hamiltonian operator is 
\be 
\wh = \dis\sum_{v_a,v_b} V_{v_a\;v_b}(2) A^\dagger(f_2 v_b) 
A(f_2 v_a)\;.
\label{eq.gue1}
\ee  
Here, $f_2=\{1^2\}$ and $v_r$'s denote irreps of the groups in the
subgroup chain of $U(N)$ that supply the labels for a complete
specification of any two-particle state; 
similarly, for any $m$, the states are $\l|f_m
v_m\ran$. Note that $A^\dagger$ and $A$ in Eq.
(\ref{eq.gue1}) are normalized two-particle  creation and destruction
operators, respectively and $V_{v_a\;v_b}(2)$ are two-particle matrix
elements. The EGUE(2) ensemble in $m$-particle spaces, with matrix
dimension $N_m = d_f(N,m)$, is generated by the $\wh$ operator in Eq.
(\ref{eq.gue1}) with GUE representation  in two-particle spaces and then
propagating it to the $m$-particle spaces using the direct product structure
of the $m$-particle states \cite{Ko-05}. With the two-particle matrix
elements $V_{v_a\;v_b}(2)$ [the $V(2)$ matrix being complex hermitian]
drawn from a GUE, $V_{v_a\;v_b}(2)$ are independent Gaussian variables
with zero center and variance given by,
\be
\overline{V_{v_a\;v_b}(2) V_{v_c\;v_d}(2)}=\lambda^2
\delta_{v_a v_d} \delta_{v_b v_c}\;. 
\label{eq.gue2}
\ee
Here, $\lambda^2 N_2$ is the ensemble averaged two-particle variance. As
in \cite{Ko-05}, the $U(\Omega) \leftrightarrow SU(\Omega)$
correspondence is used and therefore we use $U(\Omega)$ and $SU(\Omega)$
interchangeably when there is no confusion. Important step in the
analytical  study of EGUE(2) is the unitary decomposition of $\wh$ in
terms of the $SU(N)$ tensors $B(g_\nu \omega_\nu)$ with $g_\nu = \{0\},
\{21^{N-2}\}$ and $\{2^21^{N-4}\}$,
\be
B(g_\nu \omega_\nu) = \dis\sum_{v_a, v_b} \lan f_2 v_a \overline{f_2}
\overline{v_b} \mid g_\nu \omega_\nu \ran A^\dagger(f_2 v_a) 
A(f_2 v_b)\;,
\label{eq.gue21}
\ee
where $\overline{f}$ is the irrep conjugate to $f$ and $\lan f_2 v_a
\overline{f_2} \overline{v_b} \mid g_\nu \omega_\nu \ran$ is a $SU(N)$ 
Wigner  coefficient. For $f_2=\{1^2\}$ we have, 
$\overline{f_2}=\{1^{N-2}\}$ and it also contains a phase factor as 
discussed in \cite{Ko-05}.  Then we have $\wh = \sum_{g_\nu, \omega_\nu}
W(g_\nu \omega_\nu) B(g_\nu \omega_\nu) $. A significant property of  the
expansion coefficients $W$'s is that they are also independent Gaussian
random variables, just as $V$'s, with zero center  and variance given by
$\overline{W(g_\nu \omega_\nu) W(g_\mu \omega_\mu)} =  \lambda^2
\delta_{g_\nu g_\mu} \delta_{\omega_\nu  \omega_\mu}$. Using Wigner-Eckart
theorem, the matrix elements of $B$'s in $f_m$ space can be decomposed
into  a reduced matrix element and a $SU(\Omega)$  Wigner coefficient.
Using this and the expansion of $\wh$ in terms of $B$'s,  exact analytical
formulas are derived for the ensemble averaged spectral variances, 
cross-correlations in energy centroids and also for the cross-correlations in
spectral variances for EGUE(2) \cite{Ko-05}.  In addition, exact result
for the ensemble averaged excess parameter (this involves fourth moment)
for the density of eigenvalues is also derived \cite{Ko-05}. An
alternative derivation was given by Benet et al \cite{Be-01,Be-01b}. More
significantly,  all these results extend to EGUE($k$); i.e., EGUE generated
by $k$-body interactions. Two significant results for EGUE($k$) are: (i)
for $k \leq m < 2k$, the density of eigenvalues is semi-circular whereas
the density is Gaussian for $m >>2k$ with $m=2k$ being the transition
point; (ii) EGUE($k$)  generates non-zero cross-correlations between 
states with different particle numbers while they will be zero for GUE
representation for the $m$-particle $H$ matrices. See
\cite{Ko-05,Be-01,Be-01b,Pl-02,Ko-06a} 
for further details; cross-correlations are
defined and further explored in Chapters \ref{ch4} and \ref{ch6}
ahead.

\section{Embedded Ensembles for Spinless Boson Systems}
\label{begoe}

\begin{figure}[htp]
\centering
\includegraphics[width=2.5in,height=2.5in]{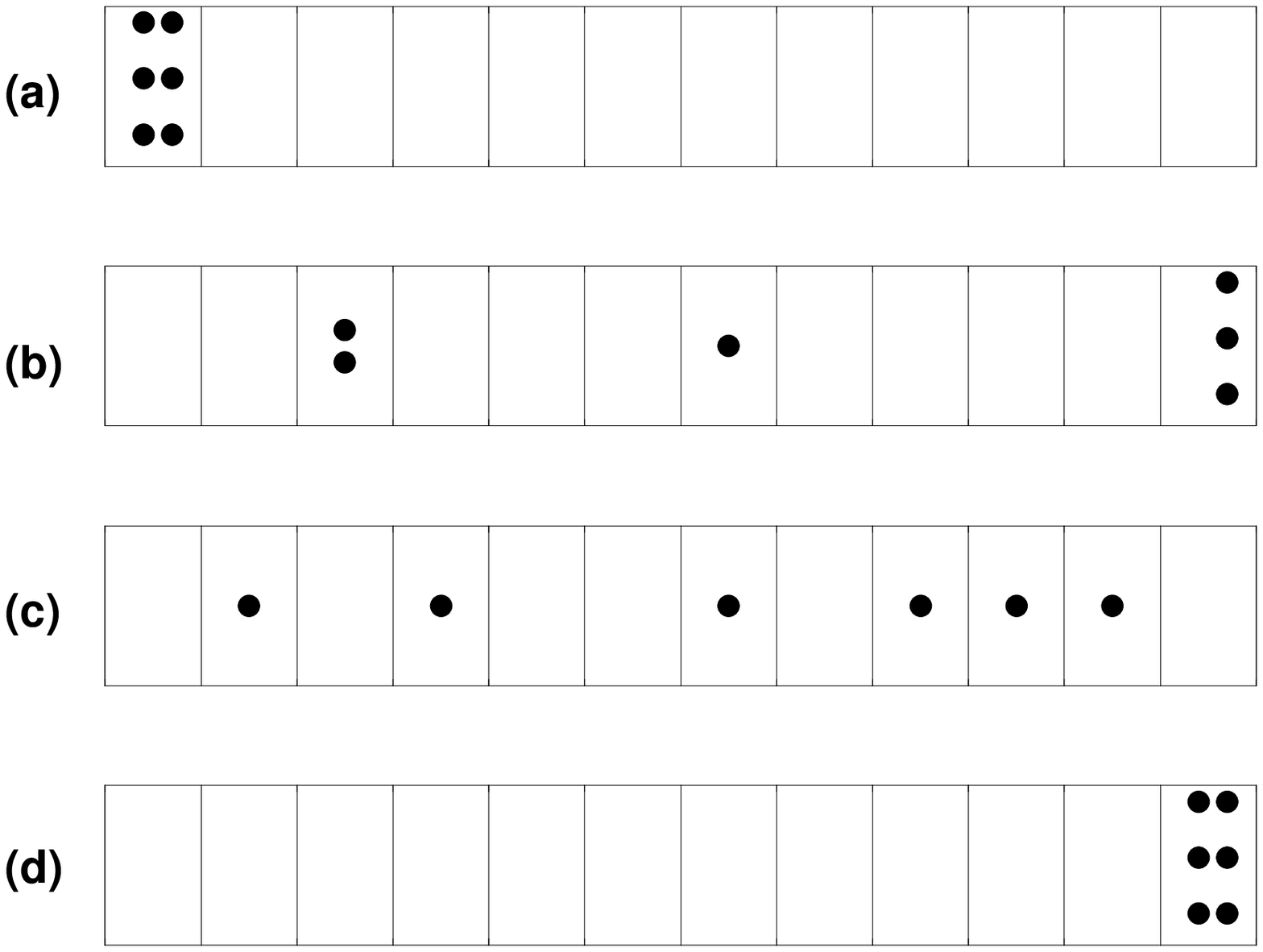}
\caption{Figure showing some configurations for the distribution of $m=6$
spinless bosons in $N=12$ single particle states. Distributing the $m$
bosons in all possible ways in $N$ single particle states generates the
$m$-particle configurations or basis states. This is similar to
distributing $m$ particles in $N$ boxes with the conditions that
occupancy of each box lies between zero and $m$ and the maximum number of
occupied boxes equals $m$.  In the figure, (a) corresponds to the basis
state $\l| (\nu_1)^6 \ran$, (b) corresponds to
the basis state $\l| (\nu_3)^2 \nu_7 (\nu_{12})^3 \ran$, (c)
corresponds to the basis state  $\l| \nu_2 \nu_4 \nu_7 \nu_9 \nu_{10}
\nu_{11} \ran$ and  (d) corresponds to the basis state $\l| (\nu_{12})^6
\ran$.}
\label{sp-boson}
\end{figure}

The BEGOE(2)/BEGUE(2) ensemble for spinless boson systems is generated
by defining the two-body Hamiltonian $H$ to be GOE/GUE in two-particle
spaces and then propagating it to many-particle spaces by using the geometry
of the many-particle spaces [this is in general valid for $k$-body
Hamiltonians, with $k < m$, generating BEGOE$(k)$/BEGUE$(k)$].  Consider
a system of $m$ spinless bosons occupying $N$  sp states $\l| \l. \nu_i
\ran \r.$, $i=1,2,\ldots,N$; see Fig. \ref{sp-boson}. Then, BEGOE(2) is
defined by the Hamiltonian operator, 
\be
\wh = \dis\sum_{\nu_i \leq \nu_j,\;\nu_k \leq \nu_l}
\dis\frac{ \lan \nu_k\;\nu_l
\mid \wh \mid \nu_i\;\nu_j\ran}{\dis\sqrt{(1+\delta_{ij})
(1+\delta_{kl})}}\;b^\dg_{\nu_k}\,b^\dg_{\nu_l}\,
b_{\nu_i}\,b_{\nu_j}\;,
\label{eq.app1a}
\ee
with the symmetries for the symmetrized two-body matrix elements 
$\lan \nu_k \; \nu_l \mid \wh \mid \nu_i \; \nu_j \ran$ being,
\be
\barr{c}
\lan \nu_k\;\nu_l \mid \wh \mid \nu_j\;\nu_i\ran = \lan \nu_k\;\nu_l
\mid \wh \mid \nu_i\;\nu_j\ran \;,\\
\lan \nu_k\;\nu_l \mid \wh \mid \nu_i\;\nu_j\ran = \lan \nu_i\;\nu_j
\mid \wh \mid \nu_k\;\nu_l\ran \;.
\earr \label{eq.app1}
\ee
The action of the Hamiltonian operator defined by Eq. (\ref{eq.app1a})
on an appropriately chosen basis states (see Fig. \ref{sp-boson} for
examples) generates the BEGOE(2) ensemble. Note that $b_{\nu_i}$ and
$b^\dg_{\nu_i}$ in Eq. (\ref{eq.app1a}) annihilate and create a boson in
the sp state $\l| \l. \nu_i \ran \r.$,  respectively.  The Hamiltonian
matrix $H(m)$ in $m$-particle spaces contains three different types of
non-zero matrix elements and explicit formulas for these are
\cite{PDPK},
\be
\lan \dis\prod _{r=i,j,\ldots} \l(\nu_r\r)^{n_r} \mid \wh \mid 
\dis\prod _{r=i,j,\ldots} \l(\nu_r\r)^{n_r} \ran = 
\dis\sum_{i \geq j }\; \dis\frac{n_i\l(n_j-\delta_{ij}\r)}{\l(1+
\delta_{ij}\r)} \; \lan \nu_i
\nu_j \mid \wh \mid \nu_i \nu_j \ran \;, \nonumber
\label{eq.app2a1}
\ee
\be
\barr{l}
\lan \l(\nu_i\r)^{n_i-1} \l(\nu_j\r)^{n_j+1} 
\dis\prod_{r^\prime=k,l,\ldots} 
\l(\nu_{r^\prime}\r)^{n_{r^\prime}} \mid \wh \mid 
\dis\prod _{r=i,j,\ldots} \l(\nu_r\r)^{n_r} \ran = \\ \\
\dis\sum_{k^\prime}\; \l[ \dis\frac{n_i\l(n_j+1\r)\l(n_{k^\prime}-
\delta_{k^\prime i}\r)^2}{\l(1+\delta_{k^\prime i}\r) \l(1+\delta_{
k^\prime j}\r)}\r]^{1/2} \; \lan \nu_{k^\prime} \nu_j \mid \wh \mid
\nu_{k^\prime} \nu_i \ran \;,
\earr \label{eq.app2}
\ee
\be
\barr{l}
\lan \l(\nu_i\r)^{n_i+1} \l(\nu_j\r)^{n_j+1} \l(\nu_k\r)^{n_k-1}
\l(\nu_l\r)^{n_l-1} \dis\prod _{r^\pr=m,n,\ldots} 
\l(\nu_{r^\pr}\r)^{n_{r^\pr}} \mid \wh \mid 
\dis\prod _{r=i,j,\ldots} \l(\nu_r\r)^{n_r} \ran = \\ \\
\l[ \dis\frac{n_k\l(n_l-\delta_{kl}\r)\l(n_i+1\r)\l(n_j+1+
\delta_{ij}\r)}{\l(1+\delta_{ij}\r) \l(1+\delta_{kl}\r)} \r]^{1/2}\; 
\lan \nu_i \nu_j \mid \wh \mid \nu_k \nu_l \ran \;.\nonumber
\earr \label{eq.app2a2}
\ee
Note that all other $m$-particle matrix elements are zero due to two-body
selection rules. In the second equation in Eq. (\ref{eq.app2}), $i \neq j$ and
in the third equation, four combinations are possible: (i) $k=l$, $i=j$, $k \neq
i$; (ii) $k=l$, $i \neq j$, $k \neq i$, $k \neq j$; (iii) $k \neq l$, $i=j$, $i
\neq k$, $i \neq l$; and (iv) $i \neq j \neq k \neq l$.  BEGOE(2) for spinless
boson systems is defined by Eqs. (\ref{eq.app1}) and (\ref{eq.app2}) with the
$H$ matrix in two-particle spaces represented by GOE($v^2$)  [see Eq.
(\ref{eq.bpp3}) and discussion below Eq. (\ref{eq.bpp4}) 
for GOE$(v^2)$]. Now the $m$-boson BEGOE(2) Hamiltonian
matrix ensemble is denoted by $\{H(m)\}$, with $\{H(2)\}$ being a GOE. Note that
the $H(m)$ matrix dimension is $d_b(N,m) = \binom{N+m-1}{m}$ and the number of
independent matrix elements is $d_b(N,2)[d_b(N,2)+1]/2$. The subscript `$b$' in
$d_b(N,m)$ stands for `bosons'. Using Eqs. (\ref{eq.app1}) and (\ref{eq.app2})
with GOE representation for $H$ in two-particle spaces, we have developed a
computer code for constructing BEGOE(2) ensemble. The extension of BEGOE(2) code
to construct BEGOE(1+2) incorporating mean-field one-body part is
straightforward. It is important to stress that,  unlike for fermionic EE, there
are only a few BEE  investigations \cite{PDPK, Ag-01, Ag-02, Ch-03, Ch-04}.

Firstly, it is important to mention that, unlike fermion systems,  for
interacting spinless boson systems with $m$ bosons in $N$ sp orbitals, 
dense limit defined by $m \to \infty$, $N  \to \infty$ and $m/N  \to
\infty$ is also possible as $m$ can be greater than $N$ for bosons. 
Also the results for bosons can be obtained from those for  fermions by
using $N \to -N$ symmetry \cite{KP-80}. It is now well understood that
BEGOE(2) [also BEGUE(2)] generates in  the dense limit, eigenvalue
density close to a Gaussian \cite{KP-80,PDPK,Ag-02}. Also the ergodic property
is found to be valid in the dense limit with sufficiently large $N$
\cite{Ch-03}; there are deviations for small $N$ \cite{Ag-01}.
Similarly, for BEGOE(1+2),  as the strength $\lambda$ of the two-body
interaction increases, there is Poisson to GOE transition in  level
fluctuations \cite{Ch-03} and with further
increase in $\lambda$, there is Breit-Wigner to Gaussian transition in
strength functions \cite{Ch-04}. For BEGUE($k$), exact analytical
results for the lowest two moments of the two-point function have been
derived by Agasa et al \cite{Ag-02}.  Level fluctuations and
wavefunction structure in interacting  boson systems are also studied
using interacting boson models of atomic nuclei \cite{Al-91,Al-93,Ca-00}
and a  symmetrized two coupled rotors model \cite{Bo-98a,Bo-98b} and the results
are understood in terms of random matrix theory. In
addition, using random interactions in interacting boson models, there
are several studies on the generation of regular structures in boson
systems with random interactions \cite{Ku-97,BF-00,Ko-04,Yo-09}.
Finally, there are also studies on thermalization in finite quantum
systems using boson systems and here also random matrix theory plays an
important role; see\cite{Ri-09,Sa-10,Sa-10a,Ol-09,Ma-11b} and references
therein. 

\section{Preview}

\begin{figure}[htp]
\centering
\includegraphics[width=4.5in,height=5.5in]{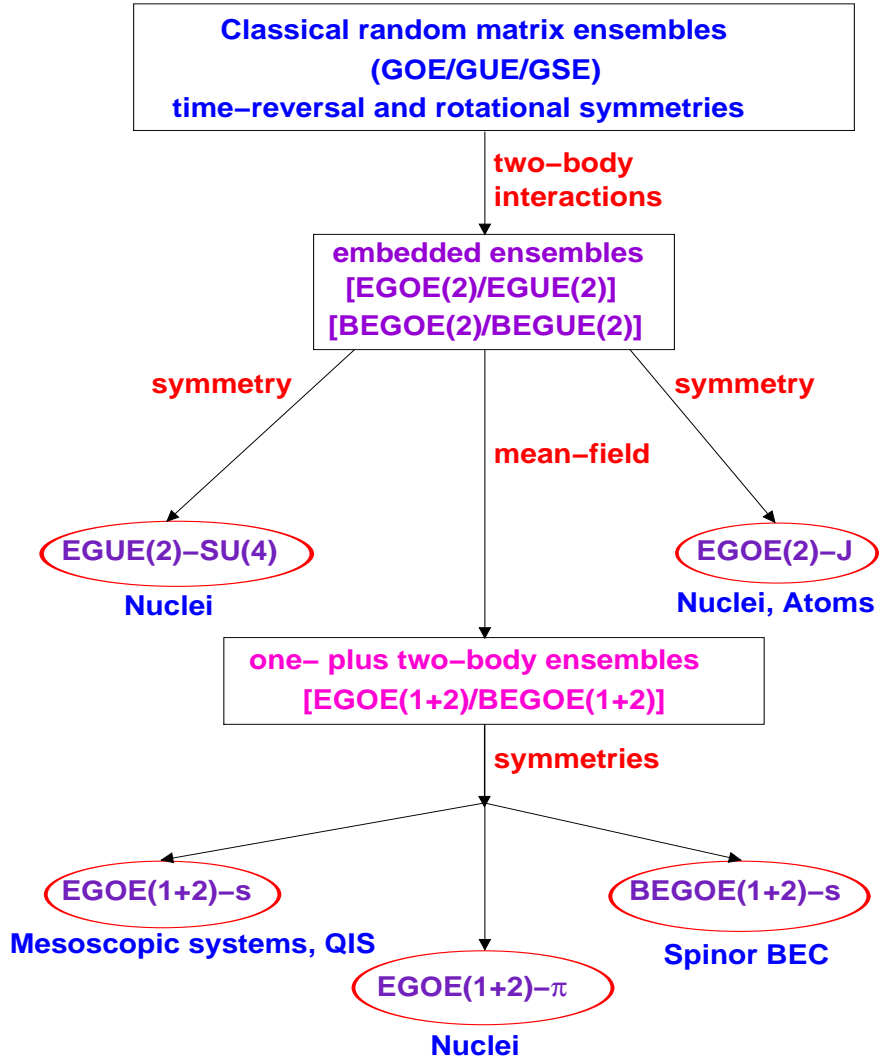}
\caption{Figure showing the information content of various random
matrix  ensembles. Also shown are the areas in which embedded ensembles
with various symmetries are relevant. Here, $\cs$ denotes two-particle
spin, $SU(4)$ denotes spin-isospin supermultiplet symmetry, $\pi$
denotes parity and $J$ denotes total angular-momentum. Note that the
symplectic ensembles EGSE/BEGSE and the  one plus two-body unitary
ensembles EGUE(1+2)/BEGUE(1+2) are not shown as there are no studies of
these ensembles till today.}
\label{tbre}
\end{figure}

Besides the mean-field and the two-body character, realistic Hamiltonians also
carry a variety of symmetries. In many applications of EGEs, generic properties
of EGE for spinless fermions are `assumed'  to extend to symmetry subspaces.
More importantly,  there are several  properties of real systems that require
explicit inclusion of symmetries and they are defined by a variety of Lie
algebras. For example, spin $S$ is a good quantum number for atoms and quantum
dots, angular-momentum $J$ and parity $\pi$ are good quantum numbers for nuclei
and so on.  Therefore, it is more appropriate to study EE with good symmetries
and symmetries in principle  provide a systematic classification of EE. Figure
\ref{tbre} shows some EE/BEE with symmetries.  It is well acknowledged that
these extended ensembles are notoriously difficult and to derive their generic
properties is thus, quite complicated. As stated by French \cite{Fr-80}:  {\it
$\ldots$ For most purposes, the resulting embedded GOE (or EGOE) is very
difficult to deal with, but by good luck, we can use it to study the questions
we have posed and the answers are different from, and much more enlightening
than, those which would come from GOE.}  EGEs operating in many-particle spaces
generate forms for distributions of various physical quantities with respect to
energy and other quantum numbers. The separation of the energy evolution of
various observables into a smoothed and a fluctuating part (with fluctuations
following GOE/GUE/GSE) provides the basis for statistical spectroscopy
\cite{KH-10,Ko-89,Fr-82}.  In statistical spectroscopy, methods are developed to
determine various moments defining the distributions (predicted by EGEs) for the
smoothed parts (valid in the chaotic region) without recourse to many-body
Hamiltonian construction [this part of statistical spectroscopy is also often
refered to as spectral distribution  theory or spectral distribution methods].

The aim of the present thesis is to identify and systematically analyze many
different physically relevant EGEs with symmetries by considering a variety of
quantities and measures that are important for finite interacting quantum
systems. Numerical as well analytical study of these more general ensembles is
challenging due to the complexities of group theory and also due to large matrix
dimensions for $m \geq 10$.  It is useful to mention that many diversified
methods like numerical Monte-Carlo methods, binary correlation approximation,
trace propagation, group theory and perturbation theory are used to derive
generic properties of EE \cite{MF-75,Be-01,Ko-05,Ko-07,Pa-09}. Towards this end,
we have obtained large number of new results for embedded ensembles and in
particular for EGOE(1+2)-$\cs$, EGUE(2)-$SU(4)$, EGOE(1+2)-$\pi$,
BEGOE(1+2)-$\cs$ and EGOE(2)-$J$ ensembles. In addition, derived are formulas
for several fourth order traces that are needed in the analysis of EGOE's and
also in the applications of spectral distribution theory generated by EGEs. We
have also obtained further evidence for EGOE representation of nuclear
Hamiltonian matrices. Results of the present thesis together with earlier
investigations establish that embedded Gaussian ensembles can be used gainfully
to study a variety of problems in many-body quantum physics and this includes
some of the new areas of research in physics such as quantum information science
(QIS) and the thermodynamics of isolated finite interacting quantum systems. It
should be noted that some of the work in the present thesis is also reviewed in
\cite{Go-11}.

Before going further, for completeness, we mention that, besides the
EE(BEE)'s that will be described in detail in Chapters
\ref{ch2}-\ref{ch8}, there a few other EE(BEE)'s that have received
limited attention in the literature. They are: (i) EGOE invariant under
particle-hole symmetry, called random quasi-particle ensembles
\cite{Jo-98,Ki-07}, (ii) a fixed Hamiltonian plus EGOE called $K$+EGOE
\cite{Ko-01} and similarly displaced TBRE \cite{Ve-02,Co-82} where a
constant is added to all the two-body interaction matrix elements, (iii)
EGOE with a partitioned GOE in the two-particle spaces, called $p$-EGOE
\cite{Ko-99,Fr-83}, (iv) in mesoscopic systems such as quantum dots, 
randomness of the sp states induces randomness in the two-body part of
the Hamiltonian and these then give rise to  induced-TBRE depending on
the underlying space-time symmetries as well as on the features of the
two-body interaction \cite{Al-05}, and (v) BEGOE(1+2) with orbital
angular-momentum $L$, denoted as  BEGOE(1+2)-$L$ or BTRBE-$L$, for bosons
in $sp$ orbits \cite{Ku-00}  and $sd$ orbits \cite{BF-01} and also
BEGOE(2) with $SO(N_1) \oplus SO(N_2)$ symmetry in IBM \cite{Ko-04}. Now
we will give a preview of the thesis. 

Results for transitions in eigenvalue and wavefunction structure in one plus
two-body random matrix ensembles with spin [EGOE(1+2)-$\cs$]  are given in
Chapter \ref{ch2}. Chapter \ref{ch3} gives the results for pairing correlations
generated by  EGOE(1+2)-$\cs$. Spectral properties of embedded Gaussian unitary
ensemble of random matrices generated by two-body interactions  with Wigner's
$SU(4)$ symmetry [EGUE(2)-$SU(4)$] are derived and discussed  in Chapter
\ref{ch4}. It is important to mention that EGUE(2)-$SU(4)$ is introduced for the
first time in the present thesis. Chapter \ref{ch5} gives the results for
density of states and parity ratios for one plus two-body random matrix
ensembles with parity [EGOE(1+2)-$\pi$].  Spectral properties of one plus
two-body random matrix ensembles for boson systems with spin [BEGOE(1+2)-$\cs$] 
are presented in Chapter \ref{ch6}. Although BEGOE(1+2)-$\cs$ was known in
literature before, it is explored in detail for the first time in this thesis.
Chapter \ref{ch7} gives the results for higher order averages, derived by
extending Mon and French's binary correlation method to two-orbits, required in
many applications. In addition, given also are some results for the traces
needed for the embedded Gaussian orthogonal ensemble of two-body interactions
with angular-momentum $J$ symmetry [EGOE(2)-$J$] for fermions in a single
$j$-shell. Chapter \ref{ch8} gives a comprehensive analysis of the structure of
$H$ matrices to establish EGOE structure of nuclear shell model $H$-matrices.
Finally, Chapter \ref{ch9} gives conclusions and future outlook. Before turning
to Chapter \ref{ch2}, we would like to add that there will be some unavoidable
repetition in Chapters \ref{ch2}-\ref{ch8} as they deal with different embedded
ensembles with applications in different physical systems. 

\chapter{EGOE(1+2)-$\cs$: Transition Markers}
\label{ch2}

\section{Introduction}
\label{ch2-int}

First non-trivial but at the same time important (from the point of view of its
applications) embedded ensembles are EE(2)-$\cs$ and EE(1+2)-$\cs$ with spin
degree of freedom, for a system of interacting fermions. In the last decade, the
GOE version, the embedded Gaussian  orthogonal ensemble of one plus two-body
interactions with spin degree of freedom [EGOE(1+2)-$\cs$],  has received
considerable attention. Both numerical \cite{Ko-06,Al-06}  and analytical
\cite{Ka-00,Kk-02} methods for analyzing and applying this ensemble have been
developed. Using these, several results are obtained and briefly they are as
follows: (i) fixed-$(m,S)$ density of levels is established, using numerical
results, to be Gaussian \cite{Ko-06,Ja-01,Ka-00}; (ii) lower order 
cross-correlations in spectra with different $(m,S)$ are studied both numerically and
analytically and they are found to be larger compared to those for the spinless
fermion systems \cite{Ko-06,Ko-06a}; (iii) ground-state (gs) spin structure
investigated using second and fourth moments established that with random
interactions there is preponderance of $S=0$ ground states \cite{Ka-00,Kk-02}; 
(iv) delay in Stoner instability in itinerant magnetic systems due to random
interactions has been established and thus with random interactions much
stronger exchange interaction is needed for gs magnetization in irregular
quantum dots \cite{Ja-00,Ja-01}; and (v) it is shown that the odd-even staggering in
the gs energies of nm-scale metallic grains, attributed normally due to
mean-field orbital energy effects or coherent pairing effects, can also come
from purely  random two-body Hamiltonians \cite{Pa-02,Kk-02}.  Thus, although
the gs structures generated by EGOE(1+2)-$\cs$ and also some results in the
strong-coupling region have been investigated in some detail, the important
question of chaos or transition markers generated by the ensemble hasn't yet
been investigated in any detail. It should be stressed that the chaos markers
form the basis \cite{Ko-03} for statistical spectroscopy
\cite{Ko-01,Fr-82,Kar-94,Fl-99,Fr-06} and also the BW to Gaussian transition
plays an important role in characterizing  multi-partite entanglement and
fidelity decay relevant for QIS
\cite{Mo-06,Br-08,Pi-07}. Our purpose in this chapter is to establish that the
EGOE(1+2)-$\cs$ ensemble exhibits three chaos markers  just as the  EGOE(1+2)
for spinless fermion systems and more importantly, by deriving the exact formula
for the propagator of the spectral variances, the spin dependence of the markers
is explained. These results, derived for the first time using an  ensemble with
additional symmetry (besides particle number), provide much stronger basis for
statistical (nuclear and atomic) spectroscopy. In addition, as recognized only
recently, entanglement and  strength functions essentially capture the same
information about eigenvector structure and therefore the change in the form
($\delta$-function to BW to Gaussian) of the strength functions in different
regimes defined by the chaos markers determines entanglement properties in
multi-qubit systems \cite{Mo-06,Br-08,Pi-07,Me-05}. Similarly, the chaos marker
$\lambda_d$ discussed in the present chapter allows us to define a region of
thermalization in finite interacting quantum systems modeled by EGOE and
thermalization in generic isolated quantum systems has applications in QIS as
emphasized in some  recent papers  \cite{Ri-08,Ca-09,De-91,Sr-94}.  All the
results presented in this chapter are published in \cite{Ma-10}.

\section{EGOE(1+2)-$\cs$ Ensemble: Preliminaries}
\label{c2s1}

Let us begin with a system of $m$ ($m > 2$)  fermions  distributed say in
$\Omega$ number of sp orbitals each with spin  $\cs=\spin$ so that the number of
sp states $N=2\Omega$. The sp states are denoted by $\l.\l| i,m_\cs=\pm
\spin\r.\ran$ with $i=1,2,\ldots,\Omega$ and similarly the  two-particle
antisymmetric states are denoted by $\l.\l|(ij)s,m_s\r.\ran$ with $s=0$ or
$1$. For one plus two-body Hamiltonians preserving $m$-particle spin $S$, the
one-body Hamiltonian is $\whh(1) = \sum_{i=1,2,\ldots,\Omega}\, \epsilon_i
\hat{n}_i$. Here the orbitals $i$ are doubly degenerate, $\hat{n}_i$ are number
operators  and $\epsilon_i$ are sp energies [it is in principle possible to
consider $\whh(1)$ with off-diagonal energies $\epsilon_{ij}$]. Similarly the
two-body Hamiltonian $\wv(2)$ is defined by the two-body matrix elements
$\lambda_s\,V^s_{ijkl}=\lan (kl)s,m_s \mid \wv(2) \mid (ij)s,m_s\ran$ 
with the two-particle spins $s=0$ and $1$. These matrix elements are independent
of  the $m_s$ quantum number. Note that the $\lambda_s$ are constants and for
$s=1$,  only $i \neq j$ and $k \neq l$ matrix elements exist.  Thus
$\wv(2)=\lambda_0 \wv^{s=0}(2) + \lambda_1 \wv^{s=1}(2)$ and  the $V$ matrix in
two-particle spaces is a direct sum matrix with the $s=0$  and $s=1$ space
matrices having dimensions $\Omega(\Omega+1)/2$ and $\Omega(\Omega-1)/2$,
respectively. Now, EGOE(2)-$\cs$ for a given $(m,S)$ system is generated  by
defining the two parts of the two-body Hamiltonian to be independent GOE's [one
for $\wv^{s=0}(2)$ and other for $\wv^{s=1}(2)$] in two-particle spaces and then
propagating the $V(2)$ ensemble $\{\wv(2)\}=\lambda_0 \{\wv^{s=0}(2)\} +
\lambda_1 \{\wv^{s=1}(2)\}$ to the $m$-particle spaces with a given spin $S$ by
using the geometry (direct product structure), defined by $U(2\Omega) \supset
U(\Omega) \otimes SU(2)$ algebra (see Appendix \ref{c3a1} and Chapter
\ref{ch3}),  of  the $m$-particle spaces. Then EGOE(1+2)-$\cs$ is defined by the
operator
\be
\{\wh\}_{\mbox{EGOE(1+2)-\cs}} = \whh(1) + \lambda_0\, \{\wv^{s=0}(2)\} +
\lambda_1\, \{\wv^{s=1}(2)\}\;,
\label{eq.def1}
\ee
where $\{\wv^{s=0}(2)\}$ and $\{\wv^{s=1}(2)\}$ in two-particle spaces are
GOE(1)  and $\lambda_0$ and $\lambda_1$ are the strengths of the $s=0$ and
$s=1$ parts of $\wv(2)$, respectively. From now onwards we drop the ``hat'' 
symbol over $H$, $h$ and $V$ operators when there is no confusion. 

The mean-field one-body Hamiltonian $h(1)$ in Eq. (\ref{eq.def1}) is a 
fixed one-body operator defined by the sp energies $\epsilon_i$ with average
spacing $\Delta$  (it is possible to draw the $\epsilon_i$'s from the
eigenvalues of a random  ensemble \cite{Ja-01} or from the center of a GOE
\cite{Al-00a}).  Without loss  of generality we put $\Delta=1$ so that
$\lambda_0$ and $\lambda_1$ are in the units of $\Delta$. Thus,
EGOE(1+2)-$\cs$ in Eq. (\ref{eq.def1}) is defined by the five  parameters
$(\Omega, m, S, \lambda_0, \lambda_1)$. The action of the Hamiltonian
operator defined by Eq. (\ref{eq.def1}) on an appropriately chosen
fixed-($m,S$) basis states generates the EGOE(1+2)-$\cs$ ensemble in ($m,S$)
spaces.   The $H$ matrix dimension $d_f(\Omega,m,S)$  for a given 
$(\Omega,m,S)$, i.e.,
number of levels in the $(m,S)$ space [with each of them being $(2S+1)$-fold
degenerate], is
\be
d_f(\Omega,m,S)=\dis\frac{(2S+1)}{(\Omega+1)} { \Omega+1 \choose
m/2+S+1} {\Omega+1 \choose m/2-S}\;,
\label{eq.def2}
\ee
satisfying the sum rule $\sum_S\;(2S+1)\;d_f(\Omega,m,S)= {N \choose m}$. Note
that the subscript `$f$' in Eq. (\ref{eq.def2}) stands for `fermions'. For
example for $\Omega=m=8$, the dimensions are 1764, 2352, 720, 63, and 1 for
$S=0$, 1, 2, 3, and 4, respectively. Similarly for $\Omega=m=10$, the dimensions
are 19404, 29700, 12375, 1925, 99, and 1 for $S=0-5$ and for $\Omega=m=12$, they
are 226512, 382239, 196625, 44044, 4214, 143, and 1 for $S=0-6$. It is useful to
note that for the EGOE(1+2)-$\cs$ ensemble three group structures are relevant
and they are $U(\Omega) \otimes SU(2)$,  $\sum_{S=0,1} O(N_{2,S}) \oplus$ and
$\sum_S O(N_{m,S}) \oplus$, $m>2$. Here $N_{m,S}=d_f(\Omega,m,S)$, the symbol
$\oplus$ stands for direct sum and $O(r)$ is the orthogonal group in $r$
dimensions.  The $U(\Omega) \otimes SU(2)$ algebra defines the embedding. The
EGOE(2) ensemble has orthogonal invariance with respect to the  $\sum_{S=0,1}
O(N_{2,S}) \oplus$ group acting in two-particle spaces. However it is not
invariant under the  $\sum_S O(N_{m,S}) \oplus$ group for $m>2$. This group is
appropriate if GOE  representation for fixed-$(m,S)$ $H$ matrices is employed;
i.e., there is an independent GOE for each $(m,S)$ subspace. 

Given the sp energies $\epsilon_i$ and the two-body matrix elements
$V^s_{ijkl}$, the many-particle Hamiltonian matrix for a given $(m,S)$ can be
constructed either using the $M_S$ representation and a spin ($S$) projection
operator \cite{Ko-06} or directly in a good $S$ basis using angular-momentum
algebra  \cite{Al-06}. The former is equivalent to employing the algebra
$U(2\Omega) \supset U(\Omega) \oplus U(\Omega)$ and the latter corresponds to
$U(2\Omega) \supset U(\Omega) \otimes SU(2)$. Just as in the earlier papers by
our group
\cite{Ko-06}, we have employed the $M_S$ representation  for constructing the
$H$ matrices and the ${\hat S}^2$ operator for projecting states with good $S$. 
Then the dimension of the basis space is $\cd(M_S^{min}) = \sum_S \,
d_f(\Omega,m,S)$; $M_S^{min}=0$ for $m$ even and $1/2$ for $m$ odd. For example,
for $\Omega=m=8$ we have $\cd(M_S^{min})=4900$, for $\Omega=8,m=6$,
$\cd(M_S^{min})=3136$ and for  $\Omega=m=10$ we have $\cd(M_S^{min})=63404$. It
is important to note that here the construction of the $m$-particle $H$ matrix
reduces to the  problem of EGOE(1+2) for spinless fermion systems and hence Eqs.
(\ref{eq.bpp1a})- (\ref{eq.bpp3}) of Chapter \ref{ch1}  will apply.   From the
dimensions given above, it is clear that numerical calculations will be
prohibitive for $m \geq 10$ even on best available computers. Therefore, most of
the numerical investigations are restricted to $m \leq 8$. For properties
related to a few lowest eigenvalues it is possible to go beyond $m=8$
\cite{Pa-02}. Now, before presenting the results for the three chaos markers
generated by EGOE(1+2)-$\cs$,  we will consider the  ensemble averaged
fixed-$(m,S)$ density of levels and present the exact formula for its variance.

\section{Gaussian Level Densities and Ensemble Averaged Spectral Variances}
\label{c2s2}

\subsection{Gaussian form for fixed-$(m,S)$ eigenvalue densities}
\label{c2s2p1}

\begin{figure}
\centering
\includegraphics[width=5.5in,height=3.5in]{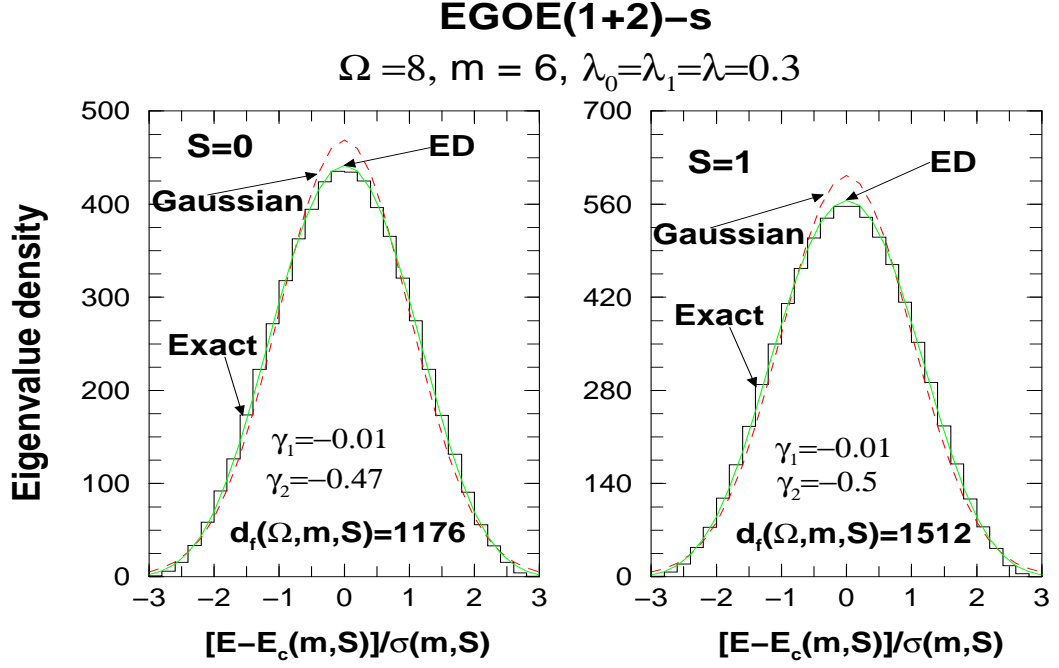}
\caption{Ensemble averaged eigenvalue density for a 20 member
EGOE(1+2)-$\cs$ ensemble with $\Omega=8$, $m=6$ and spins $S=0$ and $1$. The
dashed curves give Gaussian representation and the continuous curves give
Edgeworth corrected Gaussians [$\rho_{ED}$ in Eq. (\ref{eq.gau1})].  Values
of the skewness and excess parameters are also given in the figure. In the
plots, the densities for a given spin are normalized to the dimension
$d_f(\Omega,m,S)$. Note that the $E_c(m,S)$ and $\sigma(m,S)$ are fixed-$(m,S)$ 
energy centroids and spectral widths, respectively of the eigenvalue
densities. See text for further details.}
\label{rho}
\end{figure}

Using the $M_S$ representation,  we have numerically constructed
the $H$ matrix in large number of examples and by diagonalizing them obtained 
the ensemble averaged eigenvalue (level) densities $\overline{\rho^{m,S} (E)} = 
\lan \delta(\wh-E) \ran^{m,S}$. In general, given $m$-particle space it is
possible to decompose it into subspaces $\Gamma$ such that the $H$ preserving
$\Gamma$ symmetry will be a direct sum matrix of $H$ matrices for each $\Gamma$
subspace (as $H$ will not connect states with different $\Gamma$). Then, the
one-point function $\overline{\rho^{m,\Gamma}(E)}$,  the ensemble averaged
density of eigenvalues over each $\Gamma$ subspace, is given by
\be
\overline{\rho^{m,\Gamma}(E)} = \lan \delta(\wh-E) \ran^{m,\Gamma}\;.
\label{eq.eden}
\ee
For EGOE(1+2)-$\cs$, $\Gamma$ denotes the $m$-particle spin $S$.  For
EGUE(2)-$SU(4)$ ensemble discussed in Chapter \ref{ch4}, $\Gamma$ represents a
$m$-particle $SU(4)$ irrep $F_m$. Similarly, 
for EGOE(1+2)-$\pi$, ensemble discussed in
Chapter \ref{ch5}, $\Gamma =\pi$ and for BEGOE(1+2)-$\cs$ ensemble discussed in
Chapter \ref{ch6}, $\Gamma=S$.  Note that the trace of an operator $\co$ over a
fixed-$(m,S)$ space is defined  by $\lan\lan \co \ran\ran^{m,S} =
(2S+1)^{-1}\sum_\alpha\, \lan m,S,\alpha \mid \co \mid  m,S,\alpha \ran$ and
similarly $(m,S)$ space average is  $\lan \co \ran^{m,S} =
[d_f(\Omega,m,S)]^{-1} \lan\lan \co \ran\ran^{m,S}$.  From now onwards, we drop
the ``overline'' over $\rho$ when there is no confusion. Results are shown for
$\Omega=8$ and $m=6$ with $S=0$ and $1$ and $\lambda_0=\lambda_1=\lambda=0.3$ 
in Fig. \ref{rho}.  In these calculations and also for all other calculations
reported in this chapter, we have chosen the sp energies to be
$\epsilon_i=i+1/i$ with  $i=1,\,2,\ldots,\Omega$ just as in many of the earlier
papers \cite{Fl-96a,Fl-96b,Ko-02,Ko-06}. Note that the second term ($1/i$) in
$\epsilon_i$ has been added, as discussed first in  \cite{Fl-96a,Fl-96b}, to
avoid the degeneracy of  many-particle states for small $\lambda$. 
To construct the eigenvalue
density, we first make the centroids $E_c(m,S)$ of  all the members of the
ensemble to be zero and variance $\sigma^2(m,S)$ to be  unity,  i.e., for each
member we change the eigenvalues $E$ to the standardized  variables $\we =
[E-E_c(m,S)]/\sigma(m,S)$. Note that the parameters $E_c(m,S)$ and
$\sigma^2(m,S)$ depend also on $\Omega$. But for convenience,  we shall drop
$\Omega$ in $E_c(m,S)$ and $\sigma^2(m,S)$ throughout this chapter. Then, using
a bin-size $\Delta \we=0.2$, histograms for  $\rho^{m,S}(E)$ are generated. The
calculated results are compared with both the Gaussian  ($\rho_\cg$) and
Edgeworth (ED) corrected Gaussian $(\rho_{ED})$ forms \cite{St-87},
\be
\barr{rcl}
\rho_{\cg}(\we) & = &
\dis\frac{1}{\sqrt{2\pi}}\exp
\l(-\dis\frac{\we^2}{2}\r) \;,\\ \\
\rho_{ED}(\we) & = &
\rho_{\cg}(\we)
\l\{1+\l[{\dis\frac{\gamma_1}{6}}He_3\l(\we\r)\r]+
\l[{\dis\frac{\gamma_2}{24}}He_4\l(\we\r) +
{\dis\frac{\gamma_1^2}{72}} He_6\l(\we\r) \r]\r\}\;.
\earr \label{eq.gau1}
\ee
Here $\gamma_1$ is the skewness and $\gamma_2$ is the excess parameter. 
Similarly, $He$ are the Hermite polynomials: $He_3(x)=x^3-3x$,
$He_4(x)=x^4-6x^2+3$, and $He_6(x)= x^6-15x^4 + 45x^2-15$. From the results in
Fig. \ref{rho},  it is seen that the agreement between the exact and ED
corrected Gaussians is excellent. Further numerical examples are given in
\cite{Ko-06,Ja-01} up to $m=8$. It has been well established that the  ensemble
averaged eigenvalue density takes Gaussian form in the case of spinless fermion
as well as boson systems \cite{MF-75,Ko-01,Be-01,Ch-03}. Combining these with
the numerical results for the fixed-$(m,S)$ level densities, it can be concluded
that the Gaussian form is generic for the embedded ensembles extending to those
with good quantum numbers. This is further substantiated by the analytical
results for the ensemble averaged $\gamma_2(m,S)$ as discussed in Section
\ref{c2a1} ahead.  We will present the analytical formula for the ensemble
averaged spectral variances $\overline{\sigma^2(m,S)}$; i.e., for the  variance
of $\overline{\rho^{m,S}(E)}$ in Sec. \ref{c2s2p2}. 

It is important to point out that the variances  $\overline{\sigma^2(m,S)}$
propagate in a simple manner \cite{Pa-78,Qu-75} from the corresponding three
defining space variances, the variance in one-particle space
$\sigma^2\l(1,\spin\r)$ and the two  two-particle variances
$\overline{\sigma^2(2,s)} = \lambda_s^2 [d_f(\Omega,m,S)+1]$, $s=0,\;1$. Thus
the  $(m,S)$ space variances are a linear combination of these three basic
variances with the multiplying factors being simple functions of ($\Omega$, $m$,
$S$). These functions are called variance propagators as they carry the variance
information from the defining space to the final ($m,S$) spaces and it is easy
to derive formulas for them as given in Sec. \ref{c2s2p2}. For example, the
variances generated by the two-body part of the Hamiltonian for  $\lambda_0^2 =
\lambda_1^2 = \lambda^2$ are of the form $\overline{\sigma^2_{V(2)}(m,S)} = 
\lambda^2 \; P(\Omega,m,S)$.  The variance propagator $P(\Omega,m,S)$, given by
Eq. (\ref{eq.den9}) ahead,  determines much of the  behavior of the transitions
in  eigenvalue and wavefunction structure as discussed ahead. 

\subsection{Propagation formulas for ensemble averaged spectral 
variances}
\label{c2s2p2}

Let us start with the fixed-$(m,S)$ energy centroids $E_c(m,S) =  \lan H
\ran^{m,S}$ for a one plus two-body Hamiltonian $H = h(1) + V(2) =
h(1)+[\lambda_0 V^{s=0}(2) + \lambda_1 V^{s=1}(2)]$. The operator generating
$\lan H \ran^{m,S}$ will be a  polynomial, in the scalar operators $\hat{n}$
and $\hat{S}^2$, of maximum body rank 2. A two-body operator is said to be
of body rank 2, a three-body operator of body rank 3 and so on
\cite{MF-75}.  Note that  $\hat{n}$ is a one-body operator and $\hat{S}^2$
is a one plus  two-body operator. Hence only $\hat{n}$, $\hat{n}^2$ and
$\hat{S}^2$ are operators of maximum body rank 2 (for example, the operator
$\hat{n}\hat{S}^2$ is of maximum body rank 3). Then, $E_c(m,S) = a_0 +
a_1\,m +  a_2\,m^2 + a_3\,S(S+1)$. Solving for the $a_i$'s in terms of 
$E_c$ for $m\leq 2$, we obtain the well-known propagation formula for the
energy centroids \cite{Pa-78},
\be
\barr{rcl}
E_c(m,S) & = & \l[\lan h(1) \ran^{1,\spin}\r] \;m + 
\lambda_0 \lan\lan V^{s=0}(2)
\ran\ran^{2,0}\;\dis\frac{P^0(m,S)}{4\Omega(\Omega+1)} \\
& + & \lambda_1
\lan\lan V^{s=1}(2) \ran\ran^{2,1}\;\dis\frac{P^1(m,S)}
{4\Omega(\Omega-1)}\;; \nonumber
\earr \label{eq.den1a1}
\ee
\be
\barr{rcl}
P^0(m,S) & = & \l[ m(m+2) - 4S(S+1)\r]\,,\\
P^1(m,S) & = & \l[ 3m(m-2) + 4S(S+1)\r]\,, 
\earr \label{eq.den1}
\ee
\be
\barr{rcl}
\lan h(1) \ran^{1,\spin} & = & \Omega^{-1}\;\dis\sum_{i=1}^\Omega\;
\epsilon_i\,, \;\;
\lan\lan V^{s=0}(2) \ran\ran^{2,0} = \dis\sum_{i
\leq j}\;V^{s=0}_{ijij}\;,\;\;
\lan\lan V^{s=1}(2) \ran\ran^{2,1} = \dis\sum_{i <
j}\;V^{s=1}_{ijij}\;.\nonumber
\earr \label{eq.den1a2}
\ee
Trivially the ensemble average of $E_c$ from the $V(2)$ part will be zero.
However, the covariances in the energy centroids generated by the two-body
part $H(2)=V(2)$ of $H$ are non-zero,
\be
\barr{l}
\overline{\lan H(2) \ran^{m,S}\lan H(2) \ran^{m^\pr,S^\pr}} = \\
\dis\frac{\lambda^2_0}{16\Omega(\Omega+1)}P^0(m,S)\,P^0(m^\pr,S^\pr) + 
\dis\frac{\lambda^2_1}{16\Omega(\Omega-1)}P^1(m,S)\,P^1(m^\pr,S^\pr)\;.
\earr \label{eq.den2}
\ee

The spectral variances $\sigma^2(m,S)=\lan H^2 \ran^{m,S} - [\lan H
\ran^{m,S}]^2$ are generated by an operator that is a polynomial, in the scalar
operators $\hat{n}$ and $\hat{S}^2$, of maximum body rank 4. This gives
$\sigma^2(m,S)=\sum_{p=0}^4\,a_p\,m^{p} + \sum_{q=0}^2\,b_q\,m^{q} \, S(S+1) +
c_0\,[S(S+1)]^2$. The nine parameters $(a_i,b_i,c_i)$  can be written in terms
of $\epsilon_i$ and the two-body matrix elements  $V^{s=0,1}_{ijkl}$  using the
embedding algebra $U(N) \supset U(\Omega) \otimes SU(2)$. The final result is
given by Eq. (\ref{eq.vv1}) of Appendix \ref{c2a3} (this is derived using the
results in \cite{He-74}).  We have carried out the ensemble average of
$\sigma^2_H(m,S)$ over EGOE(1+2)-$\cs$ ensemble assuming that $h(1)$ is fixed
and the final result is as follows. Firstly, the ensemble averaged variance is,
\be
\overline{\sigma_H^2(m,S)} = \sigma_{h(1)}^2(m,S)+
\overline{\sigma^2_{V(2)}(m,S)}\;.
\label{eq.den21}
\ee
The propagation formula for $\sigma_{h(1)}^2$ is simple,
\be
\sigma_{h(1)}^2(m,S) = \dis\frac{(\Omega+2)m(\Omega-m/2)-2\Omega
S(S+1)}{(\Omega-1)(\Omega+1)} \; \sigma_{h(1)}^2\l(1,\spin\r)\;.
\label{eq.den3}
\ee
The two parts  $V^{s=0}(2)$ and $V^{s=1}(2)$ of $V(2)$ will have a  scalar
part, an effective one-body part and an irreducible two-body part denoted by
$V^{s,\nu}(2)$, with $\nu=0$, 1, and 2, respectively with respect to $U(N)
\supset U(\Omega) \otimes SU(2)$ algebra. The two $\nu=0$ parts generate the
centroids and they can be identified from Eq. (\ref{eq.den1}). As the $\nu$
decomposition is an orthogonal  decomposition, we have 
\be
\overline{\sigma^2_{V(2)}(m,S)} = \dis\sum_{s=0,1} \lambda_s^2
\dis\sum_{\nu=1,2}
\overline{\lan [V^{s,\nu}(2)]^2\ran^{m,S}} \;.
\label{eq.den4}
\ee
As seen from Eq. (\ref{eq.vv1}), for evaluating $\overline{ \lan
[V^{s,\nu=1}(2)]^2\ran^{m,S}}$ we need
$\sum_{i,j}\overline{\lambda_{i,j}^2(s)}$ where the $\lambda(s)$'s are the 
so called induced one-particle matrix elements generated by $V^s$,
\be
\barr{rcl}
\lambda_{i,i}(s) & = &  \dis\sum_j\;V_{ijij}^s\;(1+\delta_{ij})
\;-\;(\Omega)^{-1} \;\dis\sum_{k,l}\;V^s_{klkl}\;(1+\delta_{kl})\;, \\ \\
\lambda_{i,j}(s) & = & \dis\sum_k\;\dis\sqrt{(1+\delta_{ki})
(1+\delta_{kj})}\,V^s_{kikj}\;\;\;\mbox{for}\;\;\;i \neq j\;. 
\earr \label{eq.den5}
\ee
Similarly for evaluating $\overline{\lan [V^{s,\nu=2}(2)]^2\ran^{m,S}}$,  we
need $\overline{\lan [V^{s,\nu=2}(2)]^2\ran^{2,s}}$. Firstly,  applying the
fact that the $V^s$ matrix elements are independent Gaussian random
variables with zero center and variance  unity (except for the diagonal
matrix elements it is 2) and simplifying using Eq. (\ref{eq.den5}),  we
obtain
\be
\barr{rcl}
\dis\sum_{i,j}\overline{\lambda_{i,j}^2(0)} & = & (\Omega-1)
(\Omega+2)^2\;, \\ \\
\dis\sum_{i,j}\overline{\lambda_{i,j}^2(1)} & = & (\Omega-1)
(\Omega-2)(\Omega+2)\;.
\earr \label{eq.den6}
\ee
Also, $\overline{\lan [V^s(2)]^2 \ran^{2,s}} = [d_f(\Omega,2,s)+1]$.  This
along with Eqs. (\ref{eq.vv1}), (\ref{eq.den1}) and
(\ref{eq.den6}) will give $\overline{\lan [V^{s,\nu=2}(2)]^2\ran^{2,s}}$,
\be
\barr{rcl}
\overline{\lan [V^{s=0,\nu=2}(2)]^2\ran^{2,0}} & = &
\dis\frac{1}{2}(\Omega-1)(\Omega+2)\;, \\ \\
\overline{\lan [V^{s=1,\nu=2}(2)]^2\ran^{2,1}} & = & \dis\frac{(\Omega-3) 
(\Omega^2+\Omega+2)} {2(\Omega-1)}\;.
\earr \label{eq.den7}
\ee
Substituting the results in Eqs. (\ref{eq.den1}), (\ref{eq.den4}), 
(\ref{eq.den6}), and (\ref{eq.den7}) in  Eq. (\ref{eq.vv1}) gives the
final result,
\be
\barr{rcl}
\overline{\sigma^2_{V(2)}(m,S)} & = & 
\dis\frac{\lambda_0^2}{\Omega(\Omega+1)/2}
\l[\dis\frac{\Omega+2}{\Omega+1} Q^1(\{2\}:m,S) +
\dis\frac{\Omega^2+3\Omega+2}{\Omega^2+3\Omega}\,Q^2(\{2\}:m,S)\r] \\ 
& + & \dis\frac{\lambda_1^2}{\Omega(\Omega-1)/2}
\l[\dis\frac{\Omega+2}{\Omega+1} Q^1(\{1^2\}:m,S) +
\dis\frac{\Omega^2+\Omega+2}{\Omega^2+\Omega}\,Q^2(\{1^2\}:m,S)\r]\;;\nonumber
\earr \label{eq.den8a1}
\ee
\be
\barr{rcl} 
Q^1(\{2\}:m,S) & = & \l[(\Omega+1) P^0(m,S)/16\r]\,\l[m^x(m+2)/2 + 
\lan S^2\ran\r]\,,\\ 
Q^2(\{2\}:m,S) & = & \l[\Omega (\Omega+3) P^0(m,S)/32\r]\,\l[m^x(m^x+1) -
\lan S^2\ran\r]\,,\nonumber
\earr \label{eq.den8a2}
\ee
\be
\barr{rcl}
Q^1(\{1^2\}:m,S) & = & \dis\frac{(\Omega-1)}{16(\Omega-2)}
\l[(\Omega+2)\,P^1(m,S)
\,P^2(m,S) \r. \\ 
& + & \l. 8\Omega (m-1)(\Omega-2m+4) \lan S^2\ran\r]\;, 
\earr \label{eq.den8}
\ee
\be
\barr{rcl} 
Q^2(\{1^2\}:m,S)  & = & \dis\frac{\Omega}{8(\Omega-2)} 
\l.[(3\Omega^2 -7\Omega +6)
(\lan S^2\ran)^2 \r. \\ 
& + & 3m(m-2)m^x(m^x-1)(\Omega+1)(\Omega+2)/4 \\ 
& + & \l. \lan S^2\ran \l\{-m m^x (5\Omega-3)(\Omega+2)+
\Omega(\Omega-1)(\Omega+1)(\Omega+6)\r\}\r]\;,\\ 
P^2(m,S) & = & 3 m^x(m-2)/2 - \lan S^2\ran\;,\;\;\;
m^x=\l(\Omega-\dis\frac{m}{2}\r)\;. \nonumber
\earr \label{eq.den8a3}
\ee
Note that the $\nu=1$ terms (they correspond to the $Q^1$'s) are 
$1/\Omega^2$ times smaller as compared to the $\nu=2$ terms (they correspond
to the $Q^2$'s). Therefore in the dilute limit defined by $\Omega
\rightarrow \infty$, $m \rightarrow \infty$, $m/\Omega \rightarrow 0$ and $m
>> S$, the $V^{s=0,1:\nu=2}$ parts determine the variances
$\sigma^2_H(m,S)$. As a result, formula for the ensemble averaged variances
given in \cite{Kk-02} is same as the sum of the two $\nu=2$  terms in Eq.
(\ref{eq.den8}). 

\begin{figure}
\centering
\includegraphics[width=5.5in,height=4in]{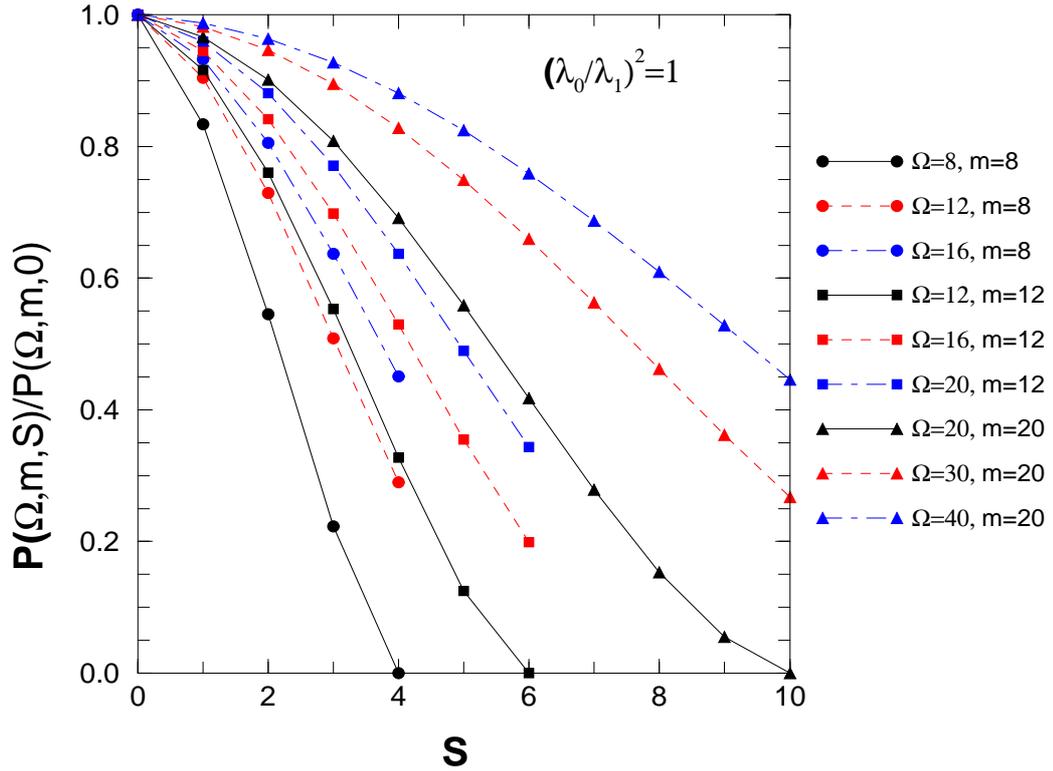} 
\caption{Variance propagator $P(\Omega,m,S)$ vs $S$ for 
different values of $\Omega$ and $m$. Eq. (\ref{eq.den9}) gives the 
formula for $P(\Omega,m,S)$.}
\label{var}
\end{figure}

In most of the numerical examples discussed in the remaining part of the present
chapter (except in Sec. \ref{c2s8}) we employ $\lambda_0 = \lambda_1 = \lambda$ 
and for this $\overline{\sigma^2_{V(2)}(m,S)}$ takes the form
\be
\barr{l}
\overline{\sigma^2_{V(2)}(m,S)}  \stackrel
{\lambda_0 = \lambda_1 = \lambda}{\longrightarrow} 
\lambda^2 \;P(\Omega,m,S)\;;\\
P(\Omega,m,S)  =  \dis\frac{1}{\Omega(\Omega+1)/2}
\l[\dis\frac{\Omega+2}{\Omega+1} Q^1(\{2\}:m,S) +
\dis\frac{\Omega^2+3\Omega+2}{\Omega^2+3\Omega}\,Q^2(\{2\}:m,S)\r] \\ \\
+  \dis\frac{1}{\Omega(\Omega-1)/2}
\l[\dis\frac{\Omega+2}{\Omega+1} Q^1(\{1^2\}:m,S) +
\dis\frac{\Omega^2+\Omega+2}{\Omega^2+\Omega}\,Q^2(\{1^2\}:m,S)\r]\;.
\earr \label{eq.den9}
\ee
Note that we are showing $\Omega$ explicitly in the formula for the variance
propagator $P(\Omega,m,S)$ as $\Omega$  plays an important role in
determining the transition markers. Figure \ref{var} shows a plot of
$P(\Omega,m,S)/P(\Omega,m,0)$ vs $S$ for various values of $m$ and
$\Omega$.  As seen from Fig. \ref{var}, $P(\Omega,m,S)$ decreases with spin
and this plays an important role in understanding the properties
of EGOE(1+2)-$\cs$ as will be seen in the following sections. Now, we will
discuss the results for transition markers generated by EGOE(1+2)-$\cs$.

\section{Poisson (or close to Poisson) to GOE Transition in Level
Fluctuations}
\label{c2s4}

Fluctuations in the eigenvalues of a fixed-($m,S$) spectrum derive from the two
and higher point correlation functions. For example, the two-point  function is
given by Eq. (\ref{eq.rho}) with $\Gamma=S$ and $\Gamma^\pr=S^\pr$. 
The commonly used Dyson-Mehta $\Delta_3$ statistic
is an exact two-point  measure while variance $\sigma^2(0)$ of the nearest
neighbor spacing distribution (NNSD) is essentially a two-point measure
\cite{Br-81}.  Note that, due to a convention as stated in the
footnote 14 of  \cite{Br-81}, the variance of the NNSD is $\sigma^2(0)$, the
second nearest $\sigma^2(1)$ etc. In all the discussion in this Sec. and all
other remaining Secs. \ref{c2s5}-\ref{c2s7} (except Sec. \ref{c2s8}),  we use
$\lambda_0=\lambda_1=\lambda$, i.e., we employ EGOE(1+2)-$\cs$ Hamiltonian,
\be
H_\lambda = h(1)+\lambda[ V^{s=0}(2)+V^{s=1}(2)]\;. 
\label{eq.ham8}
\ee
The NNSD and $\Delta_3$ statistics show Poisson character in general
\cite{Re-06}  for very small values of $\lambda$ due to the presence of many
good quantum numbers defined by  $h(1)$.  As the value of $\lambda$
increases, there is delocalization in the Fock space, i.e., the eigenstates
spread over all the basis states leading  to complete mixing of the basis
states. Hence, one expects GOE behavior for large $\lambda$ values. 

\begin{figure}
\begin{center}
{\bf\large $\;\;\;\;\;\;$EGOE(1+2)-$\cs$}
\end{center}
    \centering
    \subfigure
    {
        \includegraphics[width=2.75in,height=3.5in]{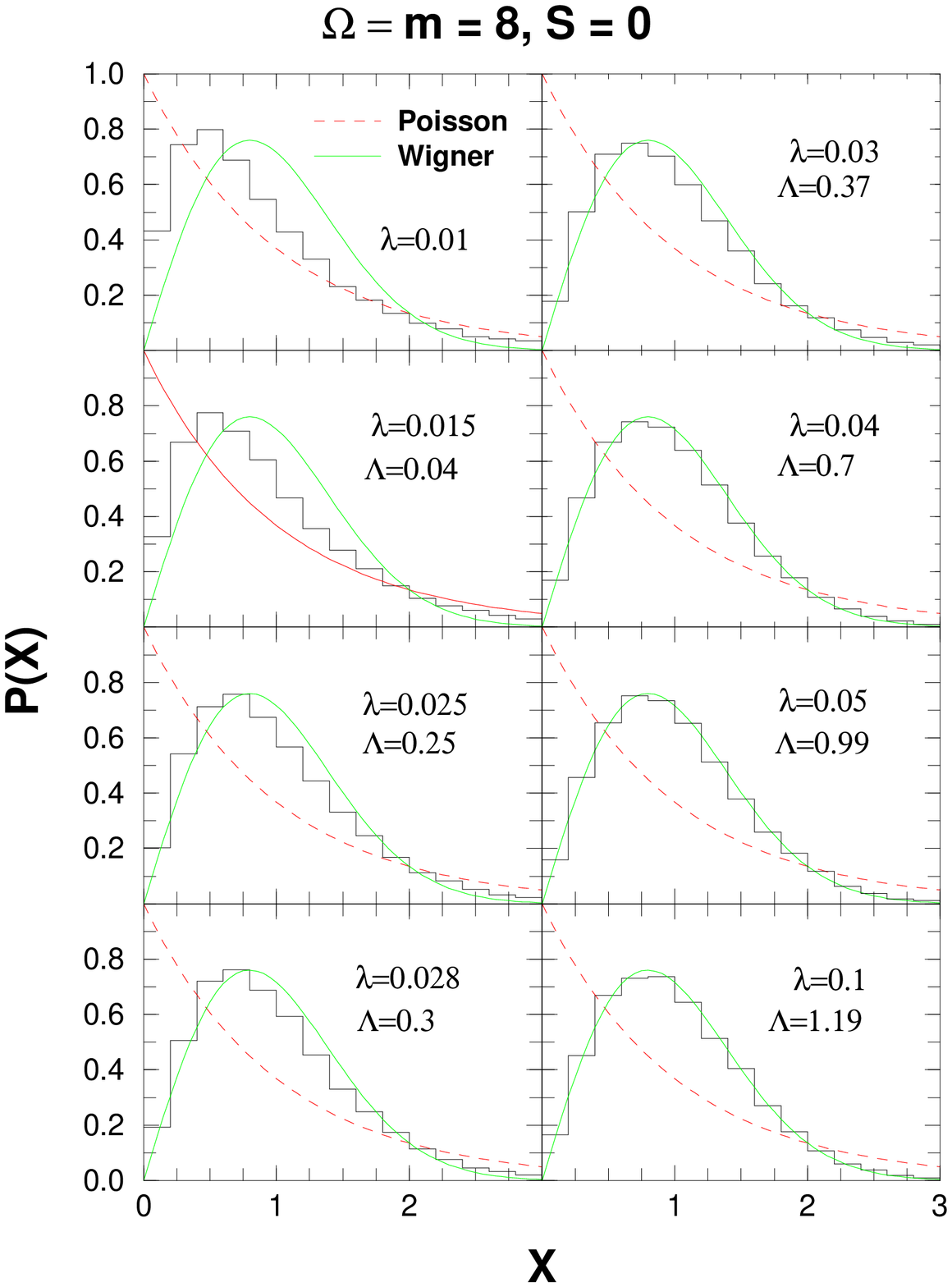}
        \label{nnsds0}
    }
    \subfigure
    {
        \includegraphics[width=2.75in,height=3.5in]{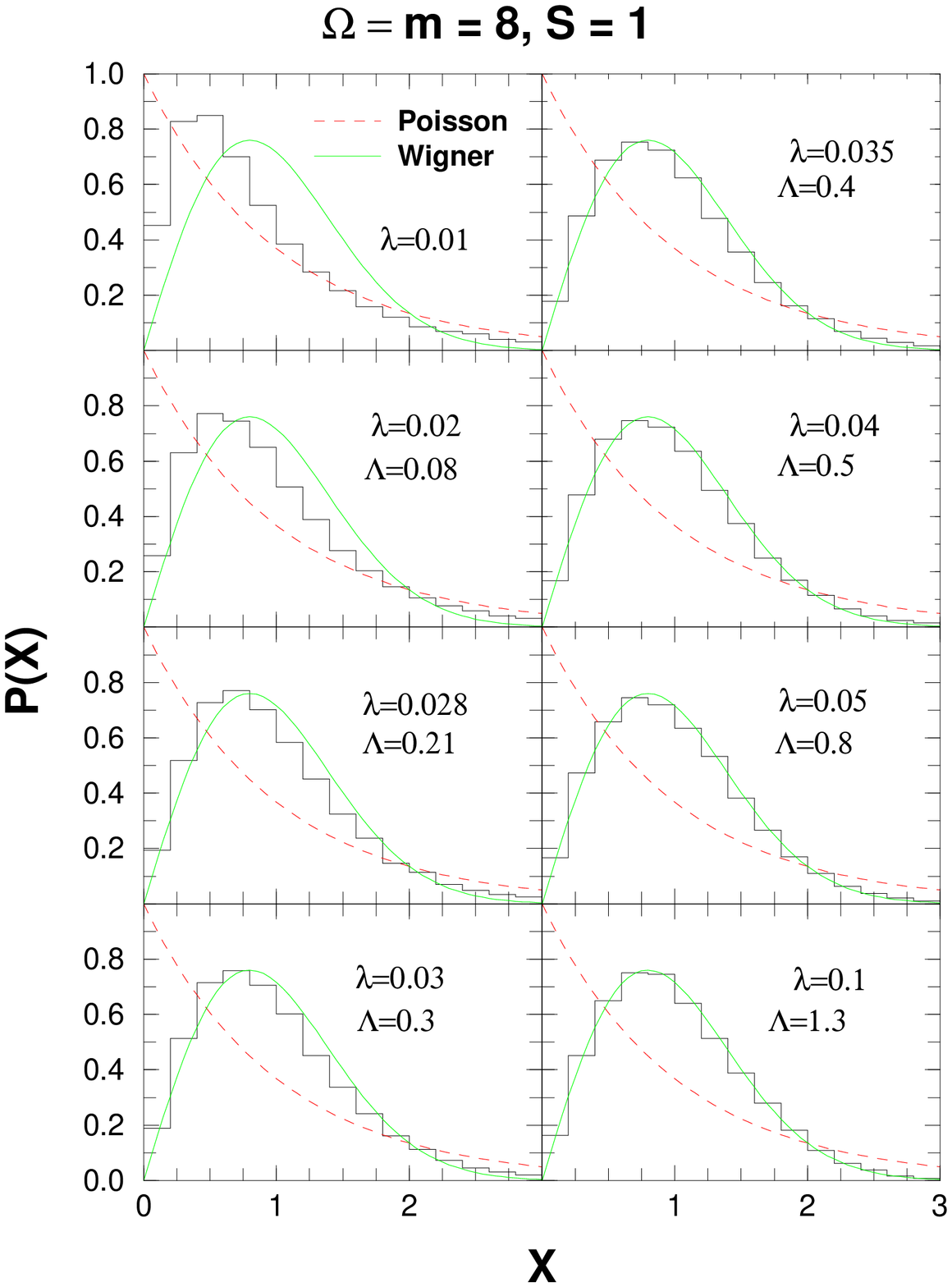}
        \label{nnsds1}
    }
    \\
    \subfigure
    {
        \includegraphics[width=2.75in,height=3.5in]{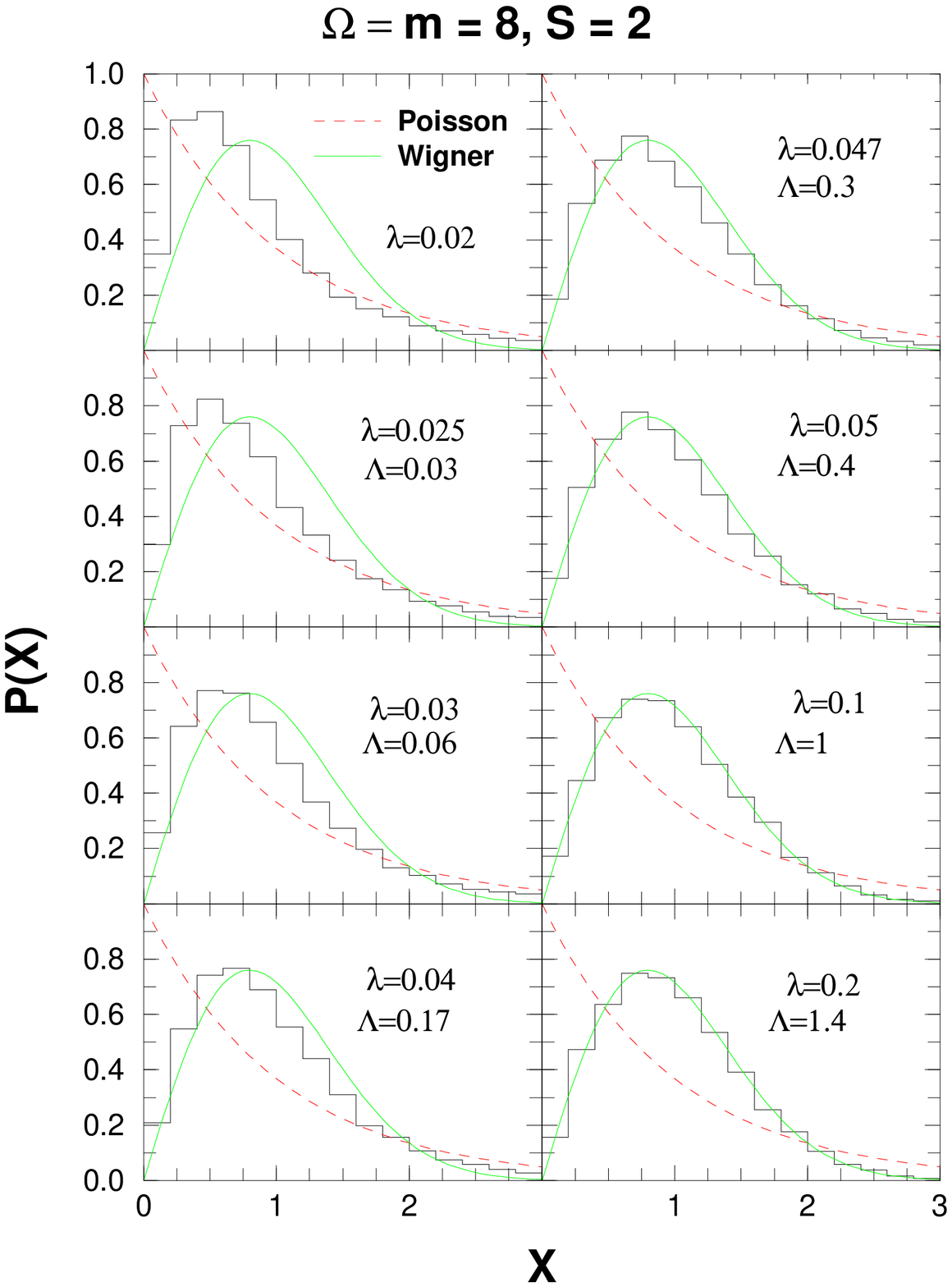}
        \label{nnsds2}
    }
    \caption{NNSD for a 20 member EGOE(1+2)-$\cs$ ensemble with $\Omega=m=8$
    and spins $S=0$, $1$, and $2$, respectively. Calculated NNSD are compared
    to the Poisson and Wigner (GOE) forms. Values of the interaction
    strength $\lambda$ and the transition parameter $\Lambda$ are given in 
    the figure. The chaos marker $\lambda_c$ corresponds to $\Lambda=0.3$. 
    Bin-size for the histograms is $0.2$. As discussed in the text, for very
    small values of $\lambda$, the NNSD, for the sp spectrum employed 
    in the calculations, is not strictly a Poisson. Therefore, the
    $\Lambda$ values are not given for $\lambda=0.01$ for spins $S=0$ and
    $1$ and for $\lambda=0.02$ for spin $S=2$.}
    \label{nnsd}
\end{figure}

\begin{figure}
\begin{center}
{\bf\large EGOE(1+2)-$\cs$}
\end{center}
\centering
\begin{tabular}{cc}
\includegraphics[width=2.75in,height=3in]{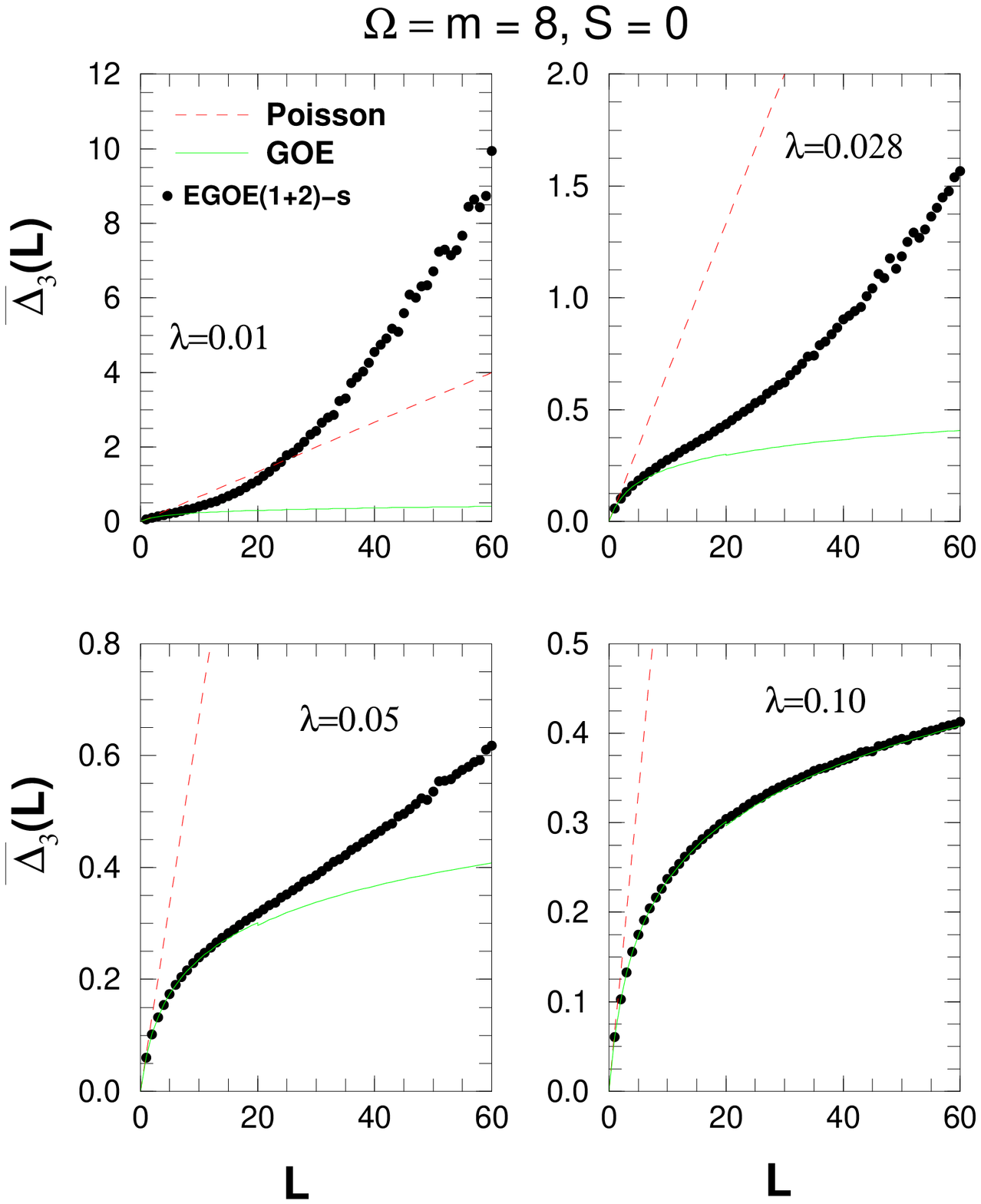} &
\includegraphics[width=2.75in,height=3in]{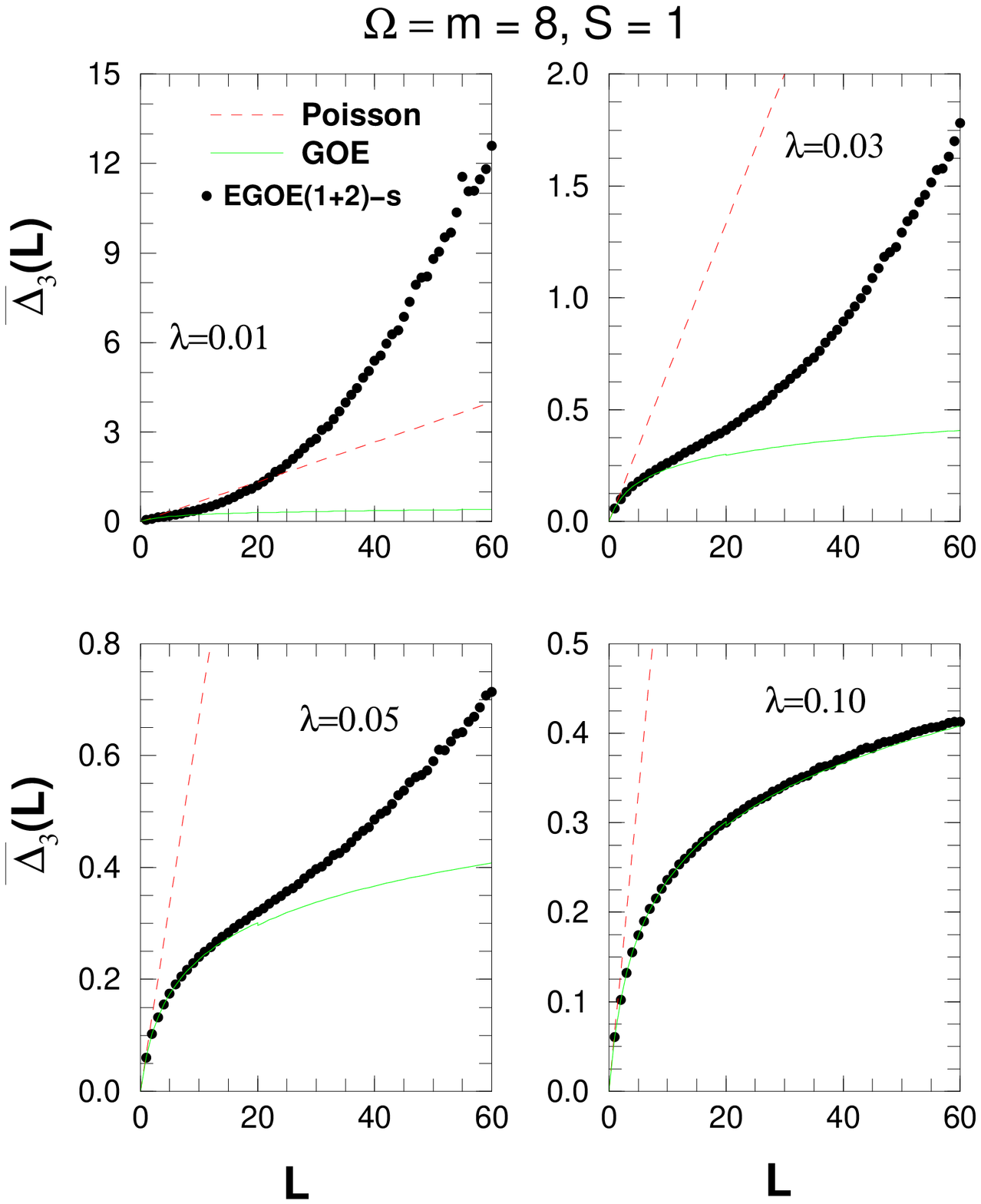}
\end{tabular}
\caption{$\overline{\Delta_3}(L)$ vs $L$ for a 20 member EGOE(1+2)-$\cs$
ensemble with $\Omega=m=8$ and spins $S=0$ and $1$. Calculated results are
compared with the Poisson and GOE forms.}
\label{delta3}
\end{figure}

For a 20 member EGOE(1+2)-$\cs$ ensemble with $\Omega=m=8$ and spins
$S=0,\;1$, and $2$, we have constructed NNSD and $\Delta_3$ for various
$\lambda$ values changing from 0.01 to 0.3. In the calculations: (i) the
spectrum for each member of the ensemble is unfolded using ED corrected
Gaussian for the eigenvalue density so that the average spacing is unity; 
(ii) we drop $5$\% of the levels from the two spectrum ends; (iii) with this
we have constructed the ensemble averaged NNSD histograms and calculated
their variances $\sigma^2(0)$; (iv) for the $\Delta_3$ statistic, overlap
interval of 0.5 (for the unfolded spectrum) is used and
$\overline{\Delta_3}(L)$ for $L \leq 60$   are calculated following Ref.
\cite{Bo-83}; $L$ is the energy interval,  measured in units of average
level spacing, over which $\Delta_3$ is calculated. Results for NNSD and
$\Delta_3$ statistic are shown in Figs. \ref{nnsd} and \ref{delta3},
respectively. As mentioned in Sec. \ref{c2s2p1}, 
in our calculation the mean-field
Hamiltonian is of a special form defined by the sp energies
$\epsilon_i=i+1/i$. For this Hamiltonian, it is easy to see that in the
dilute regime, the majority of many-body eigenvalues approach a perturbed
picket-fence spectrum. Away from the dilute limit, the spectrum is not
picket-fence and deviates from Poisson as can be seen from Figs. \ref{nnsd}
and \ref{delta3}. However, if we had used sp energies drawn from the center
of a GOE or from the eigenvalues of an irregular system, the fluctuations
will be generically Poisson \cite{Re-06}. Therefore we call the transition
seen in Figs. \ref{nnsd} and \ref{delta3},  Poisson to GOE transition and it
should be kept in mind that, the sp spectrum we have  chosen gives level
fluctuations that are close to Poisson but not strictly Poisson for
$\lambda=0$. For further discussion we focus on the  NNSD and its variance
$\sigma^2(0)$.

As we increase $\lambda$, NNSD changes rapidly from a form close to Poisson
to a  form close to  that of GOE (Wigner distribution) as seen from Fig.
\ref{nnsd}. However, the complete convergence to GOE form is very slow.
Therefore, although the transition to GOE in level fluctuations is not a
phase transition, we can still define a transition point $\lambda =
\lambda_c$ where Poisson-like fluctuations start changing to GOE character 
and we  need a criterion to determine $\lambda_c$. For this purpose we
employ $\sigma^2(0)$ given by a simple  2$\times$2 random matrix model for
Poisson to GOE transition \cite{KS-99} as used in some of the earlier
studies \cite{Ch-03}. In this model, in terms of a transition parameter
$\Lambda$ ($\Lambda$ is mean squared admixing GOE matrix element divided by
the square of the mean spacing $D_0$ of the Poisson spectrum), $\sigma^2_{P
\to  GOE}(0:\Lambda) = (8\Lambda + 2) / [\pi (\Psi(-0.5,0,2\Lambda))]^2 -
1$. Here $\Psi$ is the Kummer function. It can be argued that the transition
to GOE is nearly complete for $\Lambda \sim 0.3$ which corresponds to NNSD
variance $\sigma^2(0) = 0.37$. A plot of $\sigma^2_{P \to  GOE}(0:\Lambda)$
vs $\Lambda$ \cite{KS-99} shows that the variance decreases fast from
Poisson value  $\sigma^2(0)=1$ up to $\Lambda \sim 0.37$ and then converges
slowly to the GOE value $\sigma^2(0)=0.27$. For the NNSD that are
constructed for various EGOE(1+2)-$\cs$ examples, the calculated
$\sigma^2(0)$ are used to deduce, from the 2$\times$2 matrix formula, the
values of $\Lambda$.  In Fig. \ref{nnsd}, the values of the $\Lambda$
parameter are given for different $\lambda$ values and it is seen that the
transition point $\lambda_c$ is $0.028$, $0.030$, and $0.047$ for $S=0,\;1$,
and $2$, respectively. In Fig. \ref{delta3}, we show
$\overline{\Delta_3}(L)$  vs $L$ for some values of $\lambda$ and clearly
there is a transition to GOE statistics. It should be stressed that one
expects the $\lambda_c$ needed to approach GOE statistics for
$\overline{\Delta_3}(L)$ to scale as $L^{1/2}$ \cite{Gu-89}  although the
scaling of $\lambda_c$ with other parameters $(m,S,\Omega)$ will be same for
any given $L$. In the present example, up to $L = 20$, the $\lambda_c$
deduced from NNSD could be considered as the transition point for
$\overline{\Delta_3}(L)$. However the $L$ dependence of $\lambda_c$  is not
probed further in the present chapter.  

For a qualitative understanding of the variation of $\lambda_c$ with spin
$S$, it is plausible to use the same arguments used for spinless fermion
systems and they are based on perturbation theory \cite{Ja-97}. As $\lambda$
is increased from zero, the $m$-particle states generated by $h(1)$ will be
mixed by $V(2)$ and in lowest-order perturbation the first stage of mixing
will be between states that are directly coupled by the two-body
interaction. Poisson to GOE transition occurs when $\lambda$ is of the order
of the spacing $\Delta_c$ between the $m$-particle states that are directly
coupled by the two-body interaction. Given the two-particle spectrum span to
be $B_2$ and the number of fixed-($m,S$) states directly coupled by the
two-body interaction to be $K(\Omega,m,S)$, we have $\Delta_c(\Omega,m,S)
\propto B_2/K(\Omega,m,S)$ and therefore, $\lambda_c \propto
B_2/K(\Omega,m,S)$. Using the $h(1)$ spectrum, it is easy to see that $B_2
\propto \Omega$. Following the arguments in \cite{Ja-01} (see also
\cite{Kk-02}), the spectral variances generated by $V(2)$ can be written as
$\overline{\sigma^2_{V(2)}(m,S)} \approx  \lambda^2 K(\Omega,m,S)$ and
applying Eq. (\ref{eq.den9}) gives   $K(\Omega,m,S) \approx P(\Omega,m,S)$.
With this, we have  
\be 
\lambda_c(S) \propto
\dis\frac{\Omega}{P(\Omega,m,S)}\;. 
\label{eq.sd2} 
\ee 
From the results in Fig. \ref{var} for $P(\Omega,m,S)$, 
it is clear that $\lambda_c$ should 
increase with spin $S$. For $\Omega=m=8$, Eq. (\ref{eq.sd2}) and the formula
for $P(\Omega,m,S)$ gives $P(8,8,S=1)/P(8,8,S=0)=0.834$ and 
$P(8,8,S=2)/P(8,8,S=0)=0.55$. These and the result $\lambda_c(S=0)=0.028$
from Fig. \ref{nnsd} will give $\lambda_c(S=1)=0.034$ and
$\lambda_c(S=2)=0.05$. These predictions are close to the numerical results
shown in Fig. \ref{nnsd}. Therefore Eq. (\ref{eq.sd2}) gives a good
qualitative understanding of the $\lambda_c(S)$ variation with $S$. In the
dilute limit (sometimes also called asymptotic limit), as defined just 
after Eq. (\ref{eq.den8}), it is easily seen that $P(\Omega,m,S) \to
m^2\Omega^2$ and hence $\lambda_c \to 1/m^2\Omega$. Thus we recover the
result known  \cite{Ja-97} for spinless fermion systems as a limiting case.

\section{Breit-Wigner to Gaussian Transition in Strength Functions}
\label{c2s5}

Wavefunction structure is understood usually in terms of strength functions
[$F_k(E)$] and information entropy [$S^{info}(E)$]. Both of these are basis
dependent. In our (also by all others \cite{Ja-01,Ka-00,Al-06,Pa-02,Ko-06}) 
construction of the $H$ matrices, the basis states chosen are eigenstates of
both $\hat{h}(1)$ and $\hat{S^2}$  operators (we drop $M_S^{min}$ everywhere
although all the states have $M_S=M_S^{min}$). Given the mean field $h(1)$
basis states (denoted by $\l| k\ran$) expanded in the $H$ eigenvalue ($E$) 
basis,
\be
\l|k,S,M_S\ran=\dis\sum_E C_{k,S}^{E,S} \l|E,S,M_S\ran \;, 
\label{eq.wf1}
\ee
the strength functions $F_{k,S}(E,S)$ and information entropy
$S^{info}(E,S)$ are defined by,
\be
\barr{rcl}
F_{k,S}(E,S) & = & \dis\sum_{E^\pr}\l|C_{k,S}^{E^\pr,S}\r|^2
\;\delta(E-E^\pr)
= \l|\cac_{k,S}^{E,S}\r|^2\;d_f(\Omega,m,S)\;\rho^{m,S}(E)\;,\\
S^{info}(E,S) & = & - \dis\frac{1}{d_f(\Omega,m,S)\;\rho^{m,S}(E)}
\dis\sum_{E^\pr}\dis\sum_{k} \l|C_{k,S}^{E^\pr,S}\r|^2 \ln 
\l|C_{k,S}^{E^\pr,S}\r|^2\;\delta(E-E^\pr)\;,
\earr \label{eq.wf2}
\ee
where $\l|\cac_{k,S}^{E,S}\r|^2$ denotes the average of $|C_{k,S}^{E,S}|^2$
over the eigenstates with the same energy $E$. The strength functions give
the spreading of the basis states over the eigenstates. For $\lambda=0$, the
strength functions will be $\delta$-functions at the $h(1)$ eigenvalues. As
$\lambda$ increases from zero, the strength functions first change from 
$\delta$-function form to BW form at $\lambda = \lambda_\delta$  where
$\lambda_\delta$ is very small; see Eq. (\ref{eq.wf5}) ahead. The BW form,
with $\Gamma_{BW}$ denoting the spreading width, is defined by,
\be
F_{k,BW}(E) = \dis\frac{1}{2\pi}\;\dis\frac{\Gamma_{BW}}{(E-\xi_k)^2+
{\Gamma_{BW}^2}/4}\;.
\label{eq.wf3}
\ee
The energies $\xi_k=\lan \phi_k \mid H \mid \phi_k \ran$ are the diagonal
matrix  elements of $H$ and they are the basis state energies.  Information
entropy $S^{info}$ is a measure of complexity or chaos in wavefunctions and the
GOE value for $\exp[S^{info}(E,S)]$ is $0.48\,d_f(\Omega,m,S)$ independent of
$E$.  Our purpose is to investigate the change in $F_{k,S}(E,S)$ and
$S^{info}(E,S)$ as we change $\lambda$. In the present Sec. we consider strength
functions and in the next Sec. information entropy.

\begin{figure}
\begin{center}
{\bf\large EGOE(1+2)-$\cs$}
\end{center}
\centering
\begin{tabular}{cc}
\includegraphics[width=2.75in,height=3.5in]{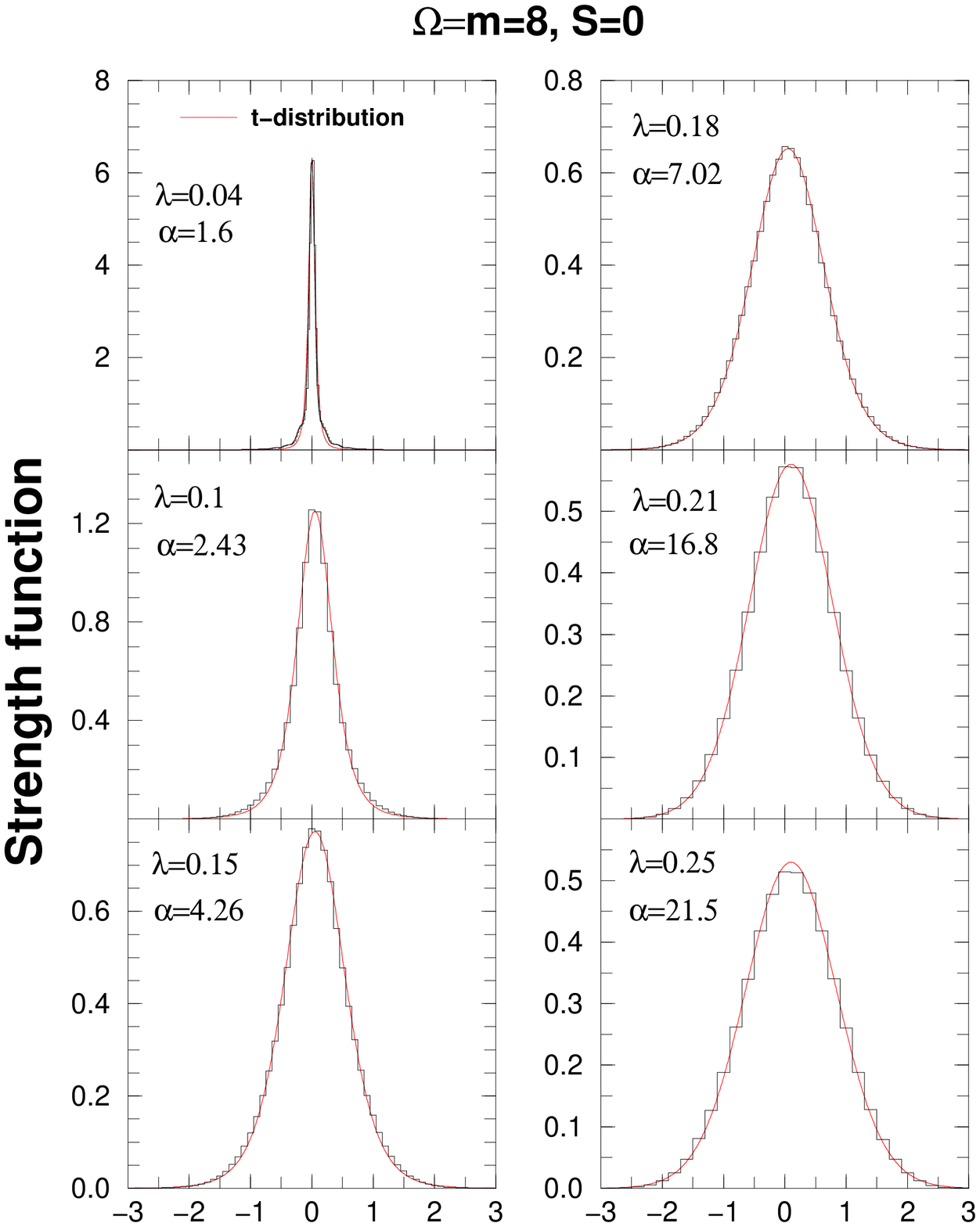} &
\includegraphics[width=2.75in,height=3.5in]{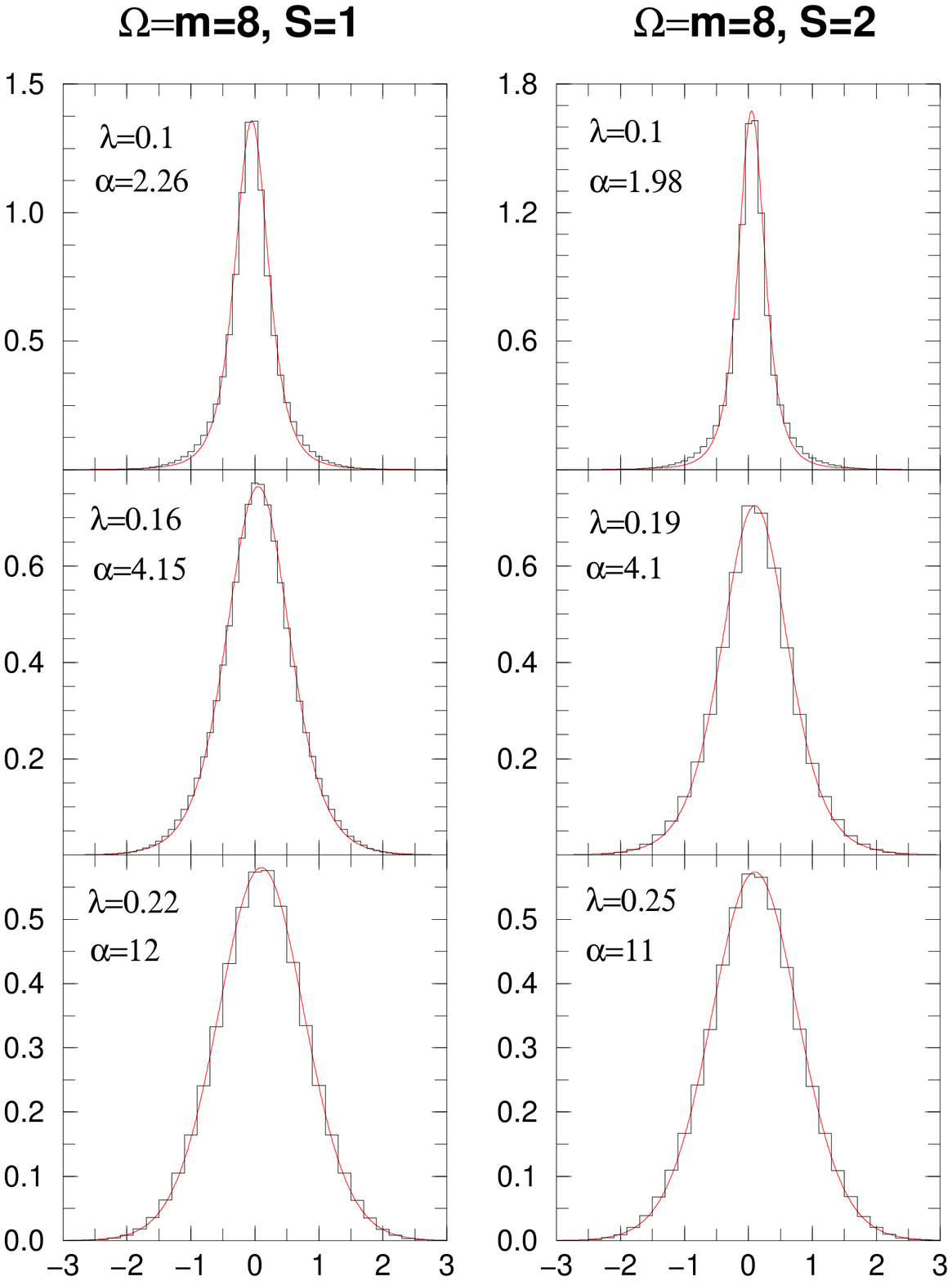}
\end{tabular}
\begin{center}
{\bf\large $\;\;\;\;\;\;[E-E_c(m,S)]/\sigma(m,S)$}
\end{center}
\caption{Strength functions as a function of $\lambda$ for a
20 member  EGOE(1+2)-$\cs$ ensemble. Calculations (histograms) are for a
$\Omega=m=8$ system  with spins $S=0$, $1$ and $2$. Note that the widths
$\sigma_{F_k}(m,S)$ of  the strength functions are different from the
spectral widths $\sigma(m,S)$. Continuous curves in the figures correspond 
to the $t$-distribution given  by Eq. (\ref{eq.wf4}). See text for details.}
\label{fke1}
\end{figure}

Figure \ref{fke1} shows strength functions as a function of $\lambda$ for 8
particles in 8 sp levels ($\Omega=m=8$) with spins $S=0$, $1$, and $2$. The
centroids ($\epsilon$) of the $\xi_k$ spectra are same as that of the
eigenvalue ($E$) spectra but their widths are different.  In the
calculations, $E$ and $\xi_k$ are zero centered for each member and scaled
by the width of the  eigenvalue spectrum. The new energies are called
$\widehat{E}$ and $\widehat{E}_k$, respectively.  For each member
$|C_{k,S}^{E,S}|^2$ are summed  over the basis states in the energy window
$\widehat{E}_k \pm \Delta_k$ and then the ensemble averaged
$F_k(\widehat{E},S)$ vs $\widehat{E}$ are constructed as histograms. We have
chosen $\Delta_k=0.025$ for $\lambda < 0.1$ and beyond this $\Delta_k=0.1$.
In the plots $\int F_k(\widehat{E},S) d\widehat{E}=1$. Clearly, strength
functions  exhibit transition from BW to Gaussian form. To describe this
transition,  a simple linear interpolation of BW and Gaussian forms, with
three parameters,  as employed in \cite{Ch-04} could be used. However, an
alternative form is given by the one-parameter $t$-distribution well known
in statistics and it is used in \cite{An-04}. In the following we employ the
$t$-distribution.

Student's $t$-distribution, with a shape parameter  $\alpha$, such that
$\alpha=1$ gives BW and $\alpha \rightarrow \infty$ gives Gaussian, is a
good interpolating function for BW to Gaussian transition and it is given
by, 
\be
F_k^{stud}(E,S:\alpha,\beta)\,dE = 
\dis\frac{(\alpha\beta)^{\alpha-1/2}\Gamma(\alpha)}
{\sqrt{\pi}\Gamma(\alpha-1/2)}\;\dis\frac{dE}{\l[(E-E_k)^2+\alpha\beta\r]
^\alpha}\;.
\label{eq.wf4}
\ee
Note that the $\Gamma$ function in Eq. (\ref{eq.wf4}) shall not be confused with
the $\Gamma$ notation used to denote subspaces; see Eq. (\ref{eq.eden}).
The parameter $\beta$ defines the energy spread and hence, it is determined
by the variance of the strength function $\sigma_{F_k}^2$, i.e.,
$\beta=\sigma_{F_k}^2(2\alpha-3)/\alpha$ for $\alpha >1.5$. For $\alpha \leq
1.5$, the spreading width determines the parameter $\beta$. Numerical
results for the strength functions are compared with the best-fit
$F_k^{stud}(E,S)$ and they are shown as continuous curves in Fig. \ref{fke1}
along with the values of the parameter $\alpha$. Although only the results
for $S=0,\;1,\;2$ and $\widehat{E}_k=0$ are shown in the figures, we have
also performed calculations for $S=0$ with  $\widehat{E}_k=\pm 0.5$. As seen
from the figures, the fits are excellent over a wide range of $\lambda$
values.  The parameter $\alpha$ rises slowly up to $\lambda_F$, then it
increases sharply (for $\alpha >16$ the curves are indistinguishable from
Gaussian). Following \cite{An-04}, the criterion $\alpha \sim 4$ defines the
transition point $\lambda_F$. From the results in Fig. \ref{fke1} it is seen
that the transition point $\lambda_F$ is $0.15$ and  $0.16$  for $S=0$ and
$1$, respectively. In addition, $\lambda_F=0.19$ for $S=2$ (for
$\lambda=0.075$ and $0.15$, $\alpha=1.69$ and $2.73$, respectively). 
Similarly for $S=0$ and $\widehat{E}_k=\pm 0.5$, the $\lambda_F$ value is
$0.16$. Thus  $\lambda_F$ increases slowly with $\widehat{E}_k$. 

For a qualitative understanding of the variation of $\lambda_F$ with spin
$S$, we consider the spreading width ${\cal \Gamma}(S)$ and the inverse
participation ratio (IPR)  $\zeta(S)$. First, Fermi golden rule gives 
$\Gamma_{BW}(S) = 2\pi\lambda^2/\overline{D(S)}$ with
$\overline{D(S)}=\Delta_c(\Omega,m,S)$ as  established in \cite{Ge-97}.
Therefore, using Eq. (\ref{eq.sd2})  gives $\Gamma_{BW}(S) 
\propto 2\pi\lambda^2
P(\Omega,m,S)/\Omega$. Similarly,  $\zeta(S) \sim \Gamma_{BW}(S)/\Delta_m(S)$
with $\Delta_m(S)$ being the average  spacing of the $m$-particle fixed-$S$
spectrum. The total spectrum span considering only $h(1)$ is $B_m \propto m
\Omega$ and therefore  $\Delta_m(S) \propto m\Omega/d_f(\Omega,m,S)$.  In the BW
domain, $\Gamma_{BW}(S)$ and $\zeta(S)$ should be such that 
(i)  $\Gamma_{BW}(S)< f_0
B_m$  and (ii) $\zeta(S) >>1$ where $f_0<1$. Condition (i) gives, $\lambda^2
< C_0 m\Omega^2/P(\Omega,m,S)$ and condition (ii) gives,  $\lambda^2 >> B_0
m\Omega^2/P(\Omega,m,S)\,d_f(\Omega,m,S)$. Note that the constants  $C_0$ and 
$B_0$
are positive. Therefore,
\be
\barr{l}
\dis\sqrt{\dis\frac{B_0\; m\Omega^2}{P(\Omega,m,S)\,d_f(\Omega,m,S)}} << 
\lambda <
\dis\sqrt{\dis\frac{C_0 \;m\Omega^2}{P(\Omega,m,S)}} 
\Rightarrow \;\;
\lambda_F(S) \propto \dis\sqrt{\dis\frac{m\Omega^2}{P(\Omega,m,S)}}\;.
\earr \label{eq.wf5}
\ee
This equation shows that just as $\lambda_c$, the marker $\lambda_F$ is
essentially determined by the variance propagator $P(\Omega,m,S)$. Also as
$\lambda$ increases from zero, the BW form sets in fast as $d_f(\Omega,m,S)$ is
usually  very large. From the results in Fig. \ref{var}, it is clear that
$\lambda_F$ should  increase with $S$. This prediction is close to the
numerical results shown in Fig. \ref{fke1}.  Equation (\ref{eq.wf5}) with the
result $\lambda_F(S=0)=0.15$ gives  $\lambda_F(S=1)=0.16$ and
$\lambda_F(S=2)=0.2$. Therefore Eq. (\ref{eq.wf5})  gives a good qualitative
understanding of $\lambda_F(S)$ variation with $S$ just as for
$\lambda_c(S)$. In the dilute limit with $P(\Omega,m,S) \to m^2\Omega^2$, we
have $\lambda_F \to 1/\sqrt{m}$ and thus reducing to the  result known
\cite{Ja-02} for spinless fermion systems. 

\section{Information Entropy and Duality Marker}
\label{c2s6}

\begin{figure}
\centering
\includegraphics[width=4in,height=5in]{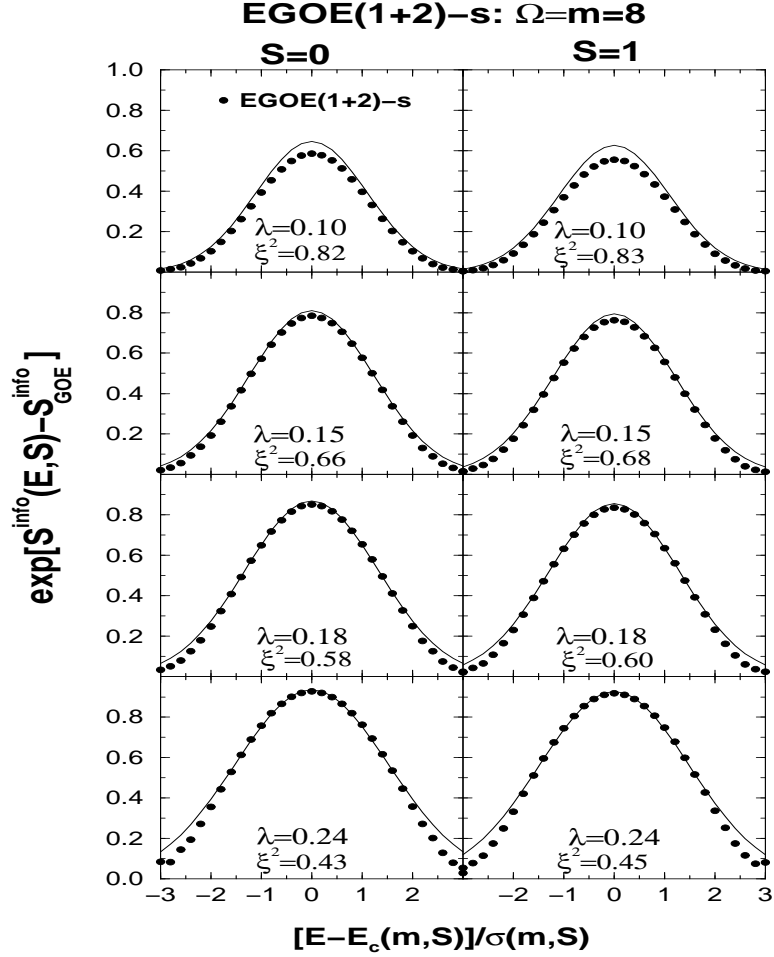}
\caption{$\exp[S^{info}(E,S)-S^{info}_{GOE}]$  for a 20 member
EGOE(1+2)-$\cs$ ensemble with $\Omega=m=8$ and spins $S=0$ and $1$ for
different $\lambda$ values. Values averaged over bin-size $0.2$ are shown 
as filled circles. The continuous curves correspond to Eq. (\ref{eq.inf1a}).
See text for details.}
\label{info}
\end{figure}

Figure \ref{info} shows information entropy $S^{info}(E,S)$ as a function of
$E$ for a 20 member EGOE(1+2)-$\cs$ ensemble with spins $S=0$ and $1$ and
for different $\lambda$ values. These results are compared with the
EGOE(1+2) formula for $S^{info}$ given in \cite{KS-01} (strictly valid only
for  $\lambda > \lambda_F$) by replacing the fixed-$m$ variances by 
fixed-($m,S$) variances,
\begin{subequations}
\be 
\exp(S^{info}(E,S)-S^{info}_{GOE})\stackrel{EGOE(1+2)-\cs}{\rightarrow}
\sqrt{1-\xi^2} \exp \l( \dis\frac{\xi^2}{2}\r) \exp \l(-
\dis\frac{\xi^2\we^2}{2}\r) \;;
\label{eq.inf1a}
\ee
\be
\xi^2=1-\dis\frac{\sigma^2_{\mbox{off-diagonal}}(m,S)}
{\sigma^2(m,S)} \sim
\dis\frac{\sigma^2_{h(1)}(m,S)}
{\sigma^2_{h(1)}(m,S)+\sigma^2_{V(2)}(m,S)}\;.
\label{eq.inf1b}
\ee 
\end{subequations}
Note that $\we$ is defined just before Eq. (\ref{eq.gau1}) and $\xi$  is a
correlation coefficient. The results given by Eq. (\ref{eq.inf1a}) are
compared with the numerical results in  Fig. \ref{info}. It is seen that the
numerical  results for  $\lambda \geq \lambda_F$ are described well by the
EGOE formula. There are  deviations at the tails because the result given by
Eq. (\ref{eq.inf1a}) assumes Gaussian form for the strength functions while
in practice there will be corrections to the Gaussian form. Thus results of
EGOE(1+2) extend to EGOE(1+2)-$\cs$ with parameters calculated in ($m,S$)
spaces. Similar analysis was done for number of principal components or IPR
in \cite{Ko-06}. 

For the EGOE(1+2)-$\cs$ Hamiltonian, two asymptotic natural basis  emerge and
they are (i) the non-interacting basis defined by  $\lambda_0=\lambda_1=0$
and (ii) the infinite interaction strength basis  defined by
$\lambda_0=\lambda_1=\infty$. In principle two more basis defined by
$\lambda_0=0, \lambda_1 = \infty$ and $\lambda_0=\infty, \lambda_1 = 0$
are possible but they are not considered in the present section. Therefore 
just as in the previous discussion we put $\lambda_0=\lambda_1=\lambda$. An
important question is \cite{Ja-02,An-04}: 
is there a point $\lambda =\lambda_d \geq \lambda_F$
where quantities defining wavefunction properties like entropy, strength
functions, temperature etc. are basis independent? To
examine this question, we compare $S^{info}(E,S)$ in $\lambda=0$ and
$\lambda=\infty$ basis by varying $\lambda$. In the $\lambda=0$ basis,
$S^{info}(E,S)$ is determined by Eq. (\ref{eq.inf1a}) with the  correlation
coefficient $\xi^2=\xi_0^2$ defined in Eq. (\ref{eq.inf1b}). Similarly, in
the  $\lambda=\infty$ basis, Eq. (\ref{eq.inf1a}) applies  with
$\xi^2=\xi_\infty^2 =\sigma^2_{V(2)}(m,S)/[\sigma^2_{h(1)}(m,S)+
\sigma^2_{V(2)}(m,S)]$; note that $\sigma^2_{V(2)}(m,S)$ depends on
$\lambda^2$. Therefore we can determine $\lambda_d$ by using the condition
that $\xi_0^2=\xi_\infty^2$ (this is equivalent to the condition that
the spreadings produced by $h(1)$ and $V(2)$ are equal).  Then we have
$\xi^2=\xi_0^2=\xi_\infty^2=0.5$ at $\lambda = \lambda_d$; see \cite{An-04}
for more details. Further, it can be argued  that the duality region
(defined by $\lambda \sim \lambda_d$) corresponds to the  thermodynamic
region for finite quantum systems and this will be discussed in Sec. \ref{c2s7}.

\begin{figure}
\centering
\includegraphics[width=4in,height=5in]{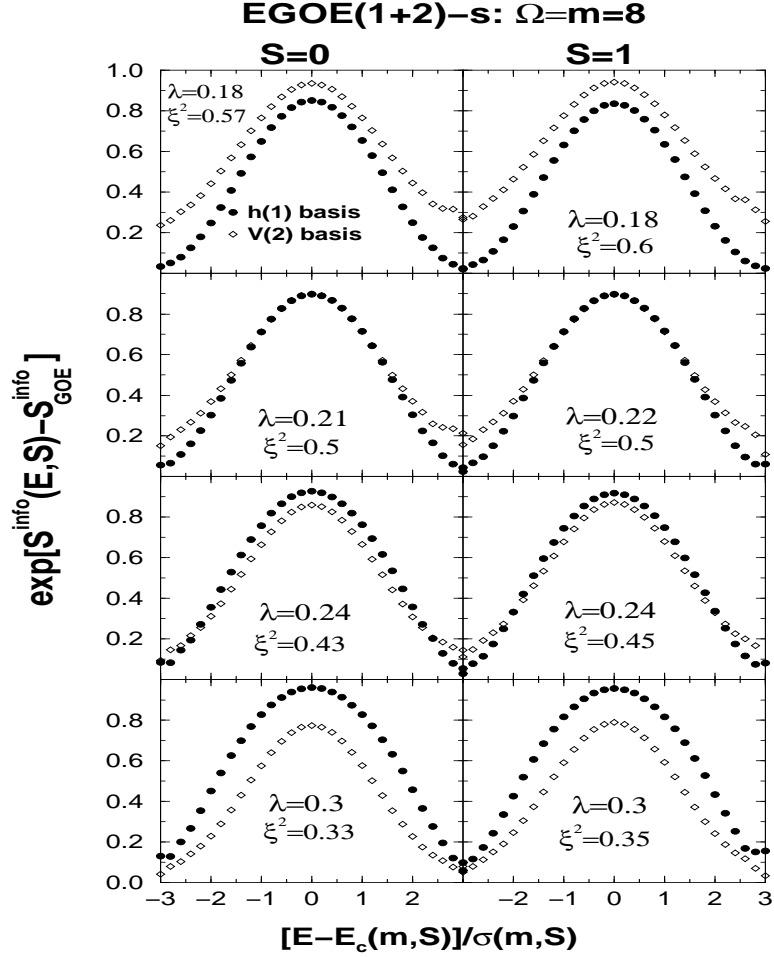}
\caption{$\exp[S^{info}(E,S)-S^{info}_{GOE}]$ in $h(1)$ and $V(2)$ basis for
a 20 member EGOE(1+2)-$\cs$ ensemble with $\Omega=m=8$ and spins  $S=0$ and
$1$ for different $\lambda$ values. Results averaged over bin-size  $0.2$
are shown  as circles; filled circles correspond to $h(1)$ basis and open
circles correspond to $V(2)$ basis. The $\xi^2$ values defined by Eq.
(\ref{eq.inf1b}) are also given in the figure. The duality point $\lambda_d$
corresponds to  $\xi^2=0.5$. See text for details.}
\label{lamd}
\end{figure}

Figure \ref{lamd} shows numerical results for the information entropy in the
$h(1)$ and $V(2)$ basis for a 20 member EGOE(1+2)-$\cs$ ensemble with
$\Omega=m=8$ and spins $S=0$ and $1$ for different $\lambda$ values ranging
from $\lambda=0.18$ to $0.3$.  It is seen from Fig. \ref{lamd} that the
duality marker  $\lambda_d=0.21$ for spin $S=0$ and $0.22$ for $S=1$. For
$\lambda$ values below and above $\lambda_d$ clearly there are differences
in $S^{info}(E,S)$ in the two basis.  The $S^{info}(E,S)$ values in the
$h(1)$ basis are smaller compared to those in the $V(2)$  basis for $\lambda
< \lambda_d$. The two entropies coincide at $\lambda=\lambda_d$ and beyond
that, $S^{info}$ in the $h(1)$ basis is comparatively larger. For a
qualitative understanding of the variation of $\lambda_d$ with $S$, we use
the criterion that around the duality region, spreadings produced by $h(1)$
and $V(2)$ are equal. This leads to the condition,
\be
\sigma^2_{h(1)}(m,S) = \lambda^2_d\;P(\Omega,m,S)\;.
\label{eq.inf2}
\ee
To determine $\sigma^2_{h(1)}(m,S)$, we consider a uniform spectrum
with $\Delta=1$. This gives, $\sigma^2_{h(1)}(1,\spin)=(\Omega^2-1)/12$.
Then, using Eq. (\ref{eq.den3}), 
\be
\sigma^2_{h(1)}(m,S)=\ch(\Omega,m,S)=\dis\frac{1}{12}\l[m(\Omega+2)
(\Omega-m/2)-2\Omega S(S+1)\r]\;.
\label{eq.inf3}
\ee
Combining this with Eqs. (\ref{eq.den9}) and (\ref{eq.inf2}) will give 
finally
\be
\lambda_d(S) \propto \dis\sqrt{\dis\frac{\ch(\Omega,m,S)}
{P(\Omega,m,S)}}\;.
\label{eq.inf4}
\ee
Eq. (\ref{eq.inf4}) with the result $\lambda_d(S=0)=0.21$ gives 
$\lambda_d(S=1)=0.22$ and $\lambda_d(S=2)=0.24$. These predictions are close
to the numerical results shown in Fig. \ref{lamd}. Therefore Eq.
(\ref{eq.inf4}) gives a good qualitative understanding of $\lambda_d(S)$
variation with $S$. In the dilute limit, simplifying the $\ch$ and $P$
factors, we have $\lambda_d \to 1/\sqrt{m}$ and this is the result for
spinless fermion systems \cite{An-04}. This also shows that in the dilute 
limit $\lambda_d$ and $\lambda_F$ have same scale. However these scales
differ parametrically  as $m$ approaches $\Omega$ (for $m>\Omega$ one has to
consider holes) and $S \gazz m/4$. In this situation there is strong spin 
dependence for the ratio $\lambda_d/\lambda_F$ as seen from Eqs.
(\ref{eq.wf5}), (\ref{eq.inf3}),  and (\ref{eq.inf4}) that  give
$\lambda_d/\lambda_F \propto  \sqrt{\frac{[m(\Omega+2) (\Omega-m/2)-2\Omega
S(S+1)]}{m \Omega^2}}$. Thus the  variance propagator determines the
behavior of the three transition markers $\lambda_c$, $\lambda_F$, and 
$\lambda_d$.

\section{Occupancies, Single-particle Entropy and Thermodynamic Region}
\label{c2s7}

A very important question for isolated finite interacting particle systems is
the following \cite{Ho-95,Fl-97,Be-01a,Ko-01,Ri-08}: in the chaotic domain will
there be a point or a region where thermalization occurs; i.e., will there be a
region where different definitions of entropy, temperature, specific heat, and
other thermodynamic variables give the same results (as valid for infinite
particle systems)? Toward answering this question within EGOE(1+2)-$\cs$, we
consider three different entropies, i.e., thermodynamic entropy defined by the
eigenvalue density, information entropy and sp entropy defined by the
occupancies of the sp orbitals. Before comparing these three different entropies
for various values of $\lambda$, now let us first consider occupancies in some
detail. 

\begin{figure}
\begin{center}
{\bf\large $\;\;\;\;\;\;$EGOE(1+2)-$\cs$}
\end{center}
\centering
\begin{tabular}{cc}
\includegraphics[width=3.5in,height=4in]{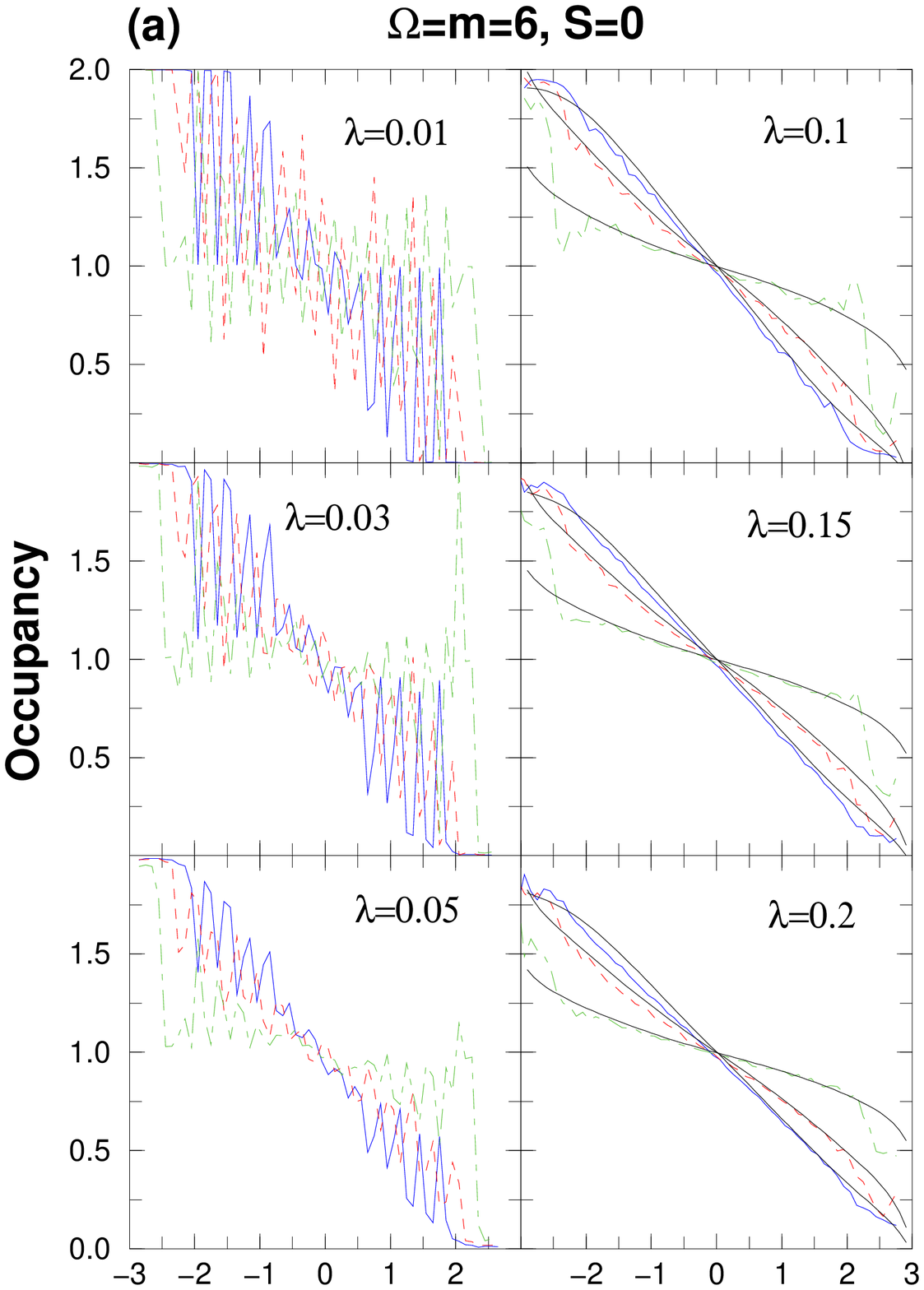} &
\includegraphics[width=1.75in,height=4in]{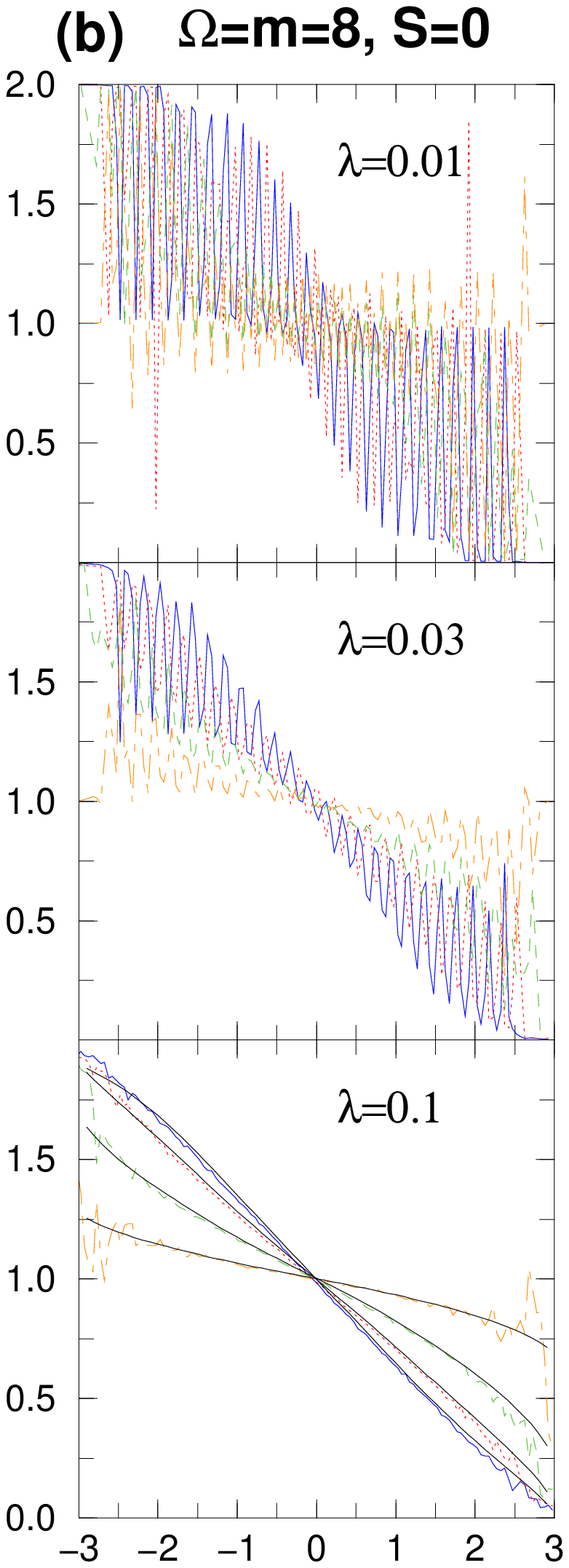}
\end{tabular}
\begin{center}
{\bf\large $\;\;\;\;\;\;\;\;[E-E_c(m,S)]/\sigma(m,S)$}
\end{center}
\caption{Occupation numbers as a function of
$\we=[E-E_c(m,S)]/\sigma(m,S)$. (a) For a 200 member EGOE(1+2)-$\cs$
ensemble with $\Omega=m=6$ and spin $S=0$, shown are the results for the
three lowest sp levels (solid blue, dashed red and dot-dashed green,  
respectively). They are  compared with the EGOE smoothed form (black) given by
Eq. (\ref{eq.appl2}) for $\lambda > \lambda_c = 0.05$. (b) For a 20 member
EGOE(1+2)-$\cs$ ensemble with $\Omega=m=8$ and spin $S=0$, shown are the
results for the four lowest sp levels (solid blue, dotted red, dashed green
and dot-dashed orange, respectively). They are compared with the EGOE
smoothed
form (black) given by Eq. (\ref{eq.appl2}) for $\lambda = 0.1$. For this
system, $\lambda_c=0.028$. Note that for the results in the figures,
occupancies are averaged over bin-size 0.1 for $\Omega=m=6$ and  $0.05$ for
$\Omega=m=8$, respectively. See text for further details.}
\label{occu}
\end{figure}

Occupation probability for a sp orbital $i$ is given by the expectation
value of $n_i$, i.e., $\lan n_i \ran^{m,S,E}$. It is possible to write this
as a ratio of two densities,
\be
\barr{rcl}
\lan n_i \ran^{m,S,E} & = & \dis\frac{\lan n_i \delta(H-E) \ran^{m,S}}
{\lan \delta(H-E) \ran^{m,S}} \\ \\
& = & \lan n_i\ran^{m,S}\dis\frac{\rho^{m,S}_{n_i}(E)}{\rho^{m,S}(E)}\;.
\earr \label{eq.appl1}
\ee 
As $n_i$ is a positive-definite operator, the occupancy density
$\rho^{m,S}_{n_i}(E)$ can be represented by a probability density with
moments $M_p(n_i)=\lan n_i H^p \ran ^{m,S}/\lan n_i \ran^{m,S}$. The
corresponding lower order central moments define Edgeworth corrected
Gaussian form for $\rho^{m,S}_{n_i}(E)$. For $\lambda > \lambda_c$, 
fluctuations follow GOE and hence $\lan n_i \ran^{m,S,E}$ take a smoothed 
form
and they can be written as the ratio of the smoothed forms for the densities
in Eq. (\ref{eq.appl1}). As the fixed-$(m,S)$ eigenvalue density is
Gaussian,  the fixed-$(m,S)$ occupancy densities also follow Gaussian form
(as discussed ahead, this is verified by calculating the excess parameter).
Therefore,
\be
\lan n_i \ran^{m,S,E}\;\;\; \stackrel{\lambda \geq \lambda_c}
{\longrightarrow} \;\;\;
\lan n_i\ran^{m,S}\dis\frac{\rho^{m,S}_{n_i:\cg}(E)}{\rho^{m,S}_{\cg}(E)}\;.
\label{eq.appl2}
\ee
Section \ref{c3s3} gives extensions of Eq. (\ref{eq.appl2}) to pairing 
Hamiltonian with further discussion on expectation value densities.
Figure \ref{occu}(a) shows occupation numbers for a 200 member 
EGOE(1+2)-$\cs$ ensemble with $\Omega=m=6$ and spin $S=0$ as a function of
$E$ for various $\lambda$ values. Results are shown for  the lowest three sp
orbitals.  As discussed in Sec. \ref{c2s8} (see also Fig. \ref{m6n6} ahead),  
$\lambda_c = 0.05$ and
$\lambda_F=0.18$ for the present example. It is clearly seen from Fig.
\ref{occu}(a) that the fluctuations are large for $\lambda < \lambda_c$ as
expected.  Beyond this, the occupancies start taking a smoothed form. The
numerical results for $\lambda >> \lambda_c$ are compared with the  smoothed
form given by Eq. (\ref{eq.appl2}). Here Edgeworth corrections are added to
the Gaussian densities. For example, for $\lambda=0.1$, the difference
between the occupancy density  centroids and the energy centroids (in units
of the spectral widths) are $-0.4,\;-0.29$, and $-0.12$ for  the sp orbitals
$1$, $2$, and $3$, respectively.   Similarly the occupancy density widths (in
units of the spectral widths) are $0.91,\;0.96$, and $0.99$ and $\gamma_2$ 
values are $-0.39,\;-0.43$, and $-0.4$ for  the sp orbitals $1$, $2$, and
$3$,
respectively. Note that $|\gamma_1| \sim 0$ in all the cases.  For the
eigenvalue density, the excess parameter $\gamma_2(m,S)=-0.38$. 
Agreement between Eq. (\ref{eq.appl2}) and the numerical results is
excellent except  at the spectrum ends as here the states are not
sufficiently complex. We have also verified this for $S=1$ and $S=2$
examples. Therefore in the $\lambda < \lambda_c$ region, fluctuations being
large (they follow Poisson), smoothed forms are not meaningful. On the other
hand, in the chaotic domain defined by $\lambda > \lambda_c$, occupation
probabilities take a smoothed form as the fluctuations here follow GOE 
(hence they
are small). The smoothed form is well  described by Eq. (\ref{eq.appl2}). It
is interesting to note that the fluctuations even in the gs region
are small for $\lambda >> \lambda_F$. All these conclusions are also
verified for a 20 member EGOE(1+2)-$\cs$ ensemble with $\Omega=m=8$ and
$S=0$ and some of these results are shown in Fig. \ref{occu}(b).

\begin{figure}
\centering
\includegraphics[width=5.5in,height=5in]{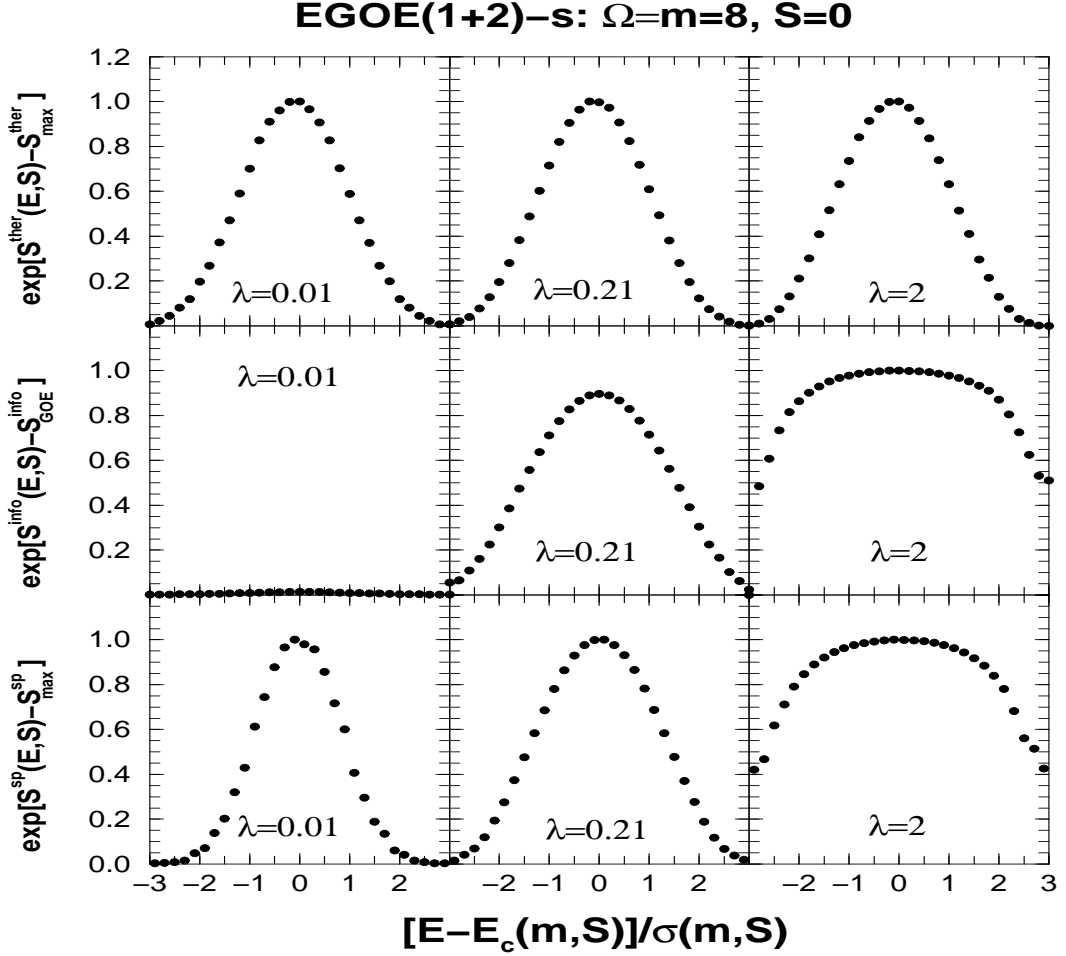}
\caption{Thermodynamic entropy $\exp[S^{ther}(E,S)-S^{ther}_{max}]$,
information entropy  $\exp[S^{info}(E,S)-S^{info}_{GOE}]$ and single-particle 
entropy  $\exp[S^{sp}(E,S)-S^{sp}_{max}]$ vs $\we = [E-E_c(m,S)] / 
\sigma(m,S)$  for a 20 member EGOE(1+2)-$\cs$ ensemble with $\Omega=m=8$ and
$S=0$ for different $\lambda$ values. Entropies averaged over bin-size $0.2$
are shown  as filled circles. Note that for $\lambda=0.01$, 
$\exp[S^{info}(E,S)-S^{info}_{GOE}]$ is close to zero for all $\we$ values. 
See text for details.}
\label{entr}
\end{figure}

Given the fractional occupation probabilities $f_i(E,S)=\frac{1}{2}\lan n_i
\ran^{m,S,E}$, the sp entropy $S^{sp}(E,S)$ is defined by,
\be
S^{sp}(E,S) = - \dis\sum_i 2 \l\{f_i(E,S) \ln f_i(E,S) + [1-f_i(E,S)] 
\ln [1-f_i(E,S)] \r\}\;.
\label{eq.appl3}
\ee 
To establish that the $\lambda=\lambda_d$ region corresponds to the
thermodynamic region, we will compare the thermodynamic entropy 
$S^{ther}(E)=\ln \rho^{m,S}(E)$ and the information entropy defined by Eqs.
(\ref{eq.wf2}) and  (\ref{eq.inf1a}) with the sp entropy for different
$\lambda$ values just as it was done before for EGOE(1+2) and the nuclear
shell-model examples \cite{Ho-95,Ko-02}. For $\Omega=m=8$ and $S=0$ system
with 20 members, we show in Fig. \ref{entr} results for
$\lambda=\lambda_d=0.21$, $\lambda=0.01 << \lambda_d$ and  $\lambda=2 >>
\lambda_d$. Note that $\exp[S^{ther}(E,S)-S^{ther}_{max}]  \longrightarrow
\exp-\frac{1}{2} {{\we}^2}$ for all $\lambda$ values as the eigenvalue
density is a Gaussian essentially independent of $\lambda$.  Similarly, Eq.
(\ref{eq.inf1a}) gives the formula for $\exp[S^{info}(E,S) - 
S^{info}_{GOE}]$. We have also verified that the extension of the EGOE(1+2) 
formula for the sp entropy \cite{Ko-02} with centroids and variances
replaced  by fixed-$(m,S)$ centroids and variances is a good approximation
for fixed-$(E,S)$ sp entropy and then the formula is,
\be
\exp[S^{sp}(E,S)-S^{sp}_{max}] =
\exp -\frac{1}{2} \xi^2 {{\we}}^2\;.
\label{eq.appl4}
\ee
For the three examples shown in Fig. \ref{entr}, $\xi^2=0.998$, $0.5$, and
$0.039$ for $\lambda=0.01$, $0.21$, and $2$, respectively. It is clearly seen
from Fig. \ref{entr} that the three entropies differ as we go away from
$\lambda=\lambda_d$ and at $\lambda=\lambda_d$ they all look similar, i.e.,
as stated in \cite{Ho-95} ``the thermodynamic entropy defined via the global
level density or in terms of occupation numbers behaves similar to the
information entropy.'' Therefore, $\lambda=\lambda_d$ region can be 
interpreted as the thermodynamic region in the sense that all different
definitions of entropy coincide in this region.

\section{Some Results for $\lambda_0 \neq \lambda_1$}
\label{c2s8}

\begin{figure}
\centering
\includegraphics[width=5.5in,height=5.5in]{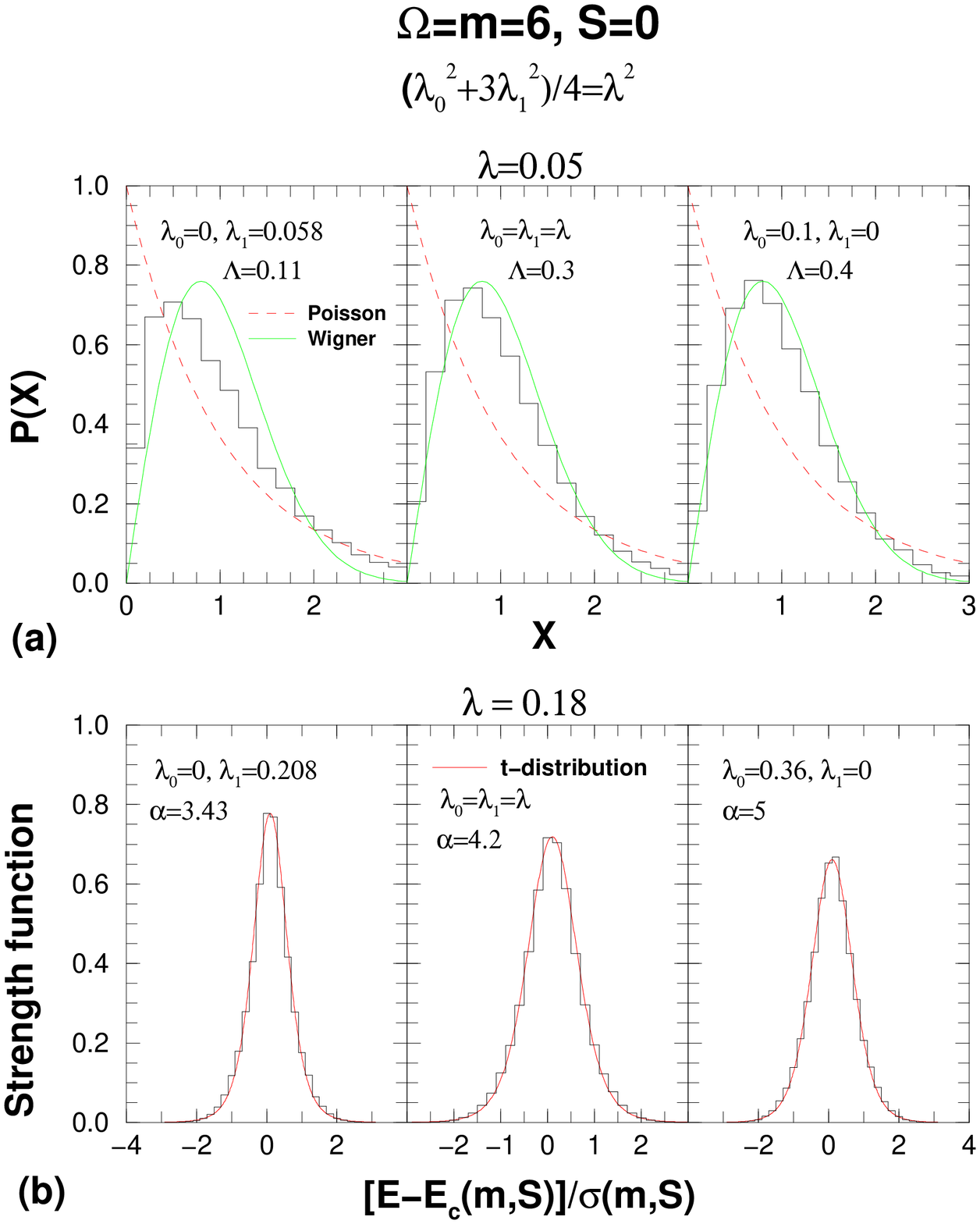}
\caption{Variation in the nearest neighbor spacing 
distributions  $P(X)$ and the strength functions $F_k(E,S)$ with   
$\lambda_0$ and $\lambda_1$, for a 200 member  EGOE(1+2)-$\cs$ ensemble with
$\Omega=m=6$ and spin $S=0$. Calculations are carried out with the
constraint $(\lambda_0^2 + 3 \lambda_1^2)/4 = \lambda^2$ and the results are
shown for (a) $\lambda=0.05$ and (b) $\lambda=0.18$ with three different
values for  $\lambda_0$.  Strength functions are shown for 
$\we_k=0$. Histograms, with bin-size $0.2$, are the calculated results. See
text for details.}
\label{m6n6}
\end{figure}

All the discussion in Secs. \ref{c2s4}-\ref{c2s7} is restricted to 
$\lambda_0=\lambda_1=\lambda$
in Eq. (\ref{eq.def1}) i.e., for equal strengths of the $s=0$ and $s=1$ parts
of the interaction. However, for completeness, here we present some results 
for the change in the eigenvalue and wavefunction structure for 
$\lambda_0^2 \neq  \lambda_1^2$.  To investigate this, we have examined NNSD
and strength functions by fixing the value for the ensemble averaged 
two-particle spectral variance $\sigma^2_{V(2)}(2)$ generated by the
two-body part of $H$ and then varying $\lambda_0$ ($\lambda_1$). The
two-particle spectral variance  for $\Omega >> 1$ is
$\sigma^2_{V(2)}(2)=(\lambda_0^2 + 3 \lambda_1^2)/16$.  Therefore we
have considered the following Hamiltonian,
\be
H_{(\lambda_0,\lambda_1:\lambda)} =
h(1)+\lambda_0 V^{s=0}(2) + \lambda_1 V^{s=1}(2)\;;\;\;\;\;
(\lambda_0^2 + 3\lambda_1^2)/4 = \lambda^2\;,
\label{eq.lam1}
\ee
and carried out calculations for various fixed values of $\lambda$ and
varying $\lambda_0$ ($\lambda_1$) with the constraint $(\lambda_0^2 +
3\lambda_1^2)/4 =  \lambda^2$. For a 200 member EGOE(1+2)-$\cs$ ensemble
defined by  $H_{(\lambda_0,\lambda_1:\lambda)}$ with $\Omega=m=6$ and 
$S=0$, results are presented in Fig. \ref{m6n6} for NNSD and strength
functions. In the calculations, we have chosen  $\lambda=0.05$ for NNSD and
$\lambda=0.18$ for the strength functions.  For the choice
$\lambda_0=\lambda_1=\lambda$, they correspond to $\lambda_c$ and
$\lambda_F$, respectively for the $\Omega=m=6$ and  $S=0$ system. This is
clearly seen in Fig. \ref{m6n6}. Results are also shown for the two extreme
choices $\lambda_0=0$, $\lambda_1=\sqrt{4/3}\lambda$ and
$\lambda_0=2\lambda$, $\lambda_1=0$. For $\lambda_0=0$, the NNSD is closer
to Poisson while for $\lambda_1=0$, NNSD is much closer to the Wigner form.
Similarly, for $\lambda_0=0$, the strength function is more closer to BW
while for $\lambda_1=0$, it is closer to Gaussian.  We can easily infer
these changes in the structures from the propagator  ratio 
$R_{(\lambda_0,\lambda_1:\lambda)}(\Omega,m,S) = 
\sigma^2_{(\lambda_0,\lambda_1:\lambda)}(m,S)/[\lambda^2\,P(\Omega,m,S)]$. 
Note that $\sigma^2_{(\lambda_0,\lambda_1:\lambda)}(m,S)$ is same as
$\overline{\sigma^2_{V(2)}(m,S)}$ given by Eq. (\ref{eq.den8}). For our
example with $\Omega=m=6$ and $S=0$,  we have
$R_{(\lambda_0,\lambda_1:\lambda)}(\Omega,m,S) = 0.93$, $0.94$, $1$, $1.1$,
$1.22$ for $\lambda=0.05$ and $\lambda_0=0$, $0.02$, $0.05$,  $0.075$, and
$0.1$, respectively. Therefore for $\lambda_0 < 0.05$,  we have
$R_{(\lambda_0,\lambda_1:\lambda)}(\Omega,m,S)<1$ and this implies [as seen
from Eq. (\ref{eq.sd2})] that the level fluctuations change from
Poisson-like  to GOE as the value of $\lambda_0$ is increased from
$\lambda_0=0$ as seen in Fig. \ref{m6n6}(a). Similarly,
$R_{(\lambda_0,\lambda_1:\lambda)}(\Omega,m,S) = 0.93$, $0.95$, $1$, $1.07$,
$1.22$ for $\lambda=0.18$ and $\lambda_0=0$,  $0.1$, $0.18$,  $0.25$, and
$0.36$, respectively. Therefore for $\lambda_0 < 0.18$,  we have
$R_{(\lambda_0,\lambda_1:\lambda)}(\Omega,m,S)<1$ and this implies [as seen
from Eq. (\ref{eq.wf5})] that the strength functions change from BW to
Gaussian form as the value of $\lambda_0$ is increased from $\lambda_0=0$ as
seen in Fig. \ref{m6n6}(b). Thus we can conclude that the general structure
of the transitions, as discussed in Fig. \ref{tmark}, remains same even for 
$\lambda_0^2 \neq  \lambda_1^2$. We have also made calculations by varying
$\lambda_0$ and $\lambda_1$ without any constraint. Here also the variance 
propagator gives predictions for the changes in NNSD and strength functions
and we have verified these predictions in some examples. Before summarizing the
results on transition markers, we now present the results for the excess 
parameter $\gamma_2(m,S)$ for EGOE(1+2)-$\cs$.

\section{Results for $\gamma_2(m,S)$ for EGOE(1+2)-$\cs$}  
\label{c2a1}

Towards providing a basis for the Gaussian form for the eigenvalue
density generated by EGOE(1+2)-$\cs$, we derive first the exact formula for
$\gamma_2(m,S)$ for a general $h(1)$ operator and then discuss 
$\gamma_2(m,S)$ for EGOE(2)-$\cs$. Given $h(1)=\sum_i
\epsilon_i n_i$, the $\gamma_2(m,S)$ is defined by  the fourth central
moment $\lan \widetilde{h}^4(1)\ran^{m,S}$ and the variance or the second
central moment $\lan \widetilde{h}^2(1)\ran^{m,S}$. Note that, 
\be
\widetilde{h}(1) =\dis \sum_{i=1}^\Omega \widetilde{\epsilon}_i n_i
\;;\;\;\;\;
\widetilde{\epsilon}_i = \epsilon_i-\dis\frac{1}{\Omega}\sum_{i=1}^\Omega 
\epsilon_i\;.
\label{eq.c2.app2}
\ee
To derive the formula for the fourth moment,  we will decompose first
$\widetilde{h}^2(1)$ into one and two body parts and apply Eq. (\ref{eq.vv1}). 
The one-body  part of
$\widetilde{h}^2(1)$ is defined by the sp energies $\tilde{\epsilon}_i ^2$
and  the two-body matrix elements $V_{ijij}^s=2\tilde{\epsilon}_i
\tilde{\epsilon}_j$  with all other matrix elements being zero; note that $i
\neq j$ for $s=1$. Then the $\lambda$'s and other averages in Eqs.
(\ref{eq.den5})-(\ref{eq.den7}) are,
\be
\barr{rcl}
\lambda_{i,i}^{s=0} & = & - \lambda_{i,i}^{s=1} = 2 \widetilde{\epsilon}_i^2 -
\dis\frac{2}{\Omega} \dis\sqrt{X}\;, \;\;\;\;
\lan V(2) \ran^{2,s} = \dis\frac{2(-1)^s}{\Omega[\Omega+(-1)^s]} 
\dis\sqrt{X}\;, \\ \\
\lan [V(2)]^2 \ran^{2,s} & = & \dis\frac{4}{\Omega[\Omega+(-1)^s]}\l[
X + (-1)^s Y \r]\;, 
\earr \label{eq.c2.app3}
\ee
\be
\barr{rcl}
\lan [V^{s:\nu=2}(2)]^2 \ran^{2,s} & = & \dis\frac{4(-1)^s}{[\Omega+(-1)^s]
[\Omega+2(-1)^s]}Y + 
\dis\frac{4(\Omega^2+3(-1)^s\Omega+3)}{\Omega [\Omega+(-1)^s]^2
[\Omega+2(-1)^s]}X \;;\\ \\
X & = & \l(\dis\sum_k \tilde{\epsilon}_k^2\r)^2\;,\;\;\;Y=\dis\sum_k 
\tilde{\epsilon}_k^4\;.\nonumber
\earr \label{eq.c2.app3a1}
\ee
Using Eq. (\ref{eq.vv1}) with Eq. (\ref{eq.c2.app3}), 
the final propagation formulas are,
\be
\barr{l}
\l[\lan \widetilde{h}^2(1) \ran^{m,S}\r]^2 = \dis\frac
{\l[m(m-2\Omega)(\Omega+2)+4\Omega S(S+1)\r]^2}{4\Omega^2(\Omega^2-1)^2}X\;,
\\ \\
\lan \widetilde{h}^4(1) \ran^{m,S} = \l[\lan \widetilde{h}^2(1) \ran^{m,S}
\r]^2 - \dis\frac{12\ch(\Omega,m,S)}{\Omega^2(\Omega^2-1)}
(X-\Omega Y)   \nonumber
\earr \label{eq.c2.app4a1}
\ee
\be
\barr{l}
+ 
\dis\frac{\l[-m (m-2 \Omega) \{-4 (\Omega+1)+m (\Omega+4)\}+4\Omega
(2\Omega-3 m+2)S(S+1)\r]}{\Omega^2(\Omega+1)(\Omega-1)(\Omega-2)}
(X-\Omega Y)
\\ \\
+ \dis\frac{1}{2 \Omega^2 (\Omega^2-1) (\Omega-2)^2 (\Omega+2)}
\l[m (m-2 \Omega) \{-4 (\Omega+1)+m (\Omega+4)\}^2 \r. \\ \\ \l.+8 \Omega 
S(S+1) \{2 (\Omega+1) (3\Omega+2)+m^2
(3 \Omega+8)-m (3 \Omega^2+16\Omega+12)\} 
\r. 
\earr \label{eq.c2.app4}
\ee
\begin{flushleft}
\be
\barr{l}
\l. +16 \Omega^2 \{S(S+1)\}^2\r]
(X-\Omega Y) \\ \\ +
\dis\frac{4}{\Omega^2(\Omega-1)(\Omega+3)} Q^2(\{2\}:m,S) 
\lan [V^{s=0:\nu=2}(2)]^2 \ran^{2,0} \\ \\ +
\dis\frac{4}{\Omega^2(\Omega+1)(\Omega-3)} Q^2(\{1^2\}:m,S) 
\lan [V^{s=1:\nu=2}(2)]^2 \ran^{2,1}\;.  \nonumber
\earr \label{eq.c2.app4a2}
\ee
\end{flushleft}
Note that $\ch$ is defined in Eq. (\ref{eq.inf3}) and $Q$'s are  defined in
Eq. (\ref{eq.den8}) respectively. Using Eq. (\ref{eq.c2.app4}), we can
calculate $\gamma_2(m,S)$ for any set of $\epsilon_i$'s and $(\Omega,m,S)$
where,
\be
\gamma_2(m,S)= \dis\frac{\lan \widetilde{h}^4(1)\ran^{m,S}}{\l[\lan
\widetilde{h}^2(1)\ran^{m,S}\r]^2} - 3\;.
\label{eq.c2.app1}
\ee
Expanding the expression, by combining Eqs. (\ref{eq.c2.app4}) and 
(\ref{eq.c2.app1}), for $\gamma_2(m,S)$ in powers of $1/\Omega$ and
retaining terms up to $1/\Omega$, we have
\be
\gamma_2(m,S) = \dis\frac{\gamma_2(1,\spin)}{m} -
\dis\frac{\{m(m-4)+4S(S+1)\}\{5\gamma_2(1,\spin)+6\}}{2m^2\Omega}
+O\l(\dis\frac{1}{\Omega^2}\r)\;.
\label{eq.c2.app5}
\ee
Therefore, for the $h(1)$ operators with $|\gamma_2(1)| \sim 1$, the excess
parameter $\gamma_2(m,S) \to 0$ for sufficiently large $m$ and also the spin
dependence is weak. Therefore $h(1)$ operators in general generate Gaussian
eigenvalue densities for large $m$ values. With $S=m/2$ and $N=2\Omega$, 
Eq. (\ref{eq.c2.app5}) reduces to,
\be
\gamma_2(m,S) = \dis\frac{\gamma_2(1,\spin)}{m}
+\dis\frac{1}{N}\l[\dis\frac{\{5\gamma_2(1,\spin)+6\}}{m} - 
\l\{5\gamma_2\l(1,\spin\r)+6\r\}\r]\;.
\label{eq.app6}
\ee
This is same as the result that follows from the exact formula for
$\gamma_2(N,m)$ for spinless fermion systems \cite{Fr-06}.

Turning to two-body interactions, first it should be mentioned that a
formalism for obtaining exact results for $\gamma_2(m,S)$ for a given $V(2)$
is given in \cite{Karw-95,Pl-96} and also they can be obtained via a
subtraction procedure using the formulation discussed in \cite{Wo-86}. As
seen from \cite{Karw-95}, the analytical result for $\gamma_2(m,S)$  is
complicated and contains too many terms.  Therefore it is not easy to derive
an easy to understand 
analytical formula for $\overline{\gamma_2(m,S)}$  for EGOE(2)-$\cs$.
However, an analytical understanding is possible in the dilute limit.  Then,
as argued in \cite{Pl-97}, the spin dependence of $\gamma_2(m,S)$ will be
weak and the first correction is of the form $C_0 [1+4S(S+1)/m^2]$ where
$C_0$ is a constant. Strikingly, Eq. (\ref{eq.c2.app5}), for the $h(1)$
operator, also gives the same result. Then one can conclude that
EGOE(2)-$\cs$ gives Gaussian eigenvalue densities. Combining the analytical
results given by Eqs. (\ref{eq.c2.app4}) and (\ref{eq.c2.app5})  for
$\gamma_2(m,S)$ for the $h(1)$ operator and the asymptotic
result for a general two-body Hamiltonian preserving spin given in 
\cite{Pl-97}, it is plausible to argue that the eigenvalue
density for EGOE(1+2)-$\cs$ will be in general of Gaussian form.

\section{Summary}
\label{c2s9}

\begin{figure}[htp]
\centering
\includegraphics[width=5.5in,height=4.5in]{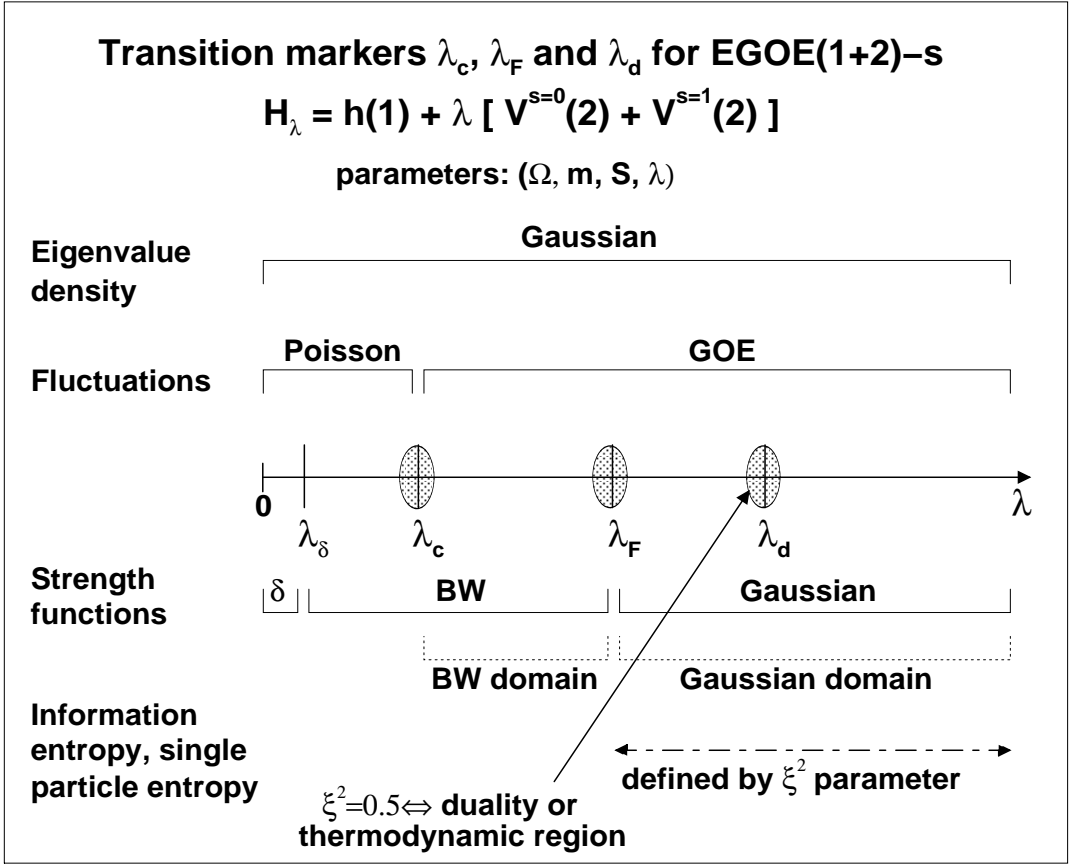}
\caption{Transition (chaos) markers for EGOE(1+2)-$\cs$.  Results in the figure
are obtained using the Hamiltonian $H_\lambda$ given in  Eq. (\ref{eq.ham8}),
i.e.,  $\lambda_0 = \lambda_1 = \lambda$ in Eq. (\ref{eq.def1}). Note that 
$\lambda$ is in the units of the average spacing $\Delta$ of the sp levels
defining $h(1)$. As discussed in the text, strength functions take
$\delta$-function form (denoted by $\delta$ in the figure) for $\lambda <
\lambda_\delta$ with $\lambda_\delta <<  \lambda_c$ and they start taking BW
form as $\lambda$ crosses  $\lambda_\delta$.  The BW domain is defined by
$\lambda_c <\lambda < \lambda_F$ and here  the strength functions take BW form
and the fluctuations are GOE. Similarly, in the Gaussian domain, defined by
$\lambda > \lambda_F$,  the strength functions take Gaussian form and the
fluctuations are GOE. Also in this region, the information entropy and
single-particle  entropy are defined by the $\xi^2$ parameter given in Eq.
(\ref{eq.inf1b}).  The basic structure of the transitions remains same even for
$\lambda_0^2 \neq \lambda_1^2$  as discussed in Sec. \ref{c2s8}. See text for
further details.}
\label{tmark}
\end{figure}

In summary, we have presented in Secs. \ref{c2s4}-\ref{c2s7},  a comprehensive
set of calculations for the  changes in level fluctuations, strength functions,
information entropy and occupancies as a function of the $\lambda$ parameter in
EGOE(1+2)-$\cs$ Hamiltonian given by Eq. (\ref{eq.def1}) with $\lambda_0 =
\lambda_1 = \lambda$.  The final results are summarized in Fig. \ref{tmark} (the
basic structure of the transitions remains same even for $\lambda_0^2 \neq
\lambda_1^2$ as discussed briefly in Sec. \ref{c2s8}). In addition, we have
derived the exact formula for the ensemble averaged fixed-($m,S$) spectral
variances and for the $V(2)$ part it is of the form $\lambda^2 P(\Omega,m,S)$. 
We have demonstrated that the variance propagator $P(\Omega,m,S)$  in Eq.
(\ref{eq.den9}) gives a good explanation for the spin dependence of the Poisson
to GOE and BW to Gaussian crossover points $\lambda_c$ and $\lambda_F$ for level
fluctuations and strength functions, respectively, and similarly for the duality
or thermodynamic region marked by  $\lambda_d$ (obtained from  information
entropy and occupancies). The three chaos markers $\lambda_c$, $\lambda_F$ and
$\lambda_d$ in terms of  $P(\Omega,m,S)$ are given by Eqs. (\ref{eq.sd2}),
(\ref{eq.wf5}), and (\ref{eq.inf4}), respectively. As seen from   Fig.
\ref{var}, $P$ decreases  with $S$ and using this in  Eqs. (\ref{eq.sd2}),
(\ref{eq.wf5}), and (\ref{eq.inf4}), establishes that the $\lambda_c$,
$\lambda_F$ and $\lambda_d$ values will increase with $S$ (as $S=m/2$
corresponds to  spinless fermions, it may be possible to investigate further,
using EGOE(1+2)-$\cs$, the recent claim by Papenbrock and  Weidenm\"{u}ller
\cite{Pa-05} that symmetries are responsible for chaos in nuclear shell-model).
Thus, introduction of the spin quantum number preserves the general structures,
generated by spinless fermion EGOE(1+2) ensemble, although the actual values of
the markers vary with the $m$-particle spin $S$.  
It should be emphasized that the first example for the transition
markers exhibited by EGOE(1+2) with additional good quantum number besides the
particle number $m$ are derived and presented using EGOE(1+2)-$\cs$ in this
chapter.

The transition markers as described in Fig. \ref{tmark} provide a basis for
statistical spectroscopy. For example, for $\lambda \geq \lambda_c$ as GOE
fluctuations are small they can be ignored. Then the smoothed eigenvalue
densities will be Gaussian. Similarly the strength functions and other related
distributions will take BW or  Gaussian form. Using these it is possible to
derive distributions (with respect to the energy eigenvalues)  for various
spectroscopic observables  \cite{Ko-01,Ko-03,Br-81,Fl-99,Kar-94} and employ them
in applications in nuclear and atomic physics. Moreover if the system is in the
thermodynamic region, then Gaussian form can be used for the strength
functions (or partial densities) defined over sub-spaces generated by any
symmetry algebra \cite{Ko-03} and this will allow one to study goodness of group
symmetries \cite{Pa-78,Wo-86}. 

\chapter*{}

\thispagestyle{empty}

\chapter{EGOE(1+2)-$\cs$: Pairing Correlations}
\label{ch3}

\section{Introduction}
\label{ch3-int}

Pairing correlations play very important role in finite interacting Fermi
systems such as nuclei \cite{Ho-07,Ka-00}, small metallic grains
\cite{Pa-02,Al-08}, quantum dots \cite{Lu-01,Al-05} and so on.  The
EGOE(1+2)-$\cs$ discussed in Chapter \ref{ch2} 
provides a model for  understanding general structures
generated by pairing correlations \cite{Pa-02,Al-05}. We adopt an algebraic
approach to pairing rather than the BCS approach. Our purpose in this
chapter is to study first the pairing symmetry in the space defined by
EGOE(1+2)-$\cs$ and then the measures for  pairing, using EGOE(1+2)-$\cs$
ensemble, that are of interest for nuclei (see \cite{Ho-07}), quantum dots and
small metallic grains (see \cite{Al-08}).  In the space defined by
EGOE(1+2)-$\cs$ ensemble, pairing symmetry is defined by the algebra
$U(2\Omega)\supset Sp(2\Omega) \supset SO(\Omega)\otimes  SU_S(2)$. Starting
with the details of this algebra we show that the state density generated by the
pairing Hamiltonian  will be a highly skewed distribution. In contrast,
the partial densities over pairing subspaces follow Gaussian form and the
propagation formulas for their centroids and variances, defined over subspaces
given by the algebra $U(2\Omega)\supset Sp(2\Omega) \supset SO(\Omega)\otimes 
SU_S(2)$, are derived.  Pair transfer strength  sum as a function of excitation
energy (for fixed $S$), a statistic for onset of chaos, is shown to follow, for
low spins, the form derived for  spinless fermion systems. The parameters
defining this form are easy to calculate using propagation equations. In
addition, we consider a quantity in terms of gs energies, giving
conductance peak spacings in mesoscopic systems at low temperatures, and study
its distribution over EGOE(1+2)-$\cs$ by including both  pairing and exchange
interactions. All the results presented in
this chapter are published in \cite{Ma-09,Ma-10c}.

\section{$U(2\Omega)\supset Sp(2\Omega) \supset SO(\Omega)\otimes SU_S(2)$ 
Pairing Symmetry}
\label{c3s1}

Pairing algebra to be discussed here is presumably familiar to others.
However to our knowledge  the details presented here are not reported
elsewhere (for a short related discussion see \cite{Fl-64}). Note that, 
we drop the ``hat'' symbol over $\wh$, $\whh$ and $\wv$ 
when there is no confusion as in Chapter \ref{ch2}.

\begin{figure}[tp]
\centering
\includegraphics[width=5.5in,height=3.5in]{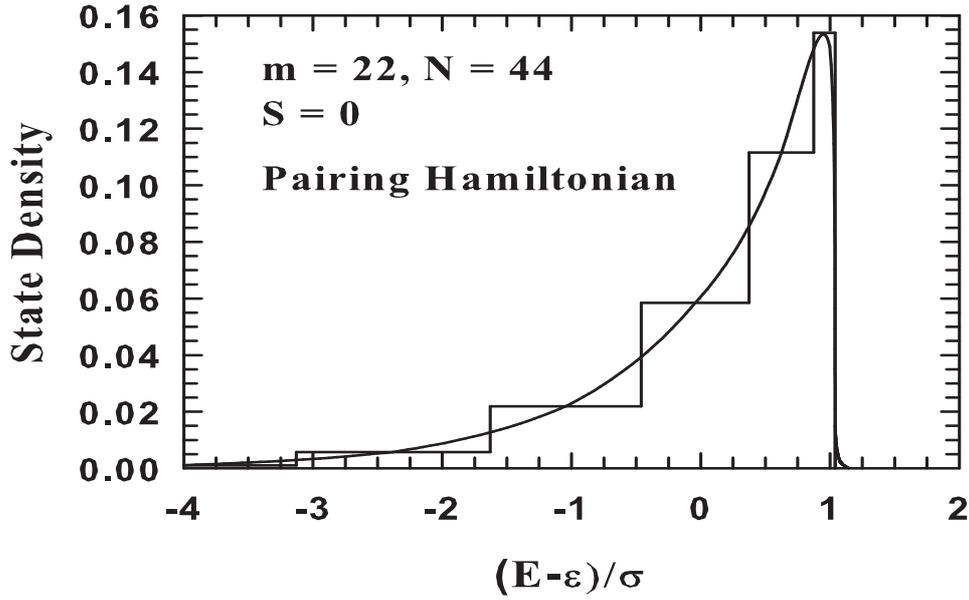}
\caption{State density for the pairing Hamiltonian $H=-H_p$ for a system
of 22 fermions in $\Omega=22$ orbits ($N=44$) and total spin $S=0$. In the
histogram, $\rho(E)$ for a given $\we=(E-\epsilon)/\sigma$  is plotted with
$\we$ as center with width given by $\Delta\we=\Delta E/\sigma$ (see Eq.
(\ref{eq.dens}) and the following discussion). The smooth curve is obtained by
joining the center points to  guide the eye. A similar plot was shown before by 
Ginocchio \cite{Gi-80} but for a system of identical fermions in a large 
single-$j$ shell.}
\label{c3f1}
\end{figure}

Consider $m$ fermions distributed in $\Omega$ number of sp 
levels each with spin $\cs=1/2$. Therefore total number of sp states
is $N=2\Omega$ and they are denoted by  $a^\dagger_{i,\cs=\spin,m_\cs} 
\l|0\ran =  \l.\l| i,\cs=\spin,m_\cs=\pm \spin\r.\ran$ with
$i=1,2,\ldots,\Omega$. Similarly, 
$$
\dis\frac{1}{\sqrt{1+\delta_{i,j}}} \l(a_{i,\cs=\spin}^\dagger 
a_{j, \cs=\spin}^\dagger\r)^s_{m_s} 
\l|0\ran = \l|\l(i,\cs=\spin;j,\cs=\spin\r)\;s,\;m_s
\ran
$$ 
denotes two-particle antisymmetric states with the two-particle in the
levels $i$ and $j$ and the two-particle  spin $s=0$ or $1$. From now on we
will drop the index $\cs=\spin$ for simplicity and then the two-particle
antisymmetric states, in spin coupled representation, are 
$$
\l.\l| (i,j)\;s,\; m_s \r.\ran = \dis\frac{1}{\sqrt{1+\delta_{i,j}}}
\l(a_i^\dagger a_j^\dagger\r)^s_{m_s} \l|0\ran\;.
$$ 
In constructing EGOE(1+2)-$\cs$, only spin invariant Hamiltonians are
considered. Thus the $m$-particle states carry good spin($S$) quantum
number \cite{Ko-06,Al-06}.  Now the pair creation operator $P_i$ for the
level $i$ and the generalized  pair creation operator (over the $\Omega$
levels) $P$ are
\be
P =  \dis\frac{1}{\sqrt{2}} \dis\sum_i \l(a^\dagger_i 
a^\dagger_i\r)^0 = \dis\sum_i P_i\;,\;\;\;P^{\dagger}=-\dis\frac{1}
{\sqrt{2}
}
\dis\sum_i(\tilde{a_i}\tilde{a_i})^0 \;.
\label{ch3.eq.npa1}
\ee
In Eq. (\ref{ch3.eq.npa1}), $\tilde{a}_{i,\cs=\spin, m_\cs}=(-1)^{\spin+m_\cs} 
a_{i,\cs=\spin,-m_\cs}$. Therefore in the space defining EGOE(1+2)-$\cs$, 
the pairing Hamiltonian $H_p$ and its two-particle matrix elements are,
\be
\barr{l}
H_p = P^2 \; = \; PP^\dagger\;,\\
\lan (k,\ell)\; s, \; m_s \mid H_p 
\mid (i,j)\; s^\pr, \; m_{s^\pr} \ran = \delta_{s,0}\,\delta_{i,j} \, 
\delta_{k,\ell}\,\delta_{s,s^\pr}\,\delta_{m_s,m_{s^\pr}} \;.
\earr \label{ch3.eq.npa2}
\ee
Note that the two-particle matrix elements of $H_p$ (also true for $H$) are
independent of the $m_s$ quantum number. With this, we will proceed to
identify and analyze the pairing algebra. Firstly,  it is easily seen that
the 4$\Omega^2$ number of one-body operators $u_{\mu}^{r}(i,j) =
(a_i^{\dagger}\tilde{a_j})_{\mu}^{r}$, $r=0,1$ generate $U(2\Omega)$
algebra; see Appendix \ref{c3a1}. They satisfy the following 
commutation relations,
\be
\barr{rcl}
\l[ u_{\mu}^{r}(i,j), u_{\mu^{\pr}}^{r^{\pr}}(k,l) \r]_{-} =
\dis\sum_{r^{\pr\pr}} (-1)^{r+r^{\pr}}\lan
r\;\mu\;r^{\pr}\;\mu^{\pr}\;|\;r^{\pr\pr}\;\mu^{\pr\pr}\ran\sqrt
{(2r+1)(2r^{\pr}
+1)}\\ \\
\times \l\{\barr{ccc} r & r^{\pr} & r^{\pr\pr} \\ 1/2 & 1/2 &
1/2 \earr\r\}\;\l[ u_{\mu^{\pr\pr}}^{r^{\pr\pr}}(k,j)\delta_{il}-
(-1)^{r+r^{\pr}
+r^{\pr\pr}}u_{\mu^{\pr\pr}}^{r^{\pr\pr}}(i,l)\delta_{jk}\r] \;.
\earr \label{ch3.eq.pai2}
\ee
Here, $\lan \ldots \mid \ldots \ran$ are CG coefficients and $\l\{ \ldots 
\ldots \ldots \r\}$ are $6j$-symbols.
The $U(2\Omega)$ irreducible representations  (irreps) are denoted
trivially by the particle number $m$ as they must be antisymmetric irreps
$\{1^m\}$. The $2\Omega(\Omega-1)$  number of  operators $V_{\mu}^r(i,j)$,
\be
V_{\mu}^r(i,j) = \sqrt{(-1)^{r+1}}\l[
u_{\mu}^{r}(i,j)-(-1)^ru_{\mu}^{r}(j,i)\r]\;;\;\;\;\;i>j\,,\;\;r=0,\;1
\label{ch3.eq.npa4}
\ee
along with the $3\Omega$ number of operators $u_{\mu}^1(i,i)$ form 
$Sp(2\Omega)$ subalgebra of $U(2\Omega)$ and this follows from the results
in \cite{Ko-06b}. 
Using anti-commutation relations for fermion creation and destruction
operators and carrying out angular-momentum algebra \cite{Ed-74}, we have
\be
\barr{l}
\l[ (a_i^{\dagger}\tilde{a_j})^k(a_j^{\dagger}\tilde{a_i})^k\r]^0 \\\\ =
(-1)^k
\sqrt{\dis\frac{2k+1}{2j+1}}(a_i^{\dagger}\tilde{a_i})^0-\dis\sum_{k^{\pr}}
\chi\l\{\barr{ccc} 1/2 & 1/2 & k \\ 1/2 & 1/2 & k \\ k^{\pr} & k^{\pr} &
0 \earr\r\} 
[(a_i^{\dagger}a_j^{\dagger})^{k^{\pr}}(\tilde{a_j}\tilde{a_i})
^{k^{\pr}}]^0 \;, 
\earr \label{ch3.eq.pai5}
\ee
\be
\l[ (a_i^{\dagger}\tilde{a_j})^k(a_i^{\dagger}\tilde{a_j})^k\r]^0 = -
\chi\l\{\barr{ccc} 1/2 & 1/2 & k \\ 1/2 & 1/2 & k \\ 0 & 0 &
0 \earr\r\} (a_i^{\dagger}a_i^{\dagger})^0(\tilde{a_j}\tilde{a_j})^0 \;.
\nonumber
\label{ch3.eq.pai5a1}
\ee
Here, $\chi\l\{ \ldots \ldots \ldots \r\}$ are $9j$ coefficients (they are not 
$9j$-symbols). Note that 
\be
\barr{rcl}
\chi\l\{\barr{ccc} 1/2 & 1/2 & k \\ 1/2 & 1/2 & k \\ s & s &
0 \earr\r\} & = & \sqrt{\dis\frac{2s+1}{4}} \;\;\;\;\mbox{for}\; k=0 \;,\\
& = & \sqrt{\dis\frac{2s+1}{3}}\l[ \frac{3}{2}-s(s+1)\r]\;\;\;\;
\mbox{for}\;k=1 \;.
\earr \label{ch3.eq.pai6}
\ee
We will show that the irreps of $Sp(2\Omega)$ algebra are
uniquely labeled by the seniority quantum number `$\v$' discussed in the
context of identical particle pairing in nuclear structure \cite{Ta-93} and
they in turn define the eigenvalues of $H_p$. 
The quadratic Casimir 
operators of the $U(2\Omega)$ and $Sp(2\Omega)$ algebras are \cite{Ko-06b},
\be
\barr{rcl}
C_2[U(2\Omega)] & = & \dis\sum_{i,j,r}u^{r}(i,j)\cdot u^{r}(j,i) \;, \\ \\
C_2[Sp(2\Omega)] & = & 2\dis\sum_i u^1(i,i)\cdot
u^1(i,i)+\dis\sum_{i>j,r}V^r(i,j)\cdot V^r(i,j) \;.
\earr \label{ch3.eq.npa5}
\ee
Simplifying 
these expressions using relations in Eqs. (\ref{ch3.eq.pai5}) and 
(\ref{ch3.eq.pai6}) [with $\hat{n}$ being the number operator], we have
\be
\barr{l}
C_2[U(2\Omega)] =  2\hat{n}\Omega-2\dis\sum_i
P_iP_i^{\dagger}-\dis\sum_{i\neq j,s} \; \sqrt{2s+1} \; [s(s+1)-1]
\l[ \l(a_i^{\dagger}a_j^{\dagger}\r)^s \l(\tilde{a_j}\tilde{a_i} \r)^s \r]^0 \;,
\\ \\
C_2[Sp(2\Omega)] = (2\Omega+1)\hat{n}-6\dis\sum_iP_iP_i^{\dagger}-4
\dis\sum_{i>j}(P_iP_j^{\dagger}+P_jP_i^{\dagger}) 
\\ \\ 
- \dis\sum_{i\neq j,s}
\sqrt{2s+1} \; [s(s+1)-1] \; \l[ \l(a_i^{\dagger}a_j^{\dagger} \r)^s
\l(\tilde{a_j} \tilde{a_i}\r)^s \r]^0 \;,
\\ \\
\Rightarrow
C_2[U(2\Omega)] - C_2[Sp(2\Omega)] = 4\;PP^{\dagger} - \hat{n} \;.
\earr \label{ch3.eq.npa6}
\ee
It is also seen that the operators $P$, $P^{\dagger}$ and $P_0$ form 
$SU(2)$ algebra,
\be
[P,P^{\dagger}] = \hat{n} - \Omega = 2\;P_0\;,\;\;[P_0,P] =
P\;,\;\;[P_0,P^{\dagger}] = -P^{\dagger} \;.
\label{ch3.eq.npa7}
\ee
The corresponding spin is called quasi-spin $Q$. As $M_Q$, the $P_0$ 
eigenvalue, is $(m-\Omega)/2$, we obtain $Q=(\Omega-\v)/2$. Then, for 
$m\le \Omega$, $\v$ take values $\v=m, m-2, \ldots, 0$ or $1$. Therefore 
eigenvalues of the pairing Hamiltonian $H_p$ are given by,
\be
E_p(m,\v,S) = \lan H_p \ran^{m,\mbox{v},S} = \lan P P^\dagger 
\ran^{m,\mbox{v},S} =
\dis\frac{1}{4}(m-\v)(2\Omega+2-m-\v) \;.
\label{ch3.eq.npa8}
\ee
As $\lan C_2[U(2\Omega)]\ran^{\{1^m\}} = m(2\Omega+1-m)$, Eqs. 
(\ref{ch3.eq.npa6}) and (\ref{ch3.eq.npa8}) will give 
\be
C_2[Sp(2\Omega)] = 2\mbox{v}\l(\Omega+1-\dis\frac{\mbox{v}}{2}\r) \;.
\label{ch3.eq.npa9}
\ee
Comparing Eq. (\ref{ch3.eq.npa9}) with the general formula for the eigenvalues
of the quadratic Casimir invariant of $Sp(2\Omega)$ \cite{Wy-74}, 
it follows that the
seniority quantum number `$\v$' corresponds to totally antisymmetric 
irrep $\lan 1^{{\v}} \ran$ of $Sp(2\Omega)$. Thus $Sp(2\Omega)$ corresponds
to  $SU(2)$ quasi-spin algebra generated by ($P$, $P^{\dagger}$, $P_0$).
More explicitly,
\be
\l|\;m,\;\v,\;S,\;\alpha \ran =
\sqrt{\dis\frac{(\Omega-\v-p)!}{(\Omega-\v)!\;p!}}\;P^{p}\;
\l|\;m=\v,\;\v,\;
S,\;\alpha \ran\;;\;\;\;\;p=\dis\frac{m-\v}{2} \;.
\label{ch3.eq.npa10}
\ee
Thus the spin $S$ is generated by `$\v$' free particles and therefore 
$\v \geq 2S$. Then, for a given $(m,S)$ we have
\be
\v = m,\;m-2,\;\ldots,\;2S\;,\;\;\;\;(m\le\Omega) \;.
\label{ch3.eq.npa11}
\ee
Number of states or dimension $D(m,\v,S)$, without the $(2S+1)$ degeneracy 
factor, for a fixed-$(m, \v, S)$ is, 
\be
\barr{rcl}
D(m,\v,S) & = & d_f(\Omega,m=\v,S) - d_f(\Omega,m=\v-2,S)\;.
\earr \label{ch3.eq.npa12}
\ee
Note that the fixed-$(m,S)$ dimensions $d_f(\Omega,m,S)$ are given by Eq.
(\ref{eq.def2}).
Table \ref{c3t1} gives the reductions $m \rightarrow S \rightarrow \v$, 
$D(m,\v,S)$
and also $E_p(m,\v,S)$ for some examples. Let us point out 
$Sp(2\Omega)\supset  SO(\Omega)\otimes SU(2)$ but $SO(\Omega)$ carries no
extra information. In fact  there is one-to-one correspondence between the 
$Sp(2\Omega)$ chain and the alternative group-subgroup chain
$U(2\Omega)\supset U(\Omega)\otimes  SU(2) \supset SO(\Omega) \otimes
SU(2)$. This is verified by comparing  the  results in Table \ref{c3t1} 
with the
irrep reductions for $U(\Omega) \supset SO(\Omega)$ that are given in Appendix 
\ref{c3a1}. 
It is useful to note that Eqs.
(\ref{ch3.eq.npa8}),  (\ref{ch3.eq.npa11}) and (\ref{ch3.eq.npa12}) 
will allow one to 
construct the  state density generated by the pairing Hamiltonian $H=-H_p$.
The dimensions $d_f(\Omega,m,S)$ and $D(m,\v,S)$
along with the energy $E_p$ of $H_p$ will give the normalized density
$\rho(E)$ to be
\be
\rho_{(-H_p)}(E) = \dis\frac{D(m,\v,S)}{d_f(\Omega,m,S)\;\Delta E}
\;;\;\;\;\Delta E =
E_p(m,\v+1,S)-E_p(m,\v-1,S)=\Omega-\v+1\;.
\label{eq.dens}
\ee
Figure \ref{c3f1} gives $\rho(\we)$ vs $\we$ plot for $\Omega=22$ (i.e., 
$N=44$),  $m=22$ and $S=0$. For this system, the spectrum spread is 132 (note
that  $v_{max}=22$), centroid  $\epsilon \sim 5.7$ and width 
$\sigma \sim 6$;  note
that $\we=(E-\epsilon)/\sigma$. Clearly, it is a highly skewed distribution 
(see also the $\alpha=0$ plot in Fig. \ref{c3f4} ahead). 

\begin{table}[htp]
\caption{Classification of states in the $U(2\Omega)\supset Sp(2\Omega) 
\supset SO(\Omega)\otimes SU_S(2)$ limit for $\Omega = 6$ with $m = 0-6$ and
$\Omega=8$ with $m=6-8$. Given are ($m$, $S$, $\v$) labels, 
the corresponding dimensions $D(m,\v,S)$ and eigenvalues $E_p(m,\v,S)$. 
Note that $\sum_{\v,S}(2S+1)D(m,\v,S)={2\Omega \choose m}$ and 
$\sum_{\v} D(m,\v ,S)=d_f(\Omega,m,S)$.}
\begin{tabular}{cccccccccccc}
\toprule
$\Omega$ & $m$ & $S$ & $\v$ & $D(m,\v,S)$ & $E_p(m,\v,S)$ & $\Omega$ 
& $m$ & $S$ & $\v$ & $D(m,\v,S)$ & $E_p(m,\v,S)$\\
\midrule
$6$ & $0$ & $0$ & $0$ & $1$ & $0$ & $8$ & $6$ & $0$ & $6$ & $840$ & $0$ \\
 & $1$ & $\frac{1}{2}$ & $1$ & $6$ & $0$ & & & & $4$ & $300$ & $4$ \\
 & $2$ & $0$ & $2$ & $20$ & $0$ & & & & $2$ & $35$ & $12$ \\
 & & & $0$ & $1$ & $6$ & & & & $0$ & $1$ & $18$ \\
 & & $1$ & $2$ & $15$ & $0$ & & & $1$ & $6$ & $1134$ & $0$ \\
 & $3$ & $\frac{1}{2}$ & $3$ & $64$ & $0$ & & & & $4$ & $350$ & $4$ \\
 & & & $1$ & $6$ & $5$ & & & & $2$ & $28$ & $10$ \\
 & & $\frac{3}{2}$ & $3$ & $20$ & $0$ & & & $2$ & $6$ & $350$ & $0$ \\
 & $4$ & $0$ & $4$ & $84$ & $0$ & & & & $4$ & $70$ & $4$ \\
 & & & $2$ & $20$ & $4$ & & & $3$ & $6$ & $28$ & $0$ \\
 & & & $0$ & $1$ & $10$ & & $7$ & $\frac{1}{2}$ & $7$ & $1344$ & $0$ \\
 & & $1$ & $4$ & $90$ & $0$ & & & & $5$ & $840$ & $3$ \\
 & & & $2$ & $15$ & $4$ & & & & $3$ & $160$ & $8$ \\
 & & $2$ & $4$ & $15$ & $0$ & & & & $1$ & $8$ & $15$ \\
 & $5$ & $\frac{1}{2}$ & $5$ & $140$ & $0$ & & & 
 $\frac{3}{2}$ & $7$ & $840$ & $0$ \\
 & & & $3$ & $64$ & $3$ & & & & $5$ & $448$ & $3$ \\
 & & & $1$ & $6$ & $8$ & & & & $3$ & $56$ & $8$ \\
 & & $\frac{3}{2}$ & $5$ & $64$ & $0$ & & & $\frac{5}{2}$ & $7$ & $160$ & 
 $0$ \\
 & & & $3$ & $20$ & $3$ & & & & $5$ & $56$ & $3$ \\
 & & $\frac{5}{2}$ & $5$ & $6$ & $0$ & & & $\frac{7}{2}$ & $7$ & $8$ & 
 $0$ \\
 & $6$ & $0$ & $6$ & $70$ & $0$ & & $8$ & $0$ & $8$ & $588$ & $0$ \\
 & & & $4$ & $84$ & $2$ & & & & $6$ & $840$ & $2$ \\
 & & & $2$ & $20$ & $6$ & & & & $4$ & $300$ & $6$ \\
 & & & $0$ & $1$ & $12$ & & & & $2$ & $35$ & $12$ \\
 & & $1$ & $6$ & $84$ & $0$ & & & & $0$ & $1$ & $20$ \\
 & & & $4$ & $90$ & $2$ & & & $1$ & $8$ & $840$ & $0$ \\
 & & & $2$ & $15$ & $6$ & & & & $6$ & $1134$ & $2$ \\
 & & $2$ & $6$ & $20$ & $0$ & & & & $4$ & $350$ & $6$ \\
 & & & $4$ & $15$ & $2$ & & & & $2$ & $28$ & $12$ \\
 & & $3$ & $6$ & $1$ & $0$ & & & $2$ & $8$ & $300$ & $0$\\
 & & & & & & & & & $6$ & $350$ & $2$ \\
 & & & & & & & & & $4$ & $70$ & $6$ \\
 & & & & & & & & $3$ & $8$ & $35$ & $0$ \\
 & & & & & & & & & $6$ & $28$ & $2$ \\
 & & & & & & & & $4$ & $8$ & $1$ & $0$ \\
\bottomrule
\end{tabular}
\label{c3t1}
\end{table}

\section{Fixed-$(m,\v,S)$ Partial Densities and their Centroids and 
Variances}
\label{c3s2}

Expansion of a given $\l.\l|m,\v,S,\alpha \r.\ran$ basis state in terms of the
$H$ eigenstates $\l.\l|E; (m,S)\r.\ran$, with the expansion coefficients being
$C_{E}^{m,\v,S,\alpha}$ will allow us to define the fixed-$(m,\v,S)$ partial
densities $\rho^{m,\v,S}(E)$,
\be
\barr{l}
\rho^{m,\v,S}(E) = \lan \delta(H-E)\ran^{m,\v,S} = 
\dis\frac{1}{D(m,\v,S)}\;\dis\sum_{\alpha}\l|
C_{E}^{m,\v,S,\alpha}\r|^2 \\ \\
\Rightarrow I^{m,\v,S}(E) = D(m,\v,S) \rho^{m,\v,S}(E) = 
\dis\sum_{\alpha}
\l|C_{E}^{m,\v,S,\alpha}\r|^2 \;.
\earr \label{ch3.eq.pa8a}
\ee
Often it is convenient to use total densities $I(E)$ rather than the 
normalized densities $\rho(E)$. It is important to note that fixed-$(m,S)$ 
density of states $\rho^{m,S}(E)$ decompose into a sum of fixed-$(m,\v,S)$ 
partial densities, 
\be
\barr{rcl}
\rho^{m,S}(E) & = & \dis\sum_{\v} \;\dis\frac{D(m,\v,S)}{d_f(\Omega,m,S)}\;
\rho^{m,\v,S}(E)
\\ \\
\Rightarrow I^{m,S}(E) & = & \dis\sum_{\v} I^{m,\v,S}(E) \;. 
\earr \label{ch3.eq.pa8b}
\ee
The partial densities are defined over broken symmetry subspaces and they
are also called `strength functions' or `local density of states'
\cite{Ko-01,Ko-03}. Partial densities $\rho^{m,\v,S}(E)$ give intensity
distribution  of a given basis state $\l| m,\v,S \ran$
over the eigenstates  $\l| E \ran$,  i.e., distribution of the expansion coefficients 
$|C_{E}^{m,\Gamma}|^2$ vs $E$. 
The partial densities have same  structure as 
that for the strength functions
defined in Eq. (\ref{eq.wf2}) as partial sums over the strength functions give
partial densities.
We will also encounter partial densities in Chapter \ref{ch5}. 

\begin{figure}
\centering
\includegraphics[width = 5.5in, height = 6in]{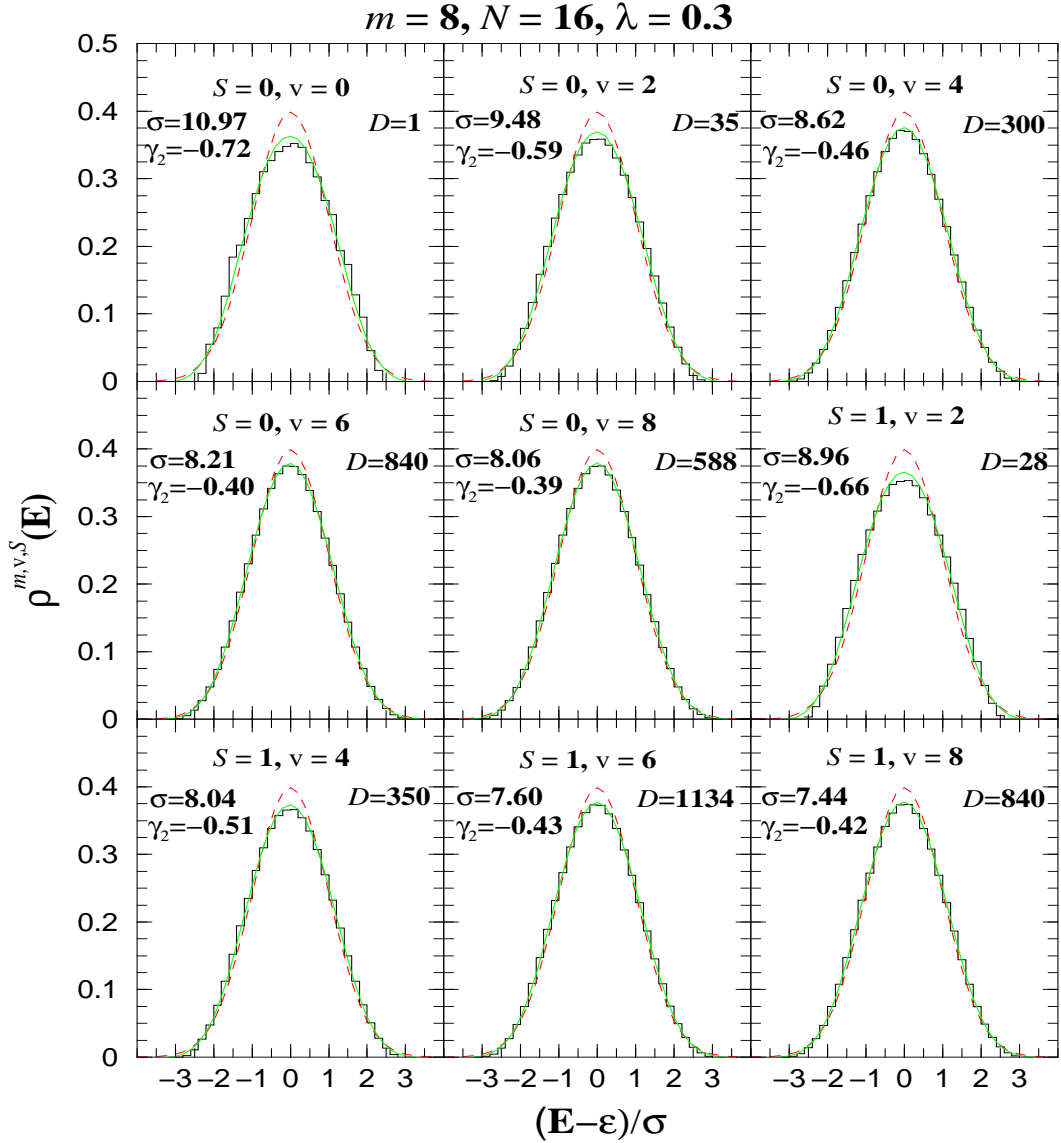}
\caption{Partial densities $\rho^{m,{\v},S}(E)$  vs $E$ for a  
EGOE(1+2)-$\cs$ ensemble $H$ defined in the text. The values of $({\v},S)$,
dimension $D$, width $\sigma$ and $\gamma_2$ for the densities are given in
the figure. Note that $\gamma_1\sim 0$ in all cases. The energies $E$ are
zero centered with respect to the centroid $\epsilon$ and scaled with the width
$\sigma$ of $\rho^{m,{\v},S}(E)$. The histograms (with $0.2$ bin size) are
exact results, dashed curves are Gaussians and the continuous curves are
Edgeworth corrected Gaussians. See text for 
further details.}
\label{c3f2}
\end{figure}

In the $\lambda > \lambda_F$ region, as discussed in Chapter \ref{ch2}, strength
functions take Gaussian form and therefore partial densities are expected to be
Gaussian in this region. Extension of this result to
EGOE(1+2)-$J$ \cite{Pa-07}  with  subspaces defined  by the pairing
Hamiltonian, i.e., fixed-$(m,\v,J)$ partial densities are Gaussian is often 
used in nuclear physics \cite{Qu-74,Qu-77}. In Fig. \ref{c3f2} 
we present tests of this 
assumption for EGOE(1+2)-$\cs$ with $J$ replaced by $S$. 
In order to discuss these results, we will
start with the EGOE(1+2)-$\cs$ Hamiltonian defined by Eq. (\ref{eq.def1}).
We choose, in all the calculations reported
in this chapter,  $\epsilon_i = i+(1/i)$, $i = 1, 2, \ldots, \Omega$ and 
$\lambda_0 = \lambda_1 = \lambda = 0.3$. Results in Chapter \ref{ch2}
confirm that $\lambda=0.3$ corresponds to strong coupling region. Before
going further, let us add that we will later consider extensions of $H$ with
the inclusion of pairing and exchange interactions (they are not random).
For a $\Omega=m=8$ system with 50 members, we have extracted the partial
densities $\rho^{m,\v,S}(E)$ in Eq. (\ref{ch3.eq.pa8a}) by numerically
constructing the $H$ matrix in  good $S$ basis and then changing it into
good $(m,\v,S)$  basis with an auxiliary diagonalization of $H_p$ in the
good $S$ basis. Results for the ensemble averaged partial densities are
shown for $S=0$ and $1$ in Fig. \ref{c3f2} 
and the results are compared with the
Gaussian ($\cg$) and ED corrected Gaussian forms given by Eq.
(\ref{eq.gau1}).
From the results in Fig. \ref{c3f2}, it is seen that the agreement between the  
exact and ED corrected Gaussians is excellent. For $\v=0$ (this is one 
dimensional) the deviations are some what larger. Similar results are also
obtained for a smaller example (these are not shown in the figure) with
$\Omega=m=6$ and $S=0,1$ and for this system we have carried out calculations
with 500 members. This shows fixed-$(m,\v,S)$ partial densities take close
to Gaussian form, just as fixed-$(m,S)$ densities, in the strong coupling
region. Thus the EGOE(1+2)-$\cs$ densities follow EGOE(1+2) even in pairing
subspaces. This is a result assumed in statistical nuclear spectroscopy
(see for example \cite{Mom-80,Fr-82}).
   
For constructing Gaussian partial $(m,\v,S)$ densities, we need
fixed-$(m,\v,S)$ centroids $E_c(m,\v,S) = \lan H \ran^{m,\v,S}$ and
variances $\sigma^2(m,\v,S) = \lan H^2\ran^{m,\v,S} - [E_c(m,\v,S)]^2$. An
important result here is, these parameters can be calculated for any
$(\Omega ,m, \v, S)$ with $m > 4$ without recourse to $H$ matrix 
construction in $(m,S)$ spaces. This derives from the fact that simple
(Casimir) propagation  is possible for  $E_c(m,\v,S)$ in terms of the
corresponding $E_c$ for $m \leq 2$ and for $\sigma^2(m,\v,S)$ in terms of 
the corresponding $\sigma^2$ for $m \leq 4$. From Table \ref{c3t1} 
one can see that
the  number of  $(m,\v,S)$ irreps $\Lambda_i$ is 5 for $m$ up to 2 and there
are 5 simple scalar operators $\widehat{C}_i$ of maximum body  rank 2,
$\widehat{C}_i = 1$, $\hat{n}$, ${\hat{n} \choose 2}$, $H_p$, and 
$\hat{S}^2$ for $i=1-5$, respectively. Note that $\lan H_p\ran^{m,\v,S}$ and
$\lan \hat{S}^2\ran^{m,\v,S}$ are  $E_p(m,\v,S)$ [see Eq. (\ref{ch3.eq.npa8})]
and $S(S+1)$, respectively.   More remarkable is that, for $m \leq 4$, the
number of  $(m,\v,S)$ irreps $\Upsilon_i$ is $14$ as seen from Table \ref{c3t1}
and
also the available simple scalars $\widehat{\cac}_i$ of maximum body rank 4
are exactly $14$. These are  $\widehat{\cac}_i=1$,  $\hat{n}$, ${\hat{n}
\choose 2}$, ${\hat{n} \choose 3}$, ${\hat{n} \choose 4}$, $H_p$,
$\hat{n}H_p$, ${\hat{n} \choose 2} H_p$,  $(H_p)^2$, $H_p \hat{S}^2$,
$\hat{S}^2$, $\hat{n}\hat{S}^2$,  ${\hat{n} \choose 2} \hat{S}^2$ and
$(\hat{S}^2)^2$ for $i=1-14$, respectively. Therefore, the spectral
variances over $(m,v,S)$ spaces propagate simply and they will be
linear combinations of the eigenvalues of the 14 operators above. 
The constants in the
expansion will follow from the variances for $m \leq 4$. 
Then, fixed-$(m,\v,S)$ energy centroids with $\cas^2=S(S+1)$, 
$X(m,S)=m(m+2)-4S(S+1)$ and $Y(m,S)=m(m-2)-4S(S+1)$, are given by
\be
\barr{l}
\lan H \ran^{m,\v,S} = E_c(m,v,S) = a_0 + a_1 m + a_2 \dis\binom{m}{2} + 
a_3\cas^2 + a_4 E_p(m,v,S) \\ \\
\Rightarrow  E_c(m,v,S) = \dis\frac{1}{2}(m-1)(m-2)\;
E_c(0,0,0) - m(m-2)\;E_c(1,1,\frac{1}{2})
\earr \label{ch3.eq.pa8p}
\ee
\be
\barr{l}
+ \dis\frac{1}{8 \Omega} \l[ -8 E_p(m,v,S)+\Omega X(m,S) \r]
E_c(2,2,0) + \dis\frac{1}{\Omega}\;E_p(m,v,S)\;E_c(2,0,0) \\
\\
+ \dis\frac{1}{8} \l[4m(m-2) - Y(m,S)\r]\;E_c(2,2,1) \;. \nonumber
\earr \label{ch3.eq.pa8pa1}
\ee
Similarly, fixed-$(m,\v,S)$ spectral variances are
\be
\barr{l}
\lan H^2 \ran^{m,\v,S} = \dis\frac{1}{24} (m-1)(m-2)(m-3)(m-4)\;\lan 
H^2\ran^{0,0,0} \\ \\
- \dis\frac{1}{6} m(m-2)(m-3)(m-4)\; \lan H^2\ran^{1,1,\frac{1}{2}} \\ \\
+ \dis\frac{1}{16\Omega} (m-3)(m-4)\l[\Omega X(m,S)-8E_p(m,v,S)\r]\;
\lan H^2\ran^{2,2,0} \\ \\
+ \dis\frac{1}{2\Omega} (m-3)(m-4)E_p(m,v,S)\; \lan H^2
\ran^{2,0,0} \\  \\
+  \dis\frac{1}{16} (m-3)(m-4)[3m(m-2)+4\cas^2]\;\lan H^2
\ran^{2,2,1} \\  \\
- \dis\frac{(m-2)(m-4)}{12(\Omega-1)} \l[(\Omega-1) X(m,S)-
12E_p(m,v,S)\r] \;\lan H^2\ran^{3,3,\frac{1}{2}} \\  \\
- \dis\frac{1}{\Omega-1} (m-2)(m-4)E_p(m,v,S)\;\lan H^2
\ran^{3,1,\frac{1}{2}} \\  \\
- \dis\frac{1}{12} (m-2)(m-4)[m(m-4)+4\cas^2]\;\lan H^2
\ran^{3,3,\frac{3}{2}} \\ \\
+ \dis\frac{1}{192(\Omega-2)(\Omega-1)}\l[96\l\{E_p(m,v,S)\r\}^2+ 
24 \l\{-(\Omega-1)m^2+2(\Omega+1)m \r.\r. 
\\ \\
+\l. 4(\Omega-1)\cas^2 - 4(\Omega+2)\r\}E_p(m,v,S) \\ \\
\l. + (\Omega-1)(\Omega-2) X(m,S) Y(m,S)\r]\;
\lan H^2\ran^{4,4,0} \\ \\
+ \dis\frac{1}{8\Omega(\Omega-2)}E_p(m,v,S)\\ \\
\times \l[\Omega\{m(m-2)-4\cas^2 +8\} 
-8\l\{E_p(m,v,S)+m-2\r\}\r]\; \lan H^2 \ran^{4,2,0} 
\earr \label{ch3.eq.pa8}
\ee
\be
\barr{l} 

\begin{figure}
\centering
\includegraphics[width = 5.5in, height = 6in]{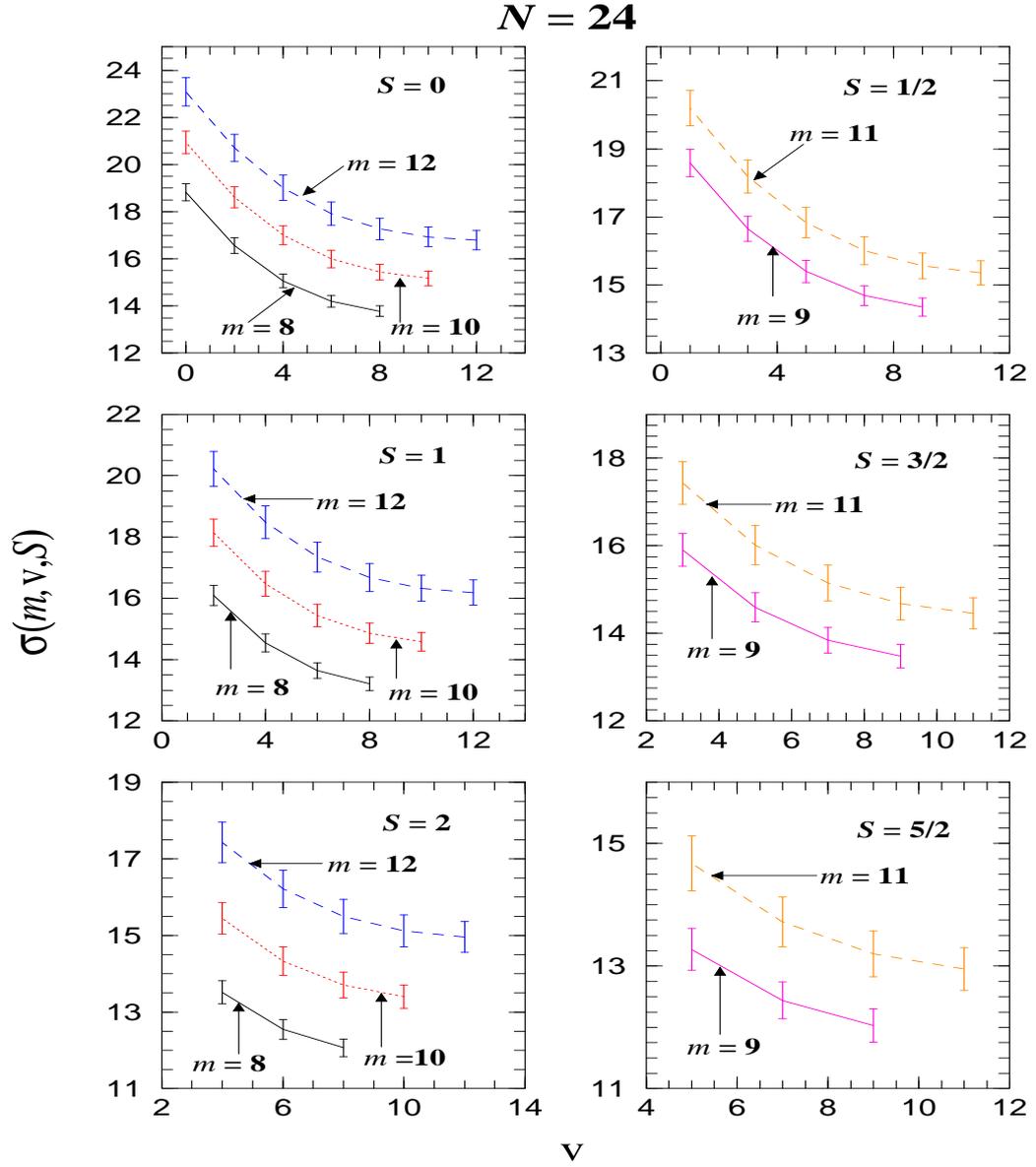}
\caption{Ensemble averaged widths $\sigma(m,{\v},S)$ vs `$\v$' for  
EGOE(1+2)-$\cs$ ensembles with  $\Omega=12$ and $(m,S)$ values as given in
the figures. Shown also in the figures are the r.m.s. deviations 
(over the ensemble)
in the widths as  error bars. For $m=12$, the results are shifted by one
unit to avoid overlapping of the error bars. See text for details.} 
\label{c3f3}
\end{figure}

+ \dis\frac{1}{2\Omega(\Omega-1)} E_p(m,v,S)
\l\{E_p(m,v,S)+m-\Omega-2\r\}\; \lan H^2\ran^{4,0,0} \\ \\
+ \dis\frac{\l[3m(m-6)+4\cas^2+24\r]}{128(\Omega -2)} \\ \\
\times \;\l[(\Omega-2) X(m,S)-16E_p(m,v,S)\r]\;
\lan H^2\ran^{4,4,1}\\ \\
+  \dis\frac{1}{8(\Omega-2)}E_p(m,v,S) \l[3m(m-6)+
4\cas^2+24\r]\; \lan H^2\ran^{4,2,1} \\ \\
+ \dis\frac{1}{384} \l[16(\cas^2)^2 + 40m^2\cas^2 -240m\cas^2 + 
288\cas^2 \r.\\ \\
\l. + 5m(m-2)(m-4)(m-6) \r]\;\lan H^2\ran^{4,4,2} \;.\nonumber
\earr \label{ch3.eq.pa8aa}
\ee
Using EGOE(1+2)-$\cs$ computer codes, it is easy to construct, even for
large $\Omega$ values, the input averages $\lan H
\ran^{m,\v,S}$, $m \leq 2$ for centroids and $\lan H^2\ran^{m,\v,S}$, 
$m \leq 4$ for variances propagation. For a 100 member ensemble with
$\Omega=12$ and $m$ changing from $8$ to $12$, we have calculated, for
three lowest spins (i.e., for even $m$, with $S=0$, $1$ and $2$ and  odd $m$
with $S=\frac{1}{2}$, $\frac{3}{2}$ and $\frac{5}{2}$), the ensemble
averaged variances using Eq. (\ref{ch3.eq.pa8}). 
We use the EGOE(1+2)-$\cs$ Hamiltonian defined by Eq. (\ref{eq.def1})
with  $\lambda=0.3$. The ensemble
averaged centroids  do not change with `$\v$' as expected and therefore we
will discuss the structure of variances. The results are shown in Fig.
\ref{c3f3}.
It is observed that  as the `$\v$' value increases from $2S$ to $m$, there
is decrease in the  variances. However the dimensions increase as `$\v$'
increases. For example, for $S=0$ and $m=10$ the widths and dimensions are
$(\sigma,D)=  (20.93,1)$, $(18.6,77)$, $(17,1638)$, $(16,14014)$,
$(15.44,55055)$, and  $(15.17,99099)$ for $\v=0,2,\ldots,10$. The decrease
in variances with increasing `$\v$' is necessary for the gs to be
dominated by $\v=0$, i.e., by pairing structure. As we shall discuss later,
this indeed happens. Going beyond the averages, we have also calculated the
variation over the ensemble for both centroids and variances as they will
give information about fluctuations and ergodicity
\cite{Br-81,Be-01,Ko-07}. We have calculated the ensemble variances for
these, say $\cav^2[E_c(m,\v,S)]$ and $\cav^2[\sigma^2(m,\v,S)]$,
respectively and then the  corresponding scaled widths 
$\Delta_c(m,\v,S)=\cav[E_c(m,\v,S)]/\{\overline{\sigma^2(m,\v,S)}\}^{1/2}$
and $\Delta_s(m,\v,S)=\cav[\sigma^2(m,\v,S)]/\overline{\sigma^2(m,\v,S)}$.
It is observed that $\Delta_c$ varies from $5-7$\% for $m=8$, $7-9$\% for
$m=9$, $8-10$\% for $m=10$, $9-13$\% for $m=11$ and $10-14$\% for $m=12$.
Thus centroid fluctuations are large just as the situation with EGOE for
spinless fermion systems \cite{Br-81,Be-01}. However the variance
fluctuations, as given by $\Delta_s$ are small, $\lazz 5$\%. Therefore the
widths are $\sigma(m,\v,S) \sim \l[\overline{\sigma^2(m,\v,S)}\r]^{1/2} 
\{1 \pm \frac{\Delta_s}{2}\}$. In Fig. \ref{c3f3} 
shown also are the fluctuations in widths.

\section{Expectation Values $\lan P P^\dagger\ran^E$ of the Pairing Operator
as Signature of Chaos}
\label{c3s3}

A series of studies in the past, using Gamow-Teller, electric quadrupole and
magnetic dipole transition operators, have established that transition  strength
sums can be considered as a statistic able to distinguish between regular  and
chaotic motion \cite{Ko-99a,Go-01,Go-03}. Moreover, for EGOE(1+2) for spinless
fermions in the strong coupling region, it is well understood that the strength
sums vary with the excitation energy as ratio of two Gaussians
\cite{FKPT,KS-00,Ko-01,Ko-03}. This result was derived using the fact that
(proved using the so-called  binary correlation approximation) the transition
strength densities, transition strengths multiplied by the state densities at
the two energies involved, for EGOE(1+2) with $\lambda > \lambda_F$,  take
bivariate Gaussian form and hence, being the marginal densities, the strength
sum densities (see ahead for the definition) will be Gaussian; see Chapter \ref{ch7} for
transition strength densities. It is now well
established that the  EGOE(1+2) (but not the GOE) provides a good description of
strength sums in nuclear shell-model in the chaotic domain
\cite{Ko-99a,Go-01,Go-03}.  Our interest is in calculating the expectation value
of $PP^\dagger$ over fixed-$(m,S)$ spaces, which is a measure of the pairing
correlations, and this is nothing but the strength sum for pair removal,
\be
\lan PP^\dagger\ran^{m, S, E} = \lan m, S, E \mid P P^\dagger \mid m, S, E 
\ran = \dis\sum_{E_f} \l|\lan m-2, S, E_f \mid P^\dagger \mid m, S, E\ran 
\r|^2 \;.
\label{ch3.eq.pa9}
\ee
Recently Horoi and Zelevinsky \cite{Ho-07} re-emphasized, in the
context of pairing correlations in nuclei, the importance of $\lan PP^\dagger\ran^E$
measure. Given a transition operator  $\co$, in terms of the
transition strength sum density $\rho_{\coo}(E)$, the
expectation value $\lan \co^\dagger \co \ran^E$ is
\be
\lan \coo \ran^E = \dis\frac{\l< \coo \delta\l(H-E\r)\r>}{\rho(E)}
= \lan \coo \ran\;\rho_{\coo}(E)/\rho(E) \;.
\label{ch3.eq.pa12}
\ee
As stated before, the normalized  $\coo$-density $\rho_{\coo}$ also takes 
Gaussian form for EGOE(1+2) with $\lambda > \lambda_F$ and it is defined by
the centroid $\epsilon_{\coo} = \lan \coo H \ran / \lan \coo \ran$  and
variance $\sigma_{\coo}^2= \lan \coo H^2 \ran / \lan \coo \ran
-\epsilon_{\coo}^2$. Similarly, skewness $\gamma_1(\coo)$ and excess 
$\gamma_2(\coo)$ for 
the $\coo$-density are defined. The normalization factor $\lan \coo\ran$ is
the average value of $\coo$ over the complete space [in our examples this
is fixed-$(m,S)$ space]. Therefore the ensemble averaged strength sum
density reduces to the ratio of two Gaussians or two ED corrected
Gaussians \cite{FKPT,KS-00,Ko-01,Ko-03},
\be
\barr{rcl}
\dis\frac{\lan \coo \ran^E}{\lan \coo \ran} & \stackrel{{\mbox{EGOE(1+2)}}} 
\longrightarrow & \rho_{\coo;{\cal G}}(E) / \rho_{\cal G}(E) \longrightarrow
\rho_{\coo; ED}(E) / \rho_{ED}(E) \;.
\earr \label{ch3.eq.pa13}
\ee 
We will now test how well the EGOE(1+2) theory given by Eq. (\ref{ch3.eq.pa13})
extends to systems with spin, i.e., for EGOE(1+2)-$\cs$ in the strong
coupling regime  and for the operator $\co=P^\dagger$. Note that in
applying Eq. (\ref{ch3.eq.pa13}),  all the averages and the densities will be
over fixed-$(m,S)$ spaces. As $H_p=PP^\dagger$ generates a highly skewed
distribution for density of states, a priori it is expected that Eq.
(\ref{ch3.eq.pa13}) may not be a good statistical formula for $\lan
PP^\dagger\ran^{m,S,E}$. Now we will investigate this using three numerical
examples.

\begin{figure}
\centering
\includegraphics[width = 5.5in, height = 7in]{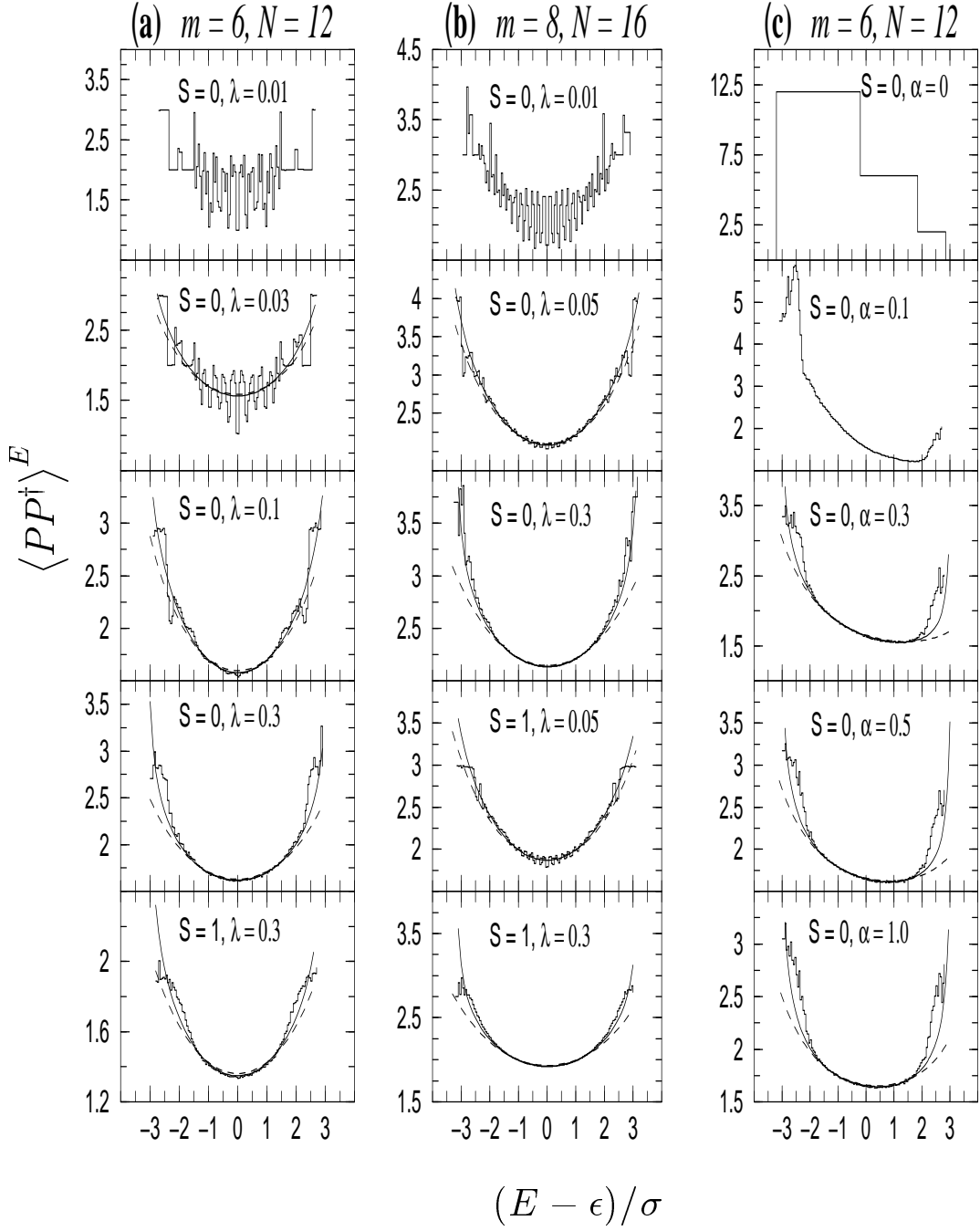}
\caption{Ensemble averaged pairing expectation value  $\lan P
P^\dagger\ran^{m,S,E}$  vs $E$ for  3 different EGOE(1+2)-$\cs$ examples.
(a) For various  values of $\lambda$ in Eq. (\ref{eq.def1}) with
$\Omega=m=6$ and $S=0,1$. (b)  For various values of $\lambda$ in Eq.
(\ref{eq.def1}) with $\Omega=m=8$ and $S=0,1$. (c) For various values of
$\alpha$ in  Eq. (\ref{ch3.eq.expec1}) with $\Omega=m=6$ and $S=0$. Results are
compared with the  EGOE(1+2) formula given by Eq. (\ref{ch3.eq.pa13}), using
Gaussian (dashed curves) and Edgeworth corrected Gaussian (solid curves)
forms. The energies $E$ are zero centered with respect to the centroid
$\epsilon$ and scaled with the width $\sigma$ of $\rho^{m,S}(E)$. See text
for details.} 
\label{c3f4}
\end{figure}

Just as in Section \ref{c3s2}, first we have used the random EGOE(1+2)-$\cs$
Hamiltonian defined in Eq. (\ref{eq.def1}) and calculated $\lan P
P^\dagger\ran^{m,S,E}$ for various values of the  $\lambda$ parameter using a
500 member ensemble for 6 fermions ($m=6$) in 6 orbits ($\Omega=6)$ and total
spins $S=0$ and $1$. Results are shown in Fig. \ref{c3f4}(a). Numerical results
are compared with the EGOE(1+2) formula given by  Eq. (\ref{ch3.eq.pa13}) both
with and without ED corrections.  For $\lambda=0.1$, we have
$\epsilon_{PP^\dagger} \sim 0$, $|\gamma_1(PP^\dagger)| \sim 0$,  
$\hat{\sigma}_{PP^\dagger} =  \sigma_{PP^\dagger}/\sigma_H \sim 1.07$,
$\gamma_2(PP^\dagger) \sim -0.47$   and $\lan PP^\dagger\ran^{m,S} \sim 1.71$
for $S=0$. Similarly, for $\lambda=0.3$,  $\gamma_2(PP^\dagger) \sim -0.55$ for
$S=0$ and $\sim -0.63$ for $S=1$. Large values of $\gamma_{2}(PP^\dagger)$ imply
that ED corrections are important  in the examples considered and this is
clearly seen in Fig. \ref{c3f4}(a).  The average value of $PP^\dagger$ follows 
easily from the centroid formula given by Eq. (\ref{ch3.eq.pa8p}),
\be
\lan PP^\dagger \ran^{m,S} = \dis\frac{2}{\Omega+1}\l\{ \dis\frac{m(m+2)}{8} -
\dis\frac{S(S+1)}{2}\r\}
\label{ch3.eq.expec}
\ee
and this has been used to verify numerical calculations. As expected, the
EGOE(1+2) smoothed form is not a good approximation to the exact  results in
the case of regular motion. Here there are large fluctuations due to
approximate good quantum numbers and the level fluctuations will be close
to  that of Poisson.  However, as $\lambda$ increases and after the onset of
chaos, in our example  for $\lambda \gazz 0.1$, the interacting particle 
system is chaotic, giving a smoothed form for pair transfer strength sums 
(with fluctuations following GOE). This behavior is clearly seen in Fig. 
\ref{c3f4}(a).
To strengthen these observations, calculations are repeated for a 50 member
EGOE(1+2)-$\cs$ ensemble with  $\Omega=m=8$ and  total spins $S=0$ and $1$.
The results are shown in  Fig. \ref{c3f4}(b).  
In this example, for  $\lambda \gazz
0.05$  [note that for EGOE(1+2) there is scaling by $\sim  1/(m^2 \Omega)$],
the EGOE(1+2) form is in good agreement with numerical results. For
$\lambda=0.05$, we have  $\epsilon_{PP^\dagger} \sim 0$,
$|\gamma_1(PP^\dagger)| \sim 0$,    $\hat{\sigma}_{PP^\dagger} \sim 1.06$,
$\gamma_2(PP^\dagger) \sim -0.33$  and $\lan PP^\dagger\ran^{m,S}\sim 2.22$
for $S=0$. Similarly, $\gamma_2(PP^\dagger) \sim -0.37$  and   $\lan
PP^\dagger\ran^{m,S}\sim 2$ for $S=1$. For $\lambda=0.3$, we have
$\gamma_2(PP^\dagger)  \sim -0.44$ for  $S=0$ and $\sim -0.47$ for $S=1$.
Thus, as seen from Figs. \ref{c3f4}(a) and \ref{c3f4}(b), 
pair expectation values follow, in the
chaotic domain the simple EGOE(1+2)  law given by Eq. (\ref{ch3.eq.pa13}). Also
it is seen from the figures that at low energies the pair expectation value
is large (but still much smaller than that for the pure pairing Hamiltonian)
and then decreases as we go to the center (after that it will again increase
as the space is finite). This trend is easily understood from the fact that
$\hat{\sigma}_{PP^\dagger} > 1$. Also expectation values in the  gs 
domain for $S=0$ are always larger than for $S=1$ and this is
consistent with previously known results \cite{Ho-07}. Thus random
interactions, even in the chaotic domain, exhibit strong pairing
correlations in the gs region and they decrease as we go up in the
energy. Perhaps this explains the preponderance of $0^+$ ground states seen
in nuclear shell-model examples \cite{Zel-04}. 

Going further, to understand the interplay between random interactions and
pairing, calculations are carried out for $\lan PP^\dagger\ran^{m,S,E}$
using  the Hamiltonian, 
\be
H = \alpha\l[ \{V^{s=0}\}+ \{V^{s=1}\}\r]+ \l[-H_p/\Omega\r] \;,
\label{ch3.eq.expec1}
\ee
which explicitly contains the pairing part. Here we divide $H_p$ by $\Omega$
so that the pairing gap (the gap between $\v=0$ and $\v=2$ states generated by
$H_p$) is unity. Therefore the parameter $\alpha$ in Eq. (\ref{ch3.eq.expec1})
is the strength of the random part of the Hamiltonian in units of the
pairing gap. Using a 500 member EGOE(1+2)-$\cs$ ensemble, with $H$ given by
Eq. (\ref{ch3.eq.expec1}), for $\Omega=m=6$ and $S=0$,  pair transfer
strength sums are calculated as a function of energy for  various $\alpha$
values. Results are shown in Fig. \ref{c3f4}(c). For $\alpha=0$, 
we have pure pairing
Hamiltonian and this generates a staircase function. As the value of the
strength of the random part increases to $\alpha > 0.3$, there is a
transition to chaotic domain with $\lan P P^\dagger\ran^{m,S,E}$ vs $E$ taking a
smoothed form (fluctuations being small and tending to that of GOE). The
smooth behavior observed for  $\alpha \geq 0.5$ is explained to some extent 
by Eq. (\ref{ch3.eq.pa13}). For better description  we use an expression (its
derivation being straightforward) based on partial $(m,\v,S)$-densities,
\be
\lan P P^\dagger\ran^{m,S,E} = \dis\sum_{\v}
\dis\frac{I^{m,\v,S}_{ED}(E)}{I^{m,S}_{ED}(E)}\;\lan P P^\dagger
\ran^{m,\v,S}\;.
\label{ch3.eq.expec2}
\ee
Note that the formula for $\lan P P^\dagger\ran^{m,\v,S}$ is given by Eq. 
(\ref{ch3.eq.npa8}) and $I^{m,S}_{ED}(E)$ is sum of $I^{m,\v,S}_{ED}(E)$. 
Following  Section \ref{c3s2}, 
we have constructed the partial densities appearing in
Eq. (\ref{ch3.eq.expec2}) as ED corrected Gaussians. The results obtained with
these are shown in Fig. \ref{c3f4}(c). 
It is seen that the agreements even at the
spectrum ends are good (without partitioning the expectation values are found
to be much larger than the exact results).  It can be concluded from Fig. 
\ref{c3f4}(c)
that for $\alpha$ of the order 0.5 times the pairing gap, pairing effects
get washed out and the structure of the expectation values is well explained
by the EGOE(1+2) smoothed formula (\ref{ch3.eq.expec2}). It is plausible that
unlike Eq. (\ref{ch3.eq.pa13}) that has worked well for the Hamiltonian defined
by Eq. (\ref{eq.def1}), the partitioned version given by Eq.
(\ref{ch3.eq.expec2}) should be used for the Hamiltonian defined by Eq.
(\ref{ch3.eq.expec1}) as this explicitly involves $H_p$, i.e., 
a regular part (as
already discussed, $H_p$  produces highly skewed density of states).

Partial densities give information about the composition, in terms of the `$\v$'
quantum number, of the wavefunctions for a given $E$. Note that
$f(\v)=I^{m,\v,S}(E)/I^{m,S}(E)$ gives the fractional intensity of states with a
given `$\v$' in the eigenstate with energy $E$; see Eqs. (\ref{ch3.eq.pa8a}) 
and (\ref{ch3.eq.pa8b}). For the Hamiltonian in Eq. (\ref{ch3.eq.expec1}) with
$\alpha=0.3$, for $\hat{E}=-3$, the $f(\v)$ for $\v=0,2,4$ and 6 are 16\%, 34\%,
33\%, and 17\%, respectively. However for the  random Hamiltonian given by Eq.
(\ref{eq.def1}) with $\lambda=0.3$, the $f(\v)$ values are 7\%, 33\%, 42\%, and
18\%.  Thus in the gs domain, although the pair  expectation values are enhanced
(see Fig. \ref{c3f4}),  the wavefunctions have  relatively small strength for
$\v=0$ states, i.e., they are not close to pure $H_p$ eigenstates. This result
is consistent with the nuclear shell-model results with random interactions,
possessing $J$-symmetry, presented in \cite{Zh-04}. Thus, some essential
features of EGOE(1+2)-$J$ are reproduced by EGOE(1+2)-$\cs$. 

\section{Distribution of $\Delta_2=E_{gs}^{(m+1)} + E_{gs}^{(m-1)} - 2\;
E_{gs}^{(m)}$ With Pairing and Exchange Interactions}
\label{c3s4}

\subsection{Brief introduction to mesoscopic systems}
\label{c3s4p1}

Mesoscopic systems are intermediate between microscopic systems (like nuclei and
atoms) and macroscopic bulk matter. Quantum dots and ultrasmall metallic grains
are good examples of mesoscopic systems whose transport properties can be
measured \cite{Im-97,Jan-01}. When the electron's phase
coherence length is comparable to or larger than the system size, the system is
called mesoscopic. As the electron phase is preserved in mesoscopic systems,
these are ideal to observe new phenomenon governed by the laws of quantum
mechanics not observed in macroscopic conductors. Also, the transport properties
of mesoscopic systems are readily measured with almost all system parameters
(like the shape and size of the system, number of electrons in the system and
the strength of coupling with the leads) under experimental control. The phase
coherence length increases rapidly with decreasing temperature. For system size
$\sim 100\;\mu$m, the system becomes mesoscopic below $\sim 100$ mK.  

Quantum dots are artificial devices obtained by confining a finite number of
electrons to regions with diameter $\sim 100$ nm by electrostatic potentials.
Typically it consists of $10^9$ real atoms but the number of mobile electrons is
much lower, $\sim 100$. Their level separation is $\sim 10^{-4}$ eV.  If the
transport in the quantum dot is dominated by electron scattering from
impurities, the dot is said to be diffusive and if the transport is dominated by
electron scattering from the structure boundaries, then dot is called ballistic.
The coupling between a dot and its leads is experimentally controllable. When
the dot is strongly coupled to the leads, the electron motion is classical and
the dot is said to be open. In isolated or closed quantum dots, the coupling is
weak and conductance occurs only by tunneling. Also the charge on the closed dot
is quantized and they have discrete excitation spectrum. The tunneling of an
electron into the dot is usually blocked by the classical Coulomb repulsion of
the electrons already in the dot. This phenomenon is called Coulomb blockade.
This repulsion can be overcome by changing the gate voltage. At appropriate gate
voltage, the charge on the dot will fluctuate between $m$ and $m+1$ electrons
giving rise to a peak in the conductance. The oscillations in conductance as a
function of gate voltage are called Coulomb blockade oscillations. At
sufficiently low temperatures, these oscillations turn into sharp peaks. In
Coulomb blockade regime $kT << \Delta << E_c$, the tunneling occurs through a
single resonance in the dot. Here, $T$ is the temperature, $\Delta$ is the mean
single particle level spacing and $E_c$ is the charging energy. 
Ultrasmall metallic grains are small pieces of metals of size $\sim 2-10$ nm.
The level separation for nm-size metallic grains is smaller than in quantum dots
of similar size and thus experiments can easily probe the Coulomb blockade
regime in quantum dots. Also, some of the phenomena observed in nm-size metallic
grains are strikingly similar to those seen in quantum dots suggesting that
quantum dots are generic systems for exploring physics of small coherent
structures \cite{Gu-98,Al-00a}.

Although the quantum dots contain many electrons, their properties cannot be
obtained by using thermodynamic limit. The description of transport through a
quantum dot at low temperatures in terms of local material constants breaks down
and the whole structure must be treated as a single coherent entity. The quantum
limits of electrical conduction are revealed in quantum dots and conductivity
exhibits statistical properties which reflect the presence of one-body chaos,
quantum interference and electron-electron interaction. The transport properties
of a quantum dot can be measured by coupling it to leads and passing current
through the dot. The conductance through the dots displays mesoscopic
fluctuations as a function of gate voltage, magnetic field and shape
deformation. The techniques used to describe these fluctuations include
semiclassical methods, random matrix theory and supersymmetric methods
\cite{Al-00a}. 

Mesoscopic fluctuations are universal dictated only by a few basic symmetries of
the system. It is now widely appreciated that the universal conductance
fluctuations  are intimately related to the universal statistics of finite
isolated  quantum systems whose classical analogs are chaotic
\cite{Ko-01,Ko-03,Pa-07}. In describing transport through these coherent
systems, we are interested in quantum manifestations of classical chaos. 
Scattering of electrons from impurities or irregular boundaries leads to single
particle dynamics that are mostly chaotic. RMT describes the statistical
fluctuations in the universal regime i.e., at energy scales below the Thouless
energy $E=g\Delta$, $g$ is the Thouless conductance. In this universal regime
RMT addresses questions about statistical behavior of eigenvalues and
eigenfunctions rather than their individual description. We consider a closed
mesoscopic system (quantum dot or small metallic grain) with chaotic single
particle dynamics and with large Thouless conductance $g$. Such a structure is
described by an effective Hamiltonian which comprises of a mean field and
two-body interactions preserving spin degree of freedom. For chaotic isolated
mesoscopic systems, randomness of single particle energies leads to randomness
in effective interactions that are two-body in nature. Hence it is important to
invoke the ideas of embedded ensembles to understand and also predict properties
of these systems theoretically.

\begin{figure}
\centering
\includegraphics[width=4.5in,height=4in]{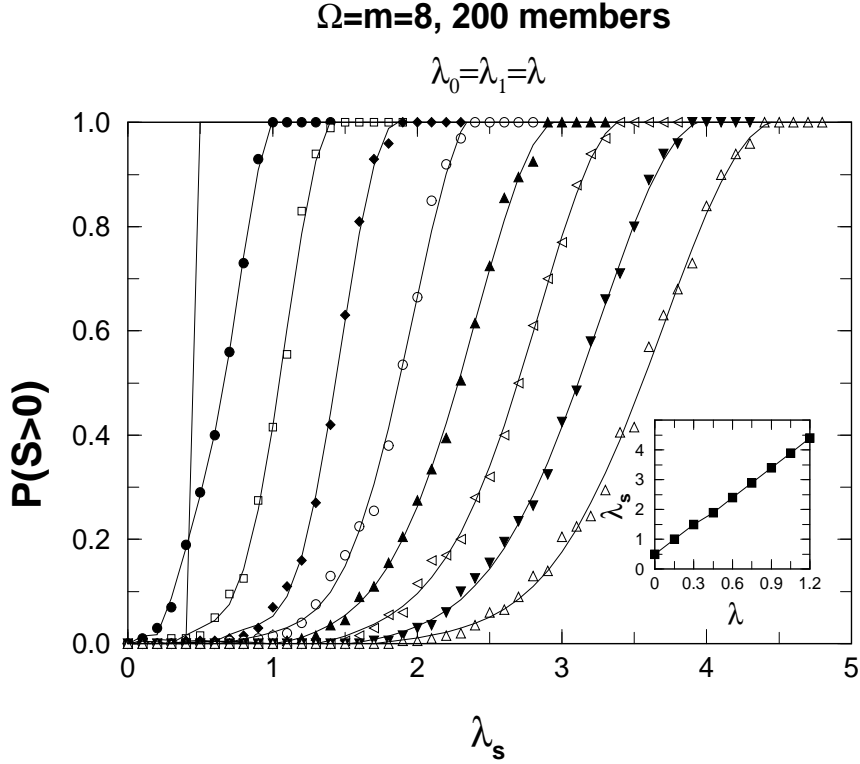}
\caption{Probability $P(S>0)$ for ground states to have $S>0$ as a  function of
exchange interaction strength $\lambda_S$ for $\lambda=0$ to $1.2$  in steps of
$0.15$; used here is $\wh(\lambda,\lambda,0,\lambda_S)$ defined by Eq.
(\ref{ham-mes}). The calculations are for $200$ member EGOE(2)-$\cs$ ensemble with
$\Omega=m=8$. Inset of figure shows the minimum exchange interaction  strength
$\lambda_S$ required for the ground states to have $S>0$ with $100$\% 
probability as a function of $\lambda$. It is seen from the results that the
probability $P(S>0)$ for gs to have $S>0$ is very small when $\lambda
> \lambda_S$ and it increases with increasing $\lambda_S$.  The results clearly
bring out the demagnetizing effect of random interaction. Similar calculations
have been performed in the past for smaller systems with $\Omega=m=6$
\cite{Ko-06,Ja-01}.}
\label{gsspin}
\end{figure}

A realistic Hamiltonian for mesoscopic systems conserves total spin $S$ and
therefore includes a mean field one-body part, (random) two-body interaction,
pairing $H_p$ and exchange interaction $\hat{S}^2$. In order to  obtain physical
interpretation of the $\hat{S}^2$ operator, we consider the space exchange or
the Majorana operator $M$ that exchanges the spatial coordinates of the
particles and leaves the spin unchanged, i.e.,
\be
M \l| i, \alpha; j, \beta \ran = \l| j, \alpha; i, \beta \ran\;.
\label{eq.maj1}
\ee 
In Eq. (\ref{eq.maj1}), labels $i,\;j$ and $\alpha,\;\beta$, respectively
denote  the spatial and spin labels. As the embedding algebra for
EGOE(1+2)-$\cs$ is $U(2\Omega) \supset U(\Omega) \otimes SU(2)$ and $\l| i,
\alpha; j, \beta \ran = (a^\dagger_{i,\alpha}a^\dagger_{j,\beta})\l|0\ran$, we
have
\be
2M=C_2\l[U(\Omega)\r] - \Omega \hat{n}\;.
\label{eq.maj2}
\ee
In Eq. (\ref{eq.maj2}), $C_2\l[U(\Omega)\r] = \sum_{i,j,\alpha,\beta}
a^\dagger_{i,\alpha} a_{j,\alpha}  a^\dagger_{j,\beta}a_{i,\beta}$ is the
quadratic Casimir invariant of the  $U(\Omega)$ group,
\be
C_2\l[U(\Omega)\r]=\hat{n}(\Omega+2)-\dis\frac{\hat{n}^2}{2} 
- \hat{S}^2\;.
\label{eq.maj3}
\ee
Combining Eqs. (\ref{eq.maj2}) and (\ref{eq.maj3}), we have finally
\be
M=-\hat{S}^2-\hat{n}\l( \dis\frac{\hat{n}}{4}-1\r)\;.
\label{maj}
\ee
Therefore, the interaction generated by the $\hat{S}^2$ operator is the 
exchange interaction with a number dependent term. This number dependent term 
becomes important when the particle number $m$ changes. The $H$ operator 
for isolated 
mesoscopic systems in universal regime has the form (with $\lambda_p$ and 
$\lambda_S$ being positive),
\be
\{\wh(\lambda_0,\lambda_1,\lambda_p,\lambda_S)\} = \whh(1) + 
\lambda_0\, \{\wv^{s=0}(2)\} + \lambda_1\, \{\wv^{s=1}(2)\} -
\lambda_p H_p -\lambda_S \hat{S}^2\;.
\label{ham-mes}
\ee
The constant part arising due to charging energy $E_c$ that depends on the 
number of fermions in the system can be easily incorporated in our model when 
required. For more details on two-body ensembles and mesoscopic systems see
\cite{Gu-98,Al-00a,Ko-01,Mi-00}. Before proceeding further, it is important to
mention that, with the  analytical formula for the propagator $P(\Omega,m,S)$
given by Eq. (\ref{eq.den9}),  EGOE(1+2)-$\cs$ generates odd-even staggering in
gs energies and also explains  preponderance of gs with spin
$0$ ($m$ even) for mesoscopic systems in a simple way. In other words, random
interaction disfavor magnetized ground states; see Fig. \ref{var}. It is
important to mention that even with the best available computing facilities, it
is not yet feasible to numerically study the properties of large systems
($\Omega>>10$) modeled by EGOE(1+2)-$\cs$.  As the minimum spin gs is
favored by random interactions, the Stoner transition will be delayed in
presence of a strong random two-body part in the Hamiltonian. The standard
Stoner picture of ferromagnetism in itinerant systems is based  on the
competition between one-body kinetic energy  [$h(1)$ in Eq. (\ref{ham-mes})] and
the exchange  interaction ($\hat{S}^2$). The probability $P(S>0)$ for the gs to
be with $S>0$ (for $m$ even) is studied as a function of $\lambda$ in Eq.
(\ref{ham-mes}) with $\lambda_p =0$ and the
results are given in Fig. \ref{gsspin}.  Thus EGOE(1+2)-$\cs$ also explains the
strong bias for low-spin ground states and the delayed gs
magnetization by random two-body interactions.

\subsection{Conductance peak spacing $(\Delta_2)$ distribution}
\label{c3s4p2}

Coulomb blockade oscillations yield detailed information about the energy and
wavefunction statistics of mesoscopic systems. We consider a closed mesoscopic
system and study the distribution $P(\Delta_2)$ of spacing $\Delta_2$ between
two neighboring conductance peaks at temperatures less than the average level
spacing. Also our focus is in the strong interaction regime [$\lambda_0 =
\lambda_1 = \lambda \geq 0.3$ in Eq. (\ref{ham-mes})] and we use fixed sp 
energies $\epsilon_i$. 
The spacing $\Delta_2$ between the peaks in conductance as a function of the
gate voltage for $T << \Delta$ is second derivative of gs energies
with respect to the number of particles, 
\be 
\Delta_2 = E_{gs}^{(m+1)} + E_{gs}^{(m-1)} - 2\;E_{gs}^{(m)}\;.
\label{ch3.eq.del1}
\ee
In Eq. (\ref{ch3.eq.del1}), $E_{gs}^{(m)}$ is the gs energy for a $m$
fermion system. The distribution $P(\Delta_2)$ has been used in the study
of the distribution of conductance peak spacings in chaotic quantum dots 
\cite{Al-05,Al-00,Al-01,AW-01}.

Let us first consider non-interacting spinless finite Fermi systems i.e.,
$H=h(1)$ and say the sp energies are $\epsilon_i;\;i=1,2,\ldots, N$.
Then Eq. (\ref{ch3.eq.del1}) gives, by applying Pauli principle,
$\Delta_2=\epsilon_{m+1} -\epsilon_m$, irrespective of whether $m$ is even or
odd. For chaotic systems it is possible to consider sp energies drawn  from GOE
eigenvalues \cite{Al-05,Al-00,Al-01,AW-01}. Therefore $P(\Delta_2)$  corresponds
to GOE spacing distribution $P_W(\Delta_2)$ - the Wigner distribution. However
recent experiments showed that $P(\Delta_2)$ is a Gaussian in many situations
\cite{Pat-98}. This calls for inclusion of two-body interactions and hence the
importance of EGOE(1+2) (in \cite{Al-00,Al-01,AW-01} this is called RIMM) in the
study of conductance fluctuations in mesoscopic systems. It was shown by
Alhassid et al \cite{Al-00,Al-01,AW-01}  that EGOE(1+2) indeed generates
Gaussian form for $P(\Delta_2)$.

\begin{figure}
\centering
\includegraphics[width=5in,height=3.5in]{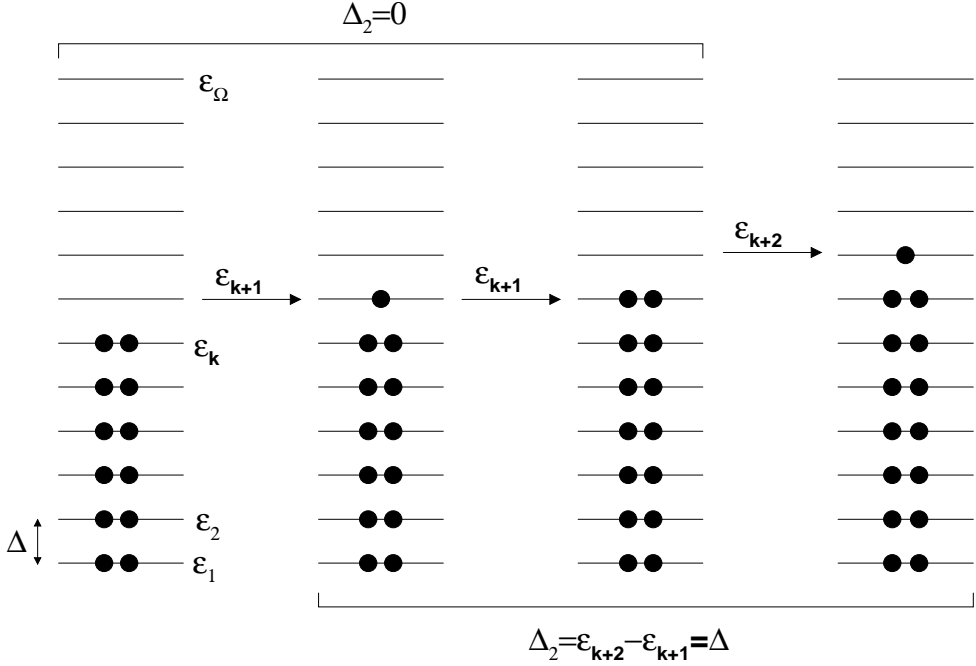}
\caption{Figure showing $\Delta_2$ values for systems with spin degree of 
freedom. For even-odd-even transitions, $\Delta_2=0$ and for odd-even-odd 
transitions, $\Delta_2=\Delta$. See text for details.}
\label{nip}
\end{figure}

As discussed in Sec. \ref{c3s4p1}, 
Hamiltonian for interacting electron systems  conserves total  spin $S$  and
thus it is important to consider sp levels that are doubly degenerate; i.e.,
spin degree of freedom should be included in $H$. Again, we start with
non-interacting finite  Fermi systems with sp energies $\epsilon_i$,
$i=1,2\ldots,\Omega$ and drawn from a GOE; total number of sp states
$N=2\Omega$. In this scenario $\Delta_2$ depends on whether $m$ is odd
or even. For $m$ odd, say $m=2k+1$, the $(m-1)$ fermion gs energy
$E_{gs}^{(m-1)}=2\sum_{i=1}^k \epsilon_i$, $E_{gs}^{(m)}=
E_{gs}^{(m-1)}+\epsilon_{k+1}$ and $E_{gs}^{(m+1)}=E_{gs}^{(m-1)}+
2\;\epsilon_{k+1}$ resulting in $\Delta_2=0$. Similar analysis for even $m=2k$
yields $\Delta_2= \epsilon_{k+1}-\epsilon_k$; note that $E_{gs}^{(m)}=
2\sum_{i=1}^k \epsilon_i$, $E_{gs}^{(m-1)}= E_{gs}^{(m)}-\epsilon_{k}$ and
$E_{gs}^{(m+1)}= E_{gs}^{(m)}+\epsilon_{k+1}$. For odd $m$,  $\Delta_2$
corresponds to  even-odd-even transition and $P(\Delta_2)$ is a delta
function. For even $m$, we have odd-even-odd transitions with $P(\Delta_2)$
following Wigner distribution. Figure \ref{nip} gives a pictorial illustration
for $\Delta_2$ calculation for systems with spin.
Therefore, by applying Pauli principle and using Eq. (\ref{ch3.eq.del1}) gives
$\Delta_2=0$ for $m$ odd and $\Delta_2 = \epsilon_{k+1}-\epsilon_k$ for even
$m$ ($k=m/2$). As we need to include, for real systems, both even and odd
$m$'s,  inclusion of spin degree of freedom gives bimodal distribution for
$P(\Delta_2)$,
\be
P(\Delta_2) =\dis\frac{1}{2}\l[\delta(\Delta_2) + P_W(\Delta_2)\r]\;.
\label{ch3.eq.del12}
\ee
Convolution of this bimodal form with a Gaussian has been used in the
analysis of  data for quantum dots obtained for situations that  correspond
to weak interactions \cite{Lu-01}. This shows that spin degree of freedom
and pairing correlations are  important for mesoscopic systems.  Hence, it
is imperative to study $P(\Delta_2)$ with a  Hamiltonian that includes mean
field one-body part, (random) two-body  interaction, exchange interaction
and pairing (defined by $H_p)$.  Therefore we have carried out 
EGOE(1+2)-$\cs$ calculations using the Hamiltonian given in Eq. (\ref{ham-mes})
(with $\lambda_p$ and
$\lambda_S$ being positive)
and constructed $P(\Delta_2)$ by combining $\Delta_2$ values obtained for
both even and odd $m$ values. Before discussing these model calculations let
us mention that very recently, for small metallic grains,  $P(\Delta_2)$
results are reported in \cite{Al-08}. These authors use a $H$ consisting of
pairing and exchange interactions just as in Eq. (\ref{ham-mes}) but  with
sp energies of $h(1)$ drawn from GOE and a two-body interaction that is  a
function of $m$. More importantly a microscopic theory is used in
\cite{Al-08} to construct $P(\Delta_2)$ at finite temperatures. When the
pairing interaction is  dominant (compared to exchange interaction), the
distribution is found to be bimodal whereas the distribution becomes
unimodal for strong exchange interaction. Following our discussion in the
previous sections, here we present results for the distribution of
$\Delta_2$  defined by Eq. (\ref{ch3.eq.del1}) with two values for $m$  and 
using $H$ defined by Eq. (\ref{ham-mes}). We use fixed $h(1)$ as in the
previous sections and  $\lambda = 0.3$. Therefore our focus is in the
strong  interaction regime.  Though our calculations are restrictive and the
model is simpler, we will show that they  reproduce all the essential
features of $P(\Delta_2)$ reported in \cite{Al-08}. 

\begin{figure}
\centering
\includegraphics[width = 5.5in, height = 6in]{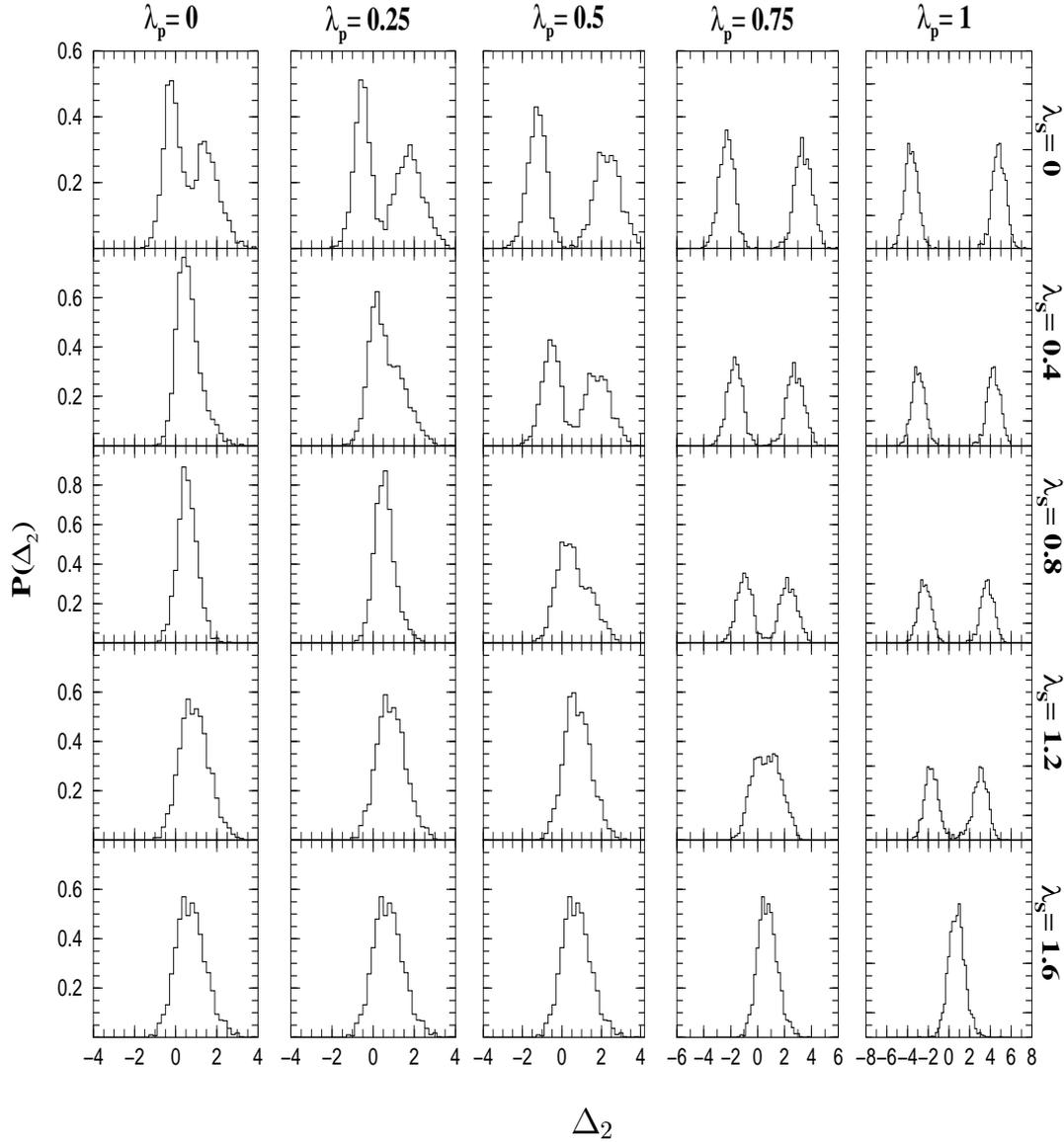}
\caption{$P(\Delta_2)$ vs $\Delta_2$ for various values of the pairing
strength $\lambda_p$ and exchange interaction strength $\lambda_S$ for the
EGOE(1+2)-$\cs$ system  defined in the text. The distributions $P(\Delta_2)$
are constructed (with bin size 0.2) by combining the results for $\Delta_2$
with $m=4$ and $5$. See text for further details.} 
\label{c3f5}
\end{figure}

Using 1000 member EGOE(1+2)-$\cs$ with $H$ defined by Eq. (\ref{ham-mes}), 
gs energies are calculated for $\lambda=0.3$ and for various values of
$\lambda_p$ and $\lambda_S$ by diagonalizing the  Hamiltonian in good spin basis
for $\Omega=6$ and $m=3, 4, 5$ and $6$. Then $\Delta_2$ is computed using Eq.
(\ref{ch3.eq.del1}) for $m=4$ and $5$ and combining these, normalized histograms
for $P(\Delta_2)$ are  constructed. Results in Fig. \ref{c3f5} show that strong
pairing correlations ($\lambda_S =0$) give rise to bimodal form for
$P(\Delta_2)$ with the two modes well separated. Increasing the exchange 
interaction reduces the separation between the two  parts and they overlap when
exchange interaction is dominant and pairing is weak. In other words,  pairing
correlations help distinguish between $m$ even and $m$ odd in Eq.
(\ref{ch3.eq.del1}). These conclusions are close to the results in Fig. 1 of
\cite{Al-08}. A qualitative understanding of these results follows from the
centroids $\lan \Delta_2\ran$ of $P(\Delta_2)$ for each $m$ generated by $H_p$
and $\hat{S}^2$ terms in $H$. When pairing is relatively stronger 
($\lambda_p>>\lambda_S$), gs has minimum spin and thus $\v=0(1)$ for
$m$ even(odd) and when pairing is weaker ($\lambda_S>>\lambda_p$),  gs
has maximum spin ($S=m/2$) and thus $\v=m$. Using the pairing eigenvalues 
$E_p(m,\v,S)$ given by Eq. (\ref{ch3.eq.npa8}), it is easily seen  that for weak
pairing, $\lan \Delta_2\ran=-\lambda_S/2$ for both $m$ even and odd and for
strong pairing,  $\lan \Delta_2\ran = (\Omega+1)\lambda_p-3/2\lambda_S$ and 
$-\Omega\lambda_p+3/2\lambda_S$  for even $m$ and odd $m$, respectively.
Therefore, for fixed $\lambda_S$, spacing between the  peaks for $m=4$ and $m=5$
increases with sufficiently large $\lambda_p$ values as seen in Fig. \ref{c3f5}.

\begin{figure}
\centering
\includegraphics[width=5in,height=3.5in]{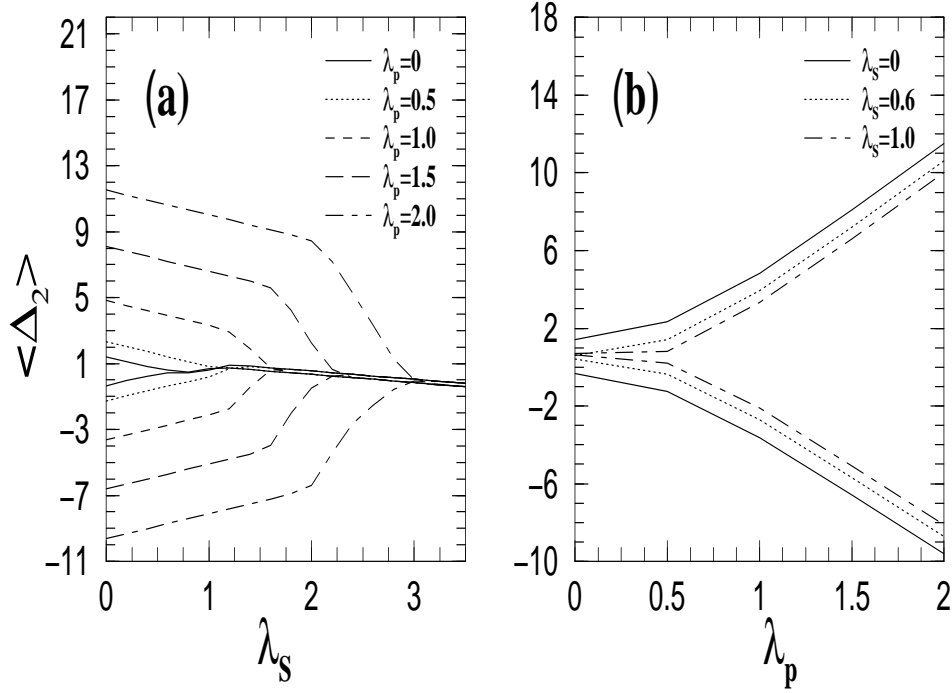}
\caption{Average peak spacing $\lan \Delta_2 \ran$ (a) as a  function of 
exchange interaction strength $\lambda_S$ for several values of  pairing 
strength $\lambda_p$ and (b) as a function of $\lambda_p$ for several
values of $\lambda_S$, for a 1000 member ensemble with $\Omega=6$. 
The curves in the upper part correspond to $m=4$ ($3 \rightarrow 4  
\rightarrow 5$) and those in the lower part to $m=5$ ($4 \rightarrow 5  
\rightarrow 6$) in Eq. (\ref{ch3.eq.del1}). See text for details.}
\label{del2}
\end{figure}

Figure \ref{del2}(a) shows the variation of average peak spacing with exchange
interaction strength $\lambda_S$ for several $\lambda_p$ values. The curves in
the upper part correspond to $m=4$ and those in the lower part to $m=5$. As the
exchange strength increases, the average peak spacing  $\lan \Delta_2\ran$ is
almost same for odd-even-odd and even-odd-even  transitions. Value of average
peak spacing and its variation with  $\lambda_S$  is different for odd-even-odd 
and even-odd-even transitions when pairing correlations are strong. The curve
for fixed value of $\lambda_p$ can be divided into two linear regions whose
slopes can be determined considering only exchange interactions i.e.,
$E_{gs}=C_0-\lambda_S\;S\;(S+1)$.  For weak exchange interaction strength,
gs spin is $0$($1/2$) for $m$ even(odd) and thus for this linear
region, $\lan\Delta_2\ran/\lambda_S \propto -3/2$($3/2$). The linear region
where exchange interactions are dominant,  $\lan\Delta_2\ran/\lambda_S\propto
-1/2$ as gs spin is $m/2$. Figure \ref{del2}(b) shows the variation of
average peak spacing with pairing strength for several $\lambda_S$ values. It
clearly shows that the separation between the distributions becomes larger with
increasing $\lambda_p$.  These results are in good agreement with the
numerically obtained results for the $P(\Delta_2)$ variation as a function of
$\lambda_p$ and $\lambda_S$ in Fig. \ref{c3f5}.  Thus, EGOE(1+2)-$\cs$ with $H$
defined in Eq. (\ref{ham-mes}) explains the interplay between  exchange
(favoring ferromagnetism) and pairing (favoring superconductivity) interaction
in the Gaussian domain as  expected in mesoscopic systems and can be used for
investigating transport properties of mesoscopic systems.

\section{Summary}
\label{c3s5}

Going beyond the results reported in Chapter \ref{ch2} for the random matrix
ensemble EGOE(1+2)-$\cs$, in the present chapter, further results are
presented with focus on pairing correlations. Firstly, in the space defined
by EGOE(1+2)-$\cs$ ensemble, pairing symmetry  defined by the algebra
$U(2\Omega)\supset Sp(2\Omega) \supset SO(\Omega)\otimes  SU_S(2)$ is
identified and some of its properties are discussed. Using numerical
calculations it is shown that in the strong coupling limit, partial
densities defined over pairing subspaces are close to Gaussian form and
propagation formulas for their centroids and variances are derived. As a
part of understanding pairing correlations in finite Fermi systems, we have
shown that pair transfer strength sums (used in nuclear structure) as a
function of excitation energy (for fixed $S$), a statistic for onset of
chaos (used in nuclei \cite{Ho-07}), follows, for low spins, the form
derived for  spinless fermion systems i.e., it is close to a ratio of
Gaussians. This is demonstrated using three detailed  examples. Going
further, we have considered a quantity in terms of gs energies,
giving conductance peak spacings in mesoscopic systems at low temperatures,
and studied its distribution over EGOE(1+2)-$\cs$ by including both 
pairing and exchange interactions. We have shown that the random matrix
model reproduces the main results that are observed recently in a realistic
calculation for small metallic grains. Finally, 
results  reported in this chapter establish that EGOE(1+2)-$\cs$ 
can be used as a random matrix model for
studying pairing correlations in finite quantum systems. 

\chapter{EGUE(2)-$SU(4)$: Group Theoretical Results}
\label{ch4}

\section{Introduction}
\label{int}

Spin-isospin $SU(4)$ supermultiplet scheme for nuclei was introduced by
Wigner \cite{Wi-37} and there is good evidence for the goodness of this
symmetry in some parts of the nuclear chart
\cite{Pa-78,Va-00,Na-01,Va-05,Ko-07a,Wu-99,Va-07} and also more recently
there is new interest in $SU(4)$ symmetry for heavy N $\sim$ Z nuclei
\cite{Va-00,Na-01,Va-05,Ko-07a}.  Therefore, it is important to define and
study EGE's generated by random two-body interactions with $SU(4)$ symmetry 
[EGUE(2)-$SU(4)$]. Given $m$ fermions (nucleons) in $\Omega$ number of sp
orbitals with spin  and isospin degrees of freedom, for $SU(4)$ scalar 
Hamiltonians, the symmetry algebra is $U(4\Omega)\supset  U(\Omega)\otimes
SU(4)$ and all the states within an $SU(4)$ [but not $U(\Omega)$] 
irrep will be degenerate in energy. In the past, applying
Wigner-Racah algebra of the embedding algebra $U(2\Omega)\supset 
U(\Omega)\otimes SU(2)$ some analytical results are derived for
EGUE(2)-$\cs$; see Appendix \ref{egue2} for some details. Going beyond the
spin ensemble (discussed in Chapters \ref{ch2}, \ref{ch3} and Appendix
\ref{egue2}), our purpose in the present chapter is to define EGUE(2)-$SU(4)$,
develop analytical formulation for solving the ensemble and derive analytical
formulas, for the lower order moments of the one-point (density of
eigenvalues) and two-point (defining level fluctuations) functions,  for some
simple class of $SU(4)$ irreps. In addition,  analytical formulation
developed in the chapter allows one to consider all these, numerically, for
any $SU(4)$ [or $U(\Omega)$] irrep. Using these, studied are: ensemble
averaged  spectral variances, expectation values of the quadratic Casimir
invariant of $SU(4)$ algebra, four periodicity in the gs energies and lower
order cross-correlations in energy centroids and spectral variances generated
by this ensemble. Before proceeding further, let us mention
that a preliminary report of some of the results in this chapter is given in
\cite{Ma-09a} and all the details are published in the long
paper \cite{Ma-10a}.

\section{Preliminaries of $U(4\Omega)\supset U(\Omega)\otimes SU(4)$ 
Algebra}
\label{su4bk}

Although all the results in this section are well-known  \cite{Pa-78, He-69,
He-74a}, we will discuss these here for completeness  and also for introducing
various quantities and  notations used in the  reminder of the
chapter\footnote{We use different notations in this chapter for mathematical
ease}.

\subsection{Generators of $U(\Omega)$ and $SU(4)$ algebras}
\label{gen}

Let us begin with $m$ fermions distributed in $4\Omega$ number of sp
states.  Then the spectrum generating algebra is $U(4\Omega)$. Associating
two quantum numbers $i$ ($i$-space) and $\alpha$ ($\alpha$-space) to each 
sp state, the sp states are denoted by  $\l|i,\alpha\ran$, where 
$i=1,2,\ldots,\Omega$ and $\alpha=1,2,3,4$. In nuclear applications, the 
$i$-space corresponds to the orbital space and the $\alpha$-space
corresponds  to the spin($\cs$)-isospin($\ct$) space, then $\l|\alpha\ran =
\l|m_\cs,m_\ct\ran = \l|\spin,\spin\ran$, $\l|\spin,-\spin\ran$,
$\l|-\spin,\spin\ran$ and  $\l|-\spin,-\spin\ran$, respectively.  From now
on in this section we will present results both in single state
representation defined by $\l|i,\alpha\ran$ states and also in the
spin-isospin representation defined by $\l|i; \cs=\spin, m_\cs ;
\ct=\spin,m_\ct\ran$ states. For the EGUE(2)-$SU(4)$ ensemble,  the former
will suffice. However the later (spin-isospin) representation  is useful for
understanding the physical relevance of the ensemble.  In the single state
representation, the $(4\Omega)^2$  number of operators $C_{i\alpha;j\beta}$
generate $U(4\Omega)$ algebra and with respect to this algebra, all the $m$
fermion states transform as the irrep $\{1^m\}$. In terms of the creation
operators $a^\dagger_{i,\alpha}$ and the annihilation operators
$a_{j,\beta}$, the generators $C_{i\alpha;j\beta}$ and their commutation
relations are,
\be
C_{i\alpha;j\beta} = a^\dagger_{i,\alpha}a_{j,\beta}\;;\;\;\;\;
\l[ C_{i\alpha;j\beta}, C_{k\alpha^\pr;l\beta^\pr}\r] =
C_{i\alpha;l\beta^\pr}\delta_{jk}\delta_{\beta\alpha^\pr} -
C_{k\alpha^\pr;j\beta}\delta_{li}\delta_{\beta^\pr\alpha}\;.
\label{ch4.eq.1}
\ee
It is possible to define commuting unitary transformations in the $i$-space
and $\alpha$-space separately and then we have $U(\Omega)$ and $U(4)$
algebras describing unitary transformations in the two respective  spaces.
With this we have the direct product group-subgroup structure
$U(4\Omega)\supset U(\Omega) \otimes U(4)$.  We can easily write the
generators $A_{ij}$ and $B_{\alpha\beta}$ for the $U(\Omega)$ and  $U(4)$
algebras, respectively, using the fact that the generators of  $U(\Omega)$ are
scalars in $\alpha$-space and similarly the $U(4)$ generators in the
$i$-space,
\be
A_{ij} = \dis\sum_{\alpha=1}^{4} C_{i\alpha;j\alpha}\;,\;\;\;\;
B_{\alpha\beta} = \dis\sum_{i=1}^{\Omega} C_{i\alpha;i\beta}\;.
\label{ch4.eq.2}
\ee
Their commutation relations can be derived using Eq. (\ref{ch4.eq.1}) by
summing over the appropriate indices,
\be
\barr{rcl}
\l[ A_{ij}, A_{kl}\r] = A_{il} \delta_{jk} - A_{kj} \delta_{li}\;,\\
\l[ B_{\alpha\beta}, B_{\alpha^\pr\beta^\pr}\r] = B_{\alpha\beta^\pr}
\delta_{\beta\alpha^\pr} - B_{\alpha^\pr\beta} \delta_{\alpha\beta^\pr}\;.
\earr \label{ch4.eq.3}
\ee
Also the $A$'s commute with the $B$'s. Instead of $U(4)$, it is possible to
consider $SU(4)$ by making the generators $B$'s traceless [see Eq. 
(\ref{ch4.eq.15}) ahead].

In the orbital $\times$ spin-isospin realization of the $U(4\Omega)\supset
U(\Omega)\otimes SU(4)$ algebra, $SU(4)$ corresponds to the Wigner's
supermultiplet algebra \cite{Wi-37}. In this physically relevant
spin-isospin  representation, the $SU(4)$ generators can be written in 
terms of the one-body operators $\caa_{ij;\mu_\cs,\mu_\ct}^{s,t}$ where,
\be
\barr{l}
\caa_{ij;\mu_\cs,\mu_\ct}^{s,t} = \l(a^\dagger_{i}\tilde{a}_{j}\r)_
{\mu_\cs,\mu_\ct}^{s,t}  \\ \\
= \dis\sum_{m_\cs(m^\pr_\cs),m_\ct(m^\pr_\ct)}\,
\lan \spin m_\cs\; \spin m^\pr_\cs \mid s\, \mu_\cs \ran\,
\lan \spin m_\ct\; \spin m^\pr_\ct \mid t\, \mu_\ct \ran\,
a^\dagger_{i;\spin,m_\cs;\spin,m_\ct}
\tilde{a}_{j;\spin,m^\pr_\cs;\spin,m^\pr_\ct}\;.
\earr \label{ch4.eq.4}
\ee
Note that $\tilde{a}_{j;\spin,\mu_\cs;\spin,\mu_\ct} = 
(-1)^{1+\mu_\cs+\mu_\ct} a_{j;\spin,-\mu_\cs;\spin,-\mu_\ct}$.  The
operators $\caa_{ij;\mu_\cs,\mu_\ct}^{s,t}$ generate $U(4\Omega)$ algebra.
Similarly, the operators $\caa_{ij}^{0,0}$ ($\Omega^2$ in number) and 
$\sum_i \caa_{ii;\mu_\cs,\mu_\ct}^{s,t}$ (16 in number) generate the 
$U(\Omega)$ and $U(4)$ algebras, respectively.  The 16 generators of $U(4)$
can be written in terms of the number operator $\hat{n}$, the three spin
generators $S_\mu^1$, the three isospin generators $T_\mu^1$ and the nine
components $(\sigma\tau)_{\mu,\mu^\pr}^{1,1}$ of the Gamow-Teller operator 
$\sigma\tau$. Dropping the number operator, we obtain the $SU(4)$ algebra. 
Given a one-body operator $\co$, it can be expressed in terms of the
creation and annihilation operators,
\be
\co = \dis\sum_{i,j,m_\cs,m_\ct,m^\pr_\cs,m^\pr_\ct} \lan i; \spin, m_\cs;
\spin, m_\ct \mid \co \mid j; \spin, m^\pr_\cs; \spin, m^\pr_\ct\ran
a^\dagger_{i; \spin, m_\cs; \spin, m_\ct} a_{j; \spin, m^\pr_\cs; \spin,
m^\pr_\ct}\;.
\label{ch4.eq.5}
\ee
Starting with Eq. (\ref{ch4.eq.5}), applying the angular-momentum algebra 
\cite{Ed-74} and using Eq. (\ref{ch4.eq.4}), will give \cite{Ko-06b}
\be
\barr{rcl}
\hat{n} & = & 2\dis\sum_i \caa_{ii;0,0}^{0,0}\;,\;\;\;\;
S^1_\mu = \dis\sum_i \caa_{ii;\mu,0}^{1,0}\;, \\ \\
T^1_\mu & = & \dis\sum_i \caa_{ii;0,\mu}^{0,1}\;,\;\;\;\;
(\sigma\tau)_{\mu,\mu^\pr}^{1,1} = \dis\sum_i \caa_{ii;\mu,\mu^\pr}^{1,1}
\;.
\earr \label{ch4.eq.6}
\ee
Commutation relations for the $SU(4)$ generators in the spin-isospin 
(sometimes called spherical) representation are,
\be
\barr{rcl}
\l[ S^1_\mu,S^1_{\mu^\pr}\r] & = & -\sqrt{2}\;\lan 1\;\mu\;1\;\mu^\pr 
\mid\;1\;\mu+\mu^\pr \ran \; S^1_{\mu+\mu^\pr} \;, \\ \\
\l[ T^1_\mu,T^1_{\mu^\pr}\r] & = & -\sqrt{2}\;\lan 1\;\mu\;1\;\mu^\pr 
\mid\;1\;\mu+\mu^\pr \ran \; T^1_{\mu+\mu^\pr} \;, \\ \\
\l[ S^1_\mu,(\sigma\tau)^{1,1}_{\mu^\pr,\mu^{\pr\pr}}\r] & = & -\sqrt{2}\;
\lan 1\;\mu\;1\;\mu^\pr \mid\;1\;\mu+\mu^\pr \ran \; (\sigma\tau)^{1,1}
_{\mu+\mu^\pr,\mu^{\pr\pr}} \;, \\ \\
\l[ T^1_\mu,(\sigma\tau)^{1,1}_{\mu^\pr,\mu^{\pr\pr}}\r] & = & -\sqrt{2}\;
\lan 1\;\mu\;1\;\mu^{\pr\pr} \mid\;1\;\mu+\mu^{\pr\pr} \ran \; 
(\sigma\tau)^{1,1}_{\mu^\pr,\mu+\mu^{\pr\pr}} \;, \\ \\
\l[ (\sigma\tau)^{1,1}_{\mu_1,\mu_2},(\sigma\tau)^{1,1}_{\mu_3,\mu_4}\r]
& = & \sqrt{2}\;(-1)^{\mu_1+1} \lan 1\;\mu_2\;1\;\mu_4
\mid\;1\;\mu_2+\mu_4\ran\;\delta_{\mu_1,-\mu_3}\;T^1_{\mu_2+\mu_4}
\\  \\ & + &
\sqrt{2}\;(-1)^{\mu_2+1} \lan 1\;\mu_1\;1\;\mu_3
\mid\;1\;\mu_1+\mu_3\ran\;\delta_{\mu_2,-\mu_4}\;S^1_{\mu_1+\mu_3}\;.
\earr \label{ch4.eq.7}
\ee
Now we will consider the quadratic Casimir invariants ($C_2$) of 
$U(\Omega)$ and $SU(4)$ and their physical interpretation. However  we will
not consider here the cubic ($C_3$) and quartic ($C_4$)  invariants of
$SU(4)$ although they are needed for some purposes  as discussed ahead; see
for example \cite{Pa-78} for $C_3$ and $C_4$  operators. 

\subsection{Quadratic Casimir operators of $U(\Omega)$ and $SU(4)$ and the
Majorana operator}
\label{cas-maj}

In the $\l|i,\alpha \ran$ representation it is easy to write down   the
quadratic Casimir invariant of $U(4)$,
\be
C_2\l[ U(4)\r] = \dis\sum_{\alpha,\beta} B_{\alpha,\beta}
B_{\beta,\alpha}
= 4\hat{n} + \dis\sum_{i,j,\alpha,\beta}
a^\dagger_{i,\alpha}a^\dagger_{j,\beta}a_{j,\alpha}a_{i,\beta}\;.
\label{ch4.eq.12}
\ee
The operator $C_2\l[ U(4)\r]$ commutes with  the generators
$B_{\alpha,\beta}$ or equivalently with $\hat{n}$, $S^1_\mu$, $T^1_\mu$ and
$(\sigma\tau)^{1,1}_{\mu,\mu^\pr}$.  Just as $C_2\l[ U(4)\r]$, the quadratic
Casimir invariant of  $U(\Omega)$ is,
\be
C_2\l[U(\Omega)\r] = \dis\sum_{i,j} A_{ij}A_{ji}
= \hat{n}\;\Omega - \dis\sum_{i,j,\alpha,\beta}
a^\dagger_{i,\alpha}a^\dagger_{j,\beta}a_{j,\alpha}a_{i,\beta}\;.
\label{ch4.eq.11}
\ee
Combining Eqs. (\ref{ch4.eq.12}) and (\ref{ch4.eq.11}) we have
\be
C_2\l[U(\Omega)\r] + C_2\l[ U(4)\r] = \hat{n}\l(\Omega + 4\r)\;.
\label{ch4.eq.13}
\ee
It is also easy to see that the $C_2\l[SU(4)\r]$ can be written in 
terms of $C_2\l[ U(4)\r]$ and  $C_2\l[U(\Omega)\r]$,
\be
\barr{rcl}
C_2\l[ SU(4)\r] & = & \dis\sum_{\alpha,\beta} B_{\alpha,\beta}^\pr
B_{\beta,\alpha}^\pr\;;\;\; 
B_{\alpha,\beta}^\pr = B_{\alpha,\beta} - \dis\frac{\mbox{Tr} (B)}{4}
\delta_{\alpha,\beta}\;,\;\;\mbox{Tr} (B) = \dis\sum_\alpha
B_{\alpha,\alpha} \\
& = & C_2\l[ U(4)\r] - \dis\frac{\hat{n}^2}{4} \\
& = & -\l[ C_2\l[U(\Omega)\r] - \hat{n}(\Omega+4) + \dis\frac{\hat{n}^2}{4} 
\r]\;.
\earr \label{ch4.eq.15}
\ee
In the angular-momentum coupled representation, $C_2\l[ SU(4)\r] = S^2 + 
T^2 +(\sigma \tau)\cdot (\sigma \tau)$. In order to obtain a physical 
interpretation for $C_2\l[ SU(4)\r]$, we will consider the  space exchange
or the Majorana operator $\whm$, with strength $\kappa$,  that exchanges the
spatial coordinates of the particles (the index $i$) and leaves the index
$\alpha$ (equivalently spin-isospin  quantum numbers) unchanged. Then
\cite{Pa-78},
\be
\whm \l| i,\alpha ; j,\beta \ran = \kappa \l| j,\alpha; i,\beta \ran\;.
\label{ch4.eq.18}
\ee
As $\l| i,\alpha; j,\beta \ran =a^\dagger_{i,\alpha}a^\dagger_{j,\beta} \l|
0 \ran$, Eq. (\ref{ch4.eq.18}) gives, with $\kappa$ a constant,
\be
\barr{rcl}
\whm & = & \dis\frac{\kappa}{2} \l[ \dis\sum_{i,j,\alpha,\beta}
\l( a^\dagger_{j,\alpha} a^\dagger_{i,\beta}\r)
\l( a^\dagger_{i,\alpha} a^\dagger_{j,\beta}\r)^
\dagger\r] \\ \\
& = & \dis\frac{\kappa}{2} \l[\dis\sum_{i,j} \l( \dis\sum_{\alpha}
a^\dagger_{j,\alpha} a_{i,\alpha}\r)
\l( \dis\sum_{\beta}
a^\dagger_{i,\beta} a_{j,\beta}\r)
- \Omega \dis\sum_{j,\alpha} a^\dagger_{j,\alpha}
a_{j,\alpha} \r] \\ \\
& = & \dis\frac{\kappa}{2} \l\{ C_2\l[U(\Omega)\r] - \Omega \hat{n}\r\}\;.
\earr \label{ch4.eq.19}
\ee
Eqs. (\ref{ch4.eq.15}) and (\ref{ch4.eq.19}) allow us to write the $\whm$ 
operator in terms of $C_2\l[ SU(4)\r]$. Then, we have
\be
\whm = \kappa\l\{2\hat{n} \l( 1+\dis\frac{\hat{n}}{16}\r) -
\spin C_2\l[ SU(4)\r] \r\}\;.
\label{ch4.eq.21}
\ee
Using Eq. (\ref{ch4.eq.21}) one can identify the $SU(4)$ [or $U(4)$] irrep for 
gs, assuming that the Hamiltonian is represented by the Majorana operator.
Towards this end, now we will consider the $SU(4)$ and $U(\Omega)$ irreps 
and the reduction of the $SU(4)$ irreps to $(S,T)$.

\subsection{$SU(4)$ and $U(\Omega)$ irreps and identification of the ground 
state $U(\Omega)$ or $SU(4)$ irreps}
\label{gsirp}

With $m$ fermions in $4\Omega$ sp states, we can decompose the basis space
with dimension $\binom{4\Omega}{m}$  into irreps of $U(4)$ [or $SU(4)$] and
$U(\Omega)$ and further the $U(4)$  irreps into ($S,T$). Firstly, the $U(4)$
irreps are represented by the Young tableaux (see Fig. \ref{young}) 
or the partitions $\{F\}$, 
\be
\{F\} = \l\{F_1,F_2,F_3,F_4\r\}\;,\;\;\;\;
F_1\geq F_2\geq F_3\geq F_4 \geq 0\;,\;\;\;\;m=\sum_{i=1}^4
F_i\;.
\label{ch4.eq.maj1}
\ee
Note that $F_\alpha$ are the eigenvalues of $B_{\alpha\alpha}$ defined in
Eq. (\ref{ch4.eq.2}).  As the total $m$-particle wavefunctions are
antisymmetric,  the $U(\Omega)$ irreps $\{f\}$ are uniquely defined by
$\{F\}$ and $\{f\}=\{\widetilde{F}\}$ (alternatively $\{F\} = 
\{\widetilde{f}\}$) which is obtained by changing rows to columns in the
Young tableaux $\{F\}$; see for example \cite{Pa-78,Wy-70,Ham-62}.  Due to
this symmetry  constraint,  $F_j\leq \Omega$, $j=1,2,3,4$ and  $f_i\leq 4$,
$i=1,2,\ldots,\Omega$. Given the $U(4)$ irrep $\{F\}$, the corresponding
$SU(4)$ irrep $\{F^\pr\}$, which is three rowed Young tableaux, can be 
defined by
\be
\{F^\pr\}=\{F^\pr_1,F^\pr_2,F^\pr_3\}=\{F_1-F_4,F_2-F_4,F_3-F_4\}\;.
\label{ch4.eq.n1}
\ee 

The $\{F\}\to(S,T)$ reductions can be obtained using group theoretical
methods \cite{Wy-70,Ham-62}. Alternatively a physically intuitive procedure,
easy to implement on a machine, is as follows. First, the $\{F\}\to(S,T)$
reductions for a symmetric $U(4)$ irrep $\{F\}=\{F_1,0,0,0\}$ can be
obtained by distributing $m=F_1$ identical bosons in the four spin-isospin
orbitals labeled by $\l|m_\cs m_\ct\ran$.  From these distributions, the $S_z$
and $T_z$  eigenvalues $M_S=\sum_i m_i(m_s)_i$ and $M_T=\sum_i m_i(m_t)_i$
and the corresponding degeneracies $d(m:M_S,M_T)$ follow easily. Here $m_i$
are  the number of bosons in the $i$th orbit with $m_\cs=(m_\cs)_i$ and 
$m_\ct=(m_\ct)_i$. Let us denote the number of times ($S,T$) appears in a
given $\{F\}$ by $D(\{F\}:S,T)$. It is easy to see that $D(\{m,0,0,0\}:S,T)$
is given by the double difference, 
\be
\barr{l}
D(\{m,0,0,0\}:S,T) = d(m:M_S=S,M_T=T)-d(m:M_S=S,M_T=T+1)\\
- d(m:M_S=S+1,M_T=T)+d(m:M_S=S+1,M_T=T+1)\;. 
\earr \label{ch4.eq.n2}
\ee
Carrying out this exercise on a machine for many $m$ values, we obtain 
the following (well known in literature) general result,
\be
\{m,0,0,0\} \to (S,T)=\;\l(\dis\frac{m}{2},\dis\frac{m}{2} \r),
\l(\dis\frac{m}{2}-1,\dis\frac{m}{2}-1 \r),\ldots,(0,0)\;\mbox{or}\;
\l(\spin,\spin\r)\;.
\label{ch4.eq.8}
\ee
It is important to note that here $D(\{m,0,0,0\}:S,T)=1$ for all allowed 
$(S,T)$ values (i.e., multiplicity is unity). The reductions for a general
$U(4)$ irrep $\{F\}=\{F_1,F_2,F_3,F_4\}$ follow by writing $\{F\}$ as a
determinant involving only totally symmetric irreps with the multiplication
of the elements in the determinant replaced by outer products. Then we have 
\cite{Wy-70,Ja-81}  
\be
\{F\}
= \l| \caf_{ij}\r|\;,\;\;\;\; \caf_{ij} = \{F_i+j-i,0,0,0\}\;;
\;\;\;\{0\}=1\;,\;\;\;\;\{-x,0,0,0\}=0\;.
\label{ch4.eq.8a}
\ee
Substituting the dimensions for symmetric irreps in the above determinant
gives the dimension formula for $U(4)$ irreps,
\be
d_4(\{F\}) = |d_{ij}|,\;d_{ij}={F_i+j-i+3 \choose 3}\;. 
\label{ch4.eq.n3}
\ee
Also the corresponding $(S,T)$ values and their multiplicities can be 
obtained by substituting the $(S,T)$ values for $\caf_{ij}$ in the 
determinant in Eq. (\ref{ch4.eq.8a}) and evaluating the determinant 
by  applying
angular-momentum coupling rules. Note that $d_4(\{F\}) = \sum_{S,T} (2S+1)
(2T+1) D(\{F\}:S,T)$. In carrying out the algebra we can exploit the
equivalence between $SU(4)$ and $U(4)$  irreps and employ  just 3 rowed
$U(4)$ irreps. This procedure is used in constructing Tables \ref{tabnew}
and \ref{tab21}.  For a realistic system such as the atomic nucleus, given
the $\Omega$   value and the number of valence nucleons $m$, we can
enumerate all the allowed $U(4)$ or $SU(4)$ irreps using Eqs.
(\ref{ch4.eq.maj1}) and (\ref{ch4.eq.n1}). Table \ref{tabnew} 
gives all the possible
$U(4)$ irreps for $\Omega=10$ and $m=0-6$ along with their spin-isospin
structure. 

\begin{table}[tp]
\caption{$m \to \{F\}\to(S,T)$ reductions for $\Omega=10$ and $m=0-6$. 
In the table, $r$ in $(S,T)^r$ gives the multiplicity of the irrep $(S,T)$.}
\begin{center}
\begin{tabular}{ccc}
\toprule
$m$ & $\l\{F_1,F_2,F_3,F_4\r\}$ & $(S,T)$ \\ 
\midrule
$0$ & $\{ 0 , 0 , 0 , 0 \}$ &  $( 0 , 0 )$ \\
$1$ & $\{ 1 , 0 , 0 , 0 \}$ &  $(\frac{ 1 }{2},\frac{ 1 }{2}) $ \\
$2$ & $\{ 1 , 1 , 0 , 0 \}$ &  $( 1 , 0 ) $,$( 0 , 1 ) $ \\
  & $\{ 2 , 0 , 0 , 0 \}$ &  $( 1 , 1 )$,$( 0 , 0 )$ \\
$3$  & $\{ 1 , 1 , 1 , 0 \}$ &  $(\frac{ 1 }{2},\frac{ 1 }{2}) $ \\
 & $\{ 2 , 1 , 0 , 0 \}$ &  $(\frac{ 3 }{2},\frac{ 1 }{2}) $,
 $(\frac{ 1 }{2},\frac{ 3 }{2}) $,$(\frac{ 1 }{2},\frac{ 1 }{2}) $ \\
 & $\{ 3 , 0 , 0 , 0 \}$ &
 $(\frac{ 3 }{2},\frac{ 3 }{2}) $,$(\frac{ 1 }{2},\frac{ 1 }{2}) $ \\
$4$  & $\{ 1 , 1 , 1 , 1 \}$ &  $( 0 , 0 ) $ \\
 & $\{ 2 , 1 , 1 , 0 \}$ &  $( 1 , 1 ) $,$( 1 , 0 ) $,$( 0 , 1 ) $
 \\
 & $\{ 2 , 2 , 0 , 0 \}$ &
 $( 2 , 0 ) $,$( 1 , 1 ) $,$( 0 , 2 ) $,$( 0 , 0 ) $ \\
 & $\{ 3 , 1 , 0 , 0 \}$ &
 $( 2 , 1 ) $,$( 1 , 2 ) $,$( 1 , 1 ) $,$( 1 , 0 ) $,
 $( 0 , 1 ) $ \\
 & $\{ 4 , 0 , 0 , 0 \}$ &  $( 2 , 2 ) $,$( 1 , 1 ) $,$( 0 , 0 ) $
 \\
$5$  & $\{ 2 , 1 , 1 , 1 \}$ & $(\frac{ 1 }{2},\frac{ 1 }{2}) $ \\
 & $\{ 2 , 2 , 1 , 0 \}$ & $(\frac{ 3 }{2},\frac{ 1 }{2}) $,
 $(\frac{ 1 }{2},\frac{ 3 }{2})$,$(\frac{ 1 }{2},\frac{ 1 }{2}) $ \\
 & $\{ 3 , 1 , 1 , 0 \}$ & $(\frac{ 3 }{2},\frac{ 3 }{2}) $,
 $(\frac{ 3 }{2},\frac{ 1 }{2}) $,$(\frac{ 1 }{2},\frac{ 3 }{2}) $,
 $(\frac{ 1 }{2},\frac{ 1 }{2})$ \\
 & $\{ 3 , 2 , 0 , 0 \}$ & $(\frac{ 5 }{2},\frac{ 1 }{2}) $,
 $(\frac{ 3 }{2},\frac{ 3 }{2})$,$(\frac{ 3 }{2},\frac{ 1 }{2})$,
 $(\frac{ 1 }{2},\frac{ 5 }{2}) $,$(\frac{ 1 }{2},\frac{ 3 }{2}) $,
 $(\frac{ 1 }{2},\frac{ 1 }{2}) $ \\
 & $\{ 4 , 1 , 0 , 0 \}$ & $(\frac{ 5 }{2},\frac{ 3 }{2}) $,
 $(\frac{ 3 }{2},\frac{ 5 }{2}) $,$(\frac{ 3 }{2},\frac{ 3 }{2}) $,
 $(\frac{ 3 }{2},\frac{ 1 }{2}) $,$(\frac{ 1 }{2},\frac{ 3 }{2}) $,
 $(\frac{ 1 }{2},\frac{ 1 }{2}) $ \\
 & $\{ 5 , 0 , 0 , 0 \}$ & $(\frac{ 5 }{2},\frac{ 5 }{2}) $,
 $(\frac{ 3 }{2},\frac{ 3 }{2}) $,$(\frac{ 1 }{2},\frac{ 1 }{2}) $ \\
$6$  & $\{ 2 , 2 , 1 , 1 \}$ & $( 1 , 0 ) $,$( 0 , 1 ) $ \\
 & $\{ 2 , 2 , 2 , 0 \}$ & $( 1 , 1 ) $,$( 0 , 0 ) $ \\
 & $\{ 3 , 1 , 1 , 1 \}$ &  $( 1 , 1 ) $,$( 0 , 0 ) $ \\
 & $\{ 3 , 2 , 1 , 0 \}$ & $( 2 , 1 ) $,$( 2 , 0 ) $,$( 1 , 2 ) $,
 $( 1 , 1 )^ 2 $,$( 1 , 0 ) $,$( 0 , 2 ) $,$( 0 , 1 ) $ \\
 & $\{ 3 , 3 , 0 , 0 \}$ & $( 3 , 0 ) $,$( 2 , 1 ) $,$( 1 , 2 ) $,
 $( 1 , 0 ) $,$( 0 , 3 ) $,$( 0 , 1 ) $ \\
 & $\{ 4 , 1 , 1 , 0 \}$ & $( 2 , 2 ) $,$( 2 , 1 ) $,$( 1 , 2 ) $,
 $( 1 , 1 ) $,$( 1 , 0 ) $,$( 0 , 1 ) $ \\
 & $\{ 4 , 2 , 0 , 0 \}$ & $( 3 , 1 ) $,$( 2 , 2 ) $,$( 2 , 1 ) $,
 $( 2 , 0 ) $,$( 1 , 3 ) $,$( 1 , 2 ) $,$( 1 , 1 )^ 2 $,$( 0 , 2 ) $
 ,$( 0 , 0 ) $ \\
 & $\{ 5 , 1 , 0 , 0 \}$ & $( 3 , 2 ) $,$( 2 , 3 ) $,$( 2 , 2 ) $,
 $( 2 , 1 ) $,$( 1 , 2 ) $,$( 1 , 1 ) $,$( 1 , 0 ) $,
 $( 0 , 1 ) $ \\
 & $\{ 6 , 0 , 0 , 0 \}$ & $( 3 , 3 ) $,$( 2 , 2 ) $,$( 1 , 1 ) $,
 $( 0 , 0 ) $ \\ 
\bottomrule
\end{tabular}
\end{center}
\label{tabnew}
\end{table}

\begin{table}[ht]
\caption{$U(4)$ and $U(\Omega)$ irreps $F_m$ and $f_m$,
respectively, with the smallest value for $\lan
C_2[SU(4)] \ran^{\tilde{f}_m}$ for a given $(m,T_z)$ value in the
($2p1f$)-shell [$\Omega=10$]. For the results in the Table, isospin
$T=|T_z|$.}
\begin{center}
\begin{tabular}{cccccccc}
\toprule
$m$ & $|T_z|$ & $F_m=\tilde{f}_m$ & $f_m$ & $m$ & $|T_z|$ & 
$F_m=\tilde{f}_m$ & $f_m$\\ 
\midrule
$4 $ & $ 0 $ & $\{ 1 , 1 , 1 , 1 \}$ & 
$\l\{ 4 \r\}$ & $9 $ & $\frac{ 1 }{2}$ & 
$\{ 3 , 2 , 2 , 2 \}$ &  $\l\{ 4,4,1 \r\}$\\
 & $ 1 $ & $\{ 2 , 1 , 1 , 0 \}$ & $\l\{ 3,1 \r\}$ &  
 & $\frac{ 3 }{2}$ & $\{ 3 , 3 , 2 , 1 \}$ &
 $\l\{ 4,3,2 \r\}$\\
 & $ 2 $ & $\{ 2 , 2 , 0 , 0 \}$ & 
 $\l\{ 2,2 \r\}$ &  & $\frac{ 5 }{2}$ & 
 $\{ 4 , 3 , 1 , 1 \}$ &
 $\l\{ 4,2,2,1 \r\}$\\
$5 $ & $\frac{ 1 }{2}$ & $\{ 2 , 1 , 1 , 1 \}$ &
 $\l\{  4,1\r\}$ &  & $\frac{ 7 }{2}$ & 
 $\{ 4 , 4 , 1 , 0 \}$ &
 $\l\{ 3,2,2,2 \r\}$\\
 & $\frac{ 3 }{2}$ & $\{ 2 , 2 , 1 , 0 \}$ &
 $\l\{  3,2\r\}$ &  & $\frac{ 9 }{2}$ & 
 $\{ 5 , 4 , 0 , 0 \}$ &
 $\l\{ 2,2,2,2,1 \r\}$\\
 & $\frac{ 5 }{2}$ & $\{ 3 , 2 , 0 , 0 \}$ &
 $\l\{  2,2,1\r\}$ & $10 $ & $ 0 $ & 
 $\{ 3 , 3 , 2 , 2 \}$ & 
$\l\{ 4,4,2 \r\}$\\
$6 $  & $ 0 $ & $\{ 2 , 2 , 1 , 1 \}$ & 
$\l\{ 4,2 \r\}$ &  & $ 1 $ & $\{ 3 , 3 , 2 , 2 \}$ & 
 $\l\{ 4,4,2 \r\}$\\
 & $ 1 $ & $\{ 2 , 2 , 1 , 1 \}$ & 
 $\l\{ 4,2 \r\}$ &  & $ 2 $ & $\{ 4 , 3 , 2 , 1 \}$ & 
 $\l\{ 4,3,2,1 \r\}$\\
 & $ 2 $ & $\{ 3 , 2 , 1 , 0 \}$ & 
 $\l\{ 3,2,1 \r\}$ &  & $ 3 $ & $\{ 4 , 4 , 1 , 1 \}$ & 
 $\l\{ 4,2,2,2 \r\}$\\
 & $ 3 $ & $\{ 3 , 3 , 0 , 0 \}$ & 
 $\l\{ 2,2,2 \r\}$ &  & $ 4 $ & $\{ 5 , 4 , 1 , 0 \}$ & 
 $\l\{ 3,2,2,2,1 \r\}$\\
$7 $ & $\frac{ 1 }{2}$ & $\{ 2 , 2 , 2 , 1 \}$ &
 $\l\{ 4,3 \r\}$ &  & $ 5 $ & $\{ 5 , 5 , 0 , 0 \}$ & 
 $\l\{ 2,2,2,2,2 \r\}$\\
 & $\frac{ 3 }{2}$ & $\{ 3 , 2 , 1 , 1 \}$ &
 $\l\{ 4,2,1 \r\}$ & $11 $ & $\frac{ 1 }{2}$ & 
 $\{ 3 , 3 , 3 , 2 \}$ & 
 $\l\{ 4,4,3 \r\}$\\
 & $\frac{ 5 }{2}$ & $\{ 3 , 3 , 1 , 0 \}$ &
 $\l\{ 3,2,2 \r\}$ &  & $\frac{ 3 }{2}$ & 
 $\{ 4 , 3 , 2 , 2 \}$ & 
 $\l\{ 4,4,2,1 \r\}$\\
 & $\frac{ 7 }{2}$ & $\{ 4 , 3 , 0 , 0 \}$ &
 $\l\{ 2,2,2,1 \r\}$ &  & $\frac{ 5 }{2}$ & 
 $\{ 4 , 4 , 2 , 1 \}$ & 
 $\l\{ 4,3,2,2 \r\}$\\
$8 $  & $ 0 $ & $\{ 2 , 2 , 2 , 2 \}$ & 
$\l\{ 4,4 \r\}$ &  & $\frac{ 7 }{2}$ & 
$\{ 5 , 4 , 1 , 1 \}$ & 
 $\l\{ 4,2,2,2,1 \r\}$\\
 & $ 1 $ & $\{ 3 , 2 , 2 , 1 \}$ & 
 $\l\{ 4,3,1 \r\}$ &  & $\frac{ 9 }{2}$ & 
 $\{ 5 , 5 , 1 , 0 \}$ & 
 $\l\{ 3,2,2,2,2 \r\}$\\
 & $ 2 $ & $\{ 3 , 3 , 1 , 1 \}$ & 
 $\l\{ 4,2,2 \r\}$ &  & $\frac{ 11 }{2}$ & 
 $\{ 6 , 5 , 0 , 0 \}$ & 
 $\l\{ 2,2,2,2,2,1 \r\}$\\
 & $ 3 $ & $\{ 4 , 3 , 1 , 0 \}$ & 
 $\l\{ 3,2,2,1 \r\}$ & & & & \\
 & $ 4 $ & $\{ 4 , 4 , 0 , 0 \}$ & 
 $\l\{ 2,2,2,2 \r\}$ & & & & \\ 
\bottomrule
\end{tabular}
\end{center}
\label{tab21}
\end{table}

Assuming that the Majorana operator is the Hamiltonian with $\kappa$ in Eq.
(\ref{ch4.eq.21}) negative, we can identify the $SU(4)$ irreps labeling gs as
follows. Using the formulas for the eigenvalues of $C_2\l[ U(4)\r]$ and 
$C_2\l[ U(\Omega)\r]$,  
\be
\barr{l}
\lan C_2\l[ U(4)\r] \ran^{\{F\}} = \dis\sum_{i=1}^4 F_i(F_i+5-2i)\;,\;\;\;\;
\lan C_2\l[ U(\Omega)\r]\ran^{\{f\}} = \dis\sum_{i=1}^\Omega 
f_i(f_i+\Omega+1-2i)\;,
\earr \label{ch4.eq.17}
\ee
and applying Eq. (\ref{ch4.eq.21}), we can order the $SU(4)$ irreps. For
physical systems, generally, the $U(\Omega)$ (spatial) irrep for the ground
states should
be the most symmetric one. The symmetric irrep, as seen from Eq.
(\ref{ch4.eq.17}), will have the largest eigenvalue for $C_2\l[ U(\Omega)\r]$.
From Eqs. (\ref{ch4.eq.19}) and (\ref{ch4.eq.21}), then it follows that the  
$SU(4)$ irrep for gs should be the one with the lowest eigenvalue for $C_2\l[
SU(4)\r]$ and these eigenvalues can be obtained by combining Eq.
(\ref{ch4.eq.15}) with Eq. (\ref{ch4.eq.17}). Now, for a given  ($m,T_z$)  with
$T=|T_z|$ and $T_z=$(N-Z)$/2$ for a nucleus with N neutrons and Z protons,
enumerating $\{F\} \to (S,T)$ reductions, we can determine the $U(4)$ 
irreps labeling gs, by applying Eq.  (\ref{ch4.eq.21}) with $\kappa$ negative.
In Table \ref{tab21},  $U(4)$ and $U(\Omega)$  irreps for gs are listed for
$\Omega=10$ and $m=4-11$ for all $T_z$ values. As it is well known and also
seen from Table \ref{tab21}, for the Majorana operator or equivalently for
the $SU(4)$ invariant Hamiltonians, for N=Z even-even ($m=4r$), N=Z odd-odd
($m=4r+2$) and N=Z$\pm 1$ ($m=4r \pm 1$) odd-A nuclei, the $U(\Omega)$
irreps for the gs,  with lowest $T$, are $\{4^r\}$, $\{4^r,2\}$, $\{4^r,1\}$
and $\{4^r,3\}$ with spin-isospin structure (see Table \ref{tabnew})  being
$(0,0)$, $(1,0) \oplus (0,1)$, $(\spin,\spin)$ and $(\spin,\spin)$,
respectively.  For convenience, we introduce the notation $\{f_m^{(p)}\}$
where
\be
\{f_m^{(p)}\} =\{4^r,p\}\;;\;\;\;m=4r+p\;\;\mbox{and}\;\; 
p=\mbox{mod}(m,4)
\label{ch4.eq.fmp}
\ee
and this is used in the reminder of the chapter. We shall see ahead that for
the special $U(\Omega)$ irreps in  Eq. (\ref{ch4.eq.fmp}), analytical
formulas are much simpler than for a general $U(\Omega)$ irrep.

Having described some of the essential properties of the $U(\Omega)  \otimes
SU(4)$ algebra, now we will introduce the EGUE(2)-$SU(4)$ random matrix
ensemble and analyze in some detail its properties. From now on we denote
the irreps $\{f\}$ and $\{F\}$ as $f$ and $F$, respectively when there is no 
confusion.

\section{Definition and Basic Properties of EGUE(2)-$SU(4)$}
\label{base}

\subsection{Definition of EGUE(2)-$SU(4)$}
\label{su4-def}

\begin{figure}[ht]
\centering
\includegraphics[width=4.5in,height=5in, angle=-90]{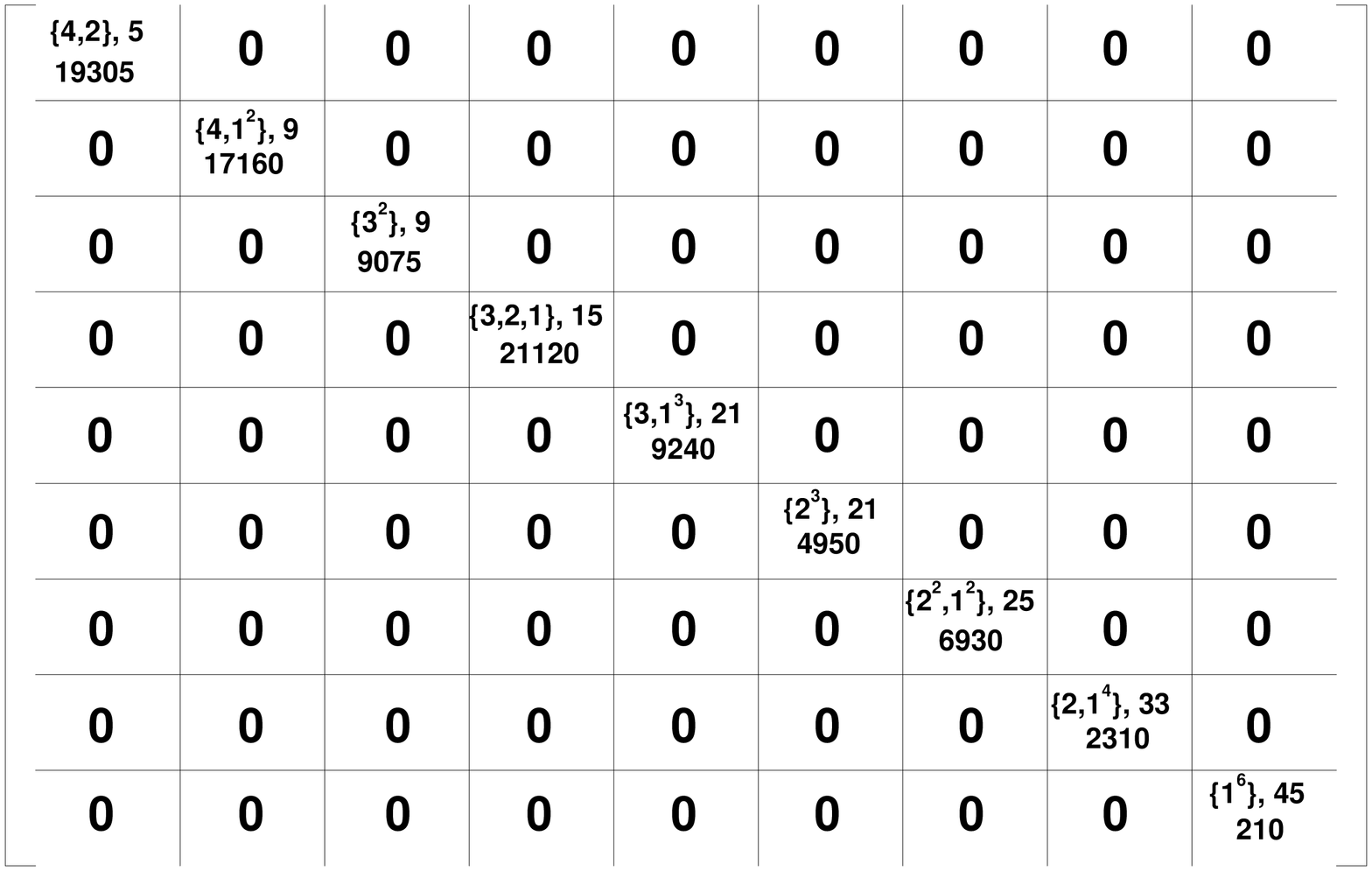} 
\caption{Direct sum matrix structure for a $SU(4)$ scalar Hamiltonian. 
The example in the figure is for $m=6$ particles in  $\Omega=10$ sp
orbitals.  The $U(\Omega)$ irreps and the corresponding eigenvalues for the
quadratic  Casimir invariant of $SU(4)$ along with the  dimensions for the
diagonal  blocks are shown in the figure. For example, for the block that
corresponds to  the irrep $f_m=\{ 3,2,1 \}$, we have $\lan C_2[SU(4)]
\ran^{\tilde{f}_m} = 15$ and $d_\Omega=21120$. As shown in the figure
(with all the off-diagonal blocks having all matrix elements zero), 
$H(m)=\sum_{f_m} H_{f_m}(m)  \oplus$ and  for each diagonal block, 
we have a EGUE(2)-$SU(4)$ matrix ensemble labeled by $(m,f_m)$.}
\label{hmatrix}
\end{figure}

Let us begin with the normalized two-particle states $\l.\l|f_2 v_2; F_2 
\beta_2\r.\ran$ where the $U(4)$ irreps $F_2=\{1^2\}$ and $\{2\}$ and the 
corresponding $U(\Omega)$ irreps $f_2$ are $\{2\}$ (symmetric) and $\{1^2\}$
(antisymmetric), respectively.  Similarly $v_2$ are the additional quantum
numbers that belong to $f_2$ and $\beta_2$ belong to $F_2$. As $f_2$
uniquely defines $F_2$, from now on we will drop $F_2$ unless they are
explicitly needed and also we will use the $f_2 \leftrightarrow F_2$
equivalence whenever needed. With $A^\dagger(f_2 v_2 \beta_2)$ and $A(f_2
v_2 \beta_2)$ denoting creation and annihilation  operators for the
normalized two-particle states, a general two-body Hamiltonian  operator
$\wh$ that is $SU(4)$ scalar can be written as
\be
\wh= \wh_{\{2\}} + \wh_{\{1^2\}} =
\dis\sum_{f_2, v_2^i, v_2^f, \beta_2; f_2=\{2\}, \{1^2\}}
\;H_{f_2 v_2^i v_2^f}(2)\;
A^\dagger(f_2 v^f_2 \beta_2)\,A(f_2 v^i_2 \beta_2)\;.
\label{ch4.eq.22}
\ee
In Eq. (\ref{ch4.eq.22}), the two-body matrix elements  $H_{f_2 v_2^i v_2^f}
(2)= \lan f_2 v^f_2 \beta_2 \mid \wh \mid f_2 v^i_2 \beta_2\ran$ are
independent  of the $\beta_2$'s.  The uniform summation over $\beta_2$ in 
Eq. (\ref{ch4.eq.22}) ensures that $\wh$ is $SU(4)$ scalar and therefore it will
not connect  states with different $f_2$'s.  However, $\wh$ is not a
$SU(4)$ invariant operator.   Just as the two-particle states, we can denote
the $m$-particle states by  $\l| f_m  v_m^f \beta_m^F \ran$;
$F_m=\widetilde{f}_m$. Action of $\wh$ on these states
generates states that are degenerate with respect to $\beta_m^F$ but not
$v_m^f$. Therefore for a given $f_m$, there will be  $d_\Omega(f_m)$ 
number of levels each with $d_4(\tilde{f}_m)$  number of degenerate states. 
Formula for the dimension $d_\Omega(f_m)$ is \cite{Wy-70}, 
\be
d_\Omega(f_m) = \dis\prod_{i<j=1}^\Omega \dis\frac{f_i-f_j+j-i}{j-i}\;,
\label{ch4.eq.28a}
\ee
where, $f_m=\{f_1,f_2,\ldots\}$. Equation (\ref{ch4.eq.28a}) also gives
$d_4(F_m)$  with the product ranging from $i=1$ to $4$ and replacing $f_i$
by $F_i$. As $\wh$ is a $SU(4)$ scalar,  the $m$-particle $H$ matrix will be
a direct sum of matrices with each of them labeled by the $f_m$'s with
dimension $d_\Omega(f_m)$. Thus 
\be
H(m)=\dis\sum_{f_m}\;H_{f_m}(m) \oplus \;.
\label{ch4.eq.hm}
\ee
Figure \ref{hmatrix} shows an example for Eq. (\ref{ch4.eq.hm}). As seen from 
Eq. (\ref{ch4.eq.22}),  the $H$ matrix in two-particle spaces is a direct sum 
of two matrices $H_{f_2}(2)$, one in  the  $f_2=\{2\}$ space and the other
in  $\{1^2\}$ space. Similarly,  for the $6$ particle example shown in Fig.
\ref{hmatrix}, there are $9$ $f_m$'s  and therefore the $H$ matrix is a 
direct sum of $9$ matrices. It should be noted that the  matrix elements of
$H_{f_m}(m)$ matrices receive contributions from  both $H_{\{2\}}(2)$ and
$H_{\{1^2\}}(2)$.

Embedded random matrix ensemble EGUE(2)-$SU(4)$ for a $m$ fermion systems 
with a fixed $f_m$, i.e., $\{H_{f_m}(m)\}$, is generated 
by the ensemble of $H$ operators given in  Eq. (\ref{ch4.eq.22}) with 
$H_{\{2\}}(2)$ and $H_{\{1^2\}}(2)$ matrices replaced by independent GUE 
ensembles of random matrices,
\be
\{H(2)\} = \{H_{\{2\}}(2)\}_{\mbox{GUE}} \oplus \{H_{\{1^2\}}(2)\}_{\mbox{GUE}}
\;.
\label{ch4.eq.ensm11}
\ee
Random variables defining the real and imaginary parts of the matrix
elements of $H_{f_2}(2)$ are  independent Gaussian variables with zero
center and variance given by (with bar representing ensemble average),
\be
\overline{H_{f_2 v_2^1 v_2^2}(2)\;H_{f_2^\pr v_2^3 v_2^4}(2)}
= \delta_{f_2 f_2^\pr} \delta_{v_2^1 v_2^4} \delta_{v_2^2 v_2^3}\,
(\lambda_{f_2})^2\;.
\label{ch4.eq.23}
\ee
Also, the independence of the $\{H_{\{2\}}(2)\}$ and $\{H_{\{1^2\}}(2)\}$ 
GUE ensembles imply,
\be
\barr{l}
\overline{\l[H_{\{2\} v_2^1 v_2^2}(2)\r]^P\;
\l[H_{\{1^2\} v_2^3 v_2^4}(2)\r]^Q} = 0 \;\;\;\mbox{for}\;\;\;P
\;\;\;\mbox{or}\;\;\;Q\;\;\;\mbox{odd}\;,
\\ \\
= \l\{\;\overline{\l[H_{\{2\} v_2^1 v_2^2}(2)\r]^P}\;\r\}\;\;
\l\{\;\overline{\l[H_{\{1^2\} v_2^3 v_2^4}(2)\r]^Q}\;\r\}
\;\;\;\mbox{for}\;\;P \;\;\mbox{and}\;\;Q\;\;\mbox{even} \;.
\earr \label{ch4.eq.indep}
\ee
Action of $\wh$ defined by Eq. (\ref{ch4.eq.22}) on $m$-particle basis states 
with a fixed $f_m$, along with Eqs. (\ref{ch4.eq.ensm11})-(\ref{ch4.eq.indep}) 
generates EGUE(2)-$SU(4)$ ensemble $\{H_{f_m}(m)\}$; it is labeled by the 
$U(\Omega)$ irrep $f_m$ with matrix dimension $d_{\Omega}(f_m)$.

\subsection{Matrix structure}
\label{stru}

For a better understanding of the size of the EGUE(2)-$SU(4)$ matrices and
the number of independent matrix elements they contain,  let us consider the
example of $8$ fermions in $N=24$ sp states. For spinless fermion systems,
we have EGUE(2) with a two-particle GUE of dimension $276$ and the number of
independent variables [denoted by $i_2(0)$] is $76176$.  These generate the
$m$ fermion EGUE(2) ensemble with $H$ matrices of dimension $d(8)=735471$.
For fermions with spin symmetry,  we have EGUE(2)-$\cs$ with $\Omega=12$.
This ensemble is generated by  independent GUE's in two-particle spin $s=0$
and $s=1$  spaces with dimensions $78$ and $66$, respectively.  Then the
number of independent variables [denoted by $i_2(2)$]  for this system is
$10440$. The $H$ matrix dimensions for EGUE(2)-$\cs$ ensembles for the $8$
particle system with spins $S=0,\;1,\;2,\;3$ and $4$ are $d(8,S)=70785$,
$113256$, $51480$, $9009$, and $495$, respectively. Going further, with
$SU(4)$ symmetry we have EGUE(2)-$SU(4)$ ensembles with $\Omega=6$. These
ensembles are  generated by two independent GUE's in $f_2=\{2\}$ and
$\{1^2\}$ spaces with dimensions $21$ and $15$ respectively.  Then the
number of independent variables [denoted by $i_2(4)$] for this system is 
$666$. The $H$ matrix dimensions for EGUE(2)-$SU(4)$ ensembles for the $8$
particle system with $f_8=\{2^2,1^4\}$, $\{2^3,1^2\}$, $\{2^4\}$,
$\{3,1^5\}$, $\{3,2,1^3\}$, $\{3,2^2,1\}$, $\{3^2,1^2\}$, $\{3^2,2\}$,
$\{4,1^4\}$, $\{4,2,1^2\}$, $\{4,2^2\}$, $\{4,3,1\}$, and $\{4^2\}$ are $15$,
$105$, $105$, $21$, $384$, $1050$, $1176$, $1470$, $315$, $2430$, $2520$,
$4410$, and $1764$, respectively. Thus $i_2$  will be considerably reduced as
the symmetry increases (with fixed $N$),  i.e., $i_2(4) << i_2(2) << i_2(0)$.
Similarly the $H$ matrix dimensions decrease as we go from EGUE(2) to
EGUE(2)-$\cs$ to EGUE(2)-$SU(4)$. For further insight, let us consider the
fraction of independent matrix elements $\cii(m,f_m)$, for $m >> 2$ for the
EGUE(2)-$SU(4)$ ensemble,  defined as the  ratio of $i_2(4)$ to the total
number (without counting the  hermitian conjugates) of matrix elements,  
\be
\cii(m,f_m)=\dis\frac{i_2(4)}{[d_\Omega(f_m)]^2}\;. 
\label{ch4.eq.su4-1} 
\ee 
Similarly, for EGUE(2) and EGUE(2)-$\cs$ ensembles, we can define the 
fraction of independent matrix elements as $\cii(m)=i_2(0)/[d(m)]^2$ and
$\cii(m,S)=i_2(2)/[d(m,S)]^2$,  respectively. In our above example,  for
EGUE(2), EGUE(2)-$\cs$ with $S=0$ and EGUE(2)-$SU(4)$ with $f_8=\{4^2\}$, we
have $\cii=1.4\times10^{-7}$, $2\times10^{-6}$, and $2\times10^{-4}$,
respectively. Therefore the $H$ matrices with more symmetry are
characterized by relatively large fraction of independent matrix
elements.   

Due to the two-body selection rules, many of the $m$-particle matrix
elements of the EGUE(2)'s will be zero. In order to understand the sparse
nature of the EGUE matrices we employ the sparsity index $\cS$ with
$\cS^{-1}$  defined as the ratio of number of $m$-particle states  that are
directly  coupled by the two-body interaction to the $m$-particle matrix
dimension. The number of many-particle states that are coupled by the
two-body interaction, i.e., the connectivity factor $K(m,f_m)$, is given by
the spectral variances; see Chapter \ref{ch2} and \cite{Ja-97}. 
Therefore, for the EGUE(2)-$SU(4)$ ensemble,
\be
\cS^{-1}(m,f_m)=\dis\frac{K(m,f_m)}{d_\Omega(f_m)}\;.
\label{ch4.eq.spar}
\ee
Similarly, $\cS^{-1}(m)=K(m)/d(m)$ for EGUE(2) and $\cS^{-1}(m,S)
=K(m,S)/d(m,S)$ for EGUE(2)-$\cs$. Formulas for the $K(m)$ and $K(m,S)$ are
given in (\cite{Fl-96a}, \cite{Ko-05}) and (\cite{Ko-07}, Chapter \ref{ch2}), 
respectively. For
EGUE(2)-$SU(4)$,  given the two-particle variances to be
$\lambda^2_{f_2}=\lambda^2$, the variances $\overline{\lan \wh^2
\ran^{m,f_m}}$ in $m$-particle space are $\sigma^2(m,f_m)=\lambda^2
K(m,f_m)$ with $K(m,f_m)$ propagating the  two-particle variances to
$m$-particle spaces. Results in Table \ref{tabw}  ahead give formulas for
the variance propagator $K(m,f_m)$ for  the $U(\Omega)$ irreps $f_m^{(p)}$.
For example,    $K(m=4r,f_m=\{4^r\}) =
r(\Omega-r+4)\l\{2r(2\Omega-2r+9)-\Omega-8\r\}$, and 
$K(m=4r+1,f_m=\{4^r,1\}) = r(\Omega-r+4)\l\{4r(2\Omega-2r+7)+2 \Omega
-15\r\}/2$. For the $8$ particle example (with $N=24$) considered before, 
the connectivity factors $K$ are  $4284$, $1440$, and $864$,  respectively 
for EGUE(2), EGUE(2)-$\cs$ with $S=0$ and EGUE(2)-$SU(4)$ with
$f_8=\{4^2\}$.  These give $\cS^{-1} = 5.8\times 10^{-3}$, $0.02$, and 
$0.49$, respectively for these ensembles.  Therefore as symmetry increases,
in general, the many-particle EGUE matrices will become more dense.
Consequences of this will be discussed further in Section \ref{num2}.

\subsection{Matrix construction} 
\label{cons}

Before proceeding to the analytical formulation, we will briefly outline a
method for numerical  construction of EGUE(2)-$SU(4)$ ensemble for a given
($\Omega,m,f_m$). Consider $m$ fermions in $\Omega$ number of sp orbitals
each four-fold degenerate. Then in the spin-isospin representation, the sp
states are denoted by $\l|i;\spin,m_\cs;\spin,m_\ct\ran$ as discussed
before, where $i=1,2,\ldots,\Omega$. We arrange  the sp states in such a way
that the first $\Omega$ states have $(m_\cs,m_\ct)=(\spin,\spin)$,
$\Omega+1$ to $2\Omega$ sp states have $(m_\cs,m_\ct)=(\spin,-\spin)$,
$2\Omega+1$ to $3\Omega$ sp states have $(m_\cs,m_\ct)=(-\spin,\spin)$ and
$3\Omega+1$ to $4\Omega$ sp states have $(m_\cs,m_\ct)=(-\spin,-\spin)$. In
this single state representation we denote the sp states as $\l| k_r \ran$,
$r=1,2\ldots,4\Omega$. Now distributing in all possible ways the $m$
fermions in these $4\Omega$ sp states will generate the $m$-particle
configurations $\cm =[m(k_1)$, $m(k_2)$, $\ldots$,  $m(k_{4\Omega})]$, with 
$m(k_r) = 0$ or $1$ and $\sum_{r=1}^{4\Omega} m(k_r) = m$. The corresponding
$(M_S,M_T)$ values are $M_S = [ \sum_{r1=1}^\Omega m(k_{r1}) +
\sum_{r2=\Omega+1}^{2\Omega} m(k_{r2}) - \sum_{r3=2\Omega+1}^{3\Omega}
m(k_{r3}) - \sum_{r4=3\Omega+1}^{4\Omega} m(k_{r4}) ]/2$ and $M_T=
[\sum_{r1=1}^\Omega m(k_{r1}) - \sum_{r2=\Omega+1}^{2\Omega} m(k_{r2}) +
\sum_{r3=2\Omega+1}^{3\Omega} m(k_{r3}) - \sum_{r4=3\Omega+1}^{4\Omega}
m(k_{r4})]/2$. The $m$-particle $H$ matrix in the basis defined by $\cm$'s
with $(M_S^{min},M_T^{min})=(0,0)$ will  contain states with all $(S,T)$
values for even $m$ and similarly with  $(M_S^{min},M_T^{min}) =
(\spin,\spin)$ for odd $m$. The dimension of this basis space, called
$\cd(M_S^{min},M_T^{min})$, is $\sum_{f_m}d_\Omega(f_m) 
\sum_{S,T}\,D(\tilde{f}_m:S,T)$. In the $(st)$ coupled representation  the
two-particle  matrix elements of $\wh$ are 
$$
\lan (i,j) s, m_s, t, m_t \mid
\wh \mid (k,l) s^\pr, m_{s^\pr}, t^\pr, m_{t^\pr} \ran \;.
$$ 
As the $SU(4)$ irreps $\{2\} \rightarrow (st)=(11) \oplus (00)$ and $\{1^2\}
\rightarrow (10) \oplus (01)$, it is easy to put these matrix elements in
one to one correspondence with the  two-body matrix elements $H_{f_2 v_2^i
v_2^f}(2)$ in Eq. (\ref{ch4.eq.22}). Applying angular-momentum algebra, it is
then possible to transform these matrix elements into  two-body matrix
elements $\lan k_c k_d | \wh |k_a k_b\ran$ in the single state
representation.  Then the construction of the $m$-particle $H$ matrix in the
$\cm$-basis with $(M_S^{min},M_T^{min})$  defined above reduces to the
problem of EGUE(2) for spinless fermion systems. The construction of EGUE(2)
for spinless fermion systems on a machine is straightforward.
For instance, the dimensions of the matrices
$\cd(M_S^{min}=0,M_T^{min}=0)$  for $m=6$, $8$ and $12$, with $\Omega=6$, 
are $17000$, $79875$, and $263844$, respectively. On the other hand, the
total  $m$-particle matrix dimensions are $d(6)=134596$, $d(8)=735471$, and
$d(12)=2704156$. Therefore,  the  $\cm$-basis formulation reduces the matrix
dimensions considerably.

After constructing this matrix, it is possible to generate the $H$ matrix
defined over a fixed $f_m$ space, for some special $f_m$'s  easily, using the
$C_2[SU(4)]$ operator as the projection operator;  eigenvalues of $C_2[SU(4)]$
will in general have degeneracies with respect to $f_m$. Some of the special
irreps that can be identified uniquely by $C_2[SU(4)]$ are the following: (a)
for $m=4r$, the irreps  $\{4^r\}$, $\{4^{r-1},3,1\}$ and $\{4^{r-1},2^2\}$ with
eigenvalues  $0$, $8$, and $12$, respectively; (b)  for $m=4r+2$, the irreps
$\{4^r,2\}$ and $\{4^{r-1},3,2,1\}$  with eigenvalues $5$, and $15$,
respectively; (c) for $m=4r+1$, the irreps  $\{4^r,1\}$, $\{4^{r-1},3,2\}$ and
$\{4^{r-1},3,1^2\}$ with eigenvalues $3$, $9$, and $13$, respectively; and (d)
for $m=4r+3$, the irreps  $\{4^r,3\}$, $\{4^{r},2,1\}$, and $\{4^{r-1},3^2,1\}$
with eigenvalues $3$, $9$, and $13$, respectively. For convenience, we denote
these special irreps by $f_m^s$. It should be noted that $f_m^{(p)}$ belong to
$f_m^s$. For the $C_2[SU(4)]$ operator, the $m$-particle matrix in the
$\cm$-basis can be constructed by identifying the two-particle matrix elements,
in single state representation,  using Eqs. (\ref{ch4.eq.12})-(\ref{ch4.eq.15}).
Diagonalizing this matrix gives a direct sum of unitary matrices and the unitary
matrix that corresponds to a given $f_m^s$ can be identified from the
eigenvalues of $C_2[SU(4)]$. Applying the unitary transformation defined by this
unitary matrix, the $m$-particle $H$ matrix with $(M_S,M_T)=(0,0)$  for even $m$
[$(M_S,M_T)=(\spin,\spin)$ for odd $m$] can be transformed to the basis with
good $f_m^s$. This method can  be successfully implemented on a machine for the
irreps $f_m^{s}$. Results  in Section \ref{su4bk} are sufficient for
constructing EGUE(2)-$SU(4)$ for these irreps. It is important to note that the
$C_2[SU(4)]$  alone will not suffice to identify the matrices corresponding to
all the $f_m$'s. To distinguish them, we need to construct the $m$-particle
matrices for the cubic and quartic Casimir invariants of  $SU(4)$ algebra and
these are more complicated. Numerical investigations of  EGUE(2)-$SU(4)$ by
matrix construction are impractical as the dimensions $\cd(M_S^{min},M_T^{min})$
are prohibitively large (even for $\Omega=6$ and $m=6$, $\cd=17000$). Therefore
our focus in this chapter is in developing analytical formulation for solving
the EGUE(2)-$SU(4)$ ensemble  (Secs. \ref{wralg} and \ref{anres} and Appendix
\ref{c4a2})  and using this we have carried out some numerical  investigations
(Secs. \ref{num1} and \ref{num2}).

\section{$U(4\Omega)\supset U(\Omega)\otimes SU(4)$ Wigner-Racah 
Algebra for Solving EGUE(2)-$SU(4)$}
\label{wralg}

Analytical solutions for EGUE(2)-$SU(4)$ follow, as discussed before for
EGUE($k$) and EGUE(2)-$\cs$ (see Sec. \ref{egue1} and Appendix \ref{egue2}), 
from the tensorial
decomposition of the  $\wh$ operator [equivalently $A^\dagger A$ in Eq.
(\ref{ch4.eq.22})] with respect to $U(\Omega) \otimes SU(4)$. 
As $\wh$ is a $SU(4)$ scalar, it transforms as the irrep
$\{0\}$ with respect to the $SU(4)$ algebra.  However with respect to
$SU(\Omega)$, the tensorial characters, in the Young tableaux notation, for
$f_2=\{2\}$ are $\bF_\nu=\{0\}$, $\{21^{\Omega-2}\}$ and $\{42^{\Omega-2}\}$
with $\nu=0,1$, and 2, respectively. Note that $\bF_\nu$ follow from the
Kronecker product of the $U(\Omega)$ irreps $\{2\}$ and $\{2^{\Omega-1}\}$
as $A^\dagger$ and $A$ transform as these irreps.  Similarly for
$f_2=\{1^2\}$,  $\bF_\nu= \{0\}$, $\{21^{\Omega-2}\}$ and $\{2^2
1^{\Omega-4}\}$ with $\nu=0,1,2$, respectively. Then we can introduce unitary
tensors $B$'s,
\be
\barr{l}
B(f_2 \bF_\nu \omega_\nu) =\\
\dis\sum_{v_2^i,v_2^f, \beta_2}\,
A^\dagger(f_2
v^f_2 \beta_2)\, A(f_2 v^i_2 \beta_2)\, \lan f_2
v_2^f\;\overline{f_2}\,\overline{v_2^i} \mid \bF_\nu \omega_\nu\ran
\lan F_2 \beta_2\;\overline{F_2}\,\overline{\beta_2} \mid \{0\} 0\ran\;,
\label{ch4.eq.24}
\earr
\ee
and expand $\wh$ in terms of these tensor operators.  In Eq. (\ref{ch4.eq.24}),
$\lan f_2 --- \ran$ are $SU(\Omega)$  Wigner coefficients  and $\lan F_2 ---
\ran$ are $SU(4)$ Wigner coefficients.  Some properties of the Wigner
coefficients are discussed in Appendix \ref{c4a1}. 
Note that in Eq. (\ref{ch4.eq.24}),
irreps $\overline{f_2}$ are complex conjugate of the irreps $f_2$
\cite{But-81}. For example, for the $U(\Omega)$ irrep $f=\{2^r\}$, the irrep
that corresponds to $\overline{f}$ is $\{2^{\Omega-r}\}$. Similarly,
$\overline{f} = \{4^{\Omega-r}\}$ for $f=\{4^r\}$, $\overline{f} =
\{4^{\Omega-r-2},2,1\}$ for $f=\{4^r,3,2\}$ and so on.  Using the
orthonormal properties of the Wigner coefficients appearing in Eq.
(\ref{ch4.eq.24}) and the action of operators $A$ and $A^\dg$ on the vaccum and
two-particle states respectively, 
it can be proved that the tensors $B$'s are orthonormal with
respect to the traces over fixed $f_2$ spaces,
\be
\lan\lan B(f_2 \bF_\nu \omega_\nu) B(f^\pr_2 \bF^\pr_\nu \omega^\pr_\nu)
\ran\ran^{f_2} = \delta_{f_2 f^\pr_2} \delta_{\bF_\nu \bF^\pr_\nu} 
\delta_{\omega_\nu \omega^\pr_\nu} \;.
\label{ch4.eq.nn1}
\ee
Expanding $\wh$ in terms of $B$'s will give the expansion coefficients 
$W$'s,
\be
\wh=\dis\sum_{f_2, \bF_\nu, \;\omega_\nu}\;W(f_2 \bF_\nu \omega_\nu)\,
B(f_2 \bF_\nu \omega_\nu)\;,
\label{ch4.eq.25}
\ee
and they can be written in terms of the $H(2)$ matrix elements using Eq.
(\ref{ch4.eq.nn1}),
\be
\barr{rcl}
W(f_2 \bF_\nu \omega_\nu) & = & \lan\lan \wh \;
B(f_2 \bF_\nu \omega_\nu) \ran\ran^{f_2} \\ 
& = & \dis\sum_{v_2^i,v_2^f}\, \dis\sqrt{d_4(\widetilde
{f}_2)}
\,\lan f_2
v_2^f\;\overline{f_2}\,\overline{v_2^i} \mid \bF_\nu \omega_\nu\ran
\,H_{f_2 v_2^i v_2^f}(2)\;.
\earr \label{ch4.eq.nn2}
\ee
Now the most significant result is that the $W$'s are also independent 
Gaussian variables just as $H(2)$'s with ensemble averaged variances 
given by,
\be
\overline{W(f_2 \bF_\nu \omega_\nu)W(f^\pr_2 \bF^\pr_\nu \omega^\pr_\nu)}
=\delta_{f_2 f^\pr_2} \delta_{\bF_\nu \bF^\pr_\nu} \delta_{\omega_\nu
\omega^\pr_\nu} \, (\lambda_{f_2})^2 d_4(F_2)\,.
\label{ch4.eq.26}
\ee
Above result is derived using Eq. (\ref{ch4.eq.23}) and (\ref{ch4.eq.nn2}).
As we will see ahead, Eq. (\ref{ch4.eq.26}) and the $(m,f_m)$-space matrix
elements  of $H$ as given by the Wigner-Eckart theorem applied  using
$SU(\Omega) \otimes SU(4)$ Wigner-Racah algebra, will completely solve
EGUE(2)-$SU(4)$. 

Analysis of the random matrix ensemble EGUE(2)-$SU(4)$ involves construction
of the one-point function $\overline{\rho^{m,f_m}(E)}$, the ensemble
averaged density of eigenvalues  given by Eq. (\ref{eq.eden}) with $\Gamma=f_m$
and the two-point and other higher point functions defining fluctuations.
The two-point function is given by,
\be
S^{m, \Gamma:m^\pr, \Gamma^\pr}(E,E^\pr) = \overline{\rho^{m,\Gamma}(E)
\rho^{m^\pr , \Gamma^\pr}(E^\pr)} - \l\{\overline{\rho^{m,\Gamma}(E)}\r\}
\;\;
\l\{\overline{\rho^{m^\pr ,\Gamma^\pr}(E^\pr)}\r\}\;, 
\label{eq.rho}
\ee
with $\rho^{m,\Gamma}(E)$ defining fixed-$(m,\Gamma)$ density of eigenvalues. 
The two-point function $S^{m \Gamma:m^\pr \Gamma^\pr}$ generates  
`self-correlations' when $m=m^\pr$ and $\Gamma=\Gamma^\pr$ and 
`cross-correlations'
between states with $m \neq m^\pr$ and/or $\Gamma \neq \Gamma^\pr$. For
EGUE(2)-$SU(4)$ ensemble, $\Gamma = f_m$. In Chapter \ref{ch6},  $\Gamma$
corresponds to the $m$-particle spin $S$. 
Therefore, for EGUE(2)-$SU(4)$, 
the
two-point function $S^{m f_m:m^\pr f_{m^\pr}}$ generates self-correlations
when $m=m^\pr$ and $f_m=f_{m^\pr}$ and cross-correlations between states
with $m = m^\pr$ and $f_m \neq f_{m^\pr}$ and also between states with  $m
\neq m^\pr$ and  $f_m \neq f_{m^\pr}$. It should be emphasized that with
$m=m^\pr$ it is possible to have $f_m \neq f_{m^\pr}$ and this should not be
confused as $f_m = f_{m^\pr}$ (confusion may arise if one substitutes the
numerical  value for $m=m^\pr$). Towards  deriving the forms for the one
and two-point functions (discussion of higher point functions is beyond the
scope of  the present thesis), the moment approach is adopted and the lower
order moments are analyzed. By definition, all odd moments of
$\overline{\rho^{m,f_m}(E)}$ will vanish and therefore the lower order
moments of interest are the ensemble averaged spectral variances
$\overline{\lan \wh^2 \ran^{m,f_m}}$ and the fourth moment $\overline{\lan
\wh^4 \ran^{m,f_m}}$ giving the excess parameter  $\gamma_2(m,f_m)$ where,
\be
\gamma_2(m,f_m) = \l[\; \overline{\lan \wh^2 \ran^{m,f_m}}\;\r]^{-2}\;
\l[\; \overline{\lan \wh^4 \ran^{m,f_m}}\;\r] - 3\;.
\label{ch4.eq.gam2}
\ee
Similarly the two lower order normalized bivariate moments of the two-point 
function are $\Sigma_{rr}$, $r=1,\; 2$ give the covariances in energy centroids
and spectral variances respectively. The formulas for these are given by,
\be
\barr{rcl}
\Sigma_{11}(m,\Gamma;m^\pr,\Gamma^\pr) & = & 
\dis\frac{\overline{\lan H \ran^{m,\Gamma}
\lan H \ran^{m^\pr,\Gamma^\pr}}}{\l[ \; 
\l\{ \overline{\lan H^2 \ran^{m,\Gamma}} \r\}\;\;
\l\{ \overline{\lan H^2 \ran^{m^\pr,\Gamma^\pr}} \r\}\;\r]^{1/2}}\;, \\ \\
\Sigma_{22}(m,\Gamma;m^\pr,\Gamma^\pr) & = & 
\dis\frac{\overline{\lan [H]^2 \ran^{m,\Gamma} 
\lan H^2 \ran^{m^\pr,\Gamma^\pr}}}{\l\{ \overline{\lan H^2 \ran^{m,\Gamma}} \r\}
\;\;
\l\{ \overline{\lan H^2 \ran^{m^\pr,\Gamma^\pr}}\r\}} - 1 \;,
\earr \label{eq.den10}
\ee
with $\Gamma=f_m$ and $\Gamma^\pr=f_{m^\pr}$ for EGUE(2)-$SU(4)$.  
For $m=m^\pr$ and $f_m=f_{m^\pr}$, the $\Sigma_{11}$ and $\Sigma_{22}$ 
give the first two terms in the normal mode decomposition of the level 
motion in the ensemble \cite{Br-81,PDPK} and hence they are of importance.
Similarly for ($m = m^\pr$, $f_m \neq f_{m^\pr}$) and ($m \neq m^\pr$,  
$f_m \neq f_{m^\pr}$), the $\Sigma_{11}$ and $\Sigma_{22}$ are important as 
they generate non-zero cross-correlations that are zero if the $m$-particle 
$H$ matrices for each $f_m$ are represented by independent GUE's.

In order to derive the analytical results for the moments of the 
one and two-point functions, 
the basic quantity that is needed is the ensemble averaged covariance 
between any two $m$-particle matrix elements of $H$, i.e.,
\be
\barr{l}
\overline{H_{f_m v_m^i v_m^f}\,H_{f_{m^\pr} v_{m^\pr}^i v_{m^\pr}^f}}
\\ \\ =
\overline{\lan f_m F_m v_m^f \beta \mid \wh \mid f_m F_m v_m^i \beta\ran
\lan f_{m^\pr} F_{m^\pr} v_{m^\pr}^f \beta^\pr \mid \wh \mid f_{m^\pr}
F_{m^\pr} v_{m^\pr}^i \beta^\pr \ran }\;.
\earr \label{ch4.eq.cv1}
\ee
Using the expansion given by Eq. (\ref{ch4.eq.25}) and applying Eq.
(\ref{ch4.eq.26}) for the ensemble average of the product of two $W$'s,
$\overline{H\;H}$ reduces to the matrix elements of the unit tensors $B$'s.
Wigner-Eckart theorem in $SU(\Omega)$ and $SU(4)$ spaces will give  
\cite{He-74a},
\be
\barr{l}
\lan f_m F_m v_m^f \beta \mid B(f_2 \bF_\nu \omega_\nu) \mid f_m F_m
v_m^i \beta \ran  \\ \\
= \dis\sum_{\rho}\lan f_m \mid\mid
B(f_2 \bF_\nu) \mid\mid f_m\ran_\rho\,
\lan f_m v_m^i \bF_\nu \omega_\nu \mid f_m v_m^f\ran_\rho \\  \\
= \dis\frac{1}{\sqrt{d_\Omega(f_2) d_4(\tilde{f}_2)}}
\;\dis\sum_\rho\;\lan f_m \mid\mid\mid
B(f_2 \bF_\nu) \mid\mid\mid f_m\ran_\rho\,
\lan f_m v_m^i \bF_\nu \omega_\nu \mid f_m v_m^f\ran_\rho\;; \\ \\
\lan f_m \mid\mid\mid
B(f_2 \bF_\nu) \mid\mid\mid f_m\ran_\rho\,=
\dis\sum_{f_{m-2}}\;F(m)\, \dis\frac{\cn_{f_{m-2}}}{\cn_{f_m}}\;
\dis\frac{U(f_m \overline{f_2} f_m f_2; f_{m-2} \bF_\nu)_\rho}{U(f_m
\overline{f_2} f_m f_2; f_{m-2} \{0\})} \;,
\earr \label{ch4.eq.27}
\ee
where the summation is over the multiplicity index $\rho$ and this arises
as  $f_m \otimes \bF_\nu$ gives in general more than once the irrep $f_m$.
In Eq. (\ref{ch4.eq.27}), $F(m)=-m(m-1)/2$ and $U(---)$ are the $SU(\Omega)$
Racah coefficients. Similarly, the standard double-barred matrix elements
(called reduced matrix elements) are changed into triple-barred matrix
elements in Eq. (\ref{ch4.eq.27}) for convenience.  The formula for the
dimension $d_\Omega(f_m)$ is given by Eq. (\ref{ch4.eq.28a}) and  the dimension
$\cn_{f_m}$ of $f_m$ with respect to the $S_m$ group is \cite{Wy-70},
\be
\cn_{f_m} = \dis\frac{m!\dis\prod_{i<k=1}^r (\ell_i-\ell_k)}
{\ell_1!\;\ell_2!\ldots\ell_r!}\;;\;\;\;\;\ell_i=f_i+r-i \;.
\label{ch4.eq.28b}
\ee
Here, $r$ denotes total number of rows in the Young tableaux for $f_m$.
Correlations generated by EGUE(2)-$SU(4)$ between states with $(m,f_m)$
and $(m^\pr,f_{m^\pr})$ follow from the covariances between the $m$-particle
matrix elements of $H$. Applying Eqs. (\ref{ch4.eq.cv1}), (\ref{ch4.eq.25}), 
(\ref{ch4.eq.26}) and (\ref{ch4.eq.27}) in that order, the final expression for
$\overline{H\;H}$ is,
\be
\barr{l}
\overline{H_{f_m v_m^i v_m^f}\,H_{f_{m^\pr} v_{m^\pr}^i v_{m^\pr}^f}}
\\ \\ =
\dis\sum_{f_2, \bF_\nu,\; \omega_\nu} \; \dis\frac{(\lambda_{f_2})^2}
{d_\Omega(f_2)}\;
\dis\sum_{\rho,\rho^\pr}\; \lan f_m \mid\mid\mid B(f_2 \bF_\nu)
\mid\mid\mid
f_m\ran_\rho  \; 
\lan f_{m^\pr} \mid\mid\mid B(f_2 \bF_\nu) \mid\mid\mid f_{m^\pr}
\ran_{\rho^\pr} \\ \\
\times
\lan f_m v_m^i\;\bF_\nu \omega_\nu
\mid f_m v_m^f\ran_\rho\;
\lan f_{m^\pr} v_{m^\pr}^i\;\bF_\nu \omega_\nu \mid f_{m^\pr}
v_{m^\pr}^f\ran_{\rho^\pr}\,.
\earr \label{ch4.eq.29}
\ee
In the following section, we will consider $\overline{\lan
\wh^2\ran^{m,f_m}}$ and $\overline{\lan \wh^r \ran^{m,f_m} \;  \lan \wh^r
\ran^{m^\pr,f_{m^\pr}}}$; $r=1,\;2$. It is important to mention here that in
evaluating these moments, the Wigner coefficients appearing in Eq.
(\ref{ch4.eq.29}) will eventually disappear due to the orthonormal properties of
these coefficients [see Eqs. (\ref{ch4.eq.prp6}) and (\ref{ch4.eq.prp7})]  and
therefore the final results for these moments will involve only the
$SU(\Omega)$ Racah coefficients given in Eq. (\ref{ch4.eq.27}). In Appendix
\ref{c4a2}, we will consider  $\overline{\lan \wh^4\ran^{m,f_m}}$ and the
algebra here is more complicated giving additional Racah coefficients than
in Eq. (\ref{ch4.eq.27}).

From now onwards, we drop the ``hat'' symbol over the $H$ operator when there
is no confusion.

\section{Exact Expressions for Spectral Variances, Lower Order 
Cross-correlations and Analytical Results for Lowest $U(\Omega)$ Irreps}
\label{anres}

\subsection{Covariances in energy centroids $\overline{\lan H \ran^{m,f_m} 
\; \lan H \ran^{m',f_{m^\pr}}}$}
\label{ansg11}

Firstly the ensemble averaged energy centroid $\overline{\lan H \ran^{m,f_m}} =
0$ by the definition of EGUE(2)-$SU(4)$ ensemble. As $\lan H \ran^{m,f_m}$ is
the trace of $H$ (divided by the dimensionality)  in $(m,f_m)$ space, only
$\bF_\nu = \{0\}$ will generate this.  Therefore for $\overline{\lan H\ran
\;\lan H\ran}$, the Wigner  coefficients in Eq. (\ref{ch4.eq.29}) and the ratio
of the $U$-coefficients in Eq. (\ref{ch4.eq.27})  will be unity. Then trivially,
\be
\barr{rcl}
\overline{\lan H \ran^{m,f_m} \; \lan H \ran^{m',f_{m^\pr}}} & = &  F(m)
\,F(m^\pr)\;\dis\sum_{f_2}\,\dis\frac{(\lambda_{f_2})^2}{d_\Omega(f_2)}
\dis\sum_{f_{m-2}} \dis\frac{\cn_{f_{m-2}}}{\cn_{f_m}}\;
\dis\sum_{f_{m^\pr-2}} \dis\frac{\cn_{f_{m^\pr-2}}}{\cn_{f_{m^\pr}}} \\
& = & \dis\sum_{f_2} \dis\frac{\l(\lambda_{f_2}\r)^2}{d_\Omega(f_2)}
\;P^{f_2}
(m,f_m)\;P^{f_2}(m^\pr,f_{m^\pr})\;,
\earr \label{ch4.eq.31}
\ee
where,
\be
P^{f_2}(m,f_m)=F(m)\dis\sum_{f_{m-2}} \dis\frac{\cn_{f_{m-2}}}
{\cn_{f_m}}\;.
\label{ch4.eq.32}
\ee
Table \ref{pf2} gives the expression for $P^{f_2}(m,f_m)$ for the irreps
$f_m^{(p)}$. It is possible to derive Eq. (\ref{ch4.eq.31}) using the trace
propagation formula for the energy  centroids \cite{Pa-78},  
\be
E_c(m,f_m) = \lan H \ran^{m,f_m} = a_0 + a_1 m + a_2 m^2 + a_3 \lan C_2
\l[SU(4)\r]\ran^{\tilde{f}_m} \nonumber
\ee
\be
\barr{rcl}
\Rightarrow
E_c(m,f_m) 
& = & \dis\frac{3m^2+12m-4\lan C_2\l[SU(4)\r]\ran^{\tilde{f}_m}}{16}
\lan H\ran^{2,\{2\}} \\ \\
& + & \dis\frac{5m^2-20m+4\lan C_2
\l[SU(4)\r]\ran^{\tilde{f}_m}}{16} \lan H\ran^{2,\{1^2\}}\;.
\earr \label{ch4.eq.cent}
\ee
Note that $\lan C_2\l[SU(4)\r]\ran^{\tilde{f}_m} =  \lan
C_2\l[U(4)\r]\ran^{\tilde{f}_m} - m^2/4$ with   $\lan
C_2\l[U(4)\r]\ran^{\tilde{f}_m}$ given by Eq. (\ref{ch4.eq.17}). We have 
verified that Eq. (\ref{ch4.eq.cent})  reproduces the results given in  Table
\ref{pf2}.

\begin{table}[htp]
\caption{$P^{f_2}(m,f_m)$ for $f_m=f_m^{(p)}=\{4^r,p\}$ 
and $f_2=\{2\}$ and $\{1^2\}$. See Eq. (\ref{ch4.eq.32}) for the definition 
of $P^{f_2}(m,f_m)$.}
\begin{center}
\begin{tabular}{ccc}
\toprule
 & $\;\;\;\;\;\;\;\;\;\;\;\;\;\;\;\;\;\;\;\;\;\;\;\;\;\;\;\;\;\;\;\;\;
\;\;\;\;\;\;\;\;P^{f_2}(m,f_m^{(p)})$ & \\ \cmidrule{2-3} 
$f_m^{(p)}$ & $f_2=\{2\}$ & $f_2=\{1^2\}$ \\ 
\midrule 
$\{4^r\}$ & $-3r(r+1)$ & $-5r(r-1)$ \\ \\
$\{4^r,1\}$ & $-\dis\frac{3r}{2}(2r+3)$ & $-\dis\frac{5r}{2}(2r-1)$\\ \\
$\{4^r,2\}$ & $-(3r^2+6r+1)$ & $-5r^2$ \\ \\
$\{4^r,3\}$ & $-\dis\frac{3}{2}(r+2)(2r+1)$ & $-\dis\frac{5r}{2}(2r+1)$\\ 
\bottomrule
\end{tabular}
\end{center}
\label{pf2}
\end{table}

\subsection{Spectral variances $\overline{\lan H^2 \ran^{m,f_m}}$}
\label{anvar}

Writing $\overline{\lan H^2 \ran^{m,f_m}}$ explicitly in terms of the 
$m$-particle $H$ matrix elements,
\be
\overline{\lan H^2 \ran^{m,f_m}} = \dis\frac{1}{d_\Omega(f_m)} 
\dis\sum_{v_m^1 ,v_m^2}
\, \overline{H_{f_m v_m^1 v_m^2}\,H_{f_m v_m^2 v_m^1}}\;,
\label{ch4.eq.33}
\ee
and then applying Eqs. (\ref{ch4.eq.27}) and (\ref{ch4.eq.29}) and the 
orthonormal properties of the $SU(\Omega)$ Wigner coefficients (see Appendix
\ref{c4a1})  lead to
\be
\barr{rcl}
\overline{\lan H^2 \ran^{m,f_m}} & = & \dis\sum_{f_2} \; \dis\frac
{(\lambda_{f_2})^2}{d_\Omega(f_2)}
\dis\sum_{\nu=0,1,2}\,\dis\sum_\rho\;\l|\lan f_m \mid\mid\mid
B(f_2,\bF_\nu) \mid\mid\mid f_m \ran_\rho\r|^2 \\ \\
& = & \dis\sum_{f_2} \;\dis\frac{(\lambda_{f_2})^2}{d_\Omega(f_2)}
\dis\sum_{\nu=0,1,2}\cq^\nu(f_2:m,f_m)\;.
\earr \label{ch4.eq.34}
\ee
The functions $\cq^\nu(f_2:m,f_m)$ involve $SU(\Omega)$ $U$-coefficients and
the explicit expression is,
\be
\barr{l}
\cq^\nu(f_2:m,f_m)=
\l[F(m)\r]^2 \dis\sum_{f_{m-2},f^\pr_{m-2}}
\dis\frac{\cn_{f_{m-2}}}{\cn_{f_m}}\dis\frac
{\cn_{f^\pr_{m-2}}}{\cn_{f_m}}X_{UU}(f_2;f_{m-2},f^\pr_{m-2};\bF_\nu)\;; \\
X_{UU}(f_2;f_{m-2},f^\pr_{m-2};\bF_\nu) = \\
\dis\sum_{\rho}\dis\frac
{U(f_m,\overline{f_2},f_m,f_2;
f_{m-2},\bF_\nu)_\rho U(f_m,\overline{f_2},f_m,f_2;f^\pr_{m-2},\bF_\nu)
_\rho}{U(f_m,\overline{f_2},f_m,f_2;f_{m-2},\{0\})
U(f_m,\overline{f_2},f_m,f_2;f^\pr_{m-2},\{0\})}\;.
\earr \label{ch4.eq.35}
\ee
Tabulations  for $X_{UU}$ (also for $Y_{UU}$ defined ahead) or equivalently
$SU(\Omega)$ $U$-coefficients, though in a complex form, are available in
\cite{He-74a}. However to gain insight into the spectral variances and the
cross-correlations $\Sigma_{rr}$, we derive analytical results by
restricting  ourselves to the physically relevant (in nuclear structure; see
Section \ref{su4bk}) irreps $f_m^{(p)}$. 

\begin{figure}[ht]
\centering
\includegraphics[width=4.5in,height=6.5in]{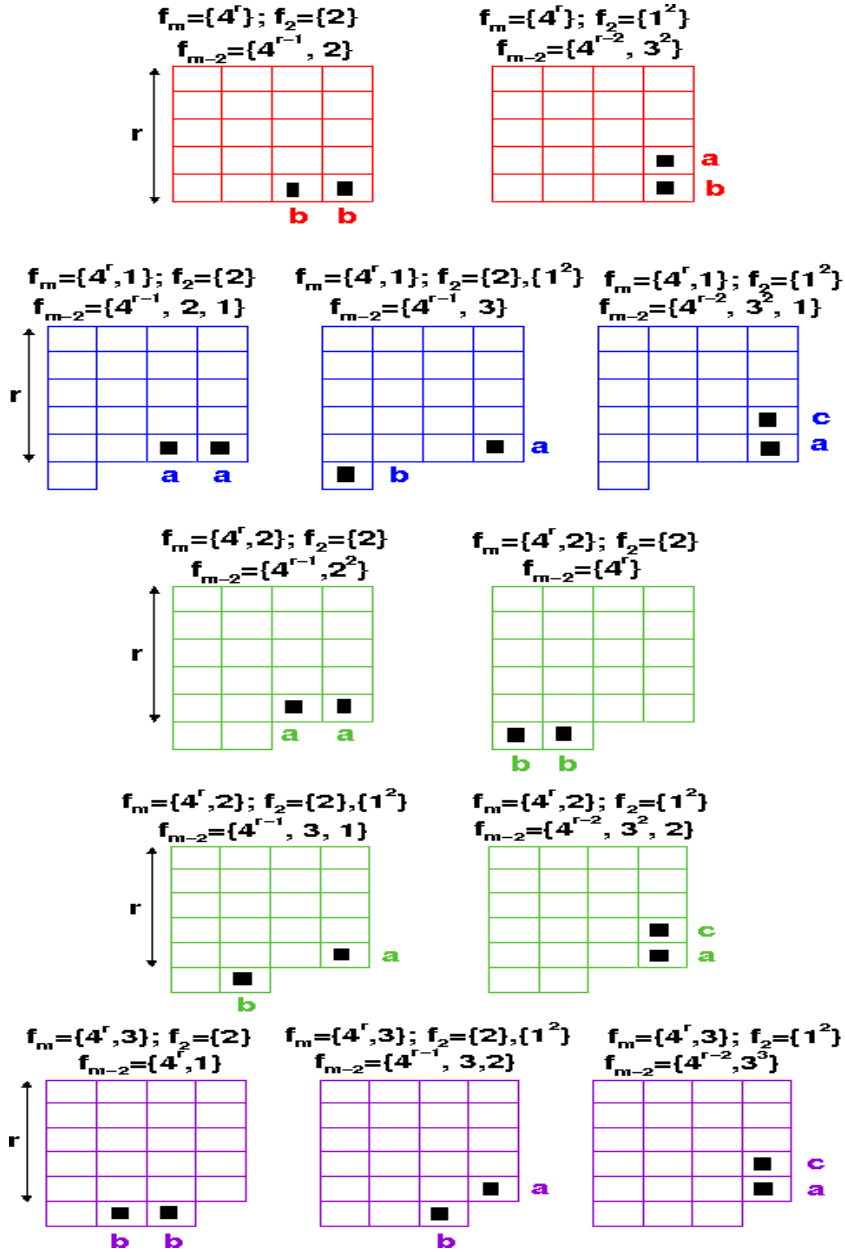} 
\caption{Schematic representation of the Young tableaux
$f_m=f_m^{(p)}=\{4^r,p\}$ with $p=0$, $1$, $2$ and $3$. Shown are  the boxes
with filled squares denoted by $a$, $b$ and $c$ whose removal  from the
irrep $f_m$ generates the irreps $f_{m-2}$ by action of  $\overline{f_2}$
where $f_2=\{2\}$ and $\{1^2\}$.}
\label{young}
\end{figure}

Summation over the multiplicity index $\rho$ appearing in Eq. (\ref{ch4.eq.35})
[also Eq. (\ref{ch4.eq.39}) ahead] arises naturally in applications to physical
problems as all the physically relevant results should be independent of
$\rho$ which is a label for equivalent $SU(\Omega)$ irreps.  Hecht derived
formulas for the sums in $X_{UU}$ (also $Y_{UU}$ defined ahead) in the
context of  spectral distribution methods in nuclei \cite{He-74a}.
Tabulations for $X_{UU}(f_2;f_{m-2},f^\pr_{m-2};\bF_\nu)$ are collected  in
Table \ref{tab1} and they are given in terms of the so-called axial
distances $\tau_{ij}$ for the boxes $i$ and $j$ in a given  Young tableaux.
Given a Young tableaux $\{f_m\}$, the axial  distance $\tau_{ij}$ between
the last box in row $i$ and the last box in  row $j$ is
$\tau_{ij}=f_i-f_j+j-i$, with $f_k$ being the number of boxes in the row
$k$. The $f_{m-2}$ irreps are obtained by removing the two-particle
symmetric ($f_2=\{2\}$) or anti-symmetric ($f_2=\{1^2\}$) irreps from
$f_m$.  Figure \ref{young} shows all the allowed $f_{m-2}$'s for the irreps
$f_m^{(p)}$. In the figure, $a$ and $b$ (or $c$) denote the  last boxes in
the rows $a$ and $b$ (or $c$), respectively, that are to be removed from the
Young tableaux $\{4^r,p\}$ to obtain the allowed $f_{m-2}$ irreps for
$f_2=\{2\}$ and $\{1^2\}$. It is seen that unlike for EGUE(2)-$\cs$ studied
in \cite{Ko-07},  for the EGUE(2)-$SU(4)$ ensemble we need a much wider
variety of $X_{UU}$'s. Results in Table \ref{tab1} (also Table \ref{tab22})
for any $f_m$ are given in terms of the following functions,
\begin{subequations}
\be
\barr{rcl}
\Pi_a^{(b)} & = & \dis\prod_{i=1,2,\ldots,\Omega;i\neq a,i \neq b}\;
\l( 1-1/\tau_{ai}\r)\;, \\ \\
\Pi_b^{(a)} & = & \dis\prod_{i=1,2,\ldots,\Omega;i\neq a,i \neq b}\;
\l( 1-1/\tau_{bi}\r)\;.
\earr
\label{ch4.eq.piab}
\ee
\be
\barr{rcl}
\Pi^\pr_a & = & \dis\prod_{i=1,2,\ldots,\Omega;i\neq a}\;
\l( 1-1/\tau_{ai}\r) \;,\\ \\
\Pi^{\pr\pr}_a & = & \dis\prod_{i=1,2,\ldots,\Omega;i\neq a}\;
(1-2/\tau_{ai})\;.
\earr
\label{ch4.eq.piapr}
\ee
\be
\Pi_a^{(bc)} = \dis\prod_{i=1,2,\ldots,\Omega;i\neq a,i \neq b,
i \neq c}\; \l( 1-1/\tau_{ai}\r) \;;\;\;a \neq b \neq c\;.
\label{ch4.eq.piabc}
\ee
\end{subequations}
In \cite{He-74a}, the functions $\Pi_a^{(b)}$ and $\Pi_{b}^{(a)}$ are  called
$\Pi_a$ and $\Pi_{b}$, respectively and sometimes this ($\Pi_a$, $\Pi_{b}$)
notation is  confusing. Further, we have introduced the functions $\Pi_a^\pr$,
$\Pi_{a}^{\pr\pr}$ and $\Pi_a^{(bc)}$. These and the notation  $\Pi_a^{(b)}$ and
$\Pi_{b}^{(a)}$ simplify considerably the formulas given by Hecht \cite{He-74a}
and therefore the results in Table \ref{tab1} (also Table \ref{tab22}) are much
easier to use in practice.  Table \ref{tauab} gives  $\tau_{ab}$, $\Pi_a^{(b)}$,
$\Pi_{b}^{(a)}$,  $\Pi_{a}^\pr$, $\Pi_{a}^{\pr\pr}$ and $\Pi_a^{(bc)}$ for the
irreps  $f_m^{(p)}$ which are required for deriving analytical formulas for the
corresponding  $X_{UU}(f_2;f_{m-2},f^\pr_{m-2};\bF_\nu)$ and also 
$Y_{UU}(f_{m-2},f^\pr_{m-2};\bF_\nu)$ defined ahead. Also given in the table
are  $\cn_{f_{m-2}}/\cn_{f_m}$ obtained by simplifying Eq. (\ref{ch4.eq.28b}).
Combining the results in Tables \ref{tab1} and \ref{tauab} and  carrying out
simplifications, final formulas for $\overline{\lan H^2 \ran^{m,f_m}}$ are
obtained and they are given in Table \ref{tabw}. In principle, the operator
generating $\lan H^2 \ran^{m,f_m}$ for any two or (1+2)-body $H$, will be a 
polynomial of maximum body rank 4 in the number operator
$\hat{n}$ and the quadratic, cubic and quartic invariants of the $SU(4)$
algebra.  The expansion coefficients  in the resulting formula will involve
$\lan H^2 \ran^{m,f_m}$ with $m=0$ to $4$ \cite{Pa-73,Pa-72} and they can be
calculated by constructing  the ensemble, for a fixed $\Omega$, on a computer.
Using these inputs,  the propagation equation can be used to compute spectral
variances for any $(m,f_m)$. However Eqs. (\ref{ch4.eq.34}) and
(\ref{ch4.eq.35}) give the ensemble averaged  variances directly in terms of
$SU(\Omega)$ $U$-coefficients.

\setlength{\LTcapwidth}{6in}

\begin{longtable}{cc}
\caption{Formulas for $X_{UU}(f_2;f_{m-2},f^\pr_{m-2};\bF_\nu)$ defined in
Eq. (\ref{ch4.eq.35}). Note that 
$\{f(ab)\}\,\{f(ab)\}$ entries satisfy the $a \leftrightarrow b$ symmetry
correctly. Similarly the entries $\{f(ab)\}\,\{f(ac)\}$ are independent of
$b \leftrightarrow c$ interchange as required by the $X_{UU}$ function. See
text for details.}\\

\endfirsthead

\multicolumn{2}{c}%
{{\bfseries \tablename\ \thetable{} -- continued}} \\
\midrule
\endhead

\bottomrule
\endfoot

\bottomrule
\endlastfoot
\toprule
$\{f_{m-2}\}\,\{f^\pr_{m-2}\}$ & $X_{UU} (\{1^2\}; f_{m-2}, 
f^\pr_{m-2}; \{2,1^{\Omega-2}\})$  \\ 
\midrule
$\{f(ab)\}\,\{f(ab)\}$ & $\dis\frac{\Omega(\Omega-1)}{2(\Omega-2)} 
\l\{\l(1+\dis\frac{1}{\tau_{ab}}\r)\dis\frac{1}{\Pi_b^{(a)}}+
\l(1-\dis\frac{1}{\tau_{ab}}\r)\dis\frac{1}{\Pi_a^{(b)}}-\dis\frac{4}
{\Omega}\r\}$ \\ \\
$\{f(ab)\}\,\{f(ac)\}$ & $\dis\frac{\Omega(\Omega-1)}{2(\Omega-2)}
\l\{\dis\frac{1}{\Pi_a^{(bc)}}-\dis\frac{4}{\Omega}\r\}$ \\ \\
$\{f(cd)\}\,\{f(ab)\}$ & $-\dis\frac{2(\Omega-1)}{(\Omega-2)}$ \\
\midrule
$\{f_{m-2}\}\,\{f^\pr_{m-2}\}$ & $X_{UU} (\{1^2\}; f_{m-2}, 
f^\pr_{m-2}; \{2^2,1^{\Omega-4}\})$ \\
\midrule
$\{f(ab)\}\,\{f(ab)\}$ & $\dis\frac{\Omega}{(\Omega-2)} 
\l\{1+\dis\frac{(\Omega-1)(\Omega-2)}{2\Pi_a^{(b)}\Pi_b^{(a)}}
-\dis\frac{(\Omega-1)}{2}\r.$ \\ \\
& $\l. \times 
\l[ \l(1+\dis\frac{1}{\tau_{ab}}\r)\dis\frac{1}{\Pi_b^{(a)}}+
\l(1-\dis\frac{1}{\tau_{ab}}\r)\dis\frac{1}{\Pi_a^{(b)}}\r]\r\}$ \\ \\
$\{f(ab)\}\,\{f(ac)\}$ & $\dis\frac{\Omega(\Omega-1)}{2(\Omega-2)}
\l\{\dis\frac{2}{\Omega-1}-\dis\frac{1}{\Pi_a^{(bc)}}\r\}$ \\ \\
$\{f(cd)\}\,\{f(ab)\}$ & $\dis\frac{\Omega}{(\Omega-2)}$ \\ 
\midrule
$\{f_{m-2}\}\,\{f^\pr_{m-2}\}$ & $X_{UU} (\{2\}; f_{m-2}, 
f^\pr_{m-2}; \{2,1^{\Omega-2}\})$ \\ 
\midrule
$\{f(ab)\}\,\{f(ab)\}$ & $\dis\frac{\Omega(\Omega+1)}{2(\Omega+2)} 
\l\{\dis\frac{(\tau_{ab}-1)^2}{\tau_{ab}(\tau_{ab}+1)}
\dis\frac{1}{\Pi_b^{(a)}} + \dis\frac{(\tau_{ab}+1)^2}{\tau_{ab}
(\tau_{ab}-1)}\dis\frac{1}{\Pi_a^{(b)}}-\dis\frac{4}
{\Omega}\r\}$ \\ \\
$\{f(aa)\}\,\{f(aa)\}$ & $\dis\frac{2\Omega(\Omega+1)}{(\Omega+2)}
\l\{\dis\frac{1}{\Pi_a^{\pr}}-\dis\frac{1}{\Omega}\r\}$ \\ 
$\{f(aa)\}\,\{f(bb)\}$ & \\
$\;\;\;\;\;\;\;\;\;\;$ \text{or} & 
$-\dis\frac{2(\Omega+1)}{(\Omega+2)}$ \\
$\{f(aa)\}\,\{f(bc)\}$ & \\ 
$\{f(aa)\}\,\{f(ab)\}$ & $\dis\frac{\Omega(\Omega+1)}{(\Omega+2)}
\l\{\dis\frac{(\tau_{ab}+1)}{(\tau_{ab}-1)}
\dis\frac{1}{\Pi_a^{(b)}}-\dis\frac{2}{\Omega}\r\}$ \\ \\
$\{f(ab)\}\,\{f(ac)\}$ & $\dis\frac{\Omega(\Omega+1)}{2(\Omega+2)}
\l\{\dis\frac{(\tau_{ab}+1)}{(\tau_{ab}-1)}
\dis\frac{(\tau_{ac}+1)}{(\tau_{ac}-1)}
\dis\frac{1}{\Pi_a^{(bc)}}-\dis\frac{4}{\Omega}\r\}$ \\ \\
$\{f(ab)\}\,\{f(cd)\}$ & $-\dis\frac{2(\Omega+1)}{(\Omega+2)}$ \\ 
\midrule
$\{f_{m-2}\}\,\{f^\pr_{m-2}\}$ & $X_{UU} (\{2\}; f_{m-2}, 
f^\pr_{m-2}; \{4,2^{\Omega-2}\})$ \\ 
$\{f(ab)\}\,\{f(ab)\}$ & $\dis\frac{\Omega(\Omega+1)}{2} 
\l\{\dis\frac{1}{\Pi_a^{(b)}\Pi_b^{(a)}}+\dis\frac{2}{(\Omega+1)(\Omega+2)}
\r.$ \\ \\
& $\l.-\dis\frac{1}{(\Omega+2)}\l[\dis\frac{(\tau_{ab}-1)^2}{\tau_{ab}
(\tau_{ab}+1)}\dis\frac{1}{\Pi_b^{(a)}} + \dis\frac{(\tau_{ab}+1)^2}
{\tau_{ab}(\tau_{ab}-1)}\dis\frac{1}{\Pi_a^{(b)}}\r]\r\}$ \\ \\
$\{f(aa)\}\,\{f(aa)\}$ & $\dis\frac{\Omega}{(\Omega+2)} 
\l\{1-2(\Omega+1)\dis\frac{1}{\Pi_a^{\pr}}+\dis\frac{(\Omega+1)(\Omega+2)}
{2}\dis\frac{1}{\Pi_a^{\pr\pr}}\r\}$ \\ 
$\{f(aa)\}\,\{f(bb)\}$ & \\
$\;\;\;\;\;\;\;\;\;\;$ \text{or} & $\dis\frac{\Omega}{(\Omega+2)}$ \\
$\{f(aa)\}\,\{f(bc)\}$ & \\ 
$\{f(aa)\}\,\{f(ab)\}$ & $\dis\frac{\Omega}{(\Omega+2)}
\l\{1-\dis\frac{(\Omega+1)(\tau_{ab}+1)}{(\tau_{ab}-1)\Pi_a^{(b)}}\r\}$ \\ \\
$\{f(ab)\}\,\{f(ac)\}$ & $\dis\frac{\Omega}{(\Omega+2)}
\l\{1-\dis\frac{(\Omega+1)(\tau_{ab}+1)(\tau_{ac}+1)}{2(\tau_{ab}-1)
(\tau_{ac}-1)}\dis\frac{1}{\Pi_a^{(bc)}}\r\}$ \\ \\
$\{f(ab)\}\,\{f(cd)\}$ & $\dis\frac{\Omega}{(\Omega+2)}$ 
\label{tab1}
\end{longtable}

\begin{sidewaystable}[htp]
\caption{Axial distances $\tau_{ab}$ and the functions $\Pi_a^{(b)}$,
$\Pi_b^{(a)}$, $\Pi_a^\pr$ and $\Pi^{\pr\pr}_a$ for the  irreps 
$f_m^{(p)}$ shown in Fig. \ref{young}.  For the
situations with  $f(ac)$, $\tau_{ab} \to \tau_{ac}$, $\Pi_a^{(b)} \to
\Pi_a^{(c)}$ and  $\Pi_b^{(a)} \to \Pi_c^{(a)}$. Also for $f(bb)$,
$\Pi_a^\pr \to \Pi_b^\pr$  and $\Pi^{\pr\pr}_a\to \Pi^{\pr\pr}_b$. For
$f_2=\{1^2\}$ and for $f_m^{(p)}=\{4^r,p\}$ with $p \neq 0$, 
we need  $\Pi_a^{(bc)}$ (see Fig. \ref{young} for $a$, 
$b$ and $c$) and for the examples in the Table, we have 
$\Pi_a^{{(bc)}}=5r/[2(4+\Omega-r)]$.}
\begin{center}
\begin{tabular}{lcccccccl}
\toprule
$f_m^{(p)}$ & $f_2$ & $f_{m-2}$ & $\tau_{ab}$ &
$\dis\frac{1}{\Pi_a^{(b)}}$  & $\dis\frac{1}{\Pi_{b}^{(a)}}$ &
$\dis\frac{1}{\Pi_{a}^\pr}$ &
$\dis\frac{1}{\Pi_{a}^{\pr\pr}}$ &
$\dis\frac{\cn_{f_{m-2}}}{\cn_{f_m}}$ \\
\midrule
$\{4^r\}$ & $\{2\}$ & $\{4^{r-1},2\}$ & $-$ & $-$ & $-$ &
$\dis\frac{4+\Omega-r}{4r}$ & $\dis\frac{(4+\Omega-r)(3+\Omega-r)}
{6r(r+1)}$ & $\dis\frac{3(r+1)}{2(4r-1)}$ \\
 & & $\to f(bb)$ & & & & & & \\
 & $\{1^2\}$ & $\{4^{r-2},3^2\}$ & $+1$ & $\dis\frac{5+\Omega-r}{5(r-1)}$ &
 $\dis\frac{4+\Omega-r}{2r}$ & $-$ & $-$ & $\dis\frac{5(r-1)}{2(4r-1)}$ \\
 & & $\to f(ab)$ & & & & & & \\
$\{4^r,1\}$ & $\{2\}$ & $\{4^{r-1},2,1\}$ & $-$ & $-$ & $-$ &
$\dis\frac{4(4+\Omega-r)}{15r}$ & $\dis\frac{(4+\Omega-r)(3+\Omega-r)}
{5r(r+1)}$ & $\dis\frac{5(r+1)}{4(4r+1)}$ \\
 & & $\to f(aa)$ & & & & & & \\
 & & $\{4^{r-1},3\}$ & $+4$ & $\dis\frac{4+\Omega-r}{5r}$ &
$\dis\frac{5(\Omega-r)}{r+4}$ & $-$ & $-$ & $\dis\frac{r+4}{4(4r+1)}$ \\
 & & $\to f(ab)$ & & & & & & \\
 & $\{1^2\}$ & $\{4^{r-1},3\}$ & $+4$ & $\dis\frac{4+\Omega-r}{5r}$ &
$\dis\frac{5(\Omega-r)}{r+4}$ & $-$ & $-$ & $\dis\frac{r+4}{4(4r+1)}$ \\
 & & $\to f(ab)$ & & & & & & \\
 & & $\{4^{r-2},3^2,1\}$ & $-1$ & $\dis\frac{8(4+\Omega-r)}{15r}$ &
$\dis\frac{5(5+\Omega-r)}{24(r-1)}$ & $-$ & $-$ &
$\dis\frac{9(r-1)}{4(4r+1)}$ \\
 & & $\to f(ac)$ & & & & & & \\
\bottomrule
\end{tabular}
\label{tauab}
\end{center}
\end{sidewaystable}

\begin{sidewaystable}[htp]
\begin{center}
\text{{\bf Table \ref{tauab} -- (continued)}}
\end{center}
\begin{center}
\begin{tabular}{lcccccccl}
\toprule
$f_m^{(p)}$ & $f_2$ & $f_{m-2}$ & $\tau_{ab}$ &
$\dis\frac{1}{\Pi_a^{(b)}}$  & $\dis\frac{1}{\Pi_{b}^{(a)}}$ &
$\dis\frac{1}{\Pi_{a}^\pr}$ &
$\dis\frac{1}{\Pi_{a}^{\pr\pr}}$ &
$\dis\frac{\cn_{f_{m-2}}}{\cn_{f_m}}$ \\
\midrule
$\{4^r,2\}$ & $\{2\}$ & $\{4^r\}$ & $-$ & $-$ & $-$ & $\dis\frac
{3(1+\Omega-r)}{2(r+3)}$ & $\dis\frac{6(\Omega-r)(1+\Omega-r)}{(r+3)(r+4)}$
& $\dis\frac{(r+3)(r+4)}{6(4r+1)(4r+2)}$ \\
 & & $\to f(bb)$ & & & & & & \\
 & & $\{4^{r-1},2^2\}$ & $-$ & $-$ & $-$ & $\dis\frac{3(4+\Omega-r)}{10r}$
& $\dis\frac{3(3+\Omega-r)(4+\Omega-r)}{10r(r+1)}$ &
$\dis\frac{10r(r+1)}{3(4r+1)(4r+2)}$ \\
 & & $\to f(aa)$ & & & & & & \\
 & & $\{4^{r-1},3,1\}$ & $+3$ & $\dis\frac{4+\Omega-r}{5r}$ &
$\dis\frac{2(1+\Omega-r)}{r+3}$ & $-$ & $-$ &
$\dis\frac{5r(r+3)}{2(4r+1)(4r+2)}$ \\
 & & $\to f(ab)$ & & & & & & \\
 & $\{1^2\}$ & $\{4^{r-1},3,1\}$ & $+3$ & $\dis\frac{4+\Omega-r}{5r}$ &
$\dis\frac{2(1+\Omega-r)}{r+3}$ & $-$ & $-$ &
$\dis\frac{5r(r+3)}{2(4r+1)(4r+2)}$ \\
 & & $\to f(ab)$ & & & & & & \\
 & & $\{4^{r-2},3^2,2\}$ & $-1$ & $\dis\frac{3(4+\Omega-r)}{5r}$ &
$\dis\frac{2(5+\Omega-r)}{9(r-1)}$ & $-$ & $-$ &
$\dis\frac{15r(r-1)}{2(4r+1)(4r+2)}$ \\
 & & $\to f(ac)$ & & & & & & \\
$\{4^r,3\}$ & $\{2\}$ & $\{4^r,1\}$ & $-$ & $-$ & $-$ &
$\dis\frac{2(2+\Omega-r)}{3(r+2)}$ & $\dis\frac{(1+\Omega-r)(2+\Omega-r)}
{(r+2)(r+3)}$ & $\dis\frac{(r+2)(r+3)}{(4r+2)(4r+3)}$ \\
 & & $\to f(bb)$ & & & & & & \\
 & & $\{4^{r-1},3,2\}$ & $+2$ & $\dis\frac{4+\Omega-r}{5r}$ &
$\dis\frac{(2+\Omega-r)}{r+2}$ & $-$ & $-$ & $\dis\frac{5r(r+2)}
{(4r+2)(4r+3)}$ \\
 & & $\to f(ab)$ & & & & & & \\
 & $\{1^2\}$ & $\{4^{r-1},3,2\}$ & $+2$ & $\dis\frac{4+\Omega-r}{5r}$ &
$\dis\frac{(2+\Omega-r)}{r+2}$ & $-$ & $-$ & $\dis\frac{5r(r+2)}
{(4r+2)(4r+3)}$ \\
 & & $\to f(ab)$ & & & & & & \\
 & & $\{4^{r-2},3^3\}$ & $-1$ & $\dis\frac{4(4+\Omega-r)}{5r}$ &
$\dis\frac{(5+\Omega-r)}{4(r-1)}$ & $-$ & $-$ &
$\dis\frac{5r(r-1)}{(4r+2)(4r+3)}$ \\
 & & $\to f(ac)$ & & & & & & \\
\bottomrule
\end{tabular}
\end{center}
\end{sidewaystable}

\subsection{Cross-correlations in energy centroids  $\Sigma_{11} (m, f_m;
m^\pr, f_{m^\pr})$}
\label{ansg11-2}

Analysis of the random matrix ensembles with various symmetries 
involves construction of the one-point function $\overline{\rho^{m,\Gamma}(E)}$
given by Eq. (\ref{eq.eden})
and the two-point and other higher point functions defining fluctuations.
Covariances in energy centroids $\Sigma_{11}(m,f_m;m^\pr,f_{m^\pr})$ follow
from Eqs. (\ref{eq.den10}), (\ref{ch4.eq.31}) and (\ref{ch4.eq.34}),
\be
\Sigma_{11}(m,f_m;m^\pr,f_{m^\pr}) = \dis\frac{\dis\sum_{f_2} 
\dis\frac{\l(\lambda_{f_2}\r)^2}{d_\Omega(f_2)} 
\;P^{f_2}(m,f_m)\;P^{f_2}(m^\pr,f_{m^\pr}) }
{\l[\;\l\{\overline{\lan H^2 \ran^{m,f_m}}\r\} \;
\l\{\overline{\lan H^2 \ran^{m^\pr,f_{m^\pr}}}\r\} \;\r]^{1/2}}.
\label{ch4.eq.spcorr}
\ee
For the irreps $f_m^{(p)}$, formulas for the functions $P^{f_2}(m,f_m)$ and
the variances $\overline{\lan H^2 \ran^{m,f_m}}$ are given in Tables
\ref{pf2} and \ref{tabw}, respectively. Table \ref{tab1} gives $X_{UU}$
required for calculating the covariances for any general $f_m$. To gain some
insight  into the structure of  $\Sigma_{11} (m,f_m;m^\pr,f_{m^\pr})$, we
consider the dilute limit defined by $\Omega \to \infty$, $r>>1$ and 
$r/\Omega \to 0$. Then the variance formulas in Table  \ref{tabw} take a
simple form for all $f_m^{(p)}$,
\be
\overline{\lan H^2\ran^{m,f_m^{(p)}}} = -\dis\frac{\Omega^2}{2}\l[ 
\lambda^2_{\{2\}} P^{\{2\}}(m,f_m^{(p)}) + \lambda^2_{\{1^2\}} P^{\{1^2\}}
(m,f_m^{(p)})\r] \;.
\label{ch4.eq.var}
\ee
Combining Eqs. (\ref{ch4.eq.spcorr}) and (\ref{ch4.eq.var}), we have 
\be
\barr{l}
\Sigma_{11}(m,f_m^{(p)};m^\pr,f_{m^\pr}^{(p)}) \stackrel{\Omega \to
\infty,r>>1}{\longrightarrow} \\ \\ 
\dis\frac{4}{\Omega^4}\dis\frac
{\dis\sum_{f_2}\lambda^2_{f_2}P^{f_2}(m,f_m^{(p)})
P^{f_2}(m^\pr,f_{m^\pr}^{(p)})}
{\l[ \l\{\dis\sum_{f_2}\lambda^2_{f_2}P^{f_2}(m,f_m^{(p)})\r\}\; 
\l\{\dis\sum_{f_2}\lambda^2_{f_2}P^{f_2}(m^\pr,f_{m^\pr}^{(p)})
\r\}\r]^{1/2}}\;.
\earr \label{ch4.eq.sg11}
\ee
Thus, $\Sigma_{11}(m,f_m^{(p)};m^\pr,f_{m^\pr}^{(p)})$ will be zero as 
$\Omega \to \infty$ and there will be no cross-correlations. However for
finite $\Omega$, there will be correlations between energy centroids of 
different states and some examples  are discussed ahead.

\begin{table}[ht]
\caption{Ensemble averaged spectral variances 
$\overline{\lan H^2\ran^{m,f_m}}$ for various $f_m=f_m^{(p)}$.}
\begin{center}
\begin{tabular}{cc}
\toprule
$f_m^{(p)}$ & $\overline{\lan H^2\ran^{m,f_m^{(p)}}}$ \\ 
\midrule
$\{4^r\}$ & $\dis\frac{r(\Omega-r+4)}{2}\l[\lambda^2_{\{2\}}3(r+1)
(\Omega-r+3)+\lambda^2_{\{1^2\}} 5(r-1)(\Omega-r+5)\r]$
\\ \\
$\{4^r,1\}$ & $\dis\frac{r(\Omega-r+4)}{4}\l[  \lambda^2_{\{2\}}
\{ 6r(\Omega-r+1)+9\Omega+15\}\r.$ 
 \\
& $\l.+ \lambda^2_{\{1^2\}} 5\{ 2r(\Omega-r+5)-\Omega-9\}\r]$ \\ \\
$\{4^r,2\}$ & $\lambda^2_{\{2\}}\spin
\l[ 3r^4-6(\Omega+2)r^3+(3\Omega^2+6\Omega-5)r^2\r.$  \\
& $\l.+(\Omega+2)(6\Omega+17)r+\Omega(\Omega+1)\r]$ \\ 
& $+\lambda^2_{\{1^2\}} \dis\frac{5r}{2}(\Omega-r+4)\{
(\Omega+4)r-r^2-3\}$ \\ \\
$\{4^r,3\}$ & $\dis\frac{1}{4}\l[\lambda^2_{\{2\}}3(r+2)(\Omega-r+2)
(2r\Omega-2r^2+6r+\Omega+1)\r.$ \\
& $\l.+\lambda^2_{\{1^2\}}5r(\Omega-r+4)
(2r\Omega-2r^2+6r+\Omega-1)\r]$  \\ 
\bottomrule
\end{tabular}
\end{center}
\label{tabw}
\end{table}

\subsection{Cross-correlations in spectral variances  $\Sigma_{22} (m, f_m;
m^\pr, f_{m^\pr})$}
\label{ansg22}

Expression for $\Sigma_{22} (m, f_m; m^\pr, f_{m^\pr})$ given by Eq.
(\ref{eq.den10}) involves evaluation of 
$$
\overline{\lan H^2\ran^{m,f_m} \lan
H^2\ran^{m^\pr,f_{m^\pr}}}\;.
$$ 
As the two-body $H$ operator defined in Eq.
(\ref{ch4.eq.22}) is a sum of $H$'s in two-particle spaces defined by
$f_2=\{2\}$ and $\{1^2\}$, we have $H(2)= H_{\{2\}}(2) + H_{\{1^2\}}(2)$. 
The $H_{f_2}$'s are independent  and the variables defining the matrix
elements of $H_{f_2}$ are independent Gaussian variables with zero center
and  variance given by Eq. (\ref{ch4.eq.23}). Expanding $\overline{\lan
H^2\ran^{m,f_m} \lan H^2\ran^{m^\pr,f_{m^\pr}}}$ and using Eqs.
(\ref{ch4.eq.23}) and (\ref{ch4.eq.indep}), we obtain 
\be
\barr{l}
\overline{\lan H^2\ran^{m,f_m} \lan H^2\ran^{m^\pr,f_{m^\pr}}} \\ \\
= \overline{\lan (H_{\{2\}})^2 \ran^{m,f_m} \lan (H_{\{2\}})^2 
\ran^{m^\pr,f_{m^\pr}}} +
\overline{\lan (H_{\{1^2\}})^2 \ran^{m,f_m} \lan (H_{\{1^2\}})^2 
\ran^{m^\pr,f_{m^\pr}}} \\ \\
+ \l\{\overline{\lan (H_{\{2\}})^2 \ran^{m,f_m}}\r\} \;
\l\{\overline{\lan (H_{\{1^2\}})^2 \ran^{m^\pr,f_{m^\pr}}}\r\} 
+ \l\{\overline{\lan (H_{\{1^2\}})^2 \ran^{m,f_m}}\r\} \;\l\{\overline{
\lan (H_{\{2\}})^2 \ran^{m^\pr,f_{m^\pr}}}\r\} \\ \\
+ 4 \overline{\lan H_{\{2\}} H_{\{1^2\}} \ran^{m,f_m} \lan H_{\{1^2\}} 
H_{\{2\}} \ran^{m^\pr,f_{m^\pr}}}\;.
\earr \label{ch4.eq.sg22-1}
\ee
Similarly, expanding $\{\overline{\lan H^2 \ran^{m,f_m}}\}\;
\{\overline{\lan H^2 \ran^{m^\pr,f_{m^\pr}}}\}$ gives,
\be
\barr{rcl}
\l\{\overline{\lan H^2 \ran^{m,f_m}}\r\}\;
\l\{\overline{\lan H^2 \ran^{m^\pr,f_{m^\pr}}}\r\}
& = & 
\l\{\overline{\lan (H_{\{2\}})^2 \ran^{m,f_m}}\r\} \; 
\l\{\overline{\lan (H_{\{2\}})^2 \ran^{m^\pr,f_{m^\pr}}}\r\} \\ \\
& + &
\l\{\overline{\lan (H_{\{2\}})^2 \ran^{m,f_m}}\r\}\; 
\l\{\overline{\lan (H_{\{1^2\}})^2 \ran^{m^\pr,f_{m^\pr}}}\r\} \\ \\
& + &
\l\{\overline{\lan (H_{\{1^2\}})^2 \ran^{m,f_m}}\r\} \; 
\l\{\overline{\lan (H_{\{2\}})^2 \ran^{m^\pr,f_{m^\pr}}}\r\} \\ \\
& + &
\l\{\overline{\lan (H_{\{1^2\}})^2 \ran^{m,f_m}}\r\} \; 
\l\{\overline{\lan (H_{\{1^2\}})^2 \ran^{m^\pr,f_{m^\pr}}}\r\}\;.
\earr \label{ch4.eq.sg22-2}
\ee
Using Eqs. (\ref{ch4.eq.sg22-1}) and (\ref{ch4.eq.sg22-2}) in the expression for
$\Sigma_{22}$ given by Eq. (\ref{eq.den10}), the numerator simplifies to give
\begin{flushleft}
\be
\barr{l}
\overline{\lan H^2\ran^{m,f_m} \lan H^2\ran^{m^\pr,f_{m^\pr}}} -
\l\{\overline{\lan H^2 \ran^{m,f_m}}\r\}\;
\l\{\overline{\lan H^2 \ran^{m^\pr,f_{m^\pr}}}\r\}
\\ \\ =
\l\{ \overline{\lan (H_{\{2\}})^2 \ran^{m,f_m} \lan (H_{\{2\}})^2 
\ran^{m^\pr,f_{m^\pr}}} - \l[\;\overline{\lan (H_{\{2\}})^2 \ran^{m,f_m}}\;\r] 
\; 
\l[\;\overline{\lan (H_{\{2\}})^2 \ran^{m^\pr,f_{m^\pr}}}\;\r]\r\} \nonumber
\earr\label{ch4.eq.sg22-3a1}
\ee
\end{flushleft}
\be
\barr{l}
+\l\{ \overline{\lan (H_{\{1^2\}})^2 \ran^{m,f_m} \lan (H_{\{1^2\}})^2 
\ran^{m^\pr,f_{m^\pr}}} - 
\l[\;\overline{\lan (H_{\{1^2\}})^2 \ran^{m,f_m}}\;\r]
 \; 
\l[\;\overline{\lan (H_{\{1^2\}})^2 \ran^{m^\pr,f_{m^\pr}}}\;\r]\r\} \\ \\
+ 4 \overline{\lan H_{\{2\}} H_{\{1^2\}} \ran^{m,f_m} \lan H_{\{1^2\}} 
H_{\{2\}} \ran^{m^\pr,f_{m^\pr}}} \\ =
X_{\{2\}} + X_{\{1^2\}} + 4 X_{\{1^2\}\{2\}}\;.
\earr \label{ch4.eq.sg22-3a2}
\ee
Then, we have
\be
\Sigma_{22}(m,f_m;m^\pr,f_{m^\pr}) = \dis\frac{X_{\{2\}}+X_{\{1^2\}}
+4\,X_{\{1^2\}\{2\}}}{\l[\;\overline{\lan H^2\ran^{m,f_m}}\;\r]\;
\l[\;\overline{\lan H^2\ran^{m^\pr,f_{m^\pr}}}\;\r]}\;.
\label{ch4.eq.36}
\ee
To evaluate $X_{\{2\}}$ and $X_{\{1^2\}}$, we use Eq. (\ref{ch4.eq.25}) and
carry out the ensemble averaging over $W$'s using the fact that $W$'s are
Gaussian random variables with zero center and variance given by Eq.
(\ref{ch4.eq.26}). Then, Eq. (\ref{ch4.eq.27}) and the sum rules for $SU(\Omega)$
Wigner coefficients [see Eqs. (\ref{ch4.eq.prp6}) and (\ref{ch4.eq.prp7})]  will
give,
\be
X_{f_2} = \dis\frac{2(\lambda_{f_2})^4}{\l[d_\Omega(f_2)\r]^2} 
\dis\sum_{\nu=0,1,2}
\l[ d(\bF_\nu)\r]^{-1}\cq^\nu(f_2:m,f_m)\cq^\nu(f_2:m^\pr,f_{m^\pr})\;.
\label{ch4.eq.sg22-4}
\ee
Similarly, we have
\be
\barr{l}
X_{\{1^2\}\{2\}} = 
\dis\frac{\lambda^2_{\{2\}}\lambda^2_{\{1^2\}}}{d_\Omega(\{2\})
d_\Omega(\{1^2\})} \\ \\ \times
\dis\sum_{\nu=0,1} \l[ d(\bF_\nu)\r]^{-1} R^\nu(\{1^2\}\{2\}:m,f_m)\;
R^\nu(\{1^2\}\{2\}:m^\pr,f_{m^\pr})\;.
\earr \label{ch4.eq.38}
\ee
Note that $\cq^\nu(f_2:m,f_m)$ are defined in Eq. (\ref{ch4.eq.35}). The
functions $R^\nu(\{1^2\}\{2\}:m,f_m)$ also involve $SU(\Omega)$
$U$-coefficients and the explicit expression for $R^\nu$ is,
\be
\barr{l}
R^\nu(\{1^2\}\{2\}:m,f_m) = \l[ F(m) \r]^2 \dis\sum_{f_{m-2},f^\pr_{m-2}}
\dis\frac{\cn_{f_{m-2}}}{\cn_{f_m}}\dis\frac{\cn_{f_{m-2}^\pr}}{\cn_{f_m}}
Y_{UU}(f_{m-2},f^\pr_{m-2};\bF_\nu)\;; \\
Y_{UU}(f_{m-2},f^\pr_{m-2};\bF_\nu) = \\
\dis\sum_{\rho}
\dis\frac{U(f_m,\{1^{\Omega-2}\},f_m,\{1^2\};f_{m-2},\bF_\nu)
_\rho\,U(f_m,\{2^{\Omega-1}\},f_m,\{2\};
f_{m-2}^\pr,\bF_\nu)_\rho}{U(f_m,\{1^{\Omega-2}\},f_m,\{1^2\};f_{m-2},
\{0\})\,U(f_m,\{2^{\Omega-1}\},f_m,\{2\};f_{m-2}^\pr,\{0\})}\;.
\earr \label{ch4.eq.39}
\ee
In $Y_{UU}(f_{m-2},f^\pr_{m-2};\bF_\nu)$, $f_{m-2}$ comes from $f_m \otimes
\{1^{\Omega-2}\}$ and $f^\pr_{m-2}$ comes  from $f_m \otimes
\{2^{\Omega-1}\}$. In Eq. (\ref{ch4.eq.38}), the summation is over $\nu=0$ and
$1$ only as $\nu=2$ parts for $f_2=\{2\}$ and $\{1^2\}$ are different. Here
$d(\bF_\nu)$ are dimension of the irrep $\bF_\nu$, and we have $d(\{0\})=1$,
$d(\{21^{\Omega-2}\})=\Omega^2-1$, $d(\{42^{\Omega-2}\})=
\Omega^2(\Omega+3)(\Omega-1)/4$, and $d(\{2^21^{\Omega-4}\}) =\Omega^2
(\Omega-3) (\Omega+1)/4$. Tables for $X_{UU}(f_2; f_{m-2}, f^\pr_{m-2}; 
\bF_\nu)$ are already discussed before (see Table \ref{tab1}).  Formulas for
$Y_{UU}(f_{m-2}, f^\pr_{m-2}; \bF_\nu)$ are tabulated in Table  \ref{tab22}
and they also involve  $\tau_{ab}$, $\Pi_a^{(b)}$, $\Pi_{b}^{(a)}$, 
$\Pi_{a}^\pr$, $\Pi_{a}^{\pr\pr}$ and $\Pi_a^{(bc)}$ introduced before. 

\begin{table}[htp]
\caption{Formulas for $Y_{UU}(f_{m-2},f^\pr_{m-2};\bF_\nu)$ defined in Eq.
(\ref{ch4.eq.39}). Note that 
$\{f(ab)\}\,\{f(ab)\}$ entries satisfy the $a \leftrightarrow b$ symmetry.
See text for details.}
\begin{center}
\begin{tabular}{lc}
\toprule
$\{f_{m-2}\}\,\{f^\pr_{m-2}\}$ &  $Y_{UU} (f_{m-2}, 
f^\pr_{m-2}; \{2,1^{\Omega-2}\})$ \\ 
\midrule
$\{f(ab)\}\,\{f(ab)\}$ & $-\dis\frac{\Omega}{2}
\l[\dis\frac{(\Omega^2-1)}{(\Omega^2-4)}\r]^{1/2} 
\l\{\l(1+\dis\frac{1}{\tau_{ab}}\r)\dis\frac{1}{\Pi_a^{(b)}}\r.$ \\ \\
& $\l.+\l(1-\dis\frac{1}{\tau_{ab}}\r)\dis\frac{1}{\Pi_b^{(a)}}-
\dis\frac{4}{\Omega}\r\}$ \\ \\
$\{f(ab)\}\,\{f(ac)\}$ & $-\dis\frac{\Omega}{2}
\l[\dis\frac{(\Omega^2-1)}{(\Omega^2-4)}\r]^{1/2} \l\{ 
\l(1+\dis\frac{1}{\tau_{ac}}\r)\dis\frac{1}{\Pi_a^{(b)}}-
\dis\frac{4}{\Omega}\r\}$ \\ \\
$\{f(ab)\}\,\{f(aa)\}$ & $-\Omega
\l[\dis\frac{(\Omega^2-1)}{(\Omega^2-4)}\r]^{1/2} \l\{ 
\dis\frac{1}{\Pi_a^{(b)}}-\dis\frac{2}{\Omega}\r\}$ \\ \\
$\{f(ab)\}\,\{f(cc)\}$ & \\
$\;\;\;\;\;\;\;\;\;\;$ \text{or} & $2\l[\dis\frac{(\Omega^2-1)}
{(\Omega^2-4)}\r]^{1/2}$ \\ 
$\{f(ab)\}\,\{f(cd)\}$ & \\ 
\bottomrule
\end{tabular}
\end{center}
\label{tab22}
\end{table}

Using the results in Tables \ref{tab1} and \ref{tab22} and  simplifying 
Eqs. (\ref{ch4.eq.35}) and  (\ref{ch4.eq.39}), expressions for  $\cq^\nu(f_2:m,f_m)$
and  $R^\nu(\{1^2\}\{2\}:m,f_m)$ are derived for the irreps  $f_m^{(p)}$. It
is found that, with $P^{f_2}$ defined in Eq. (\ref{ch4.eq.32}),
\be
\barr{rcl}
\cq^{\nu=0}(f_2:m,f_m^{(p)}) & = & \l[P^{f_2}(m,f_m^{(p)}) \r]^2\;, \\
R^{\nu=0}(\{1^2\}\{2\}:m,f_m^{(p)}) & = & P^{\{1^2\}}(m,f_m^{(p)})
P^{\{2\}}(m,f_m^{(p)})\;.
\earr \label{ch4.eq.40}
\ee
The final results for $\cq^{\nu=1,2} (f_2:m,f_m^{(p)})$ and $R^{\nu=1}
(\{1^2\}\{2\}:m,f_m^{(p)})$ are given in Tables \ref{tab2} and \ref{tabr},
respectively. Formulas in these Tables are verified numerically in many
examples by directly programming Tables \ref{tab1} and \ref{tab22}. In the
dilute limit ($\Omega \to \infty$, $r>>1$, $r/\Omega \to 0$),  the cross
term $X_{\{1^2\}\{2\}}$ will be very small compared to the direct terms
$X_{f_2}$. Dominant contribution to $X_{f_2}$ comes from
$\cq^{\nu=2}(f_2:m,f_m^{(p)})$  which has the form  $-\Omega^4
P^{f_2}(m,f_m^{(p)})/4$ (while the other terms i.e.,  $\cq^{\nu=1}
(f_2:m,f_m^{(p)})$ and $R^{\nu=1} (\{1^2\}\{2\}:m,f_m^{(p)})$ have
$\Omega^2$ dependence).  Then in the dilute limit, for the irreps
$f_m^{(p)}$, simplifying the results given in Tables \ref{tab2} and
\ref{tabr}, the covariances in  spectral variances take a simple form,
\be
\barr{l}
\Sigma_{22}(m,f_m^{(p)};m^\pr,f_{m^\pr}^{(p)}) \stackrel{\Omega \to
\infty,r>>1,r/\Omega \to 0}{\longrightarrow} \\ \\
\dis\frac{8}{\Omega^4}\dis\frac
{\dis\sum_{f_2}\lambda^4_{f_2}P^{f_2}(m,f_m^{(p)})
P^{f_2}(m^\pr,f_{m^\pr}^{(p)})}
{\l\{\dis\sum_{f_2}\lambda^2_{f_2}P^{f_2}(m,f_m^{(p)})\r\}
\l\{ \dis\sum_{f_2}\lambda^2_{f_2}P^{f_2}(m^\pr,f_{m^\pr}^{(p)})\r\}}\;.
\earr \label{ch4.eq.sg22}
\ee
As $\Omega \to \infty$, $\Sigma_{22}(m,f_m^{(p)};m^\pr,f_{m^\pr}^{(p)}) \to
0$ and there will be no correlations. For finite $\Omega$, there will be
correlations between states with different or same ($m,f_m$) and examples
for these are discussed ahead. 

\setlength{\LTcapwidth}{6.5in}

\begin{longtable}{cccc}
\caption{$\cq^\nu(f_2:m,f_m)$ for $f_m=f_m^{(p)}$
and $\nu=1$ and $2$. See Eq. (\ref{ch4.eq.35}) for the definition of
$\cq^\nu$.} \\

\toprule
$f_m^{(p)}$ & $f_2$ & $\nu$ & $\cq^\nu(f_2:m,f_m^{(p)})$ \\ 
\midrule
\endfirsthead

\multicolumn{4}{c}%
{{\bfseries \tablename\ \thetable{} -- continued}} \\
\toprule
$f_m^{(p)}$ & $f_2$ & $\nu$ & $\cq^\nu(f_2:m,f_m^{(p)})$ \\ 
\midrule
\endhead

\bottomrule
\endfoot

\bottomrule
\endlastfoot
$\{4^r\}$ & $\{2\}$ & $1$ & $\dis\frac{9r(r+1)^2(\Omega-r)(\Omega+1)
 (\Omega+4)}{2(\Omega+2)}$ \\
 & & $2$ & $\dis\frac{3r\Omega(r+1)(\Omega-r+1)(\Omega-r)
 (\Omega+4) (\Omega+5)}{4(\Omega+2)}$ \\
 & $\{1^2\}$ & $1$ & $\dis\frac{25r(r-1)^2(\Omega-r)(\Omega-1)(\Omega+4)}
 {2(\Omega-2)}$ \\
 & & $2$ & $\dis\frac{5r\Omega(r-1)(\Omega+3)(\Omega+4)(\Omega-r)
 (\Omega-r-1)}{4(\Omega-2)}$ \\ \\
$\{4^r,1\}$ & $\{2\}$ & $1$ & $\dis\frac{3r(\Omega+1)}{8(\Omega+2)}
\l[-12(\Omega+4)r^3
 +12(\Omega-3)(\Omega+4)r^2\r.$ \\
 & & & $\l.+(33\Omega ^2+100\Omega-108)r+20\Omega(\Omega+4)\r]$ \\
 & & $2$ & $\dis\frac{3r\Omega}{8(\Omega+2)}(\Omega-r+1)(\Omega+4)\l[-2
 (\Omega+5)r^2\r.$ \\
 & & & $\l.+2(\Omega-2)(\Omega+5)r+\Omega(3\Omega+11)-10\r]$ \\
 & $\{1^2\}$ & $1$ & $-\dis\frac{5r(\Omega-1)}{8(\Omega-2)}
 \l[20(\Omega+4)r^3-20(\Omega^2+5\Omega+4)r^2\r.$ \\
 & & & $\l.+(25\Omega^2+132\Omega+20)r-12\Omega(\Omega+4)\r]$ \\
 & & $2$ &$\dis\frac{5r\Omega(\Omega+4)}{8(\Omega-2)} \l[2(\Omega+3)r^3
-2(2\Omega^2+5\Omega-3)r^2\r.$ \\
 & & & $\l.+(2\Omega^3+5\Omega^2+\Omega-6)r -\Omega^3-6\Omega^2+13\Omega-6)
  \r]$ \\ \\
$\{4^r,2\}$ & $\{2\}$ & $1$ & $\dis\frac{(\Omega+1)}{4(\Omega+2)}
\l[-8(3r^2+6r+1)^2+(3r+4)(6r^2+13r+1)\Omega^2\r.$ \\
 & & & $\l.-2(9r^4-79r^2-88r-2)\Omega \r]$ \\
 & & $2$ & $\dis\frac{\Omega}{4(\Omega+2)}\l[3(\Omega+4)(\Omega+5)r^4
 -6(\Omega-1)(\Omega+4)(\Omega+5)r^3\r.$ \\
 & & & $\l.+\Omega(\Omega+4)(3\Omega^2+3\Omega-56)r^2\r.$ \\
 & & & $\l.+(\Omega-1)(\Omega+4)(6\Omega^2+29\Omega+15)r\r.$ \\
 & & & $\l.+\Omega(\Omega-1)(\Omega+2)(\Omega+3)\r]$ \\
 & $\{1^2\}$ & $1$ & $-\dis\frac{5r(\Omega-1)}{4(\Omega-2)}
 \l[10(\Omega+4)r^3-10\Omega(\Omega+4) r^2\r.$ \\
 & & & $\l.+\Omega(5\Omega+38) r-3\Omega(\Omega+4)\r]$ \\
 & & $2$ & $-\dis\frac{5r\Omega(\Omega+4)(\Omega-r-1)}{4(\Omega-2)}
\l[3(\Omega-1)-r(\Omega-r-1)(\Omega+3)\r]$ \\ \\
 $\{4^r,3\}$ & $\{2\}$ & $1$ & $-\dis\frac{3(r+2)(\Omega+1)}{8(\Omega+2)}
 \l[12(r+2)(2r+1)^2-\Omega^2(12r^2+27r+8)\r.$ \\
 & & & $\l.+4(3r^3-3r^2-19r-4)\Omega\r]$ \\
  &  & $2$ & $-\dis\frac{3\Omega(r+2)(\Omega+4)(\Omega-r-1)}
{8(\Omega+2)} \l[2r^2(\Omega+5)-2\Omega r(\Omega+5)\r.$ \\
 & & & $\l.-\Omega(\Omega+3)\r]$ \\
  & $\{1^2\}$ & $1$ & $-\dis\frac{5r(\Omega-1)}{8(\Omega-2)}
  \l[20(\Omega+4)r^3-20r^2(\Omega^2+3\Omega-4)\r.$ \\
& & & $\l. +r(-5\Omega^2+12\Omega+20)-2\Omega(\Omega+4)\r]$ \\
  &  & $2$ & $-\dis\frac{5r\Omega(\Omega+4)(\Omega-r-1)}{8(\Omega-2)}
  \l[2(\Omega+3)r^2-2r(\Omega^2+\Omega-6)\r.$ \\
& & & $\l.-\Omega(\Omega-3)\r]$ 
\label{tab2}
\end{longtable}

\begin{table}[ht]
\caption{$R^{\nu=1}(\{1^2\}\{2\}:m,f_m)$ for $f_m=f_m^{(p)}$. See 
Eq. (\ref{ch4.eq.39}) for the definition of $R^{\nu=1}$.}
\begin{center}
\begin{tabular}{cl}
\toprule
$f_m^{(p)}$ & $R^{\nu=1}(\{1^2\}\{2\}:m,f_m^{(p)})$
\\ \midrule
$\{4^r\}$ &
$-\dis\frac{15r}{2}\dis\sqrt{\dis\frac{\Omega^2-1}{\Omega^2-4}}
(r^2-1)(\Omega-r)(\Omega+4)$ \\ \\
$\{4^r,1\}$ & $\dis\frac{15r}{8}
\dis\sqrt{\dis\frac{\Omega^2-1}{\Omega^2-4}}\l[
4r^3(\Omega+4)-4r^2(\Omega+4)(\Omega-1)\r.$ \\
  & $ \l. -3r(\Omega+2)^2+4\Omega(\Omega+4)\r]$ \\ \\
$\{4^r,2\}$ & $-\dis\frac{5r}{4}
\dis\sqrt{\dis\frac{\Omega^2-1}{\Omega^2-4}}\l[ -6r^3(\Omega+4)
-3\Omega(\Omega+6)\r.$ \\
 & $\l.+6r^2(\Omega-2)(\Omega+4)+r(9\Omega^2+28\Omega-8)\r]$ \\ \\
$\{4^r,3\}$ & $-\dis\frac{15r}{8}(r+2)\dis
\sqrt{\dis\frac{\Omega^2-1}{\Omega^2-4}}\l[-4r^2(\Omega+4)\r.$ \\
 & $\l.+4r(\Omega-1)(\Omega+4)+\Omega^2-4\r]$ \\ 
\bottomrule
\end{tabular}
\end{center}
\label{tabr}
\end{table}

\section{Numerical Results for Spectral Variances, Expectation Values of
$C_2[SU(4)]$ and Four Periodicity in GS}
\label{num1}

Employing the analytical formulation described in Secs. \ref{wralg} and
\ref{anres} along with the results in Table \ref{tabw} for $f_m= f_m^{(p)}$
irreps and Table \ref{tab1} for general $f_m$ irreps, numerical calculations
are carried out for $\overline {\lan H^2 \ran^ {m,f_m}}$. In our examples,
we have chosen  $\Omega=6$ and $\Omega=10$ and they correspond to nuclear
($2s1d$) and ($2p1f$) shells, respectively. Results for spectral variances
are used to analyze expectation values of $C_2[SU(4)]$ and the four
periodicity in the gs energies. Conclusions from these studies are
summarized at the end.

\subsection{Spectral variances}
\label{num1-1}

\begin{figure}[ht]
\centering
\includegraphics[width=5.5in,height=5.5in]{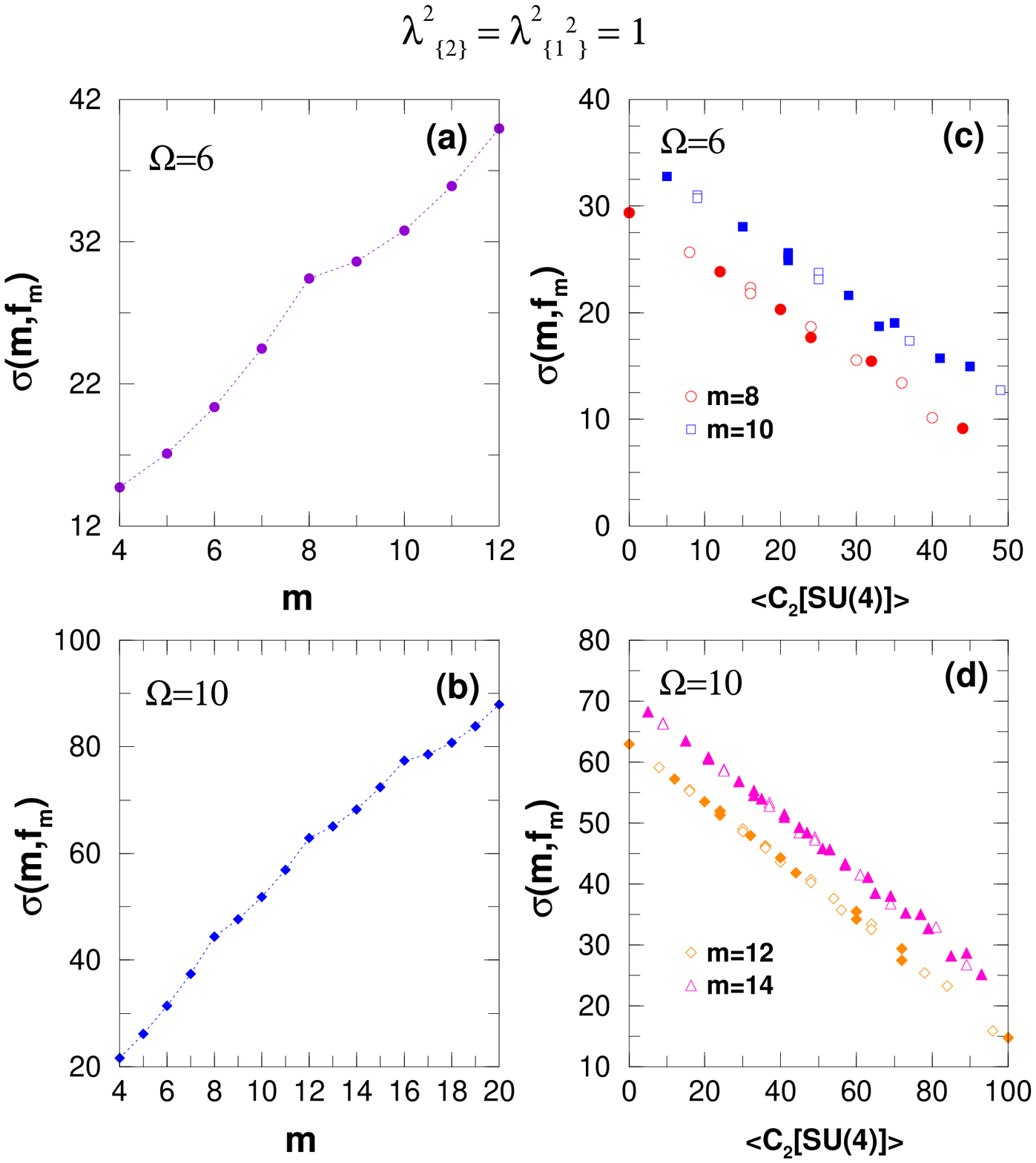} 
\caption{Variation of spectral widths $\l[\;\overline{\lan 
H^2\ran^{m,f_m}}\;\r]^{1/2}$ as a function of $m$ with fixed  $f_m$ and
similarly variation as a function of $f_m$ with fixed $m$. (a) $\Omega=6$,
$f_m=f_m^{(p)}$, (b) $\Omega=10$,  $f_m=f_m^{(p)}$, (c) $\Omega=6$ and $m=8$
and $10$, and (d)  $\Omega=10$ and $m=12$ and $14$. Note that
$f_m^{(p)}=\{4^r,p\}$ where $m=4r+p$. Similarly, instead of showing $f_m$ 
in (c) and (d) we have used  $\lan C_2[SU(4)]\ran^{\tilde{f}_m}$. We have
marked   by filled symbols in (c) and (d) the irreps $f_m$ that give
$(S,T)=(0,0)$ for  $m=4r$ systems and $(S,T)=(1,0)\oplus(0,1)$  for $m=4r+2$
systems.  See text for details.}
\label{ch4-var}
\end{figure}

Figures \ref{ch4-var}(a) and (b) show variation in the spectral widths
$\sigma(m,f_m)= [\;\overline{\lan H^2\ran^{m,f_m}}\;]^{1/2}$ as a function
of the particle number $m$ with fixed   $f_m=f_m^{(p)}$. Notice the peaks at
$m=4r$; $r=2,3,\ldots$. Except for this structure, there are no other
differences between $\{4^r\}$ and $\{4^r,2\}$ systems or equivalently
between even-even and odd-odd N=Z nuclei. Figures \ref{ch4-var}(c) and (d) show
variation in the spectral widths $\sigma(m,f_m)$ as a function of $f_m$ 
with fixed $m$ values. Results are shown for $m=8$ and $10$ for $\Omega=6$
and $m=12$ and $14$ for $\Omega=10$. In the  figures,  we have used the
physically more appropriate $\lan C_2[SU(4)] \ran^{\tilde{f}_m}$ label for
the $x$-axis instead of showing $f_m$. It is clearly seen that the variation
in the spectral widths is almost linear. Considering the eigenvalue density
to be Gaussian [extrapolating from the results known for EGUE(2), EGOE(2)
and EGOE(2)-$\cs$] and neglecting the dimension effects, energy of the
lowest state that belong to a given $f_m$ follows from the Jacquod and Stone
prescription \cite{Pa-08,Ja-01}. This gives 
\be
E_{gs}(f_m) -E_c(m,f_m) \propto -\sigma(m,f_m) \;.  
\label{eq.ch4.new}
\ee
This follows from Eq. (\ref{ch4.eq.integ}) given ahead if we restrict it to a
given $f_m$. Combining Eq. (\ref{eq.ch4.new}) 
with the results in Figs. \ref{ch4-var}(c) and
(d), we can identify the irreps that label the gs generated by 
EGUE(2)-$SU(4)$. As $\sigma(m,f_m)$ vs $\lan C_2[SU(4)] \ran^{\tilde{f}_m}$
curves are linear, clearly EGUE(2)-$SU(4)$ generates gs labeled by the
irreps that have lowest  $\lan C_2[SU(4)]\ran^{\tilde{f}_m}$.  Therefore
random interactions, which are $SU(4)$ scalar, carry the properties of
$C_2[SU(4)]$, the $SU(4)$ invariant or the Majorana force. In  Figs.
\ref{ch4-var}(c) and (d), we have marked the irreps that give $(S,T)=(0,0)$ for
$m=4r$ and $(S,T)=(1,0) \oplus (0,1)$ for $m=4r+2$ systems. If we restrict
to these irreps, the second irrep is forbidden in both cases i.e., there is a
gap between the lowest and next allowed irrep. This implies that even with
random interactions we obtain gs with $f_m=f_m^{(p)}$. We will further
substantiate this result by calculating the expectation values $\lan
C_2[SU(4)] \ran^{E}$ and also analyzing the four periodicity in $E_{gs}$.

\subsection{Expectation values $\lan C_2[SU(4)] \ran^{E}$}
\label{num1-2}

In order to examine the extent to which random interactions with $SU(4)$
symmetry carry the properties of the Majorana operator, we have studied 
expectation values (smoothed with respect to $E$)  of the quadratic Casimir
invariant of $SU(4)$ using the  Hamiltonian $H_\alpha$,   
\be 
\{H_{\alpha}\} = C_2\l[SU(4)\r] + \alpha \{H\} \;.   
\label{ch4.eq.cas1}  
\ee  
where $\{H\}$ is defined by Eq. (\ref{ch4.eq.22}) with $\lambda_{\{2\}}^2 = 
\lambda_{\{1^2\}}^2 = 1$. In order to study $\lan
C_2[SU(4)] \ran^{E}$, we decompose it in terms of $\lan C_2[SU(4)]
\ran^{\tilde{f}_m}$ (see Eq. (\ref{ch3.eq.expec2}) and \cite{Pa-78}), 
\begin{subequations}
\be 
\lan C_2[SU(4)] \ran^{E} = \dis\sum_{f_m}\;\dis\frac{  
I^{m,f_m}_\cg(E)}{I^m(E)}\,\lan C_2[SU(4)] \ran^{\tilde{f}_m}\;;
\label{ch4.eq.cas2} 
\ee
\be
I^m(E) = \dis\sum_{f_m}  I^{m,f_m}_\cg(E) = \dis\sum_{f_m}  
d_\Omega(f_m) d_4(\tilde{f}_m)\;\rho^{m,f_m}_\cg(E)\;. 
\label{ch4.eq.cas2a} 
\ee
\end{subequations}
In Eq. (\ref{ch4.eq.cas2}), $I^{m,f_m}(E)$ are partial eigenvalue densities
defined over a fixed $f_m$ space,  $I^{m,f_m}(E) = \lan \lan \delta(H-E)
\ran \ran^{m,f_m}$ and $I^m(E)$ is the total eigenvalue density.  Equation
(\ref{ch4.eq.cas2}) is exact if we remove $\cg$, as $f_m$ (equivalently 
$\tilde{f}_m$)  label the  eigenstates of $C_2[SU(4)]$. For smoothed
expectation values,  based on the $SU(4)$ partial densities that are studied
within the nuclear shell-model (with $\Omega=6$) \cite{Pa-73,Pa-72}, we assume
that $\rho^{m,f_m}(E)$ will be close  to a Gaussian ($\cg$).  Numerical
calculations of $\gamma_2$ using $H$ matrix construction as discussed in
Section \ref{cons} or using the analytical formulation  discussed in
Appendix \ref{c4a2}, will verify this assumption. However, at present both
these methods are not feasible in practice. For the Hamiltonian in Eq.
(\ref{ch4.eq.cas1}), the  centroid of $\rho^{m,f_m}_\cg(E)$ is $\lan C_2[SU(4)]
\ran^{\tilde{f}_m}$ and the  variance is $\overline{\lan H^2\ran^{m,f_m}}$. 

\begin{figure}[ht]
\centering
\includegraphics[width=5.5in,height=3in]{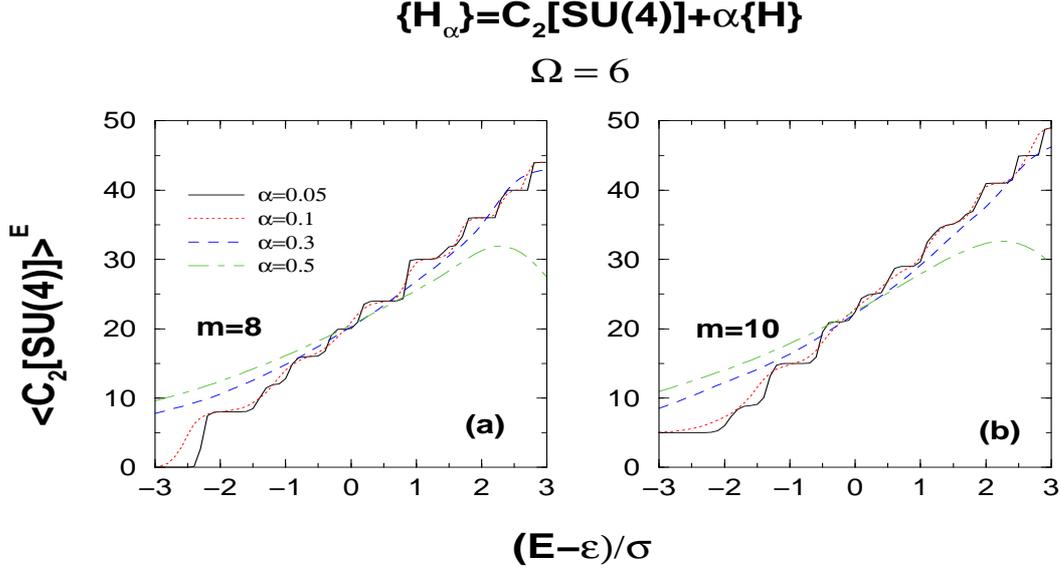} 
\caption{Expectation values of the quadratic Casimir 
invariant of $SU(4)$ as a function of excitation energy for the $H_{\alpha}$
Hamiltonian ensemble defined in Eq. (\ref{ch4.eq.cas1}). Results are shown for
four values of interaction strength $\alpha$: (a) for $m=8$ and (b) for
$m=10$. Note that the energies are zero centered with respect to the
centroid $\epsilon$ and scaled with the width $\sigma$ defined by first and
second moments of the total density of states. All the results are for
$\Omega=6$. Similar results are obtained even when we consider, in Eq.
(\ref{ch4.eq.cas2a}), the  irreps  $f_m$ that give $(S,T)=(0,0)$ for $m=8$ and
$(S,T)=(1,0)\oplus(0,1)$ for $m=10$.}
\label{cas}
\end{figure}

As an example, for $\Omega=6$ and $m=8$ and $10$, the expectation values are
calculated as a function of energy  for various values of $\alpha$ in Eq.
(\ref{ch4.eq.cas1}) and  the results are  shown in Figs. \ref{cas}(a) and (b). 
It is seen that with the increase in the strength $\alpha$, fluctuations 
decrease and the staircase form for $\alpha \sim 0$ turns into a smooth curve
for $\alpha \gazz \alpha_c = 0.3$. This conclusion remains same even when we
consider $U(\Omega)$ irreps with  $(S,T)=(0,0)$ for $m$ even  and
$(S,T)=(1,0)\oplus(0,1)$ for $m$ odd. Then the normalization for
$I^{m,f_m}_\cg(E)$ is $d_\Omega (f_m) \times d_{gs}$. Note that the 
degeneracy $d_{gs}=1$, $6$, and $4$, respectively for $m=4r$ (even-even
nuclei), $m=4r+2$ (odd-odd nuclei) and $m=4r+1$ or $4r+3$  (odd-A nuclei).
Just as for EGOE(1+2) and EGOE(1+2)-$\cs$, it is expected that the transition
point $\alpha_c \propto \Omega/K(m,f_m)$ and the  variance propagator
$K(m,f_m)$, as mentioned in Section \ref{stru},  follows from the formulas in
Table \ref{tabw} for $f_m^{(p)}$ irreps and for general irreps from Eq.
(\ref{ch4.eq.34}) and Table \ref{tab1} with
$\lambda_{\{1^2\}}^2=\lambda_{\{2\}}^2=1$. From the results in Table
\ref{tabw}, for the $f_m^{(p)}$ irreps, it follows that in the dilute limit, 
$K(m,f_m) \to m^2\Omega^2$. Thus, $\alpha_c \propto 1/m^2\Omega$ and  this
result is same as those derived before for EGOE(1+2) and EGOE(1+2)-$\cs$; see
Chapter \ref{ch2} for details. 
Therefore, with fixed $m$, $\alpha_c=0.3$ for $\Omega=6$
corresponds to $\alpha_c \sim 0.2$ for $\Omega=10$. We have verified this by
comparing the numerical results for $\Omega=6$ and $10$.

Results in Figs. \ref{cas}(a) and (b) confirm that even with random
interactions that are $SU(4)$ scalar, ground states  have lowest value for
$\lan C_2[SU(4)] \ran^{E}$ and therefore they carry the property of the
Majorana force. Also beyond a critical strength ($\alpha_c$) of the random
part in Eq. (\ref{ch4.eq.cas1}), expectation values will be smooth with 
respect to energy.

\subsection{Four-periodicity in the ground state energies}
\label{num1-3}

\begin{figure}[ht]
\centering
\includegraphics[width=5.5in,height=4.5in]{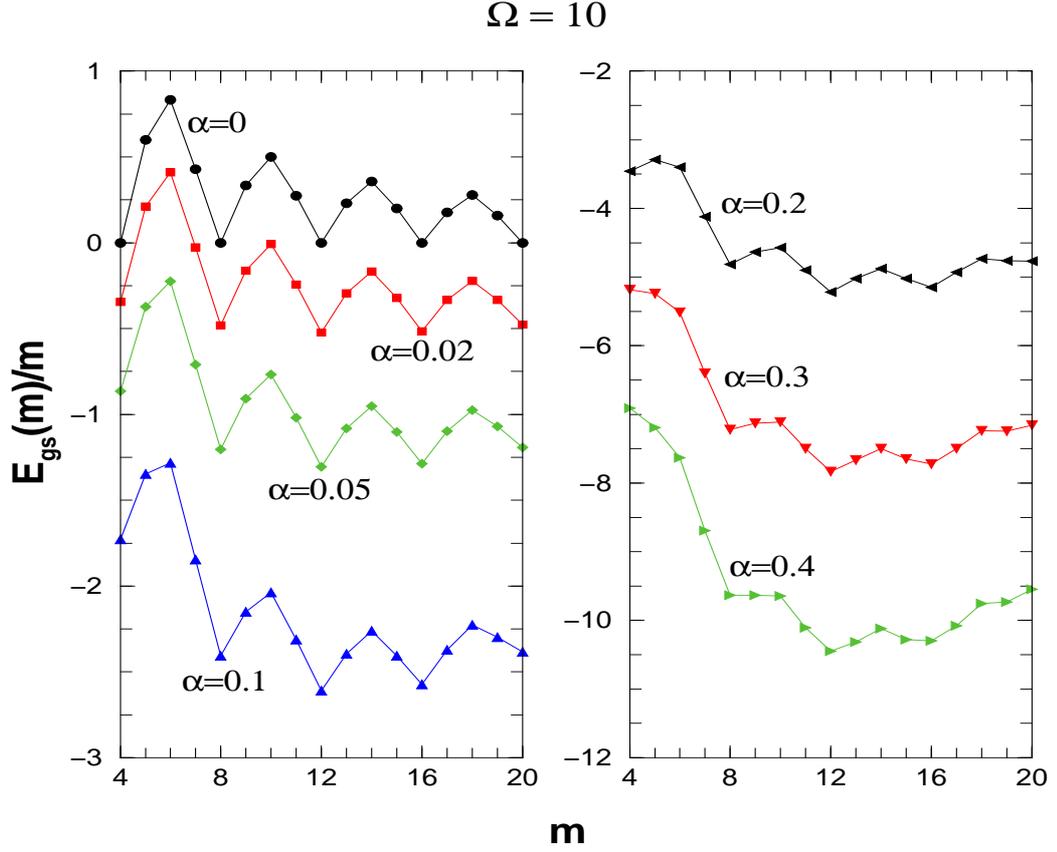} 
\caption{Ground state energy $E_{gs}(m)$ per particle $m$ as
a function of $m$ for different values of the interaction strength  $\alpha$
in Eq. (\ref{ch4.eq.cas1}). Results are shown for $\alpha \leq 0.4$. 
The variation of $E_{gs}(m)/m$ shown in the figure brings out the 
four periodicity effect in the gs energies.  See text for 
details.}
\label{gse}
\end{figure}

An evidence for effective space symmetry for nuclear ground states is
derived from the four periodicity in the gs energies $E_{gs}$ per
particle \cite{Pa-78}. An  important question is: will this feature survive
even in the presence of  random interactions.   To test this, as a  model,
we consider the Hamiltonian $H_\alpha$ in Eq. (\ref{ch4.eq.cas1}) where $\alpha$
is the strength of the random interaction with $SU(4)$ symmetry.  For the
strength $\alpha=0$, $H$ reduces to  the quadratic Casimir invariant of the
$SU(4)$ group and this, as it is well-known, produces oscillations in
$E_{gs}(m)/m$ with minima at $m=4r$  (this is called four periodicity) as seen
clearly from Fig. \ref{gse}. When the strength $\alpha$ is non-zero, for
given number of particles $m$, all the irreps $f_m$, with $(S,T)=(0,0)$ for
$m=4r$, $(S,T)=(1,0) \oplus (0,1)$ for $m=4r+2$ and $(S,T)=(\spin,\spin)$
for $m=4r+1$ and $4r+3$, contribute to the sum in Eq. (\ref{ch4.eq.cas2a}) in
generating the total  density of states. Using Eq. (\ref{ch4.eq.cas2a}), 
$E_{gs}(m)$  for a fixed $m$ is determined numerically by inverting the
integral,
\be
\dis\frac{1}{2} = \int_{-\infty}^{E_{gs}(m)} \dis\sum_{f_m} 
 d_\Omega(f_m)\;\rho^{m,f_m}_\cg(E) dE \;.
\label{ch4.eq.integ}
\ee 
This is known as ``Ratcliff procedure''
in nuclear physics literature
\cite{Ra-71,Wo-86}. We show in Fig. \ref{gse}, the variation of $E_{gs}(m)/m$ vs
$m$  for different interaction strengths $\alpha$. In the calculations,
$\Omega=10$ and $m=4-20$. It is clearly seen that the four  periodicity
produced by $C_2[SU(4)]$ is preserved by random  Hamiltonian $H_\alpha$  for
$\alpha \leq 0.2$.  The kinks in the spectral widths at $m=4r$  as a function
of $m$ as seen from Fig. \ref{ch4-var}(a) and (b) and similarly, their  
monotonic decrease with $\lan C_2[SU(4)] \ran^{\tilde{f}_m}$  as seen from
Fig. \ref{ch4-var}(c) and (d), together explain the four periodicity in the
gs energies.  

Beyond $\alpha = 0.2$, this structure starts disappearing as the
difference $\Delta$ between the centroids, produced by $C_2[SU(4)]$ for the
lowest two irreps, becomes comparable to the width of the gs irrep $\{4^r\}$;
$m=4r$. Therefore, with a regular part that is close to $C_2[SU(4)]$, random
interactions that are not too strong [$\alpha \lazz \alpha_c^\pr = 0.2$ in
Eq. (\ref{ch4.eq.cas1})] generate, in the $\Omega=10$ example, ground states that
are  spatially symmetric. Thus, $\alpha_c \sim \alpha_c^\prime$ (see Section
\ref{num1-2} for $\alpha_c$) and
therefore,  the region of onset of  smooth behavior for $\lan C_2[SU(4)]
\ran^{\tilde{f}_m}$ also marks the onset of  diminishing four periodicity
effect in the gs energies. As $\alpha_c \propto 1/m^2\Omega$, the four
periodicity effect should diminish faster for large $m$ and this is clearly
seen from Fig. \ref{gse}.

\subsection{Conclusions}
\label{num1-s}

Thus, ensemble averaged spectral variances $\overline{\lan
H^2\ran^{m,f_m}}$, expectation values $\lan C_2[SU(4)] \ran^{E}$ and the
four periodicity in $E_{gs}(m)/m$ discussed in Secs.
\ref{num1-1}-\ref{num1-3}  establish that random interactions with $SU(4)$
symmetry keep intact all the essential features of the Majorana force (see
Section \ref{cafo} for further discussion on the importance of this
result).  Therefore the EGUE(2)-$SU(4)$ and the corresponding
EGOE(2)-$SU(4)$ ensemble should be useful in nuclear structure. 

\section{Numerical Results for Correlations in Energy Centroids and Spectral
Variances}
\label{num2}

Using the results in Tables \ref{pf2}, \ref{tabw}, \ref{tab2} and \ref{tabr}
for $f_m= f_m^{(p)}$ irreps  and Tables \ref{tab1} and \ref{tab22} for
general $f_m$ irreps, the self and cross-correlations in energy centroids
and spectral variances [i.e., $\Sigma_{11}$ and $\Sigma_{22}$ in Eq.
(\ref{eq.den10})]  are calculated. See \cite{Br-81,Fl-00,PDPK} for a detailed
discussion on the significance of self-correlations (they affect level
motion in the ensemble) and  \cite{Pa-07,Ko-07,Ko-06} on the significance
of the cross-correlations (they will vanish for GE's)  generated by embedded
ensembles. Results for $\Sigma_{11}$ and $\Sigma_{22}$ are discussed in
Secs. \ref{num2-1}-\ref{num2-3} and a summary is given at the end.

\subsection{Self-correlations}
\label{num2-1}

Results for self-correlations ($m=m^\pr$, $f_m=f_{m^\pr}$)  are shown in
Table \ref{symm} for $f_m=f_m^{(p)}$ and $\Omega=6$ and $10$.  For
$\Omega=6$ we have,  $[\Sigma_{11}]^{1/2} \sim 12-28$\% and 
$[\Sigma_{22}]^{1/2} \sim 7-15$\% as $m$ changes from 6 to 12. Similarly,
for $\Omega=10$ and $m$ ranging from $12$ to $20$,  they decrease to
$10-22$\% for $[\Sigma_{11}]^{1/2}$ and $4-9$\% for $[\Sigma_{22}]^{1/2}$.
We can also infer from Table \ref{symm} that as $m$ increases, the 
self-correlations also increase. Therefore,  fluctuations in the level motion in
the ensemble increase with $m$ and as a result the ensemble averages deviate
from spectral averages with increasing $m$. This feature has been studied
before for EGOE(2) and EGOE(2)-$\cs$  \cite{Br-81,Fl-00,Le-08}.

\begin{sidewaystable}[htp]
\caption{Variation in the self-correlations in energy centroids 
($\Sigma_{11}$) and spectral variances ($\Sigma_{22}$) with symmetry. 
See text for details.} 
\begin{center}
\begin{tabular}{cccccccc}
\toprule
 & & & $\l[\Sigma_{11}\r]^{1/2}$ & & &
$\l[\Sigma_{22}\r]^{1/2}$ & \\ \cmidrule{3-5}\cmidrule{6-8}
$N$ & $m$ & EGUE(2) & EGUE(2)-$\cs$ & EGUE(2)-$SU(4)\;\;\;\;\;$ & 
 EGUE(2) & EGUE(2)-$\cs$ & EGUE(2)-$SU(4)$ \\
\midrule
$24$ & $6$ & $0.017$ & $0.043$ & $0.125$ & $0.0056$ & $0.017$ & $0.069$ \\
 & $7$ & $0.021$ & $0.055$ & $0.144$ & $0.0059$ & $0.019$ & $0.076$ \\
 & $8$ & $0.026$ & $0.066$ & $0.160$ & $0.0064$ & $0.021$ & $0.083$ \\
 & $9$ & $0.031$ & $0.081$ & $0.196$ & $0.0069$ & $0.025$ & $0.099$ \\
 & $10$ & $0.037$ & $0.094$ & $0.229$ & $0.0077$ & $0.028$ & $0.117$ \\
 & $11$ & $0.044$ & $0.112$ & $0.256$ & $0.0087$ & $0.034$ & $0.134$ \\
 & $12$ & $0.051$ & $0.128$ & $0.276$ & $0.0099$ & $0.039$ & $0.148$ \\
\midrule
$40$ & $12$ & $0.0139$ & $0.038$ & $0.105$ & $0.00222$ & $0.0079$ & $0.035$
\\
 & $13$ & $0.0157$ & $0.044$ & $0.120$ & $0.00234$ & $0.0086$ & $0.039$ \\
 & $14$ & $0.0176$ & $0.048$ & $0.134$ & $0.00247$ & $0.0093$ & $0.044$ \\
 & $15$ & $0.0196$ & $0.054$ & $0.146$ & $0.00262$ & $0.0103$ & $0.049$ \\
 & $16$ & $0.0218$ & $0.06$ & $0.156$ & $0.0028$ & $0.0112$ & $0.053$ \\
 & $17$ & $0.0241$ & $0.067$ & $0.174$ & $0.003$ & $0.0125$ & $0.061$ \\
 & $18$ & $0.0267$ & $0.073$ & $0.192$ & $0.00324$ & $0.0138$ & $0.069$ \\
 & $19$ & $0.0294$ & $0.081$ & $0.206$ & $0.00352$ & $0.0156$ & $0.078$ \\
 & $20$ & $0.0325$ & $0.088$ & $0.218$ & $0.00385$ & $0.0174$ & $0.085$ \\
\bottomrule
\end{tabular}
\end{center}
\label{symm}
\end{sidewaystable}

Further significance of the magnitude of the self-correlations  follows by
comparing the results with the corresponding ones for EGUE(2) and
EGUE(2)-$\cs$ for fixed number of sp states ($N$).  Using the analytical
formulas given in \cite{Ko-05} for EGUE(2),  $[\Sigma_{11}(m,m)]^{1/2}$ and
$[\Sigma_{22}(m,m)]^{1/2}$ are calculated  for various values of $m$ with
$N=24$ and $40$ and the results are shown in third and sixth columns of
Table \ref{symm}. Similarly, using the formulas in \cite{Ko-07} for
EGUE(2)-$\cs$, $[\Sigma_{11}(m,S;m,S)]^{1/2}$ and 
$[\Sigma_{22}(m,S;m,S)]^{1/2}$ with $S=0$ for even $m$ and $S=1/2$ for  odd
$m$ are calculated  for various values of $m$ with $N=24$ ($\Omega=12$)  and
$40$ ($\Omega=20$) and the results are shown in fourth and seventh columns of
Table \ref{symm}. It is seen from Table \ref{symm} that the magnitude of the
covariances in energy centroids and spectral variances increases by a factor
$3$ when we go from EGUE(2) $\to$ EGUE(2)-$\cs$ $\to$ EGUE(2)-$SU(4)$. 

As discussed in Section \ref{base}, the fraction of independent matrix
elements $\cii$ increases with  symmetry and also the sparsity ($\cS$)
decreases and therefore the EGUE(2)-$SU(4)$ matrices will be dense leading
to a more complete mixing of the basis states compared to EGUE(2) and
EGUE(2)-$\cs$. Therefore there is a correlation between (i) increase in
fluctuations  defined by  $\Sigma_{11}$ and $\Sigma_{22}$ and (ii) the
matrices $H_{f_m}(m)$ becoming more dense as we go from EGUE(2) $\to$
EGUE(2)-$\cs$ $\to$ EGUE(2)-$SU(4)$. See Section \ref{cafo} for further
discussion.

\subsection{Cross-correlations}
\label{num2-2}

\begin{figure}[ht]
\centering
\includegraphics[width=5.5in,height=5.5in]{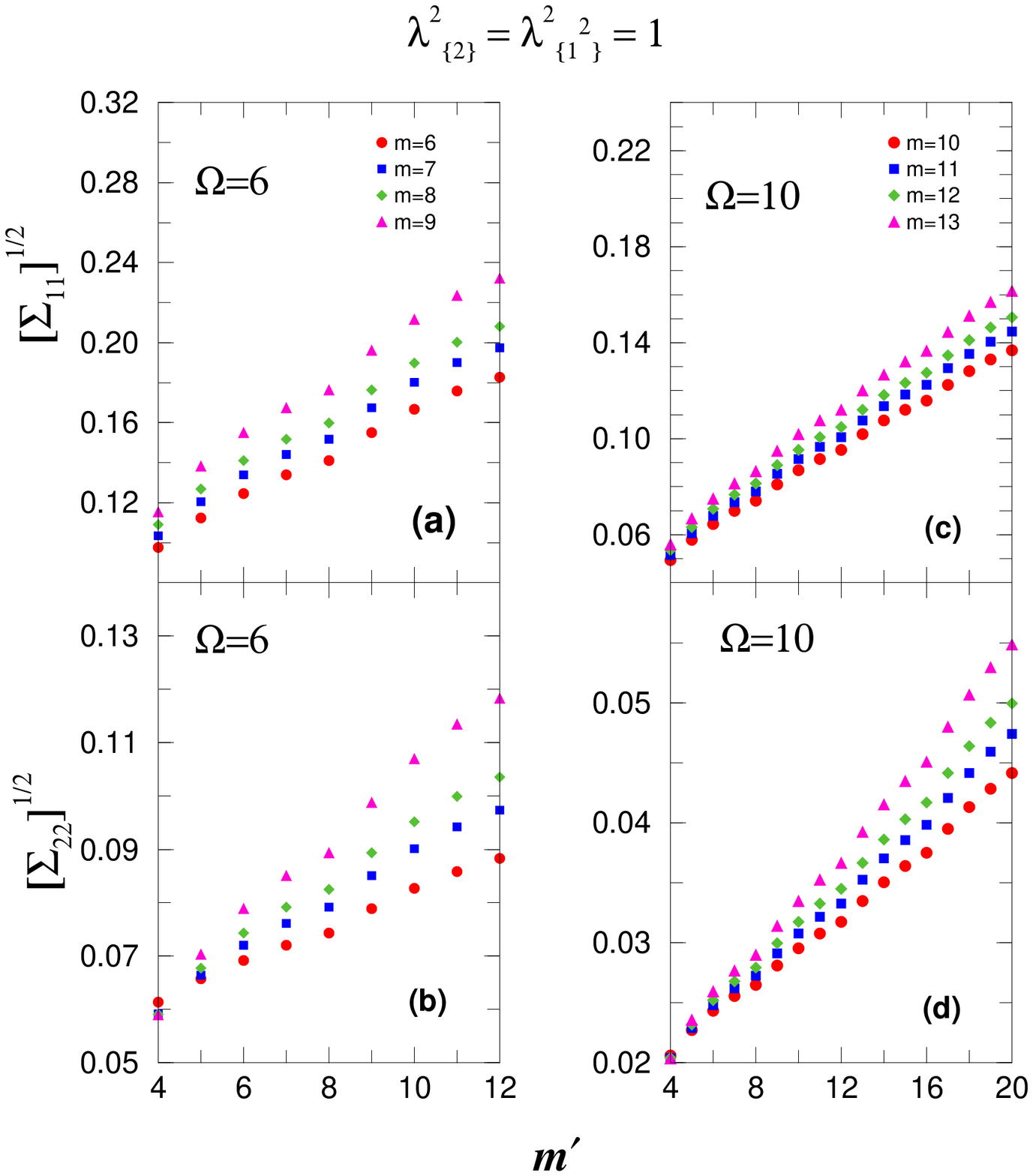} 
\caption{Self and cross-correlations in energy centroids and
spectral variances as a function of $m$ and $m^\pr$ (with fixed $f_m$ and 
$f_{m^\pr}$) for $\Omega=6$ and $\Omega=10$ examples: (a)
$\l[\Sigma_{11}(m,f_m;m^\pr,f_{m^\pr})\r]^{1/2}$ for  $\Omega=6$; (b) 
$\l[\Sigma_{22} (m,f_m;m^\pr,f_{m^\pr})\r]^{1/2}$ for $\Omega=6$; (c)
$\l[\Sigma_{11}(m,f_m;m^\pr,f_{m^\pr})\r]^{1/2}$ for $\Omega=10$; (d)
$\l[\Sigma_{22}(m,f_m;m^\pr,f_{m^\pr})\r]^{1/2}$ for $\Omega=10$. Results 
in the figure are for $f_m=f_m^{(p)}$  and $f_{m^\pr}=f_{m^\pr}^{(p)}$.  
See text for details.}
\label{corr}
\end{figure}

Results for cross-correlations in energy centroids 
$\Sigma_{11} (m, f_m; m^\pr, f_{m^\pr})$ and spectral variances
$\Sigma_{22} (m,f_m;m^\pr,f_{m^\pr})$ with $f_m=f_m^{(p)}$ as a function of
$m$ and $m^\pr$ are shown in Fig. \ref{corr} for both $\Omega=6$ and $10$.
It is seen that  $[\Sigma_{11}]^{1/2}$ and $[\Sigma_{22}]^{1/2}$ increase
almost linearly with $m$. At $m=4r$, $r=2,3,\ldots$ there is a slight dip in
$[\Sigma_{11}]^{1/2}$ as well as in $[ \Sigma_{22}]^{1/2}$. For $\Omega=6$
we have,  $[\Sigma_{11}]^{1/2}\sim 10-24$\% and $[ \Sigma_{22}]^{1/2}\sim
6-12$\%.  Similarly, for $\Omega=10$ these decrease to $5-16$\% for 
$[\Sigma_{11}]^{1/2}$ and $2-6$\% for $[\Sigma_{22}]^{1/2}$. The decrease in
$\Sigma$'s with increasing $\Omega$ is in agreement with the results
obtained for EGOE(2) for spinless fermions and EGOE(2)-$\cs$. Similarly, the
covariances in spectral variances are always smaller compared to those for
energy centroids. 

\begin{figure}[ht]
\centering
\includegraphics[width=5.5in,height=5.5in]{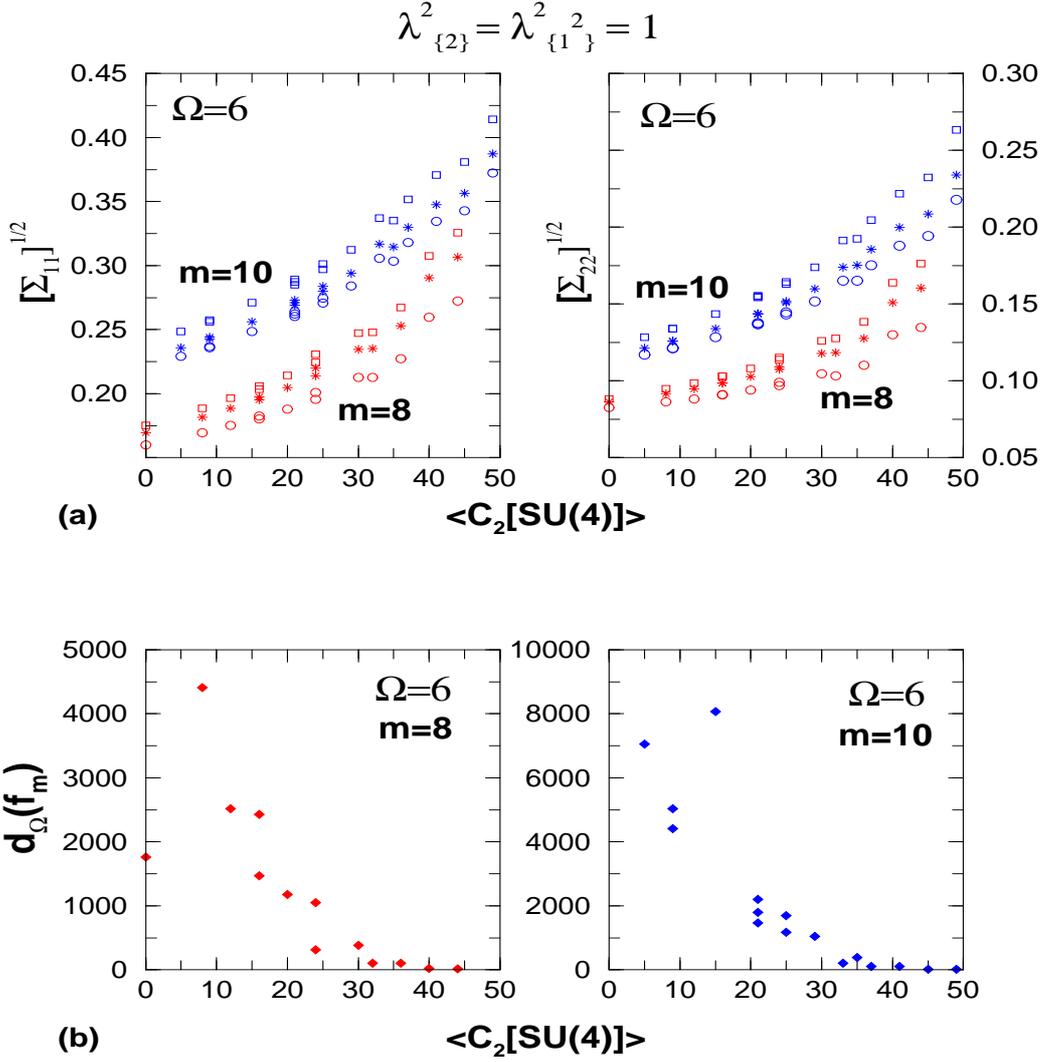} 
\caption{(a) Self and cross-correlations in energy centroids 
$\l[\Sigma_{11}(m,f_m;m^\pr,f_{m^\pr})\r]^{1/2}$ and spectral variances
$\l[\Sigma_{22}(m,f_m;m^\pr,f_{m^\pr})\r]^{1/2}$  as a function of $f_m$
and  $f_{m^\pr}$ (with fixed $m=m^\pr$). Results are shown for the first
(circle), second (star)  and fourth (square) lowest $U(\Omega)$ irreps
(ordered according to $\lan C_2[SU(4)]\ran^{\tilde{f}_m}$) with all other
irreps for $m=m^\pr=8$ (red) and  $10$ (blue) as a function of $\lan
C_2[SU(4)]\ran^{\tilde{f}_m}$. (b) Dimension $d_\Omega(f_m)$ for $m=8,\;10$
vs the  eigenvalue of $C_2[SU(4)]$ in the corresponding $SU(4)$ irrep. Note
that for a given value of the eigenvalue of $C_2[SU(4)]$ in some cases there
are more than one $f_m$ with the same eigenvalue. All results are for
$\Omega=6$.}
\label{covn6}
\end{figure}

Figures \ref{covn6}(a) and (b) show cross-correlations in energy centroids 
$\Sigma_{11}$ and spectral variances 
$\Sigma_{22}$ as a
function of $f_m$ and $f_{m^\pr}$ with fixed $m=m^\pr$. Results are shown
for the first, second and fourth lowest $U(\Omega)$ irreps, ordered
according to $\lan C_2[SU(4)] \ran^{\tilde{f}_m}$, with all other $f_m$'s
for $m=8$ and $10$ with $\Omega=6$. The correlations grow with increase in
$\lan C_2[SU(4)] \ran^{\tilde{f}_m}$.  It is important to note that there is
no correlation between variation in covariances with the variation in the
$f_m$ dimensions; see Figs. \ref{covn6}(a) and (b).

The increase in the cross-correlations with $m^\pr$ for fixed $f_m$ and
similar increase with $\lan C_2[SU(4)] \ran^{\tilde{f}_m}$ with fixed $m$,
seen from Figs. \ref{corr} and \ref{covn6}, could possibly be exploited in
deriving experimental signatures for cross-correlations. Note that the 
cross-correlations will be zero if we replace EGUE by GUE for $H_{f_m}(m)$ 
matrix.

\subsection{Results for $\lambda_{\{2\}}^2 \neq \lambda_{\{1^2\}}^2$}
\label{num2-3}

\begin{figure}[ht]
\centering
\includegraphics[width=5.5in,height=6in]{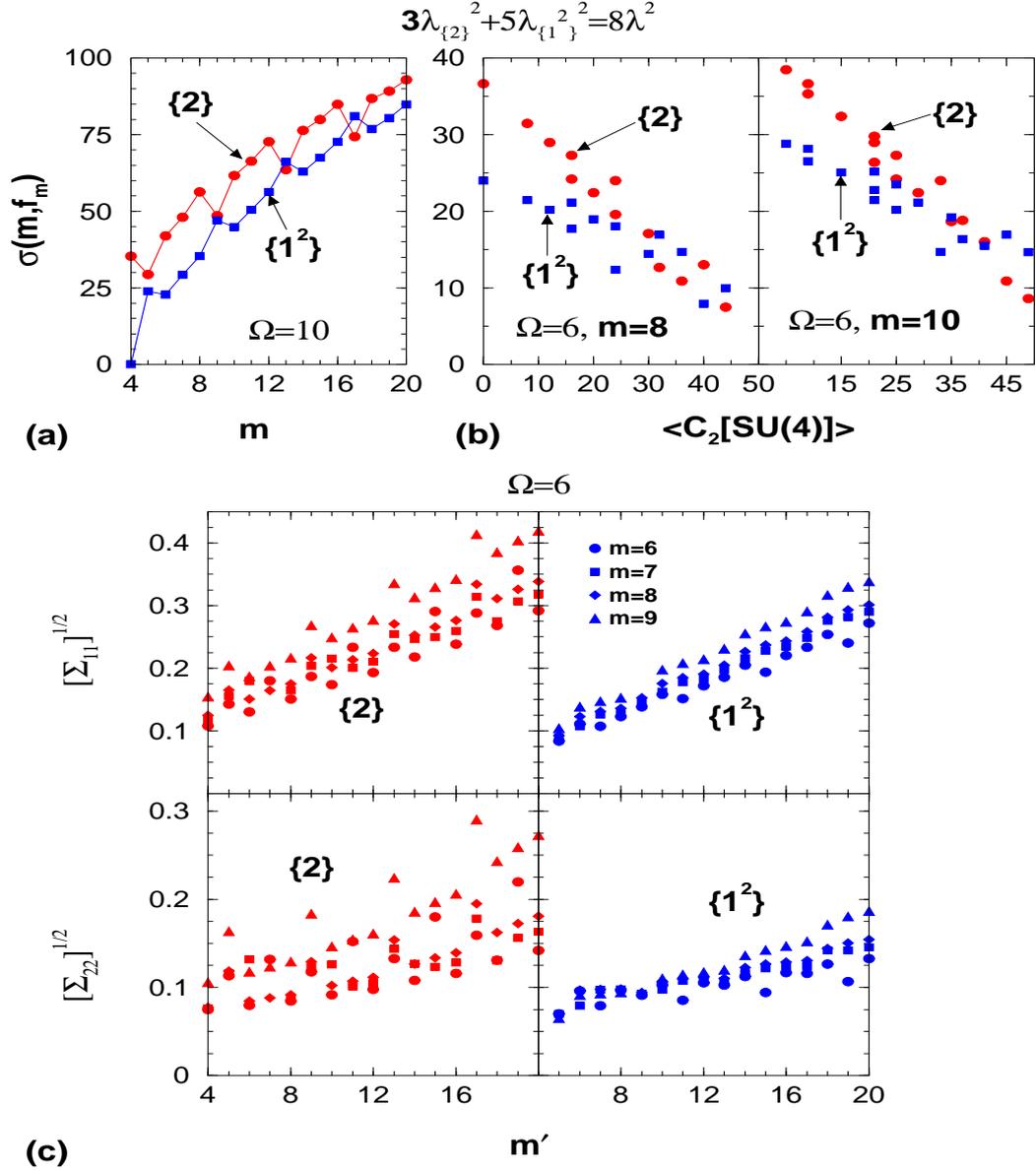} 
\caption{(a) Variation of spectral widths $\sigma(m,f_m)$ as a function of $m$
with  $f_m=f_m^{(p)}$. (b) Variation of spectral widths as a function of 
$\lan C_2[SU(4)]\ran^{\tilde{f}_m}$ for $m=8$ and $10$. In (c),   results
are shown for the covariances in energy centroids $\l[\Sigma_{11}\r]^{1/2}$
and  spectral variances $\l[\Sigma_{22}\r]^{1/2}$  for some values of $m$
and $m^\pr$ with $f_m=f_m^{(p)}$ and $f_{m^\pr}=f_{m^\pr}^{(p)}$. For the
calculations in (a), $\Omega=10$ and for (b) and (c), $\Omega=6$. Note that
in the figures  $\lambda_{\{2\}}^2=8/3$, $\lambda_{\{1^2\}}^2=0$ is denoted
as `$\{2\}$' and similarly  $\lambda_{\{2\}}^2=0$,
$\lambda_{\{1^2\}}^2=8/5$ is denoted as `$\{1^2\}$'. See text for
details.}
\label{uneq}
\end{figure}

All the discussion in the Secs. \ref{num1}, \ref{num2-1}, and \ref{num2-2} 
is restricted to $\lambda_{\{1^2\}}^2 =
\lambda_{\{2\}}^2$, i.e., for equal strengths for the symmetric and
anti-symmetric parts of the interaction. For completeness, we have studied
the variation of widths and covariances when $\lambda_{\{1^2\}}^2  \neq
\lambda_{\{2\}}^2$  by fixing the value for the ensemble averaged 
two-particle spectral variance $\sigma^2_{H(2)}(2)$ to a constant and then
varying  $\lambda_{\{2\}}$ (or equivalently $\lambda_{\{1^2\}}$).  The
two-particle spectral variance  for $\Omega >> 1$ is
$\sigma^2_{H(2)}(2)=\Omega^2[3\lambda_{\{2\}}^2 + 5 \lambda_{\{1^2\}}^2]
/16$. Therefore calculations are carried out with the constraint 
$[3\lambda_{\{2\}}^2 + 5 \lambda_{\{1^2\}}^2]=8\lambda^2$. All our previous
results correspond to $\lambda_{\{2\}}^2=\lambda_{\{1^2\}}^2=\lambda^2=1$. 
Now we will discuss some results for the extreme cases: (i) 
$\lambda_{\{1^2\}}^2=0$, $\lambda_{\{2\}}^2=8/3$ (denoted by $\{2\}$ in Fig.
\ref{uneq} and this corresponds to $H=H_{\{2\}}$) and (ii)
$\lambda_{\{1^2\}}^2=8/5$, $\lambda_{\{2\}}^2=0$ (denoted by $\{1^2\}$ in
Fig. \ref{uneq} and this corresponds to $H=H_{\{1^2\}}$).  Figure \ref{uneq}(a)
shows that the spectral widths have peaks at $m=4r$ and $m=4r+1$ for
$H_{\{2\}}$ and $H_{\{1^2\}}$, respectively. The peak for $H_{\{2\}}$ is much
larger and for $H_{\{1^2\}}$ it appears at a wrong place when compared to
the results shown in Fig. \ref{ch4-var} for  $H = H_{\{2\}} \oplus H_{\{1^2\}}$.
Similarly, it is seen from  Fig. \ref{uneq}(b) that the  variation in the
spectral widths $\sigma(m,f_m)= [\;\overline{\lan H^2\ran^{m,f_m}}\;]^{1/2}$
as a function of $f_m$ show more  fluctuations as compared to a good linear
behavior for $\lambda_{\{1^2\}}^2=\lambda_{\{2\}}^2$.  Figure \ref{uneq}(c)
shows  self and cross-correlations $\Sigma_{11} (m,f_m;m^\pr,f_{m^\pr})$ and
$\Sigma_{22} (m,f_m;m^\pr,f_{m^\pr})$ with $f_m=f_m^{(p)}$ as a function of
$m$ and $m^\pr$. Results for $H_{\{2\}}$  and $H_{\{1^2\}}$ show more
fluctuations and more importantly,  the magnitude of correlations for
$H_{\{2\}}$ is much larger and   for $H_{\{1^2\}}$ somewhat smaller compared
to the results for  $H = H_{\{2\}} \oplus H_{\{1^2\}}$. From this exercise,
we can conclude that the results for spectral widths and lower order
correlations will deviate strongly  from those reported in Secs.
\ref{num1} and \ref{num2} 
($\lambda_{\{1^2\}}^2=\lambda_{\{2\}}^2=\lambda^2$) when
$\lambda_{\{1^2\}}^2$ differs significantly from $\lambda_{\{2\}}^2$. 

\subsection{Conclusions}
\label{num2-s}

Increase in the magnitude of self-correlations in energy centroids and
spectral variances, defined by $\Sigma_{11}$ and $\Sigma_{22}$ and the
matrices $H_{f_m}(m)$ becoming more dense as we go from  EGUE(2) $\to$
EGUE(2)-$\cs$ $\to$ EGUE(2)-$SU(4)$ is an important result that deserves
more investigation. The cross-correlations increase with $m^\pr$ for fixed
$f_m$ and also with $\lan C_2[SU(4)] \ran^{\tilde{f}_m}$ with fixed $m$. For
$\lambda_{\{2\}}^2 \neq \lambda_{\{1^2\}}^2$, results for spectral widths
and lower order correlations will deviate strongly  from those with
$\lambda_{\{2\}}^2 = \lambda_{\{1^2\}}^2$ only when $\lambda_{\{1^2\}}^2$
differs significantly from $\lambda_{\{2\}}^2$. 

\section{Summary}
\label{cafo}

We have introduced  in this chapter a new embedded ensemble, EGUE(2)-$SU(4)$,
and it is defined for two-body Hamiltonians preserving $SU(4)$ symmetry for
a system of $m$ fermions in $\Omega$ number of levels each four-fold
degenerate. We have developed, for this ensemble, an analytical formulation
based on the Wigner-Racah algebra of the embedding $U(\Omega) \otimes SU(4)$
algebra.  Explicit formulas are derived for spectral variances and
covariances in energy centroids and spectral variances  for $U(\Omega)$
irreps of the type $f_m^{(p)}=\{4^r,p\}$, $p=0$, $1$, $2$ and $3$. Results 
in  Tables \ref{pf2}, \ref{tabw}, \ref{tab2} and \ref{tabr} allow one to
calculate these for any $m$ and $\Omega$. For general $U(\Omega)$ irreps
$f_m$, the analytical formulation in Secs. \ref{base}-\ref{anres} and the
formulas in the Tables \ref{tab1} and \ref{tab22} (obtained by simplifying
the tabulations due to Hecht \cite{He-74a}), allows one to carry out
numerical calculations and codes for the same are developed. The analytical
formulas in the Tables led to simple expressions for the covariances in
energy centroids and spectral variances in the dilute limit for the irreps
$f_m^{(p)}$. Using the formulation in Secs. \ref{base}-\ref{anres} and
the results in Tables \ref{pf2}-\ref{tabr},
several numerical calculations are carried out and the results are 
presented in Secs. \ref{num1} and \ref{num2} and in Figs.
\ref{ch4-var}-\ref{uneq}. Main conclusions from these are as follows:

\begin{description}

\item[(i)] Expectation values $\lan C_2[SU(4)] \ran^{E}$ studied in Section
\ref{num1-2} by constructing Gaussian partial densities with centroids given by 
$\lan C_2[SU(4)] \ran^{\tilde{f}_m}$ and variances given by $\overline{ \lan H^2
\ran^{m,f_m}}$ and similarly,  the four periodicity in the  gs energies studied
in Section \ref{num1-3}, establish that random interactions with $SU(4)$
symmetry keep intact the essential  features of the Majorana force. This
conclusion is quite similar to the result derived for EGOE(2)-$J$ (also called
TBRE some times),  the embedded ensemble with angular-momentum $J$ symmetry.
This ensemble is generated by (see
also Sec. \ref{c7s4}) random interactions that are $J$ scalar [$SO(3)$
scalar] and it is found that, for systems with even number of fermions, there is
$J^\pi=0^+$ preponderance in the ground states. This feature has been
investigated in many different ways  \cite{Zh-04,Zh-04a,Zel-04,Pa-04}. It should
be noted that the $SO(3)$ invariant operator is $J^2$ and it gives (with
$H=J^2$) $J=0$ as gs, a property  generated also by random interactions.

\item[(ii)] As shown in Section \ref{num2-1}, there is increase in the
magnitude of self-correlations in energy centroids and spectral variances,
defined by $\Sigma_{11}$ and $\Sigma_{22}$ in direct correlation with the
$H_{f_m}(m)$  matrices becoming more dense (implying stronger mixing)  as we
go from  EGUE(2) $\to$ EGUE(2)-$\cs$ $\to$ EGUE(2)-$SU(4)$. Further
investigation of this feature may provide additional justification for the
recent claim by Papenbrock and Weidenm\"{u}ller \cite{Pa-05} that symmetries
are responsible for chaos in nuclear shell-model spaces. 

\item[(iii)]  As shown in Section \ref{num2-2}, there is a significant
increase in cross-correlations with particle number  $m$ for a fixed
$U(\Omega)$ irrep $f_m$ and similarly with  $\lan C_2[SU(4)]
\ran^{\tilde{f}_m}$ for fixed $m$. This could be used as a  signature for
experimental detection of cross-correlations  generated by EGUE(2)-$SU(4)$. 

\end{description}

Finally, we conclude that the results presented in the present chapter represent
a  first detailed analytical study of an embedded ensemble with a non-trivial
symmetry that is relevant in nuclear structure.  

\chapter{EGOE(1+2)-$\pi$: Density of States and Parity Ratios}
\label{ch5}

\section{Introduction}
\label{c5s1}

Parity is an important symmetry for finite quantum systems such as nuclei and 
atoms. In this chapter, we consider EGOE that includes parity explicitly and
address three important questions related to parity in
nuclear structure. These are as follows:

\begin{description}

\item{(i)} Parity ratios of nuclear level densities is an important ingredient
in nuclear astrophysical applications. Recently, a method based on
non-interacting Fermi-gas model for proton-neutron systems has been developed
and the parity $(\pi)$ ratios as a function of excitation energy in large number
of nuclei of astrophysical interest have been tabulated \cite{Mo-07}. The method
is based on the assumption that the probability to occupy $s$ out of $N$ given 
sp states follow Poisson distribution in the dilute limit $(m
<< N, N\to \infty$ where $m$ is the number of particles). Then the ratio of the
partition functions for the $+$ve and $-$ve parity states is given by the simple
formula $Z_-/Z_+ = \tanh f$, where $f$ is average number of particles in the
$+$ve parity states. Starting with this, an iterative method is developed with
inputs from the  Fermi-Dirac distribution for occupancies including pairing
effects and the Fermi-gas form for the total level density. In the examples
studied in \cite{Mo-07}, parity ratios are found to equilibrate only around
$5-10$ MeV excitation energy. However, ab-initio interacting particle theory for
parity ratios is not yet available. 

\item{(ii)} A closely related question is about the form of the density of
states defined over spaces with fixed-$\pi$. In general, fixed-$\pi$ density of
states can be written as a sum of appropriate partial densities.  In the
situation that the form of the partial densities is determined by a few
parameters (as it is with a Gaussian or a Gaussian with one or two corrections),
it is possible to derive a theory for these parameters and using these, one can
construct fixed-$\pi$ density of states and calculate parity ratios. Such a
theory with interactions in general follows from random matrix theory
\cite{KH-10}. 

\item{(iii)} There is the important recognition in the past few years that
random interactions generate regular  structures
\cite{Zel-04,Zh-04a,Pa-07,Ho-10}. It was shown in \cite{Zh-04} that shell-model
for even-even nuclei gives  preponderance of  $+$ve parity ground states. A
parameter-free EGOE with parity has been  defined and analyzed recently by
Papenbrock and Weidenm\"{u}ller \cite{Pa-08} to address the question of
`preponderance of ground states with positive parity' for systems with even
number of fermions. They show that in the dilute limit, $+$ve parity ground
states appear with only 50\% probability. Thus, a random matrix theory
describing shell-model results is not yet available.

\end{description}

With the success of the embedded random matrix ensembles, one can argue
that the EE generated by parity preserving random interaction  may provide
generic results for the three nuclear structure quantities mentioned above. For
nuclei, the GOE versions of EE are relevant. Then, with a chaos producing
two-body interaction preserving parity in the presence of a mean-field, we have
embedded Gaussian orthogonal ensemble of one plus two-body interactions with
parity [hereafter called  EGOE(1+2)-$\pi$]. This model contains two mixing
parameters and a gap between the  $+$ve and $-$ve parity sp states and it goes
much beyond the simpler model considered in \cite{Pa-08}. In the random matrix
model used in the present chapter, proton-neutron degrees of freedom and 
angular momentum $(J)$ are not considered. Let us add that in the present
chapter for the first time a random matrix theory for parity ratios is
attempted. All the results presented in this chapter are published in
\cite{Ma-11a}. 

\section{EGOE(1+2)-$\pi$ Ensemble}
\label{c5s2}

\begin{figure}[htp]
\centering
\includegraphics[width=4in,height=2.5in]{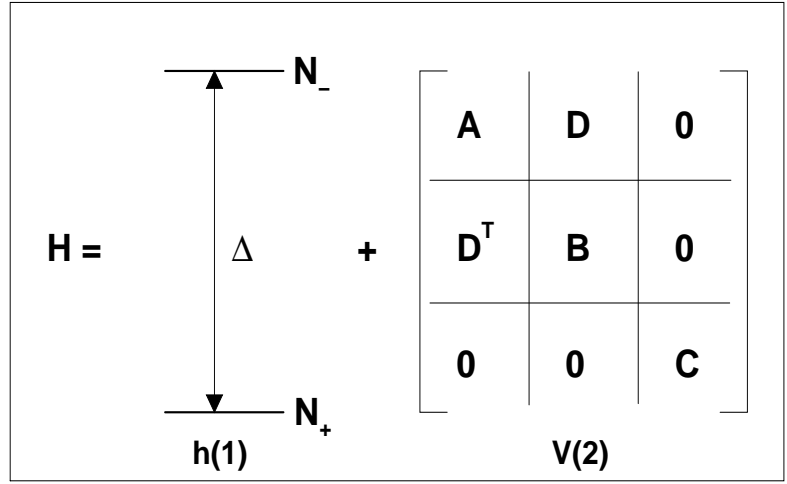}
\caption{Parity preserving one plus two-body $H$ with a sp spectrum 
defining $h(1)$ along with a schematic form of the $V(2)$ matrix. 
Dimension of the matrices A, B and C are $N_+(N_+-1)/2$, $N_-(N_--1)/2$, 
and $N_+N_-$, respectively. Note that $D^T$ is the transpose of the matrix 
$D$. See text for details.}
\label{fig1}
\end{figure}

Given $N_+$ number of positive parity sp states and  similarly $N_-$ number
of negative parity sp states, let us assume, for simplicity, that the $+$ve
and $-$ve parity states are degenerate and separated by energy $\Delta$; see
Fig. \ref{fig1}. This defines the one-body part $h(1)$ of the Hamiltonian
$H$ with $N=N_+ + N_-$ sp states. The matrix for the two-body part $V(2)$ of
$H$ [we assume $H$ is (1+2)-body] will be a $3 \times 3$ block matrix in 
two-particle spaces as there are three possible ways to generate two-particle
states with definite parity:  (i) both fermions  in $+$ve parity states;
(ii) both fermions in $-$ve parity states; (iii) one fermion  in $+$ve and
other fermion in $-$ve parity states. They will give the matrices $A$, $B$,
and $C$, respectively in Fig. \ref{fig1}. For parity  preserving interactions
only the states (i) and (ii) will be mixed and mixing matrix is $D$ in Fig.
\ref{fig1}. Note that the matrices $A$, $B$ and $C$ are symmetric square
matrices while $D$ is in  general a rectangular mixing matrix. Consider $N$
sp states arranged such that the states $1$ to $N_+$ have $+$ve parity and 
states $N_++1$ to $N$ have $-$ve parity. Then the operator form of $H$ 
preserving parity is,
\be
\barr{rcl}
H & = & h(1) + V(2)\;; \\ \\
h(1) & = & \dis\sum_{i=1}^{N_+} \epsilon_i^{(+)} \hat{n}_i^{(+)} + 
\dis\sum_{i=N_++1}^{N} \epsilon_i^{(-)} \hat{n}_i^{(-)} \;;\;\;\;\; 
\epsilon_i^{(+)}=0 \;,\;\; \epsilon_i^{(-)}=\Delta\;, \\ \\
V(2) & = & 
\dis\sum_{\barr{c}
i,j,k,l=1 \\
(i<j,\; k<\ell) \earr}^{N_+} 
\lan \nu_k \; \nu_\ell \mid V \mid \nu_i \; \nu_j
\ran a^\dagger_k \; a^\dagger_\ell \; a_j \; a_i \\ \\
& + & 
\dis\sum_{\barr{c} 
i^\pr,j^\pr,k^\pr,\ell^\pr=N_++1 \\
(i^\pr<j^\pr,\; k^\pr<\ell^\pr)
\earr}^N 
\lan \nu_{k^\pr} \; \nu_{\ell^\pr} \mid V \mid \nu_{i^\pr} \; \nu_{j^\pr}
\ran a^\dagger_{k^\pr} \; a^\dagger_{\ell^\pr} \; a_{j^\pr} \; a_{i^\pr}
\\ \\
& + & 
\dis\sum_{
i^{\pr\pr},k^{\pr\pr}=1}^{N_+}\;\; 
\dis\sum_{j^{\pr\pr},\ell^{\pr\pr}=N_++1}^N
\lan \nu_{k^{\pr\pr}} \; \nu_{\ell^{\pr\pr}} \mid V \mid \nu_{i^{\pr\pr}} 
\; \nu_{j^{\pr\pr}} \ran a^\dagger_{k^{\pr\pr}} \; a^\dagger_{\ell^{\pr\pr}}
 \; a_{j^{\pr\pr}} \; a_{i^{\pr\pr}}
\earr \label{eq.ham}
\ee
\be
\barr{rcl}
& + &
\dis\sum_{\barr{c} P,Q=1 \\ (P<Q) \earr}^{N_+}\;\;
\dis\sum_{\barr{c} R,S=N_++1 \\ (R<S) \earr}^N
\l[ \lan \nu_P \; \nu_Q \mid V \mid \nu_R \; \nu_S \ran a^\dagger_P \;
a^\dagger_Q \; a_S \; a_R + \mbox{h.c.} \r]\;. \nonumber
\earr \label{eq.hama1}
\ee
In Eq. (\ref{eq.ham}), $\nu_i$'s are sp states with $i=1,2,\ldots,N$ (the
first $N_+$ states are $+$ve parity and remaining $-$ve parity). Similarly,
$\lan \ldots \mid V \mid \ldots \ran$ are the two-particle matrix elements,
$\hat{n}_i$ are number operators and $a^\dagger_i$ and $a_i$ are creation
and annihilation operators, respectively. Note that the four terms in the RHS
of the expression for $V(2)$ in  Eq.  (\ref{eq.ham}) correspond, respectively,
to the matrices $A$, $B$, $C$ and $D$ shown in Fig. \ref{fig1}.

Many-particle states for $m$ fermions in the $N$ sp states can be obtained
by distributing $m_1$ fermions in the $+$ve parity sp states ($N_+$ in
number)  and similarly, $m_2$ fermions in the $-$ve parity sp states ($N_-$
in number) with $m=m_1+m_2$. Let us denote each distribution of $m_1$
fermions in $N_+$ sp states by $\bf{m}_1$ and similarly, $\bf{m}_2$ for
$m_2$ fermions in  $N_-$ sp states. Many-particle basis defined by
$(\bf{m}_1, \bf{m}_2)$ with $m_2$ even will form the basis for $+$ve parity
states and similarly, with $m_2$ odd  for $-$ve parity states. In the
$(\bf{m}_1, \bf{m}_2)$ basis with $m_2$ even (or odd), the $H$ matrix
construction reduces to the matrix construction for spinless  fermion
systems. The method of construction for spinless fermion systems is well
known (see Chapter \ref{ch1}) 
and therefore it is easy to construct the many-particle 
$H$ matrices in $+$ve and $-$ve parity spaces. The matrix dimensions $d_+$
for $+$ve parity and $d_-$ for $-$ve parity spaces are given by,
\be
d_+ = \dis\sum_{m_1,m_2\;(m_2\;even)} 
\dis\binom{N_+}{m_1} \dis\binom{N_-}{m_2}
\;,\;\;\;\; 
d_- = \dis\sum_{m_1,m_2\;(m_2\;odd)} 
\dis\binom{N_+}{m_1} \dis\binom{N_-}{m_2}
\;.
\label{eq.dim}
\ee
Some examples for the dimensions $d_+$ and $d_-$ are given in Table
\ref{dim}.

\begin{table}[htp]
\caption{Hamiltonian matrix dimensions $d_+$ and $d_-$ for various 
values of $(N_+,N_-,m)$.}
\begin{center}
{
\begin{tabular}{cccccccccc}
\toprule
$N_+$ & $N_-$ & $m$ & $d_+$ & $d_-$ & $N_+$ & $N_-$ & $m$ & $d_+$ & $d_-$ 
\\ 
\midrule
$6$ & $6$ & $6$ & $452$ & $472$ & $8$ & $8$ & $4$ & $924$ & $896$ \\
$7$ & $5$ & $6$ & $462$ & $462$ & $$ & $$ & $5$ & $2184$ & $2184$ \\
$7$ & $7$ & $5$ & $1001$ & $1001$ & $$ & $$ & $6$ & $3976$ & $4032$ \\
$$ & $$ & $6$ & $1484$ & $1519$ & $10$ & $6$ & $4$ & $900$ & $920$ \\
$$ & $$ & $7$ & $1716$ & $1716$ & $$ & $$ & $5$ & $2202$ & $2166$ \\
$8$ & $6$ & $5$ & $1016$ & $986$ & $$ & $$ & $6$ & $4036$ & $3972$ \\
$$ & $$ & $6$ & $1499$ & $1504$ & $6$ & $10$ & $4$ & $900$ & $920$ \\
$9$ & $5$ & $5$ & $1011$ & $911$ & $$ & $$ & $5$ & $2166$ & $2202$ \\
$$ & $$ & $6$ & $1524$ & $1479$ & $$ & $$ & $6$ & $4036$ & $3972$ \\
$5$ & $10$ & $4$ & $665$ & $700$ & $9$ & $9$ & $6$ & $9240$ & $9324$ \\
$$ & $$ & $5$ & $1501$ & $1502$ & $10$ & $8$ & $6$ & $9268$ & $9296$ \\
$$ & $$ & $$ & $$ & $$ & $10$ & $10$ & $5$ & $7752$ & $7752$ \\
$$ & $$ & $$ & $$ & $$ & $$ & $$ & $6$ & $19320$ & $19440$ \\
\bottomrule
\end{tabular}}
\end{center}
\label{dim}
\end{table}

The EGOE(1+2)-$\pi$ ensemble is defined by choosing the matrices $A$, $B$
and  $C$ to be independent GOE's with matrix elements variances $v_a^2$,
$v_b^2$, and $v_c^2$, respectively. Similarly the matrix elements of the
mixing $D$ matrix are chosen to be independent (independent of $A$, $B$ and
$C$ matrix elements) zero centered  Gaussian variables with variance
$v_d^2$. Without loss of generality we choose $\Delta=1$ so that all the
$v$'s are in $\Delta$ units. This general EGOE(1+2)-$\pi$ model will have
too many parameters ($v_a^2,v_b^2,v_c^2,v_d^2,N_+,N_-,m$) and therefore it
is necessary to reduce the number of parameters. A numerically tractable and
physically relevant (as discussed ahead) restriction is to choose the  
matrix elements variances of the diagonal blocks $A$, $B$ and $C$ to be same
and then  we have the EGOE(1+2)-$\pi$ model defined by ($N_+,N_-,m$) and the
variance parameters ($\tau$,$\alpha$) where
\be
\dis\frac{v_a^2}{\Delta^2} = \dis\frac{v_b^2}{\Delta^2} =
\dis\frac{v_c^2}{\Delta^2} = \tau^2 \;,\;\;\;\;
\dis\frac{v_d^2}{\Delta^2} = \alpha^2 \;. 
\label{eq.taual}
\ee
Thus EGOE(1+2)-$\pi$ we employ is 
\be
\barr{l}
A:\; \mbox{GOE}(0:\tau^2)\;, \;B:\; \mbox{GOE}(0:\tau^2)\;, \;C:\; 
\mbox{GOE}(0:\tau^2)\;,\;D:\;\mbox{GOE}(0:\alpha^2)\;;\\
A,\;B,\;C,\;D \mbox{\;are\;independent\;GOE's}\;.
\earr\label{eq.model}
\ee
Note that the $D$ matrix is a GOE only in the sense that the matrix elements
$D_{ij}$ are all independent zero centered Gaussian variables with variance
$\alpha^2$. In the limit $\tau^2 \rightarrow \infty$ and $\alpha = \tau$,
the model defined by Eqs. (\ref{eq.ham}), (\ref{eq.taual}) and
(\ref{eq.model})  reduces to the simpler model analyzed in \cite{Pa-08}.

Starting with the EGOE(1+2)-$\pi$ ensemble defined by Eqs. (\ref{eq.ham}),
(\ref{eq.taual}) and (\ref{eq.model}), we have numerically constructed the
ensemble in many-particle $+$ve and  $-$ve parity spaces with dimensions $d_+$
and $d_-$ given by Eq. (\ref{eq.dim}) for several values of ($N_+,N_-,m$) and
varying the parameters $\tau$ and $\alpha$. Before discussing the results of the
numerical calculations, we present the results for the energy centroids,
variances and also the shape parameters (skewness and excess) defining the
normalized fixed-$(m_1,m_2)$  partial densities $\rho^{m_1,m_2}(E) = \lan \delta
(H - E) \ran^{m_1,m_2}$. These will allow us to understand some of the numerical
results. Let us add that the fixed-$\pi$ eigenvalue densities $I_\pm(E)$ are sum
of the appropriate partial densities  as given by Eq. (\ref{eq.densty}) ahead. 
Note that the densities $I_\pm(E)$ are
normalized to $d_\pm$.

\section{Energy Centroids, Variances, Skewness and Excess Parameters 
for Fixed-$(m_1,m_2)$ Partial Densities}
\label{c5s3}

Let us call the set of $+$ve parity sp states as  unitary orbit \#1 and
similarly the set of $-$ve parity sp states as  unitary orbit \#2; 
unitary orbits notation and their significance was discussed in \cite{KH-10}. 
For convenience,
from now on, we denote the sp states by the roman letters $(i,j,\ldots)$ and
unitary orbits by greek letters $(\alpha,\beta,\ldots)$. Note that
$\alpha=1$ corresponds to the $+$ve parity unitary orbit and $\alpha=2$
corresponds to the $-$ve parity unitary orbit (with this notation, $N_1=N_+$
and $N_2=N_-$). The sp states that belong to a unitary orbit $\alpha$ are
denoted as $i_\alpha,j_\alpha,\ldots$. Propagation formulas for the energy
centroids and variances of the partial densities $\rho^{m_1,m_2}(E)$ follow
from the unitary decomposition of $V(2)$ with respect to the sub-algebra
$U(N_+) \oplus U(N_-)$ contained in $U(N)$. The operator $V(2)$ decomposes
into three parts $V(2) \to V^{[0]} + V^{[1]} +V^{[2]}$. The $V^{[0]}$
generates the energy centroids $\lan V \ran^{m_1,m_2}$, $V^{[1]}$
corresponds to the `algebraic' mean-field generated by $V$ and  $V^{[2]}$ is
the remaining irreducible two-body part. Extending the  unitary
decomposition for the situation with a single orbit for spinless fermions
(see Appendix \ref{c2a2}) and also using the detailed formulation given in 
\cite{Ch-71}, we
obtain the following formulas for the $V^{[\nu]}$'s. The $V^{[0]}$ is given
by (with $\alpha = 1,\;2$ and $\beta = 1,\;2$)
\be
\barr{rcl}
V^{[0]} & = & 
\dis\sum_{\alpha \geq \beta} 
\dis\frac{\hat{n}_\alpha (\hat{n}_\beta -
\delta_{\alpha\beta})}{(1+\delta_{\alpha\beta})}\; V_{\alpha\beta}\;;
\\ \\
V_{\alpha\alpha} & = &
\dis\binom{N_\alpha}{2}^{-1}\;\dis\sum_{i>j}V_{i_\alpha j_\alpha i_\alpha
j_\alpha}\;,\\ \\
V_{\alpha\beta} & = &
\l(N_\alpha N_\beta\r)^{-1}\;\dis\sum_{i,j}V_{i_\alpha j_\beta i_\alpha
j_\beta}\;;\;\;\;\;\alpha \neq \beta\;.
\earr \label{eq.nu0}
\ee
Then the traceless part $\widetilde{V}$ is given by
\be
\barr{l}
\widetilde{V} = V -V^{[0]} = V^{[1]} + V^{[2]} \;;\\ \\
\l( \widetilde{V}\r)_{i_\alpha j_\beta i_\alpha j_\beta} = 
V_{i_\alpha j_\beta i_\alpha j_\beta} - V_{\alpha\beta} \;,\\ \\
\l( \widetilde{V}\r)_{ijk\ell} = V_{ijk\ell} \;\;\;\; 
\mbox{\;\;for\;all\;others}\;.
\earr \label{eq.notr}
\ee
Now the $V^{[1]}$ part is
\be
\barr{l}
V^{[1]} = \dis\sum_{i_\alpha, j_\alpha} 
\widehat{\xi}_{i_\alpha j_\alpha}
a^\dagger_{i_\alpha}\;a_{j_\alpha}\;;\\ \\
\widehat{\xi}_{i_\alpha j_\alpha} = 
\dis\sum_{\beta} \dis\frac{\hat{n}_\beta -
\delta_{\alpha\beta}}{N_\beta - 2\delta_{\alpha\beta}} \; \zeta_{i_\alpha
j_\alpha} (\beta)\;,\;\;\;\; 
\zeta_{i_\alpha j_\alpha} (\beta) = 
\dis\sum_{k_\beta} \widetilde{V}_{k_\beta i_\alpha k_\beta
j_\alpha}\;.
\earr \label{eq.nu1}
\ee
It is important to stress that, with spherical ($j$) 
orbits and no radial degeneracy 
(as used in many nuclear structure studies), 
$V^{[1]}$ part will not exist. Finally, the $V^{[2]}$ part 
is as follows,
\be
\barr{rcl}
V^{[2]} & = & \widetilde{V} - V^{[1]}\;; \\ \\
{V}^{[2]}_{i_\alpha j_\beta i_\alpha j_\beta} & = &
\widetilde{V}_{i_\alpha j_\beta i_\alpha j_\beta} - \l[ 
\dis\frac{\zeta_{i_\alpha j_\alpha} (\beta)}
{N_\beta - 2\delta_{\alpha\beta}}
+ \dis\frac{\zeta_{i_\beta j_\beta} (\alpha)}
{N_\alpha - 2\delta_{\alpha\beta}}
 \r] \;,\\ \\
{V}^{[2]}_{k_\alpha i_\beta k_\alpha j_\beta} & = &
\widetilde{V}_{k_\alpha i_\beta k_\alpha j_\beta} - 
\dis\frac{\zeta_{i_\beta j_\beta} (\alpha)}
{N_\alpha - 2\delta_{\alpha\beta}}
\;;\;\;\;\; i_\beta \neq j_\beta \;,\\ \\
{V}^{[2]}_{i j k \ell} & = &
\widetilde{V}_{i j k \ell} \mbox{\;\;for\;all\;others}\;.
\earr \label{eq.nu2}
\ee
Given the $U(N) \supset U(N_+) \oplus U(N_-)$ unitary (tensorial) 
decomposition, by intuition and using Eq. (\ref{ud.eq.a3}), 
it is possible to write the
propagation formulas for the energy centroids and variances of
$\rho^{m_1,m_2}(E)$. Note
that these are essentially traces of $H$ and $H^2$ over the space defined by
the two-orbit configurations $(m_1,m_2)$; see Eqs.  (\ref{eq.cent}) and
(\ref{eq.var}) ahead. A direct approach to write the propagation formulas
for centroids and variances for a multi-orbit configuration was given in
detail  first by French and Ratcliff \cite{Fr-71}. The formula for the
variance given in \cite{Fr-71} is cumbersome  and it is realized later
\cite{Ch-71} that they can be made compact by applying group theory (see also
\cite{Ko-01,Wo-86,KH-10}).   We have adopted the group theoretical approach
for the two-orbit averages and obtained formulas. Propagation formula for
the fixed-$(m_1,m_2)$ energy centroids is,
\be
E_c(m_1,m_2) = \lan H \ran^{m_1,m_2} = m_2\;\Delta + 
\dis\sum_{\alpha \geq \beta} 
\dis\frac{m_\alpha (m_\beta -
\delta_{\alpha\beta})}{(1+\delta_{\alpha\beta})}\; V_{\alpha\beta} \;.
\label{eq.cent}
\ee
First term in Eq. (\ref{eq.cent})  is generated by $h(1)$ and is simple
because of the choice of the sp energies as shown in Fig. \ref{fig1}.
Propagation formula for fixed-$(m_1,m_2)$ variances is,
\be
\barr{l}
\sigma^2(m_1,m_2) =  
\lan H^2 \ran^{m_1,m_2} - \l[ \lan H \ran^{m_1,m_2}\r]^2 \\ \\ 
= 
\dis\sum_\alpha \dis\frac{m_\alpha\l( N_\alpha -
m_\alpha \r)}{N_\alpha \l( N_\alpha -1 \r)} \dis\sum_{i_\alpha, j_\alpha}
\l[{\xi}_{i_\alpha, j_\alpha}(m_1,m_2)\r]^2 
\earr \label{eq.var}
\ee
\be
+ \dis\sum^\pr_{\alpha,\beta,\gamma,\delta} 
\dis\frac{m_\alpha(m_\beta-\delta_{\alpha\beta})(N_\gamma -
m_\gamma)(N_\delta-m_\delta-\delta_{\gamma\delta})}
{N_\alpha(N_\beta - \delta_{\alpha\beta})
(N_\gamma - \delta_{\gamma\alpha} - \delta_{\gamma\beta})
(N_\delta - \delta_{\delta\alpha} - \delta_{\delta\beta} 
- \delta_{\delta\gamma})} \; (X) \;; \nonumber
\label{eq.vara1}
\ee
\be
{\xi}_{i_\alpha, j_\alpha}(m_1,m_2) = 
\dis\sum_{\beta} \dis\frac{m_\beta -
\delta_{\alpha\beta}}{N_\beta - 2\delta_{\alpha\beta}} \; \zeta_{i_\alpha
j_\alpha} (\beta)\;, \;\;\;\;
X = \dis\sum^\pr \l( V^{[2]}_{i_\alpha
j_\beta k_\gamma \ell_\delta} \r)^2 \;. \nonumber
\label{eq.vara2}
\ee
The `prime' over summations in Eq. (\ref{eq.var}) implies that the
summations are not free sums. Note that $(\alpha,\beta,\gamma,\delta)$ take
values $(1,1,1,1)$, $(2,2,2,2)$, $(1,2,1,2)$, $(1,1,2,2)$ and $(2,2,1,1)$.
Similarly, in the sum over $(i_\alpha,j_\beta)$, $i \leq j$ if $\alpha=\beta$
and otherwise the sum is over all $i$ and $j$. Similarly, for 
$(k_\gamma, \ell_\delta)$. Using $E_c(m_1,m_2)$ and $\sigma^2(m_1,m_2)$, 
the  fixed-parity energy centroids and spectral  variances [they define
$I_\pm(E)$] can be obtained as follows,
\be
\barr{l}
E_c(m,\pm) = \lan H \ran^{m,\pm} = \dis\frac{1}{d_\pm} \dis\sum_{
m_1,m_2}^\pr d(m_1,m_2) E_c(m_1,m_2)\;, \\ \\
\sigma^2(m,\pm) = \lan H^2 \ran^{m,\pm} - \l[ \lan H \ran^{m,\pm}\r]^2 
\;;\\ \\
\lan H^2 \ran^{m,\pm} = \dis\frac{1}{d_\pm} \dis\sum_{m_1,m_2}^\pr
d(m_1,m_2) \l[ \sigma^2(m_1,m_2) + E_c^2(m_1,m_2) \r]\;. 
\earr \label{eq.cenvar}
\ee
The `prime' over summations in Eq. (\ref{eq.cenvar}) implies that
$m_2$ is even(odd) for $+$ve($-$ve) parity.

It should be pointed out that the  formulas given by Eqs. (\ref{eq.cent}),
(\ref{eq.var}) and  (\ref{eq.cenvar}) are compact and easy to understand
compared to Eqs. (10)-(14) of \cite{Pa-08} and also those that follow from
Eqs. (129) and (133) of \cite{Fr-71} where unitary  decomposition is not
employed. We have verified Eqs. (\ref{eq.cent}) and (\ref{eq.var})  by
explicit construction of the $H$ matrices in many examples.  In principle,
it is possible to obtain a formula for the ensemble averaged variances
using  Eq. (\ref{eq.var}); the  ensemble averaged centroids derive only from
$h(1)$. Simple asymptotic formulas for ensemble  averaged variances follow
by neglecting the $\delta$-functions that appear in Eq. (\ref{eq.var}) and
replacing $(V^{[2]}_{i j k \ell})^2$ by $\tau^2$ and $\alpha^2$
appropriately. Then the final formula for the ensemble averaged
fixed-$(m_1,m_2)$ variances is,
\be
\barr{rcl}
\overline{\sigma^2(m_1,m_2)} &\approx& m \l[ \dis\sum_{\alpha=1}^2 
m_\alpha \l( N_\alpha - m_\alpha\r)  \r] \; \tau^2
+ \l[ \dis\binom{m_1}{2} \dis\binom{\wmn}{2} + 
\dis\binom{m_2}{2} \dis\binom{\wmp}{2} \r]\;\alpha^2 \\ \\
&+& \l[ \dis\binom{m_1}{2} \dis\binom{\wmp}{2} + 
\dis\binom{m_2}{2} \dis\binom{\wmn}{2} +
m_1 m_2 \wmp \wmn \r] \;\tau^2\;.
\earr \label{eq.eavar}
\ee
Here, $\wmp = N_1 - m_1$ and $\wmn = N_2 - m_2$. In Table \ref{widths}, we
compare the results obtained from Eq. (\ref{eq.eavar}) with those obtained
for various  100 member ensembles using Eq. (\ref{eq.var}) and the
agreements are quite good. Therefore, in many practical applications, one
can use Eq. (\ref{eq.eavar}). 

\begin{table}[htp]
\caption{Ensemble averaged  fixed-$(m_1,m_2)$ widths $\sigma(m_1,m_2)$  and
the total spectral width $\sigma_t$ for different $(\tau,\alpha)$ values.
For each $(\tau,\alpha)$, the $\sigma(m_1,m_2)$ are given  in the table and
they are obtained using the exact propagation formula Eq. (\ref{eq.var}) for
each member of the  ensemble. In all the calculations, 100 member ensembles
are employed.  Numbers in the bracket are obtained by using the asymptotic
formula given in Eq. (\ref{eq.eavar}). Last row for each $(N_+,N_-)$ gives
the corresponding $\sigma_t$  values. All the results are given for 6
particle systems and the dimensions  $d(m_1,m_2)$ are also given in the
table. See text for details.}
\begin{center}
{
\begin{tabular}{cccccccc}
\toprule
 & & & & \multicolumn{4}{c}{$(\tau,\alpha/\tau)$} \\ 
\cmidrule{5-8}
$(N_+,N_-)$ & $m_1$ & $m_2$ & $d(m_1,m_2)$ &
$(0.1,0.5)$ & $(0.1,1.5)$ & $(0.2,0.5)$ & $(0.2,1.5)$ \\ 
\midrule
$(8,8)$ &  0 & 6  &    28  & 1.36(1.39) & 3.21(3.21)  & 2.73(2.77) 
& 6.41(6.42)  \\
        &  1 & 5  &   448  & 1.76(1.79) & 2.70(2.72)  & 3.52(3.57) 
	& 5.41(5.44)  \\
	&  2 & 4  &  1960  & 2.05(2.09) & 2.48(2.50)  & 4.11(4.17) 
	& 4.96(5.01)  \\
	&  3 & 3  &  3136  & 2.16(2.19) & 2.42(2.45)  & 4.31(4.38) 
	& 4.84(4.90)  \\
	&  4 & 2  &  1960  & 2.05(2.09) & 2.48(2.50)  & 4.11(4.17) 
	& 4.95(5.01)  \\
	&  5 & 1  &   448  & 1.76(1.79) & 2.70(2.72)  & 3.52(3.57) 
	& 5.41(5.44)  \\
	&  6 & 0  &    28  & 1.37(1.39) & 3.21(3.21)  & 2.75(2.77) 
	& 6.42(6.42)  \\
	&    &    &        & 2.29(2.32) & 2.68(2.71)  & 4.24(4.30) 
	& 5.08(5.13)  \\
\midrule 
$(6,10)$ & 0 & 6  &  210   & 1.67(1.70) & 2.70(2.72) & 3.34(3.41) 
        & 5.41(5.44) \\
        &  1 & 5  &  1512  & 2.04(2.07) & 2.48(2.51) & 4.08(4.15) 
	& 4.97(5.02) \\
	&  2 & 4  &  3150  & 2.19(2.22) & 2.41(2.44) & 4.37(4.44) 
	& 4.82(4.88) \\
	&  3 & 3  &  2400  & 2.11(2.14) & 2.43(2.46) & 4.22(4.28) 
	& 4.86(4.91) \\
	&  4 & 2  &  675   & 1.84(1.87) & 2.60(2.62) & 3.67(3.73) 
	& 5.20(5.24) \\
	&  5 & 1  &  60    & 1.46(1.48) & 3.06(3.06) & 2.92(2.96) 
	& 6.12(6.13) \\
	&  6 & 0  &  1	   & 1.30(1.30) & 3.90(3.90) & 2.60(2.60) 
	& 7.81(7.79) \\
	&    &    &        & 2.31(2.33) & 2.65(2.67) & 4.30(4.36) 
	& 5.02(5.07) \\
\midrule 
$(10,10)$ & 0 & 6  &   210  & 1.97(2.01) & 4.16(4.19) & 3.95(4.01) 
         & 8.33(8.37) \\
         &  1 & 5  &  2520  & 2.44(2.49) & 3.63(3.66) & 4.90(4.98) 
	 & 7.25(7.32) \\
	 &  2 & 4  &  9450  & 2.76(2.81) & 3.36(3.40) & 5.53(5.61) 
	 & 6.71(6.79) \\
	 &  3 & 3  & 14400  & 2.87(2.92) & 3.28(3.32) & 5.74(5.83) 
	 & 6.56(6.64) \\
	 &  4 & 2  &  9450  & 2.76(2.81) & 3.36(3.40) & 5.53(5.61) 
	 & 6.71(6.79) \\
	 &  5 & 1  &  2520  & 2.44(2.49) & 3.63(3.66) & 4.90(4.98) 
	 & 7.25(7.32) \\
	 &  6 & 0  &   210  & 1.97(2.01) & 4.16(4.19) & 3.95(4.01) 
	 & 8.33(8.37) \\
	 &    &    &        & 2.95(2.99) & 3.54(3.57) & 5.62(5.70) 
	 & 6.83(6.91) \\
\bottomrule
\end{tabular} }
\end{center}
\label{widths}
\end{table}

In practice, fixed-$\pi$ state densities are constructed as a sum of the partial
densities $\rho^{m_1,m_2}(E)$ as discussed in Sec. \ref{c5s4s1}  ahead. Going
beyond the first two moments $M_1(m_1,m_2)=\lan H\ran^{m_1,m_2}$ and
$M_2(m_1,m_2)=\lan H^2\ran^{m_1,m_2}$, it is possible to consider the third and
fourth moments $M_3(m_1,m_2)=\lan H^3\ran^{m_1 ,m_2}$ and $M_4(m_1,m_2)=\lan
H^4\ran^{m_1,m_2}$  respectively of $\rho^{m_1,m_2}(E)$.  The skewness and
excess parameters $\gamma_1(m_1,m_2)$ and $\gamma_2(m_1,m_2)$ give information
about the shape of the partial densities and they are close to  zero implies
Gaussian form. The partial densities $\rho^{m_1,m_2}(E)$ determine the
total $+$ve and $-$ve parity state densities $I_\pm(E)$; see Eq.
(\ref{eq.densty}) ahead. By extending the binary correlation approximation to
evaluate averages over two-orbit configurations, we have derived formulas for
$\gamma_1(m_1,m_2)$ and $\gamma_2(m_1,m_2)$. All the details are discussed  in
Chapter \ref{ch7}. Exact results for skewness $\gamma_1(m,\pm)$  and excess
$\gamma_2(m,\pm)$ parameters for fixed-$\pi$ eigenvalue densities $I_\pm(E)$ are
compared with the binary correlation results in Table \ref{c5t1} and it is
clearly seen from the results in Table \ref{c5t1} that in all the examples
considered, the binary correlation results are quite close to the exact results.
In addition, the following results are inferred from the results in Chapter
\ref{ch7}.

It is seen from Eq. (\ref{eq.pty9a}),  $\gamma_1(m_1,m_2)$ will be non-zero only
when $\alpha \neq 0$ and the $\tau$ dependence is weak. Also, it is seen that
for $N_+ = N_-$, $\gamma_1(m_1,m_2) = - \gamma_1(m_2,m_1)$. Similarly,  Eq.
(\ref{eq.pty9b}) shows that for $N_+ = N_-$, $\gamma_2(m_1,m_2) =
\gamma_2(m_2,m_1)$. In the dilute limit, with some approximations as discussed
after Eq. (\ref{eq.pty9b}), the expression for $\gamma_2(m_2,m_1)$ is given by 
Eq. (\ref{eq.pty9c}). This shows that, for $\alpha << \tau$, 
$\gamma_2(m_2,m_1) = C_1/[\overline{\lan X^2 \ran^{m_1,m_2}}]^2$ with $C_1 \sim
-9 \tau^4 N^4 m^3/16$ for $m_1 = m_2 = m/2$ and $N_1 = N_2 = N$. Evaluating 
$\overline{\lan X^2 \ran^{m_1,m_2}}$ in the dilute limit then 
gives $\gamma_2 \sim
-4/m$. Similarly, for $\tau << \alpha$, we have
$\gamma_2(m_2,m_1) = C_2/[\overline{\lan D \wD \ran^{m_1,m_2}} + 
\overline{\lan \wD D \ran^{m_1,m_2}}]^2$ with $C_2 \sim - \alpha^4 N^4 m^3/16$
and this gives $\gamma_2 \sim -4/m$. Therefore, in the 
$\tau << \alpha$ and $\tau >> \alpha$ limit, the result for $\gamma_2$ is same
as the result for spinless fermion EGOE(2) \cite{Mo-73,MF-75} and this 
shows that for a
range of $(\tau,\alpha)$ values, $\rho^{m_1,m_2}(E)$ will be close to  Gaussian.
Moreover, to the extent that Eq. (\ref{eq.pty9c}) applies,  the density
$\rho^{m_1,m_2}(E)$ is a convolution of the densities generated by $X(2)$ and
$D(2)$ operators.  Let us add that the binary correlation results presented in
Chapter \ref{ch7}, with further extensions, 
will be useful in the study of partitioned
EGOE discussed in \cite{Ko-01,Ko-99}.  Now we present some numerical results.

\section{Numerical Results and Discussion}
\label{c5s4}

In order to proceed with the calculations, we need to have some idea of the
range of the parameters $(\tau, \alpha, m/N_+, N_+/N_-)$. Towards this end,
we have used realistic nuclear effective interactions in $sdfp$ \cite{sdfp}
and $fpg_{9/2}$ \cite{fpg} spaces and calculated the variances $v_a^2$,
$v_b^2$, $v_c^2$,  $v_d^2$ for these interactions. Note that it is easy to
identify  the matrices $A$, $B$, $C$ and $D$ given the interaction matrix
elements $\lan (j_1 j_2) J T \mid V \mid (j_3 j_4) JT\ran$. To calculate the
mean-squared matrix elements $v^2$'s, we put the diagonal two-particle
matrix elements to be zero and use the weight factor $(2J+1)(2T+1)$. 
Assuming that $\Delta=3$ MeV and $5$ MeV (these are reasonable values for
$A=20-80$ nuclei), we obtain $\tau \sim 0.09-0.24$ and $\alpha \sim
(0.9-1.3) \times \tau$. These deduced values of $\alpha$ and $\tau$ clearly
point out that one has to go beyond the highly restricted ensemble employed
in \cite{Pa-08} and it is necessary to consider the more general
EGOE(1+2)-$\pi$ defined in Sec. \ref{c5s2}.  Similarly, for $sdfp$ and
$fpg_{9/2}$ spaces  $N_+/N_- \sim 0.5-2.0$. Finally, for nuclei with $m$
number of valence nucleons (particles or holes) where  $sdfp$ or $fpg_{9/2}$
spaces are appropriate, usually $m \lazz N_+$ or $N_-$, whichever is lower.
Given these, we have selected the following examples: $(N_+,N_-,m) =
(8,8,4)$, $(8,8,5)$, $(10,6,4)$, $(10,6,5)$, $(6,10,4)$, $(6,10,5)$,
$(8,8,6)$, $(6,6,6)$, $(7,7,7)$ and $(7,7,6)$. To go beyond the matrix
dimensions $\sim 5000$ with $100$ members is not feasible at present with
the HPC cluster that is used for all the calculations.  Most of the
discussion in this chapter is restricted to $N = N_+ + N_- = 16$ and $m << N$
as in this dilute limit, it is possible to  understand the ensemble results
better. Following the nuclear examples mentioned above, we have chosen
$\tau=0.05,\;0.1,\;0.2,\;0.3$ and $\alpha/\tau = 0.5, \;1.0, \;1.5$. We will
make some comments on the results for other $(\tau,\alpha)$ values at
appropriate places. 

Now we will present the results for (i) the form of the  $+$ve and $-$ve
parity state densities $I_+(E)$ and $I_-(E)$, respectively, (ii) the parity
ratios $I_-(E)/I_+(E)$ vs $E$ where $E$ is the excitation energy of the
system  and (iii) the probability for $+$ve parity ground states generated
by the EGOE(1+2)-$\pi$ ensemble. 

\subsection{Gaussian form for fixed-$\pi$ state densities}
\label{c5s4s1}

\begin{figure}
    \centering
    \subfigure
    {
        \includegraphics[width=4in,height=5.5in,angle=-90]{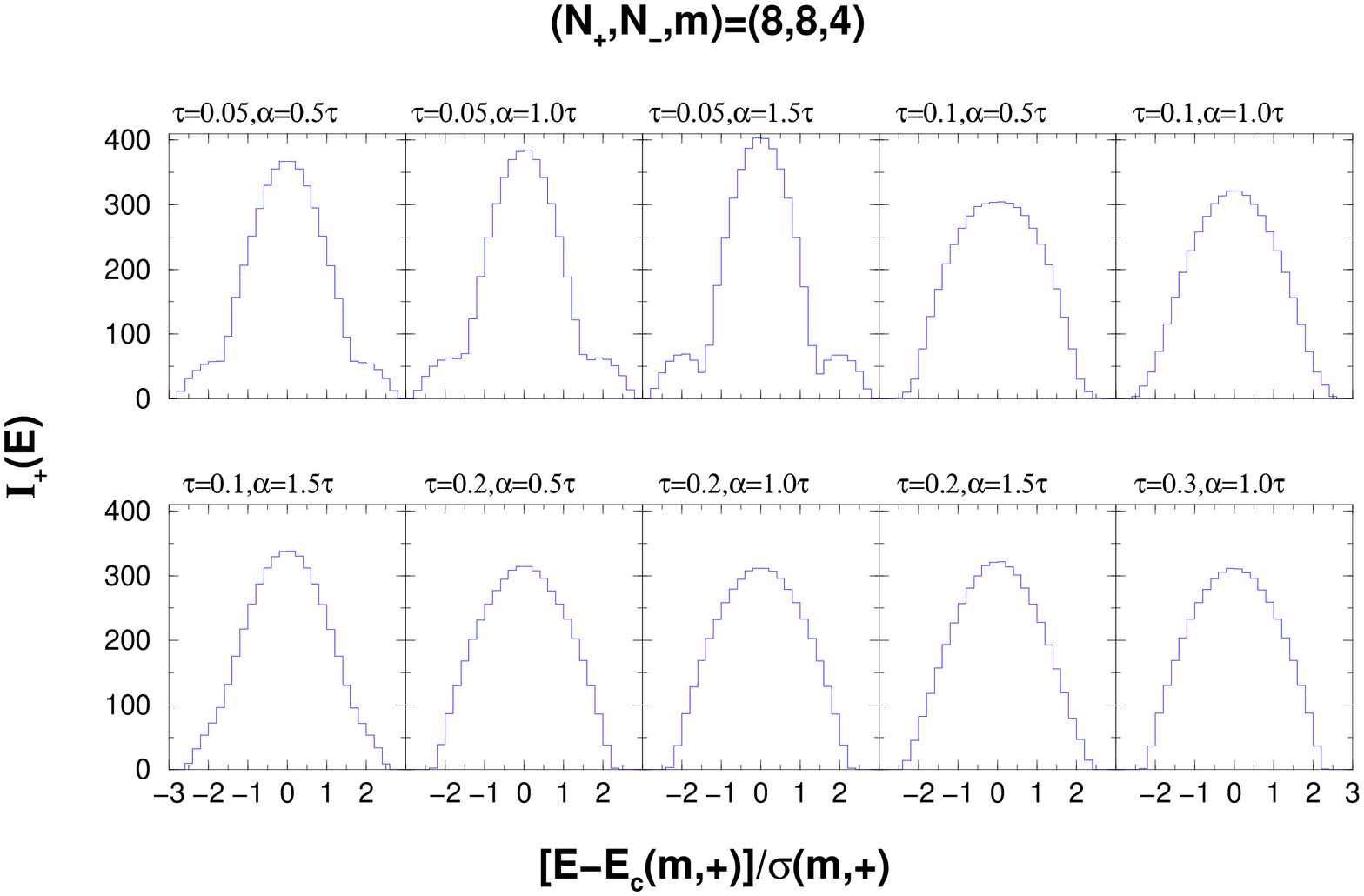}
        \label{den884-1}
    }
    \\
    \subfigure
    {
        \includegraphics[width=4in,height=5.5in,angle=-90]{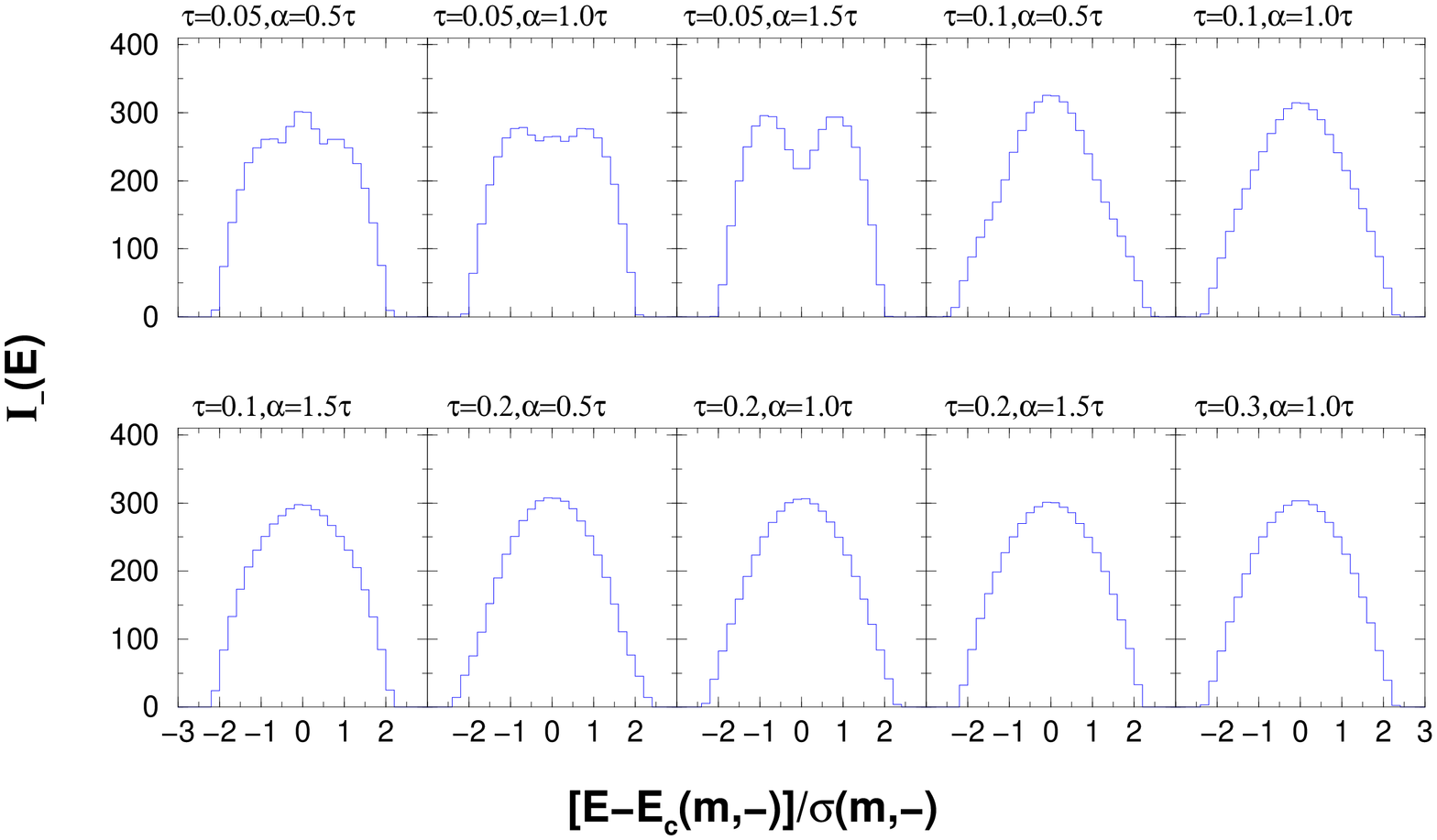}
        \label{den884-2}
    }
    \caption{Positive and negative parity state densities for
    various $(\tau,\alpha)$ values for $(N_+,N_-,m) = (8,8,4)$ system. See
    text for details.}
    \label{den884}
\end{figure}

\begin{figure}
    \centering
    \subfigure
    {
        \includegraphics[width=4.5in,height=4in]{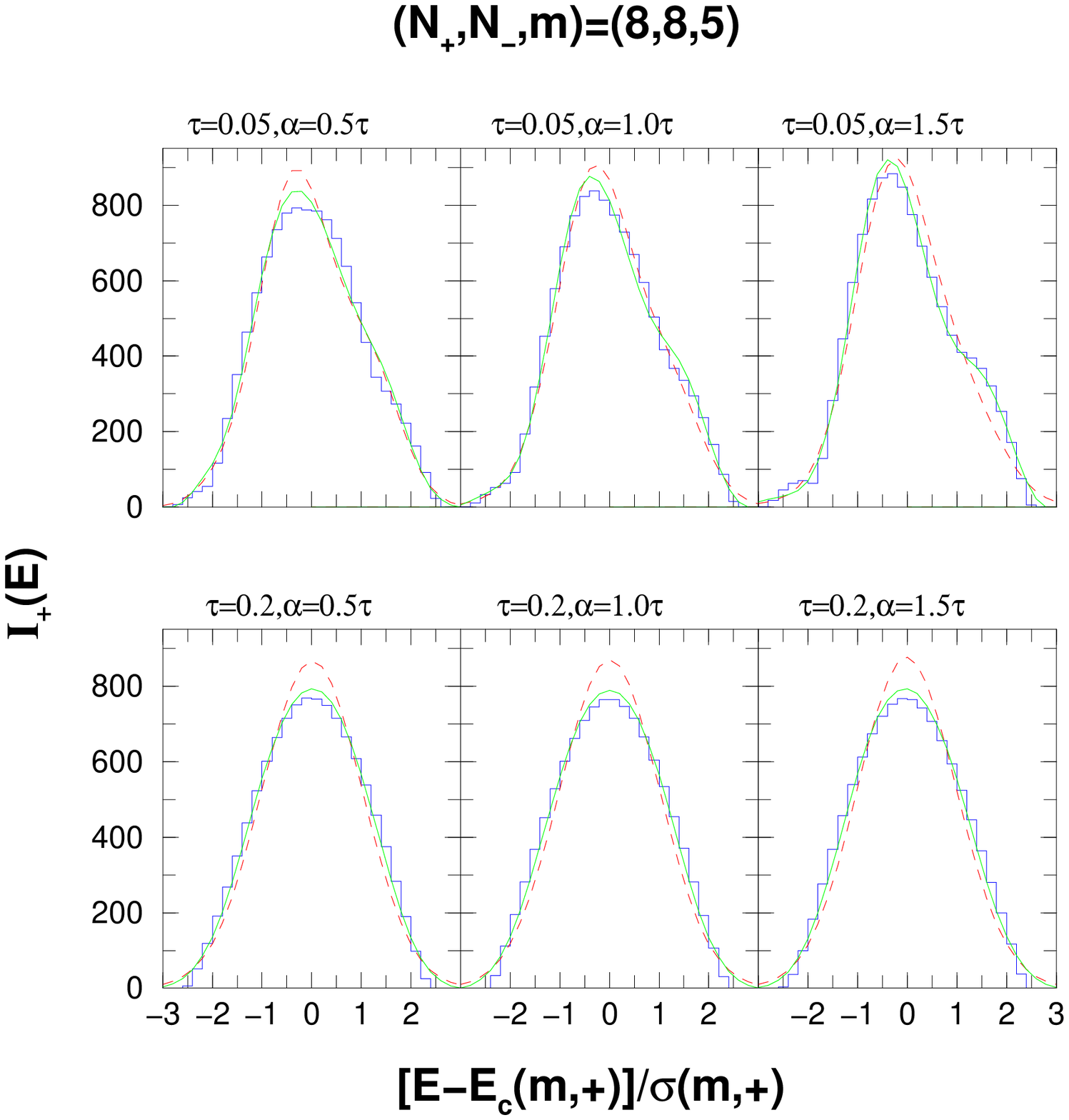}
        \label{den885-1}
    }
    \\
    \subfigure
    {
        \includegraphics[width=4.5in,height=4in]{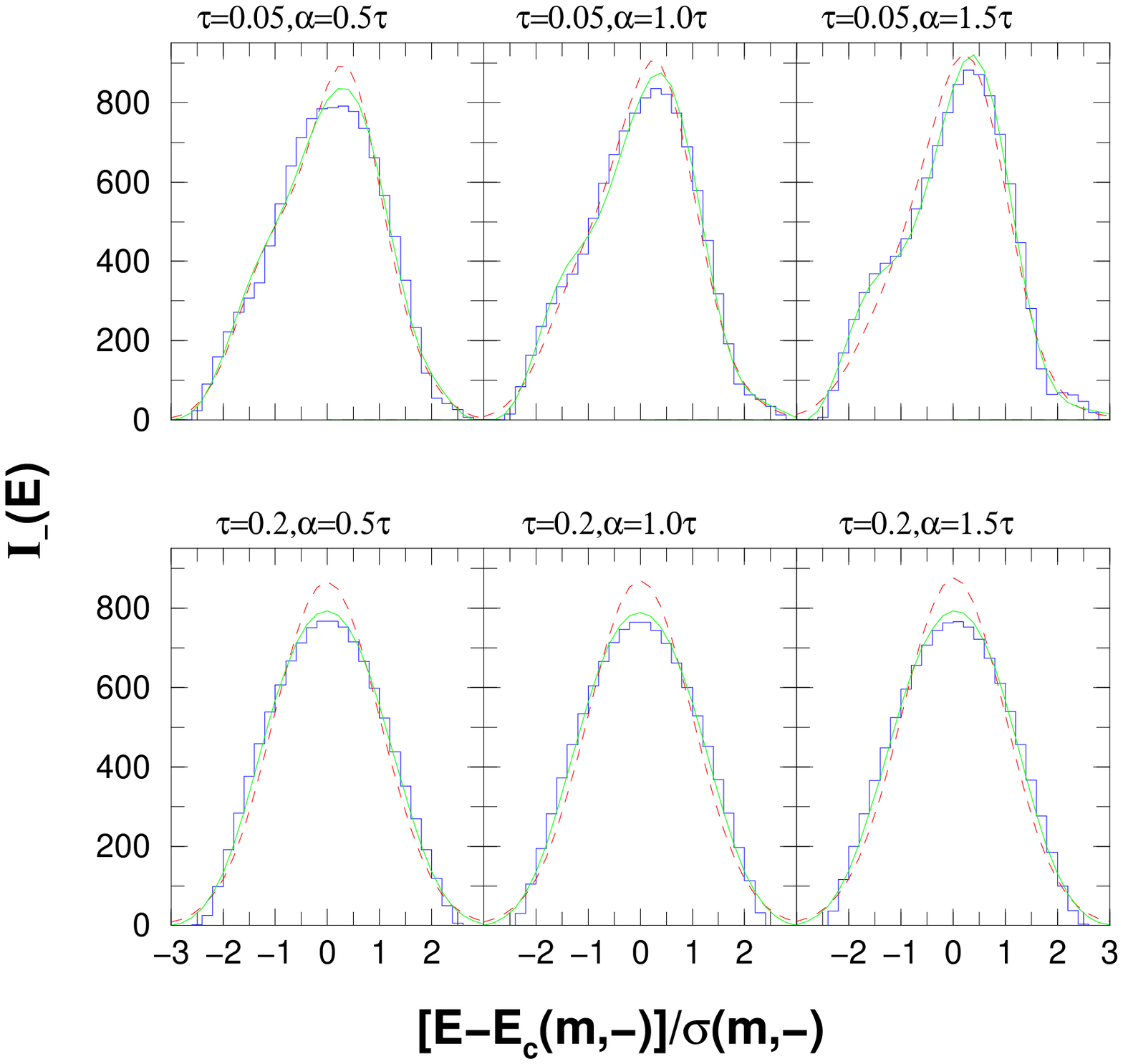}
        \label{den885-2}
    }
    \caption{Positive and negative parity state densities 
    for various $(\tau,\alpha)$ values for $(N_+,N_-,m) = (8,8,5)$ 
    system.  Histograms are numerical ensemble results. The
    dashed (red) curve corresponds to Gaussian form for $\rho^{m_1,m_2}(E)$ in
    Eq. (\ref{eq.densty}) and similarly, solid (green) curve corresponds to 
    Edgeworth corrected Gaussian form with $\gamma_1(m_1,m_2)$ and
    $\gamma_2(m_1,m_2)$ obtained using the results in Chapter \ref{ch7}. 
    See text for details.}
    \label{den885}
\end{figure}

\begin{figure}
    \centering
    \subfigure
    {
        \includegraphics[width=5.5in,height=4in]{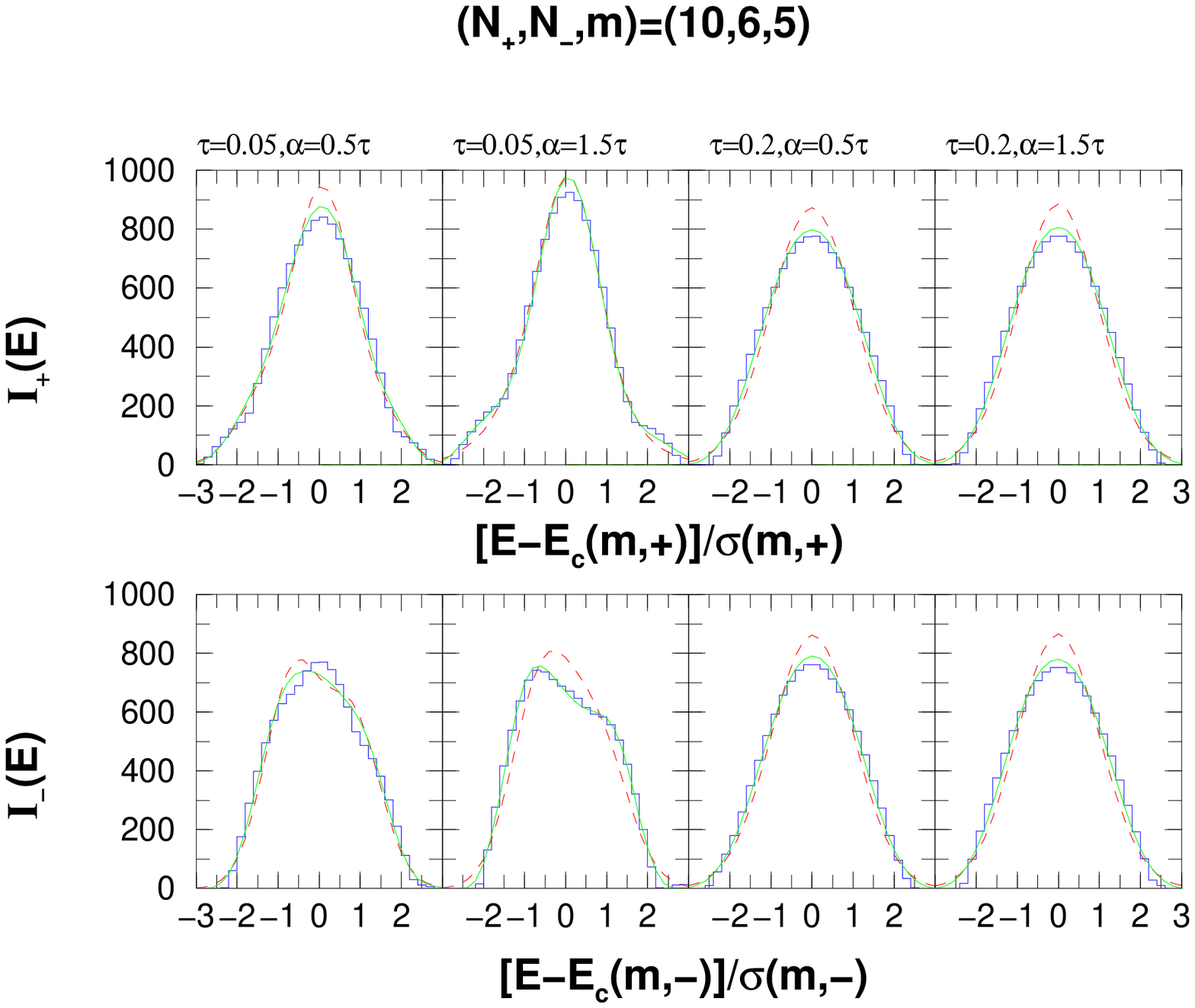}
        \label{den1065-1}
    }
    \\
    \subfigure
    {
        \includegraphics[width=5.5in,height=4in]{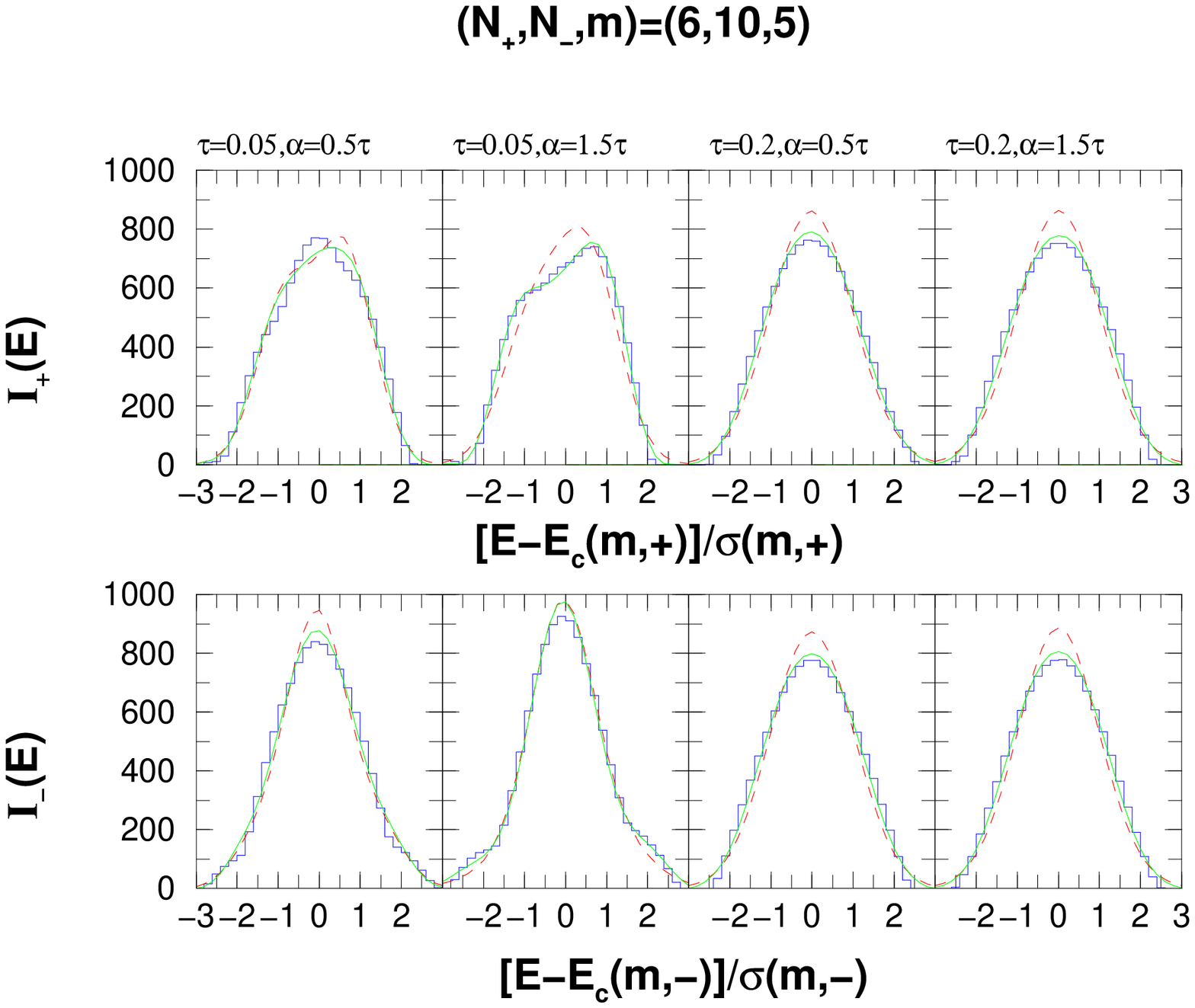}
        \label{den1065-2}
    }
    \caption{Positive and negative parity state densities 
    for various $(\tau,\alpha)$ values for $(N_+,N_-,m) = (10,6,5)$ 
    and $(6,10,5)$ systems.  Histograms are numerical ensemble results. The
    dashed (red) curve corresponds to Gaussian form for $\rho^{m_1,m_2}(E)$ in
    Eq. (\ref{eq.densty}) and similarly, solid (green) curve corresponds to 
    Edgeworth corrected Gaussian form with $\gamma_1(m_1,m_2)$ and
    $\gamma_2(m_1,m_2)$ obtained using the results in Chapter \ref{ch7}. 
    See text for details.}
    \label{den1065}
\end{figure}

Using the method discussed in Sec. \ref{c5s2}, 
we have numerically constructed in
$+$ve and $-$ve parity spaces EGOE(1+2)-$\pi$ ensembles of random matrices
consisting of 100 Hamiltonian matrices in large number of examples, i.e. for
$(N_+,N_-,m)$ and $(\tau,\alpha)$ parameters mentioned above. Diagonalizing
these matrices, ensemble averaged eigenvalue (state) densities, 
\be
\overline{I_\pm(E)} = \overline{\lan\lan\delta(H-E)\ran\ran^\pm} \;,
\label{eq.denpty2}
\ee
are constructed. From now
on, we drop the ``overline'' symbol when there is no confusion.  Results are
shown for $(N_+,N_-,m)=(8,8,4)$, $(8,8,5)$, $(10,6,5)$ and $(6,10,5)$ for
several values of $(\tau,\alpha)$ in Figs. \ref{den884}, \ref{den885} and
\ref{den1065}. To construct the fixed-parity  eigenvalue densities, we first
make the centroids $E_c(m,\pm)$ of  all the members of the ensemble to be
zero and variances $\sigma^2(m,\pm)$ to be unity, i.e., for each member we
have the  standardized  eigenvalues $\we = [E-E_c(m,\pm)]/\sigma(m,\pm)$.
Then, combining all the $\we$ and using a bin-size $\Delta \we=0.2$,
histograms for  $I_\pm(E)$ are generated. It is seen that the state
densities are multimodal for small $\tau$ values and for $\tau \geq 0.1$,
they are unimodal and close to a Gaussian.  Note that in our examples,
$\alpha=(0.5-1.5) \times \tau$.

\begin{figure}
\centering
\includegraphics[width=5in,height=5in]{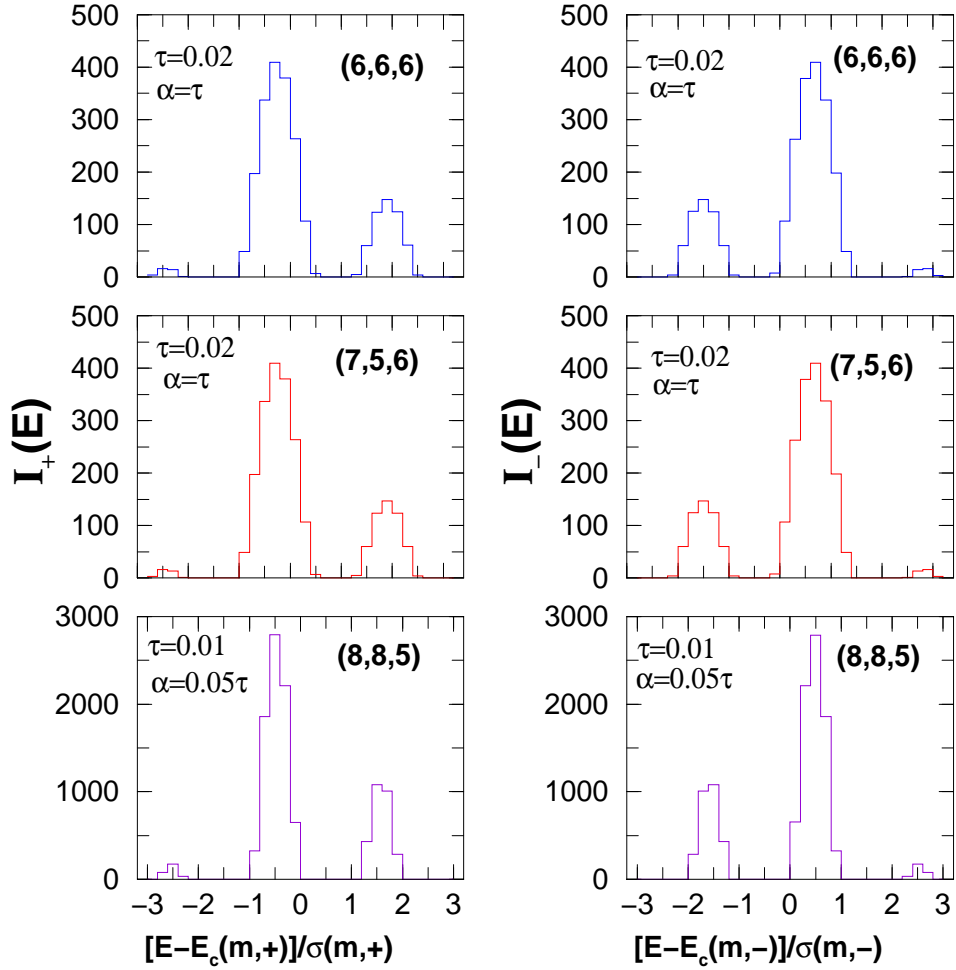}
\caption{Positive and negative parity state densities for 
some small values of $(\tau,\alpha)$. The $(N_+,N_-,m)$ values are given 
in the figures. See text for details.}
\label{smtau}
\end{figure}

For $V(2)=0$, the eigenvalue densities will be a sum of spikes  at
$0,\;2\Delta,\;4\Delta,\;\ldots$ for $+$ve parity densities and similarly at
$\Delta,\;3\Delta,\;5\Delta,\;\ldots$ for $-$ve parity densities. As we
switch on $V(2)$, the spikes will spread due to the matrices $A$, $B$ and
$C$ in Fig. \ref{fig1} and mix due to the matrix $D$. The variance
$\sigma^2(m_1,m_2)$ can be written as, 
\be
\sigma^2(m_1,m_2) = \sigma^2(m_1,m_2 \to m_1,m_2) + \sigma^2(m_1,m_2 \to m_1
\pm 2,m_2 \mp 2)\;.
\label{eq.varm1m2}
\ee
The internal variance $\sigma^2(m_1,m_2 \to m_1,m_2)$ is due to $A$, $B$ and
$C$ matrices and it receives contribution only from the $\tau$ parameter.
Similarly, the external variance $\sigma^2(m_1,m_2 \to m_1 \pm 2,m_2 \mp 2)$
is due to the matrix $D$ and  it receives contribution only from the 
$\alpha$ parameter. When we switch on $V(2)$, as the ensemble averaged
centroids generated by $V(2)$ will be zero, the positions of the spikes will
be largely unaltered. However, they will start spreading and mixing as
$\tau$ and $\alpha$  increase. Therefore, the density will be multimodal
with the modes well separated for very small $(\tau,\alpha)$ values.  Some
examples for this are shown in Fig. \ref{smtau}. As $\tau$ and $\alpha$
start increasing from zero, the spikes spread and will start overlapping for
$\sigma(m_1,m_2) \gazz \Delta$. This is the situation with $\tau=0.05$ shown
in Figs. \ref{den884}, \ref{den885} and \ref{den1065}. However, as $\tau$
increases (with $\alpha \sim \tau$), the densities start becoming unimodal
as seen from the $\tau=0.1$ and $0.2$ examples. Also, the $m$ dependence is
not strong as seen from the Figs. \ref{den884}, \ref{den885} and
\ref{den1065}. Now we will discuss the comparison of the ensemble results 
with the smoothed densities constructed using $E_c(m_1,m_2)$,
$\sigma^2(m_1,m_2)$, $\gamma_1(m_1,m_2)$ and $\gamma_2(m_1,m_2)$.

\begin{figure}
    \centering
    \subfigure
    {
        \includegraphics[width=5.5in,height=3.5in]{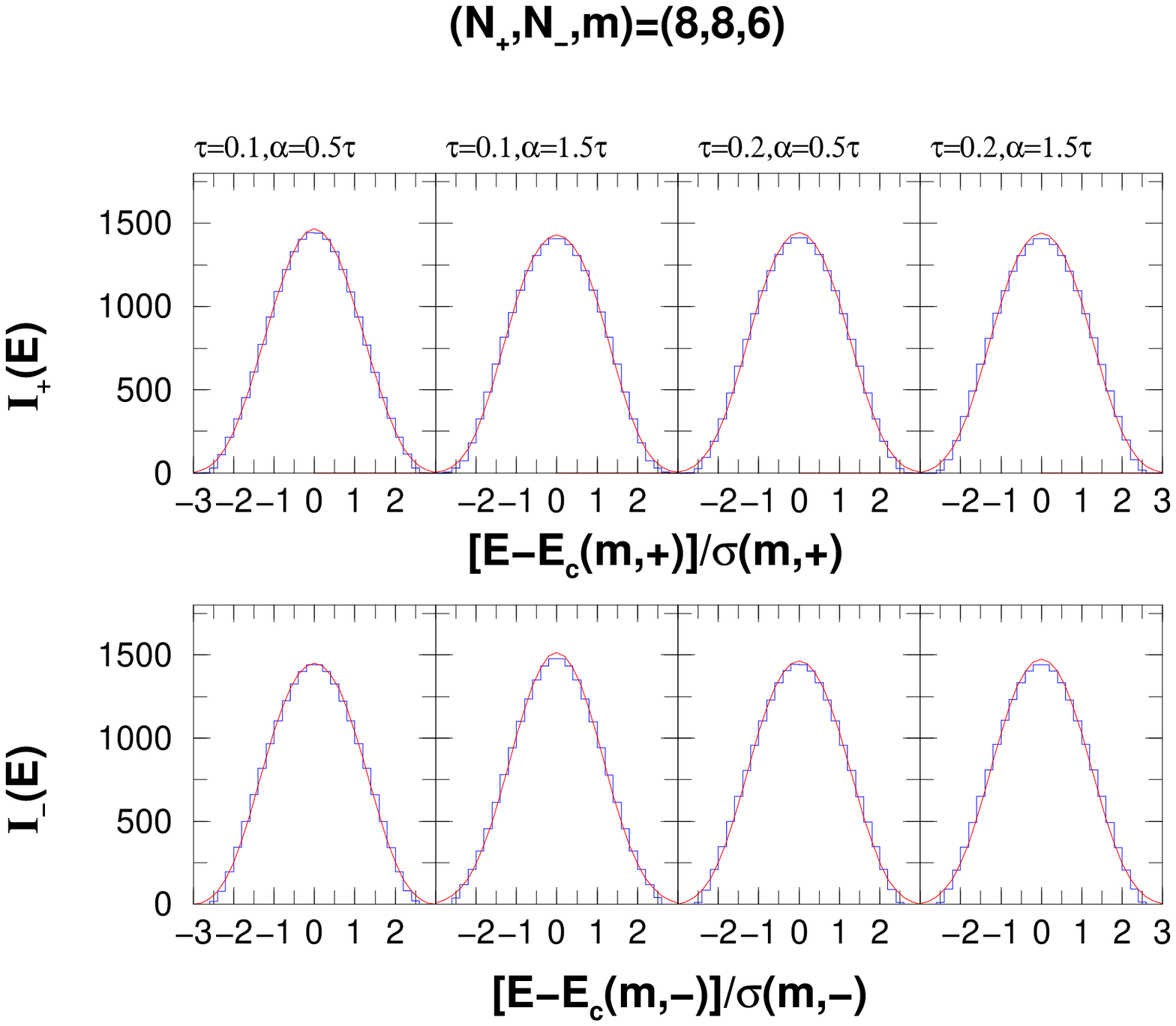}
        \label{den886-1}
    }
    \\
    \subfigure
    {
        \includegraphics[width=5.5in,height=3.5in]{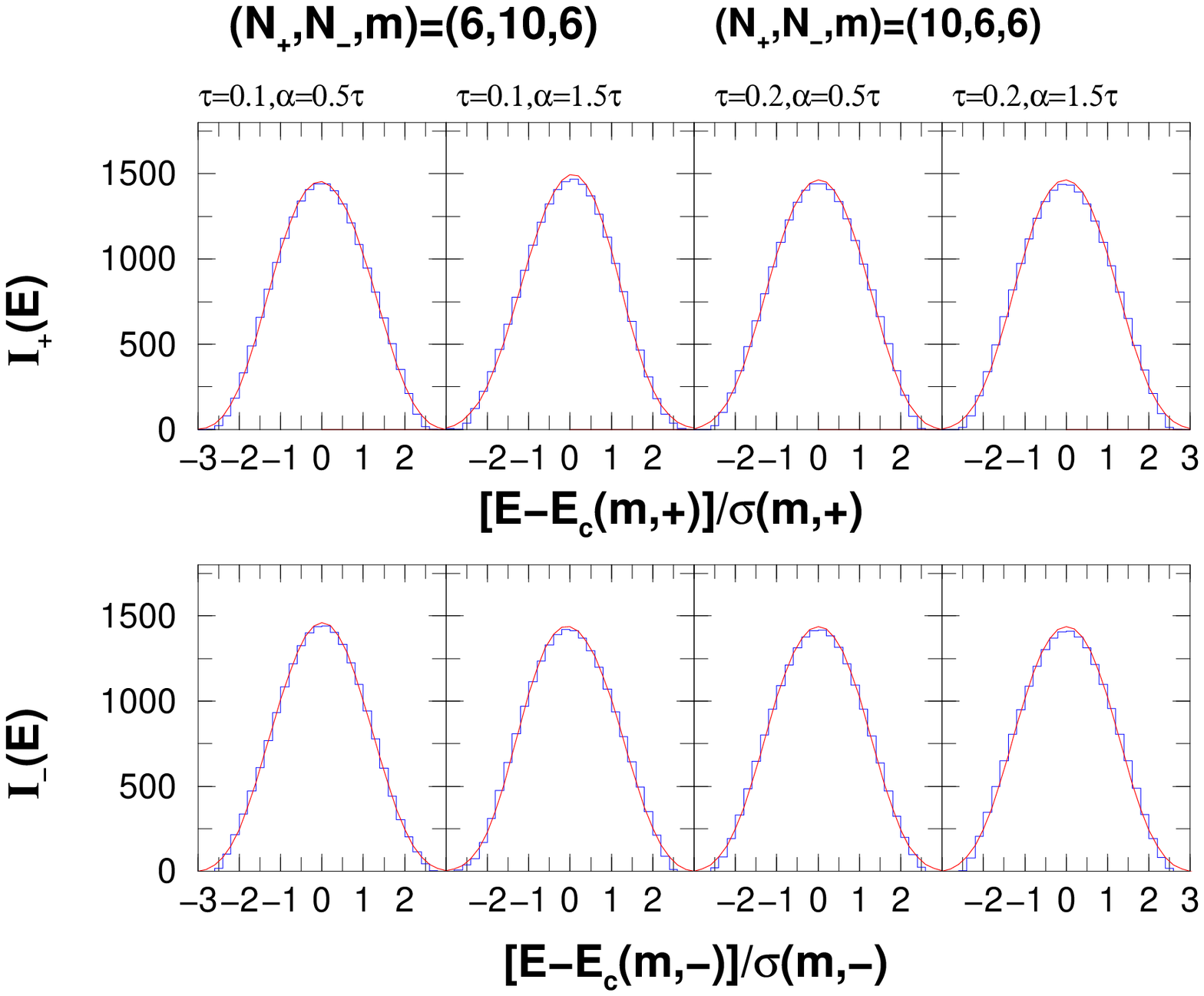}
        \label{den886-2}
    }
    \caption{Positive and negative parity state densities for
    various $(\tau,\alpha)$ values for $(N_+,N_-,m)=(8,8,6)$, $(6,10,6)$
    and $(10,6,6)$ systems. Smoothed curves (solid red lines) are obtained 
    using fixed-$(m_1,m_2)$ partial densities. See text for details.}
    \label{den886}
\end{figure}

\begin{figure}
\includegraphics[width=5.5in,height=5in]{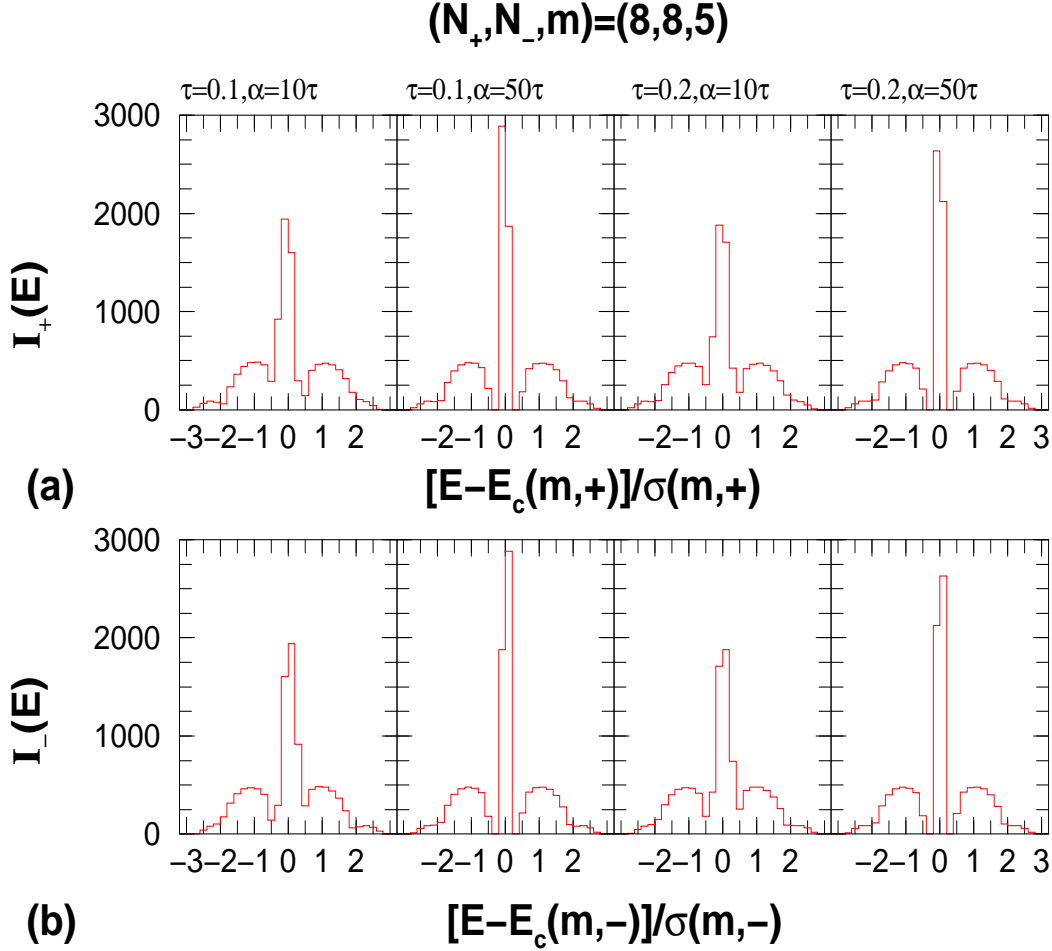}
\caption{(a) Positive and (b) negative parity state densities for
some small values of $\tau$ and large $\alpha$ values for
$(N_+,N_-,m) = (8,8,5)$ system. See text for details.}
\label{lapl}
\end{figure}

As the particle numbers in the examples shown in Figs. \ref{den884},
\ref{den885} and \ref{den1065} are small, the excess parameter
$\gamma_2^\pi(m) \sim -0.7$ to $-0.8$ (skewness parameter $\gamma_1^\pi(m)
\sim 0$ in all our examples). Therefore the densities are not very close to
a Gaussian form. It has been well established that the  ensemble averaged
eigenvalue density takes Gaussian form in the case of spinless fermion (as
well as boson) systems and also for  the embedded ensembles extending to
those with good quantum numbers; see Chapter \ref{ch2} and \cite{Ko-01,Go-11}. 
Thus, it can be anticipated that Gaussian form is 
generic for the state densities or more appropriately,  for the partial
densities $\rho^{m_1,m_2}(E)$ generated by EGOE(1+2)-$\pi$ for sufficiently
large values of $(\tau,\alpha)$.   Results for the fixed-$\pi$ densities for
$(N_+,N_-,m)=(8,8,6)$, $(6,10,6)$ and $(10,6,6)$ systems are shown in Fig.
\ref{den886}.  The smoothed $+$ve and $-$ve parity densities are a sum of
the partial  densities $\rho^{m_1,m_2}(E)$,
\be
\rho_\pm(E) = \dis\frac{1}{d_\pm} \dis\sum^\pr_{m_1,m_2} 
d(m_1,m_2) \rho^{m_1,m_2}(E)\;. 
\label{eq.parden}
\ee
Note that the summation in Eq. (\ref{eq.parden}) is over $m_2$ even for
$+$ve parity density and similarly over $m_2$ odd for $-$ve parity density.
Here $\rho_\pm(E)$ as well as $\rho^{m_1,m_2}(E)$ are normalized to unity.
However, in practice, the densities normalized to dimensions are needed and
they are denoted, as used earlier, by $I_\pm(E)$ and $I^{m_1,m_2}(E)$,
respectively,
\be
I_\pm(E) = d_\pm \rho_\pm(E) = \dis\sum^\pr_{m_1,m_2} 
I^{m_1,m_2}(E) \;;\;\;\;\; I^{m_1,m_2}(E) = d(m_1,m_2) \rho^{m_1,m_2}(E)\;.
\label{eq.densty}
\ee
We employ the Edgeworth (ED) form that includes  $\gamma_1$ and $\gamma_2$
corrections to the Gaussian partial densities $\rho_{\cg}^{m_1,m_2}(E)$. 
Then 
$$
\rho^{m_1,m_2}(E) \to \rho_{\cg}^{m_1,m_2}(E) \to \rho_{ED}^{m_1,m_2}(E)
$$ 
and in terms of the standardized variable $\we$, the ED form $\eta_{ED}(\we)$ 
is given by Eq. (\ref{eq.gau1}).
Using Eqs.
(\ref{eq.parden})  and (\ref{eq.gau1}) with exact centroids and variances
given by the propagation formulas in Sec. \ref{c5s3} and  the binary correlation
results for $\gamma_1$ and $\gamma_2$ as given by the formulas in Chapter
\ref{ch7},  the smoothed $+$ve and $-$ve parity state densities are constructed. 
We put $\eta_{ED}(\we)=0$ when  $\eta_{ED}(\we)<0$. It is clearly seen from
Fig. \ref{den886} that the sum of partial densities, with the partial
densities represented by ED corrected Gaussians, describe extremely  well
the exact fixed-$\pi$ densities.  Therefore, for the
$(\tau,\alpha)$ values in the range determined by nuclear $sdfp$ and
$fpg_{9/2}$ interactions, i.e. $\tau \sim 0.1-0.3$ and $\alpha \sim 0.5 \tau
- 2 \tau$,  the partial densities can be well represented by ED corrected
Gaussians and total densities are also close to ED corrected Gaussians.
Unlike Fig. \ref{den886}, densities in Figs. \ref{den884}, \ref{den885} and
\ref{den1065} show, in many cases, strong departures from Gaussian form.
Therefore, it is important to test how well Eq. (\ref{eq.densty}) with ED
corrected Gaussian for $\rho^{m_1,m_2}(E)$ describes the numerical results
for $I_\pm(E)$. We show this comparison for all the densities in Figs.
\ref{den885} and \ref{den1065}. It is clearly seen that the agreements  with
ED corrected Gaussians are good in all the cases. Therefore, the large
deviations from the Gaussian form for $I_\pm(E)$ arise mainly because of the
distribution of the centroids [this involves dimensions of the $(m_1,m_2)$
configurations] of the partial densities involved. It is possible that the
agreements in Figs. \ref{den885} and \ref{den1065} may become more perfect
if we employ, for the partial densities, some non-canonical forms defined by
the first four moments as given for example in \cite{Gr-95a,Te-06a}. However,
as these forms are not derived using any random matrix ensemble, we haven't
used these for the partial densities in our present investigation. In
conclusion, for the physically relevant range of $(\tau,\alpha)$ values,
the propagation
formulas for centroids and variances given by Eqs. (\ref{eq.cent}) and
(\ref{eq.var}) or alternatively with $E_c(m_1,m_2) = m_2\Delta$ and Eq.
(\ref{eq.eavar})  along with the EGOE(1+2)-$\pi$ ensemble averaged
$\gamma_1(m_1,m_2)$ and $\gamma_2(m_1,m_2)$ estimates as given in Chapter
\ref{ch7}  can be used to construct fixed-$\pi$ state densities for larger
$(N_+,N_-,m)$ systems. Finally, for a small value of $\tau$ but 
$\alpha$ very large, the densities again become multi-modal and some
examples for this are shown in Fig. \ref{lapl}. The situation here is
similar to the model discussed in \cite{Le-94}.

\subsection{Parity ratios for state densities}
\label{c5s4s2}

\begin{figure}
\centering
\includegraphics[width=4in,height=5in]{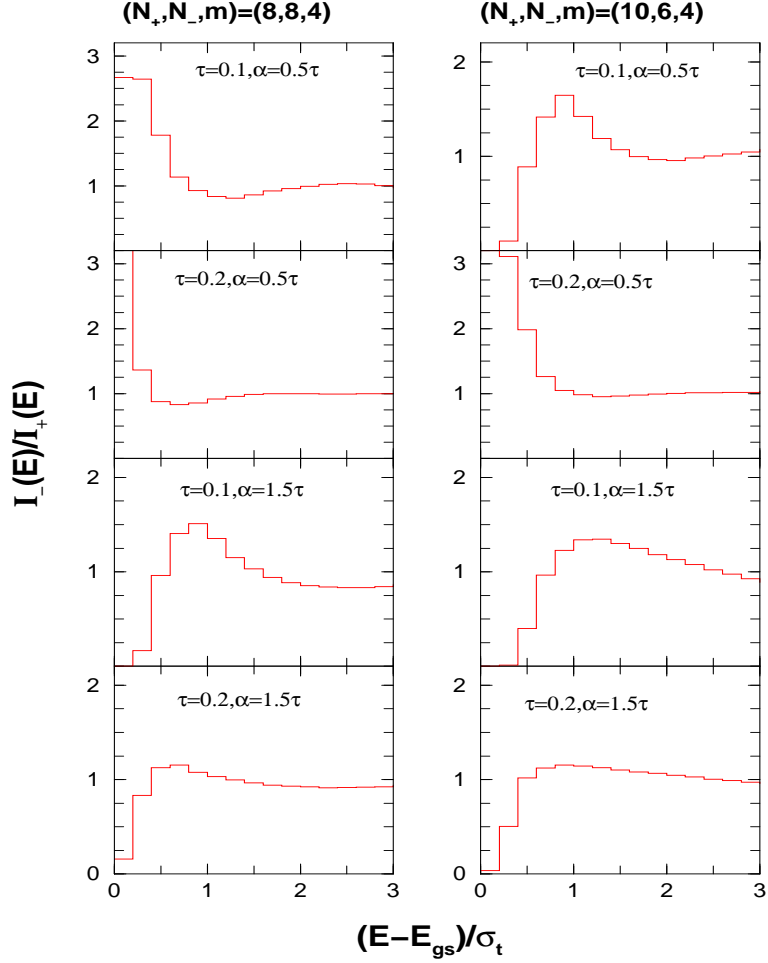}
\caption{Parity ratios for various $(\tau,\alpha)$ values for
$(N_+,N_-,m) = (8,8,4)$ and $(10,6,4)$ systems. See text for details.}
\label{pr884}
\end{figure}

\begin{figure}
    \centering
    \subfigure
    {
        \includegraphics[width=5.5in,height=4in]{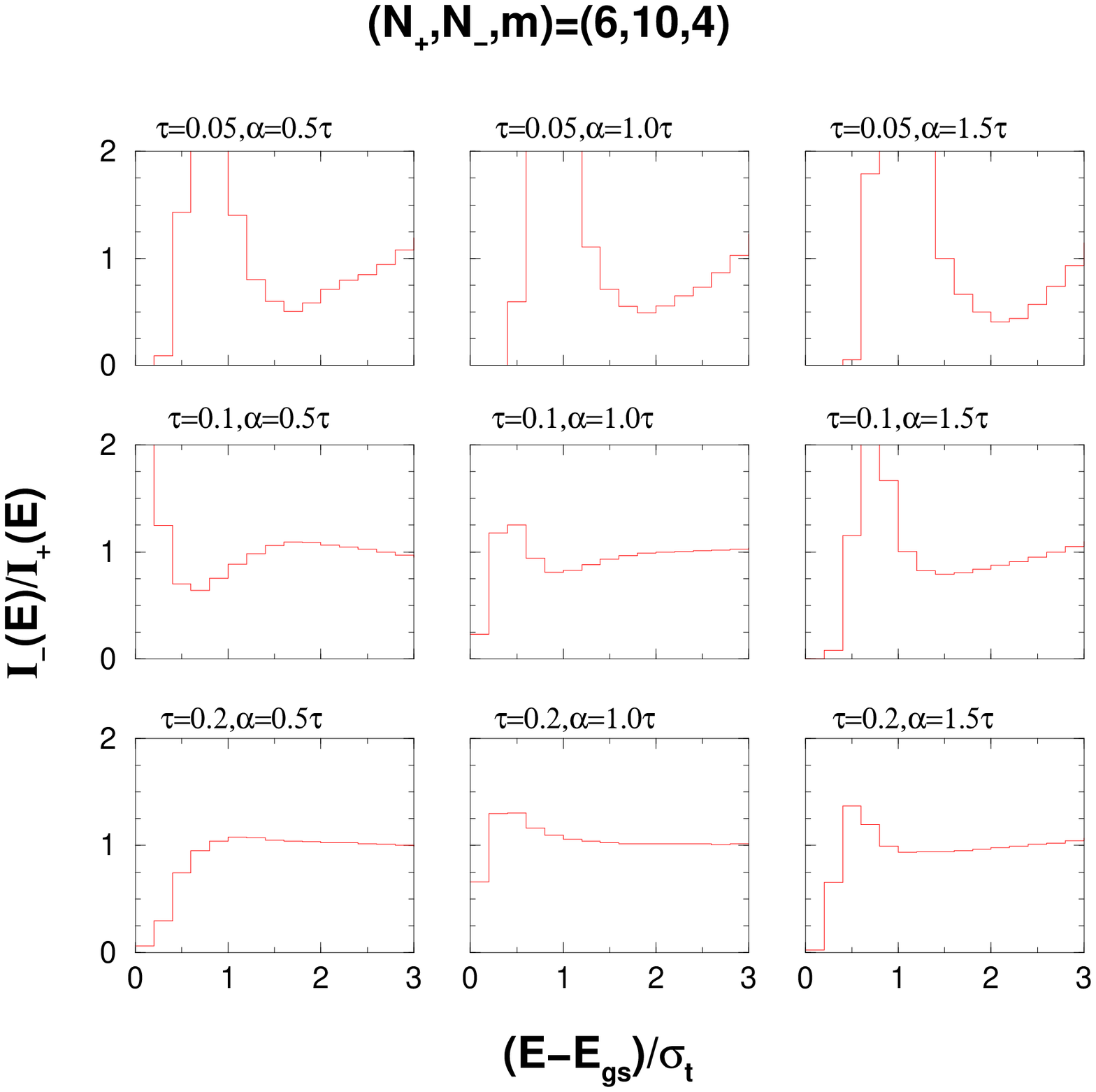}
        \label{pr610-1}
    }
    \\
    \subfigure
    {
        \includegraphics[width=5.5in,height=4in]{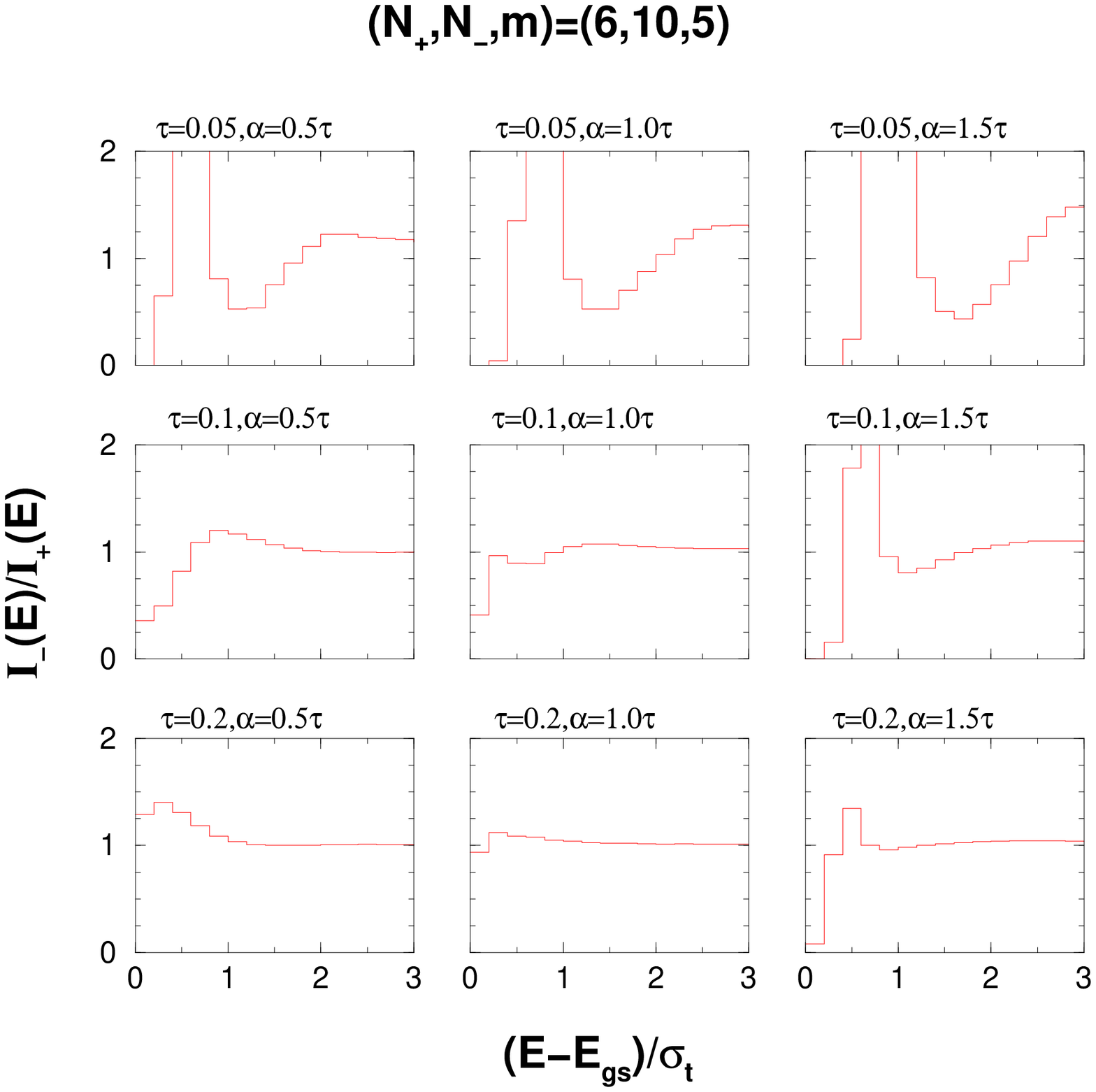}
        \label{pr610-2}
    }
    \caption{Parity ratios for various $(\tau,\alpha)$ values
    for $(N_+,N_-,m) = (6,10,4)$ and $(6,10,5)$  systems. See text for
    details.}
    \label{pr6105}
\end{figure}

\begin{figure}
\centering
\includegraphics[width=5in,height=5.5in]{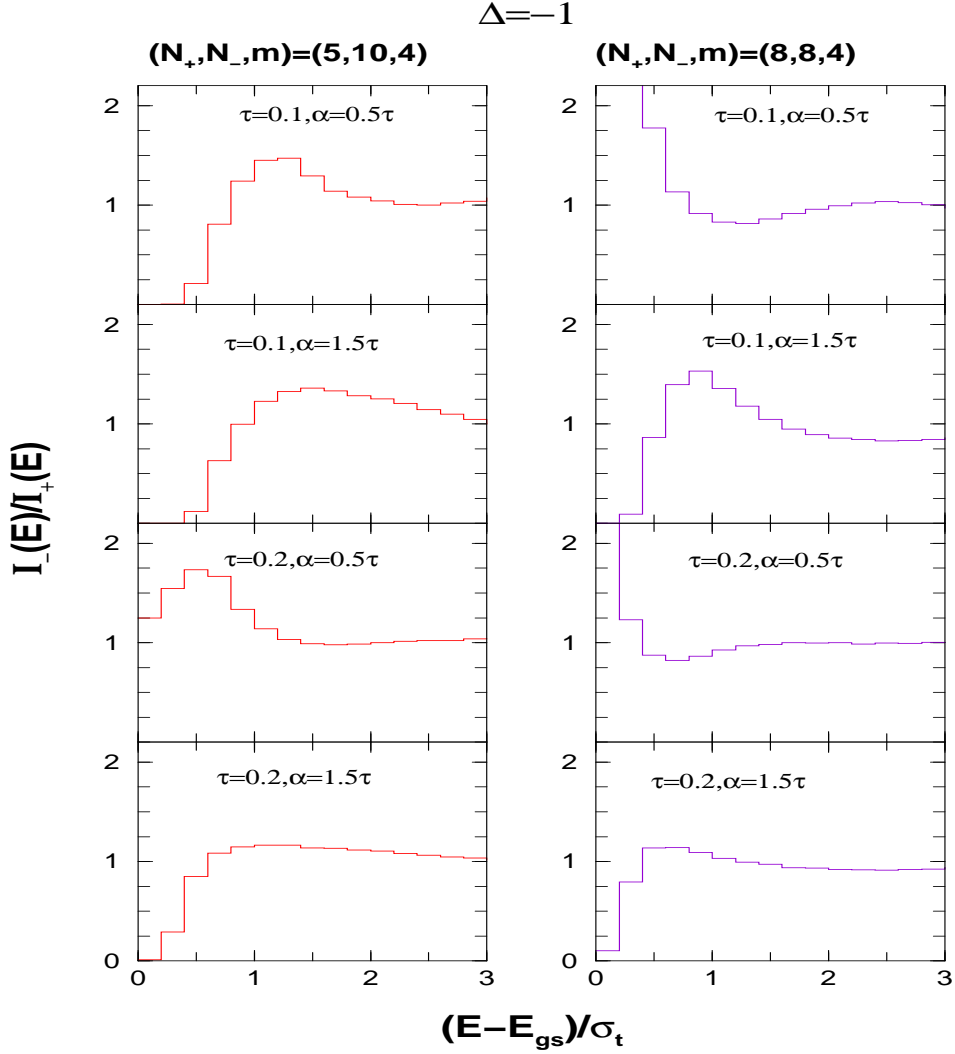}
\caption{Parity ratios for some values of $(\tau,\alpha)$ 
with $\Delta=-1$ for $(N_+,N_-,m) = (5,10,4)$ and $(8,8,4)$ systems. 
See text for details.}
\label{del-1}
\end{figure}

\begin{figure}
\centering
\includegraphics[width=5in,height=6in]{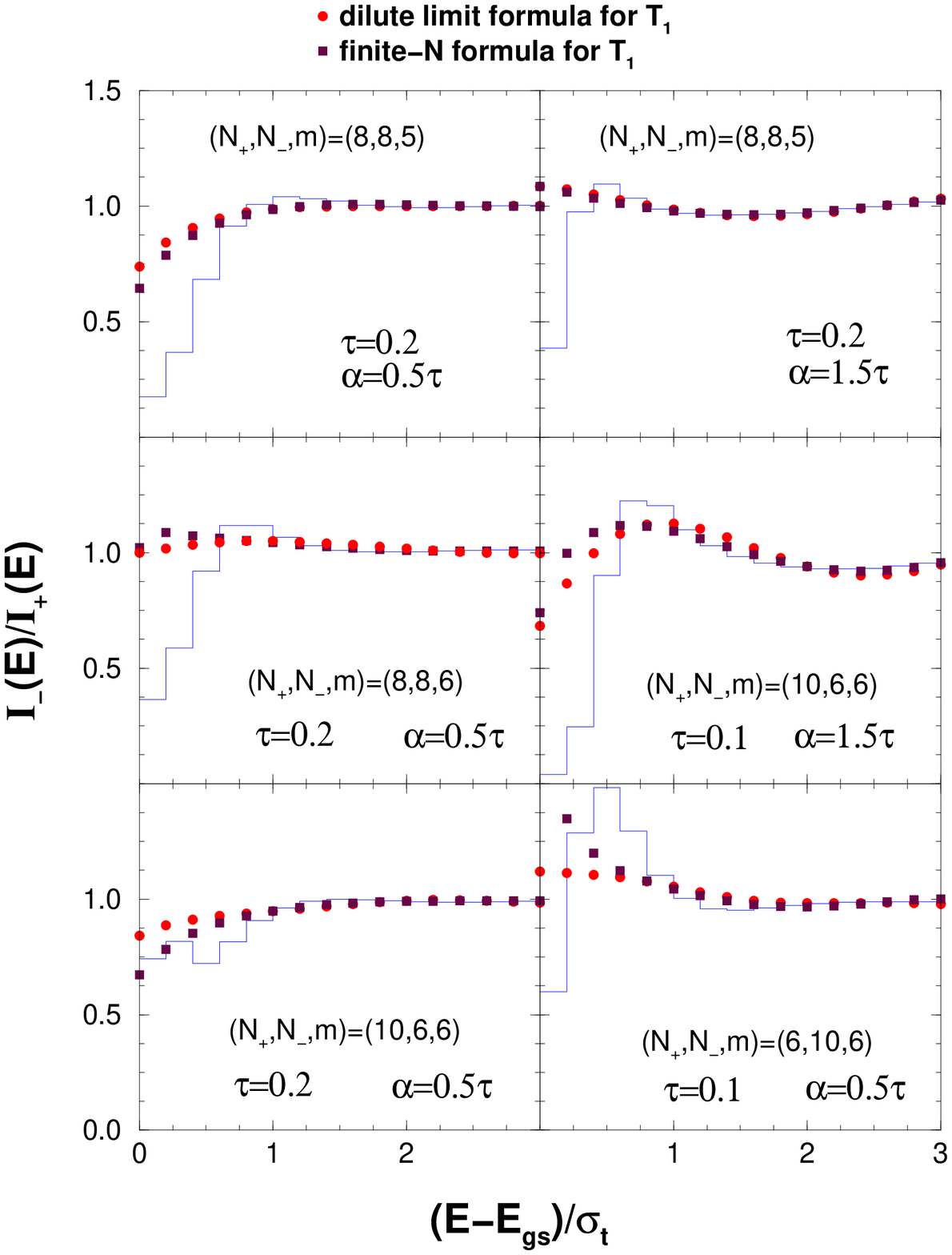}
\caption{Parity ratios for various $(\tau,\alpha)$ values and for various
$(N_+,N_-,m)$ systems. Filled circles (red) and squares (brown) are obtained
using fixed-$(m_1,m_2)$ partial densities with dilute limit formula and 
finite-$N$ formula for the functions $F(\cdots)$ given in Eqs.
(\ref{eq.b8}) and (\ref{eq.3}) respectively that are required to calculate
$T_1$ in Eq. (\ref{eq.pty6}); see Chapter \ref{ch7} and Appendix \ref{c7s1} 
for details.}
\label{pr886}
\end{figure}

As stated in the beginning of this chapter, 
parity ratio of state densities at a given
excitation energy $(E)$ is a quantity of considerable interest in nuclear
structure. For the systems shown in Figs. \ref{den884}, \ref{den885} and
\ref{den1065} and also for many other systems,  we have studied the parity
ratios and the results are shown in Figs. \ref{pr884}-\ref{pr886}. As the
parity ratios need to be  calculated at a given value of excitation energy
$E$, we measure the eigenvalues in both $+$ve and $-$ve parity spaces with
respect to the absolute gs energy $E_{gs}$ of the $N = N_+ + N_-$
system.  Thus, $E_{gs}$ is defined by taking all the $+$ve and $-$ve parity
eigenvalues of all members of the ensemble and choosing the lowest of all
these.  The gs energy can also be determined by averaging the
$+$ve and $-$ve parity gs energies over the ensemble and then the
gs energy is minimum of the two. It is seen that the results for
parity ratios are essentially independent of the choice of $E_{gs}$ and thus
we employ absolute gs energy in our calculations. We use the
ensemble averaged total ($+$ve and $-$ve eigenvalues combined) spectrum
width $\sigma_t$ of the system for scaling. The total widths $\sigma_t$ can
be calculated also  by using $E_c(m_1,m_2)$ and $\sigma^2(m_1,m_2)$.
Examples for $\sigma_t$ are shown in Table \ref{widths} and they are in good
agreement with the results obtained using the simple formula given by 
Eq. (\ref{eq.eavar}). We use the variable $\bee = (E-E_{gs})/\sigma_t$ for
calculating parity ratios. Starting with $E_{gs}$ and using a bin-size of
$\Delta\bee = 0.2$, we have calculated the number of states $I_+(\bee)$ with
$+$ve parity and also the number of states $I_-(\bee)$  with $-$ve parity 
in a given bin and then the ratio $I_-(\bee)/I_+(\bee)$ is the parity ratio.
Note that the results in  Figs. \ref{pr884}-\ref{pr886} are shown for
$\bee=0-3$ as the spectrum span is $\sim 5.5\sigma_t$. To go beyond the
middle of the spectrum,  for real nuclei, one has to include more sp levels
(also a finer splitting of the $+$ve and $-$ve parity levels may be needed)
and therefore, $N_+$ and $N_-$ change. Continuing with this, one obtains the
Bethe form for nuclear level densities \cite{KH-10}.

General observations  from Figs. \ref{pr884}-\ref{pr886} are as follows. (i)
The parity ratio $I_-(\bee)/I_+(\bee)$ will be zero up to an energy
$\bee_0$. (ii) Then, it starts increasing and becomes larger than unity at
an  energy $\bee_m$. (iii) From here on, the parity ratio decreases and
saturates quickly to unity from an energy $\bee_1$. In these examples, 
$\bee_0 \lazz 0.4$, $\bee_m \sim 1$ and $\bee_1 \sim 1.5$.  It is seen that
the curves shift towards left as $\tau$ increases. Also the position of the
peak shifts to much larger value of $\bee_m$ and equilibration gets delayed
as $\alpha$ increases for a fixed $\tau$ value.  Therefore for larger
$\tau$, the energies $(\bee_0, \bee_m, \bee_1)$ are smaller compared to
those for a smaller $\tau$. The three transition energies also depend on
$(N_+,N_-,m)$. We have also verified, as shown in Fig. \ref{del-1}, that the
general structure of the parity ratios will remain same even when we change
$\Delta \to -\Delta$ (i.e., $-$ve parity sp states below the $+$ve parity sp
states). For $(N_+,N_-,m)=(8,8,4)$ system, results for $\Delta = 1$ are
given in Fig. \ref{pr884} and they are almost same as the results with
$\Delta=-1$ given in Fig. \ref{del-1}. The general structures (i)-(iii) are
clearly seen in the numerical examples shown in \cite{Mo-07} where a method
based on the Fermi-gas model has been employed. If $\sigma_t \sim 6-8$ MeV,
equilibration in parities is expected around $E \sim 8-10$ MeV and this is
clearly seen in the examples in \cite{Mo-07}. It is also seen from Fig.
\ref{pr6105} that equilibration is quite poor for very small values of
$\tau$ and therefore comparing with the results in \cite{Mo-07}, it can be
argued that very small values of $\tau$ are ruled out for nuclei. Hence, it
is plausible to conclude that generic results for parity ratios can be
derived using EGOE(1+2)-$\pi$ with reasonably large $(\tau,\alpha)$ values.
Let us add that the interpretations in \cite{Mo-07} are based on the
occupancies of the sp orbits while in the present chapter, they are in terms of
$\tau$ and $\alpha$ parameters.

Using the smoothed $I_\pm(\bee)$, constructed as discussed in Sec.
\ref{c5s4s1},
smoothed forms for parity ratios are calculated as follows. Starting with
the absolute gs energy $E_{gs}$ and using a bin-size of
$\Delta\bee = 0.2$, $+$ve and $-$ve parity  densities in a given energy bin 
are obtained and their ratio is the parity ratio at a given $\bee$. We have
chosen the examples where $I_+$ and $I_-$ are close to Gaussians. It is seen
from Fig. \ref{pr886} that the agreement with exact results is good for
$\bee \gazz 0.5$. However, for smaller $\bee$, to obtain a good agreement
one should have a better prescription for determining the tail part of the
$\rho^{m_1,m_2}(\bee)$ distributions. Developing the theory for this is
beyond the scope of the present thesis as this requires more complete
analytical treatment of the ensemble.  

\subsection{Probability for $+$ve parity ground states}
\label{c5s4s3}

Papenbrock and Weidenm\"{u}ller used the $\tau \to \infty$, $\alpha = \tau$
limit of EGOE(1+2)-$\pi$ for several $(N_+,N_-,m)$ systems to study the
probability $(R_+)$  for $+$ve parity ground states over the ensemble
\cite{Pa-08}. As stated before, this exercise was motivated by shell-model
results with random interaction giving preponderance of  $+$ve parity ground
states \cite{Zh-04}. The numerical calculations in \cite{Pa-08} showed
considerable variation ($18-84$\%) in $R_+$. In addition, they gave a
plausible proof that in the dilute limit $[m << (N_+,N_-)]$, $R_+$ will
approach $50$\%. Combining these, they argued that the observed
preponderance of $+$ve parity ground states could be a finite size (finite
$N_+$, $N_-$, $m$) effect. For the extended EGOE(1+2)-$\pi$ considered in
the present chapter, where the $\tau \to \infty$ and $\alpha = \tau$
restriction is relaxed, as we will discuss now, $R_+$ can reach $100$\%.

\begin{figure}
\includegraphics[width=5in,height=6.5in]{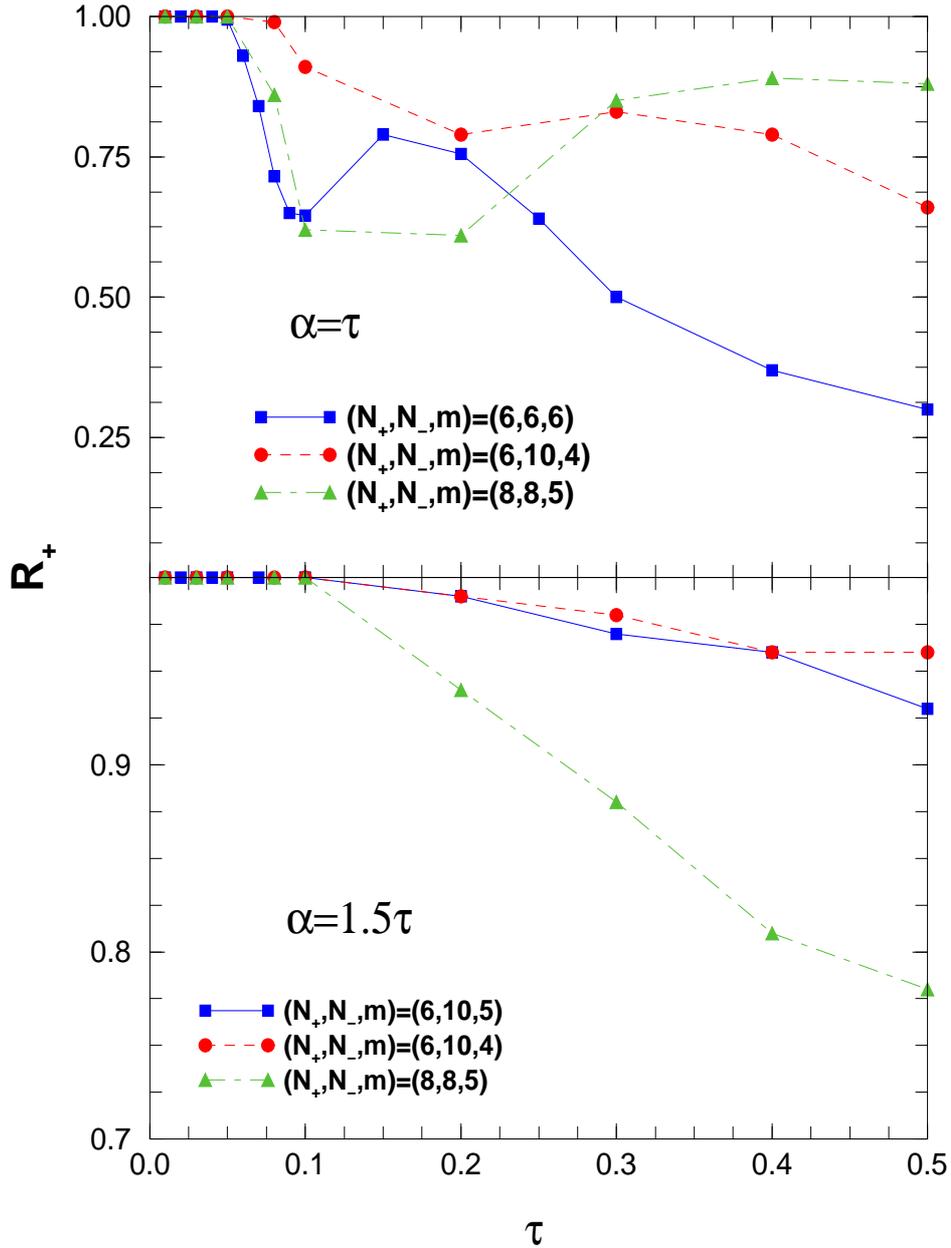}
\caption{Probability $(R_+)$ for $+$ve parity ground states 
for various $(\tau,\alpha)$ values and for various $(N_+,N_-,m)$ 
systems. See text for details.}
\label{rplus}
\end{figure}

For EGOE(1+2)-$\pi$ with $\tau \sim 0$, clearly one will get  $R_+ = 100$\%
(for even $m$ and $m << N_+,N_-$) and therefore it is of interest to study
$R_+$ variation with $(\tau,\alpha)$. We have carried out calculations using
a 200 member ensemble for $(N_+,N_-,m)=(6,6,6)$ and 100 member ensembles for
$(8,8,5)$, $(6,6,6)$,  $(6,10,4)$ and $(6,10,5)$ systems. In these
calculations, we use $\alpha=\tau$ and $1.5\tau$. The results are shown in
Fig. \ref{rplus}.  For $\alpha=\tau$, the results are as follows. For $\tau
\lazz 0.04$, we have $R_+ \sim 100$\% and then $R_+$ starts decreasing with
some fluctuations between $\tau=0.1$ and $0.2$. The origin of these
fluctuations is not clear. As $\tau > 1$ is not realistic, we have
restricted the $R_+$ calculations to $\tau \leq 1$. We see from the figure
that EGOE(1+2)-$\pi$ generates $R_+ \gazz 50$\% for $\tau \leq 0.3$
independent of $(N_+,N_-,m)$. Also, $R_+$ decreases much faster with $\tau$
and reaches $\sim 30$\% for $\tau=0.5$ for $(N_+,N_-,m)=(6,6,6)$. For $m <
(N_+,N_-)$, the decrease in $R_+$ is slower. If we increase $\alpha$, from
the structure of the two-particle $H$ matrix in Fig. \ref{fig1}, we can
easily infer that the width of the lowest $+$ve parity $(m_1,m_2)$ unitary
configuration becomes much larger compared to the lowest $-$ve parity
unitary configuration (see  Table \ref{widths} for examples). Therefore,
with increasing $\alpha$ we expect $R_+$ to increase and this is clearly
seen in Fig. \ref{rplus}. Thus $\alpha \gazz \tau$ is required for $R_+$ to
be large. A quantitative description of $R_+$ requires the construction of
$+$ve and $-$ve parity state densities more accurately in the tail region
and the  theory for this is not yet available. 

\section{Summary}
\label{c5s5}

In the present chapter, we have introduced a generalized EGOE(1+2)-$\pi$ 
ensemble for identical fermions and its construction follows from EGOE(1+2)
for spinless fermion systems. Using this generalized EE, we have not only
studied $R_+$, as it was done by Papenbrock and Weidenm\"{u}ller \cite{Pa-08}
using a simpler two-body ensemble with parity, but also studied the form of
fixed-$\pi$ state densities and parity ratios which are important nuclear
structure quantities. Numerical examples (see Figs.
\ref{den884}-\ref{den1065} and \ref{den886}), with the range of the various
parameters in the model fixed using realistic nuclear effective
interactions, are used to show that the fixed-$\pi$ state densities in
finite dimensional spaces are of Gaussian form for sufficiently large values
of the mixing parameters $(\tau,\alpha)$. The random matrix model also
captures the essential features of parity ratios as seen in the method based
on non-interacting Fermi-gas model reported in \cite{Mo-07}. We also found
preponderance of $+$ve parity ground states for $\tau \lazz 0.5$ and $\alpha
\sim 1.5 \tau$. In addition, for constructing fixed-$\pi$ Gaussian densities
we have derived an easy to understand propagation formula [see Eq.
(\ref{eq.var})] for the spectral variances  of the partial densities
$\rho^{m_1,m_2}(E)$ that generate $I_+$ and $I_-$. Similarly, for calculating
the corrections to the Gaussian forms, formulas for skewness $\gamma_1$ and
excess $\gamma_2$ of the partial densities $\rho^{m_1,m_2}(E)$ are derived
using the binary correlation approximation (see Chapter \ref{ch7} for the
formulas). The smoothed densities constructed using Edgeworth corrected
Gaussians are shown to  describe the numerical results for $I_\pm(E)$ [for
$(\tau,\alpha)$ values in the range defined by nuclear $sdfp$ and
$fpg_{9/2}$ interactions - see beginning of Sec. \ref{c5s4}] and also
the parity ratios at energies away from the gs. Numerical results
presented for parity ratios at lower energies show that a better theory for
the tails of the partial densities is needed (see Figs.
\ref{pr884}-\ref{pr886}). Thus, the results in the present chapter represent
considerable progress in analyzing EGOE(1+2)-$\pi$ ensemble going much
beyond the analysis presented in \cite{Pa-08}.

\chapter{BEGOE(1+2)-$\cs$: Spectral Properties}
\label{ch6}

\section{Introduction}
\label{c6s1}

In the present chapter, our focus is on embedded ensembles for boson systems. As
already emphasized in Chapter \ref{ch1}, unlike for fermion systems, there are
only a few BEE investigations for finite  interacting spinless boson systems
\cite{Ag-01,Ag-02,Ch-03,Ch-04}. Going beyond the embedded ensembles for spinless
boson systems, our purpose in this chapter is to introduce and analyze spectral
properties of embedded Gaussian orthogonal ensemble of random matrices for boson
systems with spin degree of freedom [BEGOE(2)-$\cs$ and also BEGOE(1+2)-$\cs$]
and for Hamiltonians that conserve the total spin of the $m$-boson systems. Here
the spin is, for example, as the $F$-spin in the proton-neutron interacting
boson model ($pn$IBM) of atomic nuclei \cite{Ca-05}. Just as the earlier BEE
studies for spinless boson systems, a major motivation for the study undertaken
in the present chapter is the possible applications of generalized BEEs to
ultracold atoms. The BEGOE(1+2)-$\cs$ with spin-$\spin$ bosons is a simple yet
non-trivial extension of BEGOE(1+2). This ensemble is useful in obtaining 
several physical conclusions, like spin dependence of the order to chaos
transition marker in level fluctuations, the spin of the gs, the spin ordering
of excited states and pairing correlations in the gs region generated by random
interactions, that explicitly require inclusion of spin degree of freedom. These
are discussed in Secs. \ref{c6s3}, \ref{c6s5} and \ref{c6s6}. 

It should be emphasized that the present chapter opens a new direction in
defining and analyzing embedded ensembles for boson systems with symmetries.
There are now many studies of spinor BEC using Hamiltonians conserving the total
spin with the bosons carrying $\cs=1$ (also higher) degree of freedom
\cite{Pe-10,Yi-07}. Also, there are several studies of the properties of a
mixture of two species of atoms which correspond to pseudospin-$\spin$ bosons
(i.e., two-component boson systems)  with $m_\cs=\pm \spin$ distinguishing the
two species; see for example \cite{Al-03,Yu-10}. However, the Hamiltonians
appropriate for these studies do not conserve the total spin (as the system does
not have true $\spin$-spins). BEE with good $M_S$ are appropriate in
understanding the statistical properties of these systems. These explorations
are beyond the scope of the present thesis. Extensions of BEGOE(1+2)-$\cs$ with
$\cs=\spin$ to boson ensembles  with integer spin $\cs=1$ and to
BEGOE(1+2)-$M_S$ are briefly discussed in Appendix \ref{c6a1} for completeness. 
All the results presented in this chapter are reported in \cite{Ma-11}.
Now, we begin with the definition and construction of BEGOE(1+2)-$\cs$.

\section{Definition and Construction of BEGOE(1+2)-$\cs$} 
\label{c6s2}

Let us consider a system of $m$ ($m>2$) bosons distributed in $\Omega$
number of sp orbitals each with spin $\cs=\spin$. Then the number of sp 
states is $N=2\Omega$. The sp  states are denoted by  $\l.\l| i,m_\cs=\pm
\spin\r.\ran$ with $i=1,2,\ldots,\Omega$ and the two-particle symmetric
states are denoted by $\l.\l|(ij)s,m_s\r.\ran$ with $s=0$ or $1$. It is
important to note that for EGOE(1+2)-$\cs$, the embedding algebra is
$U(2\Omega) \supset U(\Omega) \otimes SU(2)$ with $SU(2)$ generating spin;
see Secs. \ref{c6s5} and \ref{c6s6} ahead.  The dimensionalities of the 
two-particle
spaces with $s=0$  and $s=1$ are $\Omega(\Omega-1)/2$ and
$\Omega(\Omega+1)/2$, respectively. For one plus two-body Hamiltonians
preserving $m$-particle spin $S$, the one-body Hamiltonian is
$\whh(1)=\sum_{i=1}^\Omega\, \epsilon_i n_i$ where the orbitals $i$ are
doubly degenerate,  $n_i$ are number operators  and $\epsilon_i$ are sp
energies.  The two-body Hamiltonian $\wv(2)$
preserving $m$-particle spin $S$ is defined by the symmetrized  two-body
matrix elements  $V^s_{ijkl}=\lan (kl)s,m_s \mid \wv(2) \mid (ij)s,m_s\ran$
with $s=0,\,1$ and they are independent of the $m_s$ quantum number; note
that for $s=0$, only $i \neq j$ and $k \neq l$ matrix elements exist. Thus
$\wv(2)=\wv^{s=0}(2) + \wv^{s=1}(2)$  and the sum here is a direct sum. The
BEGOE(2)-$\cs$ ensemble for a given $(m,S)$ system is generated by first  
defining the two parts of the two-body Hamiltonian to be independent GOE(1)'s in
the two-particle spaces [one for $\wv^{s=0}(2)$ and other for
$\wv^{s=1}(2)$]. Now the $V(2)$ ensemble
defined by  $\{\wv(2)\}=\{\wv^{s=0}(2)\} + \{\wv^{s=1}(2)\}$ is propagated
to the  $(m,S)$-spaces by using the geometry (direct product structure) of
the $m$-particle spaces. By adding the
$\whh(1)$ part, the BEGOE(1+2)-$\cs$ is defined by the operator
\be
\{\wh\}_{\mbox{BEGOE(1+2)-\cs}} = \whh(1)+  \lambda_0\, \{\wv^{s=0}(2)\} +
\lambda_1\, \{\wv^{s=1}(2)\}\,.
\label{eq.begoe-s}
\ee
Here $\lambda_0$ and $\lambda_1$ are the strengths of the $s=0$ and  $s=1$
parts of $\wv(2)$, respectively.  The mean-field one-body Hamiltonian
$\whh(1)$ in Eq. (\ref{eq.begoe-s}) is defined by sp energies $\epsilon_i$
with average spacing $\Delta$. As already mentioned in Chapter \ref{ch2}, 
we put
$\Delta=1$ so that $\lambda_0$ and $\lambda_1$ are in the  units of
$\Delta$ and choose $\epsilon_i=i+1/i$. Thus BEGOE(1+2)-$\cs$ is
defined by the five  parameters $(\Omega, m, S, \lambda_0, \lambda_1)$. The
$H$ matrix dimension $d_b(\Omega,m,S)$ for a given $(m,S)$ is
\be
d_b(\Omega,m,S)=\frac{(2S+1)}{(\Omega-1)} { \Omega+m/2+S-1
\choose m/2+S+1} {\Omega+m/2-S-2 \choose m/2-S}\;,
\label{eq.msdim}
\ee
and they satisfy the sum rule $\sum_S\;(2S+1)\;d_b(\Omega,m,S)= {N+m-1 \choose
m}$.  For example: (i)   $d_b(4,10,S)=196$, $540$, $750$,  $770$, $594$ and
$286$ for spins $S=0-5$; (ii)  $d_b(4,11,S)=504$, $900$, $1100$, $1056$, $780$
and $364$ for $S=1/2-11/2$; (iii)   $d_b(5,10,S)=1176$,  $3150$, $4125$,
$3850$, $2574$ and $1001$ for $S=0-5$; (iv) $d_b(6,12,S)=13860$, $37422$,
$50050$, $49049$, $36855$, $20020$ and $6188$ for $S=0-6$; and (v) 
$d_b(6,16,S)=70785$, $198198$, $286650$, $321048$, $299880$, $235620$,
$151164$, $72675$ and $20349$ for $S=0-8$.

Given $\epsilon_i$ and $V^s_{ijkl}$, the many-particle Hamiltonian matrix
for a given ($m,S$) can be constructed using the $M_S$ representation ($M_S$
is the $S_z$ quantum number) and for spin projection the $S^2$ operator is
used as it was done for fermion systems in Chapter \ref{ch2}. 
Alternatively, it
is possible to construct the $H$ matrix directly in a good $S$ basis using
angular-momentum algebra as it was done for fermion systems in
\cite{Al-06}.  We have employed the $M_S$ representation for constructing
the $H$ matrices with $M_S=M_S^{min}=0$ for even $m$ and
$M_S=M_S^{min}=\spin$ for odd $m$ and they will contain states with all $S$
values.  The dimension of this basis space is $\cd(\Omega,m,M_S^{min}) =
\sum_S\,d_b(\Omega,m,S)$.  For example, $\cd(4,10,0)=3136$,
$\cd(4,11,\spin)=4704$,   $\cd(5,10,0)=15876$, $\cd(6,12,0)=213444$ and
$\cd(6,16,0)=1656369$. 

To construct the many-particle Hamiltonian matrix for a given $(m,S)$, 
first the  sp states $\l.\l| i,m_\cs=\pm \spin\r.\ran$ are arranged in such
a way that the first $\Omega$ states have $m_\cs=\spin$ and the remaining
$\Omega$ states have $m_\cs=-\spin$ so that the sp states are
$\l.\l|r\r.\ran=\l.\l|i=r,m_\cs=\spin\r.\ran$ for $r \leq \Omega$ and
$\l.\l| r\r.\ran=\l.\l|i=r-\Omega,m_\cs=-\spin\r.\ran$ for $r > \Omega$.
Using the direct product structure of the many-particle states,  the
$m$-particle configurations $\cm$, in occupation number representation, are
\be
\cm=\left| {\prod\limits_{r=1}^{N=2\Omega} {m_r } } \right\rangle  =
\left| {m_1, m_2, \ldots, m_\Omega, m_{\Omega+1}, m_{\Omega+2},
\ldots, m_{2\Omega} } \right\rangle,
\label{eq.ms-conf}
\ee
where $m_r \geq 0$ with $\sum_{r=1}^N{m_r}=m$ and 
$M_S=\spin\l[\sum_{r=1}^{\Omega}\,m_r - \sum_{r^\prime=
\Omega+1}^{2\Omega}\,m_{r^\prime}\r]$. To proceed further, the (1+2)-body
Hamiltonian defined by $\epsilon_i$ and $V^{s=0,1}_{ijkl}$ is converted into
the $\l.\l| i,m_\cs=\pm \spin\r.\ran$ basis. Then the sp energies 
$\epsilon_i^\pr$ with $i=1,2,\ldots,N$ are $\epsilon_i^\pr = 
\epsilon_{i+\Omega}^\pr = \epsilon_i$ for $i \leq \Omega$. Similarly,
$V^{s}_{ijkl}$ are changed to  $V_{im_i,jm_j,km_k,lm_l}={\left\langle {im_i,
jm_j } \right|\left. {V(2) } \right|\left. {km_k, lm_l} \right\rangle }$
using,
\be
\barr{rcl}
V_{i\spin,j\spin,k\spin,l\spin} & = & 
V_{i -\spin,j -\spin,k -\spin,l -\spin} = 
V^{s=1}_{ijkl} \;,\\ \\
V_{i\spin,j -\spin,k\spin,l -\spin} & = & \dis\frac{\sqrt{(1+\delta_{ij})
(1+\delta_{kl})}}{2} \l[ V^{s=1}_{ijkl}+ V^{s=0}_{ijkl}\r]\;,
\earr \label{tbme_new}
\ee
with all the other matrix elements being zero except for the symmetries, 
\be 
V_{im_i,jm_j,km_k,lm_l} = V_{km_k,lm_l,im_i,jm_j}=
V_{jm_j,im_i,lm_l,km_k} = V_{im_i,jm_j,lm_l,km_k}\;. 
\ee 
Using $(\epsilon_r^\pr, V_{im_i,jm_j,km_k,lm_l})$'s, construction of the 
$m$-particle $H$ matrix in the basis defined by Eq. (\ref{eq.ms-conf}) 
reduces to the problem of BEGOE(1+2) for spinless boson systems and hence
Eq. (\ref{eq.app2}) will give the formulas for the non-zero matrix
elements; see Sec. \ref{begoe} for details. Now
diagonalizing the $S^2$ matrix in the basis defined by  Eq.
(\ref{eq.ms-conf}) will give the unitary transformation required to  change
the $H$ matrix in $M_S$ basis into good $S$ basis.  Following this method,
we have numerically constructed BEGOE(1+2)-$\cs$ in many examples and
analyzed various spectral properties generated by this ensemble.  In
addition, we have also derived some analytical results as discussed ahead in
Secs. \ref{c6s4} and \ref{c6s6}.  These results are also used to validate the
BEGOE(1+2)-$\cs$ numerical code we have developed. In addition, we have also
verified the code by comparing the results with those \cite{Ch-10} obtained by
directly programming the operations that give Eq.
(\ref{eq.app2}). 
In this chapter, we deal with both BEGOE(2)-$\cs$ and BEGOE(1+2)-$\cs$ and the
focus is on the dense limit defined by   $m \to \infty$, $\Omega \to
\infty$, $m/\Omega  \to \infty$ and $S$ is  fixed. Now we will discuss these
results.

\section{Numerical Results for Eigenvalue Density and Level Fluctuations in
the Dense Limit}
\label{c6s3}

\begin{figure}
\centering
\includegraphics[width=5in,height=6.5in]{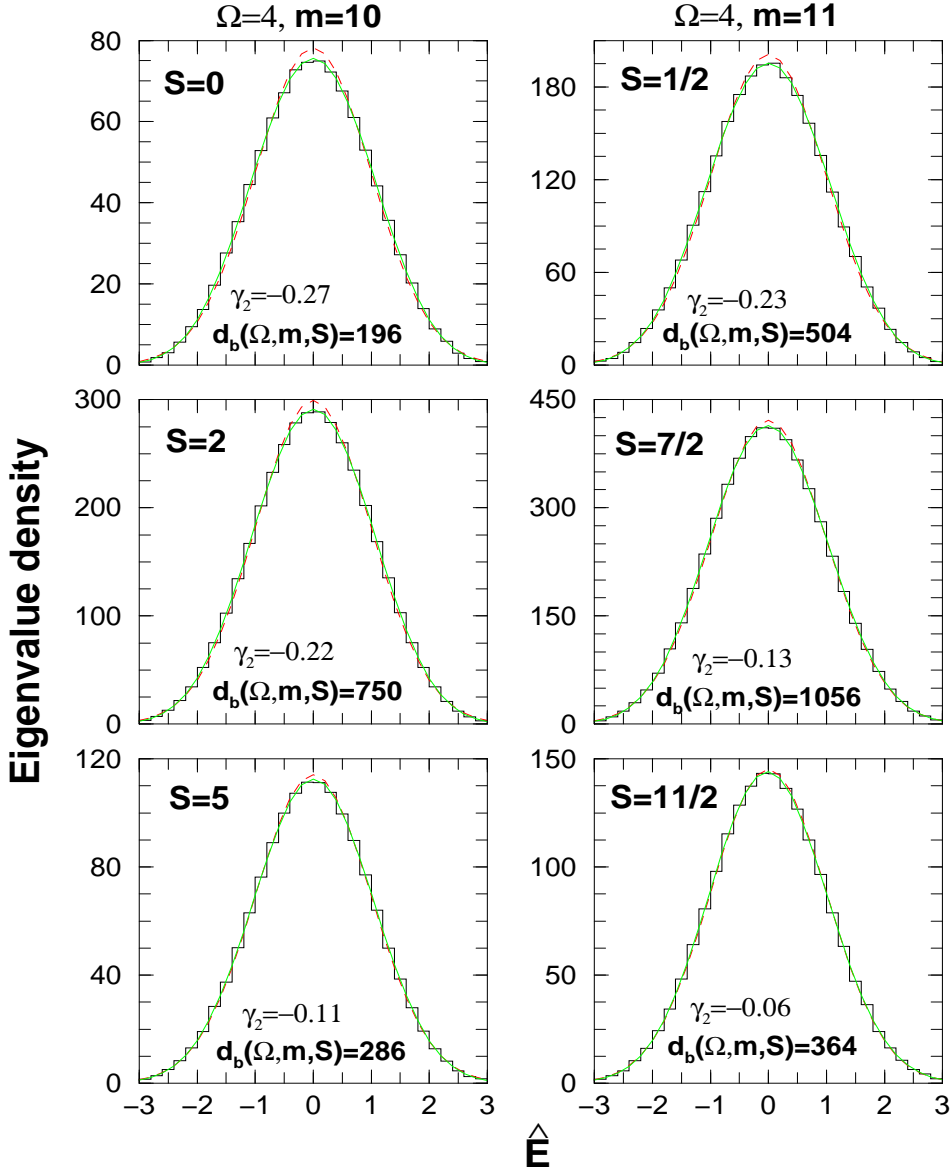}
\caption{Ensemble averaged eigenvalue density 
$\rho^{m,S}(\widehat{E})$ vs $\widehat{E}$ for BEGOE(2)-$\cs$ ensembles with
$\Omega=4$, $m=10$ and $\Omega=4$, $m=11$.  In the figure, histograms
constructed with a bin size $0.2$  are BEGOE(2)-$\cs$ results and they are
compared with Gaussian (dashed red) and Edgeworth (ED) corrected Gaussian
(solid green) forms.  The ensemble averaged values of the excess  parameter
$(\gamma_2)$ are also shown in the figure.  In the plots, the area under the
curves is normalized to the dimensions $d_b(\Omega,m,S)$. See text for further
details.}
\label{den}
\end{figure}

We begin with the ensemble averaged  fixed-($m$, $S$) eigenvalue density
$\rho^{m,S}(E)$, the one-point function for eigenvalues. First we present
the results for BEGOE(2)-$\cs$ ensemble defined by $\whh(1)=0$  
in Eq. (\ref{eq.begoe-s}) and then the Hamiltonian operator is,
\be
\{\wh\}_{\mbox{BEGOE(2)-\cs}} =  \lambda_0\, \{\wv^{s=0}(2)\} +
\lambda_1\, \{\wv^{s=1}(2)\}\,.
\label{eq.v2}
\ee
We have considered a 500 member BEGOE(2)-$\cs$ ensemble with $\Omega=4$ and 
$m=10$ and similarly a 100 member ensemble with $\Omega=4$ and  $m=11$. 
Here and in all other numerical results presented in the present 
chapter, we use
$\lambda_0 = \lambda_1 = \lambda$. In the construction of the ensemble
averaged  eigenvalue densities, the spectra of each member of the ensemble
is first zero centered and scaled to unit width (therefore the densities are
independent of the $\lambda$ parameter).  The eigenvalues are then denoted
by $\widehat{E}$. Given the fixed-($m,S$) energy centroids $E_c(m,S)$ and
spectral widths $\sigma(m,S)$, $\widehat{E}=[E-E_c(m,S)]/\sigma(m,S)$.  Then
the histograms  for the density are generated by combining the eigenvalues
$\widehat{E}$ from all the members of the  ensemble. Results are shown in
Fig. \ref{den} for a few selected $S$ values. The calculations have been
carried out for all $S$ values (the results for other $S$ values are close
to those given in the figure) and also for many other BEGOE(2)-$\cs$
examples.  It is clearly seen that the eigenvalue densities are close to
Gaussian (denoted by $\cg$ below) with the ensemble averaged skewness
($\gamma_1$) and excess ($\gamma_2$) being very small; $|\gamma_1| \sim 0$,
$|\gamma_2| \sim 0.1-0.27$.  The agreements with Edgeworth (ED) corrected
Gaussians are excellent. The ED form that includes $\gamma_1$ and $\gamma_2$
corrections is given by $\rho_{ED}$ in Eq. (\ref{eq.gau1}). 

\begin{figure}
\centering
\includegraphics[width=5in,height=6.75in]{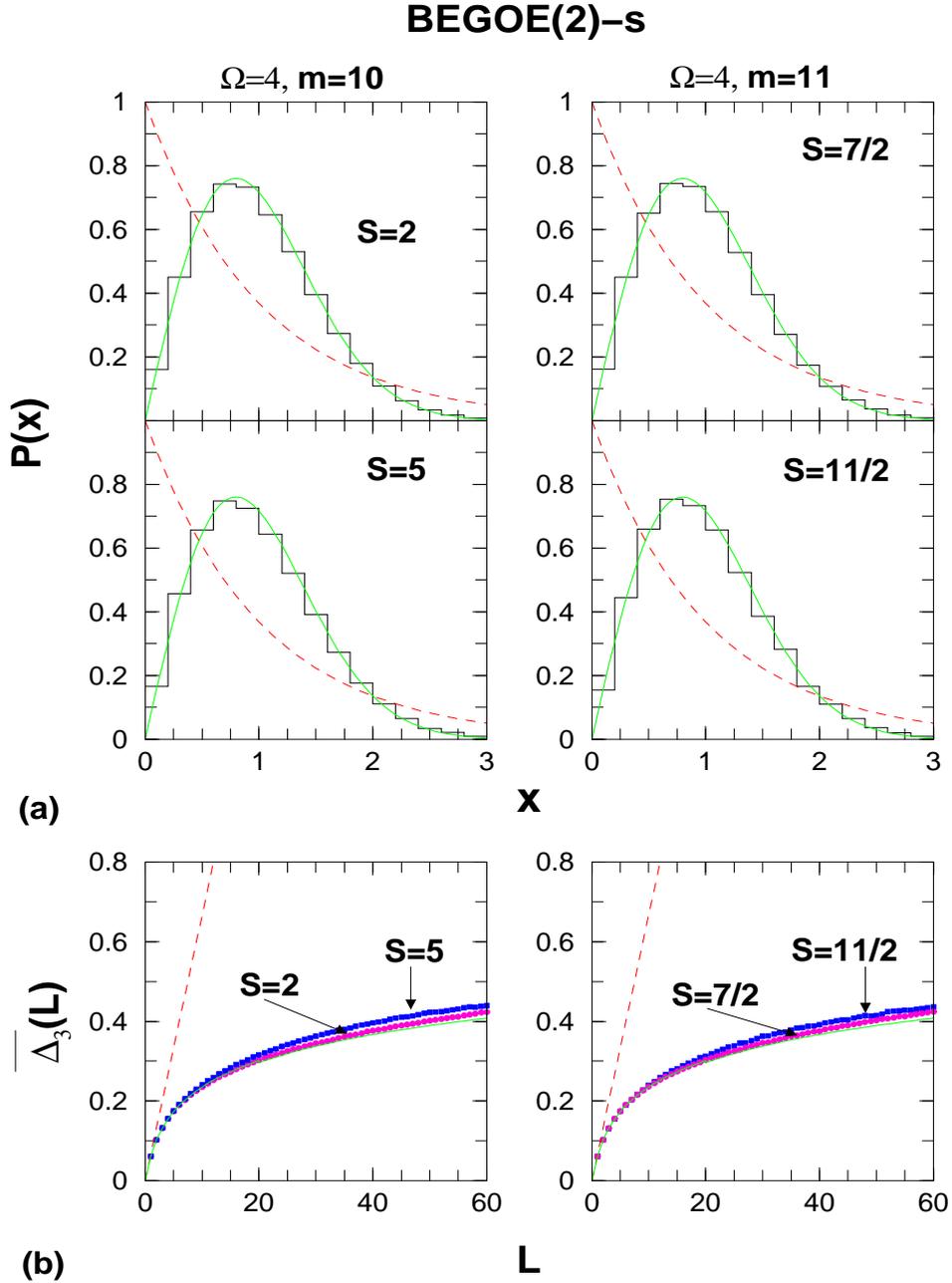}
\caption{(a) Ensemble averaged nearest neighbor spacing 
distribution (NNSD) and (b) Dyson-Mehta statistic $\overline{\Delta_3}(L)$
vs  $L$ for $L \leq 60$. Results are for the same systems considered in
Fig. \ref{den}; first column gives the results for ($\Omega=4$, $m=10$) and
the second column for ($\Omega=4$, $m=11$) systems. The NNSD histograms from
BEGOE(2)-$\cs$ are  compared with Poisson (dashed red) and  GOE (Wigner)
forms (solid green) and similarly the $\overline{\Delta_3}(L)$ results.  In
the NNSD graphs, the bin-size is $0.2$ and  $x$ is the nearest neighbor
spacing in the units  of local mean spacing. See text and Fig. \ref{den} for
further details.}
\label{nnsd-del3}
\end{figure}

For the analysis of level fluctuations (equivalent to studying the two-point
function for the eigenvalues), each spectrum in the ensemble is unfolded
using  a sixth order polynomial correction to the Gaussian and then the 
smoothed density is $\overline{\eta(\widehat{E})}=\eta_{\cg}(\widehat{E}) \{
1+ \sum_{\zeta \geq 3}^{\zeta_0}  ( \zeta !)^{-1} S_{\zeta}
He_{\zeta}(\widehat{E})\}$ with $\zeta_0=6$ \cite{Le-08,PDPK}. The
parameters $S_{\zeta}$ are determined by minimizing
$\Delta^2=\sum_{i=1}^{d_b(\Omega,m,S)} [ F(E_i)-\overline{F(E)}]^2$. The
distribution function $F(E)=\int_{-\infty}^E \eta(x) dx$ and similarly
$\overline{F(E)}$ is defined. We require that the continuous function
$\overline{F(E)}$ passes through the mid-points of the jumps in the discrete
$F(E)$ and therefore, $F(E_i) = (i-1/2)$.  The ensemble averaged
$\Delta_{RMS}$ is $\sim 3$ for $\zeta_0=3$, $\sim 1$ for $\zeta_0=4$ and
$\sim 0.8$ for $\zeta_0=6$ with some variation with respect to $S$. As
$\Delta_{RMS}\sim 0.88$ for GOE, this implies GOE fluctuations set in when
we add 6th order corrections to the asymptotic Gaussian density. Using the
unfolded energy levels of all the members of the BEGOE(2)-$\cs$ ensemble,
the nearest neighbor spacing distribution  (NNSD) that gives information
about level repulsion and  the Dyson-Mehta $\overline{\Delta_3}(L)$
statistic that gives  information about spectral rigidity are studied.
Results for the same systems used in Fig. \ref{den} are shown in Fig.
\ref{nnsd-del3} with $S=2$ and $5$ for $m=10$ and $S=7/2$ and $11/2$ for $m=11$ 
(for other spins, the results are
similar).  In the calculations, middle 80\% of the eigenvalues from each
member are employed. It is clearly seen from the figures that the NNSD are
close to GOE (Wigner) form and the widths of the NNSD are $\sim 0.288$ (GOE
value is $\sim 0.272$). The $\overline{\Delta_3}(L)$ values show some
departures from GOE  for $L \gazz 30$ for $S=S_{max}$ and this could be
because the matrix dimensions are small for $S=S_{max}$ in our examples
(also the systems considered are not strictly in the dense limit and
numerical examples with  much larger $m$ and $\Omega$ with $m >> \Omega$ are
currently not feasible). It is useful to add that $S=S_{max}$ states are
important for boson systems with random interactions  as discussed in 
Secs. \ref{c6s4}-\ref{c6s6} ahead. In conclusion, sixth order unfolding removes
essentially all the secular behavior and then the fluctuations follow 
closely  GOE. This is similar to the result known before for spinless 
boson systems \cite{Le-08,Ch-03}. 

\begin{figure}
\centering
\includegraphics[width=3in,height=6.5in]{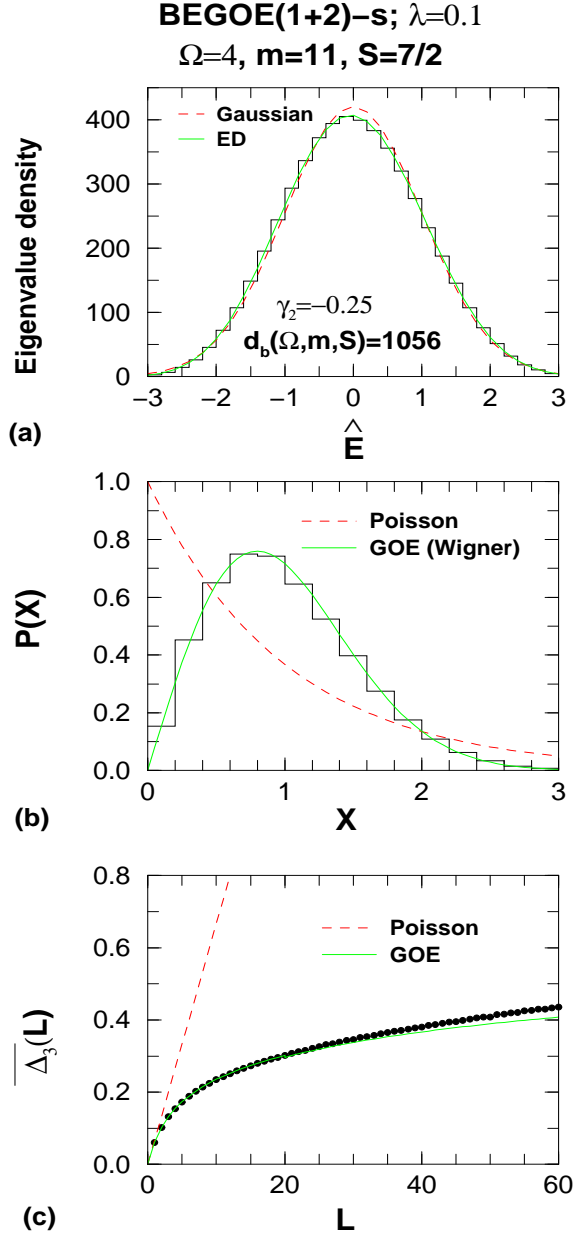}
\caption{(a) Ensemble averaged eigenvalue density 
$\rho^{m,S}(\widehat{E})$, (b) NNSD and (c) 
$\overline{\Delta_3}(L)$ vs.  $L$ for a 100 member BEGOE(1+2)-$\cs$ 
ensemble for $\Omega=4$, $m=11$ and $S=7/2$ system with
$\lambda_0=\lambda_1=\lambda=0.1$ in Eq. (\ref{eq.begoe-s}). For all other
details, see text and Figs. \ref{den} and \ref{nnsd-del3}.}
\label{n4m11}
\end{figure}

\begin{figure}
\centering
\includegraphics[width=5.5in,height=6.5in]{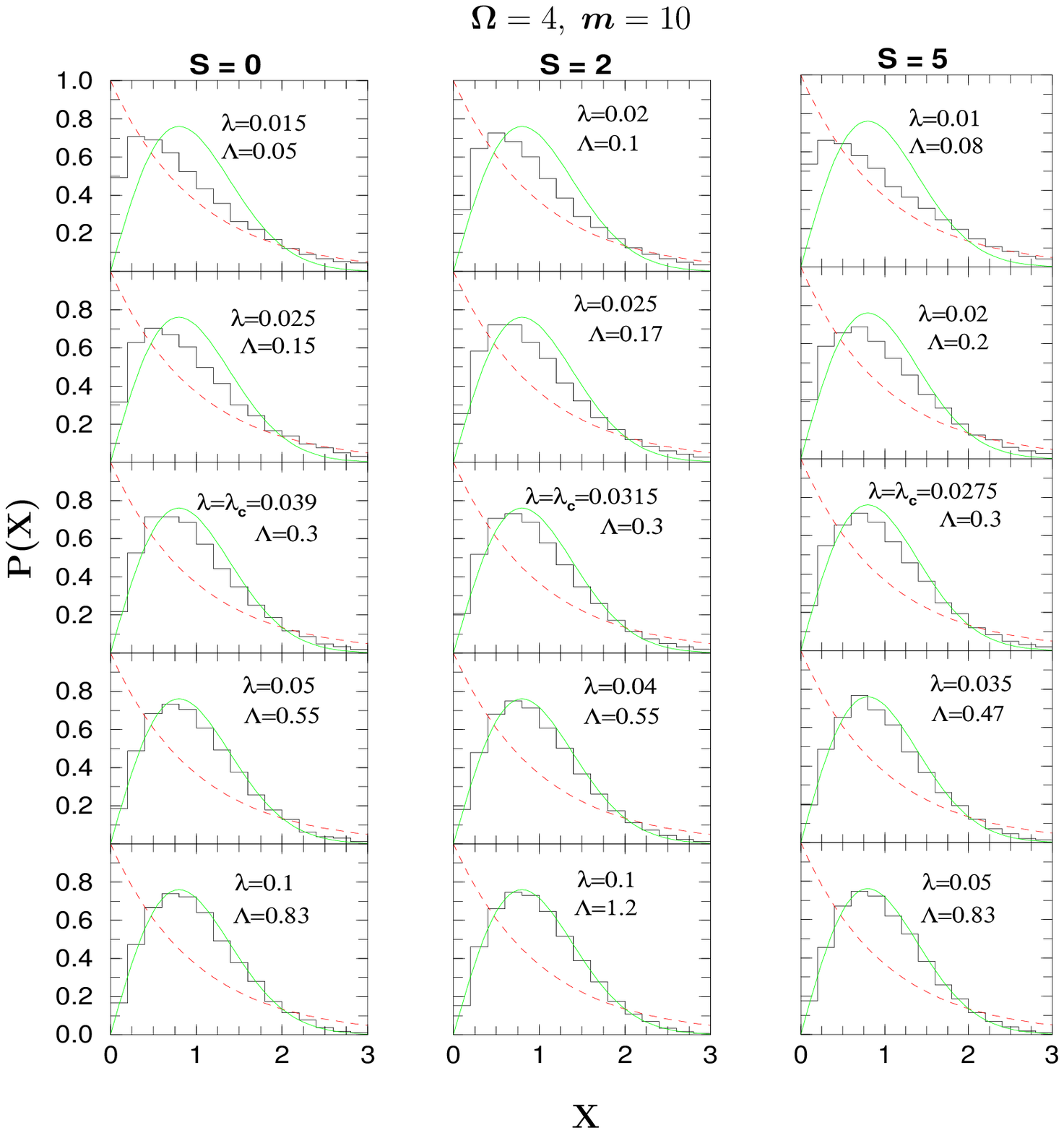}
\caption{NNSD for a 100 member BEGOE(1+2)-$\cs$ ensemble 
with $\Omega=4$, $m=10$ and spins $S=0$, $2$ and $5$.  Calculated NNSD are
compared to the Poisson (red dashed) and Wigner (GOE) (green solid) forms. 
Values of the interaction
strength $\lambda$ and the transition parameter $\Lambda$ are given in  the
figure. The values of $\Lambda$ are deduced as discussed in Chapter \ref{ch2}. 
The chaos marker $\lambda_c$ corresponds to $\Lambda=0.3$ and its values, as
shown in the figure, are $0.039,\;0.0315,\;0.0275$ for $S=0,\;2$, and 5,
respectively. Bin-size for the histograms is $0.2$.}
\label{ns-new}
\end{figure}

Going beyond BEGOE(2)-$\cs$, calculations are also carried out for
BEGOE(1+2)-$\cs$ systems using Eq. (\ref{eq.begoe-s}) with $\lambda_0 =
\lambda_1 = \lambda$. We have verified the Gaussian behavior for the
eigenvalue density for BEGOE(1+2)-$\cs$; an example is shown in  Fig.
\ref{n4m11}(a). This result is essentially  independent of $\lambda$. In
addition, we have also verified that BEGOE(1+2)-$\cs$ also generates level
fluctuations close to GOE for $\lambda \gazz 0.1$ for $\Omega=4$ and $m=10$,
$11$ systems; Figs. \ref{n4m11}(b) and \ref{n4m11}(c) 
show the results for $\lambda=0.1$ for $\Omega=4,\;m=11,\;S=7/2$ system. Going
beyond this, in  Fig. \ref{ns-new}, we show the NNSD results, for a 100
member  BEGOE(1+2)-$\cs$ ensemble with $\Omega=4,\;m=10$ and total spins
$S=0,\;2$  and $5$, by varying $\lambda$ from 0.01 to 0.1 to demonstrate
that as $\lambda$ increases from zero,  there is generically Poisson to GOE
transition. A similar study is reported in Chapter \ref{ch2} for fermion
systems. As discussed there, for very small $\lambda$, the NNSD will be
Poisson (as we use sp energies to be $\epsilon_i=i+1/i$, the $\lambda=0$
limit will not give strictly a Poisson). Moreover, as discussed in detail in
Chapter \ref{ch2}, the variance of the NNSD  can be  written in terms of a
parameter $\Lambda$ ($\Lambda$ is a parameter in a 2$\times$2 random matrix
model that generates Poisson to GOE transition)  with $\Lambda=0$ giving
Poisson, $\Lambda \gazz 1$ GOE and $\Lambda=0.3$ the transition point
$\lambda_c$ that marks the onset of GOE fluctuations. We show in Fig.
\ref{ns-new}, for each $\lambda$, the deduced value of $\Lambda$ from the
variance of the NNSD  (Fig. \ref{nnsd-del3} gives the results for $\lambda \to
\infty$).  As seen from the Fig. \ref{ns-new}, $\lambda_c=
0.039, \; 0.0315, \; 0.0275$ for $S=0$, $2$, and $5$, respectively. Thus
$\lambda_c$ decreases with increasing spin $S$ and this is opposite to the
situation for fermion systems. For a fixed $\Omega$ value, as discussed in
Chapter \ref{ch2}, 
the $\lambda_c$ is inversely proportional to $K$, where $K$ is
the number of many-particle states [defined by $h(1)$] that are directly
coupled by the two-body interaction. For fermion systems, $K$ is
proportional to the variance propagator but not for boson systems as
discussed in \cite{Ch-03}. At present, for BEGOE(1+2)-$\cs$ we don't have a
formula for $K$. However, if we use the variance propagator $Q(\Omega,m,S)$
for the boson systems [see Eq. (\ref{eq.varav1}) and Fig. \ref{prop} ahead],
then qualitatively we understand the decrease in $\lambda_c$ with increasing
spin.

Finally, it is well-known that the Gaussian form for the eigenvalue density is
generic for embedded ensembles of spinless boson (also fermion)  systems; see
Chapter \ref{ch1}. In addition,  ensemble averaged fixed-($m,S$) eigenvalue
densities for the fermion  EGOE(1+2)-$\cs$ are shown to take  Gaussian form; see
Chapter \ref{ch2}.  Hence, from the results shown in Figs. \ref{den} and
\ref{n4m11}(a), it is  plausible  to conclude that the Gaussian form is generic
for BEE (also EE)  with good  quantum numbers. With the eigenvalue density being
close to Gaussian, it is useful to derive formulas for the energy centroids and
ensemble averaged spectral variances. These in turn, as already discussed in
Chapter \ref{ch4}, will also allow us to study the lowest two moments of the
two-point function. From now on, we will drop the ``hat''  over the
operators $H$, $h(1)$ and $V(2)$ when there is no confusion.

\section{Energy Centroids, Spectral Variances and Ensemble Averaged Spectral
Variances and Covariances}
\label{c6s4}

\subsection{Propagation formulas for energy centroids and
spectral variances}
\label{c6s4s1}

Given a general (1+2)-body Hamiltonian $H=h(1)+V(2)$, which is a typical
member of BEGOE(1+2)-$\cs$, the energy centroids will be polynomials in  the
number operator and the $S^2$ operator. As $H$ is of maximum body rank 2,
the polynomial form for the energy centroids is $\lan H \ran^{m,S} =
E_c(m,S) = a_0 +a_1 m +a_2 m^2 + a_3 S(S+1)$. Solving for the $a$'s in terms
of the centroids in one and two-particle spaces, the propagation formula for
the energy  centroids is,
\be
\barr{rcl}
\lan H \ran^{m,S} = E_c(m,S) & = & \l[\lan h(1) \ran^{1,\spin}\r]\;m +
\lambda_0\,
\lan\lan V^{s=0}(2) \ran\ran^{2,0}\;\dis\frac{P^0(m,S)}{4\Omega 
(\Omega-1)} \\ \\
& + & \lambda_1\,\lan\lan V^{s=1}(2) \ran\ran^{2,1}\;\dis\frac{P^1(m,S)}{
4\Omega (\Omega+1)}\;; \nonumber
\earr \label{eq.bcenta}
\ee
\be
\barr{l}
P^0(m,S) = \l[ m(m+2) - 4S(S+1)\r]\;,\;\;
P^1(m,S) = \l[3m(m-2) + 4S(S+1)\r]\;, \\ 
\lan h(1) \ran^{1,\spin} =
\overline{\epsilon} = \Omega^{-1}\;\dis\sum_{i=1}^{\Omega}\;
\epsilon_i\;,
\earr \label{eq.bcent}
\ee
\be
\lan\lan V^{s=0}(2) \ran\ran^{2,0} = \dis\sum_{i
< j}\;V^{s=0}_{ijij}\;,\;\;\;\;
\lan\lan V^{s=1}(2) \ran\ran^{2,1} = \dis\sum_{i \leq j}\;V^{s=1}_{ijij}\;.
\nonumber
\label{eq.bcentb}
\ee
For the energy centroid of a two-body Hamiltonian [member of a
BEGOE(2)-$\cs$], the $h(1)$ part in Eq. (\ref{eq.bcent}) will be absent.

Just as for the energy centroids, polynomial form for the spectral variances
$$
\sigma_{H=h(1)+V(2)}^2(m,S) = \lan H^2 \ran^{m,S} - \l[E_c(m,S)\r]^2
$$ 
is $\sum_{p=0}^4 a_p m^p + \sum_{q=0}^2 b_q m^q S(S+1) + c_0 [S(S+1)]^2$. It
is well-known that the propagation formulas for fermion systems will give
the formulas for the corresponding boson systems by applying $\Omega
\rightarrow -\Omega$ transformation \cite{Ko-79a,KP-80,Ko-81,Cv-82,Ko-05}.
Applying this transformation to the propagation equation for the spectral
variances for fermion systems with spin given by Eq. (\ref{eq.vv1}),
we obtain the propagation equation for $\sigma_{H=h(1)+V(2)}^2(m,S)$ in
terms of inputs that contain the sp energies $\epsilon_i$
defining $h(1)$ and the two-particle matrix elements $V_{ijkl}^s$. The final
result is,
\be
\barr{l}
\sigma_{H=h(1)+V(2)}^2(m,S) = \lan H^2 \ran^{m,S} - \l[E_c(m,S)\r]^2 \\
\\
= \dis\frac{(\Omega-2)mm^\star+2\Omega \;\lan S^2 \ran}{ (\Omega-1) \Omega
(\Omega+1)}\;\;\dis\sum_i\;{\tilde{\epsilon}}_i^2 \\ \\
+ \dis\frac{m^\star P^0(m,S)}{2  (\Omega-1) \Omega
(\Omega+1)}\;\;\dis\sum_i\;{\tilde{\epsilon}}_i \lambda_{i,i}(0) \\ \\
+ \dis\frac{(\Omega-2)m^\star P^1(m,S) +
8 \Omega (m-1) \lan S^2 \ran}{2 (\Omega-1) \Omega (\Omega+1) (\Omega+2)}\;\;
\dis\sum_i\;{\tilde{\epsilon}}_i \lambda_{i,i}(1) \\ \\
+ P^{\nu=1,s=0}(m,S)\;\; \dis\sum_{i,j}\; \lambda_{i,j}^2(0)
+ P^{\nu=1,s=1}(m,S)\;\; \dis\sum_{i,j}\;\lambda_{i,j}^2(1) \\ \\
+ \dis\frac{P^2(m,S) P^0(m,S)}
{4  (\Omega-1) \Omega (\Omega+1) (\Omega+2)}\;\; \dis\sum_{i,j}\;
\lambda_{i,j}(0) \lambda_{i,j}(1) \\ \\
+ P^{\nu=2,s=0}(m,S)\,\lan \l(V^{\nu=2,s=0}\r)^2 \ran^{2,0} +
P^{\nu=2,s=1}(m,S)\,\lan \l(V^{\nu=2,s=1}\r)^2 \ran^{2,1}\;.
\earr \label{eq.bvar}
\ee
The propagators $P^{\nu,s}$'s, which are used later, are
\be
\barr{l}
P^{\nu=1,s=0}(m,S) = \dis\frac{\l[(m+2)m^\star/2 - \lan S^2 \ran\r]
P^0(m,S)}
{8 (\Omega-2) (\Omega-1) \Omega (\Omega+1) } \;,\\ \\
P^{\nu=1,s=1}(m,S) =\dis\frac{8\Omega(m-1)(\Omega+2m-4) \lan S^2 \ran +
(\Omega-2) P^2(m,S) P^1(m,S)}
{8  (\Omega-1) \Omega (\Omega+1) (\Omega+2)^2} \;,\\ \\
P^{\nu=2,s=0}(m,S) = \l[m^\star(m^\star-1) - \lan S^2 \ran\r]
P^0(m,S)/[8 \Omega (\Omega+1)]\;,\\ \\
P^{\nu=2,s=1}(m,S) = \l\{ \l[ \lan S^2 \ran \r]^2 (3\Omega^2+7\Omega+6)/2 +
3m(m-2) m^\star (m^\star+1) \r. \\ \times
\l. (\Omega-1)(\Omega-2)/8 + \l[\lan S^2 \ran/2 \r] \l[ (5\Omega+3)
(\Omega-2)m m^\star +
\Omega(\Omega-1)(\Omega+1) \r. \r. \\
\l. \l. \times (\Omega-6) \r.] \r.\} / 
\l[(\Omega-1) \Omega (\Omega+2)(\Omega+3)\r]\;; \\
P^2(m,S) = 3(m-2) m^\star/2 + \lan S^2 \ran\;,\;\;\;\;m^\star =
\Omega+m/2\;,\;\;\;\; \lan S^2 \ran =S(S+1)\;.
\earr \label{eq.bvar1}
\ee
The inputs in Eq. (\ref{eq.bvar}) are given by,
\be
\barr{l}
{\tilde{\epsilon}}_i=\epsilon_i -\overline{\epsilon} \;,\\ \\
\lambda_{i,i}(s) =  \dis\sum_j\;V_{ijij}^s\;(1+\delta_{ij})
\;-\;(\Omega)^{-1} \;\dis\sum_{k,l}\;V^s_{klkl}\;(1+\delta_{kl}) \;, \\ \\
\lambda_{i,j}(s) = \dis\sum_k\;\dis\sqrt{(1+\delta_{ki})(1+\delta_{kj})}\,
V^s_{kikj}\;\;\;\mbox{for}\;\;\;i \neq j \;,\\ \\
V^{\nu=2,s}_{ijij} =  V^s_{ijij} - \l[\lan V(2)\ran^{2,s} +
(\lambda_{i,i}(s) + \lambda_{j,j}(s))\l(\Omega-2(-1)^s\r)^{-1}\r] \;,\\ \\
V^{\nu=2,s}_{kikj} = V^s_{kikj} - \l(\Omega-2(-1)^s\r)^{-1}\,
\dis\sqrt{(1+\delta_{ki})(1+\delta_{kj})}\,\lambda^s_{i,j}
\;\;\;\mbox{for}\;\;\;i \neq j \;,\\ \\
V^{\nu=2,s}_{ijkl} = V^s_{ijkl}\;\;\;\mbox{for all other cases}\;.
\earr \label{eq.bvar2}
\ee
Eqs. (\ref{eq.bcent}) and (\ref{eq.bvar}) can be applied to individual
members of the BEGOE(1+2) ensemble. On the other hand, it is possible to use
these to obtain ensemble averaged spectral variances and ensemble averaged
covariances in energy centroids just as it was done before for fermion
systems; see Chapter \ref{ch2} for details. Now we  will consider these.

\subsection{Ensemble averaged spectral variances for BEGOE(2)-$\cs$}
\label{spvar}

In the present subsection, we restrict to $H=V(2)$ i.e., BEGOE(2)-$\cs$ and
consider BEGOE(1+2)-$\cs$ at the end.

\begin{figure}
\centering
\includegraphics[width=5in]{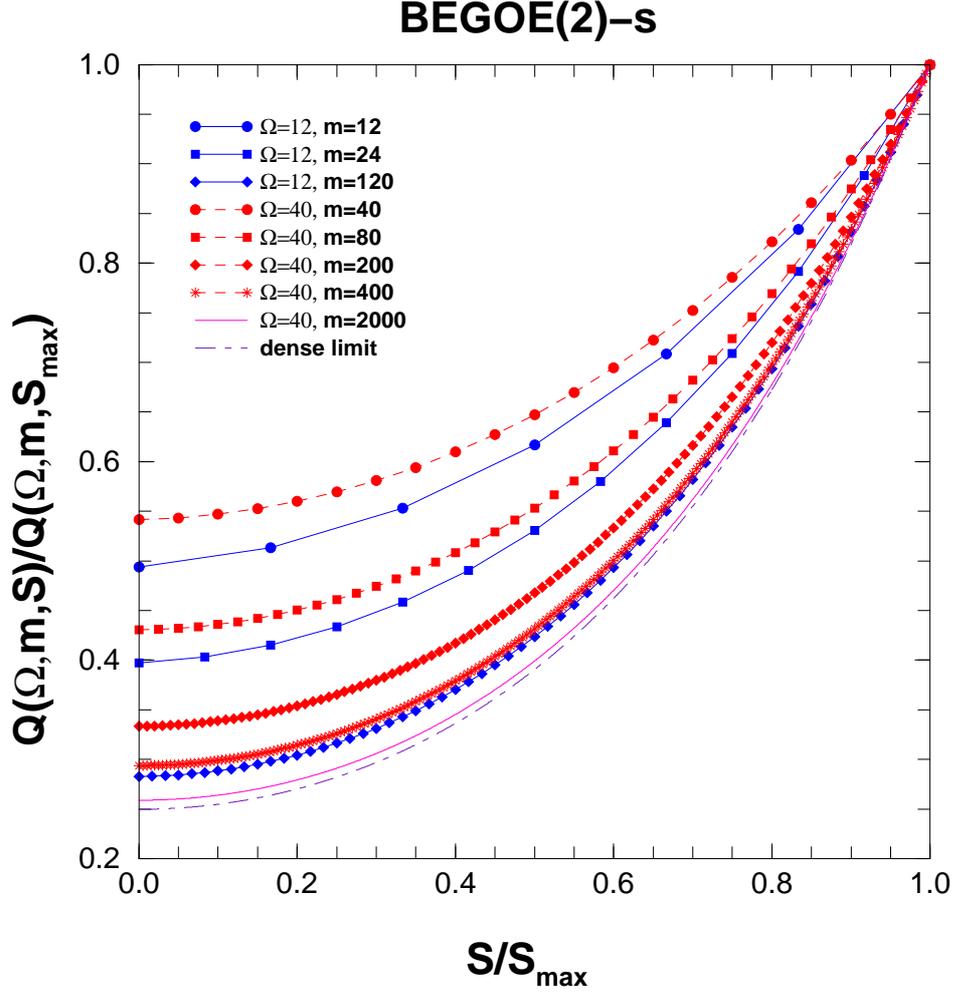}
\caption{BEGOE(2)-$\cs$ variance propagator 
Q($\Omega$,$m$,$S$)/Q($\Omega$, $m$,$S_{max}$) vs $S/S_{max}$ for various
values of $\Omega$ and $m$.   Formula for Q($\Omega$,$m$,$S$) follows from
Eqs. (\ref{eq.bvar1}), (\ref{eq.varav}) and  (\ref{eq.varav1}). Note that
the results in the figure are for $\lambda_0=\lambda_1=\lambda$ in Eq.
(\ref{eq.v2}) and therefore independent of $\lambda$. Dense limit
(dot-dashed) curve corresponds to the result given by Eq. (\ref{asymp2}) 
with $m=2000$.}
\label{prop}
\end{figure}

For the ensemble averaged spectral variances generated by $H$, only the
fourth, fifth, seventh and eighth terms in Eq. (\ref{eq.bvar}) will
contribute. Evaluating the ensemble averages of the inputs in these four
terms, we obtain,
\be
\barr{l}
\overline{\dis\sum_{i,j}\;\lambda_{i,j}^2(0)} =
\lambda_0^2 (\Omega-1)(\Omega-2)(\Omega+2) \;,\\ \\
\overline{\dis\sum_{i,j}\;\lambda_{i,j}^2(1)} =
\lambda_1^2 (\Omega-1)(\Omega+2)^2 \;,\\ \\
\overline{\lan \l(H^{\nu=2,s=0}\r)^2 \ran^{2,0}} =
\lambda_0^2 \dis\frac{(\Omega-3)(\Omega^2+\Omega+2)}{2(\Omega-1)} \;,\\ \\
\overline{\lan \l(H^{\nu=2,s=1}\r)^2 \ran^{2,1}} =
\lambda_1^2 \dis\frac{(\Omega-1)(\Omega+2)}{2}\;.
\earr \label{eq.varinp}
\ee
Note that these inputs follow from the results for EGOE(2)-$\cs$ for
fermions given in Chapter \ref{ch2} by interchanging $s=0$ with $s=1$.
Now the final expression for the ensemble averaged variances is
\be
\barr{rcl}
\overline{\sigma^2_{H}(m,S)} & = & \dis\sum_{s=0,1}
\lambda_s^2 (\Omega-1)(\Omega-(-1)^s 2)(\Omega+2)\;P^{\nu=1,s}(m,S) \\ \\
& + & \lambda_0^2 \dis\frac{(\Omega-3)(\Omega^2+\Omega+2)}{2(\Omega-1)}
\; P^{\nu=2,s=0}(m,S) \\ \\
& + & \lambda_1^2 \dis\frac{(\Omega-1)(\Omega+2)}{2}\; P^{\nu=2,s=1}(m,S)\;.
\earr \label{eq.varav}
\ee
In most of the numerical calculations, we employ $\lambda_0 = \lambda_1
= \lambda$ and then $\overline{\sigma^2_{H}(m,S)}$ takes the form,
\be
\overline{\sigma^2_{H}(m,S)} \stackrel{\lambda_0 = \lambda_1 =
\lambda}{\longrightarrow} \lambda^2 Q(\Omega,m,S)\;.
\label{eq.varav1}
\ee
Expression for the variance propagator $Q(\Omega,m,S)$ follows easily from
Eqs. (\ref{eq.bcent}), (\ref{eq.bvar1}) and (\ref{eq.varav}). In Fig.
\ref{prop}, we show a plot of $Q(\Omega,m,S)/Q(\Omega,m,S_{max})$ vs
$S/S_{max}$ for various $\Omega$ and $m$ values. It is clearly seen that the
propagator value increases as spin increases and this is just opposite to
the result for fermion systems (see Fig. \ref{var}). 
An important consequence of
this is BEGOE(2)-$\cs$ gives ground states with $S=S_{max}$ [for fermion
EGOE(2)-$\cs$, the ground states with random interactions have $S=0$; see Figs.
\ref{var} and \ref{gsspin}]. This result follows from Eq. (\ref{eq.ch4.new})
with $f_m$ replaced by $S$.

Before proceeding further, let us remark that for the BEGOE(1+2)-$\cs$
Hamiltonian $\{H\}=h(1)+\{V(2)\}$, assuming that $h(1)$ is fixed, we have
$\overline{\sigma^2_{H}} = \sigma^2_{h(1)} + \overline{\sigma^2_{V(2)}}$.
The first term $\sigma^2_{h(1)}$ is given by the first term of Eq.
(\ref{eq.bvar}) and the second term is given by Eq. (\ref{eq.varav}).
In the situation  $h(1)$ is represented by an ensemble independent of 
$\{V(2)\}$, we have to replace $\sigma^2_{h(1)}$ by
$\overline{\sigma^2_{h(1)}}$ in $\overline{\sigma^2_H}$.

\subsection{Ensemble averaged covariances in energy centroids and spectral
variances for BEGOE(2)-$\cs$}
\label{c6s4s3}

Normalized covariances in energy centroids $\Sigma_{11}$ 
and spectral variances $\Sigma_{22}$ are
defined by Eq. (\ref{eq.den10}) with $\Gamma = S$.
These define the lowest two moments of the two-point function,
$\cS^{m,S:m^\pr,S^\pr}(E,W)$; see Eq. (\ref{eq.rho}).
For  $(m,S) = (m^\pr,S^\pr)$ they will give information about fluctuations
and in particular about level motion in the ensemble \cite{PDPK}. For $(m,S)
\neq (m^\pr,S^\pr)$, the covariances (cross-correlations) are non-zero for
BEGOE while they will be zero for independent GOE representation for the $m$
boson Hamiltonian matrices with different $m$ or $S$. Note that the 
$\Omega$ value has to be  same for both $(m,S)$ and $(m^\pr,S^\pr)$ systems
so that the Hamiltonian in two-particle spaces remains same. Now we will
discuss analytical  and numerical results  for $\Sigma_{11}$ and numerical
results for $\Sigma_{22}$ for large  values of $(\Omega,m)$ and they are
obtained using the results in Secs. \ref{c6s4s1} and \ref{spvar}.

\begin{figure}
\centering
\includegraphics[width=5in]{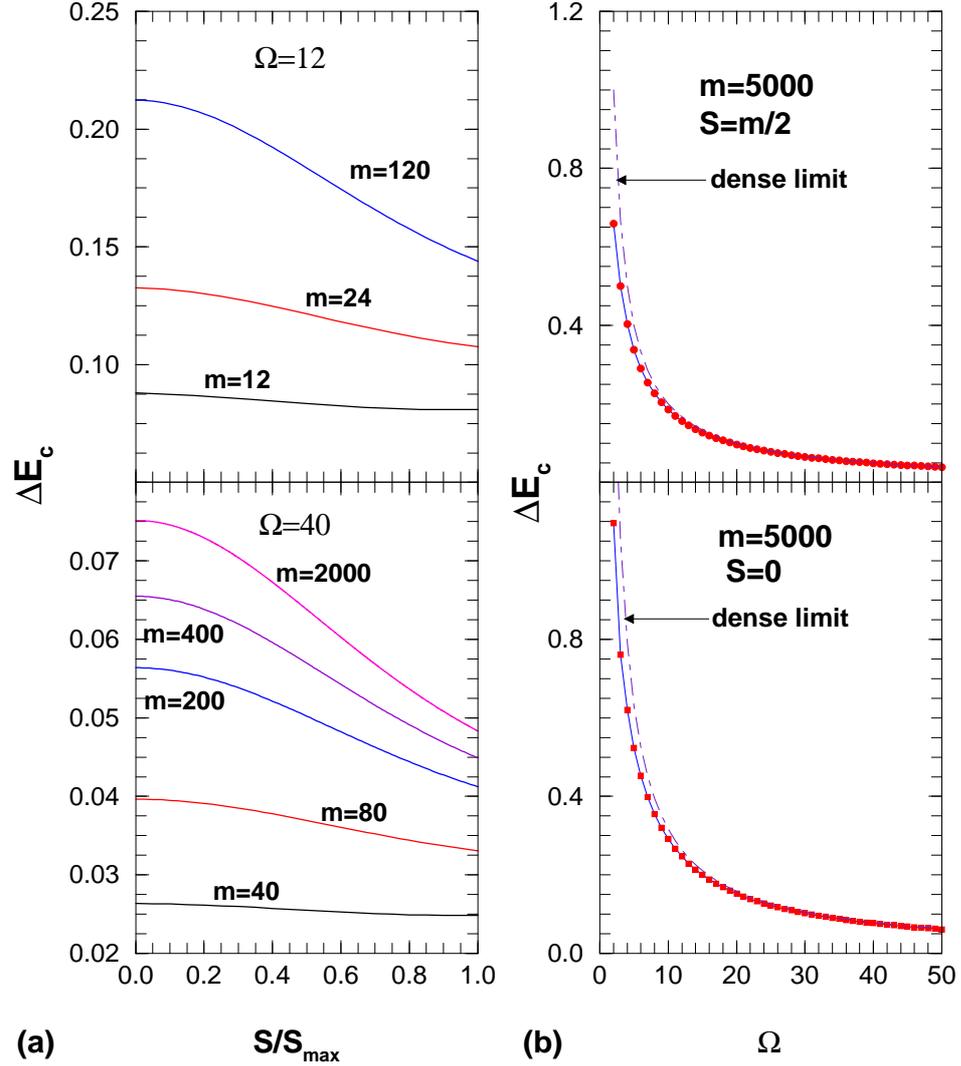}
\caption{(a) Self-correlations $\Sigma^{1/2}_{11}$ in energy
centroids, giving  width $\Delta E_c$ of the fluctuations in energy
centroids scaled to the spectrum width, as a function of spin $S$ for
different values of $m$  and $\Omega$.  (b) Self-correlations as a function
of $\Omega$ for 5000 bosons  with minimum spin $(S=0)$ and maximum spin
$(S=2500)$. Dense limit (dot-dashed) curves for 
$S=0$ and $S=m/2$ in (b) correspond to the results given by 
Eq. (\ref{asymp4}). See text for details.}
\label{sig11-s}
\end{figure}

\begin{figure}
\centering
\includegraphics[width=5.5in]{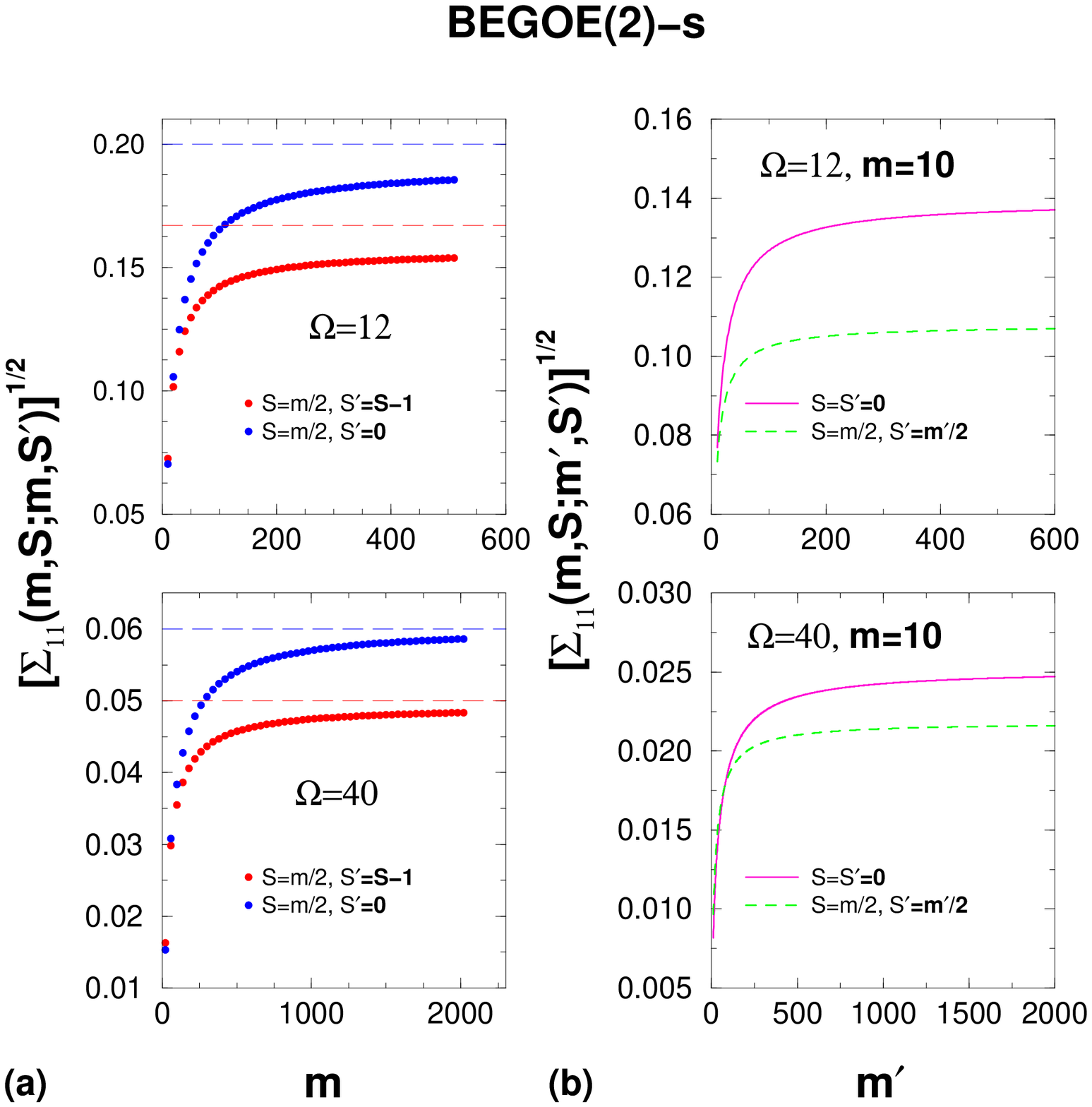}
\caption{Cross-correlations $\Sigma^{1/2}_{11}$ in energy
centroids for  various BEGOE(2)-$\cs$  systems. (a) $\Sigma^{1/2}_{11}$ vs
$m$ with $m=m^\pr$ but different spins  $(S \neq S^\pr)$. (b)
$\Sigma^{1/2}_{11}$ vs $m^\pr$ with $m=10$ and  $S=S^\pr=0$ and
$S=5,S^\pr=m^\pr/2$. The dashed lines in (a) are the dense limit results.
See text for details.}
\label{sig11-c}
\end{figure}

Trivially, the ensemble average of the energy centroids $E_c(m,S)$ will be
zero [note that $H$ is two-body for BEGOE(2)-$\cs$]; i.e., $\overline{\lan
H \ran^{m,S}}=0$. However the covariances in the energy centroids of $H$ are
non-zero and Eq. (\ref{eq.bcent}) gives,
\be
\barr{l}
\overline{\lan H \ran^{m,S}\lan H \ran^{m^\pr,S^\pr}} = \\
\dis\frac{\lambda^2_0}{16\Omega(\Omega-1)}P^0(m,S)\,P^0(m^\pr,S^\pr) +
\dis\frac{\lambda^2_1}{16\Omega(\Omega+1)}P^1(m,S)\,P^1(m^\pr,S^\pr)\;.
\earr \label{eq.eccor}
\ee
Equations (\ref{eq.varav}), (\ref{eq.varav1}) and (\ref{eq.eccor}) allow us
to calculate $\Sigma_{11}$ for any $(\Omega,m,S)$. For $m=m^\pr$ and
$S=S^\pr$, the $[\Sigma_{11}]^{1/2}$ gives the width $\Delta E_c$ of the
fluctuations in the energy centroids. In the numerical calculations, we use
$\lambda_0=\lambda_1=\lambda$ and therefore, $\Sigma_{11}$ and $\Sigma_{22}$
are independent of $\lambda$. Figure \ref{sig11-s} gives some numerical
results for $\Delta E_c$ and it is seen that : (i) for $m >> \Omega$, the
${\Delta}E_c$ is $\sim 20$\% for $S=0$ and it goes down to $\sim 15$\% for
$S=S_{max}=m/2$ for $\Omega=12$; (ii) going from $\Omega=12$ to $40$,
$\Delta E_c$ decreases to $\sim 2-7$\%; (iii) for fixed $(m,\Omega)$, there
is decrease in $\Delta E_c$ with increasing $S$ value; (iv) for fixed
$(m,S)$ and very large $m$ value, there is a sharp decrease in $\Delta E_c$
with increasing $\Omega$ up to $\Omega \sim 20$ and then it slowly converges
to zero. It is possible to understand these results and the results for
cross-correlations $[\Sigma_{11}(m,S:m^\pr,S^\pr)]^{1/2}$, with $(m,S) \neq
(m^\pr,S^\pr)$ as shown in Fig. \ref{sig11-c}, using the asymptotic
structure of $Q(\Omega,m,S)$.

Let us consider the dense limit defined by $m \rightarrow \infty$, $\Omega
\rightarrow \infty$ and $m/\Omega \rightarrow \infty$.
Firstly the $P^{{\nu},s}(m,S)$ in Eq. (\ref{eq.bvar1}) take the simpler
forms, with $\cas^2=S(S+1)$,
\be
\barr{l}
P^{\nu=1,s=0} = \dis\frac{\l(m^2 - 4\cas^2\r)^2}{32 \Omega^4}\;,\;\;\;\;
P^{\nu=1,s=1} = \dis\frac{64m^2\cas^2 \l(3m^2 + 4\cas^2\r)^2}{32 \Omega^4}\;,\\
\\
P^{\nu=2,s=0} = \dis\frac{\l(m^2 - 4\cas^2\r)^2}{32 \Omega^2}\;,\;\;\;\;
P^{\nu=2,s=1} = \dis\frac{3m^4 + 40 m^2\cas^2 + 48(\cas^2)^2}{32 \Omega^2}\;.
\earr \label{asymp1}
\ee
Using these in Eq. (\ref{eq.varav}), with $\lambda_0=\lambda_1=\lambda$, we
have
\be
\barr{l}
\overline{\sigma_H^2(m,S)} = \lambda^2 \dis\frac{\l(m^2 +
4\cas^2\r)^2}{16} \\ \\
\Rightarrow 
\l[\overline{\sigma_H^2(m,S_{max})}\r]^{-1} \overline{\sigma_H^2(m,S)} =
\l[ \dis\frac{m/(m+2)+\cas^2/\cas_{max}^2}{m/(m+2)+1} \r]^2 \;.
\earr \label{asymp2}
\ee
The dense limit result given by Eq. (\ref{asymp2}) with $m=2000$  is
compared with the exact results in Fig. \ref{prop}. Firstly, it should be
noted that for the applicability of Eq. (\ref{asymp2}), $\Omega$ should be
sufficiently large and $m >> \Omega$. Also, the result is independent of
$\Omega$. Comparing with the $\Omega = 12$ and $\Omega = 40$ results, it is
seen that the dense limit result is very close to the $\Omega = 40$ results
for $m \gazz 200$. Thus for sufficiently large value of $\Omega$ and $m
\gazz 5\Omega$, the dense limit result describes quite well the exact 
results.

Simplifying $\overline{\lan H \ran^{m,S} \lan H \ran^{m^\pr,S^\pr}}$ gives 
in the dilute limit,
\be
\barr{l}
\overline{\lan H \ran^{m,S}\lan H\ran^{m^\pr,S^\pr}} \\ \\
=  \dis\frac{\lambda^2}{16 \Omega^2}\l[\l(m^2 -4\cas^2\r)\l\{(m^\pr)^2
-4(\cas^\pr)^2\r\} + \l(3m^2 +4\cas^2\r)\l\{3(m^\pr)^2
+4(\cas^\pr)^2\r\}\r]\;.
\earr \label{asymp3}
\ee
Then $[\Sigma_{11}]^{1/2}$, with $m=m^\pr$ and $S=S^\pr$ (for
$\lambda_0=\lambda_1$) giving $\Delta E_c$, is
\be
[\Sigma_{11}]^{1/2} = \Delta E_c = \dis\frac{\dis\sqrt{2(5m^4 +8m^2\cas^2 +
16(\cas^2)^2)}}{\Omega\l(m^2+4\cas^2\r)}\;.
\label{asymp4}
\ee
Eq. (\ref{asymp4}) gives $[\Sigma_{11}]^{1/2}$ to be $\sqrt{10}/\Omega$ and
$2/\Omega$ for $S=0$ and $S=S_{max}$ and these dense limit results are well
verified by the results in Fig. \ref{sig11-s}(b).  Similarly, Eqs.
(\ref{asymp2}) and (\ref{asymp3}) will give $[\Sigma_{11}]^{1/2}$ to be
$\sqrt{6}/\Omega$ for $(m=m^\pr : S=S_{max},S^\pr=0)$ and $2/\Omega$ for
$(m=m^\pr : S=S_{max},S^\pr=S_{max}-1)$. The upper and lower dashed lines in
Fig. \ref{sig11-c}(a)  for $\Omega = 12$ (similarly for $\Omega = 40$)
correspond to these two dense limit results, respectively. It is seen that
the dense limit results are close to exact results for $\Omega  = 40$ but
there are deviations for $\Omega = 12$. Also, for $\Omega = 40$, the
agreements are good only for $m \gazz 80$ and these are similar to the
results discussed earlier with reference to Fig. \ref{prop}.

\begin{figure}
\centering
\includegraphics[width=5in]{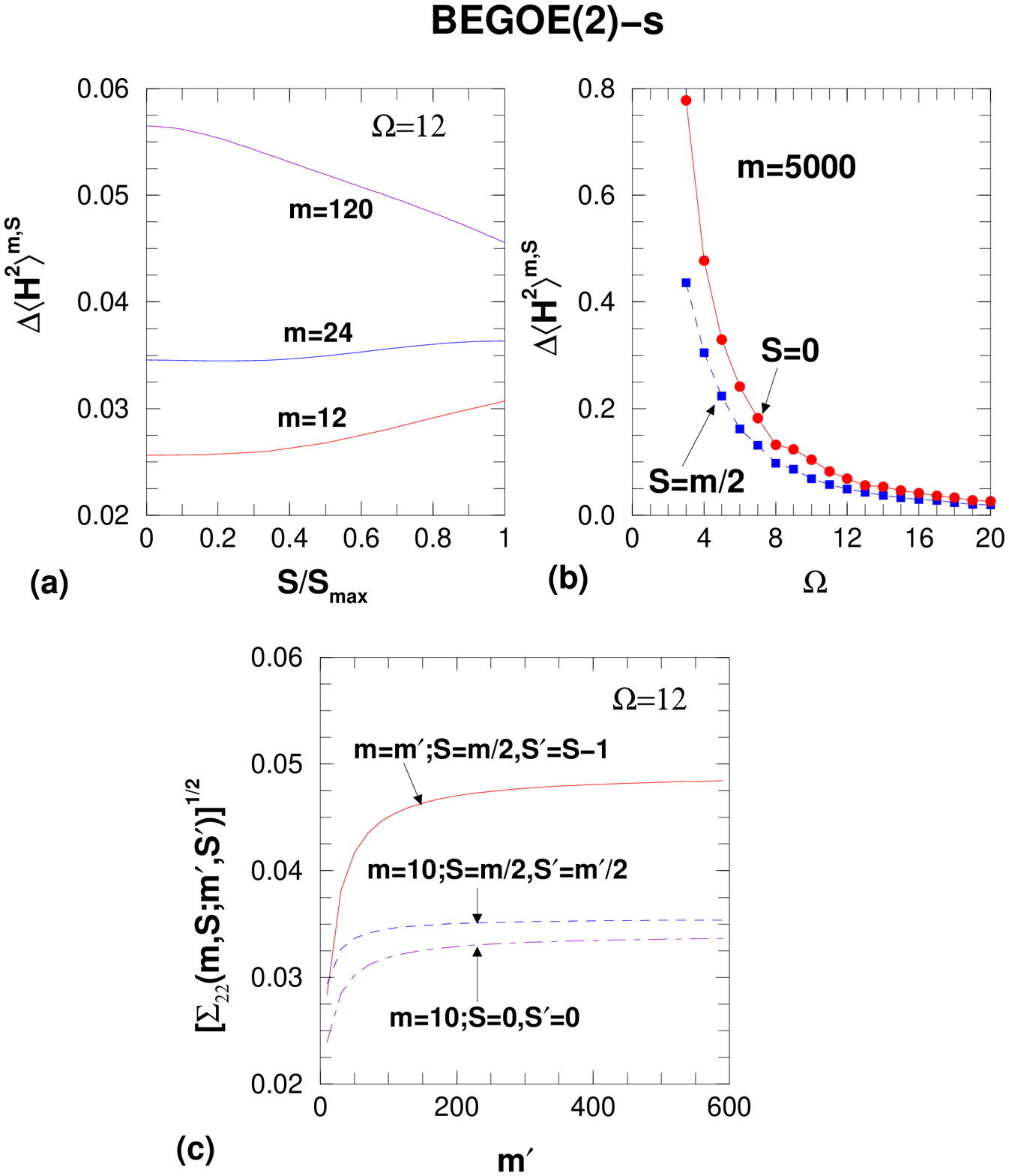}
\caption{Correlations in spectral variances
$\Sigma^{1/2}_{22}$ for various  BEGOE(2)-$\cs$ systems. (a)
Self-correlations, giving width $\Delta\lan H^2 \ran^{m,S}$  of the 
spectral variances,  as a function of spin $S$ for $m=12$, $24$ and $120$
with $\Omega=12$. (b) Self-correlations as a function of $\Omega$ for 5000
bosons with  $S=0$ and $2500$. (c) Three examples for cross-correlation in 
spectral variances with same or different particle numbers  and same or
different spins. All the results are obtained using 500 member  ensembles. 
See text for details.}
\label{sig22}
\end{figure}

Unlike for the covariances in energy centroids, we do not have at present
complete analytical formulation for the covariances in spectral variances.
However, for a given member of BEGOE(2)-$\cs$, generating numerically (on a
computer) the ensembles $\{V^{s=0}(2)\}$ and $\{V^{s=1}(2)\}$ and applying
Eqs. (\ref{eq.bcent}) and (\ref{eq.bvar}) to each member of the ensemble 
will give $\overline{\lan H^2\ran^{m.S}} =  \overline{\sigma^2(m,S)} +
\overline{\l[E_c(m,S)\r]^2}$. This procedure has been used with 500 members
and results for $\Sigma_{22}$ are obtained for various $({\Omega},m,S)$
values. For some examples, results are shown in Fig. \ref{sig22} for both
self-correlations giving the width $\Delta \lan H^2 \ran^{m,S}$ of variances
and cross-correlations $[\Sigma_{22}]^{1/2}$ with $(m,S) \neq
(m^\pr,S^\pr)$. It is seen that $[\Sigma_{22}]^{1/2}$ are always much
smaller than $[\Sigma_{11}]^{1/2}$ just as for EGOE(2) for spinless fermion
systems \cite{Ko-06a}. It is seen from Fig. \ref{sig22}(a) that for
$\Omega=12$, width of the fluctuations in the variances $\lan H^2\ran^{m,S}$
are $\sim 3-5${\%}. Similarly for large $m$, with $\Omega$ very small, the
widths are quite large but they decrease fast with increasing $\Omega$ as
seen from Fig. \ref{sig22}(b). Finally, for $\Omega=12$, the cross-correlations
are $\sim 4$\%. Finally, let us add
that it is important to identify measures involving $\Sigma_{11}$ and
$\Sigma_{22}$ that can be tested using some experiments so that
evidence for BEGOE(2) operation in real quantum systems can be
established.  

\section{Preponderance of $S_{max}=m/2$ Ground States and Natural Spin Order
: Role of Exchange Interaction}
\label{c6s5}

\subsection{Introduction to regular structures with random interactions}
\label{c6s5s1}

Johnson et al \cite{Jo-98} discovered in 1998 that the  nuclear shell-model with
random interactions generates, with high probability, $0^+$ ground states in
even-even nuclei (also generates odd-even staggering in binding energies, the
seniority pairing gap etc.) 
and similarly, Bijker and Frank \cite{BF-00} found
that the interacting boson model ($sd$IBM) of atomic nuclei [in this model, one
considers identical bosons carrying  angular-momentum $\ell=0$ (called $s$
bosons) and $\ell=2$  (called $d$ bosons)] with random interactions generates
vibrational and rotational structures with high probability. Starting with
these, there are now many studies on regular structures in many-body systems
generated by random interactions. See for example \cite{Zh-04a,Zel-04,We-09a}
for reviews on the subject and Sec. \ref{c5s4s3} for results on preponderance of
$+$ve parity ground states. 
More recently,  the effect of random interactions in
the $pn$-$sd$IBM with $F$-spin quantum number  has been studied by Yoshida et al
\cite{Yo-09}. Here, proton and neutron bosons are treated as the two components
of a spin $\spin$ boson and this spin is called $F$-spin. Yoshida et al found
that random interactions conserving $F$-spin generate  predominance of maximum
$F$-spin ($F_{max}$) ground states. It should be noted that the low-lying states
generated by $pn$-$sd$IBM correspond to those of $sd$IBM and all $sd$IBM states
will have $F=F_{max}$. Thus random interactions preserve the property that the
low-lying states generated by $pn$-$sd$IBM are those of $sd$IBM. Similarly,
using shell-model with isospin conserving interactions (here protons and
neutrons correspond to the two projections of isospin $\ct=\spin$), Kirson and
Mizrahi \cite{Ki-07} showed that random interactions generate natural isospin
ordering. Denoting the lowest energy state (les) for a given many nucleon
isospin $T$ by $E_{les}(T)$, the natural isospin ordering corresponds to
$E_{les}(T_{min}) \leq E_{les}(T_{min}+1) \leq \ldots$;  for even-even N=Z
nuclei, $T_{min}=0$.  Therefore, one can ask if  BEGOE(1+2)-$\cs$ generates a
spin ordering.

As an application of BEGOE(1+2)-$\cs$, we present here results for the
probability of gs spin to be $S=S_{max}$ and also for natural spin ordering
(NSO). Here NSO corresponds to $E_{les}(S_{max}) \leq E_{les}(S_{max}-1)
\ldots$. In this analysis, we add the Majorana force or the space  exchange
operator to the Hamiltonian in Eq. (\ref{eq.begoe-s}). Note that $S$ in
BEGOE(1+2)-$\cs$ is similar to $F$-spin in the $pn$-$sd$IBM. First we will
derive the exchange interaction and then present some numerical results.

\subsection{$U(\Omega)$ algebra and space exchange operator}
\label{c6s5s2}

In terms of boson creation ($b^\dagger$) and annihilation ($b$) operators,
the  sp states for $(\Omega)^m$ systems are $\l.\l| i,m_\cs \pm \spin\r.\ran =
b^\dagger_{i,\spin,m_\cs}\l|0\ran$ with
$i=1,2,\ldots,\Omega$. It can be easily identified that  the 4$\Omega^2$
number of one-body operators $A_{ij;\mu}^{r}$,
\be
A_{ij;\mu}^{r} =  \l(b_i^{\dagger}\tilde{b_j}\r)_{\mu}^{r}\;;\;\;\;\;
r=0,\;1\,,
\label{ch6.eq.maj1}
\ee
generate $U(2\Omega)$ algebra. In Eq. (\ref{ch6.eq.maj1}), $\tilde{b}_{i,\spin,
m_\cs}=(-1)^{\spin+m_\cs}  b_{i,\spin,-m_\cs}$.  The $U(2\Omega)$
irreducible representations are denoted trivially by the particle
number $m$ as they must be symmetric irreps $\{m\}$. The $\Omega^2$  number
of  operators $A_{ij}^{0}$ generate $U(\Omega)$ algebra and similarly there
is a $U(2)$ algebra generated by the number operator $\hat{n}$ and the spin
generators $S^1_\mu$,
\be
\barr{l}
\hat{n}=\dis\sqrt{2} \dis\sum_i A_{ii}^0\;;\;\;\;\;
S^1_\mu= \dis\frac{1}{\sqrt{2}} \dis\sum_i A_{ii;\mu}^{1}\;.
\earr \label{eq.ope}
\ee
Then we have the group-subgroup algebra $U(2\Omega) \supset U(\Omega)
\otimes SU(2)$ with $SU(2)$ generated by $S^1_\mu$. Note that
$S_0=S^1_0$, $S_+=-\sqrt{2}S^1_1$ and $S_-=\sqrt{2}S^1_{-1}$.
As the $U(2)$ irreps are
two-rowed, the $U(\Omega)$ irreps have to be two-rowed and they are labeled
by $\{m_1,m_2\}$ with $m=m_1+m_2$ and $S=(m_1-m_2)/2$; $m_1 \geq m_2 \geq
0$. Thus with respect to $U(\Omega) \otimes SU(2)$ algebra, many boson 
states are labeled by $\l| \{m_1,m_2\}, \xi \ran$ or equivalently by 
$\l| (m,S), \xi \ran$, where $\xi$ are extra labels required for a complete 
specification of the states. The quadratic Casimir  operator of the
$U(\Omega)$ algebra is,
\be
C_2[U(\Omega)] = 2\dis\sum_{i,j} A_{ij}^{0} \cdot A_{ji}^{0}
\label{ch6.eq.maj2}
\ee
and its eigenvalues are $\lan C_2[U(\Omega)] \ran^{\{m_1,m_2\}} = 
m_1(m_1+\Omega-1)+m_2(m_2+\Omega-3)$ or equivalently,
\be
\lan C_2[U(\Omega)] \ran^{(m,S)} = 
\dis\frac{m}{2}(2\Omega+m-4)+2S(S+1)\;.
\label{eq.maj4}
\ee
Note that the Casimir invariant of $SU(2)$ is $\hat{S}^2$ with eigenvalues
$S(S+1)$. Now we will show that the space exchange or the Majorana operator 
$\whm$ is simply related to $C_2\l[ U(\Omega)\r]$.

Majorana operator $\whm$ acting on a two-particle state exchanges the spatial
coordinates of the particles (index $i$) and leaves the  spin  quantum
numbers ($m_\cs$) unchanged. The operator form of $\whm$ is
\be
\whm = \dis\frac{\kappa}{2} \dis\sum_{i,j,m_\cs,m_\cs^\pr} \l(
b^\dagger_{j,m_\cs} b^\dagger_{i,m_\cs^\pr}\r) \l(
b^\dagger_{i,m_\cs} b^\dagger_{j,m_\cs^\pr}\r)^\dagger \;.
\label{eq.maj5}
\ee
Equation (\ref{eq.maj5}) gives, with $\kappa$ a constant,
\be
\whm = \dis\frac{\kappa}{2} \l\{ C_2\l[U(\Omega)\r] - \Omega \hat{n}\r\}\;.
\label{eq.maj6}
\ee
Then, combining Eqs. (\ref{eq.maj4}) and (\ref{eq.maj6}), we have
\be
\whm = \kappa \l\{\hat{n}\l( \dis\frac{\hat{n}}{4}-1\r)+\hat{S}^2\r\}\;.
\label{eq.maj7}
\ee
As seen from Eq. (\ref{eq.maj7}), exchange interaction with  $\kappa > 0$
generates gs with $S=S_{min}=0$($\spin$)  for even(odd) $m$ (this is
opposite to the result for `fermion systems' where the exchange interaction
generates gs with $S=S_{max}=m/2$ \cite{Ma-10c,Ja-01}).  Now we will study
the interplay between random interactions and the Majorana force in
generating gs spin structure in boson systems. Note that for states with
boson number fixed, $\whm \propto \hat{S}^2$ as seen from Eq.
(\ref{eq.maj7}) and therefore, from now on, we refer to $\hat{S}^2$ as the
exchange interaction just as in Chapter \ref{ch3}.

\subsection{Numerical results for $S_{max}=m/2$ ground states and natural
spin order}
\label{smax}

\begin{figure}
\centering
\includegraphics[width=5in,height=7.5in]{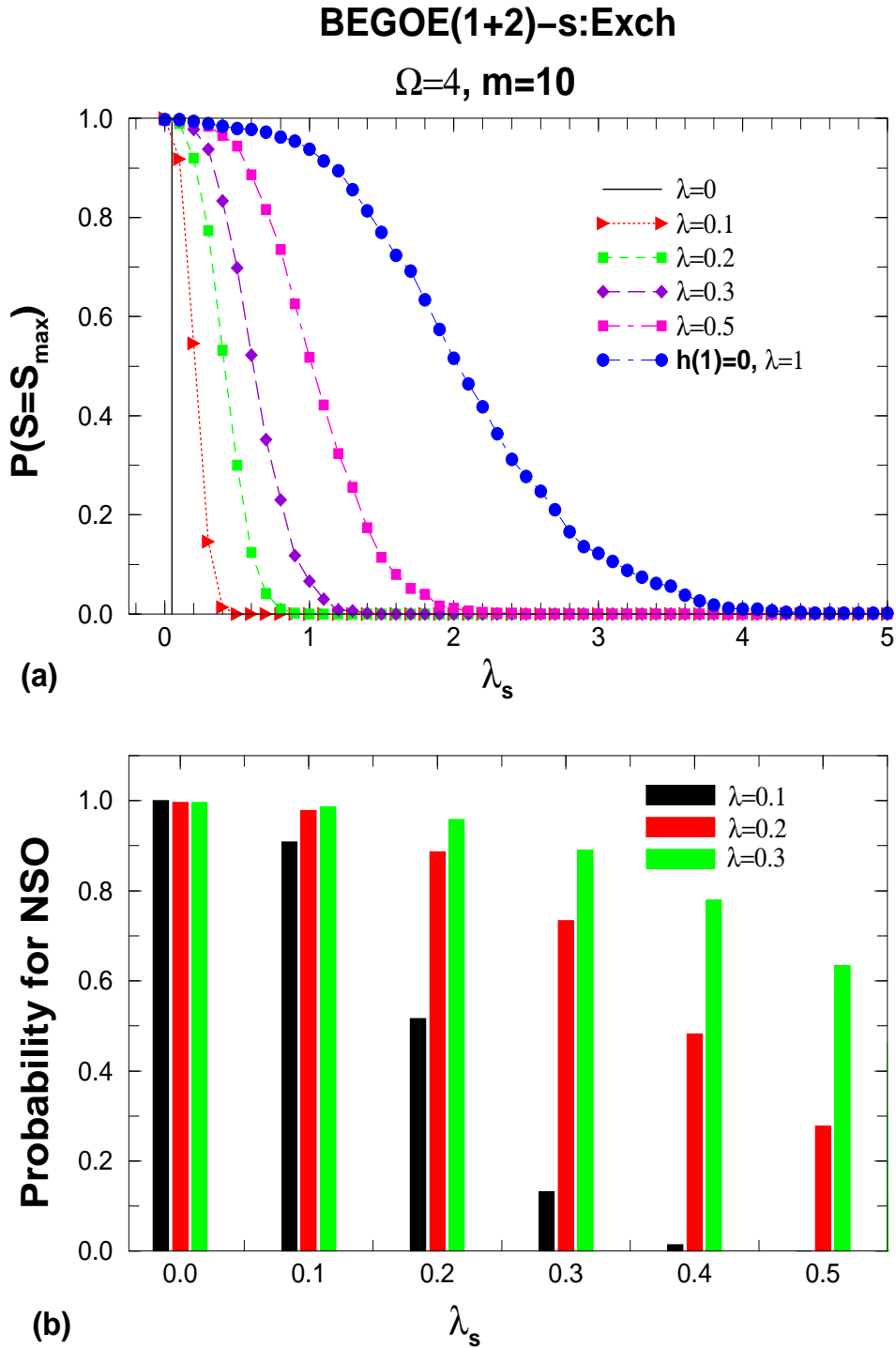}
\caption{(a) Probability  for ground  states to have spin
$S=S_{max}$ as a function of the exchange interaction strength $\lambda_S
\geq 0$. (b) Probability for natural spin order (NSO)  as a function of
$\lambda_S$.  Results are shown for a 500 member 
BEGOE(1+2)-$\cs:\mbox{Exch}$ ensemble generated by Eq. (\ref{H-exch})  for a
system with $\Omega=4$ and $m=10$. Values of the interaction strength
$\lambda$ are shown in the figure.}
\label{pofgs}
\end{figure}

In order to understand the gs structure in BEGOE(1+2)-$\cs$, we have
studied $P(S=S_{max})$, the probability for the gs to be with spin
$S_{max}=m/2$, by adding the exchange term $\lambda_S\,S^2$ with $\lambda_S
> 0$ to the Hamiltonian in Eq. (\ref{eq.begoe-s}) i.e., using
\be 
\{H\}_{\mbox{BEGOE(1+2)-\cs}:\mbox{Exch}} = h(1) + \lambda\,\l[ \,
\{V^{s=0}(2)\} + \{V^{s=1}(2)\}\, \r]\, +\lambda_S\,S^2\;. 
\label{H-exch}
\ee 
Note that the operator $S^2$ is simple in the $(m,S)$ basis. Fig.
\ref{pofgs}(a) gives probability $P(S=S_{max})$ for the ground states to  have
spin $S=S_{max}$ as a function of exchange interaction strength $\lambda_S$
for $\lambda_0=\lambda_1=\lambda=0$, $0.1$, $0.2$, $0.3$ and $0.5$ and also
for $h(1)=0$ with $\lambda=1$. Similarly, Fig.
\ref{pofgs}(b) shows the results for NSO. Calculations are  carried out for 
($\Omega=4$, $m=10$) system using a 500 member ensemble and the  mean-field
Hamiltonian $h(1)$ is as defined in Sec. \ref{c6s2}.  

\subsubsection{Preponderance of $S_{max}=m/2$ ground states}
\label{c6s5s3}

Let us begin with pure random two-body interactions. Then $h(1)=0$ in Eq.
(\ref{H-exch}). Now in the absence of the exchange interaction
($\lambda_S=0$), as seen from Fig. \ref{pofgs}(a), ground states will have
$S=S_{max}$ i.e., the probability $P(S=S_{max}) = 1$. The variance
propagator (see Fig. \ref{prop}) derived earlier gives a  simple explanation
for this by applying the Jacquod and Stone prescription given by Eq.
(\ref{eq.ch4.new}) with $f_m$ replaced by $S$ for BEGOE(1+2)-$\cs$. 
Thus pure random interactions generate preponderance of $S=S_{max}$
ground states. On the other hand, as discussed in Sec. \ref{c6s5s2}, 
the exchange
interaction acts in opposite direction by generating $S=S_{min}$ ground
states. Therefore, by adding the exchange interaction to the  $\{V(2)\}$
ensemble, $P(S=S_{max})$ starts decreasing as the strength $\lambda_S$
($\lambda_S > 0$) starts increasing. For the example considered in Fig.
\ref{pofgs}(a), for $\lambda_S > 4$, we have $P(S=S_{max}) \sim 0$. The
complete variation with $\lambda_S$ is shown in Fig. \ref{pofgs}(a) marked
$h(1) = 0$ and $\lambda = 1$.

Similarly, on the other end, for $\lambda=0$ in Eq. (\ref{H-exch}), we have
$H=h(1)$ in the absence of the exchange interaction. In this situation,  as
all the bosons can occupy the lowest sp state, gs spin $S=S_{max}$.
Therefore, $P(S=S_{max})=1$. When the exchange interaction is turned on,
$P(S=S_{max})$ remains unity until  $\lambda_S$ equals the spacing between
the lowest two sp states divided by $m$. As in our example, the sp energies
are $\epsilon_i=i+1/i$, we have $P(S=S_{max}) = 1$ for $\lambda_S < 0.05$. 
Then  $P(S=S_{max})$ drops to zero for $\lambda_S \geq 0.05$. This variation
with $\lambda_S$ is shown in Fig. \ref{pofgs}(a) marked  $\lambda = 0$. Figure
\ref{pofgs}(a) also shows the variation of $P(S=S_{max})$ with $\lambda_S$ for
several values of $\lambda$ between $0.1$ and $0.5$. It is seen that there
is a critical value ($\lambda_S^c$) of $\lambda_S$ after which $P(S=S_{max})
= 0$ and its value increases with $\lambda$. Also, the variation of
$P(S=S_{max})$ with $\lambda_S$ becomes slower as $\lambda$ increases.

In summary, results in Fig. \ref{pofgs}(a)  clearly show that with random
interactions there is preponderance of $S=S_{max}=m/2$ ground states. This is
unlike for fermions where there is preponderance of $S=S_{min}=0(\spin)$ ground
states for $m$ even(odd); see Fig. \ref{gsspin}. With the addition of the
exchange  interaction,  $P(S=S_{max})$ decreases and finally goes to zero for 
$\lambda_S \geq  \lambda_S^c$ and the value of $\lambda_S^c$ increases with
$\lambda$. We have also carried out calculations for ($\Omega=4$, $m=11$) system
using a 100 member ensemble  and the results are close to those given in Fig.
\ref{pofgs}(a).  All these explain the results given in \cite{Yo-09} where random
interactions are employed within $pn$-$sd$IBM. 

\subsubsection{Natural spin ordering}
\label{c6s5s4}

For the system considered in  Fig. \ref{pofgs}(a), for each member of the
ensemble, eigenvalue of the  lowest state for each spin $S$ is calculated
and using these, we have obtained total number of members $N_\lambda$ having
NSO as a function of  $\lambda_S$ for $\lambda=0.1, 0.2$ and $0.3$ using the
Hamiltonian given in Eq. (\ref{H-exch}). As stated in Sec. \ref{c6s5s1}, the NSO
here corresponds to (as $S=S_{max}$ is the spin of the gs of the system)
$E_{les}(S_{max}) < E_{les}(S_{max}-1) < E_{les}(S_{max}-2) < \ldots$.  The
probability for NSO is $N_\lambda/500$ and the results are shown in Fig.
\ref{pofgs}(b).  In the absence of the exchange interaction, as seen from Fig.
\ref{pofgs}(b), NSO is found in all the members  independent of $\lambda$.
Thus random interactions strongly favor NSO. The presence of exchange
interaction reduces the probability for NSO. Comparing Figs. \ref{pofgs}(a)
and (b),  it is clearly seen that with increasing exchange
interaction strength, probability for gs state spin  to be  $S=S_{max}$ is
preserved for much larger values of $\lambda_S$ (with a fixed $\lambda$)
compared to the NSO. Therefore for preserving both $S=S_{max}$ gs and the
NSO with high probability, the $\lambda_S$ value has to be small. We have
also verified this for the ($\Omega=4$, $m=11$) system. Finally, it is
plausible to argue  that the  results in Fig. \ref{pofgs}  obtained using
BEGOE(1+2)-$\cs$ are generic for boson systems with spin.  Now we will turn
to pairing in BEGOE(2)-$\cs$.

\section{Pairing in BEGOE(2)-$\cs$}
\label{c6s6}

Pairing correlations are known to be important not only for fermion systems (see
Chapter \ref{ch3}) but also for boson systems \cite{Pe-10}. An important issue
that is raised in the recent years is: to what extent random interactions carry
features of pairing. See Chapter \ref{ch3} and \cite{Zh-04a,Zel-04,Ho-07} for
some results for fermion systems. In order to address this question for boson
systems, first we will identify the pairing algebra in $(\Omega,m,S)$ spaces of
BEGOE(2)-$\cs$. Then we will consider expectation values of the pairing
Hamiltonian in the eigenstates  generated by BEGOE(2)-$\cs$ as they carry
signatures of pairing.

\subsection{$U(2\Omega)\supset [U(\Omega) \supset SO(\Omega)] \otimes 
SU_S(2)$ Pairing symmetry}
\label{c6s6s1}

In constructing BEGOE(2)-$\cs$, it is assumed that spin is a good symmetry
and thus the $m$-particle states carry spin ($S$) quantum number.  Now,
following the $SO(5)$ pairing algebra for fermions \cite{Fl-64}, it is
possible to consider pairs that are vectors in spin space. The pair creation
operators $P_{i:\mu}$ for the level $i$ and the generalized  pair creation
operators (over the $\Omega$ levels) $P_\mu$, with $\mu=-1,0,1$, in spin
coupled representation, are
\be
P_\mu =  \dis\frac{1}{\sqrt{2}} \dis\sum_i \l(b^\dagger_i 
b^\dagger_i\r)^1_\mu = \dis\sum_i P_{i:\mu}\;,\;\;\;
\l( P_\mu \r)^{\dagger}=\dis\frac{1}{\sqrt{2}}
\dis\sum_i(-1)^{1-\mu}\l(\tilde{b_i}\tilde{b_i}\r)^1_{-\mu}\;.
\label{eq.npa1}
\ee
Therefore in the space defining BEGOE(2)-$s$, the  pairing Hamiltonian $H_p$
and its two-particle matrix elements are,
\be
H_p = \dis\sum_\mu P_\mu \l( P_\mu \r)^\dagger\;,\;\;\;\lan (k\ell) s 
\mid H_p \mid (ij) s \ran = \delta_{s,1}\,\delta_{i,j} \, \delta_{k,\ell}\;.
\label{eq.npa2}
\ee
With this, we will proceed to identify and analyze the pairing algebra. It
is easy to verify that the  $\Omega(\Omega-1)/2$ number of operators
$C_{ij}=A_{ij}^{0}-A_{ji}^{0}$, $i>j$ generate a $SO(\Omega)$ subalgebra of
the $U(\Omega)$ algebra; $A_{ij:\mu}^r$ are defined in Eq. (\ref{ch6.eq.maj1}). 
Therefore we have $U(2\Omega) \supset [U(\Omega)
\supset SO(\Omega)] \otimes SU(2)$. We will show that the irreps of
$SO(\Omega)$ algebra are uniquely labeled by the seniority quantum number
$v$ and a reduced spin $\tilde{s}$  similar to the reduced isospin
introduced in the context of nuclear
shell-model \cite{Fl-52} and they in turn define the eigenvalues of $H_p$.
The quadratic Casimir operator of the $SO(\Omega)$ algebra is,
\be
C_2[SO(\Omega)] = 2\dis\sum_{i>j} C_{ij} \cdot C_{ji} \;.
\label{eq.npa5}
\ee
Carrying out angular-momentum algebra \cite{Ed-74} it can be shown that,
\be
C_2[SO(\Omega)] = C_2[U(\Omega)] - 2\,H_p - \hat{n}\;.
\label{eq.npa6}
\ee
The quadratic Casimir operator of the $U(\Omega)$ algebra is given in 
Eq. (\ref{ch6.eq.maj2}). Before discussing the eigenvalues of the pairing
Hamiltonian $H_p$, let us first consider the irreps of $SO(\Omega)$.

Given the two-rowed $U(\Omega)$ irreps $\{m_1,m_2\}$; $m_1+m_2=m$, 
$m_1-m_2=2S$, it should be clear that the $SO(\Omega)$ irreps should be of 
$[v_1,v_2]$ type and for later simplicity we use $v_1+v_2=v$ and 
$v_1-v_2=2\tilde{s}$. The quantum number $v$ is called seniority and
$\tilde{s}$ is called reduced spin; see also Appendix \ref{c3a1}. 
The $SO(\Omega)$ irreps for a given
$\{m_1,m_2\}$ can be  obtained as follows. First expand the $U(\Omega)$
irrep $\{m_1,m_2\}$ in  terms of totally symmetric irreps, 
\be
\{m_1,m_2\}=\{m_1\} \times \{m_2\}-\{m_1+1\} \times \{m_2-1\}\;. 
\label{eq.npa22}
\ee
Note that the irrep multiplication in Eq. (\ref{eq.npa22}) is a Kronecker
multiplication \cite{Ko-06c,Wy-70}.
For a totally symmetric $U(\Omega)$ irrep $\{m^\pr\}$, the $SO(\Omega)$ 
irreps are given by the well-known result
\be
\{m^\pr\} \to [v] = [m^\pr] \oplus [m^\pr-2] \oplus \ldots \oplus [0] 
\mbox{ or } [1]\;. 
\label{eq.npa23}
\ee
Finally, reduction of the Kronecker product of two symmetric $SO(\Omega)$
irreps $[v_1]$ and $[v_2]$, $\Omega>3$ into $SO(\Omega)$ irreps $[v_1,v_2]$
is given by (for $v_1 \geq v_2$) \cite{Ko-06c,Wy-70},
\be
[v_1] \times [v_2] = \dis\sum_{k=0}^{v_2}\dis\sum_{r=0}^{v_2-k}
[v_1-v_2+k+2r,k] \oplus \;.
\label{eq.npa10}
\ee
Combining Eqs. (\ref{eq.npa22}), (\ref{eq.npa23}) and (\ref{eq.npa10}) gives
the $\{m_1,m_2\} \to [v_1,v_2]$ reductions. It is easy to implement this
procedure on a computer. 

Given the space defined by $\l| \{m_1,m_2\}, [v_1,v_2], \alpha \ran$, with
$\alpha$ denoting extra labels needed for a complete specification of the
state, the eigenvalues of $C_2[SO(\Omega)]$ are \cite{Ko-06c}
\be
\lan C_2[SO(\Omega)] \ran^{\{m_1,m_2\}, [v_1,v_2]} = 
v_1(v_1+\Omega-2)+v_2(v_2+\Omega-4)\;. 
\label{eq.npa31}
\ee
Now changing $\{m_1,m_2\}$ to $(m,S)$ and $[v_1,v_2]$ to $(v,\tilde{s})$ and
using Eqs. (\ref{eq.npa6}) and (\ref{eq.maj4}) will give the formula for the
eigenvalues of the  pairing Hamiltonian $H_p$. The final result is,
\be
E_p(m,S,v,\tilde{s}) = \lan H_p \ran^{m,S,v,\tilde{s}} = 
\dis\frac{1}{4}(m-v)(2\Omega-6+m+v) + [S(S+1)-\tilde{s}(\tilde{s}+1)]\;.
\label{eq.npa8}
\ee
This is same as the result that follows from Eq. (18) of \cite{Fl-64} for
fermions by using $\Omega \to -\Omega$ symmetry; see also Eq. (\ref{eq.pa4}). 
From now on, we denote the
$U(\Omega)$ irreps by $(m,S)$ and $SO(\Omega)$ irreps by $(v,\tilde{s})$. In
Table \ref{red}, for $(\Omega,m) = (4,10),(5,8)$ and $(6,6)$ systems, given
are the $(m,S) \to (v,\tilde{s})$ reductions, the pairing eigenvalues given
by Eq. (\ref{eq.npa8}) in the spaces defined by these irreps and also the
dimensions of the  $U(\Omega)$ and $SO(\Omega)$ irreps. The dimensions 
$d_b(\Omega,m,S)$ of the $U(\Omega)$ irreps $(m,S)$ are given by Eq.
(\ref{eq.msdim}). Similarly, the dimension $\cads(v_1,v_2) \Leftrightarrow 
\cads(v,\tilde{s})$ of the $SO(\Omega)$ irreps  $[v_1,v_2]$ follow from Eqs.
(\ref{eq.npa23}) and (\ref{eq.npa10}) and they will give
\be
\barr{l}
\cads(v_1,v_2) =  \cads(v_1) \cads(v_2) - \dis\sum_{k=0}^{v_2-1} 
\dis\sum_{r=0}^{v_2-k} \cads(v_1-v_2+k+2r,k)\;;\\ \\
\cads(v) = \dis\binom{\Omega+v-1}{v} - \dis\binom{\Omega+v-3}{v-2}\;.
\earr \label{ch6.eq.dim}
\ee
Note that in general the $SO(\Omega)$ irreps $(v,\tilde{s})$ can appear more
than once in the reduction of $U(\Omega)$ irreps $(m,S)$. For example,
$(2,1)$ irrep of  $SO(\Omega)$ appears twice in the reduction of the
$U(\Omega)$ irrep $(10,1)$.   

It is useful to remark that just as the fermionic $SO(5)$ pairing algebra
for nucleons in $j$ orbits \cite{Pa-65,He-65,Fl-64}, there will be a
$SO(4,1)$ complementary pairing algebra corresponding to the  $SO(\Omega)$
subalgebra. The ten operators $P^1_\mu$,  $(P^1_\mu)^\dagger$, $S^1_\mu$ and
$\hat{n}$ form the $SO(4,1)$ algebra. Their commutation relations follow from 
the basic two commutation relations,
\be
\barr{rcl}
\l[ P^1_{\mu_1}, \l(P^1_{\mu_2}\r)^\dg \r] 
& = & - \l(\Omega + \hat{n} + 2\mu S^1_0 \r) \; 
\;\; \mbox{for} \;\; \mu_1=\mu_2=\mu \\ \\
& = & 2\dis\sqrt{2} (-1)^{\mu_2} \lan 1 \mu_1 1 -\mu_2 \mid 1 \mu_1-\mu_2 \ran 
S^1_{\mu_1-\mu_2} \;\;\; \mbox{for} \;\; \mu_1 \neq \mu_2 \;,
\earr \label{ch6.eq.so41}
\ee
\be
\barr{rcl}
\l[ \dis\sqrt{2} P^1_{\mu_1}, \l(b^\dg \tilde{b} \r)^s_{\mu_2}\r] & = & 
\dis\sqrt{6(2s+1)} (-1)^{s+1} 
\lan 1 \mu_1 s \mu_2 \mid 1 \mu_1+\mu_2 \ran
\\ \\
& \times &
\l\{ \barr{ccc} 1 & \spin & \spin
\\ \spin & 1 & s \earr \r\} \; P^1_{\mu_1+\mu_2} 
\;; \;\;\;\;s=0,\;1\;. \nonumber
\earr \label{ch6.eq.so41a}
\ee
It is possible, in principle,  to exploit this algebra to derive properties of the eigenstates 
defined by the pairing Hamiltonian. 

\subsection{Pairing expectation values}
\label{c6s6s2}

\begin{figure}[htp]
\centering
\includegraphics[height=5.5in,width=7in]{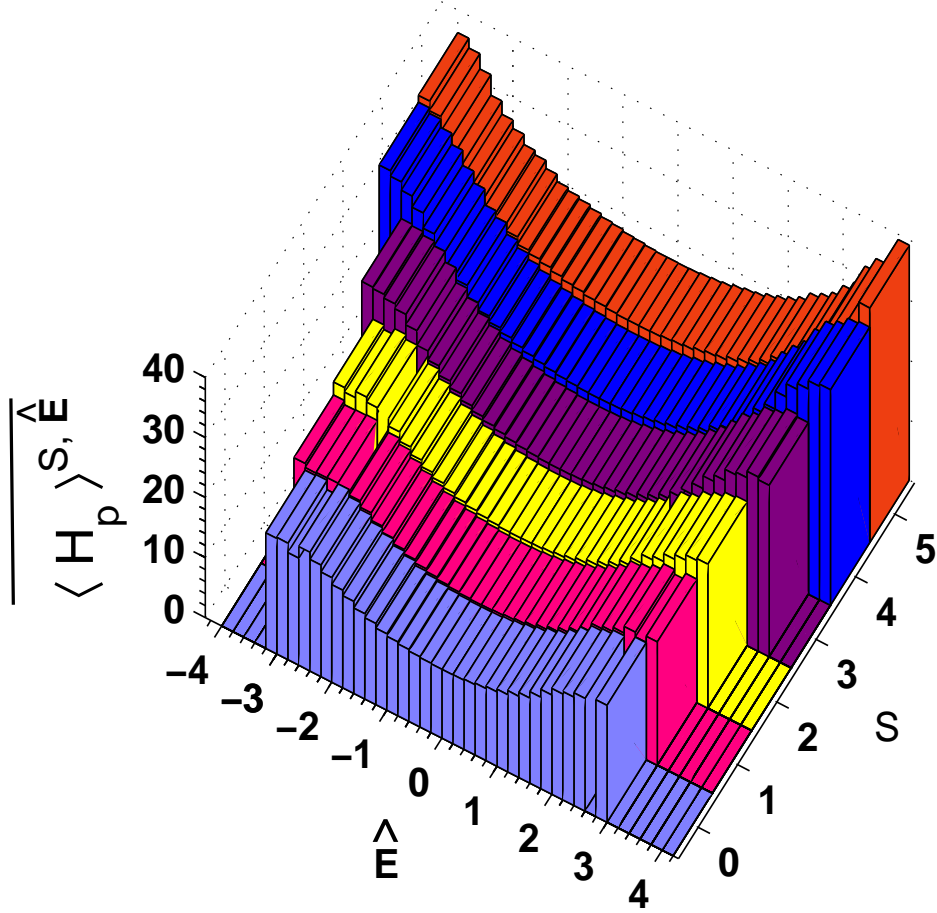}
\caption{Ensemble averaged pairing expectation values
$\overline{\lan H_p \ran^{S,\widehat{\bee}}}$ vs $\widehat{\bee}$ and $S$,
shown as a 3D  histogram, for a 500 member BEGOE(2)-$\cs$ ensemble with
$\Omega=4$ and  $m=10$. The bin-size is 0.2 for $\widehat{\bee}$. Note that
the $\widehat{\bee}$ label in this figure is different from the
$\widehat{E}$ used in Figs. \ref{den} and \ref{n4m11}(a).}
\label{pair}
\end{figure}

Pairing expectation values are defined by $\lan H_p \ran^{S,E} =  \lan m,S,E
\mid H_p \mid m,S,E \ran$ for eigenstates with energy $E$ and spin $S$
generated by a Hamiltonian $H$ for a system of $m$ bosons in $\Omega$ number
of sp orbitals (for simplicity, we have dropped $\Omega$ and $m$ labels in
$\lan H_p \ran^{S,E}$). In our analysis, $H$ is a member of BEGOE(2)-$\cs$.
As we will be comparing the results for all spins at a given energy $E$, for
each member of the ensemble the eigenvalues for all spins are zero centered
and normalized using the $m$-particle energy centroid $E_c(m)=\lan H \ran^m$
and spectrum width $\sigma(m)=[\lan H^2 \ran^m -\{E_c(m)\}^2]^{1/2}$.
Then the eigenvalues $E$ for all $S$ are changed to 
$\widehat{\bee}=[E-E_c(m)]/\sigma(m)$. Using the method described in Sec.
\ref{c6s2}, the $H_p$ matrix is constructed in good $M_S$ basis and 
transformed into
the  eigenbasis of a given $S$  for each member of the  BEGOE(2)-$\cs$
ensemble. Then the ensemble average of the diagonal elements of the $H_p$
matrix will give the ensemble averaged pairing expectation values
$\overline{\lan H_p \ran^{S,E}} \Leftrightarrow  \overline{\lan H_p
\ran^{S,\widehat{\bee}}}$.  Using this procedure for a  500 member
BEGOE(2)-$\cs$ ensemble with $\Omega=4$, $m=10$ and $S=0-5$, results for 
$\overline{\lan H_p \ran^{S,\widehat{\bee}}}$ as a  function of energy
$\widehat{\bee}$ (with $\widehat{\bee}$ as described above)  and spin $S$
are obtained and they are shown  as a 3D histogram in Fig. \ref{pair}. From
Table \ref{red}, it is seen that the maximum value of the eigenvalues
$E_p(m,S,v,\tilde{s})$ increases with spin $S$ for a fixed-$(\Omega,m)$. 
The values are $28$, $32$, $34$, $42$, $48$, and $60$ for $S=0-5$,
respectively for $\Omega=4$ and $m=10$.  Numerical results in Fig.
\ref{pair} also  show that for states near the lowest $\widehat{\bee}$
value,  $\overline{\lan H_p \ran^{S,\widehat{\bee}}}$ increases  with spin 
$S$. Thus random interactions preserve this property of the pairing
Hamiltonian in addition to generating $S=S_{max}$ ground states as discussed
in  Sec. \ref{smax}. It is useful to remark that random interactions will
not generate $S=S_{max}$ ground states with $(v,\tilde{s})=(m,m/2)$ as
required for example in the $pn$-$sd$IBM. This needs explicit inclusion of
pairing and exchange terms in the Hamiltonians defined by Eqs.
(\ref{eq.begoe-s})  and (\ref{eq.v2}).

\setlength{\LTcapwidth}{6in}

\begin{longtable}{cccccccccc}
\caption{Classification of states in the $U(2\Omega)\supset$  $[U(\Omega)
\supset  SO(\Omega)]\otimes SU_S(2)$ limit for $(\Omega,m) = (4,10),(5,8)$
and $(6,6)$.  Given are $U(\Omega)$ labels $(m, S)$ and $SO(\Omega)$ labels 
$(v, \tilde{s})$ with the corresponding dimensions  $d_b(\Omega,m,S)$ and 
$\cads(v,\tilde{s})$, respectively, and also the pairing eigenvalues  $E_p = 
E_p(m,S,v,\tilde{s})$.  Note that $\sum_{v,\tilde{s}} r 
\cads(v,\tilde{s})=d_b(\Omega,m,S)$; here $r$ denotes multiplicity of the 
$SO(\Omega)$ irreps and in the table,  they are shown only for the cases
when $r>1$.} \\

\toprule
$\Omega$ & $m$ & $(m,S)_{d_b(\Omega,m,S)}$ &
$(v,\tilde{s})^r_{\cads(v,\tilde{s})}$ & $E_p$ & $\Omega$
& $m$ & $(m,S)_{d_b(\Omega,m,S)}$ &  $(v,\tilde{s})^r_{\cads(v,\tilde{s})}$ &
$E_p$ 
\\ 
\midrule 
\endfirsthead

\multicolumn{10}{c}%
{{\bfseries \tablename\ \thetable{} -- continued}} \\
\toprule
$\Omega$ & $m$ & $(m,S)_{d_b(\Omega,m,S)}$ &
$(v,\tilde{s})^r_{\cads(v,\tilde{s})}$ & $E_p$ & $\Omega$
& $m$ & $(m,S)_{d_b(\Omega,m,S)}$ &  $(v,\tilde{s})^r_{\cads(v,\tilde{s})}$ &
$E_p$  \\ \midrule
\endhead

\bottomrule
\endfoot

\hline 
\endlastfoot
$4$ & $10$ & $( 10 ,  0 )_{196}$ & $( 2 ,  0 )_{6}$ & $28$ &
$5$ & $8$ & $( 8 ,  0 )_{490}$ & $( 0 ,  0 )_{1}$ & $24$ \\     	
 &  &  & $( 4 ,  1 )_{30}$ & $22$ &  
 &  &  & $( 2 ,  1 )_{14}$ & $19$ \\     	
 &  &  & $( 6 ,  2 )_{70}$ & $12$ &  
 &  &  & $( 4 ,  2 )_{55}$ & $10$ \\     	
 &  &  & $( 6 ,  0 )_{14}$ & $18$ &  
 &  &  & $( 4 ,  0 )_{35}$ & $16$ \\     	
 &  &  & $( 8 ,  1 )_{54}$ & $8$ &  
 &  &  & $( 6 ,  1 )_{220}$ & $7$ \\     	
 &  &  & $( 10 ,  0 )_{22}$ & $0$ &  
 &  &  & $( 8 ,  0 )_{165}$ & $0$ \\     	
 &  & $( 10 ,  1 )_{540}$ & $( 2 ,  1 )^ 2_{9}$ & $28$ &  
 &  & $( 8 ,  1 )_{1260}$ & $( 2 ,  1 )_{14}$ & $21$ \\     	
 &  &  & $( 4 ,  2 )^ 2_{25}$ & $20$ &  
 &  &  & $( 4 ,  2 )_{55}$ & $12$ \\     	
 &  &  & $( 6 ,  3 )_{49}$ & $8$ &  
 &  &  & $( 4 ,  1 )^ 2_{81}$ & $16$ \\     	
 &  &  & $( 4 ,  1 )_{30}$ & $24$ &  
 &  &  & $( 6 ,  2 )_{260}$ & $5$ \\     	
 &  &  & $( 6 ,  2 )_{70}$ & $14$ &  
 &  &  & $( 6 ,  1 )_{220}$ & $9$ \\     	
 &  &  & $( 6 ,  1 )^ 2_{42}$ & $18$ &  
 &  &  & $( 8 ,  1 )_{455}$ & $0$ \\     	
 &  &  & $( 8 ,  2 )_{90}$ & $6$ &  
 &  &  & $( 2 ,  0 )_{10}$ & $23$ \\     	
 &  &  & $( 8 ,  1 )_{54}$ & $10$ &  
 &  &  & $( 6 ,  0 )_{84}$ & $11$ \\     	
 &  &  & $( 10 ,  1 )_{66}$ & $0$ &  
 &  & $( 8 ,  2 )_{1500}$ & $( 4 ,  2 )^ 2_{55}$ & $16$ \\     	
 &  &  & $( 0 ,  0 )_{1}$ & $32$ &  
 &  &  & $( 6 ,  3 )_{140}$ & $3$ \\     	
 &  &  & $( 4 ,  0 )_{10}$ & $26$ &  
 &  &  & $( 6 ,  2 )_{260}$ & $9$ \\     	
 &  &  & $( 8 ,  0 )_{18}$ & $12$ &  
 &  &  & $( 8 ,  2 )_{625}$ & $0$ \\     	
 &  & $( 10 ,  2 )_{750}$ & $( 4 ,  2 )^ 2_{25}$ & $24$ &  
 &  &  & $( 2 ,  1 )^ 2_{14}$ & $25$ \\     	
 &  &  & $( 6 ,  3 )_{49}$ & $12$ &  
 &  &  & $( 4 ,  1 )_{81}$ & $20$ \\     	
 &  &  & $( 6 ,  2 )^ 2_{70}$ & $18$ &  
 &  &  & $( 6 ,  1 )_{220}$ & $13$ \\     	
 &  &  & $( 8 ,  3 )_{126}$ & $4$ &  
 &  &  & $( 0 ,  0 )_{1}$ & $30$ \\     	
 &  &  & $( 8 ,  2 )_{90}$ & $10$ &  
 &  &  & $( 4 ,  0 )_{35}$ & $22$ \\     	
 &  &  & $( 10 ,  2 )_{110}$ & $0$ &  
 &  & $( 8 ,  3 )_{1155}$ & $( 6 ,  3 )_{140}$ & $9$ \\     	
 &  &  & $( 2 ,  1 )_{9}$ & $32$ &  
 &  &  & $( 8 ,  3 )_{595}$ & $0$ \\     	
 &  &  & $( 4 ,  1 )^2_{30}$ & $28$ &  
 &  &  & $( 4 ,  2 )_{55}$ & $22$ \\     	
 &  &  & $( 6 ,  1 )_{42}$ & $22$ &  
 &  &  & $( 6 ,  2 )_{260}$ & $15$ \\     	
 &  &  & $( 8 ,  1 )_{54}$ & $14$ &  
 &  &  & $( 2 ,  1 )_{14}$ & $31$ \\     	
 &  &  & $( 2 ,  0 )_{6}$ & $34$ &  
 &  &  & $( 4 ,  1 )_{81}$ & $26$ \\     	
 &  &  & $( 6 ,  0 )_{14}$ & $24$ &  
 &  &  & $( 2 ,  0 )_{10}$ & $33$ \\     	
 &  & $( 10 ,  3 )_{770}$ & $( 6 ,  3 )^ 2_{49}$ & $18$ &  
 &  & $( 8 ,  4 )_{495}$ & $( 8 ,  4  )_{285}$ & $0$ \\     	
 &  &  & $( 8 ,  4 )_{81}$ & $2$ &  
 &  &  & $( 6 ,  3  )_{140}$ & $17$ \\     	
 &  &  & $( 8 ,  3 )_{126}$ & $10$ &  
 &  &  & $( 4 ,  2  )_{55}$ & $30$ \\     	
 &  &  & $( 10 ,  3 )_{154}$ & $0$ &  
 &  &  & $( 2 ,  1  )_{14}$ & $39$ \\     	
 &  &  & $( 4 ,  2 )^ 2_{25}$ & $30$ &  
 &  &  & $( 0 ,  0  )_{1}$ & $44$ \\     	
 &  &  & $( 6 ,  2 )_{70}$ & $24$ &  
$6$ & $6$ & $( 6 ,  0 )_{490}$ & $( 2 ,  0 )_{15}$ & $14$ \\     	
 &  &  & $( 8 ,  2 )_{90}$ & $16$ &  
 &  &  & $( 4 ,  1 )_{175}$ & $6$ \\     	
 &  &  & $( 2 ,  1 )^ 2_{9}$ & $38$ &  
 &  &  & $( 6 ,  0 )_{300}$ & $0$ \\     	
 &  &  & $( 4 ,  1 )_{30}$ & $34$ &  
 &  & $( 6 ,  1 )_{1134}$ & $( 2 ,  1 )^ 2_{20}$ & $14$ \\     	
 &  &  & $( 6 ,  1 )_{42}$ & $28$ &  
 &  &  & $( 4 ,  2 )_{105}$ & $4$ \\     	
 &  &  & $( 0 ,  0 )_{1}$ & $42$ &  
 &  &  & $( 4 ,  1 )_{175}$ & $8$ \\     	
 &  &  & $( 4 ,  0 )_{10}$ & $36$ &  
 &  &  & $( 6 ,  1 )_{729}$ & $0$ \\     	
 &  & $( 10 ,  4 )_{594}$ & $( 8 ,  4 )_{81}$ & $10$ &  
 &  &  & $( 0 ,  0 )_{1}$ & $20$ \\     	
 &  &  & $( 10 ,  4 )_{198}$ & $0$ &  
 &  &  & $( 4 ,  0 )_{84}$ & $10$ \\     	
 &  &  & $( 6 ,  3 )_{49}$ & $26$ &  
 &  & $( 6 ,  2 )_{1050}$ & $( 4 ,  2 )_{105}$ & $8$ \\     	
 &  &  & $( 8 ,  3 )_{126}$ & $18$ &  
 &  &  & $( 6 ,  2 )_{735}$ & $0$ \\     	
 &  &  & $( 4 ,  2 )_{25}$ & $38$ &  
 &  &  & $( 2 ,  1 )_{20}$ & $18$ \\     	
 &  &  & $( 6 ,  2 )_{70}$ & $32$ &  
 &  &  & $( 4 ,  1 )_{175}$ & $12$ \\     	
 &  &  & $( 2 ,  1 )_{9}$ & $46$ &  
 &  &  & $( 2 ,  0 )_{15}$ & $20$ \\     	
 &  &  & $( 4 ,  1 )_{30}$ & $42$ &  
 &  & $( 6 ,  3 )_{462}$ & $( 6 ,  3  )_{336}$ & $0$ \\     	
 &  &  & $( 2 ,  0 )_{6}$ & $48$ &  
 &  &  & $( 4 ,  2  )_{105}$ & $14$ \\     	
 &  & $( 10 ,  5 )_{286}$ & $( 10 ,  5  )_{121}$ & $0$ &  
 &  &  & $( 2 ,  1  )_{20}$ & $24$ \\     	
 &  &  & $( 8 ,  4  )_{81}$ & $20$ &  
 &  &  & $( 0 ,  0  )_{1}$ & $30$ \\     	
 &  &  & $( 6 ,  3  )_{49}$ & $36$ &  
 &  &  &  & \\     	
 &  &  & $( 4 ,  2  )_{25}$ & $48$ &  
 &  &  &  & \\     	 
 &  &  & $( 2 ,  1  )_{9}$ & $56$ &  
 &  &  &  & \\     	
 &  &  & $( 0 ,  0  )_{1}$ & $60$ &  
 &  &  &  &    	
\label{red} 
\end{longtable}

For a given spin $S$, the pairing expectation values as a function of $E$
are expected, for two-body ensembles,  to be given by a ratio of expectation
value density Gaussian (the first two moments given by $\lan H_p H
\ran^{m,S}$ and $\lan H_p H^2 \ran^{m,S}$) and the eigenvalue density 
Gaussian with normalization given by $\lan H_p \ran^{m,S}$ and this itself 
will be a Gaussian; see Chapters \ref{ch2} and \ref{ch3} for details.  
Let us denote the expectation value density centroid by
$E_c(m,S:H_p)$ and width by $\sigma(m,S:H_p)$. Then  the ratio of Gaussians
[see Eq. (\ref{ch3.eq.pa12})] will give 
\be 
\barr{rcl} 
\overline{\lan H_p
\ran^{S,\widehat{E}}} & = & \dis\frac{\lan H_p \ran^{m,S}}
{\widehat{\sigma}(m,S)}  \exp{\dis\frac{\widehat{\epsilon}^2(m,S)}
{2\l[1-\widehat{\sigma}^2(m,S)\r]}} \\ \\
& \times &
\exp\l\{\dis\frac{(\widehat{\sigma}^2(m,S)-1)}{2\widehat{\sigma}^2(m,S)} \l[
\widehat{E} - \dis\frac{\widehat{\epsilon}(m,S)}
{1-\widehat{\sigma}^2(m,S)}\r]^2\r\}\;. 
\earr \label{eq.ratio} 
\ee 
Here, $\widehat{\epsilon}(m,S)=\{ E_c(m,S:H_p)-E_c(m,S)\}/
\sigma(m,S)$, $\widehat{\sigma}(m,S) = \sigma(m,S:H_p)/\sigma(m,S)$ and 
$\widehat{E}=[\sigma(m)/\sigma(m,S)]\{\widehat{\bee}-\ce\}$;
$\ce=[E_c(m,S)-E_c(m)]/\sigma(m)$. The
Gaussian form given by Eq. (\ref{eq.ratio}) is clearly  seen in Fig.
\ref{pair} and this also gives a quantitative description of the results.
Note that in our example, $\widehat{\epsilon}(10,S) = 0.001$, $0.001$,
$0.001$, $0.002$, $0.002$, $0.003$ and $\widehat{\sigma}(10,S) = 1.045$,
$1.047$, $1.053$, $1.062$, $1.073$, $1.082$, respectively for $S=0-5$.

\section{Summary}
\label{c6s7}

In the present chapter, we have introduced the BEGOE(1+2)-$\cs$ ensemble and a
method for constructing BEGOE(1+2)-$\cs$ for numerical calculations has been
described. Numerical examples are used to show that, like the spinless
BEGOE(1+2), the spin BEGOE(1+2)-$\cs$ ensemble also generates Gaussian
density of  states in the dense limit. Similarly, BEGOE(2)-$\cs$ exhibits
GOE level  fluctuations. On the other hand, BEGOE(1+2)-$\cs$ exhibits
Poisson to GOE transition as the interaction strength $\lambda$ is increased
and the transition marker $\lambda_c$ is found to decrease with increasing
spin. Moreover, ensemble averaged covariances in energy centroids and
spectral variances for BEGOE(2)-$\cs$ between spectra with different
particle numbers and spins are studied using the propagation formulas
derived for the  energy centroids and spectral variances. For $\Omega=12$
systems, the cross-correlations in energy centroids are  $\sim 15$\% and
they reduce to $\sim 4$\% for spectral variances. We have also derived the
exact formula for the ensemble averaged fixed-$(m,S)$ spectral variances and
demonstrated that the variance propagator gives a simple explanation for the
preponderance of spin $S=S_{max}$ ground states generated by random
interactions as in $pn$-$sd$IBM. It is also shown, by including exchange
interaction $\hat{S}^2$ in BEGOE(1+2)-$\cs$, that random interactions
preserving spin symmetry strongly favor NSO (just  as with isospin in 
nuclear shell-model). These results  are comprehensive and
give a mathematical foundation for the results in \cite{Yo-09}. 
In addition, we have identified the pairing
$SO(\Omega)$ symmetry and showed using numerical examples that random
interactions exhibit pairing correlations in the gs region and also they
generate a Gaussian form for the variation of the pairing expectation values
with respect to energy. 

\chapter*{}

\thispagestyle{empty}

\chapter{Higher Order Traces and their Applications}
\label{ch7}

\section{Introduction}
\label{c7int}

Embedded ensembles operating in many-particle spaces generate forms for
distributions of various physical quantities with respect to energy and other
quantum numbers; several examples for these are already discussed in Chapters
\ref{ch2}-\ref{ch6}. The separation of the energy evolution of various
observables into a smoothed and a fluctuating part  provides a basis for
statistical spectroscopy. In statistical spectroscopy, methods are developed to
determine various moments defining the distributions (predicted by EGEs) for the
smoothed parts (valid in the chaotic region) without recourse to many-particle
Hamiltonian construction.  Parameters defining many of the important spectral
distributions, generated by EGEs, involve traces of product of four (or even
more) two-body (or one-body or a mixture of one and two-body) operators
\cite{Mom-80,KH-10}. For
example, they are required for calculating nuclear structure matrix elements for
$\beta$ and  $0\nu-\beta\beta$ decay and also for establishing Gaussian density
of states generated by various extended two-body ensembles.

Propagation formulas for the moments $M_r = \lan  H^r  \ran^m$,  $r=3,4$ and
also for traces over multi-orbit configurations for a given one plus two-body 
Hamiltonian $H = h(1) + V(2)$ follow from the results, derived using
diagrammatic methods, given in \cite{Wo-86,No-72,Ay-74,Po-75,Ch-78,Karw-95} many
years back. These results extend to traces of product of four operators each of
maximum body-rank 2. From now on, we refer to these traces as fourth order
traces or averages. The propagation formulas derived using diagrammatic methods 
contain very large  number of complicated terms (in particular for fourth order
averages) and carrying out analytically ensemble averaging of all these terms is
proved to be impractical (we are not aware if anyone was successful in the
past). Some idea of the difficulty in carrying out simplifications can be seen
from the attempt in \cite{Pl-97}. Ensemble averages from trace propagation
formulas is  feasible for the second order moments and  we have already
presented examples for these in Chapters  \ref{ch2}, \ref{ch5} and \ref{ch6}. An
alternative is to program the exact formulas and evaluate the moments
numerically for  each member of EGE's by considering say 500 members in
two-particle spaces. However, as pointed out  by Ter\'{a}n and Johnson
\cite{Te-06}  in their most recent  attempt in this direction, these
calculations for the fourth order averages are time consuming if not
impractical. All the problems with the exact formulas have been emphasized in
\cite{KH-10}. Because of these (in future with much faster computers it may be
possible to use the exact formulas), we have adopted the binary correlation
approximation, first used by Mon and French \cite{Mo-73,MF-75} and later by
French et al \cite{FKPT,To-86} for deriving formulas for ensemble averaged
traces and they are good in the  dilute limit. All the ``basic'' binary
correlation results for averages over one orbit and two orbit configurations are
available in literature and for easy reference, we discuss these in Appendix
\ref{c7s1}.  Extending the binary correlation approximation method for two
different operators and for traces over two orbit configurations, we have
addressed two applications: (i) derived formulas for the skewness $\gamma_1$
and excess $\gamma_2$ parameters for EGOE(1+2)-$\pi$ ensemble in the dilute
limit; and (ii) we have derived formula for the fourth order trace defining 
correlation coefficient and sixth order traces defining the fourth order
cumulants of the bivariate transition strength density generated by the 
transition operator relevant for 0$\nu$-$\beta\beta$ decay (also $\beta$ decay).
The results for (i) and (ii) are presented in Secs. \ref{c7s2}
and \ref{c7s3}. In addition, we have derived formulas for cumulants (they also
involve fourth order traces) over $m$-particle spaces that enter into the
expansions for the energy centroids and spectral variances, up to order
$[J(J+1)]^2$, for EGOE(2)-$J$ i.e., embedded Gaussian orthogonal ensemble
generated by random two-body interactions  with angular momentum $J$ symmetry
for fermions in a single-$j$ shell. The expansions for fixed-$J$ centroids and
variances involve traces of powers of operators $H$ and $J^2$. As $H$ preserves
$J$ symmetry, we use exact methods to evaluate these traces.  More specifically,
we have derived trace propagation formulas for the bivariate moments $\lan H^P
(J^2)^Q \ran^m$, $P+Q \leq 4$ and the results are presented in Sec. \ref{c7s4}.
All the results in Secs. \ref{c7s2} and \ref{c7s4} are published in 
\cite{Ma-11a} and \cite{Ko-08}, respectively. 

\section{Application to EGOE(1+2)-$\pi$: Formulas for Skewness and Excess
Parameters}
\label{c7s2}

\begin{sidewaystable}[htp]
\caption{Exact results for skewness and excess parameters for fixed-$\pi$
eigenvalue densities $I_\pm(E)$ compared with the binary correlation
results (in the table, called `Approx'). For exact results, we have used the
eigenvalues obtained from  EGOE(1+2)-$\pi$ ensembles with 100 members. The
binary correlation results are obtained using Eqs.
(\ref{eq.pty1})-(\ref{eq.pty9}) and extension of  Eq. (\ref{eq.cenvar}). See
text for details.}
\begin{center}
\begin{tabular}{ccrrrrrrrr}
\toprule
 &  & \multicolumn{4}{c}{$\gamma_1(m,\pi)$} & 
\multicolumn{4}{c}{$\gamma_2(m,\pi)$} \\ \cmidrule{3-10}
$(N_+,N_-,m)$ & $(\tau,\alpha/\tau)$ & 
\multicolumn{2}{c}{Exact} & \multicolumn{2}{c}{Approx} &
\multicolumn{2}{c}{Exact} & \multicolumn{2}{c}{Approx} \\ \cmidrule{3-10}
 & & $\pi=+$ & $\pi=-$ & $\pi=+$ & $\pi=-$ 
 & $\pi=+$ & $\pi=-$ & $\pi=+$ &  $\pi=-$ \\
\midrule
$(8,8,4)$ 
 & $(0.05,0.5)$ & $0.01$ & $0$ & $0$ & $0$ & $-0.05$ & $-0.99$ & $-0.05$ &
 $-1.00$ \\
 & $(0.05,1.0)$ & $0.01$ & $0$ & $0$ & $0$ & $0.12$ & $-1.08$ & $0.13$ & 
 $-1.08$ \\
 & $(0.05,1.5)$ & $0.01$ & $0$ & $0$ & $0$ & $0.33$ & $-1.16$ & $0.34$ & 
 $-1.17$ \\
 & $(0.1,0.5)$  & $0$ & $0$ & $0$ & $0$ & $-0.84$ & $-0.66$ & $-0.84$ & 
 $-0.67$ \\
 & $(0.1,1.0)$  & $0$ & $0$ & $0$ & $0$ & $-0.70$ & $-0.79$ & $-0.71$ & 
 $-0.79$ \\
 & $(0.1,1.5)$  & $0$ & $0$ & $0$ & $0$ & $-0.51$ & $-0.90$ & $-0.51$ & 
 $-0.91$ \\
 & $(0.2,0.5)$  & $0$ & $0$ & $0$ & $0$ & $-0.83$ & $-0.74$ & $-0.84$ & 
 $-0.75$ \\
 & $(0.2,1.0)$  & $0$ & $0$ & $0$ & $0$ & $-0.84$ & $-0.81$ & $-0.84$ & 
 $-0.81$ \\
 & $(0.2,1.5)$  & $0$ & $0$ & $0$ & $0$ & $-0.74$ & $-0.87$ & $-0.74$ & 
 $-0.87$ \\
 & $(0.3,1.0)$  & $0$ & $0$ & $0$ & $0$ & $-0.85$ & $-0.83$ & $-0.85$ & 
 $-0.84$ \\
\bottomrule
\end{tabular}
\label{c5t1}
\end{center}
\end{sidewaystable}

\begin{sidewaystable}[htp]
\begin{center}
\text{{\bf Table \ref{c5t1} -- (continued)}}
\end{center}
\begin{center}
\begin{tabular}{ccrrrrrrrr}
\toprule
 &  & \multicolumn{4}{c}{$\gamma_1(m,\pi)$} & 
\multicolumn{4}{c}{$\gamma_2(m,\pi)$} \\ \cmidrule{3-10}
$(N_+,N_-,m)$ & $(\tau,\alpha/\tau)$ & 
\multicolumn{2}{c}{Exact} & \multicolumn{2}{c}{Approx} &
\multicolumn{2}{c}{Exact} & \multicolumn{2}{c}{Approx} \\ \cmidrule{3-10}
 & & $\pi=+$ & $\pi=-$ & $\pi=+$ & $\pi=-$ 
 & $\pi=+$ & $\pi=-$ & $\pi=+$ &  $\pi=-$ \\
\midrule
$(8,8,5)$ 
 & $(0.05,0.5)$ & $0.15$ & $-0.15$ & $0.15$ & $-0.15$ & $-0.52$ & $-0.52$ &
 $-0.52$ & $-0.52$ \\
 & $(0.05,1.0)$ & $0.16$ & $-0.16$ & $0.16$ & $-0.16$ & $-0.50$ & $-0.50$ &
 $-0.50$ & $-0.50$ \\
 & $(0.05,1.5)$ & $0.18$ & $-0.17$ & $0.18$ & $-0.18$ & $-0.46$ & $-0.46$ &
 $-0.46$ & $-0.46$ \\
 & $(0.2,0.5)$  & $-0.03$ & $0.03$ & $-0.03$ & $0.03$ & $-0.71$ & $-0.71$ &
 $-0.71$ & $-0.71$ \\
 & $(0.2,1.0)$  & $-0.01$ & $0.01$ & $-0.01$ & $0.01$ & $-0.73$ & $-0.73$ &
 $-0.74$ & $-0.74$ \\
 & $(0.2,1.5)$  & $0.02$ & $-0.02$ & $0.02$ & $-0.02$ & $-0.72$ & $-0.72$ &
 $-0.73$ & $-0.73$ \\
$(10,6,5)$ 
 & $(0.05,0.5)$ & $-0.06$ & $0.09$ & $-0.07$ & $0.09$ & $-0.26$ & $-0.76$ &
 $-0.26$ & $-0.75$ \\
 & $(0.05,1.5)$ & $-0.04$ & $0.15$ & $-0.05$ & $0.15$ & $-0.01$ & $-0.86$ &
 $-0.01$ & $-0.86$ \\
 & $(0.2,0.5)$  & $0.01$ & $-0.04$ & $0.01$ & $-0.04$ & $-0.73$ & $-0.69$ &
 $-0.73$ & $-0.69$ \\
 & $(0.2,1.5)$  & $0.01$ & $0.02$ & $0.01$ & $0.02$ & $-0.69$ & $-0.75$ &
 $-0.70$ & $-0.75$ \\
$(6,10,5)$ 
 & $(0.05,0.5)$ & $-0.09$ & $0.07$ & $-0.09$ & $0.07$ & $-0.76$ & $-0.26$ &
 $-0.75$ & $-0.26$ \\
 & $(0.05,1.5)$ & $-0.15$ & $0.05$ & $-0.15$ & $0.05$ & $-0.86$ & $-0.01$ &
 $-0.86$ & $-0.01$ \\
 & $(0.2,0.5)$  & $0.04$ & $-0.01$ & $0.04$ & $-0.01$ & $-0.68$ & $-0.73$ &
 $-0.69$ & $-0.73$ \\
 & $(0.2,1.5)$  & $-0.02$ & $-0.01$ & $-0.02$ & $-0.01$ & $-0.75$ & $-0.69$ &
 $-0.75$ & $-0.70$ \\
\bottomrule
\end{tabular}
\end{center}
\end{sidewaystable}

For the EGOE(1+2)-$\pi$ Hamiltonian, we have $H = h(1) + V(2) = h(1) + X(2)
+ D(2)$ with  $X(2) = A \oplus B \oplus C$ is the direct sum of the
spreading matrices $A$, $B$ and $C$ and $D(2) = D + \wD$ is the off-diagonal
mixing matrix as defined in Chapter \ref{ch5}.  Here, $\wD$ is the transpose
of the matrix $D$. The operator form for $D$ is 
\be
D(2) = \sum_{\gamma,\delta}
v_D^{\gamma\delta} \gamma_1^\dg(2) \delta_2(2) \;,
\label{eq.ptya1}
\ee
with $\overline{[v_D^{\gamma\delta}]^2} = v_D^2$. Note that the operator form of
$X(2)$ is given by Eq. (\ref{eq.b11}) and then $v_X^2(i,j) = 
\tau^2$ with $i + j = 2$ and similarly, $v_D^2 = \alpha^2$; see Chapter
\ref{ch5} for further discussion on the $(\alpha,\tau)$ parameters. 
Using this and
the property that  $h(1)$ conserves $(m_1,m_2)$ symmetry and $X$ preserves
$(m_1,m_2)$ symmetry, we apply the results in Appendix \ref{c7s1}  and derive
formulas for $M_r(m_1,m_2)$ with $r \leq 4$. These results are good in the
dilute limit: $m_1,N_1,m_2,N_2 \to \infty$, $m/N_1 \to 0$ and $m/N_2 \to 0$ with
$m=m_1$ or $m_2$. With the sp energies defining
the mean  field $h(1)$ as in Chapter \ref{ch5}, the first moment $M_1$ of
the partial densities $\rho^{m_1,m_2}(E)$ is trivially,
\be
M_1(m_1,m_2) = \overline{\lan (h + V) \ran^{m_1,m_2}} = m_2 \;,
\label{eq.pty1}
\ee
as $\lan h^r \ran^{m_1,m_2} = (m_2)^r$ and  $\overline{\lan V
\ran^{m_1,m_2}} = 0$. Applying the results in Appendix \ref{c7s1} in different 
ways, we derive formulas for the second, third and fourth order traces giving
$M_r(m_1,m_2)$, $r=2-4$.  However, the presence of the mixing matrix $D$
makes the application involved. The second moment $M_2$ is,
\be
\barr{rcl}
M_2(m_1,m_2) & = & 
\overline{\lan (h + V)^2 \ran^{m_1,m_2}} \\ 
& = & \lan h^2 \ran^{m_1,m_2} +
\overline{\lan V^2 \ran^{m_1,m_2}} = \l(m_2\r)^2 + \overline{\lan V^2
\ran^{m_1,m_2}} \;; \nonumber
\earr \label{eq.pty2a}
\ee
\be
\overline{\lan V^2 \ran^{m_1,m_2}} =
\overline{\lan X^2 \ran^{m_1,m_2}} + \overline{\lan D\wD \ran^{m_1,m_2}} +
\overline{\lan \wD D \ran^{m_1,m_2}}\;, 
\label{eq.pty2}
\ee
\be
\barr{rcl}
\overline{\lan X^2 \ran^{m_1,m_2}} & = & \tau^2 \; \dis\sum_{i+j=2}
T(m_1,N_1,i) \; T(m_2,N_2,j)\;, \\ \\
\overline{\lan D\wD \ran^{m_1,m_2}} & = & \alpha^2 \; \dis\binom{m_1}{2}
\dis\binom{\wmn}{2}\;,\;\;\;\;
\overline{\lan \wD D \ran^{m_1,m_2}} = \alpha^2 \; \dis\binom{\wmp}{2}
\dis\binom{m_2}{2}\;.\nonumber
\earr \label{eq.pty2b}
\ee
The second line in Eq. (\ref{eq.pty2}) follows by using the fact that $X(2)$ and
$D(2)$ are independent and $D(2)$ can correlate only with $\wD(2)$. 
In Eq. (\ref{eq.pty2}), the expression for $\overline{\lan X^2
\ran^{m_1,m_2}}$ follows directly from Eq. (\ref{eq.b12}).  
The last two equations in Eq. (\ref{eq.pty2}) can be derived using Eq.
(\ref{eq.ptya1}) giving the definition of the operator $D(2)$ and  using Eqs.
(\ref{eq.b2}) and (\ref{eq.b3}) appropriately to contract the operators
$\gamma^\dg$ with $\gamma$ and  $\delta$ with $\delta^\dg$. For the
$T(\cdots)$'s in Eq. (\ref{eq.pty2}),  we use Eq. (\ref{eq.b6a}). Note that, Eq.
(\ref{eq.pty2}) gives the binary  correlation formula for 
$\overline{\sigma^2(m_1,m_2)}$. Similarly, the  third moment $M_3$ is
\be
\barr{rcl}
M_3(m_1,m_2) & = & \overline{\lan (h + V)^3 \ran^{m_1,m_2}} \\ \\ 
& = & \lan h^3 \ran^{m_1,m_2}
+ 2 \lan h \ran^{m_1,m_2} \overline{\lan V^2 \ran^{m_1,m_2}} 
+ \overline{\lan X h X \ran^{m_1,m_2}} \\ \\
& + & 
\overline{\lan D h \wD \ran^{m_1,m_2}} +
\overline{\lan \wD h D \ran^{m_1,m_2}} \\ \\
& = & \l(m_2\r)^3 + 2\; m_2 \; \overline{\lan V^2 \ran^{m_1,m_2}} +
m_2 \; \overline{\lan X^2 \ran^{m_1,m_2}} \\ \\
& + & (m_2 + 2) \; \overline{\lan D \wD \ran^{m_1,m_2}}
+ (m_2 - 2) \; \overline{\lan \wD D \ran^{m_1,m_2}} \;.
\earr \label{eq.pty3}
\ee
In Eq. (\ref{eq.pty3}), the last three terms on the 
RHS are evaluated by using the following properties of the 
operators $X$, $D$ and $\wD$,
\be
\barr{l}
X(2) \l| m_1,m_2 \ran \to \l| m_1,m_2 \ran \;,\;\;\; D(2) \l| m_1,m_2 \ran \to
\l| m_1+2,m_2-2 \ran \;, \\ \\ 
\wD(2) \l| m_1,m_2 \ran \to \l| m_1-2,m_2+2 \ran \;.
\earr \label{eq.pty3ab}
\ee
Also, the fixed-$(m_1,m_2)$ averages involving $X^2$, $V^2$, $D\wD$ and $\wD
D$ in Eq. (\ref{eq.pty3}) follow from Eq. (\ref{eq.pty2}). Now, the formula 
for the fourth moment  $M_4$ is,
\be
\barr{rcl}
M_4(m_1,m_2) & = & \overline{\lan (h + V)^4 \ran^{m_1,m_2}} \\ \\ & = &  
\lan h^4 \ran^{m_1,m_2}
+ 3 \lan h^2 \ran^{m_1,m_2} \overline{\lan V^2 \ran^{m_1,m_2}} 
+ \lan h^2 \ran^{m_1,m_2} \overline{\lan X^2 \ran^{m_1,m_2}} \\ \\ & + &
\overline{\lan D h^2 \wD \ran^{m_1,m_2}} +
\overline{\lan \wD h^2 D \ran^{m_1,m_2}} +
2 \;\overline{\lan h X h X \ran^{m_1,m_2}} \\ \\
& + & 
2 \;\overline{\lan h D h \wD \ran^{m_1,m_2}} +
2 \;\overline{\lan h \wD h D \ran^{m_1,m_2}} +
\overline{\lan V^4 \ran^{m_1,m_2}} \\ \\
& = & \l( m_2 \r)^4 + 3\; \l( m_2 \r)^2 \overline{\lan V^2 \ran^{m_1,m_2}} +
\l( m_2 \r)^2 \overline{\lan X^2 \ran^{m_1,m_2}} \\ \\
& + & (m_2 + 2)^2 \; \overline{\lan D \wD \ran^{m_1,m_2}} +
(m_2 - 2)^2 \; \overline{\lan \wD D \ran^{m_1,m_2}}  \\ \\
& + & 2\; \l( m_2 \r)^2 \overline{\lan X^2 \ran^{m_1,m_2}} + 
2 \; m_2 (m_2 + 2) \; \overline{\lan D \wD \ran^{m_1,m_2}} \\ \\
& + & 2 \; m_2 (m_2 - 2)\; \overline{\lan \wD D \ran^{m_1,m_2}}
+ \overline{\lan V^4 \ran^{m_1,m_2}} \;. 
\earr \label{eq.pty3a}
\ee
The first term in Eq. (\ref{eq.pty3a}) is trivial. The next two terms 
follow from Eq. (\ref{eq.pty2}). The terms $4-8$ in Eq. (\ref{eq.pty3a}) are
also simple and follow from Eq. (\ref{eq.pty3ab}).
The expression for $\overline{\lan V^4 \ran^{m_1,m_2}}$, which is
non-trivial, is,
\be
\barr{rcl}
\overline{\lan V^4 \ran^{m_1,m_2}} & = & 
\overline{\lan X^4 \ran^{m_1,m_2}} + 3 \;
\overline{\lan X^2 \ran^{m_1,m_2}} \l\{
\overline{\lan D \wD \ran^{m_1,m_2}} + \overline{\lan \wD D 
\ran^{m_1,m_2}} \r\} \\  \\ 
& + & \overline{\lan D X^2 \wD \ran^{m_1,m_2}} + 
\overline{\lan \wD X^2 D \ran^{m_1,m_2}} \\ \\ & + &
2 \; \overline{\lan X D X \wD \ran^{m_1,m_2}} + 2 \;
\overline{\lan X \wD X D
\ran^{m_1,m_2}} + \overline{\lan (D + \wD)^4 \ran^{m_1,m_2}} \;.
\earr \label{eq.pty5}
\ee
The formula for the first term in Eq. (\ref{eq.pty5}) follows from Eq.
(\ref{eq.b14}), 
\be
\barr{rcl}
\overline{\lan X^4 \ran^{m_1,m_2}} & = & 
2 \l\{ \overline{\lan X^2 \ran^{m_1,m_2}} \r\}^2 + T_1 \;; \\ \\
T_1 & = & \tau^4 \; \dis\sum_{i+j=2,\;t+u=2} F(m_1,N_1,i,t) \;
F(m_2,N_2,j,u) \;.
\earr \label{eq.pty6}
\ee
Combining Eqs. (\ref{eq.pty5}) and (\ref{eq.pty6}), we have,
\be
\barr{l}
\overline{\lan V^4 \ran^{m_1,m_2}} \\ \\
 = 
2 \l\{ \overline{\lan X^2 \ran^{m_1,m_2}} \r\}^2 + T_1
+ 3 \;
\overline{\lan X^2 \ran^{m_1,m_2}} \l\{
\overline{\lan D \wD \ran^{m_1,m_2}} + \overline{\lan \wD D 
\ran^{m_1,m_2}} \r\} \\  \\
+ \l\{ \overline{\lan D X^2 \wD \ran^{m_1,m_2}} + 
\overline{\lan \wD X^2 D \ran^{m_1,m_2}} \r\} 
\\ \\ 
+ 2 \; \l\{ \overline{\lan X D X \wD \ran^{m_1,m_2}} + 
\overline{\lan X \wD X D
\ran^{m_1,m_2}} \r\} + 
\overline{\lan (D + \wD)^4 \ran^{m_1,m_2}} \\ \\
= 2 \l\{ \overline{\lan X^2 \ran^{m_1,m_2}} \r\}^2
+ 3 \; \overline{\lan X^2 \ran^{m_1,m_2}} \l\{
\overline{\lan D \wD \ran^{m_1,m_2}} + \overline{\lan \wD D 
\ran^{m_1,m_2}} \r\} \\  \\
+ T_1 +  T_2 + 2\; T_3 + T_4 \;.
\earr \label{eq.pty4}
\ee
To simplify the notations, we have introduced $T_1$, $T_2$, $T_3$ and $T_4$ in
Eq. (\ref{eq.pty4}). The first and second terms  in the RHS of the last step 
in Eq. (\ref{eq.pty4}) are completely determined by Eq. (\ref{eq.pty2}). Also,
expression for $T_1$ is given in Eq. (\ref{eq.pty6}).
Now, we will evaluate the terms $T_2$, $T_3$ and $T_4$. Firstly, using Eq.
(\ref{eq.pty3ab}), we have
\be
\barr{rcl}
T_2 & = & \overline{\lan D X^2 \wD \ran^{m_1,m_2}} + 
\overline{\lan \wD X^2 D \ran^{m_1,m_2}}
\\ \\
& = & \l\{\overline{\lan D \wD \ran^{m_1,m_2}}\r\} \;
\l\{\overline{\lan X^2 \ran^{m_1-2,m_2+2}}\r\} \\ \\
& + & \l\{\overline{\lan \wD D \ran^{m_1,m_2}}\r\} \;
\l\{\overline{\lan X^2 \ran^{m_1+2,m_2-2}}\r\} \;.
\earr \label{eq.pty7a}
\ee
Formulas for the averages involving $X^2$, $D\wD$ and $\wD D$  in Eq.
(\ref{eq.pty7a}) are  given by Eq. (\ref{eq.pty2}). Using Eqs. (\ref{eq.b4}) and
(\ref{eq.b5}) appropriately to contract the  operators $D$ with $\wD$ across
operator $X$ along with the expression for $\overline{\lan X^2 \ran^{m_1,m_2}}$ 
in Eq. (\ref{eq.pty2}), we have
\be
\barr{rcl}
T_3 & = & \overline{\lan X D X \wD \ran^{m_1,m_2}} + \overline{\lan X \wD X D
\ran^{m_1,m_2}} \\ \\ 
& = & \tau^2 \; \alpha^2 \; \dis\sum_{i+j=2} \l[ 
\dis\binom{m_1-i}{2} \dis\binom{\wmn-j}{2} + \dis\binom{\wmp-i}{2}
\dis\binom{m_2-j}{2} \r] \\ \\ 
& \times & T(m_1,N_1,i) \; T(m_2,N_2,j) \;.
\earr \label{eq.pty7b}
\ee
Similarly, the expression for $T_4$ is as follows,
\be
\barr{rcl}
T_4 & = & \overline{\lan (D + \wD)^4 \ran^{m_1,m_2}} \\ \\
& = &  \overline{\lan D^2 
\wD^2 \ran^{m_1,m_2}} +
\overline{\lan \wD^2 D^2 \ran^{m_1,m_2}} + \overline{\lan D \wD D \wD 
\ran^{m_1,m_2}} \\ \\
& + & \overline{\lan \wD D \wD D 
\ran^{m_1,m_2}} + \overline{\lan D \wD^2 D 
\ran^{m_1,m_2}} + \overline{\lan \wD D^2 \wD  
\ran^{m_1,m_2}} \;. 
\earr \label{eq.pty8a}
\ee
As, in leading order, $D$ can correlate only with $\wD$, we have
\be
\barr{l}
\overline{\lan D^2 \wD^2 \ran^{m_1,m_2}} = 
\overline{\lan \crd \cbd \crdt \cbdt \ran^{m_1,m_2}} +
\overline{\lan \crd \cbd \cbdt \crdt \ran^{m_1,m_2}} \\ \\
=
\alpha^4 \; \dis\sum_{\gamma,\delta,\kappa,\eta} \;
\lan \gamma_1^\dg(2) \delta_2(2) \kappa_1^\dg(2) \eta_2(2) \delta_2^\dg(2)
\gamma_1(2) \eta_2^\dg(2) \kappa_1(2) \ran^{m_1,m_2} \\ \\
+
\alpha^4 \; \dis\sum_{\gamma,\delta,\kappa,\eta} \;
\lan \gamma_1^\dg(2) \delta_2(2) \kappa_1^\dg(2) \eta_2(2) \eta_2^\dg(2) 
\kappa_1(2) \delta_2^\dg(2) \gamma_1(2) \ran^{m_1,m_2} \\ \\
= \alpha^4 \; \dis\sum_{\gamma,\delta,\kappa,\eta} \;
\lan \gamma_1^\dg(2) \kappa_1^\dg(2) \gamma_1(2) \kappa_1(2) \ran^{m_1} \;
\lan \delta_2(2)  \eta_2(2)  \delta_2^\dg(2) \eta_2^\dg(2) \ran^{m_2} \\ \\
+  \alpha^4 \; \dis\sum_{\gamma,\delta,\kappa,\eta} \;
\lan \gamma_1^\dg(2) \kappa_1^\dg(2) \kappa_1(2) \gamma_1(2) \ran^{m_1} \;
\lan \delta_2(2)  \eta_2(2)  \eta_2^\dg(2) \delta_2^\dg(2) \ran^{m_2} \\ \\
= 2 \; \alpha^4 \; \dis\sum_{\gamma,\kappa} \;
\lan \gamma_1^\dg(2) \kappa_1^\dg(2) \kappa_1(2) \gamma_1(2) \ran^{m_1} \;
\dis\sum_{\delta,\eta} \;
\lan \delta_2(2)  \eta_2(2) \eta_2^\dg(2) \delta_2^\dg(2) \ran^{m_2} 
\\ \\
= 2\; \overline{\lan D \wD \ran^{m_1,m_2}} \;\; \overline{\lan D \wD
\ran^{m_1-2,m_2+2}} \;.
\earr \label{eq.pty8b}
\ee
In order to obtain the last step in Eq. (\ref{eq.pty8b}), the operators $\kappa^\dg\kappa$ and
$\gamma^\dg\gamma$ are contracted  
using Eq. (\ref{eq.b2}) that gives
$\binom{m_1-2}{2}$ and $\binom{m_1}{2}$ respectively. 
Similarly, contracting operators $\eta\eta^\dg$ and
$\delta\delta^\dg$ using Eq. (\ref{eq.b3}) gives $\binom{\wm_2-2}{2}$ and $\binom{\wm_2}{2}$ 
respectively. Combining these gives the last step in Eq. (\ref{eq.pty8b}).
Note that the correlated pairs of operators are represented using same color
in Eq. (\ref{eq.pty8b}). Also, the third binary pattern $\overline{\lan \crd
\crd \cbdt \cbdt \ran^{m_1,m_2}}$ is not considered as it will be $1/N_1$ or
$1/N_2$ order smaller compared to the other two binary patterns shown in Eq.
(\ref{eq.pty8b}). Similarly, we obtain
\be
\barr{rcl}
\overline{\lan \wD^2 D^2 \ran^{m_1,m_2}} & = & 
\overline{\lan \crdt \cbdt \crd \cbd \ran^{m_1,m_2}} +
\overline{\lan \cbdt \crdt \crd \cbd \ran^{m_1,m_2}} \\ \\
& = & 2\; \overline{\lan \wD D \ran^{m_1,m_2}} \;\; \overline{\lan \wD D
\ran^{m_1+2,m_2-2}} \;, \\ \\
\overline{\lan D \wD D \wD \ran^{m_1,m_2}} & = &
\overline{\lan \crd \crdt \cbd \cbdt \ran^{m_1,m_2}} +
\overline{\lan \crd \cbdt \cbd \crdt \ran^{m_1,m_2}} \\ \\
& = & \l\{ \overline{\lan D \wD \ran^{m_1,m_2}} \r\}^2 \;
+ \overline{\lan  D \wD \ran^{m_1,m_2}} \;\; \overline{\lan \wD D
\ran^{m_1-2,m_2+2}} \;, \\ \\
\overline{\lan D \wD \wD D \ran^{m_1,m_2}} & = &
\overline{\lan \crd \crdt \cbdt \cbd \ran^{m_1,m_2}} +
\overline{\lan \crd \cbdt \crdt \cbd \ran^{m_1,m_2}} \\ \\
& = & 2 \; \overline{\lan D \wD \ran^{m_1,m_2}} \;\;
\overline{\lan  \wD D \ran^{m_1,m_2}}  \;, \\ \\
\overline{\lan \wD D D \wD \ran^{m_1,m_2}} & = &
\overline{\lan \crdt \crd \cbd \cbdt \ran^{m_1,m_2}} +
\overline{\lan \cbdt \crd \cbd \crdt \ran^{m_1,m_2}} \\ \\
& = & 2 \; \overline{\lan D \wD \ran^{m_1,m_2}} \;\;
\overline{\lan  \wD D \ran^{m_1,m_2}}  \;, \\ \\
\overline{\lan \wD D \wD D \ran^{m_1,m_2}} & = &
\overline{\lan \crdt \crd \cbdt \cbd \ran^{m_1,m_2}} +
\overline{\lan \cbdt \crd \crdt \cbd \ran^{m_1,m_2}} \\ \\
& = & \l\{ \overline{\lan \wD D \ran^{m_1,m_2}} \r\}^2
+ \overline{\lan \wD D \ran^{m_1,m_2}} \;\;
\overline{\lan  D \wD \ran^{m_1+2,m_2-2}}  \;. 
\earr \label{eq.pty8c}
\ee
Combining Eqs. (\ref{eq.pty8a})-(\ref{eq.pty8c}), we have
\be
\barr{rcl}
T_4 & = & 
\l\{\; \overline{\lan D \wD \ran^{m_1,m_2}} \;\r\}^2 + 
\l\{\; \overline{\lan \wD D \ran^{m_1,m_2}} \;\r\}^2 \\ \\
& + &
\overline{\lan D \wD \ran^{m_1,m_2}} \l[ \; 2\; \overline{\lan D \wD
\ran^{m_1-2,m_2+2}} + \overline{\lan \wD D \ran^{m_1-2,m_2+2}} \; \r] \\ \\
& + & 
\overline{\lan \wD D \ran^{m_1,m_2}} \l[ \; 2\;\overline{\lan \wD D
\ran^{m_1+2,m_2-2}} + \overline{\lan D \wD \ran^{m_1+2,m_2-2}} \;\r]
\\ \\
& + &
4 \;\l\{\;\overline{\lan D \wD \ran^{m_1,m_2}} \;\r\}
\;\l\{\;\overline{\lan \wD D \ran^{m_1,m_2}}\;\r\} \;.
\earr \label{eq.pty8}
\ee
Therefore, combining Eqs. (\ref{eq.pty3a}), (\ref{eq.pty6}), (\ref{eq.pty4}),
(\ref{eq.pty7a}), (\ref{eq.pty7b}) and (\ref{eq.pty8}), the expression for
the fourth moment is,
\be
\barr{l}
M_4(m_1,m_2) 
= \l( m_2 \r)^4 + 3\; \l( m_2 \r)^2 \overline{\lan V^2 \ran^{m_1,m_2}} +
3 \; \l( m_2 \r)^2 \overline{\lan X^2 \ran^{m_1,m_2}} 
\\ \\
+ (m_2 + 2)^2 \; \overline{\lan D \wD \ran^{m_1,m_2}} +
(m_2 - 2)^2 \; \overline{\lan \wD D \ran^{m_1,m_2}} 
\\ \\
+ 2 \; m_2 (m_2 + 2) \; \overline{\lan D \wD \ran^{m_1,m_2}}
+ 2 \; m_2 (m_2 - 2)\; \overline{\lan \wD D \ran^{m_1,m_2}}
+ 2 \l\{ \overline{\lan X^2 \ran^{m_1,m_2}} \r\}^2 
\\ \\
+ 3 \; \overline{\lan X^2 \ran^{m_1,m_2}} \l\{
\overline{\lan D \wD \ran^{m_1,m_2}} + \overline{\lan \wD D 
\ran^{m_1,m_2}} \r\} \nonumber
\earr \label{eq.pty9ac1}
\ee
\be
\barr{l}
+ \tau^4 \; \dis\sum_{i+j=2,\;t+u=2} F(m_1,N_1,i,t) \;
F(m_2,N_2,j,u) \\ \\
+  \l\{\overline{\lan D \wD \ran^{m_1,m_2}}\r\} \;
\l\{\overline{\lan X^2 \ran^{m_1-2,m_2+2}}\r\} \\ \\
+ \l\{\overline{\lan \wD D \ran^{m_1,m_2}}\r\} \;
\l\{\overline{\lan X^2 \ran^{m_1+2,m_2-2}}\r\} \\ \\
+ 2\; \tau^2 \; \alpha^2 \; \dis\sum_{i+j=2} \l[ 
\dis\binom{m_1-i}{2} \dis\binom{\wmn-j}{2} + \dis\binom{\wmp-i}{2}
\dis\binom{m_2-j}{2} \r] \\ \\
\times T(m_1,N_1,i) \; T(m_2,N_2,j) 
\earr \label{eq.pty9ac}
\ee
\be
\barr{l}
+ \l\{ \overline{\lan D \wD \ran^{m_1,m_2}} \r\}^2 + 
\l\{ \overline{\lan \wD D \ran^{m_1,m_2}} \r\}^2 \\ \\
+
\overline{\lan D \wD \ran^{m_1,m_2}} \l[ \; 2\; \overline{\lan D \wD
\ran^{m_1-2,m_2+2}} + \overline{\lan \wD D \ran^{m_1-2,m_2+2}} \; \r] \\ \\
+ 
\overline{\lan \wD D \ran^{m_1,m_2}} \l[ \; 2\;\overline{\lan \wD D
\ran^{m_1+2,m_2-2}} + \overline{\lan D \wD \ran^{m_1+2,m_2-2}} \;\r]
\\ \\
+
4 \;\l\{\overline{\lan D \wD \ran^{m_1,m_2}} \r\}
\;\l\{\overline{\lan \wD D \ran^{m_1,m_2}}\r\} \;. \nonumber
\earr \label{eq.pty9ac2}
\ee
Equations (\ref{eq.pty1}), (\ref{eq.pty2}), (\ref{eq.pty3}), and
(\ref{eq.pty9ac}),
respectively give the first four non-central moments  $[M_1(m_1,m_2)$,
$M_2(m_1,m_2)$, $M_3(m_1,m_2)$ and $M_4(m_1,m_2)]$. In  Eq. (\ref{eq.pty9ac}), 
we use Eq. (\ref{eq.b6a}) for $T(\cdots)$'s  and  for $F(\cdots)$'s, we use Eq.
(\ref{eq.b8}) and also Eq. (\ref{eq.3}) in applications. 
The first four cumulants $[k_1(m_1,m_2)$, $k_2(m_1,m_2)$,
$k_3(m_1,m_2)$, $k_4(m_1,m_2)]$ can be calculated from these non-central
moments using the formulas \cite{St-87},
\be
\barr{rcl}
k_1(m_1,m_2) & = & M_1(m_1,m_2)\;, \;\;\;\;
k_2(m_1,m_2) = M_2(m_1,m_2) - M_1^2(m_1,m_2)\;, \\ 
k_3(m_1,m_2) & = & M_3(m_1,m_2) - 3\; M_2(m_1,m_2)\; M_1(m_1,m_2) + 
2\; M_1^3(m_1,m_2) \;, \\ 
k_4(m_1,m_2) & = & M_4(m_1,m_2) - 4\; M_3(m_1,m_2)\; M_1(m_1,m_2) - 
3\; M_2^2(m_1,m_2) \\ 
& + & 12\; M_2(m_1,m_2)\; M_1^2(m_1,m_2) - 6\; M_1^4(m_1,m_2) \;.
\earr \label{eq.pty9}
\ee
Then, the skewness and excess parameters are, 
\be
\gamma_1(m_1,m_2) =
\dis\frac{k_3(m_1,m_2)}{[k_2(m_1,m_2)]^{3/2}}\;,\;\;\;\;\;\;
\gamma_2(m_1,m_2) = \dis\frac{k_4(m_1,m_2)}{[k_2(m_1,m_2)]^2}\;.
\label{eq.pty9ab}
\ee 
After carrying out the simplifications using Eqs. 
(\ref{eq.pty1}), (\ref{eq.pty2}), (\ref{eq.pty3}), (\ref{eq.pty9ac}) and 
(\ref{eq.pty9}), it is easily seen that,
\be
\gamma_1(m_1,m_2) = \dis\frac{2\l[ \overline{\lan D \wD \ran^{m_1,m_2}} - 
\overline{\lan \wD D \ran^{m_1,m_2}}\r]}
{\l\{ \overline{\lan D \wD \ran^{m_1,m_2}} + 
\overline{\lan \wD D \ran^{m_1,m_2}} + 
\overline{\lan X^2 \ran^{m_1,m_2}}\r\}^{3/2}}\;.
\label{eq.pty9a}
\ee
Thus, $\gamma_1$ will be non-zero only when $\alpha \neq 0$ and the $\tau$
dependence appears only in the denominator. Also, it is seen that for $N_+ =
N_-$, $\gamma_1(m_1,m_2) = - \gamma_1(m_2,m_1)$.  The expression for
$\gamma_2$ is more cumbersome. Denoting  $\cd = \overline{\lan D \wD
\ran^{m_1,m_2}}$,  $\bd = \overline{\lan \wD D \ran^{m_1,m_2}}$ and  $\cx =
\overline{\lan X^2 \ran^{m_1,m_2}}$ for brevity, we have
\be
\barr{l}
\gamma_2(m_1,m_2) + 1 = \dis\frac{T_1 + T_2 + 2\;T_3 + T_4 + 
\l( \bd + \cd \r) \l(4 -\cx \r) - 2 \;\l( \bd + \cd \r)^2}
{\l\{ \bd + \cd + \cx \r\}^2} \;.
\earr \label{eq.pty9b}
\ee
The formulas for $T$'s, $\cd$, $\bd$ and $\cx$ given before together with
Eq. (\ref{eq.pty9b}) show that, for $N_+ = N_-$, $\gamma_2(m_1,m_2) =
\gamma_2(m_2,m_1)$. With, $T_1 \sim \cx^2 + C_1$, $T_2 = T_3
\sim \cx(\bd + \cd)$ and  $T_4 \sim 3 (\bd + \cd)^2 + C_2$ which are 
good in the dilute limit ($|C1/T_1|$
and $|C_2/T_4|$ will be close to zero), we have
\be
\gamma_2(m_1,m_2) = \dis\frac{C_1 + C_2 + 4\;(\bd + \cd)}
{\l\{ \bd + \cd + \cx \r\}^2} \;.
\label{eq.pty9c}
\ee
Note that $C_1$ and $\cx$ depend only on $\tau$. Similarly, $C_2$ and
$(\bd,\cd)$ depend only on $\alpha$.  The $(\bd+\cd)$ term in the numerator
will contribute to $\gamma_2(m_1,m_2)$ when $\tau = 0$ and $\alpha$ is very
small. The approximation $T_2 = T_3 \sim \cx(\bd + \cd)$ is crucial in
obtaining the numerator in Eq. (\ref{eq.pty9c}) with no cross-terms
involving the  $\alpha$ and $\tau$ parameters. With this, we have $k_4$ to
be the sum of $k_4$'s coming from $X(2)$ and $D(2)$ matrices [note that, as
mentioned before,  $X(2) = A \oplus B \oplus C$ and $D(2) = D + \wD$]. 

To test the accuracy of the formulas for $M_r$ given by Eqs.
(\ref{eq.pty1}), (\ref{eq.pty2}), (\ref{eq.pty3}) and (\ref{eq.pty9ac}), 
the binary correlation results for
$\gamma_1(m,\pm)$ and $\gamma_2(m,\pm)$  are compared with exact results
obtained using the eigenvalues from EGOE(1+2)-$\pi$ ensembles with 100
members for several values of $(N_+,N_-,m)$ and $(\tau,\alpha)$ parameters
in Table \ref{c5t1}. Extension of Eq. (\ref{eq.cenvar}) along with  the
results derived for $M_r(m_1,m_2)$ will give the binary correlation results
for $\gamma_1(m,\pm)$ and $\gamma_2(m,\pm)$. It is clearly seen from the
results in the Table that in all the examples considered, the binary
correlation results are quite close to the exact results. Similar agreements
are also seen in many other examples which are not shown in the table. 

\section{Application to $\beta\beta$ Decay: Formulas for the Bivariate
Correlation Coefficient and Fourth Order Cumulants for the Transition Strength
Density} 
\label{c7s3}

\subsection{Transition matrix elements and bivariate strength densities}
\label{c7s3s1}

Given a transition operator $\co$, the transition matrix elements are given by
$| \lan f \mid \co \mid i \ran|^2$, with $i$ and $f$ being the initial and final
states. These are also generally called transition strengths. Operation of EGEs
in many-particle spaces will lead to a theory for the smoothed part of
transition strengths and the fluctuations in the locally renormalized strengths
follow Porter-Thomas form for systems in the chaotic region.  The transition
matrix elements are needed in many applications. Examples are one-particle
transfer \cite{Po-91},  E2 and M1 transition strengths in nuclei \cite{Ha-82a}, 
dipole strengths in atoms \cite{Fl-98}, beta-decay \cite{Ma-07},   giant dipole
resonances \cite{Ma-98} and  problems involving time-reversal non-invariance and
parity \cite{FKPT,To-00}. Here, our focus is on $0\nu-\beta\beta$ decay. 
Half-life for 0$\nu$ double-beta decay (NDBD), for the 0$^+_i$ gs  of a initial
even-even nucleus decay to the 0$^+_f$ gs of the final even-even nucleus, with a
few approximations, is given by \cite{El-02}
\be
\barr{rcl}
\l[ T_{1/2}^{0\nu}(0^+_i \to 0^+_f) \r]^{-1} & = & G^{0\nu}
\l| M^{0\nu} \r|^2 \dis\frac{\lan m_\nu \ran^2}{m_e^2} \;,\\ \\
M^{0\nu} & = & M^{0\nu}_{GT} - \dis\frac{g_V^2}{g_A^2} 
M^{0\nu}_{F} = \lan 0^+_f \mid\mid \co(2:0\nu) \mid\mid 0^+_i \ran \;,
\earr \label{eq.bbd1}
\ee
where $\lan m_\nu \ran$ is effective neutrino mass and the $G^{0\nu}$ is an
accurately  calculable phase space integral \cite{Bo-92,Doi-93}.  Similarly
$g_A$ and $g_V$ are the weak axial-vector and vector coupling constants
($g_A/g_V = 1$ to $1.254$). The nuclear matrix elements $M_{GT}$ and $M_F$ are
matrix elements of Gamow-Teller and Fermi like two-body operators respectively.
Forms for them will follow from the closure approximation which is well
justified for NDBD  \cite{El-02}. As seen from Eq. (\ref{eq.bbd1}), the  NDBD
half-lives are generated by the two-body transition operator  $\co(2:0\nu)$. An
experimental value of (bound on) $T_{1/2}^{0\nu}$ will give a value for (bound
on) neutrino mass via Eq. (\ref{eq.bbd1}) provided we know the value of
$|M^{0\nu}|^2$ generated by the NDBD two-body  transition operator 
$\co(2:0\nu)$, connecting the ground states of the initial  and final even-even
nuclei involved. 

Transition strengths multiplied by the eigenvalue densities at the two energies
involved define the transition strength densities. With EGOE(1+2) operating in
the Gaussian domain, it was established in the past that transition strength
densities follow close to bivariate Gaussian form for spinless fermion systems
and for operators that preserve particle number with the additional assumption
that the transition operator and the Hamiltonian operator can be represented by
independent EGOEs.  With extensions of these results (without a EGOE basis), 
the bivariate Gaussian form is used in practical applications. Our purpose is to
establish that for the $0\nu-\beta\beta$ decay (also for $\beta$ decay),
transition strength densities are close to bivariate Gaussian form and also to
derive a formula for the bivariate correlation coefficient. We will address
these two important questions so that the EGOE results can be applied to
formulate a theory for calculating 0$\nu$-$\beta\beta$  transition  matrix
elements \cite{Ko-08a}. With space \#1 denoting  protons and similarly space \#2
neutrons, the general form of the transition operator $\co$ is,
\be
\co(k_\co) = \dis\sum_{\gamma,\delta} v_\co^{\gamma\delta}(k_\co)\; 
\gamma_1^\dg(k_\co) \delta_2(k_\co) \;; \;\;\;\; k_\co=2 \; 
\mbox{for NDBD}\;.
\label{eq.9}
\ee
Therefore, in order to derive the form for the transition strength densities
generated by $\co$, it is necessary to deal with two-orbit configurations
denoted by $(m_1,m_2)$, where $m_1$ is the number of particles in the first
orbit (protons) and $m_2$ in the second orbit (neutrons).   Now, the
transition strength density  $I_\co(E_i,E_f)$ is
\be
\barr{l}
I_\co^{(m_1^f, m_2^f), (m_1^i, m_2^i)}(E_i,E_f) \\ \\ =  
I^{(m_1^f, m_2^f)}(E_f) \l|\lan (m_1^f, m_2^f) E_f 
\mid \co \mid (m_1^i, m_2^i) E_i \ran \r|^2
I^{(m_1^i, m_2^i)}(E_i) \;,
\earr \label{eq.bbd3}
\ee
and the corresponding bivariate moments are  
\be
\wtM_{PQ}(m_1^i,m_2^i) =   \overline{\lan \co^\dagger(k_\co)
H^Q(k_H) \co(k_\co) H^P(k_H) \ran^{m_1^i,m_2^i}} \;.
\label{eq.bbd4}
\ee
Note that $\wtM$ are in general non-central and non-normalized moments.  The
general form of the operator $H(k_H)$ is given by Eq. (\ref{eq.b11}) and it
preserves $(m_1^i,m_2^i)$'s.  However,  $\co$ and its hermitian conjugate
$\co^\dg$ do not preserve $(m_1,m_2)$ i.e.,   $\co(k_\co) \l| m_1, m_2 \ran =
\l| m_1 + k_\co, m_2 - k_\co \ran$ and  $\co^\dg(k_\co) \l| m_1, m_2 \ran = \l|
m_1 - k_\co, m_2 + k_\co \ran$. Thus, given a $(m_1^i,m_2^i)$ for an initial
state,  the $(m_1^f,m_2^f)$ for the final state generated by the action of $\co$
is uniquely defined and therefore,  in the bivariate moments defined in Eq.
(\ref{eq.bbd4}), only the initial $(m_1^i,m_2^i)$ is specified. For
completeness, let us mention that  given the marginal centroids $(\epsilon_i ,
\epsilon_f)$, widths  $(\sigma_i, \sigma_f)$ and the bivariate correlation
coefficient  $\zeta_{biv}$, the normalized bivariate Gaussian is defined by,
\be
\barr{l}
\rho_{\mbox{biv}-\cg;\co}(E_i , E_f) = \rho_{\mbox{biv}-\cg;\co}
(E_i,E_f;\epsilon_i,\epsilon_f,
\sigma_i,\sigma_f,\zeta_{biv}) 
\\ \\
= \dis\frac{1}{2\pi \sigma_i \sigma_f \sqrt{(1-\zeta_{biv}^2)}} 
\\ \\
\times \exp-\dis\frac{1}{2(1-\zeta_{biv}^2)}
\l\{\l(\frac{E_i - \epsilon_i}{\sigma_i}\r)^2 
- 2\zeta_{biv} \l(\frac{E_i-\epsilon_i}{\sigma_i}
\r)\l(\frac{E_f-\epsilon_f}{\sigma_f}\r) + \l(\frac{E_f-\epsilon_f}{
\sigma_f}\r)^2\;\r\}\;.
\earr \label{eq.bivg}
\ee

\subsection{Formulas for the bivariate moments}
\label{c7s3s2}

Using binary correlation approximation, we derive formulas for  the first four
moments $\wtM_{PQ}(m_1^i,m_2^i)$, $P+Q \leq 4$ of
$I_\co^{(m_1^f, m_2^f), (m_1^i, m_2^i)}(E_i,E_f)$ for any $k_\co$ by
representing $H(k_H)$ and $\co(k_\co)$ operators by independent EGOEs and
assuming $H(k_H)$ is a $k_H$-body operator preserving $(m_1,m_2)$'s. Note
that the ensemble averaged $k_H$-particle matrix elements of $H(k_H)$ are
$v_H^2(i,j)$ with $i+j=k_H$ [see Eq. (\ref{eq.b11})] 
and similarly the ensemble average of 
$(v_\co^{\gamma \delta})^2$ is $v_\co^2$. From now on, we use $(m_1^i,m_2^i) = 
(m_1,m_2)$. Using Eq. (\ref{eq.9}) and  applying
the basic rules given by Eqs. (\ref{eq.b2}) and (\ref{eq.b3}), we have 
\be 
\barr{rcl}
\wtM_{00}(m_1,m_2) & = & \overline{\lan \co^\dg(k_\co) \co(k_\co) 
\ran^{m_1,m_2}} \\ \\
& = & \dis\sum_{\gamma, \delta} \;
\overline{\l\{v_\co^{\gamma\delta}\r\}^2} \; 
\lan \delta_2^\dg(k_\co) \gamma_1(k_\co) \gamma_1^\dg(k_\co) \delta_2(k_\co)
\ran^{m_1,m_2} \\ \\
& = & v_\co^2 \; \dis\binom{\wm_1}{k_\co} \; \dis\binom{m_2}{k_\co} \;.
\earr \label{eq.9aa}
\ee
Trivially, $\wtM_{10}(m_1,m_2)$ and $\wtM_{01}(m_1,m_2)$ will be zero as 
$H(k_H)$ is represented by EGOE$(k_H)$. Thus,
$\wtM_{PQ}(m_1,m_2)$ are central moments. Moreover, by definition, all the
odd-order moments, i.e., $\wtM_{PQ}(m_1,m_2)$ with 
$\mod(P+Q,2) \neq 0$, will be zero. Now, the $\wtM_{11}$ is
given by,
\be
\barr{rcl}
\wtM_{11}(m_1,m_2)  & = & \overline{\lan \co^\dagger(k_\co) H(k_H) 
\co(k_\co) H(k_H) \ran^{m_1,m_2}} \\ \\
& = & 
v_\co^2 \;\;
\dis\sum_{\alpha_1, \beta_1, \alpha_2, \beta_2, \gamma_1,
\delta_2; \; i+j = k_H} v_H^2(i,j) \\ \\
& \times &
\lan \gamma_1^\dagger(k_\co) \alpha_1(i) \beta_1^\dagger(i) \gamma_1(k_\co) 
\beta_1(i) \alpha_1^\dagger(i) \ran^{m_1} 
\\ \\
& \times &
\lan \delta_2(k_\co) \alpha_2(j) \beta_2^\dagger(j) \delta_2^\dagger(k_\co) 
\beta_2(j) \alpha_2^\dagger(j) \ran^{m_2} \;.
\earr \label{eq.9a}
\ee
Then, contracting over the $\gamma^\dg\gamma$ and $\delta\delta^\dg$ 
operators, respectively in
the first and second traces in Eq. (\ref{eq.9a}) using Eqs. (\ref{eq.b4}) 
and (\ref{eq.b5}) appropriately, we have
\be
\barr{rcl}
\wtM_{11}(m_1,m_2) & = & v_\co^2 \;\; \dis\sum_{i+j=k_H} v_H^2(i,j) \;
\dis\binom{\wm_1-i}{k_\co} \dis\binom{m_2-j}{k_\co} \\ \\
& \times & T(m_1,N_1,i) \; T(m_2,N_2,j) \;.
\earr \label{eq.9b}
\ee
Note that the formulas for the functions $T(\cdots)$'s appearing in 
Eq. (\ref{eq.9b}) 
are given by Eqs. (\ref{eq.b6a}), (\ref{eq.b6b}) and (\ref{eq.5b}). Similarly,
the functions $F(\cdots)$'s appearing ahead are given by Eqs. 
(\ref{eq.b8}) and (\ref{eq.3}). For the marginal variances, we have
\be
\barr{rcl}
\wtM_{20}(m_1,m_2)  & = &  \overline{\lan \co^\dagger(k_\co) \co(k_\co)
H^2(k_H) \ran^{m_1,m_2}} \\ \\
& = & \wtM_{00}(m_1,m_2) \; \overline{\lan H^2(k_H) \ran^{m_1,m_2}} \;,
\\ \\
\wtM_{02}(m_1,m_2)  & = &  \overline{\lan \co^\dagger(k_\co) H^2(k_H) 
\co(k_\co) \ran^{m_1,m_2}} \\ \\
& = & \wtM_{00}(m_1,m_2) \; \overline{\lan H^2(k_H) \ran^{m_1+k_\co, 
m_2-k_\co}} 
\;.
\earr \label{eq.9ab}
\ee
In Eq. (\ref{eq.9ab}), the ensemble averages of $H^2(k_H)$ are given by Eq.
(\ref{eq.b12}). Now, the correlation coefficient $\zeta_{biv}$ is
\be
\zeta_{biv}(m_1,m_2) = \dis\frac{\wtM_{11}(m_1,m_2)}{\dis\sqrt{\wtM_{20}(m_1,m_2)
\; \wtM_{02}(m_1,m_2)}} \;.
\label{eq.11}
\ee
Clearly, $\zeta_{biv}$ will be independent of $v_\co^2$. 

Proceeding further, we derive formulas for the fourth order moments 
$\wtM_{PQ}$,  $P+Q = 4$. The results are as follows. Firstly, for $(PQ) =
(40)$ and $(04)$, we have 
\be
\barr{rcl}
\wtM_{40}(m_1,m_2) & = & \overline{\lan \co^\dagger(k_\co) \co(k_\co)
H^4(k_H) \ran^{m_1,m_2}} \\ \\
& = & \wtM_{00}(m_1,m_2) \; \overline{\lan H^4(k_H) \ran^{m_1,m_2}} 
\;, \\ \\
\wtM_{04}(m_1,m_2) & = & \overline{\lan \co^\dagger(k_\co) 
H^4(k_H) \co(k_\co) \ran^{m_1,m_2}} \\ \\
& = & \wtM_{00}(m_1,m_2) \; \overline{\lan H^4(k_H) 
\ran^{m_1+k_\co,m_2-k_\co}} \;.
\earr \label{eq.10}
\ee
In Eq. (\ref{eq.10}), the ensemble averages of $H^4(k_H)$ are 
given by Eq. (\ref{eq.b14}). For $(PQ) = (31)$, we have
\be
\barr{rcl}
\wtM_{31}(m_1,m_2) & = & \overline{\lan \co^\dagger(k_\co) H(k_H) \co(k_\co) 
H^3(k_H) \ran^{m_1,m_2}} \\ \\
& = & \overline{\lan \co^\dagger(k_\co) \cbh \co(k_\co) 
\cbh \crh \crh \ran^{m_1,m_2}} \\ \\
& + & \overline{\lan \co^\dagger(k_\co) \cbh \co(k_\co) 
\crh \cbh \crh \ran^{m_1,m_2}} \\ \\
& + & \overline{\lan \co^\dagger(k_\co) \cbh \co(k_\co) 
\crh \crh \cbh \ran^{m_1,m_2}} \;.
\earr \label{eq.11a}
\ee
Note that in Eq. (\ref{eq.11a}), we use the same color 
for the binary correlated pairs of operators. First and last terms on RHS of Eq.
(\ref{eq.11a}) are simple and this gives,
\be 
\barr{l}
\wtM_{31}(m_1,m_2) = 2\; \overline{\lan H^2(k_H) \ran^{m_1,m_2}} \;
\wtM_{11}(m_1,m_2) \\ \\
+ \overline{\lan \co^\dagger(k_\co) \cbh \co(k_\co) 
\crh \cbh \crh \ran^{m_1,m_2}} \\ \\
= 2\; \overline{\lan H^2(k_H) \ran^{m_1,m_2}} \;
\wtM_{11}(m_1,m_2) 
+ v_\co^2 \; \dis\sum_{i+j=k_H, t+u=k_H} \; v_H^2(i,j) \; v_H^2(t,u) \;
\\ \\ \times \dis\binom{m_2-j}{k_\co}
\; \dis\binom{\wm_1-i}{k_\co} \; F(m_1,N_1,i,t) \; F(m_2,N_2,j,u) \;.
\earr \label{eq.12}
\ee
Similarly, we have
\be 
\barr{l}
\wtM_{13}(m_1,m_2)  = \overline{\lan \co^\dagger(k_\co) H^3(k_H) \co(k_\co) 
H(k_H) \ran^{m_1,m_2}} \\ \\
= \overline{\lan \co^\dagger(k_\co) \cbh \cbh \crh \co(k_\co) 
\crh \ran^{m_1,m_2}} \\ \\
+ \overline{\lan \co^\dagger(k_\co) \cbh \crh \cbh \co(k_\co) 
\crh \ran^{m_1,m_2}} \\ \\
+ \overline{\lan \co^\dagger(k_\co) \cbh \crh \crh \co(k_\co) 
\cbh \ran^{m_1,m_2}} \\ \\ 
=  2\; \overline{\lan H^2(k_H) \ran^{m_1+k_\co,m_2-k_\co}} \;
\wtM_{11}(m_1,m_2) \\ \\
+ v_\co^2 \; 
\dis\sum_{i+j=k_H, t+u=k_H}\; v_H^2(i,j) \; v_H^2(t,u) 
\; G(t,u) 
\\ \\
\times \dis\binom{\wm_1-k_\co-t+i}{i} \; \dis\binom{m_1+k_\co-t}{i}
\; \dis\binom{\wm_2-u+k_\co+j}{j} \; \dis\binom{m_2-k_\co-u}{j} \;;\\ \\
G(t,u) = \dis\binom{\wm_1-t}{k_\co} \dis\binom{m_2-u}{k_\co} T(m_1,N_1,t) 
\;  T(m_2,N_2,u)\;.
\earr\label{eq.13}
\ee
Finally, $\wtM_{22}(m_1,m_2)$ is given by,
\be
\barr{l}
\wtM_{22}(m_1,m_2) = \overline{\lan \co^\dagger(k_\co) H^2(k_H) \co(k_\co) 
H^2(k_H) \ran^{m_1,m_2}} \\ \\
= \overline{\lan \co^\dagger(k_\co) \cbh \cbh \co(k_\co) 
\crh \crh \ran^{m_1,m_2}} \\ \\
+ \overline{\lan \co^\dagger(k_\co) \cbh \crh 
\co(k_\co) 
\cbh \crh \ran^{m_1,m_2}} \\ \\
+ \overline{\lan \co^\dagger(k_\co) \cbh \crh \co(k_\co) 
\crh \cbh \ran^{m_1,m_2}} \\ \\
=  \wtM_{00}(m_1,m_2) \; \overline{\lan H^2(k_H)
\ran^{m_1+k_\co,m_2-k_\co}} \;\; \overline{\lan H^2(k_H) \ran^{m_1,m_2}} \\ \\
+ v_\co^2 \; \dis\sum_{i+j=k_H, t+u=k_H}\; v_H^2(i,j) \; v_H^2(t,u) \;
\dis\binom{\wm_1-i-t}{k_\co} \; \dis\binom{m_2-u-j}{k_\co} \\ \\
\times \l[ F(m_1,N_1,i,t) \; F(m_2,N_2,j,u) \r. 
\\ \\ \l. + 
T(m_1,N_1,i) \; T(m_1,N_1,t)\; T(m_2,N_2,j) \; T(m_2,N_2,u) \r] \;.
\earr \label{eq.14}
\ee
Given the $\wtM_{PQ}(m_1,m_2)$, the normalized central moments $M_{PQ}$ are
$M_{PQ}=\wtM_{PQ}/\wtM_{00}$. 

\subsection{Numerical results for bivariate correlation coefficient and fourth
order cumulants}
\label{c7s3s3}

Firstly, the scaled moments $\widehat{M}_{PQ}$ are 
\be
\widehat{M}_{PQ} = \dis\frac{M_{PQ}(m_1,m_2)}
{\l[M_{20}(m_1,m_2)\r]^{P/2} \l[
M_{02}(m_1,m_2)\r]^{Q/2}}\;;\;\;\;\;
P+Q \geq 2\;.
\label{eq.20}
\ee
Now the fourth order cumulants are \cite{St-87},
\be
\barr{l}
k_{40}(m_1,m_2) = \widehat{M}_{40}(m_1,m_2) - 3\;, 
k_{04}(m_1,m_2) = \widehat{M}_{04}(m_1,m_2) - 3\;, \\
k_{31}(m_1,m_2) = \widehat{M}_{31}(m_1,m_2) - 3\; 
\widehat{M}_{11}(m_1,m_2)\;, \\
k_{13}(m_1,m_2) = \widehat{M}_{13}(m_1,m_2) - 3\; 
\widehat{M}_{11}(m_1,m_2)\;, \\
k_{22}(m_1,m_2) = \widehat{M}_{22}(m_1,m_2) - 2\; 
\widehat{M}_{11}^2(m_1,m_2) -1 \;. 
\earr \label{eq.21}
\ee

\begin{table}[htp] 
\caption{Correlation coefficients $\zeta_{biv}(m_1,m_2)$ for some nuclei with
$k_\co = 2$ as appropriate for $0\nu-\beta\beta$ decay operator. 
Note that space \#1 is for protons and space \#2 for neutrons. 
The configuration spaces corresponding to $N_1$ or
$N_2 = 20$, 22, 30, 32, 44 and 58 are $r_3f$, $r_3g$, $r_4g$, $r_4h$, $r_5i$,
and $r_6j$, respectively with $f$ = $^1f_{7/2}$, $g$ = $^1g_{9/2}$, 
$h$ = $^1h_{11/2}$, $i$ = $^1i_{13/2}$, $j$ = $^1j_{15/2}$, 
$r_3$ = $^1f_{5/2}$ $^2p_{3/2}$ $^2p_{1/2}$, $r_4$ =
$^1g_{7/2}$ $^2d_{5/2}$ $^2d_{3/2}$ $^3s_{1/2}$, 
$r_5$ = $^1h_{9/2}$ $^2f_{7/2}$ $^2f_{5/2}$ $^3p_{3/2}$
$^3p_{1/2}$ and $r_6$ = $^1i_{11/2}$ $^2g_{9/2}$ $^2g_{7/2}$ $^3d_{5/2}$ 
$^3d_{3/2}$ $^4s_{1/2}$. See text for details.}
\begin{center}
\begin{tabular}{cccccc}
\toprule
Nuclei & $N_1$ & $m_1$ & $N_2$ & $m_2$ & $\zeta_{biv}(m_1,m_2)$ \\ 
\midrule
$^{76}_{32}$Ge$_{44}$  & $22$ & $4 $ & $22$ & $16$ & $0.64$ \\ \\
$^{82}_{34}$Se$_{48}$  & $22$ & $6 $ & $22$ & $20$ & $0.6$ \\ \\
$^{100}_{42}$Mo$_{58}$ & $30$ & $2 $ & $32$ & $8 $ & $0.57$ \\\\
$^{128}_{52}$Te$_{76}$ & $32$ & $2 $ & $32$ & $26$ & $0.62$ \\\\
$^{130}_{52}$Te$_{78}$ & $32$ & $2 $ & $32$ & $28$ & $0.58$ \\\\
$^{150}_{60}$Nd$_{90}$ & $32$ & $10$ & $44$ & $8 $ & $0.72$ \\\\
$^{154}_{62}$Sm$_{92}$ & $32$ & $12$ & $44$ & $10$ & $0.76$ \\\\
$^{180}_{74}$W$_{106}$ & $32$ & $24$ & $44$ & $24$ & $0.77$ \\\\
$^{238}_{92}$U$_{146}$ & $44$ & $10$ & $58$ & $20$ & $0.83$ \\
\bottomrule
\end{tabular}
\label{corr-coef}
\end{center}
\end{table}

\begin{sidewaystable}[htp] 
\caption{Cumulants $k_{PQ}$, $P+Q =4$ for  some nuclei listed in Table
\ref{corr-coef}.  The  numbers in the brackets are for the strict dilute limit
as explained in the text.  Just as in the construction of Table \ref{corr-coef},
we use $v_H^2(i,j)$  independent of $(i,j)$. See Table \ref{corr-coef} and text
for details.}
\begin{center}
\begin{tabular}{cccccccccc}
\toprule
Nuclei & $N_1$ & $m_1$ & $N_2$ & $m_2$ & $k_{40}$ & $k_{04}$ & $k_{13}$
& $k_{31}$ & $k_{22}$ \\ 
\midrule
$^{100}_{42}$Mo$_{58}$ & $30$ & $2 $ & $32$ & $8 $ & $-0.45(-0.39)$ 
& $-0.42(-0.38)$ & $-0.24(-0.23)$ & $-0.26(-0.25)$ & $-0.20(-0.22)$ \\\\
$^{150}_{60}$Nd$_{90}$ & $32$ & $10$ & $44$ & $8 $ & $-0.27(-0.22)$ 
& $-0.29(-0.23)$ & $-0.22(-0.18)$ & $-0.20(-0.17)$ & $-0.19(-0.18)$ \\\\
$^{154}_{62}$Sm$_{92}$ & $32$ & $12$ & $44$ & $10$ & $-0.24(-0.18)$ 
& $-0.25(-0.18)$ & $-0.19(-0.15)$ & $-0.18(-0.15)$ & $-0.17(-0.15)$ \\\\
$^{180}_{74}$W$_{106}$ & $32$ & $24$ & $44$ & $24$ & $-0.19(-0.08)$ 
& $-0.20(-0.08)$ & $-0.17(-0.08)$ & $-0.15(-0.08)$ & $-0.15(-0.08)$ \\\\
$^{238}_{92}$U$_{146}$ & $44$ & $10$ & $58$ & $20$ & $-0.18(-0.13)$ 
& $-0.18(-0.13)$ & $-0.15(-0.11)$ & $-0.15(-0.11)$ & $-0.13(-0.11)$ \\
\bottomrule
\end{tabular}
\label{cumu}
\end{center}
\end{sidewaystable}

Assuming $v_H^2(i,j)$ defining $H(2)$ are independent of $(i,j)$ so that
$\zeta_{biv}$ is independent of $v_H^2$, we have calculated the value of
$\zeta_{biv}$ with $k_\co = 2$ for several $0\nu-\beta \beta$ decay candidate
nuclei using Eq. (\ref{eq.11}) along with Eqs. (\ref{eq.9aa}), (\ref{eq.9b}),
(\ref{eq.9ab}) and (\ref{eq.b12}). For the function $T(\cdots)$, we use Eq.
(\ref{eq.b6a}). Note that $v_H^2(i,j)$ correspond to the variance of
two-particle matrix elements from the p-p $(i=2,j=0)$, n-n $(i=0,j=2)$ and p-n
$(i=1,j=1)$ interactions.
Results are given in   Table \ref{corr-coef}. It is seen that $\zeta_{biv} \sim
0.6-0.8$. It is important to mention that $\zeta_{biv} = 0$ for GOE.  Therefore,
the transition strength density will be narrow in $(E_i,E_f)$ plane. In order to
establish the bivariate Gaussian form for the $0\nu -\beta\beta$ decay
transition strength density, we have examined $k_{PQ}$, $P+Q =4$. For a good
bivariate Gaussian, $|k_{PQ}| \lazz 0.3$. Using Eqs. (\ref{eq.9aa}),
(\ref{eq.9b}), (\ref{eq.9ab}), (\ref{eq.10}), (\ref{eq.12})-(\ref{eq.21}) along
with Eqs. (\ref{eq.b12}) and (\ref{eq.b14}), we have calculated the cumulants
$k_{PQ}(m_1,m_2)$, $P+Q = 4$. These involve $T(\cdots)$ and $F(\cdots)$
functions. For set \#1 calculations in Table \ref{cumu}, we use Eq.
(\ref{eq.b6a}) for  $T(\cdots)$ and Eq. (\ref{eq.3}) for $F(\cdots)$. For the
set \#2 calculations, shown in `brackets' in Table \ref{cumu}, we use Eq.
(\ref{eq.b6b}) for  $T(\cdots)$, Eq. (\ref{eq.b8}) for $F(\cdots)$ and replace
everywhere $\binom{\wm_i+r}{s} \to \binom{N_i}{s}$  for any $(r,s)$ with
$i=1,2$. Then we have the strict dilute limit. We show in Table \ref{cumu},
bivariate cumulants for five heavy nuclei for both sets of calculations and they
clearly establish that bivariate Gaussian is a good approximation. We have also
examined this analytically in the dilute limit with $N_1,N_2 \to \infty$ and
assuming $v_H^2(i,j)$ independent of $(i,j)$. With these, we have expanded
$k_{PQ}$ in powers of  $1/m_1$ and $1/m_2$ using Mathematica. It is seen that
all the  $k_{PQ}$, $P+Q =4$ behave as,
\be
k_{PQ} = - \dis\frac{4}{m_1} + O\l( \dis\frac{1}{m_1^2} \r)
+ O\l( \dis\frac{m_2^2}{m_1^3} \r) + \ldots \;.
\label{eq.22}
\ee
Therefore, for $m_1 >> 1$ and $m_2 << m_1^{3/2}$, the strength density
approaches bivariate Gaussian form in general. It is important to recall that
the strong dependence on $m_1$ in Eq. (\ref{eq.22}) is due to the nature of the
operator $\co$ i.e., $\co(k_\co) \l| m_1, m_2 \ran = \l| m_1 + k_\co, m_2 -
k_\co \ran$. Thus, we conclude that bivariate Gaussian form is a good
approximation for $0\nu-\beta\beta$ decay transition strength densities. With
this, one can apply the formulation given in \cite{Ko-08a} with the bivariate
correlation coefficient $\zeta_{biv}$ given by Eqs. (\ref{eq.11}), 
(\ref{eq.9ab}) and (\ref{eq.9b}). The values given by the
two-orbit binary correlation theory  for $\zeta_{biv}$ 
can be  used as starting values in practical calculations.

For completeness, we have also calculated the correlation coefficient
and fourth order moments for the transition operator relevant for $\beta$ decay
and the results presented in Table \ref{cumu-beta} confirm that bivariate
Gaussian form is a good approximation for $\beta$ decay transition strength
densities. These results justify the assumptions made in \cite{Ko-95}. 

\begin{sidewaystable}[htp] 
\caption{Correlation coefficients $\zeta_{biv}(m_1,m_2)$ and  cumulants
$k_{PQ}$, $P+Q =4$ for  some nuclei relevant for $\beta$ decay [$k_\co = 1$ in
Eq. (\ref{eq.9})]. The first four nuclei in the table are relevant for $\beta^-$
transitions, next four nuclei are relevant for electron capture and the last two
nuclei are relevant for $\beta^+$ transitions.  The  numbers in the brackets for
$k_{PQ}$ are for the strict dilute limit as in Table \ref{cumu}. We
assume  $v_H^2(i,j)$ are independent of $(i,j)$ just as in the calculations for
generating Tables \ref{corr-coef} and \ref{cumu}. See caption to Table 
\ref{corr-coef} for other details.}
\begin{center}
\begin{tabular}{ccccccccccc}
\toprule
Nuclei & $N_1$ & $m_1$ & $N_2$ & $m_2$ & $\zeta_{biv}(m_1,m_2)$ & $k_{40}$ & $k_{04}$ & $k_{13}$ & $k_{31}$ & $k_{22}$ \\ 
\midrule
$^{62}_{27}$Co$_{35}$ & $20$ & $7 $ & $30$ & $15$ & $0.72$ & $-0.26(-0.18)$ & $-0.27(-0.18)$ & $-0.24(-0.16)$ & $-0.23(-0.16)$ & $-0.22(-0.16)$ \\\\
$^{64}_{27}$Co$_{37}$ & $20$ & $7 $ & $30$ & $17$ & $0.73$ & $-0.27(-0.16)$ & $-0.27(-0.16)$ & $-0.24(-0.15)$ & $-0.23(-0.15)$ & $-0.21(-0.15)$ \\\\
$^{62}_{26}$Fe$_{36}$ & $20$ & $6 $ & $30$ & $16$ & $0.72$ & $-0.28(-0.18)$ & $-0.28(-0.18)$ & $-0.24(-0.16)$ & $-0.24(-0.16)$ & $-0.22(-0.16)$ \\\\
$^{68}_{28}$Ni$_{40}$ & $20$ & $8 $ & $30$ & $20$ & $0.72$ & $-0.27(-0.14)$ & $-0.27(-0.14)$ & $-0.24(-0.13)$ & $-0.23(-0.13)$ & $-0.21(-0.13)$ \\\\
$^{65}_{32}$Ge$_{33}$ & $36$ & $5 $ & $36$ & $4 $ & $0.55$ & $-0.45(-0.41)$ & $-0.46(-0.42)$ & $-0.35(-0.33)$ & $-0.34(-0.32)$ & $-0.34(-0.34)$ \\\\
$^{69}_{34}$Se$_{35}$ & $36$ & $7 $ & $36$ & $6 $ & $0.66$ & $-0.36(-0.29)$ & $-0.34(-0.30)$ & $-0.28(-0.25)$ & $-0.28(-0.25)$ & $-0.27(-0.25)$ \\\\
$^{73}_{36}$Kr$_{37}$ & $36$ & $9 $ & $36$ & $8 $ & $0.72$ & $-0.28(-0.23)$ & $-0.28(-0.23)$ & $-0.24(-0.20)$ & $-0.24(-0.20)$ & $-0.23(-0.20)$ \\\\
$^{77}_{38}$Sr$_{39}$ & $36$ & $11$ & $36$ & $10$ & $0.76$ & $-0.24(-0.19)$ & $-0.24(-0.19)$ & $-0.21(-0.17)$ & $-0.21(-0.17)$ & $-0.20(-0.17)$ \\\\
$^{85}_{42}$Mo$_{43}$ & $36$ & $15$ & $36$ & $14$ & $0.79$ & $-0.20(-0.14)$ & $-0.21(-0.14)$ & $-0.19(-0.13)$ & $-0.18(-0.13)$ & $-0.17(-0.13)$ \\\\
$^{93}_{46}$Pd$_{47}$ & $36$ & $19$ & $36$ & $18$ & $0.80$ & $-0.19(-0.11)$ & $-0.19(-0.11)$ & $-0.18(-0.10)$ & $-0.17(-0.10)$ & $-0.16(-0.10)$ \\
\bottomrule
\end{tabular}
\label{cumu-beta}
\end{center}
\end{sidewaystable}

\section{EGOE(2)-$J$ Ensemble: Structure of Centroids and Variances for Fermions in a Single-$j$ Shell}
\label{c7s4}

\subsection{Definition and construction of EGOE(2)-$J$}
\label{c7s4s1}

Shell-model corresponds to $m$ fermions occupying sp  $j$-orbits
$j_1,j_2,\ldots$ and interacting via a two body interaction  $H=V(2)$ that
preserves total $m$-particle angular momenta $J$.  For simplicity we restrict to
identical nucleons and degenerate sp energies. Firstly, the $V(2)$ matrix
$[V(2)]$ in two-particle spaces is a direct sum of matrices, $[V(2)]=
[V^{J_{12}}(2)] \oplus [V^{J^\pr_{12}}(2)]  \oplus [V^{J^{\pr \pr}_{12}}(2)]
\oplus \ldots$  where $J_{12}$ are two-particle angular momenta. Now the
$[V^{J_{12}}(2)]$ matrices are represented by GOE, i.e., $V(2)$ in two-particle
spaces is a direct sum of GOE's.  Let us consider the example of
$j=(7/2,5/2,3/2,1/2)$, i.e., the nuclear $2p1f$ shell. Here $J_{12}=0-6$ and the
corresponding matrix dimensions are $4$, $3$, $8$, $5$, $6$, $2$, and $2$,
respectively. This gives $94$ independent matrix elements for the $\{V(2)\}$
ensemble and they are chosen to be Gaussian variables with zero center and
variance unity (variance of the diagonal elements being 2); see 
Eq. (\ref{eq.bpp3}). The EGOE(2)-$J$
ensemble in $m$-particle spaces is then generated by propagating  this 
$\{V(2)\}$ ensemble to a given $(m,J)$ space by using the shell-model geometry,
i.e., by the algebra $U(N) \supset SO_J(3)$ with a suitable sub-algebra in
between, where $N=\sum_i\;(2j_i+1)$. Then, the $m$-particle $H$ matrix elements are
linear combinations of two-particle matrix elements with the expansion
coefficients being essentially fractional parentage coefficients. For the
$(2p1f)^{m=8}$ system, the dimensions $D(m,J)$ are 347, 880, 1390, 1627, 1755,
1617, 1426, 1095, 808, 514, 311, 151, 73, 22, 6 for $J=0$ to $14$,
respectively.  As the shell-model geometry is complex, EGOE(2)-$J$ is
mathematically a difficult ensemble. In the case of a single-$j$ shell,
$J_{12}=0,2,4,\ldots,(2j-1)$ and  $\{V^{J_{12}}(2)\}$ are one dimensional.  In
general, nuclear shell-model codes can be used to construct EGOE(2)-$J$ 
\cite{Br-81,Zel-04,Zh-04,Pa-07}.

For a $(j)^{m}$ system with $H$'s preserving angular momentum  
$J$ symmetry, the operator form for a two-body $H$ is,
\be 
H =\;\dis\sum_{J_2=\mbox{even},M_2}\;V_{J_2} \; A\l(j^2;J_2M_2\r) 
\l[A\l(j^2;J_2M_2\r)\r]^{\dagger} \;,
\label{eq.egj1}
\ee
where $V_{J_2}=\lan (j^2)J_2 M_2 \mid H \mid (j^2)J_2 M_2\ran$ are independent 
of $M_2$ and $J_2=0,2,4,\ldots$, $(2j-1)$. The operator $A(j^2;J_2M_2)$ 
creates a two-particle state. The 
EGOE(2)-$J$ ensemble for the $(j)^m$ system 
is generated by assuming $V_{J_2}$'s to be
independent Gaussian random variables with zero center and variance unity,
\be
\rho_{V_{J_{2a}},V_{J_{2b}},\ldots}(x_a,x_b,\ldots)dx_a dx_b \ldots =
\rho_{V_{J_{2a};\cg}}(x_a)\;\rho_{V_{J_{2b};\cg}}(x_b)\ldots \;dx_a dx_b \ldots
\label{eq.egj2}
\ee 
One simple way to construct the EGOE(2)-$J$ ensemble in $m$-particle spaces
with a fixed-$J$  value is as follows. Consider the $N=(2j+1)$ sp
states $|jm\rangle$, $m=-j,-j+1,\ldots,j$. Now distributing $m$ fermions in
the $|jm\rangle$ orbits in all possible ways will give the configurations
$[m_{\nu}]=|n_{\nu_1},n_{\nu_2},\ldots,n_{\nu_m}\rangle$ where
$(\nu_1,\nu_2,\ldots,\nu_m)$ are the filled orbits so that $n_{\nu_i}=1$. We
can select configurations such that $M=\sum_{i=1}^{m}n_{\nu_i}m_{\nu_i}=0$
for even $m$ and $M=1/2$ for odd $m$. The number of  $[m_{\nu}]$'s for $M=0$,
with $m$ even, is $D(m,M=0)=  \sum_{J=0}^{J_{max}} d(m,J)$ and similarly for
odd $m$, $D(m,M=1/2) = \sum_{J=1/2}^{J_{max}} d(m,J)$. Converting $V_{J_2}$
into the $\l.\l|jm \r.\ran \l.\l|jm^\pr \r.\ran$ basis will give,
\be 
\barr{rcl}
V_{m_1,m_2,m_3,m_4} & = & \langle jm_3jm_4|V| jm_1jm_2\rangle \\ \\
&=& 2\;\dis\sum_{J_2=\mbox{even},M_2} \langle j m_1 j m_2|J_2M_2\rangle 
\langle j m_3 j m_4|J_2M_2\rangle V_{J_2} \;,
\earr \label{eq.egj3}
\ee
where $M_2=m_1+m_2=m_3+m_4$. The $V$ matrix in the $[m_{\nu}]$ basis follows
easily from the formalism used for  EGOE(2) for spinless fermion systems when
we use $V_{m_1,m_2,m_3,m_4}$ matrix  elements; see Chapter \ref{ch1} for
details. Starting with the $J^2$ operator and writing its one and two-body 
matrix elements  in the $\l.\l| jm \r.\ran \l.\l|jm^\pr \r.\ran$  basis, 
it is possible to construct
the $J^2$ matrix in the $[m_{\nu}]$ basis. Diagonalizing  this matrix will
give (with $M_0=0$ for even $m$ and  $1/2$ for odd $m$) the $C$-coefficients
in
\be 
\l. \l| (j)^m \alpha J M_0 \r.\ran = \dis\sum_{[m_{\nu}]}\;
C_{[m_{\nu}]}^{\alpha J}\;\phi_{[m_{\nu}]}
\label{eq.egj4}
\ee
and we can identify the $J$-value of the eigenfunctions by using the 
eigenvalues $J(J+1)$. With this, the $H$ matrix in the
$|(j)^m \alpha J M_0\rangle$ basis is 
\be
\langle (j)^m \beta J M_0|H|(j)^m \alpha J M_0 \rangle 
=\dis\sum_{[m_{\nu}]_i,[m_{\nu}]_f}\;\dis C_{[m_{\nu}]_i}^{\alpha J}
\;C_{[m_{\nu}]_f}^{\beta J} \;\; \langle \phi_{[m_{\nu}]_f} \mid
V \mid \phi_{[m_{\nu}]_i} \rangle \;.
\label{eq.egj5}
\ee
The above procedure can be implemented on a computer easily. In our study we
analyze  EGOE(2)-$J$ without explicitly constructing the $H$ matrices in the
$m$-particle spaces. In particular, we analyze the structure of fixed-$J$ 
energy
centroids $E_c(m,J)$ and spectral variances $\sigma^2(m,J)$ 
for $(j)^m$ systems. 

Exact formulas for $E_c(m,J)$ and $\sigma^2(m,J)$ can be obtained from the
results in \cite{Ja-79,Ja-79a,Wo-86,Ve-81,Ve-82,Ve-84,No-72}. However, 
they are
too complicated and computationally extensive.  An alternative is to use the
bivariate Edgeworth form for $\rho(E,M)$ and seek expansions for the centroids
and variances. The expansion coefficients then will involve fourth order traces
over fixed-$m$ spaces. We will derive the expansions in Sec. \ref{c7s4s3}. 
Trace
propagation formulas for the expansion coefficients are given in Sec.
\ref{c7s4s3p2}. Finally, in Sec. \ref{c7s4s4}, we will discuss the structure of
$E_c(m,J)$ and $\sigma^2(m,J)$  for $(j)^m$ systems. 

\subsection{Expansions for centroids $E_{c}(m,J)$ and variances 
$\sigma^2(m,J)$}
\label{c7s4s3}

Firstly, fixed-$J$ averages of a $J$ invariant operator 
$\co$ follow from fixed-$M$ averages using,
\be
\barr{rcl}
\lan \co \ran^{m,J} & = & \dis\frac{\lan\lan \co\ran\ran^{m,M=J} -
\lan\lan \co \ran\ran^{m,M=J+1}}{\cd(m,M=J)-\cd(m,M=J+1)} \\ \\ \\
& \simeq & \l[-\l.\dis\frac{\partial \cd(m,M)}{\partial M}
\r|_{M=J+1/2}\r]^{-1}
\l[-\l.\dis\frac{\partial \lan\lan \co \ran\ran^{M}}{\partial M} 
\r|_{M=J+1/2}\r] .
\earr \label{eq.fj2}
\ee
Here, $\cd(m,M)$ is fixed-$M$ dimension. We use an expansion for the bivariate
distribution $\rho^{H,m}(E,M)$ and obtain the expansion for various quantities
in Eq. (\ref{eq.fj2}). Applying this to $H$ and $H^2$ operators, we have derived
expansions to order $[J(J+1)]^2$ for $E_c(m,J)$ and $\sigma^2(m,J)$. Now we
present these results.

The operators $H$ and $J_z$ whose eigenvalues are $E$ and $M$, respectively,
commute and therefore the bivariate moments of $\rho^{H,m}(E,M)$ are just
$M_{rs}(m) = \lan H^r J_z^s \ran^m$; note that nuclear effective
Hamiltonians are all $J$ invariant. Now some important results are: (i)
$M_{rs}(m)=0$ for $s$ odd and therefore all the cumulants 
$k_{rs}(m)=0$ for $s$ odd; (ii)
the marginal densities $\rho(E)$ and $\rho(M)$ are close to Gaussian, the
first one is a result of the fact that nuclear $H$'s can be represented by
two-body random matrix ensembles giving $k_{40}(m) \sim -4/m$ and the second
as $J_z$ is a one-body operator giving $k_{04}(m) \sim -1/m$; (iii) the
correlation coefficient $\zeta_{biv}(m)=k_{11}(m)=0$ and hence the bivariate
Gaussian in $(E,M)$ is just $\rho_\cg(E)\rho_\cg(M)$; (iv) random matrix
representation of $H$ shows that $k_{22}(m) \sim -2/3m$ in the dilute limit 
and this follows from the results in Eqs. (\ref{eq.fj10}), (\ref{eq.fj36}) 
and (\ref{eq.fj45}); (v) as $k_{rs}(m) \sim 1/m$ for
$r+s=4$, one can assume further that $k_{rs}(m) \sim 1/[m^{(r+s-2)/2}]$.
With (i)-(v), it is possible to use bivariate ED expansion for
$\rho(E,M)$ and the system parameter that decides the convergence of the
expansion is the particle  number $m$; see \cite{Ko-01,Ko-84,St-87}.
The expansion for $\eta(E,M)$ up to order $1/m^2$ follow from Eq. (12) and 
Table 2 of \cite{Ko-84}. 
Using these and noting that $\we=He_1(\we)$ and $\we^2-1=He_2(\we)$, 
the traces $\lan\lan (\wh)^p \ran\ran^{m,M}$, 
$p=0,\;1,\;2$ are given by
\be
\barr{rcl}
\lan\lan \wh \ran\ran^{m,M} & = & \cd_\cg(m,M)
\l\{\dis\frac{k_{12}(m)}{2}He_2(\whm) + \dis\frac{k_{14}(m)}{24}
He_4(\whm) \r. \\ \\ 
& + & \l. \dis\frac{k_{04}(m)
k_{12}(m)}{48}He_6(\whm) + O\l(\dis\frac{1}{m^{5/2}}\r)\r\} \;,
\\ \\
\lan\lan \wh^2 -1 \ran\ran^{m,M}  & = &  \cd_\cg(m,M) \l\{
\dis\frac{k_{22}(m)}{2} He_2(\whm) + 
\dis\frac{[k_{12}(m)]^2}{4} He_4(\whm) \r. 
\\ \\
& + & \dis\frac{k_{24}(m)}{24} He_4(\whm) +
\dis\frac{k_{14}(m)k_{12}(m)}{24}He_6(\whm) \\ \\
& + & \l. \dis\frac{k_{22}(m) k_{04}(m)}{48} He_6(\whm) +
\dis\frac{k_{04}(m) [k_{12}(m)]^2}{96} He_8(\whm) \r.
\\ \\
& + & \l. O\l(\dis\frac{1}{m^3}\r) \r\} \;,
\earr \label{eq.fj17a}
\ee
\be
\barr{rcl}
\cd(m,M) & = & \cd_\cg(m,M)\l\{He_0(\whm)+
\dis\frac{k_{04}(m)}{24} He_4(\whm) \r. 
\\ \\
& + & \l. \dis\frac{k_{06}(m)}{720} He_6(\whm) + 
\dis\frac{[k_{04}(m)]^2}{1152} He_8(\whm) +
O\l(\frac{1}{m^3}\r) \r\} \;.\nonumber
\earr \label{eq.fj17a-1}
\ee
Here we have used the results that $\int He_r(\we) He_s(\we) \eta_\cg(\we) d\we
=r! \; \delta_{rs}$ and $\whm=M/\sigma_{J_z}(m)$ with $\sigma^2_{J_z}(m) =\lan
J_z^2 \ran^m$.

Using Eqs. (\ref{eq.fj2}) and (\ref{eq.fj17a}) and carrying out some 
tedious algebra (and also verified using Mathematica)
will give the following expansions to order $[J(J+1)]^2$,
\be
\barr{rcl}
D(m,J) & \simeq & \dis\binom{N}{m} \; \dis\frac{(2J+1)}
{\sqrt{8\pi}\sigma_{J_z}^3(m)}\,
\exp-\dis\frac{(J+1/2)^2}{2 \sigma_{J_z}^2(m)} \\ \\
& \times & \l[1+\dis\frac{k_{04}(m)}{24}
\l\{\l[\dis\frac{J(J+1)}{\sigma_{J_z}^2(m)}\r]^2-10
\dis\frac{J(J+1)}{\sigma_{J_z}^2(m)}+15\r\}\r]\;,
\earr \label{eq.fj6a}
\ee
\be
\barr{l}
\lan \wh \ran^{m,J} =  \dis\frac{E_c(m,J) -E_c(m)}{\sigma(m)} 
\\ \\ 
= \l[\dis\frac{k_{12}(m)}{2} \l(-3 + \dis\frac{1}{4\sigma^2_{J_z}(m)}
\r) + \dis\frac{k_{14}(m)}{8}\l(5-\dis\frac{5}{6\sigma^2_{J_z}(m)}
+ \dis\frac{1}{48\sigma^4_{J_z}(m)}\r) \r. \\ \\
+ \l. \dis\frac{k_{04}(m) k_{12}(m)}{4} \l(-5 +
\dis\frac{5}{4\sigma^2_{J_z}(m)} -
\dis\frac{1}{24\sigma^4_{J_z}(m)}\r) \r] 
\\ \\ 
+ \dis\frac{J(J+1)}{\sigma_{J_z}^2(m)} \l\{\dis\frac{k_{12}(m)}{2}
+ \dis\frac{k_{14}(m)}{12} \l(-5 +\dis\frac{1}{4\sigma^2_{J_z}(m)}
\r) \r.  \\ \\ \l. 
+ \dis\frac{k_{04}(m) k_{12}(m)}{4} \l(5
-\dis\frac{1}{3\sigma^2_{J_z}(m)}\r)\r\} \\ \\ +
\dis\frac{[J(J+1)]^2}{\sigma_{J_z}^4(m)}\l\{\dis\frac{k_{14}(m)}{24}
- \dis\frac{k_{04}(m) k_{12}(m)}{6} \r\} \\ \\
\simeq \l[-\dis\frac{3 k_{12}(m)}{2} \r] + \dis\frac{k_{12}(m)}{2}
\dis\frac{J(J+1)}{\sigma_{J_z}^2(m)} +
\l\{\dis\frac{k_{14}(m)}{24} - \dis\frac{k_{04}(m) k_{12}(m)}{6}
\r\} \dis\frac{[J(J+1)]^2}{\sigma_{J_z}^4(m)}\;, 
\earr \label{eq.fj6b}
\ee
\be
\barr{l}
\dis\frac{\sigma^2(m,J)}{\sigma^2(m)}  =  \lan \wh^2\ran^{m,J} -
\l(\lan\wh\ran^{m,J}\r)^2 
\\ \\ 
= \l[1 -\dis\frac{3 k_{22}(m)}{2} + 
\dis\frac{3 [k_{12}(m)]^2}{2} + \dis\frac{5 k_{24}(m)}{8} - 
\dis\frac{5 k_{14}(m) k_{12}(m)}{2} \r. \\ \\
- \l. \dis\frac{5 k_{22}(m) 
k_{04}(m)}{4} + \dis\frac{15 k_{04}(m)[k_{12}(m)]^2}{4} \r] +
\l[\dis\frac{J(J+1)}{\sigma^2_{J_z}(m)} +
\dis\frac{1}{4\sigma^2_{J_z}(m)} \r]\; 
\\ \\ 
\l\{\dis\frac{k_{22}(m)}{2} 
- [k_{12}(m)]^2 - \dis\frac{5 k_{24}(m)}{12} + 
\dis\frac{5k_{14}(m) k_{12}(m)}{2} - 5 k_{04}(m) [k_{12}(m)]^2 \r. 
\\ \\ 
+ \l. \dis\frac{5 k_{22}(m) k_{04}(m)}{4} \r\} + \l[\dis\frac{J(J+1)}
{\sigma^2_{J_z}(m)} + \dis\frac{1}{4\sigma^2_{J_z}(m)} \r]^2 \;
\l\{\dis\frac{k_{24}(m)}{24} - \dis\frac{k_{14}(m)k_{12}(m)}{3} \r. 
\\ \\
- \l. \dis\frac{k_{22}(m) k_{04}(m)}{6} + 
\dis\frac{5 k_{04}(m) [k_{12}(m)]^2}{6} \r\}  
\earr \label{eq.fj6c} 
\ee
\be
\barr{l}
\simeq \l[1 -\dis\frac{3 k_{22}(m)}{2} \r] + \l[\dis\frac{k_{22}(m)}{2} \r]
\; \dis\frac{J(J+1)}{\sigma^2_{J_z}(m)} 
+ \l\{\dis\frac{k_{24}(m)}{24} - \dis\frac{k_{22}(m) k_{04}(m)}{6} 
\r\} \;\l[\dis\frac{J(J+1)}{\sigma^2_{J_z}(m)}\r]^2 \;. \nonumber
\earr \label{eq.fj6c-1}
\ee
The last step in Eq. (\ref{eq.fj6b})  follows from the assumption
that $\sigma_{J_z}^2(m) >> 1$.  Similarly, in the last step in Eq. 
(\ref{eq.fj6c}), assuming that 
$\sigma_{J_z}^2(m) >> 1$, we have neglected $1/4\sigma^2_{J_z}(m)$ 
terms and so also the terms with squares and products of cumulants 
that are expected to be small.
Note that the expansions to order $J(J+1)$ were given before \cite{Kk-02} 
and the terms with $[J(J+1)]^2$ are new. From now on, we use the last forms in
Eqs. (\ref{eq.fj6b}) and (\ref{eq.fj6c}) and apply them to $(j)^m$ systems in
the present section.
To proceed further, we need to define and evaluate the bivariate cumulants 
$k_{rs}(m)$. 

Bivariate cumulants $k_{rs}(m)$ are defined in terms of the bivariate moments 
$\lan \widetilde{H}^r J_z^s \ran^m$ with
$\widetilde{H}=H-\lan H \ran^m$,
\be 
\barr{rcl}
k_{04}(m) & = & \dis\frac{\lan J_z^4 \ran^m}{\sigma^4_{J_z}(m)}-3\;,\;\;\;
k_{12}(m) = \dis \frac{\lan
\widetilde{H} J_z^2 \ran^m}{\sigma(m)\sigma_{J_z}^2(m)}\;,  \\ \\
k_{14}(m) & = & \;\dis\frac{\lan {\widetilde{H}J_z^4} \ran^m}{\sigma(m)
\sigma_{J_z}^4(m)} - \frac{6\; \lan \widetilde{H}
 J_z^2 \ran^m}{\sigma(m) \sigma_{J_z}^2(m)}\;, \\ \\
k_{22}(m) & = & \dis\frac{\lan \widetilde{H}^2 J_z^2 \ran^m}{
\sigma_{J_z}^2(m)\sigma^2(m)}-1 \;, 
\earr \label{eq.fj17}
\ee
\be
\barr{rcl}
k_{24}(m) & = & \dis\frac{\lan \widetilde{H}^2 J_z^4\ran^m}{\sigma_{J_z}^4(m) \;
\sigma^2(m)} - \dis\frac{\lan J_z^4\ran^m}{\sigma_{J_z}^4(m)} 
- 6\; \dis\frac{\lan \widetilde{H}^2 J_z^2\ran^m}{\sigma_{J_z}^2(m)
\;\sigma^2(m)} \\ \\
& - & 6\; \dis\frac{\l[\lan \widetilde{H} J_z^2\ran^m\r]^2}
{\sigma_{J_z}^4(m) \;\sigma^2(m)} + 6 \;. \nonumber
\earr \label{eq.fj17aa}
\ee 
Note that, $\sigma^2(m)= \lan \widetilde{H}^2 \ran^m$. 

\subsection{Propagation equations for bivariate cumulants $k_{rs}(m)$ 
for $(j)^m$ systems}
\label{c7s4s3p2}

To begin with, let us mention that the tensorial decomposition of the $H$  
and $J^2$  operators with
respect to the $U(N)$, $N=2j+1$, algebra will be useful for deriving 
propagation equations for $k_{rs}(m)$. For the single-$j$ shell situation, 
the $H$ operator is defined by the two-body matrix elements $V_{J_2} = 
\lan (j)^2 J_2 \mid H \mid (j)^2 J_2 \ran$ with $J_2$ being even taking 
values $0,2,\ldots,2j-1$.
Using the results in Appendix \ref{c2a2}, unitary decomposition for 
the operators $H$ and $J^2$ are,
\be 
\barr{l} 
H = H^{\nu=0} + H^{\nu=2} \;;\\ \\
H^{\nu=0} = \dis\frac{\hat{n}(\hat{n}-1)}{2}\; \overline{V}
\;,\;\;\;\;
\overline{V} = \dis\binom{N}{2}^{-1}\;\dis\sum_{J_2}\,
(2J_2 +1)\, V_{J_2} \;, \\ \\
H^{\nu=2} \Longleftrightarrow V^{\nu=2}_{J_2} = V_{J_2} - 
\overline{V} 
\;. 
\earr \label{eq.fj8}
\ee
\be
\barr{rcl}
J^2 & = & (J^2)^{\nu=0} + (J^2)^{\nu=2}\;; \\ \\
(J^2)^{\nu=0} & = &  \dis\frac{\hat{n}(N-\hat{n})}{N(N-1)}\,j(j+1)(2j+1)\;, 
\\ \\
(J^2)^{\nu=2} & \Longleftrightarrow & (J^2)^{\nu=2}_{J_2} = 
J_2(J_2+1) - (2j-1)(j+1)\;.
\earr \label{eq.fj8a}
\ee
To proceed further, we write the cumulants defined in Eq. (\ref{eq.fj17}) 
in terms of $H^{\nu=2}$ and $(J^2)^{\nu=2}$. For
this purpose, we  use the equalities $\lan H^p J_z^2 \ran^m = \lan H^p
J^2\ran^m/3$ and  $\lan H^p J_z^4 \ran^m = \lan H^p (J^2)^2\ran^m/5 - 
\lan H^p (J^2) \ran^m/15$. Then the formulas are,
\be
\barr{rcl}
k_{12}(m) & = & \dis\frac{\lan H^{\nu=2}(J^2)^{\nu=2}\ran^m} 
{3\; \sigma(m) \sigma_{J_z}^2(m)} 
\;, \\ \\
k_{14}(m) & = &
\dis\frac{1}{\sigma(m)\;\sigma_{J_z}^4(m)} \l\{\dis\frac{1}{5} 
\lan (J^2)^{\nu=2}
(J^2)^{\nu=2} H^{\nu=2}\ran^m \r.
\\ \\
& - & \l. \l[\dis\frac{4}{5} \sigma^2_{J_z}(m)
+\dis\frac{1}{15} \r] \lan (J^2)^{\nu=2} H^{\nu=2}\ran^m \r\} 
\\ \\
& \simeq & \dis\frac{1}{24\;\sigma_{J_z}^8(m)} 
\l\{\dis\frac{1}{5} \lan (J^2)^{\nu=2}
(J^2)^{\nu=2} H^{\nu=2}\ran^m \r.
\\ \\
& - & \l. \dis\frac{4}{5} \sigma^2_{J_z}(m) \;\; 
\lan (J^2)^{\nu=2}
H^{\nu=2}\ran^m \r\} \;, 
\earr \label{eq.fj9}
\ee
\be
\barr{rcl}
k_{22}(m) & = & \dis\frac{\lan (H^{\nu=2})^2(J^2)^{\nu=2}
\ran^m}{3 \sigma^2_{J_z}(m)\,\sigma^2(m)} \;, \\ \\
k_{24}(m) & = & \dis\frac{9}{5} - 
\dis\frac{1}{5\;\sigma_{J_z}^2(m)} - \dis\frac{
\lan J_z^4 \ran^m}{\sigma_{J_z}^4(m)} +
\dis\frac{\lan ((J^2)^{\nu=2})^2(H^{\nu=2})^2 \ran^m}
{5\;\sigma_{J_z}^4(m) \;
\sigma^2(m)} 
\\ \\
& - & \dis\frac{2\,\l[\lan (J^2)^{\nu=2}H^{\nu=2} \ran^m \r]^2}{3\;
\sigma_{J_z}^4(m)\;\sigma^2(m)} - 
\dis\frac{\lan (J^2)^{\nu=2}(H^{\nu=2})^2
\ran^m}{15\; \sigma_{J_z}^4(m) \;\sigma^2(m)} \\ \\
& - &
\dis\frac{4\;\lan (J^2)^{\nu=2}(H^{\nu=2})^2 \ran^m}{5
\;\sigma_{J_z}^2(m) \; \sigma^2(m)} \;.\nonumber
\earr \label{eq.fj9-1}
\ee
Simple trace propagation formulas that follows from the results in Appendix 
\ref{c2a2} are as follows,
\be
\barr{l}
\sigma_{J_z}^2(m) = \lan J_z^2 \ran^m = \dis\frac{1}{3} 
\lan (J^2)^{\nu=0}
\ran^m = \dis\frac{m(N-m)}{N(N-1)}\; \dis\frac{1}{3} \; j(j+1)(2j+1)
\;, \\ \\
\lan J_z^4 \ran^m = \dis\frac{9}{5} \sigma_{J_z}^4(m) - \dis\frac{1}{5}
\sigma_{J_z}^2(m) + \dis\frac{1}{5} \lan (J^2)^{\nu=2} (J^2)^{\nu=2} \ran^m 
\;;\\ \\
\lan X^{\nu=2} Y^{\nu=2} \ran^m 
= \dis\frac{m(m-1)(N-m)(N-m-1)}{N(N-1)(N-2)(N-3)}\,
\dis\sum_{J_2}
(2J_2+1) X^{\nu=2}_{J_2} \;Y^{\nu=2}_{J_2}\;.
\earr \label{eq.fj10}
\ee
Note that for $\sigma^2(m) = \lan H^{\nu=2} H^{\nu=2} \ran^m$ is given by 
$X=Y=H$ in last equality in Eq. (\ref{eq.fj10}). Similarly,  
$\lan H^{\nu=2} (J^2)^{\nu=2} \ran^m$ and $\lan (J^2)^{\nu=2} (J^2)^{\nu=2} 
\ran^m$ are given by $X=H, \;Y=J^2$ and $X=Y=J^2$, respectively.
From  now on, we use the symbols 
$m^\times = (N-m)$ and
$[X]_{r} = X(X-1) \ldots (X-r+1)$, $X = m, N, m^\times$.
Then, the propagation equation for $\lan (J^2)^{\nu=2} (J^2)^{\nu=2} 
H^{\nu=2} \ran^m$ is \cite{Ko-01}, 
\be
\lan (J^2)^{\nu=2} (J^2)^{\nu=2} H^{\nu=2}\ran^m = \dis\frac{[m]_3
[m^\times]_3}{ [N]_6} A + \l[\dis\frac{[m]_2[m^\times]_4 + [m]_4
[m^\times]_2}{[N]_6}\r] B \;,
\label{eq.fj36} 
\ee 
where 
\be 
\barr{rcl}
A & = & \dis\sum_{\Delta}\,(-1)^{\Delta} (2\Delta +1)^{-1/2}
[\beta^{\Delta}((J^2)^{\nu=2})]^2 \beta^{\Delta}(H^{\nu=2}) \;,\\ \\
B & = & \lan\lan (J^2)^{\nu=2} (J^2)^{\nu=2} H^{\nu=2}\ran\ran^2
= \dis\sum_{J_2} (2J_2+1)\;\l[(J^2)^{\nu=2}_{J_2}\r]^2\;V^{\nu=2}_{J_2}\;.
\earr \label{eq.fj361}
\ee
Note that $\Delta$ symbol in Eq. (\ref{eq.fj361}) should not be confused with
`$\Delta$' used in Chapters \ref{ch2}-\ref{ch6}.
In Eq. (\ref{eq.fj361}), the term 
$A$ is more complicated involving particle-hole matrix elements 
$\beta^{\Delta}$ of the $(J^2)^{\nu=2}$ and $H^{\nu=2}$ operators. 
The $\beta^{\Delta}$ for a $\nu=2$ operator $\bv$, in the example 
of a single $j$ shell is
\be
\beta^{\Delta}(\bv) =-2 \dis\sum_{J_2 =\;even}\; (-1)^\Delta
\dis\sqrt{2\Delta +1} \;\;(2J_2+1)\; \l\{\barr{ccc} j & j & J_2 
\\ j & j &
\Delta \earr\r\} \;\bv_{J_2}\;.
\label{eq.fj37}
\ee
For $j >>1$, $(J^2)^{\nu=2}$ can be approximated as
\be
(J^2)^{\nu=2}_{J_2} \simeq -2j(j+1)(2j+1) \l\{\barr{ccc} j & j & 
J_2 \\
j & j & 1 \earr\r\}\;.
\label{eq.fj38}
\ee
Substituting this in Eqs. (\ref{eq.fj37}) will give,
\be
\beta^{\Delta}[(J^2)^{\nu=2}] = 2j(j+1)(2j+1) \dis\sqrt{2\Delta +1} \;
(-1)^\Delta \l[\dis\frac{1}{3} \delta_{\Delta,1} + (-1)^{\Delta + 1}
\l\{\barr{ccc} j & j & \Delta \\ j & j & 1 \earr\r\} \r]\;.
\label{eq.fj39}
\ee
Now $A$ in Eq. (\ref{eq.fj36}) takes a simple form,
\be
\barr{rcl}
A & = & -8 [j(j+1)(2j+1)]^2 \dis\sum_{J_2} (2J_2+1)\;V_{J_2}^{\nu=2} 
\;\; X_{J_2} \;; \\ \\
X_{J_2} & = & \dis\sum_{\Delta} (2\Delta +1)\;
\l\{\barr{ccc} j & j & J_2 \\ j & j & \Delta \earr\r\}\;
\l[\dis\frac{1}{3} \delta_{\Delta,1} + (-1)^{\Delta + 1}
\l\{\barr{ccc} j & j & \Delta \\ j & j & 1 \earr\r\} \r]^2 \\ \\
& = & \dis\frac{1}{3} \l\{\barr{ccc} j & j & J_2 \\ j & j & 1 
\earr\r\}
+ 2 \l\{\barr{ccc} j & j & 1 \\ j & j & 1 \earr\r\}
\l\{\barr{ccc} j & j & J_2 \\ j & j & 1 \earr\r\} +
\l\{\barr{ccc} j & j &J_2 \\ j & 1 & j \\ 1 & j & j \earr\r\} \\ \\
& = & - \dis\frac{(J^2)^{\nu=2}_{J_2}}{6Y_j} + \dis\frac{
(J^2)^{\nu=2}_{J_2} [j(j+1)-1]}{Y_j^2} +
\dis\frac{ (J^2)^{\nu=2}_{J_2}[(J^2)^{\nu=2}_{J_2}+2]}{4Y_j^2} \;,
\earr \label{eq.fj40}
\ee
where $Y_j=j(j+1)(2j+1)$. Above simplifications are obtained using the results
given in \cite{Ed-74,Br-94} for angular-momentum recoupling coefficients.
Going further, Eq. (\ref{eq.fj36}) will give 
the expression for $\lan (J^2)^{\nu=2}(H^{\nu=2}
)^2\ran^m$ with $A$ and $B$ defined by
\be
\barr{rcl}
A & = & \dis \sum_{\Delta} \frac{(-1)^{\Delta}}{(2\Delta +1)^{\frac{1}{2}}}
\l[\beta^{\Delta}(V^{\nu=2})\r]^2\,\beta^{\Delta}((J^2)^{\nu=2}) \;,
\\ \\
B & = & \dis\sum_{J_2} (2J_2+1)(V^{\nu=2}_{J_2})^2\,(J^2)^{\nu=2}_{J_2}\;.
\earr \label{eq.fj45}
\ee
Using the expression for $\beta^{\Delta}$ for a $\nu=2$ operator from
Eq. (\ref{eq.fj37}) and the simple formula for $\beta^\Delta((J^2)^{\nu=2})$
given by Eq. (\ref{eq.fj39}), the term $A$ in Eq. (\ref{eq.fj45}) simplifies to,
\be
\barr{rcl}
A & = & 8j(j+1)(2j+1) \dis\sum_{J_2,J_2^\pr} (2J_2+1)(2J_2^\pr+1) 
V_{J_2}^{\nu=2} V_{J_2'}^{\nu=2}
\\ \\ 
& \times & \l[\l\{\barr{ccc} j & j & J_2 \\ j & j & 1 
\earr\r\} \l\{\barr{ccc} j & j & J_2^\pr \\ j & j & 1 \earr\r\} - 
\l\{\barr{ccc} J_2^\pr & j & j \\ j & J_2 & 1 \earr\r\}^2 \r] 
\earr \label{eq.fj46}
\ee
\be
\barr{rcl}
& = & \dis\frac{2}{j(j+1)(2j+1)}\l[
\dis\sum_{J_2}(2J_2+1)V_{J_2}^{\nu=2}(J^2)_{J_2}^{\nu=2}\r]^2 
\\ \\
& - & 2\dis\sum_{J_2}(2J_2+1)(V_{J_2}^{\nu=2})^2 J_2(J_2+1)\;.
\earr \label{eq.fj46-1}
\ee
Most complicated is the $k_{24}(m)$ cumulant that involves $\lan
(J^2)^{\nu=2}(J^2)^{\nu=2}  (H^{\nu=2})(H^{\nu=2})\ran^m$. Equations (69)
and (70) in \cite{Wo-86} give a formula for this trace in a complex form.
After carrying out the simplification of these equations  and correcting
errors at many places, it is seen that there will be 9 terms  in the
propagation equation. Table \ref{c7tj} gives the final result.  
We have verified
that the results in Table  \ref{c7tj} are correct by replacing  
$(J^2)^{\nu=2}$ with
$H^{\nu=2}$ and then comparing with the formulas given  in \cite{No-72}. Results
that are simple as in Table \ref{c7tj} for $k_{24}(m)$ for multi-$j$
shell situation are not yet available and because of this, 
we have restricted our discussion to single-$j$ shell in this section. For
multi-$j$ shell with realistic sp energies, the EGOE(1+2)-$J$ is also called
realistic TBRE (RTBRE) \cite{Fl-00}.

\begin{table}[pt]
\caption{Propagation equation for $\lan\lan (J^2)^{\nu=2} (J^2)^{\nu=2}
H^{\nu=2} H^{\nu=2} \ran\ran^m$. Column 2 gives the input trace in a symbolic
form and the corresponding expressions are given in the footnote. Column
3 gives the corresponding  propagators. Multiplying the terms in column 
2 with corresponding ones in column 3 and summing gives the propagation 
formula. Note that $N=2j+1$.}
{\begin{center} \begin{tabular}{@{}ccc@{}} \toprule  
term & Input Trace & Propagator \\
\midrule
$\#1$ & $J^1H^1H^1J^1$ & ${N-8 \choose m-2}+{N-8 \choose m-6}+4{N-8 \choose m-4}$
\\ \\
$\#2$ & $J^1J^2H^2H^1$ & $2{N-8 \choose m-4}+\frac{4}{N}\l\{
{N-8 \choose m-3}+{N-8 \choose m-5 }\r\}-\frac{24}{N}{N-8 \choose m-4}$\\ \\
$\#3$ & $J^1H^2H^2J^1$ & ${N-8 \choose m-4}+\frac{4}{N}\l\{{
N-8 \choose m-3}+{N-8 \choose m-5}\r\}-\frac{8}{N}{N-8 \choose m-4}$\\ \\
$\#4$ & $J^1\beta^2_J\beta^2_HH^1$ & $-4{N-8 \choose m-3}-4{N-8 \choose m-5}+
8{N-8 \choose m-4}$ \\ \\
$\#5^a$ & $J^1\beta^2_H\beta^2_HJ^1$ & $-2{N-8 \choose m-3}-2{N-8 \choose m-5}$ 
\\ \\
$\#6^a$ & $H^1\beta^2_J\beta^2_JH^1$ & $-2{N-8 \choose m-3}-2{N-8 \choose m-5}$ 
\\ \\
$\#7$ & $\beta^1_H\beta^1_H\beta^1_J\beta^1_J$ & $3{N-8 \choose m-4}$\\ \\
$\#8$ & $\beta^1_J\beta^2_J\beta^2_H\beta^1_H$ & $4{N-8 \choose m-3}+
4{N-8 \choose m-5}-8{N-8 \choose m-4}$ \\ \\
$\#9^a$ & $\beta^1_H\beta^2_J\beta^2_J\beta^1_H$ & $-8{N-8 \choose m-4}$ \\
\midrule
\end{tabular} \label{c7tj} \end{center}}
\noindent ${\#1} = \dis \lan \lan  \l[ (J^2)^{\nu=2} \r]^2
(H^{\nu=2})^2 \ran \ran^{m=2}$,$\;\;\;\;{\#2} = \dis\l[ \lan \lan
(J^2)^{\nu=2} H^{\nu=2} \ran \ran^{m=2} \r]^2$ \\ \\
${\#3} = \dis \lan \lan \l[(J^2)^{\nu=2}\r]^2 \ran
\ran^{m=2} \; \lan \lan (H^{\nu=2})^2 \ran \ran^{m=2} $\\ \\
${\#4} = \dis\sum_{\Gm, \Delta} (2\Gm +1)
\l\{\barr{ccc} j
& j & \Gm \\ j  & j & \Delta \earr\r\}
(J^2)^{\nu=2}_{\Gm}\beta^{\Delta}((J^2)^{\nu=2})
\beta^{\Delta}(H^{\nu=2}) V_{\Gm}^{\nu=2}$
\\ \\
${\#7} = \dis \sum_{\Delta}\frac{1}{(2\Delta
+1)} \;\l[\beta^{\Delta}(H^{\nu=2})\r]^2 \; \l[\beta^{\Delta}((J^2)^{\nu=2})
\r]^2$ 
\\ \\
${\#8} = \dis\sum_{\Delta}
\beta^{\Delta}((J^2)^{\nu=2}) \; \beta^{\Delta}(H^{\nu=2}) \;
\dis\sum_{\Gm_2,\Gm_3}(2\Gm_2+1) (2\Gm_3+1)
\l\{\barr{ccc} \Delta &
\Gm_2 & \Gm_3 \\ j & j & j \earr\r\}^2 (J^2)^{\nu=2}_{\Gm_2}\,
V^{\nu=2}_{\Gm_3}$
\\ \\
$^a$Terms $J^1\beta^2_H\beta^2_HJ^1$ and
$H^1\beta^2_J\beta^2_JH^1$ follow from appropriate permutations
of $(J^2)^{\nu=2}$ and $H^{\nu=2}$ in the
$J^1\beta^2_J\beta^2_HH^1$ expression. Similarly
$\beta^1_H\beta^2_J\beta^2_J\beta^1_H$ follows by 
appropriate permutations in the $\beta^1_J\beta^2_J\beta^2_H\beta^1_H$ 
expression.
\end{table}

\subsection{Structure of centroids and variances}
\label{c7s4s4}

\subsubsection{Centroids $E_c(m,J)$}

In the dilute limit with $m \rightarrow \infty$, $N \rightarrow
\infty$ and $m/N \rightarrow 0$, the centroids $E_c(M,J)$ given by 
Eq. (\ref{eq.fj6b}) take a simple form.
Firstly, the constant term in the expansion for $E_c(M,J)$ is [after
simplifying $k_{12}(m)$ and $k_{14}(m)$],
\be
E_c(m) - 3\sigma(m) \dis\frac{k_{12}(m)}{2} \simeq  
\dis\frac{m^2}{N^2} 
\dis\sum_{J_2}\,(2J_2+1) V_{J_2} \;.
\label{eq.egj6}
\ee
Similarly, the $J(J+1)$ term is
\be
\sigma(m) \dis\frac{k_{12}(m)}{2 \sigma_{J_z}^2(m)} \simeq 
\dis\frac{3}{2 
[j(j+1)]^2 N^2} \dis\sum_{J_2} (2J_2+1)\; V_{J_2} \; 
(J^2)^{\nu=2}_{J_2}\;.
\label{eq.egj7}
\ee
More remarkable is that the $[J(J+1)]^2$ term $\frac{\sigma(m)\,k_{14}(m)}
{24\,
\sigma_{J_z}^4(m)} - \frac{\sigma(m) k_{04}(m)
k_{12}(m)}{6\sigma_{J_z}^4(m)}$ also takes a simple form.
The results in Sec. \ref{c7s4s3p2} will give 
the expression for the first term as,
\be
\barr{rcl}
\sigma(m)\;\dis\frac{k_{14}(m)}{24\;\sigma_{J_z}^4(m)} & = & \dis\sum_{J_2}
(2J_2+1)\;V_{J_2}^{\nu=2} \;\;S_{J_2} \;; \\ \\
S_{J_2} & \simeq & \dis\frac{9}{40 m^2 (N-m)^2 N^2 [j(j+1)]^4} \l\{
3\l[(J^2)^{\nu=2}_{J_2}\r]^2 (N-2m)^2 \r. \\ \\
& & \l. -4 (J^2)^{\nu=2}_{J_2}\;j(j+1)\;\l[2N^2 -2Nm + 2m^2\r] \r\}\;.
\earr \label{eq.egj8}
\ee
Similarly, we can write the expression for
$\frac{\sigma(m)k_{04}(m)k_{12}(m)}{6\sigma_{J_z}^4(m)}$ and in the dilute
limit this reduces exactly to the second piece in the expression for $S_{J_2}$
in Eq. (\ref{eq.egj8}). Therefore, in the dilute limit, the term
multiplying $[J(J+1)]^2$ in the $E_c(m,J)$ expansion is,
\be
\barr{l}
\dis\frac{\sigma(m)}{\sigma_{J_z}^4(m)}\;\l\{\dis\frac{k_{14}(m)}{24}
-  \dis\frac{k_{04}(m) k_{12}(m)}{6}\r\} = \dis\sum_{J_2}
(2J_2+1)\;V_{J_2}^{\nu=2} \;\;R_{J_2} \;; \\ \\
R_{J_2} \sim
\dis\frac{9 (N-2m)^2}{40 m^2 (N-m)^2 N^2 [j(j+1)]^4} \l\{3\l[J_2(J_2+1)
-2j(j+1)\r]^2\r\} \;.
\earr \label{eq.egj9}
\ee
It is already pointed out in \cite{Kk-02} that the constant term and the
$J(J+1)$ term as given by Eqs. (\ref{eq.egj6}) and (\ref{eq.egj7}) are same as
those derived by Mulhall et al \cite{Mu-00,Mu-00a} using statistical mechanics 
approach that is
completely different from the present moment method approach. For EGOE(2)-$J$
ensemble, Eqs. (\ref{eq.egj6})-(\ref{eq.egj9}) will give the distribution of the
centroids over the ensemble as discussed in \cite{Mu-00a}. 
More remarkable is that the $[J(J+1)]^2$
term given by Eq. (\ref{eq.egj9})  is also very close to the formula given by
Mulhall \cite{Mu-00a}. These results confirm that the approximations used in
\cite{Mu-00,Mu-00a} are equivalent to the proposition that $\rho(E,M)$ is a
Edgeworth corrected bivariate Gaussian as assumed in the present approach. The
equivalence of Mulhall et al approach with the moment method approach in the
dilute limit is further substantiated by the expansion for fixed-$M$
occupancies; the results are given in Appendix \ref{c7a3}.

\begin{figure}[htp]
\includegraphics[width=5in,height=2in]{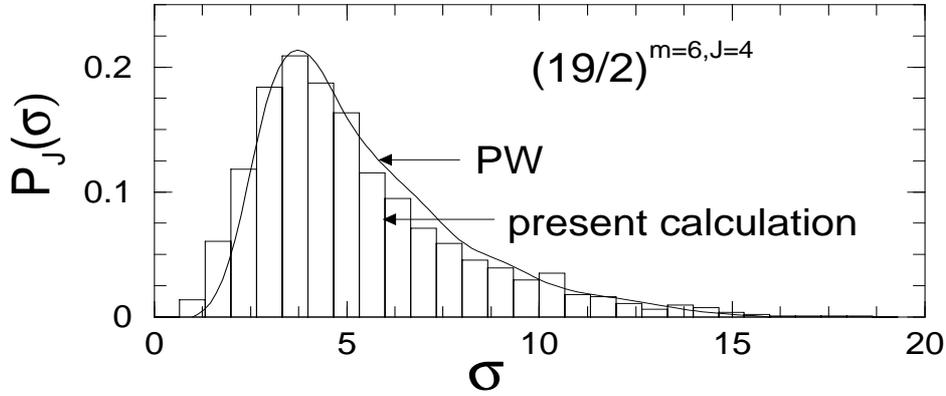}
\caption{Probability distribution for widths $\sigma$ for EGOE(2)-$J$ 
ensemble; see text for details.} 
\label{dist-wid}
\end{figure}

\subsubsection{Variances $\sigma^2(m,J)$}

In the dilute limit, simplifying $k_{22}(m)$ and $\sigma^2(m)$ will give
\be
\barr{rcl}
\sigma^2(m,J) & = & \dis\frac{m^2}{N^2}\dis\sum_{J_2}(2J_2+1)
(V_{J_2}^{\nu=2})^2 \\ \\
& + & \dis\frac{3\;J(J+1)}{2\;N^2[j(j+1)]^2}\dis\sum_{J_2}(2J_2+1)
(V_{J_2}^{\nu=2})^2(J^2)^{\nu=2} \;.
\earr \label{eq.fj789}
\ee
However to add $[(J(J+1)]^2$ correction, we need to simplify $k_{24}(m)$ and
this is quite cumbersome. 
A quantity of interest, as pointed out by Papenbrock and Weidenm\"{u}ller 
\cite{Pa-04} (PW) is the probability  distribution for the spectral widths
$\sigma=\{\lan H^2 \ran^{m,J}\}^{1/2}=[\sigma^2(m,J)+E_c^2(m,J)]^{1/2}$ over the
EGOE(2)-$J$ ensemble. To compare with PW results, we  have generated a
EGOE(2)-$J$ ensemble for $(\frac{19}{2})^{m=6}$ system with  2500 members, i.e
we have used 2500 sets of $V_{J_2}$'s. Using the formalism described in Secs.
\ref{c7s4s3} and \ref{c7s4s3p2} 
we have calculated the bivariate  cumulants $k_{rs}(m)$. For our
example $\sigma_{J_z}(m)=12.124$ and $k_{04}(m)=-0.229$.  The ensemble averaged
cumulants $\overline{k_{12}(m)},\;\overline{k_{14}(m)} \sim 0$  as expected.
However  $\overline{k_{22}(m)} =-0.053$ and $\overline{k_{24}(m)}=-0.114$.  With
these, it is clear that the expansions to order  $[J(J+1)]^2$ are needed.
Equation
(\ref{eq.fj6c}) is found to be good for $J < 30$.  We have  calculated $\langle
H^2 \rangle^{m,J}$ for each member of the ensemble and then $P_J({\sigma})$ vs
$\sigma$ histograms are constructed for various $J$  values. Results for  $J=4$ 
are shown in Fig. \ref{dist-wid}. The calculated histogram is in good agreement
with the exact curve given by PW \cite{Pa-04}; in \cite{Pa-04} a completely
different formalism is used.  Though not shown in Fig. \ref{dist-wid}, we
have noticed that for  $J=0$, the widths  given by the exact results (they are
reported in \cite{Pa-04}) are  somewhat larger than the numbers given by the
present formalism.  This could be because $J=0$ is at one extreme end of the
Edgeworth expansion and therefore, the truncation to $1/m^2$ terms may not be
adequate. 

\section{Summary}
\label{c7s5}

To summarize, by extending the binary correlation approximation method for two
different operators and for traces over two-orbit configurations, we have
derived formulas for $\gamma_1$ and $\gamma_2$ parameters for EGOE(1+2)-$\pi$
ensemble. Note that EGOE(1+2)-$\pi$ is defined by the embedding algebra $U(N)
\supset U(N_+) \oplus U(N_-)$ with the Hamiltonian breaking the symmetry
in a particular way as discussed in Chapter \ref{ch5}. In addition, we have
derived formula for the fourth order trace defining  correlation coefficient of
the bivariate transition strength of the transition operator relevant for
0$\nu$-$\beta\beta$ decay. We have also derived the formulas for the fourth
order cumulants in order to establish bivariate Gaussian form of the transition
strength densities. Here also the embedding algebra is $U(N) \supset U(N_p)
\oplus U(N_n)$ with the Hamiltonian preserving the symmetry and the transition
operator breaking the symmetry in a particular way. Going further, we have
considered an application to EGOE(2)-$J$ for fermions in a single-$j$ shell.
Here the embedding algebra is $U(2j+1) \supset SO_J(3)$.  Expansions to order
$[J(J+1)]^2$ for energy centroids $E_c(m,J)$ and spectral variances
$\sigma^2(m,J)$ are obtained. Formulas are derived for fixed-$m$ bivariate
cumulants and they are used to show the expansion to order $[J(J+1)]^2$ explain
the structure of fixed-$J$ centroids and variances. These results are important
in the subject of regular structures generated by random interactions.

\chapter{Hamiltonian Matrix Structure}
\label{ch8}

\section{Introduction}
\label{c8s1}

In Chapters \ref{ch2}-\ref{ch7}, our focus is in analyzing extended embedded
ensembles for isolated finite interacting quantum systems. 
As discussed in Chapter \ref{ch1}, the classical GOE is universally regarded as
the model for fluctuation properties of generic chaotic quantum systems. However
for a complete statistical description of systems such as nuclei and atoms, as
the interactions for these systems are two-body, as already emphasized in the
previous chapters, EGOE
is expected to be most appropriate. On the other hand, banded random matrix
ensembles (BRME) \cite{Wi-55,Wi-57,Ca-90,Ca-93,Fy-91,Fy-92} are also employed 
by some groups. One can infer the appropriateness of GOE, BRME or
EGOE representation, for describing statistical properties, by analyzing
eigenvalue densities, strength functions, chaos measures such as information
entropy, transition strength distributions, expectation values of operators,
level and strength fluctuations and so on \cite{Br-81,Ko-01,Fl-99,Go-01,Go-04}.
However, an important question is:  is it possible to infer the random matrix
structure by directly examining the Hamiltonian matrix itself. 

Some of the earlier studies of GOE and EGOE structure of nuclear shell-model
Hamiltonian matrices were due to Gervois \cite{Ge-68}, French and Wong
\cite{Fr-70,Fr-71a,Wo-72} and Bohigas and Flores \cite{Bo-71}. Similarly, for
atoms, they were due to  Rosenzweig and Porter \cite{Ro-60} and Parikh
\cite{Pa-78a}.  In many of these studies, the matrix dimensions are quite small
(in most cases they are 10-50 dimensional). More recently (in the 90's) there
has been renewed interest in examining Hamiltonian  matrices in nuclear and
atomic  examples as it is now possible to  construct much larger size matrices 
and more importantly, because random matrix theory has been well established in
the 80's.   For example,  characteristic  features, in terms of GOE and EGOE, of
the shell-model Hamiltonian matrix of $^{22}$Na, $(J^\pi T)=(2^+0)$ with
dimension $d=307$, were studied by French et al  \cite{FKPT,To-86}. Similarly,
Zelevinsky et al analyzed GOE and BRME structure of  $^{28}$Si, $(J^\pi
T)=(2^+0)$ and $(0^+0)$ shell-model Hamiltonian matrices, with $d=3276$ and
$839$, respectively \cite{Ze-96}.  On the other hand, Flambaum et al
\cite{Fl-94,Gr-95} using $ls$ coupling studied, in terms of GOE and BRME,  the
structure of Hamiltonian matrices of CeI, $J^\pi=4^\pm$, with dimension $d=260$
for odd parity and $276$ for even parity. Similarly,  both $ls$ and $jj$
coupling schemes were investigated for Hamiltonian matrices of CeI,
$J^\pi=4^\pm$; PrI, $J^\pi=11/2^\pm$, with dimension for PrI being $d=887$ and
$737$ for odd and even parities, respectively, by Cummings and collaborators
\cite{Cu-01}. Going beyond these studies,  our purpose in this chapter is to
carry out a comprehensive analysis of the structure of nuclear  Hamiltonian
matrices, with two shell-model examples  ($^{22}$Na, $^{24}$Mg), by employing
all the measures, for GOE, BRME and EGOE, that are introduced in the literature
at various times. For comparison, we have employed SmI atomic example, as this
appears to be, from the past analysis \cite{An-03,An-05}, the best atomic
example for EGOE. All the results presented in this chapter are published in
\cite{Ma-10b}.

\section{Matrix Structure by Visualization}
\label{c8s2}

With the advances in computer graphics, it is now possible to visualize  the
coarse grained structure of the $H$ matrices. Given the many-particle matrix
elements $H_{ij}$ (in $J^\pi T$ basis for nuclei and in $J^\pi$ basis for
atoms) one can make a plot of the squares of the matrix elements  $H_{ij}^2$
as a function of $(i,j)$. In general, there are many choices for the indices
$i$. Most commonly employed one for $i$ are the basis states indices 
defined by the ordering of the basis states as used in the shell-model 
codes. Note that the Hamiltonian operator is one plus two body, i.e.,
$H=h(1)+V(2)$ and the basis states used for constructing the $H$ matrix are 
the eigenstates of  the one-body part [$h(1)$] of $H$. This exercise has
been carried out: (i)  for lanthanide atoms CeI and PrI by Flambaum et al
\cite{Fl-94} and Cummings et al \cite{Cu-01}; (ii) for the  lanthanide atoms
NdI, PmI and SmI by Angom and Kota \cite{An-05a}; (iii) for $^{28}$Si
nucleus  by Zelevinsky et al  \cite{Ze-96}.  Alternatively it is also
possible to plot the same as a function of the basis state energies
$e_i=H_{ii}$.  Physically the basis states indices do not carry any 
significant information. However the basis state energies $e_i$'s give the
location of the corresponding strength functions [see Eq. (\ref{eq.strn})
ahead] and hence they are more meaningful. This exercise has been carried
out  for the  lanthanide atoms NdI, PmI and SmI by Angom and Kota
\cite{An-05}.  Following this, for visualization we plot $H_{ij}$ as a
function of $(e_i,e_j)$. 
 
\begin{figure}[htp]
\centering
\includegraphics[width=5.5in,height=5in]{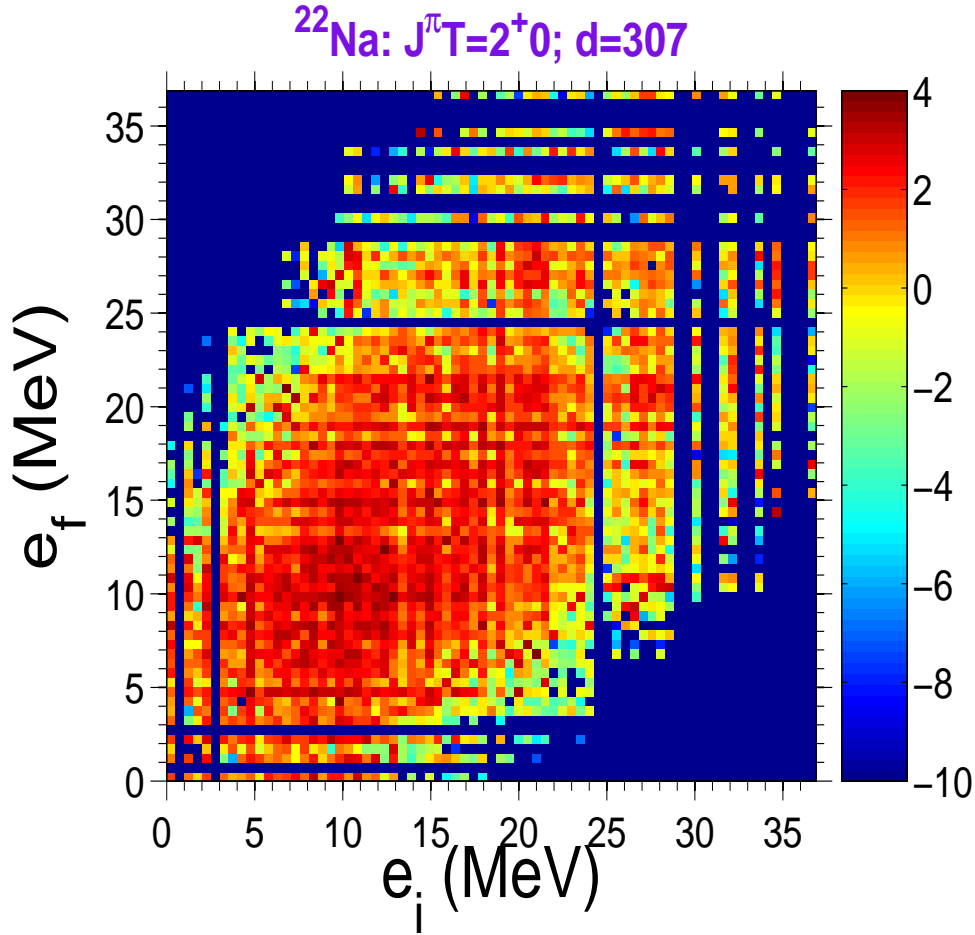}
\caption{Intensity plot showing natural logarithm of  the 
squares of the off-diagonal matrix elements $|\lan f \mid H \mid i \ran|^2$ 
(whose value is determined by the color with scale as indicated in the
figure) as a function of the single-particle basis state energies $e_i=\lan
i \mid H \mid i \ran$ and $e_f =\lan f \mid H \mid f \ran$ for  $^{22}$Na
nucleus. Note that diagonal matrix elements $\lan i \mid H \mid i \ran$ are
put to zero in calculating $\sum_{i,f}\,|\lan f \mid H \mid i  \ran|^2$ in a
given bin. Bin-size is $0.5 \times 0.5$. All matrix elements  are in MeV
units  and $d$ stands for the matrix dimension. Calculation  used Kuo
interaction with $^{17}$O sp energies; see  \cite{Ko-98} for 
details.}
\label{na-mat}
\end{figure}

\begin{figure}[htp]
\centering
\includegraphics[width=5.5in,height=5in]{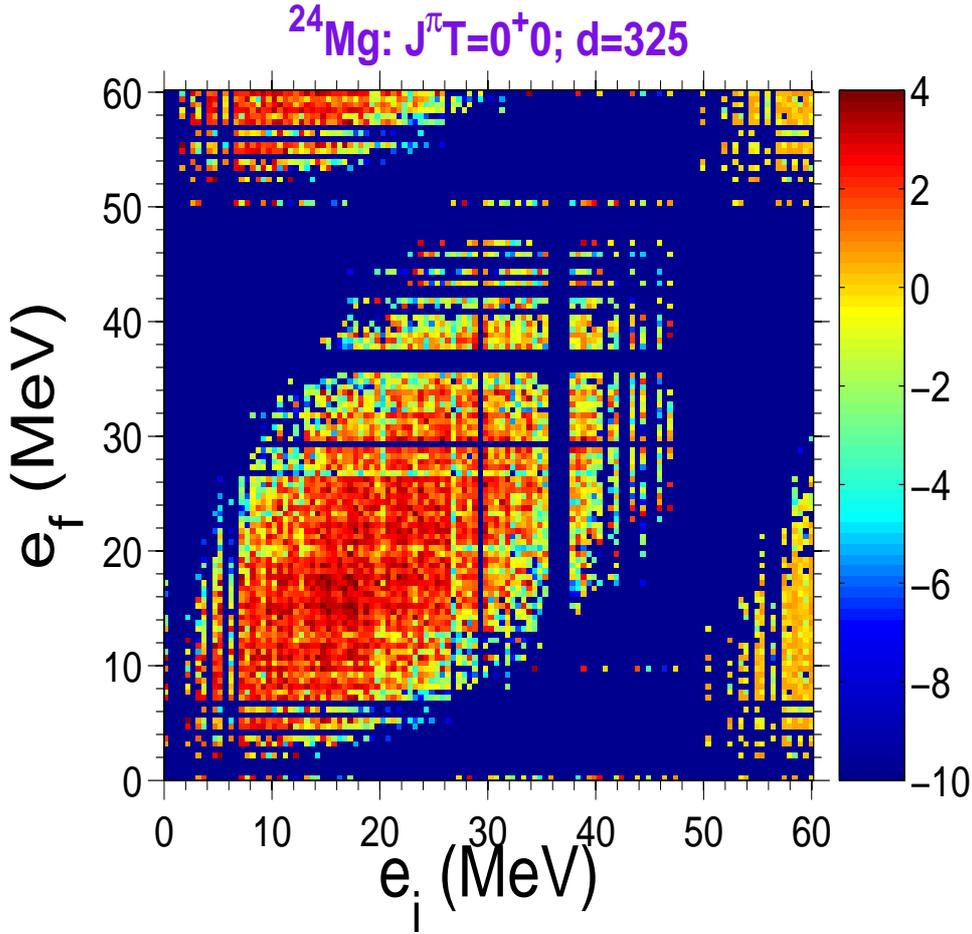}
\caption{Same as Fig. \ref{na-mat} but for $^{24}$Mg. 
Bin-size is $0.5\times 0.5$. All matrix elements are in MeV units. 
Calculation used Kuo interaction with $^{17}$O sp energies;  
see \cite{Ko-02} for details.}
\label{mg-mat}
\end{figure}

\begin{figure}[htp]
\centering
\includegraphics[width=5.5in,height=5in]{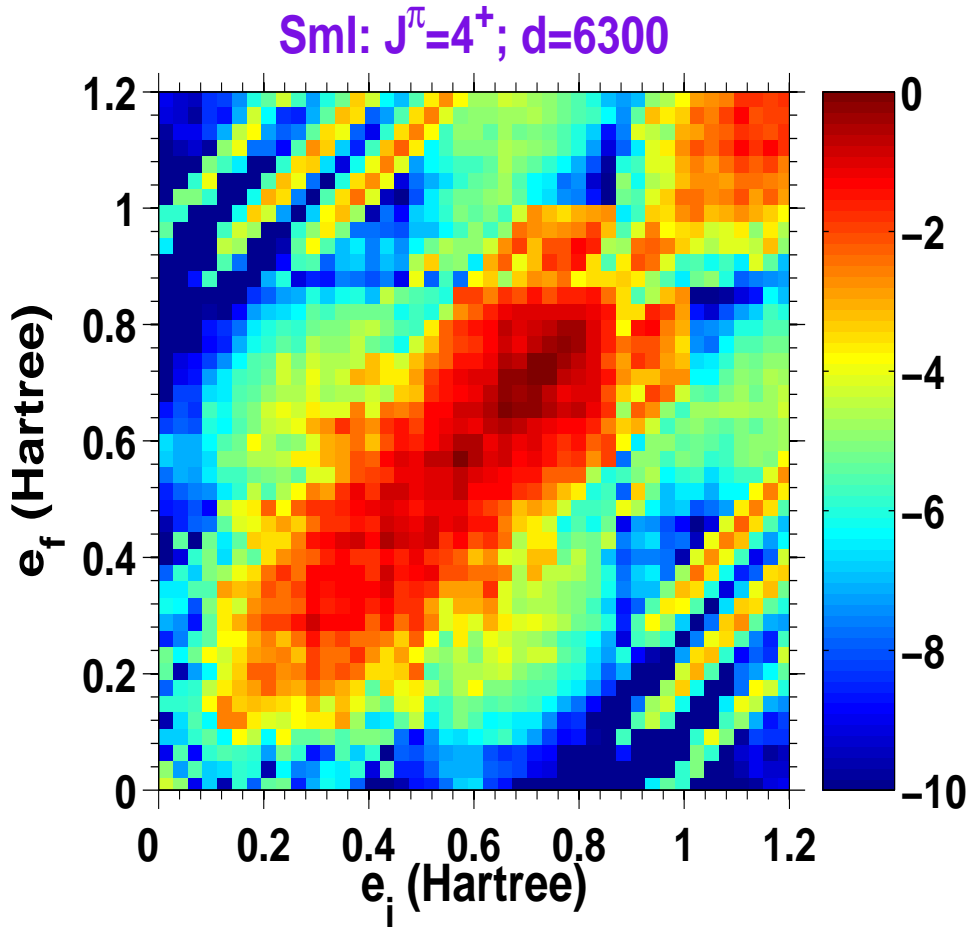} 
\caption{Same as Fig. \ref{na-mat} but for SmI. The bin-size 
used is $0.03 \times 0.03$. The matrix construction was discussed in
\cite{An-05} and these results are used to construct the color plot. Note
that all the matrix elements are in Hartree units.}
\label{sm-mat}
\end{figure}

We show in Figs. \ref{na-mat} and \ref{mg-mat} for $^{22}$Na ($J^\pi
T=2^+0$, $d=307$) and $^{24}$Mg ($J^\pi T=0^+0$, $d=325$) nuclei,
respectively, the plot of squares of matrix elements $H^2_{kl}$, by
averaging over an area in the $e_k-e_l$ plane, as a function of the basis
state energies $(e_k , e_l)$. For details of the nuclear shell-model
calculations for $^{22}$Na and $^{24}$Mg see Refs. \cite{Ko-98} and
\cite{Ko-02}, respectively. In the plots, we employ a color code for better
visualization.  Similarly, in Fig. \ref{sm-mat}, $H$ matrix plot for SmI
($J^\pi=4^+$, $d=7325$) atom is shown. Only the first 6300 basis states are
taken into consideration in the plot as discussed in Ref. \cite{An-05} and
unlike in \cite{An-05}, we have used a color code for the plot for better
visualization. For $^{22}$Na and  $^{24}$Mg examples, the matrix is more
spread compared to that for SmI. This is because, unlike in SmI example
(also in many other atomic examples as discussed in \cite{An-05}), in
the nuclear shell-model all excitations within the model space are taken
into account. Figs. \ref{na-mat} and \ref{mg-mat} show a sparse, band-like
structure with block structure within the band. As seen from Fig.
\ref{sm-mat} for SmI $H$ matrix, there are prominent diagonal blocks and 
streaks of large matrix elements parallel to the diagonal but far away from
the diagonal. Also, there is a sparse, band like structure with block
structures within the bands. It is seen from Figs. \ref{na-mat},
\ref{mg-mat} and \ref{sm-mat} that, in general, strictly speaking the $H$
matrices  are neither GOE nor banded. Also, from the visualization in Figs.
\ref{na-mat}, \ref{mg-mat} and \ref{sm-mat}, it is not possible to infer 
the two-body selection rules which form the basis for EGOE description. 

In order to bring out the two-body selection rules clearly,  we consider the
following representation. In the nuclear shell-model, for $m$ fermions
distributed over $r$ sub shells  with total angular momentum $j_l$,
$l=1,2,\ldots,r$, the many-particle states are  labeled by the spherical
configurations ${\bf m} =(m_1,m_2,\ldots,m_r)$, total angular momentum $J$,
isospin $T$ and the multiplicity label $\alpha$.  Note that $m=\sum_{i=1}^r
m_i$. The $m$-particle basis states $\l|{\bf m} \alpha JT\ran$ can be
ordered according to the ${\bf m}$'s. Then the $H$ matrix will contain
diagonal blocks which couple the states within same spherical configuration
and off-diagonal blocks  which couple states with different spherical
configurations. As the interaction is two-body in nature, there will be only
two types of  off-diagonal blocks containing non-zero matrix elements which
can mix configurations differing in position of one or two particles. All
other off-diagonal blocks will contain zero matrix elements. For visual
demonstration of this result,  we have shown in Chapter \ref{ch1} in 
Fig. \ref{m8-block}, a plot of the
$H$ matrix for the $^{24}$Mg example displaying the structure due to the 
two-body selection rules.  For this nucleus, there are $8$ valence nucleons
occupying the three spherical orbits  ($1d_{5/2}$, $1d_{3/2}$, $2s_{1/2}$).
Therefore the spherical configurations are $(m_1,m_2,m_3)$ with $m_1$ number
of nucleons in $1d_{5/2}$ orbit, $m_2$ in $1d_{3/2}$  orbit and $m_3$ in
$2s_{1/2}$ orbit. There are $33$ configurations generating the $325$
dimensional $(J^\pi T=0^+0)$ $H$ matrix. Their dimensions are $35$, $34$, 
$28$, $27$, $23$, $20$, $19$, $15^2$, $14$, $12$, $10$, $9^2$, $7^2$, $5$, 
$4^4$, $3$,  $2^6$ and $1^5$; here $d^n$ means there are $n$ number of
configurations with dimension $d$. The configurations are ordered such that
the block matrices start from the maximum size (35 $\times$ 35) and go  to
the minimum (1 $\times$ 1). Figure \ref{m8-block} clearly shows the diagonal
blocks and the off-diagonal blocks that involve  change of occupancy of one
and two nucleons, respectively.  The regions that correspond to all other
off-diagonal blocks are  forbidden by the two-body selection rules. A
similar figure was given earlier in \cite{Pa-05} for $^{28}$Si with $(J^\pi
T)=(0^+0)$. Although Fig. \ref{m8-block} brings out clearly the structure
due to two-body selection rules, it will not give any further insight into
the EGOE structure of the matrix.   Therefore, for a quantitative
understanding of GOE, BRME and EGOE structures of the matrices, we employ
various measures introduced in the literature for the structure of these
ensembles. Now we will turn to this analysis.

\section{Analysis in Terms of GOE and BRME}
\label{c8s3}

Hamiltonian matrices, prior to the actual diagonalization, are analyzed  for
the nuclear and atomic examples using measures defining GOE and BRME. To
ascertain the GOE character, the distribution of off-diagonal elements is
studied. As discussed in Sec. \ref{c8s2}, the 3D matrix plots show a banded
structure. In order to quantify the banded structure, we calculate the
bandwidth and the sparsity parameters.

\subsection{GOE structure: distribution of the off-diagonal matrix elements}
\label{c8s3s1}

\begin{figure}[htp]
\centering
\includegraphics[width=3in,height=4.5in]{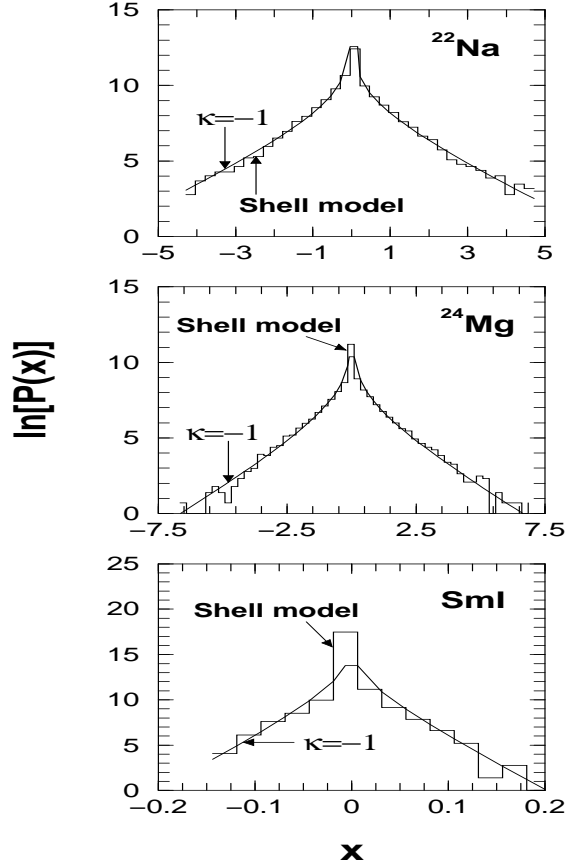}
\caption{Plot showing distribution $P(x)$ of the off-diagonal  matrix
elements for $^{22}$Na, $^{24}$Mg and SmI. Note that $P(x)$ gives  number of
$x=\widetilde{H}_{kl}$ in a given energy bin. The bin-size in the figures is
$0.25$  for $^{22}$Na and $^{24}$Mg and $0.025$ for SmI. In the figure,
exact results are shown as histograms and the best fits  $P_{\kappa=-1}(x)$
are shown as continuous curves. The function $P(x)$ is normalized to
$d(d-1)$, the number of off-diagonal matrix elements.  Finally, note that
the plots are for $\ln[P(x)]$ vs $x$. The units for $x$ are MeV for
$^{22}$Na and $^{24}$Mg and Hartree for SmI.}
\label{dist-na}
\end{figure}

Figure \ref{dist-na} shows the probability densities $P(x)$ for the
off-diagonal matrix elements  $x=\widetilde{H}_{kl}=H_{kl}$, $k\neq l$ for
$^{22}$Na, $^{24}$Mg and SmI examples. Figure shows that there are large
number  of small matrix elements and almost half of these are zeroes. This
also implies the leading role of the diagonal matrix elements in forming the
spectrum which we will discuss in detail in Sec. \ref{c8s4}. 
For GOE, $P(x)$ should
be a Gaussian. However, for large $\widetilde{H}_{kl}$, it was found that
the distribution $P(x)$ is well  described by the Porter-Thomas ($P-T$)
distribution \cite{Fl-94}, 
\be 
P_{P-T}(x) =\dis\frac{1}{2\sqrt{\pi w \l|x\r|}}
\exp\l(-\dis\frac{\l|x\r|}{w}\r)\;;\;
x=\widetilde{H}_{kl},\; 
w=\sqrt{\overline{\widetilde{H}_{kl}^2}}\;.
\label{eq.h1}
\ee  
But, the agreement is not good when $\widetilde{H}_{kl} \sim 0$. A better
proposition is to use a generalized $P-T$ distribution as suggested first by
Zelevinsky et al \cite{Ze-96},
\be
P_{\kappa}(x) = \spin \l[ (2x_0)^{\kappa+1}
\Gamma(\kappa+1)\r]^{-1} \l|x\r|^{\kappa} \exp\l(-\dis\frac
{\l|x\r|}{2x_0}\r)\;.
\label{eq.h2}
\ee
Note that $x_0=w/2(\kappa+1)$ and $w$ is given in Eq. (\ref{eq.h1}). Equation
(\ref{eq.h2}) is found to explain the distribution of the off-diagonal 
elements in  the nuclear examples considered in \cite{Ze-96}. Our examples
substantiate this further as discussed below. Note that $\kappa=-1/2$ in Eq.
(\ref{eq.h2}) will  give Eq. (\ref{eq.h1}). 

For $^{22}$Na, $^{24}$Mg and SmI examples that correspond to Figs.
\ref{na-mat}, \ref{mg-mat}, and \ref{sm-mat}, 
respectively, we have carried
out fits to Eq. (\ref{eq.h2}) with  $\kappa=0$, $-1/2$, $-1$ and $-2$ and
found that there is good agreement for $\kappa=-1$ but not for the other
values. The fits with $\kappa=-1$ are shown in Fig. \ref{dist-na} as
continuous curves. In the fits to Eq. (\ref{eq.h2}), a small region around
$x=0$ is not considered as $P_\kappa(x)$ will not be regular at $x=0$ for
$\kappa \leq -1$. Note that the deviations are larger for SmI example as
compared to the nuclear examples. Therefore, our two nuclear examples (to
some extent, even the atomic example) are in conformity with the conclusion
of Zelevinsky et al. They state \cite{Ze-96}: {\it Eq. (\ref{eq.h2})
implies that the normally distributed quantities in the realistic cases are
not the off-diagonal matrix elements themselves as would be the case in
canonical random matrix ensembles but rather some quantities resembling
square roots of them}. As  they have argued, it is possible that the
multipole-multipole form of the nuclear interactions could be the physical
reason for this. Hence, it is clear that GOE is not an appropriate
representation for the nuclear (also atomic) Hamiltonian matrices. 

\subsection{BRME structure: bandwidths and sparsity}
\label{c8s3s2}

As seen from Figs. \ref{na-mat}, \ref{mg-mat} and \ref{sm-mat}, the $H$
matrices have a band-like structure. We calculate a measure for sparsity and
also the energy bandwidths for testing the BRME representation for the
nuclear and atomic Hamiltonian examples.  
 
\begin{figure}[htp]
\centering
\includegraphics[width=5.5in,height=6in]{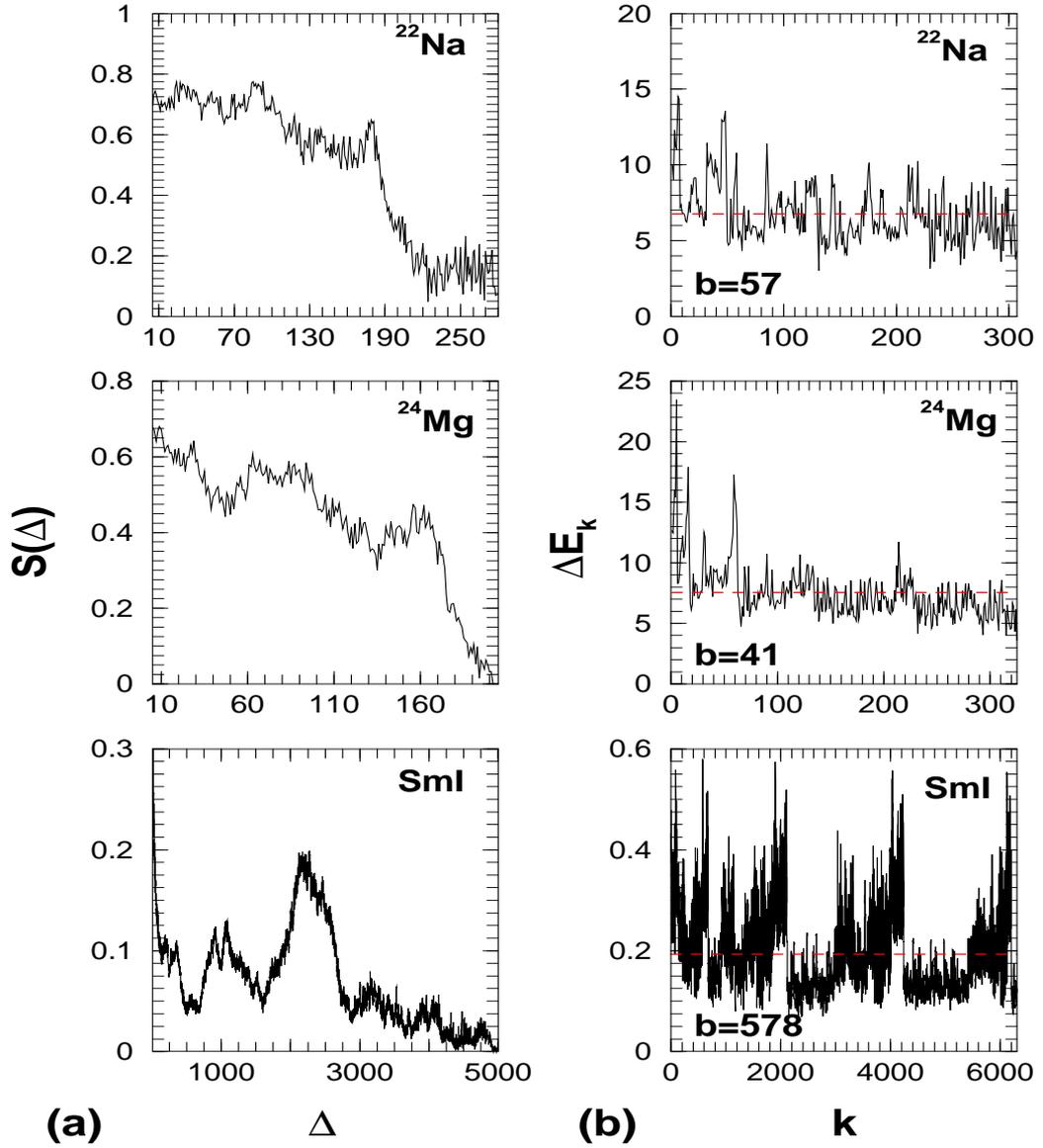}
\caption{(a) Sparsity $S(\Delta)$ defined by Eq. (\ref{eq.h5}) as a function
of  $\Delta=|k-l|$ for $^{22}$Na, $^{24}$Mg and SmI matrices. Results are
shown for $\Delta \geq 5$. Note that for 
calculating sparsity, all the matrix elements whose absolute value is  $\geq
10^{-5}$ and $10^{-8}$, respectively for the nuclear and atomic examples 
are taken as non-zero. (b) Energy bandwidths $\Delta E_k$  defined
by Eq. (\ref{eq.h3}) as a function of state index $k$ along with mean values
of $\Delta E_k$ (dashed lines) for $^{22}$Na, $^{24}$Mg and SmI matrices.
The units for $\Delta E_k$ are MeV for $^{22}$Na and  $^{24}$Mg and Hartree
for SmI. The values of the bandwidths $b$ are shown in the figures.}
\label{band}
\end{figure}

Gribankina et al \cite{Gr-95} and Cummings et al \cite{Cu-01} defined the 
sparsity $S$ as a function  of $\Delta$, the difference in the indices of
the basis states connected by the Hamiltonian, as a measure for band-like
structure. The definition of $S$ is,
\be
S(\Delta) = \dis\frac{\mbox{number of}\; |H_{kl}| \neq 0}
{\mbox{number of all}\; H_{kl}}\;,\;\;|k-l|=\Delta\;.
\label{eq.h5}
\ee
For a BRME of bandwidth $b$, the sparsity $S=1$ for $\Delta
\leq b$ and zero for $\Delta > b$, thus it is a step function. Figure
\ref{band}(a) shows the results for $S(\Delta)$ for the three examples
$^{22}$Na, $^{24}$Mg  and SmI.  In the nuclear examples, $S(\Delta)$
essentially decreases as a function of $\Delta$ with  approximate linear
dependence on $\Delta$  up to $\Delta \sim 150$ and then falls sharply to
zero. However, there are sizeable fluctuations in $S$ as a function of 
$\Delta$. On the other hand, for SmI, the structure is quite different with
large fluctuations  and a peak at  $\Delta \sim 2250$. The latter may be due
to the large off-diagonal streaks seen in Fig. \ref{sm-mat}. Thus,
$S(\Delta)$ shows clear deviation from band-like structure in all the
examples. This is further substantiated by the energy bandwidths for the
basis states and we will turn to this now.

Energy bandwidth $b$ gives the energy interval in which the basis states are
strongly mixed. The energy bandwidths $\Delta E_k$ for each basis state $k$
are defined as \cite{Fe-91},
\be
\Delta E_k^2 = \dis\frac{\dis\sum_{l}(H_{kk}-H_{ll})^2
|H_{kl}|^2}{\dis\sum_{l\neq k}|H_{kl}|^2}\;.
\label{eq.h3}
\ee
In Fig. \ref{band}(b) we show the results for $\Delta E_k$ for the
$^{22}$Na, $^{24}$Mg and SmI examples.  The value of the average bandwidth
$b$ is given as the ratio of the mean value of $\Delta E_k$ and the mean
level spacing of the unperturbed energy levels $D$, i.e., $b=\overline{\Delta
E_k}/D$; in general $b$ can be energy dependent as $D$ can be defined as the
local mean spacing of the energy  levels. The values of $b$ are given in
Fig. \ref{band}(b) and for all the three examples, $b$ is smaller than the 
matrix dimension $d$ by a factor of $\sim 4$.  The number $b$ can also be
calculated by fitting the mean squared matrix elements to the simple
exponential ansatz \cite{Fy-91,Fy-92},
\be
\lan H_{kl}^2 \ran_{|k-l|=\Delta} = H_0^2\,\exp\l(-\dis\frac
{\Delta}{b}\r)\;.
\label{eq.h4}
\ee
The values obtained using Eq. (\ref{eq.h4}) are almost same as those 
obtained using Eq. (\ref{eq.h3}). Significant observation from the figures
is as follows. For a BRME, the bandwidth  $\Delta E_k$ should be independent
of $k$. However, there are significant fluctuations in the energy bandwidths
in the nuclear examples and quite large fluctuations in SmI example. Note
that it is impossible to reach a banded form even by reordering the basis
states and this is due to the two-body selection rules.  By combining the
results for sparsity $S(\Delta)$ and the energy bandwidths $\Delta E_k$
shown in Fig. \ref{band}, we can conclude that BRME is not a good
representation for the $H$ matrices. 

\section{Analysis Using Measures for EGOE Structure}
\label{c8s4}

Going further, we analyze three measures for quantifying the EGOE structure
of the $H$ matrices for the two nuclear examples and the one atomic example
in the present section.    

\subsection{Correlations between diagonal matrix elements and eigenvalues}
\label{c8s4s1}

Large number of numerical calculations in the past in the context of
statistical nuclear spectroscopy have clearly indicated \cite{Fr-83,Ko-89}
that the joint probability distribution $\rho(E,e_k)$ of the diagonal matrix
elements $e_k$ and eigenvalues $E$ is a bivariate Gaussian for EGOE.
Therefore the marginal densities $\rho(E)$ and $\rho(e_k)$ will be close to
Gaussians with same centroids but different widths. In addition, the widths
of the conditional densities $\rho(E|e_k)$ will be independent of $e_k$.
These results were used to derive a formula for the chaos measures, the
number of principal components and information entropy in wavefunctions for
embedded ensembles \cite{KS-01}. The close to Gaussian form of $\rho(E)$ and
$\rho(e_k)$ imply that the eigenvalues $E$ and  the diagonal elements of the
$H$ matrix  (or equivalently the basis state energies) will be correlated. 
Flambaum et al examined, for CeI, eigenvalue spectrum vs the spectrum
generated  by $e_k$ \cite{Fl-94}. They found a close correlation between the
two spectra. 

\begin{figure}[htp]
\centering
\includegraphics[width=4in,height=5in]{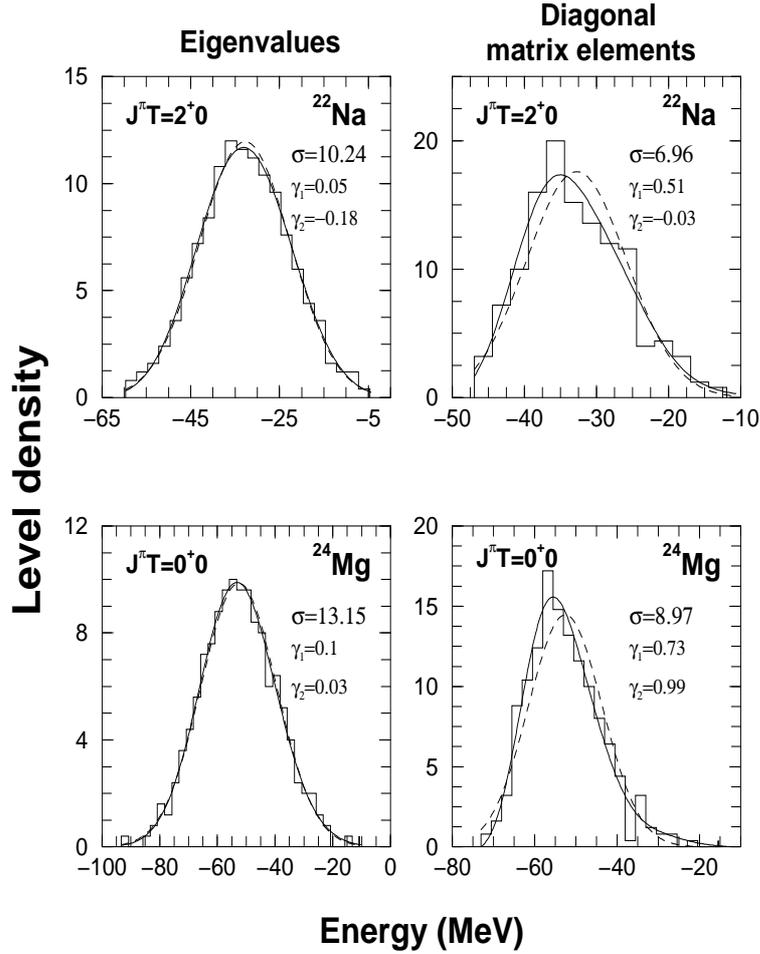}
\caption{Plot showing density of eigenvalues and density of  diagonal matrix
elements for the Hamiltonian matrices of $^{22}$Na and $^{24}$Mg. Values of
the widths $\sigma$, skewness $\gamma_1$ and  excess $\gamma_2$ are  given
in the figures. The units for $\sigma$ are MeV. The centroid $E_c=-32.77$
MeV for $^{22}$Na and $-52.59$ MeV for $^{24}$Mg. Histograms are the exact
results with bin size $2.5$ MeV for all the examples. The  dashed curves are
the Gaussians with centroid $E_c$ given above and width $\sigma$ whose value
is given in the figure. Similarly continuous curves are Edgeworth corrected
Gaussians defined in Eq. (\ref{eq.gau1}).}
\label{den-n}
\end{figure}

\begin{figure}[htp]
\centering
\includegraphics[width=4in,height=2.5in]{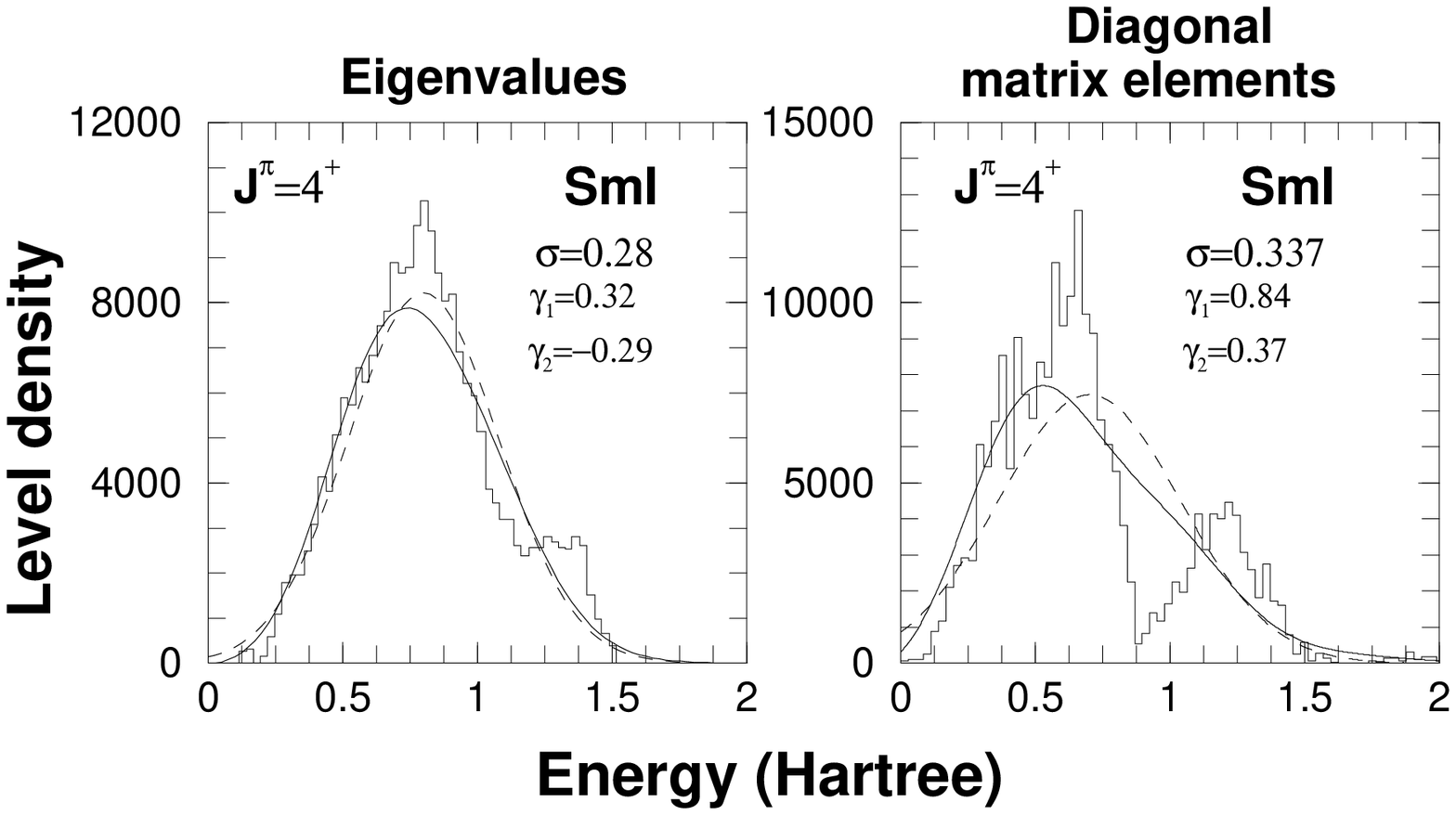}
\caption{Same as Fig. \ref{den-n} but for SmI matrix. The units for $\sigma$
are Hartree. The centroid is $E_c=0.7$ Hartree (all energies are given with
respect to the lowest energy). Histograms are the exact results with
bin size $0.028$ Hartree. The eigenvalue density for SmI is constructed by
scaling appropriately the data taken from Fig. 3 in \cite{An-05}.} 
\label{den-a}
\end{figure}

There is recent interest in the topic of correlations between eigenvalues
and diagonal matrix elements and several examples from nuclei and also
random matrices have been discussed in \cite{Sh-08,Yo-09a}. We show in Figs.
\ref{den-n} and \ref{den-a}, density of eigenvalues and density of diagonal
matrix elements for the Hamiltonian matrices of $^{22}$Na, $^{24}$Mg and
SmI.  The distributions are compared with the Gaussian form $\rho_\cg(\wx)$ 
and
the Edgeworth (ED) corrected Gaussian form $\rho_{ED}(\wx)$; see
Eq.(\ref{eq.gau1})  for definitions.
Here, $\wx=(x-x_c)/\sigma$ where $x_c$ is the centroid and
$\sigma$ is the width of the distribution of $x$. It is clearly seen
that the eigenvalue distributions for the two nuclear examples are quite
close to  $\rho_\cg(\wx)$ while the  densities of the diagonal matrix elements
are, with some deviations,  close to $\rho_{ED}(\wx)$. However, there are
stronger deviations from $\rho_{ED}(\wx)$ for the SmI example, both for the
eigenvalue density and the density of the diagonal elements. Here, the
eigenvalue density has a secondary peak and the density of diagonal matrix
elements displays a stronger bimodal form.  Results in Fig. \ref{den-n}
reconfirm that in the nuclear examples, the eigenvalues and  the  diagonal
matrix elements of the $H$ matrix are highly correlated and their
distributions are close to Gaussian forms. However, there are stronger
deviations from this behavior for the SmI example.  

\subsection{Fluctuations in the basis states spreading widths}
\label{c8s4s2}

\begin{figure}[htp]
\centering
\includegraphics[width=4in,height=5in]{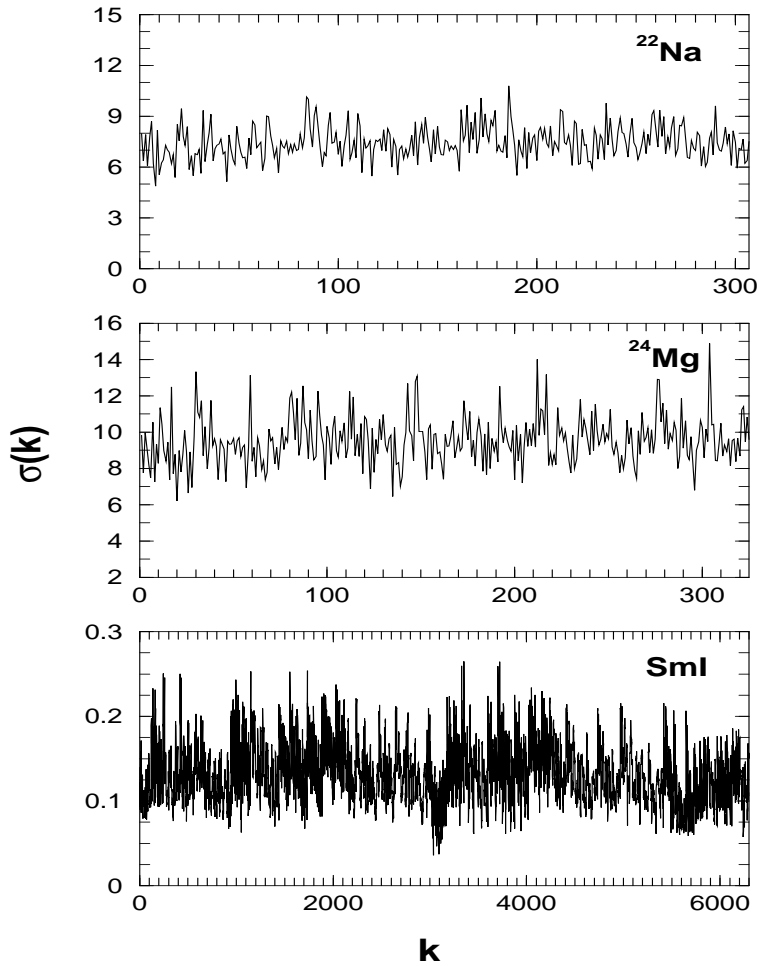}
\caption{Plot showing the variation of width $\sigma(k)$ with the basis
state index $k$ for $^{22}$Na, $^{24}$Mg and SmI matrices. The units for
$\sigma(k)$ are  MeV for $^{22}$Na and $^{24}$Mg and Hartree for SmI.} 
\label{wid}
\end{figure}

Going beyond the diagonal matrix elements, it is also useful to consider the
basis state widths $\sigma(k)$ where 
\be
\sigma^2(k)=\lan k \mid
H^2\mid k \ran-e_k^2=\sum_{l\neq k}|\lan l \mid H \mid k \ran|^2\;.
\label{eq.h66}
\ee
It should be noted that $\sigma(k)$ are the widths of the strength functions
$F_k(E)$ and similarly $e_k$ are their
centroids. Given the mean field $h(1)$ basis  states (denoted by $\l|
k\ran$) expanded in the $H$ eigenvalue ($E$)  basis, $\l|k\ran=\sum_E\,
C_{k}^{E} \l|E\ran$,  the strength functions $F_{k}(E)$ are defined by,
\be
F_{k}(E) = \dis\sum_{E^\pr}\l|C_{k}^{E^\pr}\r|^2 
\;\delta(E-E^\pr)
= \l|\cac_{k}^{E}\r|^2\;I(E)\;.
\label{eq.strn}
\ee
In Eq. (\ref{eq.strn}), $\l|\cac_{k}^{E}\r|^2$ denotes the  average of
$|C_{k}^{E}|^2$ over the eigenstates with the same energy $E$ and all the
quantities are defined over good $JT$ (nuclei) or $J$ (atoms) spaces;  the
strength functions over good spin spaces are also  defined in Chapter \ref{ch2}.
The strength functions define the spreading of the basis states over the
eigenstates and for EGOE the spreadings are of Gaussian form in the strong
coupling limit; see \cite{Fr-83,Ko-01} and Chapter \ref{ch2}. Also as stated
above, the bivariate Gaussian form of $\rho(E,e_k)$ implies $\sigma^2(k)$ should
be constant i.e., they are independent of $k$. We show in Fig. \ref{wid},
results for $\sigma(k)$ vs $k$ for $^{22}$Na, $^{24}$Mg and  SmI  matrices. It
is seen that the basis state widths $\sigma(k)$ are almost  constant apart from
small fluctuations in the nuclear examples.  This result is in agreement with
several  previous  numerical calculations \cite{Ze-96,Fr-83,Ko-08}.  Writing
$\sigma(k) = \overline{\sigma(k)}(1 \pm \delta)$, it is seen that the relative
rms deviation of the fluctuations from the mean values is  $14$\% and $15$\%
(i.e., $\delta =0.14$ and $0.15$, respectively) and the mean values 
$\overline{\sigma(k)}=7.5$ MeV and  $9.6$ MeV, respectively for $^{22}$Na and
$^{24}$Mg  matrices.  For SmI, $\overline{\sigma(k)}=0.14$ Hartree and
$\delta=0.25$. Therefore, the fluctuations in $\sigma(k)$ are much  larger for
SmI as compared to those for the nuclear examples. 

It is possible to estimate the magnitude of the fluctuations in $\sigma(k)$. Say
there are $K$ number of many-particle states that are directly coupled by the
two-body interaction. The connectivity factor $K$ also defines the spectral
variances; see \cite{Fl-96a,Fl-96b,Ja-01} and Chapter \ref{ch2}.  Assuming that
the coupling matrix elements are independent Gaussian random variables with zero
centroid and variance $v^2$, we have $\overline{\sigma^2(k)}=v^2 K$. Now  the
relative rms fluctuations in the $\sigma^2(k)$ are given by $\sqrt{2/K}$.
Therefore, $\sigma(k) \sim  [\;\overline{\sigma^2(k)}\;]^{1/2} (1 \pm
1/\sqrt{2K})$ giving $\delta$ defined above to be $1/\sqrt{2K}$.  For embedded
ensembles for spinless fermion systems with $m$ fermions in  $N$ sp states, the
connectivity factor $K \sim m(m-1)(N-m)(N-m-1)/4$  \cite{Fl-96a,Fl-96b}.  For
example, for $6$ fermions in $12$ sp states ($N=12,\;m=6$), $\delta \sim 0.05$. 
Going to embedded ensembles for fermion systems with spin  ($\cs=\spin$) degree
of freedom  and assuming that the variances of the  matrix elements in the
two-particle spin $s=0$ and $s=1$ channels to be $v_s^2$, we can relate $v_s$ to
$v$ by demanding that the two-particle spectral variance in both the models is
same. This gives $v_s^2=v^2/4$ for large $N$. Using this scaling and the  result
for the connectivity factor $K(S) = K(\Omega,m,S) = P(\Omega,m,S)$  given in
Chapter \ref{ch2}, we obtain $\delta \sim 0.1$  for $6$ fermions in $6$ sp
orbits (so that $N=12$) with total spin $S=0$.  Therefore, going from embedded
ensembles for spinless fermion systems to systems with spin, relative rms
fluctuations in the basis states variances change from $5$\% to $10$\% (see also
Table \ref{symm}).  We expect the EGOE results for nuclei with $JT$ symmetry to
be larger than that of the  embedded ensembles for spin systems and this
explains the results in Fig. \ref{wid} for the nuclear examples.

\subsection{Structure of the two-body part of the Hamiltonian in 
the eigenvalue basis}
\label{c8s4s3}

In general, it is possible to examine the $H$ matrices in different bases.
For example for $(2s1d)$ shell nuclei, the $U(24) \supset [U(6) \supset
SU(3) \supset SO_L(3)] \otimes [SU(4) \supset SU_S(2) \otimes SU_T(2)]
\supset SO_J(3) \otimes SU_T(2)$ basis \cite{El-58} will be interesting.
Similarly, Zuker et al \cite{Zu-01} examined the structure of Lanczos
tridiagonal $H$ matrices for nuclei. Unlike examining the total $H$ matrix,
it was suggested in \cite{FKPT} that it may be useful to analyze the pure
two-body   part ${\bf V}$ of $H$ [${\bf V}$ is defined by dropping the 
diagonal matrix elements of the two-body part $V(2)$] as this part is
responsible for chaos (note that the one-body part of $H$ generates Poisson
fluctuations). The two natural basis to consider are the shell-model
mean-field basis and the $H$ eigenvalue basis. The structure of ${\bf V}$ in
the mean-field basis is essentially same as that shown in Figs.
\ref{na-mat}, \ref{mg-mat} and \ref{sm-mat}.  Therefore new insight is
expected from the structure of ${\bf V}$ in the $H$ eigenvalue basis. Unlike
the mean-field basis or the $SU(3)$ basis mentioned above (or even any other
basis defined by a  group symmetry), the $H$ basis is expected to be the
least biased and also it is the most natural basis.  More importantly, EGOE
has a prediction,  as discussed ahead, for the structure of ${\bf V}$ in the
$H$ basis. As the $^{22}$Na nuclear example was discussed before
\cite{FKPT,To-86} and the SmI example showed strong deviations from EGOE
structure as discussed in Secs. \ref{c8s4s1} and \ref{c8s4s2}, 
we restrict our discussion here to $^{24}$Mg example.

\begin{figure}[htp]
\centering
\includegraphics[width=3.5in,height=3.5in]{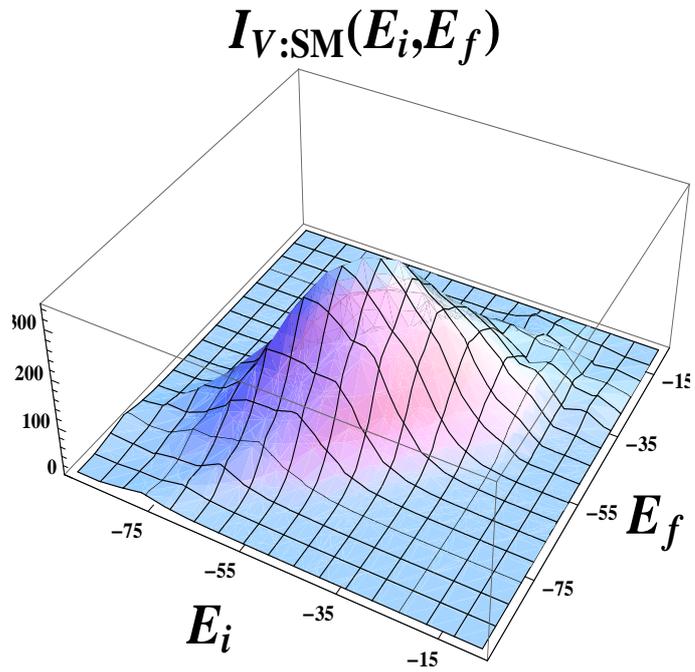}
\vskip 1.5cm
\includegraphics[width=3.5in,height=3.5in]{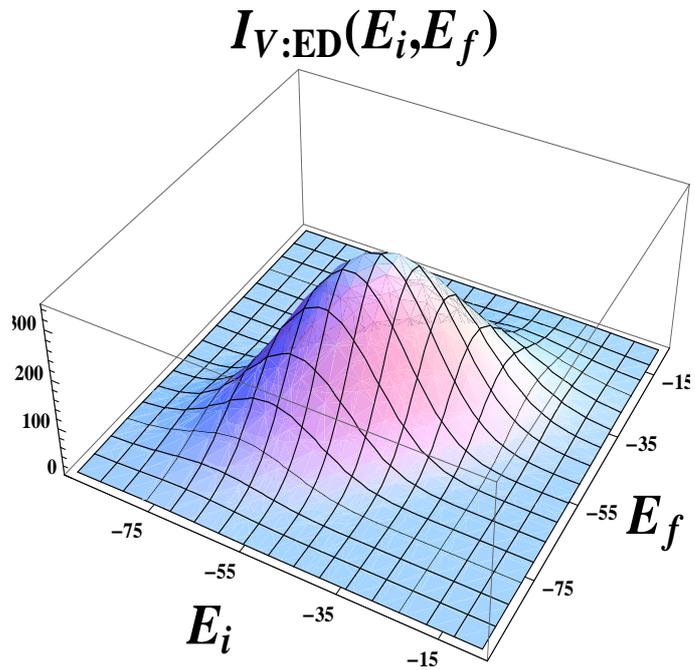}
\caption{ Plots showing the bivariate transition strength 
density for $^{24}$Mg with $(J^\pi T)=(0^+0)$. Compared are the results from
exact  shell-model (denoted by $I_{{\bf V}:SM}$ in the figure) with the
Edgeworth  corrected bivariate Gaussian $I_{{\bf V}:ED}$ in Eq.
(\ref{eq.ew9}) obtained using the bivariate cumulants given in the text.
The units for $E_i$ and $E_f$ are MeV.}
\label{vinh}
\end{figure}

For $^{24}$Mg example, starting with the matrices for ${\bf V}$ and $H$ in
the mean-field basis [$H$ operator consists of two-body  matrix elements due
to Kuo \cite{Kuo-67} defining $V(2)$ and $^{17}$O sp energies  $-4.15$ MeV,
$0.93$ MeV and $-3.28$ MeV for $1d_{5/2}$, $1d_{3/2}$ and $2s_{1/2}$ orbits
defining $h(1)$] we have constructed the matrix $\lan E_f \mid {\bf V} \mid
E_i \ran$. Using this we have analyzed the bivariate transition strength
density generated by the operator ${\bf V}$   (we put $\lan E_i \mid {\bf V}
\mid E_i\ran=0$ as discussed in \cite{FKPT} so that we are dealing with the
pure two-body part of $H$). Given the transition  operator ${\bf V}$, 
transition strength density
$I^{H,m}_{{\bf V}}(x,y)$ with the two  variables $x$ and $y$ being eigenvalues
of $H$ is $I^{H,m}_{{\bf V}}(x,y) = I^{H,m}(y) |\lan m , y \mid {\bf V} 
\mid m , x\ran|^2 I^{H,m}(x)$. 
The bivariate moments of this distribution are 
$\widetilde{M_{pq}} = \lan\lan {\bf V} H^q {\bf V} H^p\ran\ran^{m}$.
Note that the normalization factor is $M_{00}$. Starting with 
$\widetilde{M}_{pq}$, we can obtain normalized moments, the central moments,
reduced moments and also the reduced cumulants $k_{rs}$, $r+s \geq 3$. 
It is possible to write down the Edgeworth corrected bivariate Gaussian that
includes the cumulants $k_{rs}$ with $r+s=3,\;4$ \cite{FKPT,Ko-95}. 
Following the spinless EGOE results in \cite{FKPT,To-86} and the new results in
Chapter \ref{ch7}, it can be argued that EGOE gives in general close to
bivariate 
Gaussian form with Edgeworth corrections for $I^{H,m}_{{\bf V}}(x,y)$. Equation
(\ref{eq.ew9}) in Appendix \ref{c7a2} 
gives the bivariate Gaussian form with ED corrections. 
This prediction of EGOE is tested
in Fig. \ref{vinh} for $^{24}$Mg. The spectrum span for this nucleus is from
$-93.29$ to $-10.06$ MeV.  The bivariate distribution $I_{{\bf V}}$ is shown
in Fig. \ref{vinh} and it is constructed using the bin-size  5$\times$5
MeV$^2$. For comparison, we also show  the corresponding ED corrected
Gaussian distribution. The marginal centroids  $\epsilon_i,\epsilon_f$ are
equal and their value is $-50.44$ MeV.  Similarly, the marginal widths are 
$13.76$ MeV and the bivariate correlation coefficient $\zeta_{biv}=0.61$
MeV.  Thus, it is clear that the matrix can not be represented by a GOE as
$\zeta_{biv}^{GOE}=0$.  The bivariate cumulants ($k_{rs}=k_{sr}$ due to
symmetry of the ${\bf V}$  matrix) for $r+s \leq 4$ are $k_{21} =  0.035$,
$k_{30} =  0.070$, $k_{22} =  -0.092$, $k_{31} =  -0.053$ and $k_{40} = 
-0.015$. The overall normalization is $12933.25$ MeV$^2$. It is seen from
Fig. \ref{vinh}  that the r.m.s. matrix elements of ${\bf V}$ in the $H$
eigenvalue basis  are well described by the EGOE bivariate Gaussian form.
This along with the previous \cite{FKPT,To-86} $^{22}$Na example and all
other results in  Secs. \ref{c8s4s1} and \ref{c8s4s2} 
support the conjecture that EGOE is a
good representation for nuclear Hamiltonians. 

\subsection{Comments on deviations from EGOE in the atomic example}
\label{c8s4s4}

Although both the nuclear and atomic shell-model Hamiltonians include a
one-body and two-body parts, it is clearly seen that EGOE does not  describe
very well the atomic shell-model  Hamiltonian while it is good for nuclei. 
Following are some of the differences in the two systems:  (i) given the sp
orbits and the number of valence fermions, only a few configurations (that
correspond to single and double excitations  with respect to the leading
configuration) are included in atomic calculations  \cite{Fl-99,Cu-01,An-05}
whereas all configurations allowed in the model space, just as in EGOE, are
included in the nuclear examples;  (ii) for atoms, both positive and
negative parity (interwoven) sp orbits are included (and this is
necessary)   while in  nuclear examples, orbits of only one parity are
considered; (iii) the inter-configuration mixing is weak for atoms as
discussed in earlier atomic calculations \cite{Fl-99,Cu-01}; (iv) the
Coulomb interaction  is of long range while nuclear interactions are of
short range.  A simple plot of the distribution of the configuration
centroids with degeneracy given by the dimensions shows multimodal structure
for atoms (see Fig. 4 of \cite{An-05}). However, for nuclear examples it is
essentially an   unimodal distribution and this difference can be ascribed
to (i) and (ii). Random matrix model taking into account (i), (ii) and (iii)
corresponds to  partitioned EGOE \cite{Ko-01}. A simpler version of this
model shows that weak mixing between configurations generates bimodal (in
general, multimodal) forms for density of states \cite{Ko-99}.  Similarly,
in order to understand the effects due to (iv),   the model considered by
Bae et al \cite{Ot-92} may be relevant. This model includes a parameter
$\xi$, where $\xi$ is the ratio of the radius of the many-body system to the
range of the interaction.  It will be useful to examine the statistical
properties considered in the present chapter using both partitioned EGOE and
the Bae et al model. However, this analysis is beyond the scope of the
present thesis.     

\section{Summary}
\label{c8s5}

In the present chapter, a comprehensive analysis of the structure of nuclear
shell-model Hamiltonian matrices has been carried out by employing all available
measures for GOE, BRME and EGOE random matrix ensembles. To this end, considered
are $^{22}$Na and $^{24}$Mg nuclear examples and for comparison the SmI atomic
example. In the nuclear examples, the matrix sizes are $\sim 300$ and comparing
with some of the analysis carried out by Zelevinsky et al \cite{Ze-96} and
Papenbrock and Weidenm\"{u}ller \cite{Pa-07} where much larger size matrices are
used, it is clearly seen from the results in Secs. \ref{c8s2}-\ref{c8s4}  that
the present examples are adequate for bringing out all the essential features of
the nuclear shell model Hamiltonians and in particular, the EGOE structure.
Results for SmI in Secs. \ref{c8s2}-\ref{c8s4} indicate that further 
investigations are needed for establishing the extent to which EGOE can be
applied for describing statistical properties of atomic levels.  For nuclear
Hamiltonians, it is possible to argue, using chaos measures  applied to the
diagonal blocks in Fig. \ref{m8-block}, that there is a local GOE structure
(i.e., each diagonal block is close to a GOE with weak admixings between these
blocks) in the  matrices although there is a global EGOE structure
\cite{Pa-05}.  This aspect was also recognized in the earlier studies of $H$
matrices by French et al  \cite{FKPT}.  The study presented in Secs.
\ref{c8s2}-\ref{c8s4} together with the previous analysis in
\cite{Pa-07,FKPT,Ze-96} clearly  establishes that EGOE is the best random matrix
representation for nuclear shell-model Hamiltonians. 

\chapter*{}

\thispagestyle{empty}

\chapter{Conclusions and Future Outlook}
\label{ch9}

In this thesis, we have identified and systematically analyzed many  different
physically relevant EGEs with symmetries by considering a variety of quantities
and measures that are important for finite interacting quantum systems such as
nuclei, quantum dots, small metallic grains and ultracold atoms. The studies
carried out and the main results obtained in the thesis are as follows.

Finite interacting Fermi systems with a mean-field and a chaos generating
two-body interaction are modeled, more realistically, by one plus two-body
embedded Gaussian  orthogonal ensemble of random matrices with spin degree
of freedom [called EGOE(1+2)-$\cs$]. Numerical calculations are used to
demonstrate that, as $\lambda$, the strength of the interaction (measured in
the  units of the average spacing of the single particle levels defining the
mean-field), increases, generically there is Poisson to GOE transition  in
level fluctuations, Breit-Wigner to Gaussian transition in strength
functions (also called local density of states) and also a duality  region
where information entropy will be the same in both the mean-field and
interaction defined basis. Spin dependence of the transition points
$\lambda_c$, $\lambda_F$ and  $\lambda_d$, respectively, is described using
the propagator for the spectral variances and the analytical formula for the
propagator is derived. We have further established that the duality region
corresponds to a region of thermalization. For this purpose we have compared the
single particle entropy defined by the occupancies of the single particle
orbitals with thermodynamic entropy and information entropy for various
$\lambda$ values  and they are very close to each other at $\lambda =
\lambda_d$. Chaos markers play an important role in quantum information science,
statistical nuclear spectroscopy and thermalization in finite quantum systems.

EGOE(1+2)-$\cs$  also provides a model for understanding general structures
generated by pairing correlations.  In the space defined by EGOE(1+2)-$\cs$
ensemble for fermions, pairing defined by the algebra $U(2\Omega) \supset
Sp(2\Omega) \supset SO(\Omega) \otimes  SU_S(2)$ is identified and some of its
properties are derived. Using numerical calculations it is shown that
in the strong coupling limit, partial densities defined over pairing subspaces
are close to Gaussian form and propagation formulas for their centroids and
variances are derived. As a part of understanding pairing correlations in finite
Fermi systems, we have shown that pair transfer strength sums (used in nuclear
structure) as a function of excitation energy (for fixed $S$), a statistic for
onset of chaos, follows, for low spins, the form derived for  spinless fermion
systems, i.e., it is close to a ratio of Gaussians. Going further, we have
considered a quantity in terms of ground state energies, giving conductance peak
spacings in mesoscopic systems at low temperatures, and studied its distribution
over EGOE(1+2)-$\cs$ by including both  pairing and exchange interactions. This
model is shown to generate  bimodal to unimodal  transition in the distribution
of conductance peak spacings consistent with the results obtained using
realistic calculations for small metallic grains. 

For $m$ fermions in $\Omega$ number of single particle orbitals, each  four-fold
degenerate, we have introduced and analyzed in detail embedded Gaussian unitary
ensemble of random matrices generated by random two-body interactions that are
$SU(4)$ scalar [EGUE(2)-$SU(4)$]. Here, the $SU(4)$ algebra corresponds to the
Wigner's  supermultiplet  $SU(4)$ symmetry in nuclei.  Embedding algebra for the
EGUE(2)-$SU(4)$ ensemble is $U(4\Omega) \supset U(\Omega) \otimes SU(4)$.
Exploiting the Wigner-Racah algebra of the embedding algebra, analytical
expression for the ensemble average of the product of any two $m$-particle
Hamiltonian  matrix  elements is derived.   Using this, formulas for a special
class of $U(\Omega)$ irreps are derived for the
ensemble  averaged spectral  variances and also for the covariances in energy
centroids and spectral variances. On the other hand, simplifying the tabulations
available for $SU(\Omega)$ Racah coefficients, numerical calculations are
carried out for general $U(\Omega)$ irreps. Spectral variances clearly show, by
applying the so-called Jacquod and Stone prescription, that the EGUE(2)-$SU(4)$
ensemble generates ground state structure just as the quadratic Casimir
invariant $(C_2)$ of $SU(4)$. This is further corroborated by the calculation of
the expectation values of $C_2[SU(4)]$ and the four periodicity in the ground
state energies. Secondly, it is found that the covariances in energy centroids
and spectral variances increase in magnitude considerably as we go from EGUE(2)
for spinless fermions to EGUE(2) for fermions with spin to  EGUE(2)-$SU(4)$
implying that the differences in ensemble and spectral averages grow with
increasing symmetry. Also for EGUE(2)-$SU(4)$ there are, unlike for GUE,
non-zero cross-correlations in energy centroids and spectral variances defined
over spaces with different particle numbers and/or  $U(\Omega)$ [equivalently
$SU(4)$] irreps. In the dilute limit defined by $\Omega \to \infty$, $r >> 1$
and $r/\Omega \to 0$, for the $\{4^r,p\}$ irreps,  we have derived analytical
results for these correlations. All correlations are non-zero for finite
$\Omega$ and they tend to zero as $\Omega \to \infty$.

One plus two-body embedded Gaussian orthogonal ensemble of random matrices with
parity [EGOE(1+2)-$\pi$] generated by a random two-body interaction (modeled by
GOE in two particle spaces) in the presence of a mean-field, for spinless
identical  fermion systems, in terms of two mixing parameters and a gap between
the positive $(\pi=+)$ and negative $(\pi=-)$ parity single particle states is
introduced.   Numerical calculations are used to demonstrate, using realistic
values of the mixing parameters, that  this ensemble generates Gaussian form
(with corrections) for fixed parity state densities for sufficiently large
values of the mixing parameters. The random matrix model also generates many
features in parity ratios of state densities that are similar to those predicted
by a method based on the Fermi-gas model for nuclei. We have also obtained a
simple formula for the spectral variances defined over fixed-$(m_1,m_2)$ spaces
where $m_1$ is the number of fermions in the $+$ve parity single particle states
and $m_2$ is the number of fermions in the $-$ve parity single particle states.
The smoothed densities generated by the sum of fixed-$(m_1,m_2)$ Gaussians with
lowest two shape corrections describe the numerical results in many situations. 
The model  also generates preponderance of $+$ve  parity ground states for
small  values of the mixing parameters and this is a feature seen in nuclear
shell-model results.  

Turning to interacting boson systems, for $m$ number of bosons, carrying spin
($\cs=\spin$) degree of freedom, in $\Omega$ number of single particle orbitals,
each doubly degenerate, we have introduced and analyzed embedded Gaussian
orthogonal ensemble of random matrices generated by random two-body interactions
that are spin ($S$) scalar [BEGOE(2)-$\cs$]. The ensemble BEGOE(2)-$\cs$ is 
intermediate to the BEGOE(2) for spinless bosons and for bosons with spin
$\cs=1$ which is relevant for spinor BEC. Embedding algebra for the
BEGOE(2)-$\cs$ ensemble and also for BEGOE(1+2)-$\cs$ that includes the
mean-field one-body part  is $U(2\Omega) \supset U(\Omega) \otimes SU(2)$ with
$SU(2)$ generating spin.  A method for constructing the ensembles in
fixed-($m,S$) spaces has been developed. Numerical calculations show that for
BEGOE(2)-$\cs$, the  fixed-$(m,S)$ density of states is close to Gaussian and
level fluctuations follow GOE in the  dense limit. For BEGOE(1+2)-$\cs$,
generically there is Poisson to GOE transition in level fluctuations as the
interaction strength (measured in the units of the average spacing of the single
particle levels defining the mean-field) is  increased. The interaction strength
needed for the onset of the transition  is found to decrease with increasing
$S$. Propagation formulas for the fixed-$(m,S)$ space energy centroids and
spectral variances are derived for a general one plus two-body Hamiltonian
preserving spin. Derived also is the formula for the variance propagator for the
fixed-$(m,S)$ ensemble averaged spectral variances. Using these, covariances in
energy centroids and spectral variances are analyzed.  Variance propagator
clearly shows that the BEGOE(2)-$\cs$ ensemble generates ground states with spin
$S=S_{max}$. This is further corroborated by analyzing the structure of the
ground states in the  presence of the exchange interaction $\hat{S}^2$ in
BEGOE(1+2)-$\cs$.  Natural spin ordering ($S_{max}$, $S_{max}-1$, $S_{max}-2$,
$\ldots$, $0$ or $\spin$) is also observed with random interactions. Going
beyond these,  we have also introduced  pairing symmetry in the space defined by
BEGOE(2)-$\cs$. Expectation values of the pairing Hamiltonian show that random
interactions exhibit pairing correlations in the ground state region. 

Parameters defining many of the important spectral distributions (valid in the
chaotic region), generated by EGEs, involve traces of product of four two-body
operators. For example, these higher order traces are required for calculating
nuclear structure matrix elements for $\beta\beta$ decay and also for
establishing Gaussian density of states generated by various embedded ensembles.
Extending the binary correlation approximation method for two different
operators and for traces over two-orbit configurations, we have derived
formulas, valid in the dilute limit, for both skewness
and excess parameters for EGOE(1+2)-$\pi$. In addition, we have derived a
formula for the traces defining the correlation coefficient of the bivariate
transition strength distribution generated by the two-body transition operator 
appropriate for calculating 0$\nu$-$\beta \beta$ decay nuclear transition matrix
elements and also for other higher order traces required for justifying the
bivariate Gaussian form for the strength distribution. With applications in the
subject of regular structures generated by random interactions, we have also
derived expressions for the coefficients in the expansions to order $[J(J+1)]^2$
for the energy centroids $E_c(m,J)$ and spectral variances $\sigma^2(m,J)$
generated by EGOE(2)-$J$ ensemble members for the single-$j$ situation. These
also involve traces of four two-body operators. 

In order to establish random matrix structure of nuclear shell-model 
Hamiltonian matrices, we have presented a comprehensive analysis of the
structure of  Hamiltonian matrices based on visualization of the matrices in
three dimensions as well as in terms of measures for GOE, banded and embedded
random matrix ensembles. We have considered two nuclear shell-model examples,
$^{22}$Na with $J^\pi T = 2^+0$ and $^{24}$Mg with $J^\pi T = 0^+0$  and, for
comparison we have also considered  SmI atomic example with $J^\pi = 4^+$. It is
clearly established that the matrices are  neither GOE nor banded. For the EGOE
[strictly speaking, EGOE(2)-$JT$ or EGOE(2)-$J$] structure we have examined the
correlations between diagonal elements and eigenvalues, fluctuations in the
basis states variances and structure of the two-body part of the Hamiltonian in
the eigenvalue basis. Unlike the atomic example, nuclear examples show that the
nuclear shell-model Hamiltonians can be well represented by EGOE. 

In summary, in this thesis, large number of new results are obtained for
embedded ensembles EGOE(1+2)-$\cs$, EGUE(2)-$SU(4)$, EGOE(1+2)-$\pi$ and
BEGOE(1+2)-$\cs$, with EGUE(2)-$SU(4)$  introduced for the first time in this
thesis. Moreover, some results are presented for EGOE(2)-$J$ and for the first
time BEGOE(1+2)-$\cs$  has been explored in detail in this thesis. In addition, 
formulas are derived, by extending the binary correlation approximation method,
for higher order traces for embedded ensembles with  $U(N) \supset U(N_1) \oplus
U(N_2)$ embedding and some of these are needed for new applications of
statistical nuclear spectroscopy. Results of the present thesis establish that
embedded Gaussian ensembles can be used gainfully to study a variety of problems
in many-body quantum physics. 

Some of the future studies in embedded ensembles should include the following.

\begin{itemize}

\item It is important to examine the energy dependence of the transition
markers generated by EGOE(1+2)-$\cs$ and this will give new information about
onset of chaos in interacting many-particle systems as we increase the excitation
energy. In addition,  going beyond the strength functions and occupancies, the
distribution of transition strengths, generated by a general one-body transition
operator, that is a vector in the spin space should be studied. This is
important for producing a better random matrix basis for the smoothed forms for
transition strength densities.

\item Going beyond the measures employed in Chapter \ref{ch2},  new
entanglement measures, introduced in the context of quantum information science,
should be analyzed to characterize complexity in quantum many-body
systems; for disordered spin-$1/2$ lattice systems, entanglement and
delocalization are found to be strongly correlated \cite{Br-08,Pi-08}. Besides
the entanglement measures, further analysis of the thermodynamic region
generated by two-body  ensembles (defined by $\lambda_d$ in Fig. \ref{tmark})
using long-time averages of various complexity measures as discussed in
\cite{Ca-09,Ri-08} is needed. This, besides being important in QIS, should lead
to a deeper understanding of wavefunction thermalization in generic isolated
many-body quantum systems \cite{Ma-11b,Ca-09,Ri-08,Ge-00,Fl-00}.

\item It is possible to apply the Hamiltonian in Eq. (\ref{ham-mes}) with sp
energies drawn from GOE (or GUE), adding a particle number dependent term and
also by varying the interaction strength $\lambda$. Analysis with this $H$ will
generalize the results in Fig. \ref{c3f5} and also those reported in 
\cite{Al-08}. Furthermore, it will be useful to consider a generalized pairing
operator by extending Eq. (\ref{ch3.eq.npa1}) to $P = \sum_i \beta_i P_i$ where
$\beta_i$ are free parameters.

\item For evaluating $\gamma_2(m,f_m)$ for EGUE(2)-$SU(4)$, even for
$f_m^{(p)}$ irreps, the needed $SU(\Omega)$ Racah coefficients are not
available in analytical form nor there are tractable methods for their numerical
evaluation. The mathematical problem here is challenging and its solution will
establish the Gaussian form of the eigenvalue densities generated by
EGUE(2)-$SU(4)$.

\item It will also be interesting to analyze EGOE/EGUE with $SU(4)-ST$
symmetry. With this, it will be possible to understand the role of random
interactions in generating the differences in the gs  structure of even-even and
odd-odd N=Z nuclei.

\item It is important to investigate EGOE(1+2)-$\pi$ for proton-neutron
systems and then we will have four unitary orbits (two for protons and two for
neutrons). This extended EGOE(1+2)-$\pi$ model with protons and neutrons
occupying different sp states will be generated by a 10$\times$10 block matrix
for $V(2)$ in two-particle spaces with 14 independent variances. 
Therefore, parametrization of this ensemble is more complex.

\item Further extension of BEGOE(1+2) including $\cs=1$ (also spin 
$2$ etc.) degree of freedom for bosons, as discussed in Appendix \ref{c6a1}, is
relevant for spinor BEC studies \cite{Pe-10,Yi-07} and this ensemble should be
analyzed so that realistic applications of BEGOE can be attempted. 

\item Wavefunction structure should be analyzed for BEGOE(1+2)-$\cs$ and with
this, it is possible to address questions related to thermalization in finite
interacting boson systems. 

\item Binary correlation theory for EGEs with symmetries [going beyond direct
sum sub-algebra of $U(N)$] needs to be developed and then it is possible to
derive results for the excess parameter  $\gamma_2$ for EGOE(2)-$\cs$ and 
EGUE(2)-$SU(4)$ ensembles. 

\item Extensions of binary correlation approximation to
spinless boson systems and for boson systems with spin ($\cs=1/2$ and $1$) will
be interesting and may prove to be useful in ultracold atom studies.

\item Binary correlation results presented for EGOE(1+2)-$\pi$ in
Chapter \ref{ch7} should be extended further for deriving spectral properties 
of partitioned EGOE \cite{Ko-01,Ko-99}.

\item Applications of embedded ensembles to wider class of systems like
quantum dots, BEC etc. should be carried out by deriving results for physically
relevant quantities that can be confronted directly with experimental data.

\item In systems like nuclei and quantum dots, it is important to find
experimental signatures for cross-correlations (they are discussed in Chapters
\ref{ch4} and \ref{ch6} and Appendix \ref{egue2}) as they will give direct
evidence for embedded ensembles. This requires identifying measures involving
cross-correlations in lower order moments of the two-point function that can be
used in data analysis.

\item In future, it is important to analyze embedded ensembles with much larger
matrix dimensions that are needed for particle number $m \geq 10$. This requires
new numerical efforts.

\item In literature and also in this thesis, embedded ensembles with only GOE
and GUE embedding are explored. In future, embedded ensembles with GSE embedding
(EGSE) should be attempted. 

\item New efforts in developing further the Wigner-Racah algebra for
general $SU(N)$ groups are needed for more complete analytical tractability of
embedded random matrix ensembles. For example, analytical form for  the
two-point function is not yet available even for the spinless EGOE/BEGOE.

\item Starting with $U(N)$ algebra for $m$ fermions/bosons in $N$ sp
states, we have identified EEs with embedding defined by some of the $U(N)$
sub-algebras. As $U(N)$ admits very large class of sub-algebras, it is possible
to identify many more EEs that could be physically relevant and this exploration
will enrich the subject of embedded random matrix ensembles. 

\end{itemize}

\appendix
\chapter{Unitary decomposition for a one plus two-body Hamiltonian for 
spinless fermions}
\label{c2a2}

\renewcommand{\theequation}{A\arabic{equation}}
\setcounter{equation}{0}   

Let us consider a system of $m$ fermions in $N$ sp states with a (1+2)-body
Hamiltonian $H=h(1)+V(2)$ where $h(1)=\sum_i \epsilon_i\, {\hat{n}}_i$  and
$V(2)$ is defined by the two-body matrix elements $V_{ijkl}= \lan kl
\mid V(2) \mid ij \ran$.  With respect to the $U(N)$ group, the two-body
interaction $V(2)$ can be separated into scalar ($\nu=0$), effective
one-body ($\nu=1$) and irreducible two-body ($\nu=2$) parts. Then, we have
\cite{Ch-71,KS-01},
\be
\barr{l}
V^{\nu=0} = \dis\frac{\hat{n}(\hat{n} -1)}{2}
{\overline{V}}\;\;;\;\;\;\; {\overline{V}}=
\dis\binom{N}{2}^{-1} \; \dis\sum_{i<j} V_{ijij} \;,\\ \\
V^{\nu=1} = \dis\frac{{\hat{n}}-1}{N-2} \dis\sum_{i,j} \zeta_{i,j}
a^\dagger_ia_j\;\;;\;\;\zeta_{i,j}=
\l[\dis\sum_{k} V_{kikj}\r] -\l[(N)^{-1}\;\dis\sum_{r,s} 
V_{rsrs}\r] \delta_{i,j} \;,\nonumber
\earr \label{ud.eq.a1a1}
\ee
\be
V^{\nu=2} = V - V^{\nu=0} - V^{\nu=1}\;\; \Longleftrightarrow \;\;
V^{\nu=2}_{ijkl}\;\;; 
\label{ud.eq.a1}
\ee
\be
\barr{l}
V^{\nu=2}_{ijij} = V_{ijij}- \overline{V} - (N-2)^{-1} \l(\zeta_{i,i}+
\zeta_{j,j} \r)\;\;,\\  \\
V^{\nu=2}_{ijik} = V_{ijik} - (N-2)^{-1} \zeta_{j,k}\;\;\;\mbox{for}\;\; 
j \neq k \;,\\ \\
V^{\nu=2}_{ijkl} = V_{ijkl} \;\;\;\mbox{for all other cases} \;.\nonumber
\earr \label{ud.eq.a1a2}
\ee
Similar to Eq. (\ref{ud.eq.a1}), the $h(1)$ operator will have $\nu=0,1$ parts,
\be
\barr{l}
h^{\nu=0}= {\overline{\epsilon}}\;{\hat{n}}\;,\;\;\;
{\overline{\epsilon}}=(N)^{-1} \dis\sum_i \epsilon_i \;,\\
h^{\nu=1} =
\dis\sum_i\,\epsilon^1_i {\hat{n}}_i\;\;,\;\;\;\epsilon^1_i =
\epsilon_i-{\overline{\epsilon}} \;.
\earr \label{ud.eq.a2}
\ee
Then the propagation equations for the $m$-particle centroids and 
variances are,
\be
\barr{rcl}
E_c(m) & = & \lan H \ran^m =  m\,\overline{\epsilon} + 
\dis\binom{m}{2}\;\overline{V} \;,\\ \\
\sigma^2(m) & = & \lan H^2 \ran^m - \l[ E_c(m) \r]^2 \\ \\
& = & 
\dis\frac{m(N-m)}{N(N-1)} \;\;\dis\sum_{i,j}\;
\l\{\epsilon^1_i \delta_{i,j} + \dis\frac{m-1}{N-2} \zeta_{i,j}\r\}^2 \\ \\
& + & \dis\frac{m(m-1)(N-m)(N-m-1)}{N(N-1)(N-2)(N-3)} \;\lan\lan 
\l(V^{\nu=2} \r)^2 \ran\ran^2 \;.
\earr \label{ud.eq.a3}
\ee

\chapter{Exact variance formula for a given member of EGOE(1+2)-$\cs$}
\label{c2a3}

\renewcommand{\theequation}{B\arabic{equation}}
\setcounter{equation}{0}   

For completeness, we reproduce here the formula for spectral variances generated
by each member of EGOE(1+2)-$\cs$.
Given a one plus two-body Hamiltonian $H$, the fixed-$S$ 
spectral variance $\sigma^2(m,S)=\lan H^2 \ran^{m,S} - [\lan H
\ran^{m,S}]^2$ will be a fourth order polynomial in $m$ and $S(S+1)$
\cite{Fr-69,No-86}. This gives 
\be
\sigma^2(m,S)=\sum_{p=0}^4\,a_p\,m^{p} +
\sum_{q=0}^2\,b_q\,m^{q} \, S(S+1) + c_0\,[S(S+1)]^2 \;. 
\label{eq.c2a3e1}
\ee
The parameters
$(a_i,b_i,c_i)$ follow from  $\sigma^2(m,S)$ for $m \leq 4$ and to determine
these inputs one has to  construct $H$ matrices for $m$ up to 4. However an
elegant method, allowing $\sigma^2(m,S)$ to be expressed in terms of
$(\epsilon_i,V^{s=0,1}_{ijkl})$, is to use the embedding algebra $U(N)
\supset U(\Omega) \otimes SU(2)$. With respect to this algebra, as pointed
out in \cite{Ko-79,Kk-02}, $h(1)$ decomposes into a scalar $\nu=0$ part
[given by the first term in the first equation in Eq. (\ref{eq.den1})]  
and an irreducible
one-body part with $\nu=1$. The $\nu=0$ and $\nu=1$ parts transform, in
Young tableaux notation \cite{He-74}, as the 
irreps $[0]$ and $[21^{\Omega-2}]$ respectively of $U(\Omega)$.  Similarly
$V^s(2)$, $s=0,1$ decompose into $\nu=0,1$ and $2$ parts. The scalar  parts
$V^{\nu=0:s=0,1}$ can be identified from Eq. (\ref{eq.den1}) and they will not
contribute to the variances. The effective one-body parts $V^{\nu=1:s=0,1}$,
generated by $V^{s=0,1}_{ijkl}$, are defined by the  induced single particle
energies $\lambda_{i,j}(s)$ given ahead in Eq. (\ref{eq.vv1}). 
The diagonal induced
energies $\lambda_{i,i}(s)$ are  identified for the first time in
\cite{Ko-79}. However for EGOE(1+2)-$\cs$ it is possible to have
$\lambda_{i,j}(s)$, $i \neq j$. Now the irreducible two-body part
$V^{\nu=2:s=0}=V-V^{\nu=0:s=0}-V^{\nu=1:s=0}$ and similarly $V^{\nu=2:s=1}$
is defined. It should be noted that the two $\nu=0$ parts of $V(2)$
transform as the $U(\Omega)$ irrep $[0]$ and the two $\nu=1$ parts of $V(2)$
transform as the irrep $[21^{\Omega-2}]$. Similarly  $V^{\nu=2:s=0}$
transforms as the irrep $[42^{\Omega-2}]$ and the $V^{\nu=2:s=1}$ as the
irrep $[2^21^{\Omega-4}]$. Using these and the  group theory of $U(N)
\supset U(\Omega) \otimes SU(2)$ algebra as given by Hecht and Draayer
\cite{He-74}, a compact and easy to  understand expression for fixed-$S$
variances emerges, with $\cas^2=S(S+1)$, $m^x=\Omega-m/2$, 
$X(m,S)=m(m+2)-4S(S+1)$ and $Y(m,S)=m(m-2)-4S(S+1)$,
\be 
\barr{l} 
\sigma_{H=h(1)+V(2)}^2(m,S) = 
\dis\frac{(\Omega+2)m m^x -2\Omega \;\cas^2}{ \Omega (\Omega-1)
(\Omega+1)}\;\;\dis\sum_i\;{\tilde{\epsilon}}_i^2 \\\\
+ \dis\frac{m^x\;X(m,S)}{2 \Omega (\Omega-1)
(\Omega+1)}\;\;\dis\sum_i\;{\tilde{\epsilon}}_i \lambda_{i,i}(0) \\\\ 
+ \dis\frac{(\Omega+2) m^x \l[3Y(m,S)+16\cas^2\r] -
8 \Omega (m-1) \cas^2}{2 \Omega (\Omega-1) (\Omega+1) (\Omega-2)}\;\;
\dis\sum_i\;{\tilde{\epsilon}}_i \lambda_{i,i}(1) \\\\ 
+ \dis\frac{\l[(m+2) m^x /2 + \cas^2 \r] X(m,S)}
{8 \Omega (\Omega-1) (\Omega+1) (\Omega+2)}\;\; \dis\sum_{i,j}\;
\lambda_{i,j}^2(0) \\\\ 
+ \dis\frac{1}{8 \Omega (\Omega-1) (\Omega+1) (\Omega-2)^2}
\l\{8\Omega(m-1)(\Omega-2m+4)\cas^2 \r. \\ \\
+\l. (\Omega+2) \l[3(m-2)m^x/2 - \cas^2\r]\l[3Y(m,S)+8\cas^2\r] \r\}
\;\; \dis\sum_{i,j}\;\lambda_{i,j}^2(1) \\\\ 
+ \dis\frac{\l[3(m-2)m^x/2 - \cas^2\r]X(m,S)
}{4 \Omega (\Omega-1) (\Omega+1) (\Omega-2)}\;\; \dis\sum_{i,j}\;
\lambda_{i,j}(0) \lambda_{i,j}(1) \\\\
+ P_2^0(m,S)\,\lan \l(V^{\nu=2,s=0}\r)^2 \ran^{2,0} +
P_2^1(m,S)\,\lan \l(V^{\nu=2,s=1}\r)^2 \ran^{2,1}\;;\\
\earr \label{eq.vv1}
\ee
\be
\barr{l}
P_2^0(m,S) = \dis\frac{\l[m^x(m^x+1) -\cas^2\r]X(m,S)}
{8 \Omega (\Omega-1)}\;, \nonumber
\earr \label{eq.vv1a2}
\ee
\be
\barr{l}
P_2^1(m,S) = \dis\frac{1}
{\Omega(\Omega+1) (\Omega-2)(\Omega-3)} \;\;
\l\{(\cas^2)^2(3\Omega^2-7\Omega+6)/2  +  
3m(m-2) m^x (m^x-1) \r. \\ \\ 
\l. \times (\Omega+1)(\Omega+2)/8 
- \cas^2 \l[(5\Omega-3)(\Omega+2)m^x m +
\Omega(\Omega-1)(\Omega+1)(\Omega+6)\r]/2 \r\}\;, \nonumber
\earr \label{eq.vv1a1}
\ee
with
\be
\barr{l}
{\tilde{\epsilon}}_i=\epsilon_i -\overline{\epsilon} \;,\\ \\
\lambda_{i,i}(s) =  \dis\sum_j\;V_{ijij}^s\;(1+\delta_{ij})
\;-\;(\Omega)^{-1} \;\dis\sum_{k,l}\;V^s_{klkl}\;(1+\delta_{kl}) \;,\\ \\
\lambda_{i,j}(s) = \dis\sum_k\;\dis\sqrt{(1+\delta_{ki})(1+\delta_{kj})}\,
V^s_{kikj}\;\;\;\mbox{for}\;\;\;i \neq j \;,\\ \\
V^{\nu=2,s}_{ijij} =  V^s_{ijij} - \l[\lan V(2)\ran^{2,s} +
(\lambda_{i,i}(s) + \lambda_{j,j}(s))\l(\Omega+2(-1)^s\r)^{-1}\r] \;,\\ \\
V^{\nu=2,s}_{kikj} = V^s_{kikj} - \l(\Omega+2(-1)^s\r)^{-1}\,
\dis\sqrt{(1+\delta_{ki})(1+\delta_{kj})}\,\lambda^s_{i,j}
\;\;\;\mbox{for}\;\;\;i \neq j \;,\\ \\
V^{\nu=2,s}_{ijkl} = V^s_{ijkl}\;\;\;\mbox{for all other cases}\;.
\earr  \label{eq.vv2}
\ee
Equations (\ref{eq.vv1}) and (\ref{eq.vv2}) are tested, by using some members 
of the
EGOE(1+2)-$\cs$ ensemble, for all $S$ values with $m=6,7$ and 8 and also for
many different $\Omega$ values. 

\chapter*{}

\thispagestyle{empty}

\chapter{EGUE(2)-$\cs$ ensemble}
\label{egue2}

\renewcommand{\theequation}{C\arabic{equation}}
\setcounter{equation}{0}   

For $m$ fermions occupying $\Omega$ number of sp orbitals each with spin  
$\cs=\spin$ so that the number of sp states $N=2\Omega$, we consider
Hamiltonians that preserve total $m$-particle spin $S$. Then the $m$-particle 
states can be classified according to $U(N) \supset U(\Omega)
\otimes SU(2)$ algebra with $SU(2)$ generating spin $S$. The $U(\Omega)$
irrep that corresponds to spin $S$ is  $f_m=\{2^p,1^q\}$ where $m=2p+q$ and
$S=q/2$ and therefore  the $m$-particle states are denoted by $\l.\l|f_m \;
v_m M\r.\ran$;  $v_m$ are the additional quantum numbers that belong to
$f_m$ and $M$ is the $S_z$ quantum number. With this, a general two-body
Hamiltonian operator  preserving spin $S$ can be written as,
\be
\wh = \dis\sum_{f_2,v_2^i,v_2^f,m_2} V_{f_2 v_2^i v_2^f}(2) 
A^\dagger(f_2 v_2^f m_2)  A(f_2 v_2^i m_2)\;. 
\label{eq.gue3}
\ee
Here, $A^\dagger(f_2 v_2 m_2)$ and $A(f_2 v_2 m_2)$ denote creation and
annihilation operators for the normalized two-particle states and $V_{f_2
v_2^i v_2^f}(2)= \lan f_2 v^f_2 s m_2 \mid \wh \mid f_2 v^i_2 s m_2\ran$
independent of the $m_2$'s. Note that the two-particle spin $s=0$ and $1$
and the  corresponding $U(\Omega)$ irreps $f_2$ are $\{2\}$ (symmetric) and
$\{1^2\}$ (antisymmetric), respectively.  The EGUE(2)-$\cs$ ensemble for
a given $(m,S)$ is generated by the action of $\wh$ on $m$-particle basis
space with a GUE representation for the $H$ matrix in two-particle spaces.
Then, the two-particle matrix  elements  $V_{f_2 v_2^i v_2^f}(2)$ are
independent Gaussian variables with zero center and variance given by,
\be
\overline{V_{f_2 v_2^1 v_2^2}(2) V_{f_2^\pr v_2^3 v_2^4}(2)} =
\lambda_{f_2}^2 \delta_{f_2 f_2^\pr} \delta_{v_2^1 v_2^4} 
\delta_{v_2^2 v_2^3} \;.
\label{eq.gue4}
\ee
Thus $V(2)$ is a direct sum of GUE matrices for $s=0$ and $s=1$. Just as for
EGUE(2), tensorial decomposition of $\wh$ with respect to $U(\Omega) \otimes
SU(2)$ algebra gives analytical results for the spin ensemble. As $\wh$
preserves $S$, it is a scalar in spin $SU(2)$ space. However with respect to
$SU(\Omega)$, the tensorial characters  for $f_2=\{2\}$ are $\bF_\nu=\{0\}$,
$\{21^{\Omega-2}\}$ and $\{42^{\Omega-2}\}$. Similarly for $f_2=\{1^2\}$
they are $\{0\}$,  $\{21^{\Omega-2}\}$ and $\{2^2 1^{\Omega-4}\}$. Here the
unitary tensors $B$'s are 
\be
\barr{l}
B(f_2 \bF_\nu \omega_\nu) \\ = 
\dis\sum_{v_2^i,v_2^f, m_2}\, \lan f_2
v_2^f\;\overline{f_2}\,\overline{v_2^i} \mid \bF_\nu \omega_\nu\ran\,\lan s
m_2\;\overline{s}\,\overline{m_2} \mid 0 0\ran \,A^\dagger(f_2 
v^f_2 m_2)\, A(f_2 v^i_2 m_2)\;. 
\label{eq.gue4a}
\earr
\ee
In Eq. (\ref{eq.gue4a}), $\lan f_2 --- \ran$ are $SU(\Omega)$  Wigner
coefficients  and $\lan s -- \ran$ are $SU(2)$ Wigner coefficients.
Then we have $\wh(2) = \sum_{f_2, \bF_\nu, \omega_\nu} W(f_2 \bF_\nu
\omega_\nu) B(f_2 \bF_\nu \omega_\nu) $. The expansion coefficients $W$'s
are also independent Gaussian random variables, just as $V$'s,
with zero center  and variance given by 
$$
\overline{W(f_2 \bF_\nu \omega_\nu)W(f^\pr_2 \bF^\pr_\nu \omega^\pr_\nu)}
=\delta_{f_2 f^\pr_2} \delta_{\bF_\nu \bF^\pr_\nu} \delta_{\omega_\nu
\omega^\pr_\nu} \, (\lambda_{f_2})^2 (2s+1)\;.
$$ 
The $m$-particle $H$ matrix will be a direct sum matrix with the diagonal
blocks represented by $f_m$. Then $H(m)=\sum_{f_m} H_{f_m}(m) \oplus$ and
the EGUE(2)-$\cs$ is generated for each $H_{f_m}(m)$.

Using Wigner-Eckart theorem, the matrix elements of $B$'s in $f_m$ space can
be decomposed as,
\be
\barr{l}
\lan f_m v_m^f M \mid B(f_2 \bF_\nu \omega_\nu) \mid f_m v_m^i M\ran  \\ = 
\dis\sum_\rho\;\lan f_m \mid\mid B(f_2 \bF_\nu) \mid\mid f_m\ran_\rho\,
\lan f_m v_m^i \bF_\nu \omega_\nu \mid f_m v_m^f\ran_\rho \;,
\earr 
\label{eq.gue5}
\ee
where the summation is over the multiplicity index $\rho$ and this arises
as  $f_m \otimes \bF_\nu$ gives in general more than once the irrep $f_m$.
Applying Eq. (\ref{eq.gue5}) and the expansion of $\wh$ in terms of $B$'s,  
exact analytical formulas are derived for the ensemble averaged spectral
variances, cross-correlations in energy centroids and also for the 
cross-correlations in  spectral variances. In addition, the ensemble averaged
excess parameter for the fixed-$(m,S)$ density of states is  given in terms
of $SU(\Omega)$ Racah coefficients \cite{Ko-07}. For finite $m$ and  $\Omega
\to \infty$, some important results are:  (i) the
ensemble averaged variances, to the leading order, just as for the spinless
fermion systems \cite{Be-01}, are same for both EGOE(2)-$\cs$ and 
EGUE(2)-$\cs$ and this is inferred from the exact analytical formulas
available for both the ensembles [comparing Eq. (\ref{eq.den8}) with Eq. (19) 
of \cite{Ko-07}]; (ii) similarly it is seen that the cross-correlations 
in energy centroids for
EGOE(2)-$\cs$ are twice that of EGUE(2)-$\cs$  to the
leading order \cite{Ko-06a,Be-01} [as an aside,
let us point out that Eqs. (\ref{eq.den2}) and
(\ref{eq.den8}) give for EGOE(2)-$\cs$, the exact formula for the normalized
cross-correlations in the energy  centroids]; (iii) combining (ii) with the exact analytical
results for spinless fermion EGUE(2)-$\cs$ (see Sec. \ref{egue1} for details), it is
conjectured that the covariances in spectral variances for EGOE(2)-$\cs$ are twice that of
EGUE(2)-$\cs$  to the
leading order [note that for EGUE(2)-$\cs$ an analytical result is available but not for
EGOE(2)-$\cs$]; and (iv) combining the analytical results for the excess
parameter for EGOE(2) and EGOE(2)-$\cs$ (see Appendix 
\ref{c7s1} and Sec. \ref{c2a1} for
details), it is expected that the density of eigenvalues will be Gaussian for
EGUE(2)-$\cs$.

\chapter*{}

\thispagestyle{empty}

\chapter{$U(2\Omega)\supset[U(\Omega)\supset SO(\Omega)]\otimes SU(2)$
pairing symmetry}
\label{c3a1}

\renewcommand{\theequation}{D\arabic{equation}}
\setcounter{equation}{0}   

With the 4$\Omega^2$ number of one-body operators $u_{\mu}^{r}(i,j)$; 
$r=0,1$, defined in Sec. \ref{c3s1}, generating $U(2\Omega)$
algebra, it is easily seen that the
operators $u^0(i,j)$, which are $\Omega^2$ in
number, generate $U(\Omega)$  algebra. Similarly the operators 
$C(i,j)=u^0(i,j)-u^0(j,i),\; i > j$, which are $\Omega(\Omega-1)/2$ in
number, generate the $SO(\Omega)$ sub-algebra of $U(\Omega)$. The spin
operator $\hat{S}=S^1_\mu$, the number operator $\hat{n}$ and the quadratic
Casimir operators $C_2$'s of  $U(\Omega)$ and $SO(\Omega)$ are
\be
\barr{l}
S^1_\mu = \dis\frac{1}{\sqrt{2}}\,\dis\sum_{i=1}^{\Omega} 
\,  u^1_\mu(i,i)\;,
\\ \\
\hat{n}=\dis\sum_i n_i\,,\;\;\;\;n_i=\dis\sqrt{2} u^0(i,i)\;, 
\\ \\
C_2(U(\Omega))=2 \dis\sum_{i,j} u^0(i,j) u^0(j,i)\,,
\\ \\
C_2(SO(\Omega)) = 2 \dis\sum_{i > j} C(i,j) C(j,i) \;.
\earr \label{eq.pa2}
\ee
The structure of $C_2(U(\Omega))$ in terms of the number operator and the 
$\hat{S} \cdot \hat{S} = \hat{S}^2$ operator is, 
\be
\barr{rcl} 
C_2(U(\Omega)) & = &
\hat{n}\l(\Omega+2-\frac{\hat{n}}{2}\r) - 2 \hat{S}^2\;\;, 
\\ \\
\lan C_2(U(\Omega)) \ran^{m,S} & = & m\l(\Omega+2- \frac{m}{2}\r) 
- 2 S(S+1)\;.  
\earr \label{eq.pa21}
\ee 
Note that $\lan C_2(U(\Omega)) \ran^{\{f\}}=\sum_if_i(f_i+\Omega+1-2i)$.
As $U(2\Omega) \supset U(\Omega) \otimes SU(2)$ with the $SU(2)$ algebra
generating total spin $S$, the $U(\Omega)$ irreps are labeled by two 
column irreps $\{2^p 1^q\}$ with $m=2p+q\,,\,S=q/2$. As a consequence, the
$SO(\Omega)$ irreps  are  also of two column type and we will denote them
by $[2^{v_1} 1^{v_2}]$. Here, $v_S=2v_1+v_2$ is called seniority and
$\tilde{s}=v_2/2$  is called reduced spin. We also have \cite{Wy-74}
\be
\barr{rcl}
\lan C_2(SO(\Omega)) \ran^{\lan \omega\ran} & = & \dis\sum_i 
\omega_i  (\omega_i+\Omega-2i)
\\ \\
\Rightarrow \lan C_2(SO(\Omega)) \ran^{\lan 2^{v_1}\;1^{v_2}\ran} 
& = &
v_S \l(\Omega + 1 - \dis\frac{v_S}{2} \r)-2\tilde{s}(\tilde{s}+1) \;.
\earr \label{eq.pa21a}
\ee
After some commutator algebra it  can be shown that,
\be
\barr{c}
2H_p=-C_2(SO(\Omega)) + \hat{n}\l(\Omega+1-\frac{\hat{n}}{2}\r) - 2
\hat{S}^2\;,
\\ \\
\lan H_p \ran^{(m,S,v_S,\tilde{s})} = \frac{1}{4} (m-v_S)
(2\Omega+2-m-v_S) + \l[\tilde{s}(\tilde{s}+1)-S(S+1)\r]\;,
\earr \label{eq.pa4}
\ee
where the pairing Hamiltonian $H_p$ is defined by Eq. (\ref{ch3.eq.npa2}).
Classification of $U(2\Omega) \supset \l[U(\Omega)  
\supset SO(\Omega)\r]\otimes SU(2)$ states defined by $(m,S,v_S,\tilde{s})$ 
quantum numbers is needed, i.e., $(m,S) \rightarrow (v_S,
\tilde{s})$ reductions are required and they are obtained by group theory. 
Using the 
tabulations in \cite{Wy-70}, results are given in Tables \ref{c3a1t1} and 
\ref{c3a1t2} 
for: (i) $m \leq 4, \; \Omega\geq 4$ and (ii) $m=6,\;\Omega=6$ and 
$m=5-8,\;\Omega=8$, respectively.

\begin{table}[htp]
\centering
\caption{$(m,S) \rightarrow (v_S,\tilde{s})$ reductions for 
$m \leq 4$ and $\Omega \geq 4$.}
\begin{tabular}{lr}
\toprule
$(m,S)\;\;\;\;\;$ & $(v_S,\tilde{s})$ \\
\midrule
$(0,0)$ & $(0,0)$ \\ 
$(1,\frac{1}{2})$ & $(1,\frac{1}{2})$ \\
$(2,0)$ & $(2,0)$, $(0,0)$ \\
$(2,1)$ & $(2,1)$ \\
$(3,\frac{1}{2})$ & $(3,\frac{1}{2})$, $(1,\frac{1}{2})$ \\
$(3,\frac{3}{2})$ & $\{(1,\frac{1}{2})_{\Omega=4}\,;\;\;
(2,1)_{\Omega=5}\,;\;\;(3,\frac{3}{2})_{\Omega \geq 6}\}$ \\
$(4,0)$ & $(4,0)$, $(2,0)$, $(0,0)$ \\
$(4,1)$ & $\{(2,0)_{\Omega=4}\,;\;\;
(3,\frac{1}{2})_{\Omega=5}\,;\;\;(4,1)_{\Omega \geq 6}\}$, $(2,1)$ \\
$(4,2)$ & $\{(0,0)_{\Omega=4}\,;\;\;
(1,\frac{1}{2})_{\Omega=5}\,;\;\;(2,1)_{\Omega=6}\,,\;\;
(3,\frac{3}{2})_{\Omega=7}\,;\;\;(4,2)_{\Omega \geq 8}\}$ \\
\bottomrule
\end{tabular}
\label{c3a1t1}
\end{table}

\begin{table}[htp]
\centering
\caption{$(m,S) \rightarrow (v_S,\tilde{s})$ irrep reductions for 
($\Omega = 6; m = 6$) and ($\Omega = 8; m = 5-8$). Note that the dimensions $d_f(\Omega,m,S)$ of the
$(m,S)$ and $d(\Omega,v_S,\tilde{s})$ of the $(v_S,\tilde{s})$ space are given as subscripts; 
$d_f(\Omega,m,S) = \sum_{v_S,\tilde{s}} d(\Omega,v_S,\tilde{s})$.}
\begin{tabular}{llr}
\toprule
$\Omega\;\;\;\;\;\;\;\;$ & $(m,S)_{d_f(\Omega,m,S)}\;\;\;\;\;$ & 
$(v_S,\tilde{s})_{d(\Omega,v_S,\tilde{s})}$ \\
\midrule
6 & $(6,0)_{175}$ & $(6,0)_{70}$, $(4,0)_{84}$, $(2,0)_{20}$, $(0,0)_{1}$ \\\\
& $(6,1)_{189}$ & $(4,1)_{90}$, $(4,0)_{84}$, $(2,1)_{15}$ \\\\
& $(6,2)_{35}$ & $(2,1)_{15}$, $(2,0)_{20}$ \\\\
& $(6,3)_{1}$ & $(0,0)_{1}$ \\\\
8 & $(5,\frac{1}{2})_{1008}$ & $(5,\frac{1}{2})_{840}$, $(3,
\frac{1}{2})_{160}$, $(1,\frac{1}{2})_{8}$ \\\\
& $(5,\frac{3}{2})_{504}$ & $(5,\frac{3}{2})_{448}$, $(3,
\frac{3}{2})_{56}$ \\\\
& $(5,\frac{5}{2})_{56}$ & $(3,\frac{3}{2})_{56}$ \\\\
& $(6,0)_{1176}$ & $(6,0)_{840}$, $(4,0)_{300}$, $(2,0)_{35}$, $(0,0)_{1}$ 
\\\\
& $(6,1)_{1512}$ & $(6,1)_{1134}$, $(4,1)_{350}$, $(2,1)_{28}$ \\\\
& $(6,2)_{420}$ & $(4,1)_{350}$, $(4,2)_{70}$ \\\\
& $(6,3)_{28}$ & $(2,1)_{28}$ \\\\
& $(7,\frac{1}{2})_{2352}$ & $(7,\frac{1}{2})_{1344}$, 
$(5,\frac{1}{2})_{840}$,
$(3,\frac{1}{2})_{160}$, $(1,\frac{1}{2})_{8}$ \\\\
& $(7,\frac{3}{2})_{1344}$ & $(5,\frac{1}{2})_{840}$, 
$(5,\frac{3}{2})_{448}$, 
$(3,\frac{3}{2})_{56}$ \\\\
& $(7,\frac{5}{2})_{216}$ & $(3,\frac{1}{2})_{160}$, 
$(3,\frac{3}{2})_{56}$ \\\\
& $(7,\frac{7}{2})_{8}$ & $(1,\frac{1}{2})_{8}$ \\\\
& $(8,0)_{1764}$ & $(8,0)_{588}$, $(6,0)_{840}$, $(4,0)_{300}$, 
$(2,0)_{35}$, 
$(0,0)_{1}$ \\\\
& $(8,1)_{2352}$ & $(6,0)_{840}$, $(6,1)_{1134}$, $(4,1)_{350}$, 
$(2,1)_{28}$ \\\\
& $(8,2)_{720}$ & $(4,0)_{300}$, $(4,1)_{350}$, $(4,2)_{70}$ \\\\
& $(8,3)_{63}$ & $(2,0)_{35}$, $(2,1)_{28}$ \\\\
& $(8,4)_{1}$ & $(0,0)_{1}$ \\
\bottomrule
\end{tabular}
\label{c3a1t2}
\end{table}

\chapter*{}

\thispagestyle{empty}

\chapter{Some properties of $SU(\Omega)$ Wigner coefficients}
\label{c4a1}

\renewcommand{\theequation}{E\arabic{equation}}
\setcounter{equation}{0}   

Some properties of the $SU(\Omega)$ Wigner coefficients 
used in Chapter \ref{ch4} are given here and they are similar to those used for 
EGUE(2) and EGUE(2)-$\cs$ in \cite{Ko-05,Ko-07} and discussed in detail in
\cite{But-81}. Firstly (dropping the multiplicity index $\rho$ everywhere 
for simplicity),
\be
\lan f_a\,v_a\,f_b\,v_b\mid f_{ab}\,v_{ab} \ran = 
(-1)^{\phi(f_a,f_b,f_{ab})}
\lan f_b\,v_b\,f_a\,v_a\mid f_{ab}\,v_{ab} \ran\;,
\label{ch4.eq.prp1}
\ee
where $\phi$ is a function of $(f_a,f_b,f_{ab})$ that defines the phase 
for $a \to b$ interchange in the Wigner coefficient. With $\l|
\overline{f_a}\,\overline{v_a} \ran$ denoting the time-reversal partner 
(complex conjugate) of $\l| f_a\,v_a\ran$, we have
\be
\lan f_a\,v_a\,f_b\,v_b\mid f_{ab}\,v_{ab} \ran = 
\lan \overline{f_a}\,\overline{v_a}\,\overline{f_b}
\,\overline{v_b}\mid \overline{f_{ab}}\,\overline{v_{ab}} \ran\;.
\label{ch4.eq.prp2}
\ee
Similarly,
\be
\lan f_a\,v_a\,f_b\,v_b\mid f_{ab}\,v_{ab} \ran =
(-1)^{\phi(f_a,f_b,f_{ab})}\, \dis\sqrt{\dis\frac{d_\Omega(f_{ab})}
{d_\Omega(f_a)}}
\lan f_{ab}\,v_{ab}\,\overline{f_b}
\,\overline{v_b}\mid f_{a}\,v_{a} \ran\;.
\label{ch4.eq.prp3}
\ee
In addition we also have,
\be
\lan f_a\,v_a\,\overline{f_a}\,\overline{v_a}\mid\{0\}\,0 \ran = 
\dis\frac{1} {\dis\sqrt{d_\Omega(f_a)}}\;,
\label{ch4.eq.prp4}
\ee
\be
\l[\lan f_a\,v_a\,\overline{f_a}\,\overline{v_b}\mid f_{ab}\,v_{ab} 
\ran\r]^*
=  \lan f_a\,v_b\,\overline{f_a}\,\overline{v_a}\mid f_{ab}\,v_{ab} \ran\;.
\label{ch4.eq.prp5}
\ee
Orthonormal properties of the Wigner coefficients are,
\begin{subequations}
\be
\dis\sum_{v_a,v_b} 
\lan f_a\,v_a\,f_b\,v_b\mid f_{ab}\,v_{ab} \ran \;
\l[\lan f_a\,v_a\,f_b\,v_b\mid f_{cd}\,v_{cd} 
\ran\r]^*
= \delta_{f_{ab},f_{cd}} \, \delta_{v_{ab},v_{cd}}\;, 
\label{ch4.eq.prp6}
\ee
\be
\dis\sum_{f_{ab},v_{ab}} 
\lan f_a\,v_a\,f_b\,v_b\mid f_{ab}\,v_{ab} \ran \;
\lan f_a\,v_c\,f_b\,v_d\mid f_{ab}\,v_{ab} \ran
= \delta_{v_a,v_c} \, \delta_{v_b,v_d}\;. 
\label{ch4.eq.prp7}
\ee
\end{subequations}
Finally,
\be
\barr{l}
\dis\sum_{v_{ab}} \lan f_a v_a f_b v_b \mid f_{ab} v_{ab} \ran
\lan f_{ab} v_{ab} f_c v_c \mid f v \ran \\ \\  =
\dis\sum_{f_{bc},v_{bc}}\lan f_b v_b f_c v_c \mid f_{bc} v_{bc} \ran
\lan f_a v_a  f_{bc} v_{bc} \mid f v \ran U(f_a f_b f f_c;f_{ab} 
f_{bc})\;.
\earr \label{ch4.eq.prp8}
\ee

\chapter{Excess parameter $\gamma_2(m,f_m)$ in terms of $SU(\Omega)$ Racah
coefficients}
\label{c4a2}

\renewcommand{\theequation}{F\arabic{equation}}
\setcounter{equation}{0}   

The formula for $\gamma_2(m,f_m)$, given by Eq. (\ref{ch4.eq.gam2}), involves
$\overline{\lan H^4 \ran^{m,f_m}}$. As the Hamiltonian in Eq. (\ref{ch4.eq.22})
is a direct sum of matrices in  $f_2=\{2\}$ and $\{1^2\}$ spaces, we have
\be
\overline{\lan H^4 \ran^{m,f_m}} = 
\overline{\lan (H_{\{2\}}+H_{\{1^2\}})^4 \ran^{m,f_m}}\;. 
\label{ch4.eq.gam-1}
\ee
Expanding the RHS of Eq. (\ref{ch4.eq.gam-1}) using the cyclic invariance 
of the averages and applying the property that terms with odd powers of 
$H_{\{2\}}$ and $H_{\{1^2\}}$ will vanish [see Eq. (\ref{ch4.eq.indep})], 
we have
\be
\barr{l}
\overline{\lan H^4 \ran^{m,f_m}} = 
\overline{\lan (H_{\{2\}})^4 \ran^{m,f_m}} + 
\overline{\lan (H_{\{1^2\}})^4 \ran^{m,f_m}} +
4\overline{\lan (H_{\{2\}})^2 (H_{\{1^2\}})^2 \ran^{m,f_m}} \\ \\
+2\overline{\lan H_{\{2\}} H_{\{1^2\}} H_{\{2\}} 
H_{\{1^2\}}\ran^{m,f_m}}\;. 
\earr \label{ch4.eq.gam-2}
\ee
Writing $H$ in terms of the unit tensors $B$'s using Eq. (\ref{ch4.eq.25}), 
the first two terms in Eq. (\ref{ch4.eq.gam-2}) will give
\be
\barr{l}
\overline{\lan H_{f_2}^4 \ran^{m,f_m}} \\ \\= 
\dis\frac{1}{d_\Omega(f_m)} 
\dis\sum_{v_1,v_2,v_3,v_4,\bF_{\nu_1},\bF_{\nu_2},\bF_{\nu_3},\bF_{\nu_4},
\omega_{\nu_1},\omega_{\nu_2},\omega_{\nu_3},\omega_{\nu_4}}
\lan f_m v_1 \mid B(f_2 \bF_{\nu_1} \omega_{\nu_1}) \mid f_m v_2 \ran \nonumber
\earr \label{ch4.eq.gam-3a1}
\ee
\be
\barr{l}
\times
\lan f_m v_2 \mid B(f_2 \bF_{\nu_2} \omega_{\nu_2}) \mid f_m v_3 \ran
\lan f_m v_3 \mid B(f_2 \bF_{\nu_3} \omega_{\nu_3}) \mid f_m v_4 \ran
\\  \\ \times
\lan f_m v_4 \mid B(f_2 \bF_{\nu_4} \omega_{\nu_4}) \mid f_m v_1 \ran
\\  \\ \times
\overline{W(f_2 \bF_{\nu_1} \omega_{\nu_1})W(f_2 \bF_{\nu_2} \omega_{\nu_2})
W(f_2 \bF_{\nu_3} \omega_{\nu_3})W(f_2 \bF_{\nu_4} \omega_{\nu_4})}\;.
\earr\label{ch4.eq.gam-3}
\ee
Using Eq. (\ref{ch4.eq.26}), it is easy to see that the term  $\overline{\lan
H_{f_2}^4 \ran^{m,f_m}}$ will have non-zero contribution in three cases, (i)
$\delta_{\bF_{\nu_1},\bF_{\nu_2}} = 1$, 
$\delta_{\omega_{\nu_1},\omega_{\nu_2}} = 1$,
$\delta_{\bF_{\nu_3},\bF_{\nu_4}} = 1$, 
$\delta_{\omega_{\nu_3},\omega_{\nu_4}} = 1$; (ii)  
$\delta_{\bF_{\nu_1},\bF_{\nu_4}} = 1$, 
$\delta_{\omega_{\nu_1},\omega_{\nu_4}} = 1$,
$\delta_{\bF_{\nu_2},\bF_{\nu_3}} = 1$, 
$\delta_{\omega_{\nu_2},\omega_{\nu_3}} = 1$; and (iii)
$\delta_{\bF_{\nu_1},\bF_{\nu_3}} = 1$, 
$\delta_{\omega_{\nu_1},\omega_{\nu_3}} = 1$,
$\delta_{\bF_{\nu_2},\bF_{\nu_4}} = 1$, 
$\delta_{\omega_{\nu_2},\omega_{\nu_4}} = 1$. The first two cases are
equivalent due to cyclic invariance of the traces and they can be  called
direct terms whereas the third case involves cross-correlations and thus is
called the exchange term. For (i) and (ii), applying the Wigner-Eckart
theorem and carrying out simplifications using the properties of the Wigner 
coefficients (see Appendix \ref{c4a1}), the direct terms reduce to $2\l[\;
\overline{\lan H_{f_2}^2 \ran^{m,f_m}}\;\r]^2$. Similarly, for the exchange
term, reordering of the  Wigner coefficients [see Eq. (\ref{ch4.eq.prp8})]
yields an  expression in terms of a new Racah coefficient. With these, we
have
\be
\barr{l}
\overline{\lan H_{f_2}^4 \ran^{m,f_m}} = 2\l[\; \overline{\lan H_{f_2}^2
\ran^{m,f_m}}\;\r]^2 + \lambda_{f_2}^4 \l[d_4(F_2)\r]^2 d_\Omega(f_m)
\\ \\ \times
\dis\sum_{\bF_{\nu_1},\bF_{\nu_2},\rho_1,\rho_2,\rho_3,\rho_4}
\dis\frac{1}{\dis\sqrt{d_\Omega(\bF_{\nu_1}) d_\Omega(\bF_{\nu_2})}}
U(f_m \overline{f_m} f_m f_m; (\bF_{\nu_1})_{\rho_1 \rho_3}
(\bF_{\nu_2})_{\rho_2 \rho_4}) \nonumber
\earr \label{ch4.eq.gam-4a1}
\ee 
\be
\barr{l}
\times
\lan f_m \mid\mid B(f_2 \bF_{\nu_1}) \mid\mid f_m \ran_{\rho_1} 
\lan f_m \mid\mid B(f_2 \bF_{\nu_2}) \mid\mid f_m \ran_{\rho_2} 
\\ \\ \times
\lan f_m \mid\mid B(f_2 \bF_{\nu_1}) \mid\mid f_m \ran_{\rho_3} 
\lan f_m \mid\mid B(f_2 \bF_{\nu_2}) \mid\mid f_m \ran_{\rho_4}\;.
\earr \label{ch4.eq.gam-4}
\ee
In Eq. (\ref{ch4.eq.gam-4}), the multiplicity labels appearing in the new 
$U$-coefficient [this is quite different from the $U$-coefficient appearing
in Eq. (\ref{ch4.eq.27})] can be easily understood from the corresponding 
labels  in the reduced matrix elements. Similarly, we have 
\begin{subequations}
\be
\barr{l}
\overline{\lan H_{\{2\}}^2 H_{\{1^2\}}^2 \ran^{m,f_m}} = 
\l\{\overline{\lan H_{\{2\}}^2\ran^{m,f_m}}\r\} \;\;
\l\{\overline{\lan H_{\{1^2\}}^2\ran^{m,f_m}}\r\}\;,
\earr \label{ch4.eq.gam-5a}
\ee
\be
\barr{l}
\overline{\lan H_{\{2\}} H_{\{1^2\}} H_{\{2\}} H_{\{1^2\}}\ran^{m,f_m}} =
\lambda_{\{2\}}^2 \lambda_{\{1^2\}}^2 d_4(\{2\}) d_4(\{1^2\}) d_\Omega(f_m)
\\ \\ \times
\dis\sum_{\bF_{\nu_1},\bF_{\nu_2},\rho_1,\rho_2,\rho_3,\rho_4}
\dis\frac{1}{\dis\sqrt{d_\Omega(\bF_{\nu_1}) d_\Omega(\bF_{\nu_2})}}
U(f_m \overline{f_m} f_m f_m; (\bF_{\nu_1})_{\rho_1 \rho_3}
(\bF_{\nu_2})_{\rho_2 \rho_4}) 
\\ \\ \times
\lan f_m \mid\mid B(\{2\} \bF_{\nu_1}) \mid\mid f_m \ran_{\rho_1} 
\lan f_m \mid\mid B(\{1^2\} \bF_{\nu_2}) \mid\mid f_m \ran_{\rho_2} 
\\ \\ \times
\lan f_m \mid\mid B(\{2\} \bF_{\nu_1}) \mid\mid f_m \ran_{\rho_3} 
\lan f_m \mid\mid B(\{1^2\} \bF_{\nu_2}) \mid\mid f_m \ran_{\rho_4}\;.
\earr\label{ch4.eq.gam-5}
\ee
\end{subequations}
Substituting the results in Eqs. (\ref{ch4.eq.gam-4}), (\ref{ch4.eq.gam-5a})  and
(\ref{ch4.eq.gam-5}) in Eq. (\ref{ch4.eq.gam-2}) gives $\overline{\lan H^4
\ran^{m,f_m}}$. Using this and Eqs. (\ref{ch4.eq.34}) and (\ref{ch4.eq.gam2}),  we
have the analytical result for the excess parameter  $\gamma_2(m,f_m)$. This
involves $SU(\Omega)$  Racah coefficients with multiplicity labels and
evaluation of these is in general  complicated \cite{Kl-05,Kl-09}. Similarly,
evaluation of the  reduced matrix elements in Eq. 
(\ref{ch4.eq.gam-4}) is also complicated.   The only simple situation is, when
the multiplicity labels are all unity. We denote the  $U(\Omega)$ irreps 
that satisfy this as  $f_m^{(g)}$ and we have verified that one of these 
irreps is  $\{4^r\}$  where $m=4r$. For these irreps, the expression for
$\gamma_2$ is,
\be
\barr{l}
\l[\gamma_2(m,f_m^{(g)})+1\r] = \l[\;\overline{\lan H^2 \ran^{m,f_m^{(g)}}}
\;\r]^{-2}\;
\\ \\ \times \l\{
\dis\sum_{f_a,f_b=\{2\},\{1^2\}} \dis\frac{\lambda_{f_a}^2 \lambda_{f_b}^2} 
{d_\Omega(f_a) d_\Omega(f_b)}
\dis\sum_{\bF_{\nu_1},\bF_{\nu_2}} \dis\frac{d_\Omega(f_m^{(g)})}{\dis\sqrt
{d_\Omega(\bF_{\nu_1})d_\Omega(\bF_{\nu_2})}} \r.\\ \\ 
\times \l. U(f_m^{(g)} \overline{f_m^{(g)}} f_m^{(g)} f_m^{(g)}; \bF_{\nu_1} 
\bF_{\nu_2})\; \cq^{\nu_1}(f_a:m,f_m^{(g)})\; 
\cq^{\nu_2}(f_b:m,f_m^{(g)}) \r\}
\;.
\earr \label{ch4.eq.gam-6}
\ee 
The $\cq^\nu(f_2:m,f_m)$ in Eq. (\ref{ch4.eq.gam-6}) are defined by   Eq.
(\ref{ch4.eq.35}). They can be calculated  using $X_{UU}$ given in Table
\ref{tab1}. Therefore the only unknown in Eq. (\ref{ch4.eq.gam-6}) is the
$SU(\Omega)$ Racah coefficient $U(f_m^{(g)} \overline{f_m^{(g)}} f_m^{(g)}
f_m^{(g)};  \bF_{\nu_1} \bF_{\nu_2})$. There are many attempts in the past
to derive analytical formulation and  also to develop numerical methods for
evaluating general $SU(N)$ Racah coefficients 
\cite{BL-68,LB-70,Lo-70,Bl-87,BS-82,Se-88,Vi-95}. 
There are also attempts to derive analytical formulas for some simple class
of Racah coefficients; see  \cite{Vi-95,Li-90} and references therein.  In
addition,  there is a recent effort to develop a new numerical method for 
evaluating  $SU(N)$ Racah coefficients with multiplicities  \cite{Kl-05,Kl-09}.
From all the attempts we made in trying to use these results, we conclude
that further group theoretical work on $SU(N)$  Racah coefficients is needed
to be able to  derive analytical formulas for, or for evaluating
numerically,  the Racah coefficients appearing in  Eq. (\ref{ch4.eq.gam-6}).

\chapter{Further extensions of BEGOE(1+2)}
\label{c6a1}

\renewcommand{\theequation}{G\arabic{equation}}
\setcounter{equation}{0}   

For completeness, we briefly outline here extension of BEGOE(1+2) to
BEGOE(1+2)-$M_S$ and  BEGOE(1+2)-$p$; here $p$ corresponds to spin $\cs=1$
bosons and $M_S$ is the $S_z$ quantum number for spin $\cs =\spin$ bosons. We
restrict our discussion to the definition and construction of these  ensembles
using the results for spinless BEGOE(1+2) discussed in Chapter \ref{ch1}.  

\subsubsection{BEGOE(1+2)-$M_S$}

Consider a system of $m$ bosons occupying $\Omega$ number of sp orbitals each
with spin  $\cs=\spin$ so that the number of sp states $N=2\Omega$. The sp
states are denoted by $\l| \l. \nu_i, m_\cs \ran \r.$, $i=1,2,\ldots,\Omega$ and
$m_\cs = \pm \spin$. The average spacing between the $\nu_i$ states is assumed
to be $\Delta$ and between two $m_\cs$ states for a given $\nu_i$ to be  
$\Delta_{m_\cs}$. For constructing the $H$ matrix in good $M_S$ representation,
we arrange the  sp states $\l.\l| i,m_\cs=\pm \spin\r.\ran$ in such a way that
the first $\Omega$ states have $m_\cs=\spin$ and the remaining $\Omega$ states
have $m_\cs=-\spin$. Many-particle states for $m$ bosons in the $2\Omega$ sp
states, arranged as explained above, can be obtained by distributing $m_1$
bosons in the $m_\cs = \spin$ sp states ($\Omega$ in number)  and similarly,
$m_2$ fermions in the $m_\cs = -\spin$ sp states ($\Omega$ in number) with
$m=m_1+m_2$.  Thus, $M_S = (m_1 -m_2)/2$. Let us denote each distribution of
$m_1$ fermions in $m_\cs = \spin$ sp states by $\bf{m}_1$ and similarly, 
$\bf{m}_2$ for $m_2$ fermions in  $m_\cs = -\spin$ sp states. Many-particle
basis defined by $(\bf{m}_1, \bf{m}_2)$ with $m_1-m_2=2M_S$ will form the basis
for BEGOE(1+2)-$M_S$. As the two-particle $m_s$ can take values $\pm 1$ and $0$,
the two-body part of the Hamiltonian preserving $M_S$
will be $\wv(2) = \lambda_0 \wv^{m_s=0}(2) + \lambda_1 \wv^{m_s=1}(2)
+ \lambda_{-1} \wv^{m_s=-1}(2)$ with the corresponding two-particle matrix being
a direct sum matrix generated by $\wv^{m_s}(2)$.
Therefore, the Hamiltonian is
\be
\wh = \whh(1) +  \lambda_0 \l\{ \wv^{m_s=0}(2) \r\} + 
\lambda_1 \l\{ \wv^{m_s=1}(2) \r\} + \lambda_{-1} \l\{ \wv^{m_s=-1}(2) \r\} \;.
\label{eq.c6a1e1}
\ee
In Eq. (\ref{eq.c6a1e1}), the $\{\wv^{m_s}(2)\}$  ensembles in two-particle
spaces are represented  by independent GOE(1)'s [see Eq. (\ref{eq.bpp3})] and
$\lambda_{m_s}$'s are their corresponding strengths.  The action of the
Hamiltonian operator defined by Eq. (\ref{eq.c6a1e1}) on the $(\bf{m}_1,
\bf{m}_2)$ basis states with a given $M_S$ generates the BEGOE(1+2)-$M_S$
ensemble  in $m$-particle spaces. Therefore, BEGOE(1+2)-$M_S$ is defined by six
parameters $(\Omega, m, \Delta_{m_\cs}, \lambda_0, \lambda_1, \lambda_{-1})$ 
[we put $\Delta = 1$ so that $\Delta_{m_\cs}$ and $\lambda_{m_s}$'s are in the
units of $\Delta$]. In the $(\bf{m}_1, \bf{m}_2)$ basis with a given $M_S$, the
$H$ matrix construction reduces to the matrix construction for spinless  boson
systems; see Chapter \ref{ch1}. The $H$ matrix dimension for a given $M_S$ is 
$\sum_{S \geq M_S} d_b(\Omega,m,S)$. Finally, pairing can also
be introduced in this ensemble using the algebra $U(2\Omega) \supset 
SO(2\Omega) \supset SO(\Omega) \otimes SO(2)$ with $SO(2)$ generating $M_S$; 
see \cite{Ko-06c}.

\subsubsection{BEGOE(1+2)-$p$}

Let us begin with a system of $m$ bosons  distributed say in $\Omega$ number of
sp orbitals each with spin  $\cs=1$ so that the number of sp states $N=3\Omega$.
The sp states are denoted by $\l.\l| i,m_\cs \r.\ran$ with $m_\cs = 0,\pm1$
and  $i=1,2,\ldots,\Omega$.  For a one plus two-body Hamiltonians preserving
$m$-particle spin $S$, the one-body Hamiltonian $\whh(1)$ is defined by the sp
energies $\epsilon_i$; $i=1,2,\ldots,\Omega$,   with average spacing $\Delta$. 
Similarly the two-body Hamiltonian $\wv(2)$ is defined by the two-body matrix
elements $\lambda_s\,V^s_{ijkl}=\lan (kl)s,m_s \mid \wv(2) \mid
(ij)s,m_s\ran$  with the two-particle spins $s=0,1$ and $2$. These matrix
elements are independent of  the $m_s$ quantum number.  Note that the
$\lambda_s$ are parameters. For generating the many-particle states, firstly, the
sp states are arranged such that the first $\Omega$ number of sp states have
$m_\cs = 1$,  next $\Omega$ number of sp states have $m_\cs = 0$ and the
remaining $\Omega$ sp states have $m_\cs = -1$. Now, the many-particle states
for $m$ bosons can be obtained by distributing $m_1$ bosons in the $m_\cs = 1$
sp states,  $m_2$ bosons in the $m_\cs = 0$ sp states and similarly, $m_3$
bosons in the $m_\cs = -1$ sp states with $m=m_1+m_2+m_3$.  Thus, $M_S = (m_1
-m_3)$. Let us denote each distribution of $m_1$ bosons in $m_\cs = 1$ sp states
by $\bf{m}_1$, $m_2$ bosons in $m_\cs = 0$ sp states by $\bf{m}_2$ and
similarly,  $\bf{m}_3$ for $m_3$ bosons in  $m_\cs = -1$ sp states.
Many-particle basis defined by $(\bf{m}_1, \bf{m}_2, \bf{m}_3)$  will form a
basis for BEGOE(1+2)-$p$. The $V$ matrix in two-particle spaces will be a direct
sum matrix and the $V(2)$ operator is 
$\wv(2)=\lambda_0 \wv^{s=0}(2) + \lambda_1 \wv^{s=1}(2) +
\lambda_2 \wv^{s=2}(2)$ with three parameters $(\lambda_0,\lambda_1,\lambda_2)$.
Now, BEGOE(1+2)-$p$ for a given $(m,S)$ system is
generated  by defining the three parts of the two-body Hamiltonian to be
independent GOE(1)'s in two-particle spaces and then propagating the $V(2)$
ensemble $\{\wv(2)\}=\lambda_0 \{\wv^{s=0}(2)\} + \lambda_1 \{\wv^{s=1}(2)\} +
\lambda_2 \{\wv^{s=2}(2)\}$ to the $m$-particle spaces with a given spin $S$ by
using the geometry (direct product structure) of  the $m$-particle spaces. The
embedding algebra is  $U(3\Omega) \supset G \supset G1 \otimes SO(3)$ with SO(3)
generating spin S. Thus BEGOE(1+2)-$p$ is defined by the operator 
\be 
\wh =
\whh(1) + \lambda_0\, \{\wv^{s=0}(2)\} + \lambda_1\, \{\wv^{s=1}(2)\} +
\lambda_2\, \{\wv^{s=2}(2)\}\;. 
\label{eq.c6a1e2} 
\ee 
The sp levels defined by $\whh(1)$ will be triply degenerate. The action of the 
Hamiltonian operator defined by Eq. (\ref{eq.c6a1e2}) on  $(\bf{m}_1, \bf{m}_2,
\bf{m}_3)$ basis states with fixed-($m,M_S=M_S^{min}$)  generates the
ensemble in ($m,M_S$) spaces. It is important to note that the
construction of the $m$-particle $H$ matrix in fixed-($m,M_S=M_S^{min}$) spaces
reduces to the  problem of BEGOE(1+2) for spinless boson systems and hence Eqs.
(\ref{eq.app1a})- (\ref{eq.app2}) of Chapter \ref{ch1}  will apply. Then the
${\hat S}^2$ operator is used for projecting states with good $S$. Therefore,
BEGOE(1+2)-$p$ ensemble is defined by five parameters $(\Omega, m, \lambda_0,
\lambda_1, \lambda_2)$ with $\lambda_s$ in units of $\Delta$.   Finally, it is
important to mention that it is also possible to study the pairing symmetry in
the space defined by BEGOE(1+2)-$p$ ensemble. For this, there are two possible
algebras (each defining a particular type of pairing), $U(3\Omega) \supset
[U(\Omega) \supset SO(\Omega)] \otimes [U(3) \supset SO(3)]$ and $U(3\Omega)
\supset SO(3\Omega)  \supset SO(\Omega) \otimes SO(3)$ and they can be studied
in detail by extending the results for IBM-3 model in nuclear structure where
$\Omega = 6$  \cite{Ga-99,Ko-96}.  Exploiting the group chain $U(3\Omega) \supset
U(\Omega)  \otimes [U(3) \supset SO(3)]$, it is possible to write the dimension
formulas for the $H$ matrices for a given $(m,S)$ as it was done in Sec.
\ref{gsirp} for $SU(4)-ST$ reductions. 

\chapter*{}

\thispagestyle{empty}

\chapter{Basic binary correlation results}
\label{c7s1}

\renewcommand{\theequation}{H\arabic{equation}}
\setcounter{equation}{0}   

We denote a $k_H$-body operator as,
\be
H(k_H) = \dis\sum_{\alpha,\; \beta} v_H^{\alpha\beta}\; \alpha^\dg(k_H)
\beta(k_H) \;.
\label{eq.b1}
\ee
Here, $\alpha^\dg(k_H)$ is the $k_H$ particle creation operator and
$\beta(k_H)$ is the $k_H$ particle annihilation operator. Similarly,
$v_H^{\alpha\beta}$ are matrix elements of the operator $H$ in $k_H$
particle space i.e., $v_H^{\alpha\beta} = \lan k_H \beta \mid H \mid k_H 
\alpha \ran$ (it should be noted that Mon and French \cite{Mo-73,MF-75} 
used operators with
daggers to denote annihilation operators and operators without daggers to
denote creation operators). Following basic traces will be used throughout,
\be
\dis\sum_\alpha \alpha^\dg(k) \alpha(k)  = \dis\binom{\hat{n}}{k} \;\;\;\;
\Rightarrow \lan \dis\sum_\alpha \alpha^\dg(k) \alpha(k) \ran^m =  
\dis\binom{m}{k} \;.
\label{eq.b2}
\ee
\be
\dis\sum_\alpha \alpha(k) \alpha^\dg(k) = \dis\binom{N - \hat{n}}{k} 
\;\;\;\;
\Rightarrow \lan \dis\sum_\alpha \alpha(k) \alpha^\dg(k) \ran^m =  
\dis\binom{\wm}{k} \;; \;\;\;\; \wm = N-m \;.
\label{eq.b3}
\ee
\be
\barr{l}
\dis\sum_\alpha \alpha^\dg(k) B(k^\pr) \alpha(k) = 
\dis\binom{\hat{n} - k^\pr}{k} B(k^\pr) \\ \\
\Rightarrow \lan \dis\sum_\alpha \alpha^\dg(k) B(k^\pr) \alpha(k) \ran^m =  
\dis\binom{m-k^\pr}{k} B(k^\pr) \;.
\earr \label{eq.b4}
\ee
\be
\barr{l}
\dis\sum_\alpha \alpha(k) B(k^\pr) \alpha^\dg(k) = 
\dis\binom{N - \hat{n} - k^\pr}{k} B(k^\pr) \\ \\
\Rightarrow \lan \dis\sum_\alpha \alpha(k) B(k^\pr) \alpha^\dg(k) \ran^m =  
\dis\binom{\wm-k^\pr}{k} B(k^\pr) \;.
\earr \label{eq.b5}
\ee
Equation (\ref{eq.b2}) follows from the fact that the average should be zero
for $m < k$ and one for $m=k$ and similarly, Eq. (\ref{eq.b3})  follows from
the same argument except that the particles are replaced by holes. Equation
(\ref{eq.b4}) follows first by writing the $k^\pr$-body operator $B(k^\pr)$
in operator form using Eq. (\ref{eq.b1}), i.e.,
\be
B(k^\pr) = \dis\sum_{\beta,\; \gamma} v_B^{\beta\gamma}\; \beta^\dg(k^\pr)
\gamma(k^\pr) \;,
\label{eq.b5a}
\ee
and then applying the commutation relations for the fermion creation and
annihilation operators. This gives $\sum_{\beta,\gamma} v_B^{\beta\gamma}\;
\beta^\dg(k^\pr)  \sum_\alpha \alpha^\dg(k) \alpha(k) \gamma(k^\pr)$. Now
applying Eq. (\ref{eq.b2}) to the sum involving $\alpha$ gives Eq.
(\ref{eq.b4}). Eq. (\ref{eq.b5}) follows from the same arguments except one has
to assume that $B(k^\pr)$ is fully irreducible $\nu = k^\pr$ operator and
therefore, it has particle-hole symmetry. For  a general $B(k^\pr)$ operator,
this is valid only in the $N \to \infty$ limit. Therefore, this equation has to
be  applied with caution.

Using the definition of the $H$ operator in Eq. (\ref{eq.b1}), 
we have
\be
\barr{rcl}
\overline{\lan H(k_H) H(k_H) \ran^m} & = & \dis\sum_{\alpha,\;\beta}
\overline{\l\{v_H^{\alpha\beta}\r\}^2} \; \lan \alpha^\dg(k_H) \beta(k_H)
\beta^\dg(k_H) \alpha(k_H) \ran^m \\ \\
& = & v_H^2 \; \lan \dis\sum_\alpha \alpha^\dg(k_H) \l\{ 
\dis\sum_\beta \beta(k_H) \beta^\dg(k_H) \r\} \alpha(k_H) \ran^m
\\ \\
& = & v_H^2 \; T(m,N,k_H) \;.
\earr \label{eq.b6a1}
\ee
Here, $H$ is taken as EGOE$(k_H)$ with all the $k_H$ particle matrix elements
being Gaussian variables with zero center and same variance for off-diagonal
matrix elements (twice for the diagonal matrix elements). This gives 
$\overline{(v_H^{\alpha \beta})^2} = v_H^2$ to be independent of  $\alpha, \;
\beta$ labels. It is important to note that in the dilute limit, the diagonal
terms [$\alpha =\beta$ in Eq. (\ref{eq.b6a1})] in the averages are neglected
(as they are smaller by at least one power of $1/N$)  and the individual
$H$'s are unitarily irreducible. These assumptions are no longer valid for
finite-$N$ systems and hence, evaluation of averages is more complicated. In
the dilute limit, we have
\be
\barr{rcl}
T(m,N,k_H) & = &  \lan \dis\sum_\alpha \alpha^\dg(k_H) \l\{ 
\dis\sum_\beta \beta(k_H) \beta^\dg(k_H) \r\} \alpha(k_H) \ran^m
\\ \\
& = &  \dis\binom{\wm+k_H}{k_H} \; \lan \dis\sum_\alpha \alpha^\dg(k_H)
\alpha(k_H) \ran^m \\ \\ 
& = &  \dis\binom{\wm+k_H}{k_H} \; \dis\binom{m}{k_H} \;.
\earr \label{eq.b6a}
\ee
Note that, we have used Eq. (\ref{eq.b3}) to evaluate the summation over
$\beta$ and Eq. (\ref{eq.b2}) to evaluate summation over $\alpha$ in Eq.
(\ref{eq.b6a}). In the `strict' $N \to \infty$ limit, we have
\be
\barr{l} 
T(m,N,k_H) \stackrel{N \to \infty}{\to} \dis\binom{m}{k_H} \; 
\dis\binom{N}{k_H} \;.
\earr \label{eq.b6b}
\ee
In order to incorporate the finite-$N$ corrections, we have to consider the
contribution of the diagonal terms. Then, we have,
\be
\barr{l}
T(m,N,k_H) \\ \\
=  \lan \dis\sum_{\alpha \neq \beta} \alpha^\dg(k_H)
\beta(k_H) \beta^\dg(k_H) \alpha(k_H) \ran^m
+ 2 \lan \dis\sum_\alpha \alpha^\dg(k_H) \alpha(k_H) \alpha^\dg(k_H) 
\alpha(k_H) \ran^m
 \\ \\
= \lan \dis\sum_\alpha \alpha^\dg(k_H) \l\{ 
\dis\sum_\beta \beta(k_H) \beta^\dg(k_H) \r\} \alpha(k_H) \ran^m 
\\ \\
+ \lan \dis\sum_\alpha \alpha^\dg(k_H) \alpha(k_H) \alpha^\dg(k_H) \alpha(k_H)
\ran^m
\\ \\
= \dis\binom{\wm+k_H}{k_H} \; \dis\binom{m}{k_H} + \dis\binom{m}{k_H}
= \dis\binom{m}{k_H}\l[ \dis\binom{\wm+k_H}{k_H} + 1 \r]\;.
\earr \label{eq.5b}
\ee
Note that the prefactor `2' in the second term of first line in  Eq.
(\ref{eq.5b}) comes because variance of the diagonal terms is twice that of
the  off-diagonal terms. Also, the trace $\sum_\alpha \alpha^\dg(k_H)
\alpha(k_H) \alpha^\dg(k_H)  \alpha(k_H) = \sum_\alpha \alpha^\dg(k_H)
\alpha(k_H)$ as the operator $\alpha^\dg(k_H)  \alpha(k_H)$ conserves the
number of particles. Now we turn to evaluating fourth order averages.

For averages involving product of four operators of the form
$$
\lan H(k_H) G(k_G) H(k_H) G(k_G) \ran^m \;,
$$ 
with operators $H$ and $G$ independent and of body ranks $k_H$ and $k_G$
respectively, there are two possible ways of
evaluating this trace. Either (a) first contract the $H$ operators across the
$G$ operator using Eq. (\ref{eq.b5}) and  then contract the $G$ operators 
using Eq. (\ref{eq.b4}),  or (b) first contract the $G$ operators across the
$H$ operator  using Eq. (\ref{eq.b5}) and then contract the $H$ operators
using Eq. (\ref{eq.b5}). Following (a), in the dilute limit, we get
\be
\barr{l}
\overline{\lan H(k_H) G(k_G) H(k_H) G(k_G)\ran^m} \\ \\
=
\dis\sum_{\alpha,\beta} \overline{\l\{v_H^{\alpha\beta}\r\}^2} \;
\lan \alpha^\dg(k_H) \beta(k_H) G(k_G) \beta^\dg(k_H)
\alpha(k_H) G(k_G) \ran^m \\ \\
= v_H^2 \; \dis\binom{\wm+k_H-k_G}{k_H} \; \dis\sum_\alpha
\lan \alpha^\dg(k_H) G(k_G) \alpha(k_H) G(k_G) \ran^m \\ \\
= v_H^2 \; \dis\binom{\wm+k_H-k_G}{k_H} \; \dis\binom{m-k_G}{k_H} \;
\lan G(k_G) G(k_G) \ran^m \\ \\
= v_H^2 \; v_G^2 \; \dis\binom{\wm+k_H-k_G}{k_H} \; \dis\binom{m-k_G}{k_H}
\; \dis\binom{\wm+k_G}{k_G} \; \dis\binom{m}{k_G} \;.
\earr \label{eq.b7a}
\ee
Similarly, following (b), in the dilute limit, we get
\be
\barr{l}
\overline{\lan H(k_H) G(k_G) H(k_H) G(k_G) \ran^m} \\ \\
=
v_H^2 \; v_G^2 \; \dis\binom{\wm+k_G-k_H}{k_G} \; \dis\binom{m-k_H}{k_G}
\; \dis\binom{\wm+k_H}{k_H} \; \dis\binom{m}{k_H} \;.
\earr \label{eq.b7b}
\ee
The result should be independent of the preference. In other words, the
average should have the $k_H \leftrightarrow k_G$ symmetry. As seen from Eqs.
(\ref{eq.b7a}) and (\ref{eq.b7b}), this symmetry is violated except for the
trivial case of $k_H = k_G$. However, the
$k_H \leftrightarrow k_G$ symmetry is valid for `strict'  $N \to \infty$
result and also for the result incorporating finite $N$ corrections as
discussed below. In general, the final result can be expressed as,
\be
\overline{\lan H(k_H) G(k_G) H(k_H) G(k_G) \ran^m} = 
v_H^2 \; v_G^2 \; F(m,N,k_H,k_G)\;.
\label{eq.b7}
\ee
In the `strict' dilute  limit, both Eqs. (\ref{eq.b7a}) and (\ref{eq.b7b})
reduce to give result for $F(m,N,k_H,k_G)$,
\be
F(m,N,k_H,k_G) =  \dis\binom{m-k_H}{k_G} \; \dis\binom{m}{k_H} \; 
\dis\binom{N}{k_H} \; \dis\binom{N}{k_G}\;,
\label{eq.b8}
\ee
In order to obtain finite-$N$ corrections to $F(\cdots)$, we have to contract
over operators whose lower symmetry parts can not be ignored. The operator
$H(k_H)$ contains irreducible symmetry parts $\cf(s)$ denoted by 
$s=0,1,2,\ldots,k_H$
with respect to the unitary group $SU(N)$ decomposition of the operator. For
a $k_H$-body number conserving operator \cite{Ch-71,MF-75},
\be
H(k_H) = \dis\sum_{s=0}^{k_H} \; \dis\binom{m-s}{k_H-s} \; \cf(s) \;.
\label{eq.b8a}
\ee 
Here, the $\cf(s)$ are orthogonal with respect to $m$-particle averages, 
i.e., $\lan \cf(s) \cf^\dg(s^\pr) \ran^m = \delta_{ss^\pr}$. Now,
the $m$-particle trace in Eq. (\ref{eq.b7a}) 
with binary correlations  will have four parts,
\be
\barr{l}
\overline{\lan  H(k_H) G(k_G) H(k_H) G(k_G) \ran^m} \\ \\
= v_H^2 v_G^2
\dis\sum_{\alpha,\beta,\gamma,\delta} \lan \alpha^\dg(k_H) \beta(k_H)
\gamma^\dg(k_G)
\delta(k_G)  \beta^\dg(k_H) \alpha(k_H) \delta^\dg(k_G) \gamma(k_G) \ran^m 
\\ \\
+ v_H^2 v_G^2
\dis\sum_{\alpha,\gamma,\delta} \lan \alpha^\dg(k_H) 
\alpha(k_H) \gamma^\dg(k_G)
\delta(k_G)  \alpha^\dg(k_H) \alpha(k_H) \delta^\dg(k_G) \gamma(k_G) \ran^m 
\earr \label{eq.2}
\ee
\be
\barr{l}
+ v_H^2 v_G^2
\dis\sum_{\alpha,\beta,\gamma} \lan \alpha^\dg(k_H) 
\beta(k_H) \gamma^\dg(k_G)
\gamma(k_H)  \beta^\dg(k_H) \alpha(k_H) \gamma^\dg(k_G) \gamma(k_G) \ran^m
\\ \\
+ v_H^2 v_G^2
\dis\sum_{\alpha,\delta} \lan \alpha^\dg(k_H) \alpha(k_H) \delta^\dg(k_G)
\delta(k_G)  \alpha^\dg(k_H) \alpha(k_H) \delta^\dg(k_G) \delta(k_G) \ran^m 
\\ \\
= X + Y_1 + Y_2 + Z \;. \nonumber
\earr \label{eq.2-1}
\ee
Note that we have decomposed each operator into diagonal and off-diagonal parts.
We have used the condition that the variance of the diagonal matrix elements is 
twice that of the off-diagonal matrix elements in the defining spaces to convert
the restricted  summations into unrestricted summations appropriately to obtain
the four terms in the  RHS of Eq. (\ref{eq.2-1}).  Following Mon's thesis
\cite{Mo-73}  and applying unitary decomposition to $\gamma \delta^\dagger$
(also $\delta \gamma^\dagger$) in the first two terms and $\alpha \beta^\dagger$
(also $\beta \alpha^\dagger$) in the  third term we get $X$, $Y_1$ and $Y_2$. To
make things clear, we will discuss the derivation for $X$ term in detail before
proceeding further. Applying unitary decomposition to the operators
$\gamma^\dg(k_G)\delta(k_G)$ and  $\gamma(k_G)\delta^\dg(k_G)$ using Eq.
(\ref{eq.b8a}), we have
\be
X = v_H^2 \; v_G^2 \; \dis\sum_{\alpha,\beta,\gamma,\delta} \;
\dis\sum_{s=0}^{k_G} \; \dis\binom{m-s}{k_G-s}^2 \; 
\lan \alpha^\dg(k_H) \beta(k_H) \cf^\dg_{\gamma\delta}(s) 
\beta^\dg(k_H) \alpha(k_H) \cf_{\gamma\delta}(s) \ran^m \;.
\label{eq.b8ba}
\ee
Contracting the operators $\beta\beta^\dg$ across $\cf$'s using Eq.
(\ref{eq.b5}) and operators $\alpha^\dg\alpha$ across $\cf$ using Eq.
(\ref{eq.b4}) gives,
\be
X
= v_H^2 \; v_G^2 \; 
\dis\sum_{s=0}^{k_G} \; \dis\binom{m-s}{k_G-s}^2 \; 
\dis\binom{\wm+k_H-s}{k_H}
\; \dis\binom{m-s}{k_H} \; \dis\sum_{\gamma,\delta} \;
\lan \cf^\dg_{\gamma\delta}(s) \cf_{\gamma\delta}(s) \ran^m \;.
\label{eq.b8b}
\ee
Inversion of the equation,
\be
\dis\sum_{\gamma,\delta} 
\lan \gamma^\dg(k_G) \delta(k_G) \delta^\dg(k_G) \gamma(k_G) \ran^m 
= Q(m) = \dis\sum_{s=0}^{k_G} 
\dis\binom{m-s}{k_G-s}^2 \; \dis\sum_{\gamma,\delta} 
\lan \cf^\dg_{\gamma\delta}(s) \cf_{\gamma\delta}(s) \ran^m \;,
\label{eq.b8c}
\ee
gives,
\be
\barr{l}
\dis\binom{m-s}{k_G-s}^2 \; \dis\sum_{\gamma,\delta} 
\lan \cf^\dg_{\gamma\delta}(s) \cf_{\gamma\delta}(s) \ran^m \\ \\
= 
\dis\binom{m-s}{k_G-s}^2 \; \dis\binom{N-m}{s} \; \dis\binom{m}{s}\;
\l[ (k_G-s)! s! \r]^2 (N-2s+1) \\\\
\times \dis\sum_{t=0}^s \dis\frac{(-1)^{t-s}\l[ (N-t-k_G)!\r]^2}
{(s-t)! (N-s-t+1)! t! (N-t)!} Q(N-t) \;.
\earr \label{eq.b8d}
\ee
It is important to mention that there are errors in the equation given  in
Mon's thesis and we have verified Eq. (\ref{eq.b8d}) using Mathematica (Mon =
Eq. (\ref{eq.b8d})/$[(N-2s)! (s!)^2]$).  For the average required in Eq.
(\ref{eq.b8c}), we have
\be
Q(m) = \dis\sum_{\gamma,\delta} \lan \gamma^\dg(k_G) 
\delta(k_G) \delta^\dg(k_G)
\gamma(k_G) \ran^m = \dis\binom{\wm+k_G}{k_G} \; \dis\binom{m}{k_G} \;.
\label{eq.b8e}
\ee
Simplifying Eq. (\ref{eq.b8d}) using Eq. (\ref{eq.b8e}) and using the 
result in Eq. (\ref{eq.b8b}) along with the series summation
\be
\dis\sum_{t=0}^s \;\dis\frac{(-1)^{t-s} (N-t-k_G)! \; (k_G+t)!}
{(s-t)! \; (t!)^2 \; (N-s-t+1)!} = \dis\frac{k_G! (N-k_G-s)!}{(N+1-s)!} \;
\dis\binom{k_G}{s} \;\dis\binom{N+1}{s} \;,
\label{eq.ser-x}
\ee
the expression for $X$ is,
\be
\barr{rcl}
X & = & v_H^2 v_G^2 \; F(m,N,k_H,k_G) \;;  \\ \\
F(m,N,k_H,k_G) & = & \dis\sum_{s=0}^{k_G} \dis\binom{m-s}{k_G-s}^2
\dis\binom{\wm+k_H-s}{k_H} \dis\binom{m-s}{k_H} \dis\binom{\wm}{s}
\dis\binom{m}{s} \dis\binom{N+1}{s} \\ \\
& \times & \dis\frac{N-2s+1}{N-s+1}
\dis\binom{N-s}{k_G}^{-1} \dis\binom{k_G}{s}^{-1} \;.
\earr \label{eq.3}
\ee
Although not obvious, $X$ has $k_H \leftrightarrow k_G$ symmetry and we have
verified this explicitly for $k_H, k_G \leq 2$.  Similarly, 
the terms $Y_1$ and $Y_2$ are given by,
\be
\barr{rcl}
Y_1 & = & v_H^2 v_G^2 \; B(m,N,k_H,k_G) \;,\;\; \;\;
Y_2 = v_H^2 v_G^2 \; B(m,N,k_G,k_H) \;; \\ \\
B(m,N,k_H,k_G) & = & \dis\sum_{s=0}^{k_G} \dis\binom{m-s}{k_G-s}^2
\dis\binom{\wm+k_H-s}{k_H} \dis\binom{m-s}{k_H} \dis\binom{\wm}{s}
\dis\binom{m}{s} \\ \\
& \times &
\dis\frac{N-2s+1}{N-s+1}
\dis\binom{N-s}{k_G}^{-1} \dis\binom{k_G}{s}^{-1} \;.
\earr \label{eq.4}
\ee
In order to derive Eq.(\ref{eq.4}), we have used $Q(m)=\binom{m}{k_G}$ along with
the series summation,
\be
\dis\sum_{t=0}^s \;\dis\frac{(-1)^{t-s} (N-t-k_G)! \; k_G! \; t!}
{(s-t)! \; (t!)^2 \; (N-s-t+1)!} = \dis\frac{k_G! (N-k_G-s)!}{(N+1-s)!} \;
\dis\binom{k_G}{s} \;.
\label{eq.ser-y}
\ee
Note that Mon's thesis gives $\binom{m-s}{s}$ in place of $\binom{m-s}{k}$
with $k=k_H$ or $k_G$ for $X$, $Y_1$ and $Y_2$ in Eqs. (\ref{eq.3}) and
(\ref{eq.4}) and it should be a printing error.  The expressions given in 
Eqs. (\ref{eq.3}) and (\ref{eq.4}) agree with the results given in
Tomsovic's thesis \cite{To-86}. Finally, the result for $Z$ is
\be
\barr{l}
Z = v_H^2 v_G^2 
\dis\sum_{\alpha,\delta} \lan \alpha^\dg(k_H) \alpha(k_H) \delta^\dg(k_G)
\delta(k_G)  \alpha^\dg(k_H) \alpha(k_H) \delta^\dg(k_G) \delta(k_G) 
\ran^m \\ \\
= v_H^2 v_G^2 
\dis\sum_{\alpha} \lan \alpha^\dg(k_H) \alpha(k_H) \ran^m 
\dis\sum_{\delta} \lan \delta^\dg(k_G) \delta(k_G) \ran^m \\ \\
= v_H^2 v_G^2 \; C(m,N,k_H,k_G) \;; \\ \\
C(m,N,k_H,k_G) = \dis\binom{m}{k_H} \dis\binom{m}{k_G} \;.
\earr \label{eq.5}
\ee
Equation (\ref{eq.5}) is in agreement with the result in Mon's thesis with
$k_H=k_G=k$. However, it differs from the result given in Tomsovic's thesis.
For a one-body operator, obviously $Z=m^2$ and this confirms that Eq.
(\ref{eq.5}) is correct. Therefore Eqs. (\ref{eq.2})-(\ref{eq.5}) give the
final formula for the trace $\overline{\lan H(k_H) G(k_G) H(k_H) G(k_G)
\ran^{m}}$. It is easily seen that dominant contribution to the average 
$\overline{\lan  H(k_H) G(k_G) H(k_H) G(k_G) \ran^m}$ comes from the $X$ term
and therefore, in all the applications, we use 
\be
\overline{\lan  H(k_H) G(k_G) H(k_H) G(k_G) \ran^m} = X = 
v_H^2 \; v_G^2 \;  F(m,N,k_H,k_G) \;.
\label{eq.6}
\ee

An immediate application of these averages is in evaluating the fourth order
average $\lan H^4(k_H) \ran^m$. There will be three different correlation
patterns that will contribute to this average in the binary correlation
approximation (we must correlate in pairs the operators for all moments 
of order $> 2$),
\be
\barr{rcl}
\overline{\lan H^4(k_H) \ran^m} & = & 
\overline{\lan \crh \crh \cbh \cbh \ran^m} \\ \\
& + &
\overline{\lan \crh \cbh \cbh \crh \ran^m}
\\ \\ & + &
\overline{\lan \crh \cbh \crh \cbh \ran^m} 
\;.
\earr \label{eq.b10a}
\ee  
In Eq. (\ref{eq.b10a}), we denote the correlated pairs of operators by same
color in each pattern. The first two terms on the RHS of Eq. (\ref{eq.b10a})
are 
equal due to cyclic invariance and follow easily from Eq. (\ref{eq.b6a1}),
\be
\barr{rcl}
\overline{\lan \crh \crh \cbh \cbh \ran^m} & = &
\overline{\lan \crh \cbh \cbh \crh \ran^m} \\ \\
& = &
\l[ \; \overline{\lan H^2(k_H) \ran^m} \; \r]^2\;.
\earr \label{eq.b10b}
\ee
Similarly, the third term on the RHS of Eq. (\ref{eq.b10a}) follows easily 
from Eq. (\ref{eq.6}),
\be
\overline{\lan \crh \cbh \crh \cbh \ran^m} = 
v_H^4 \; F(m,N,k_H,k_H) \;. 
\label{eq.b10c}
\ee
Finally, $\overline{\lan H^4(k_H) \ran^m}$ is given by,
\be
\overline{\lan H^4(k_H) \ran^m} =  v_H^4 \; \l[ 2 \; \l\{T(m,N,k_H)\r\}^2 + 
F(m,N,k_H,k_H) \r]\;.
\label{eq.b10}
\ee
Simplifying $T(\cdots)$ and $F(\cdots)$ in `strict' $N \to\infty$ limit and
using Eqs. (\ref{eq.b6a1}) and (\ref{eq.b10}), the
excess parameter for spinless EGOE($k_H$) is,
\be
\gamma_2(m) = \dis\frac{\overline{\lan H^4(k_H) \ran^m}}
{\l[ \overline{\lan H^2(k_H) \ran^m} \r]^2} - 3 = 
\dis\frac{\dis\binom{m-k_H}{k_H}}{\dis\binom{m}{k_H}} - 1 
\stackrel{m >> k_H}{\to} -\dis\frac{k_H^2}{m} \;.
\label{eq.b10d}
\ee
Equation (\ref{eq.b10d}) was first derived in \cite{MF-75}. As seen from Eq.
(\ref{eq.b10d}),  state densities for spinless EGOE$(k_H)$ approach Gaussian
form for large $m$ and they exhibit, as $m$ increases from $k_H$, semicircle to
Gaussian transition with $m=2k_H$ being the transition point. The results for
$\overline{\lan H^2(k_H) \ran^m}$ and  $\overline{\lan H^4(k_H) \ran^m}$ easily
extend, though not obvious,  to averages over two-orbit spaces with operator $H$
having fixed body ranks in the two spaces. It is useful to mention that the
details for the two-orbit averages using binary correlation approximation are
not available in literature. Now, we will discuss the two-orbit results.

In many nuclear structure applications and also for applications to
interacting spin systems, fourth order traces over two orbit configurations
are needed.  Let us consider $m$ particles in two orbits with number of sp
states being $N_1$ and $N_2$ respectively. Now the $m$-particle space can be
divided into configurations $(m_1,m_2)$ with $m_1$ particles in the \#1 orbit
and $m_2$ particles in the \#2 orbit such that $m = m_1 + m_2$. Considering
the operator $H$ with fixed body ranks in $m_1$ and $m_2$ spaces such that
$(m_1,m_2)$ are preserved by this operators, the general form for $H$ is,
\be
H(k_H) = \dis\sum_{i+j=k_H\;;\alpha,\beta,\gamma,\delta} 
\l[ v_H^{\alpha\beta\gamma\delta}(i,j) \r] \; \alpha_1^\dg(i) 
\beta_1(i) \gamma_2^\dg(j)\delta_2(j) \;.
\label{eq.b11}
\ee
Now, it is easily seen that, in the dilute limit, 
\be
\barr{l}
\overline{\lan H^2(k_H) \ran^{m_1,m_2}} \\ \\
= 
\dis\sum_{i+j=k_H} v_H^2(i,j)\;
\dis\sum_{\alpha,\beta,\gamma,\delta} \lan \alpha_1^\dg(i) 
\beta_1(i) \gamma_2^\dg(j)\delta_2(j) \beta_1^\dg(i) 
\alpha_1(i) \delta_2^\dg(j)\gamma_2(j) \ran^{m_1,m_2}
\\ \\
=
\dis\sum_{i+j=k_H} v_H^2(i,j)\;
\dis\sum_{\alpha,\beta} \lan \alpha_1^\dg(i) 
\beta_1(i)  \beta_1^\dg(i) \alpha_1(i) \ran^{m_1} \dis\sum_{\gamma,\delta}
\lan \gamma_2^\dg(j)\delta_2(j) \delta_2^\dg(j)\gamma_2(j) \ran^{m_2} 
\\  \\
= \dis\sum_{i+j=k_H} v_H^2(i,j)\;
T(m_1,N_1,i)\;T(m_2,N_2,j) \;.
\earr \label{eq.b12}
\ee
Note that $v_H^2(i,j) = \overline{[v_H^{\alpha\beta\gamma\delta} (i,j)]^2}$
and $T$'s are defined by Eqs. (\ref{eq.b6a}) and (\ref{eq.b6b}). The ensemble
is defined such that $v_H^{\alpha\beta\gamma\delta}(i,j)$ are independent
Gaussian random variables with zero center and the variances depend only on
the indices $i$ and $j$. The formula for $\overline{\lan H(k_H)
H(k_H)  \ran^{m_1,m_2}}$ with finite $(N_1,N_2)$ corrections is,
\be
\overline{\lan H(k_H) H(k_H) \ran^{m_1,m_2}} = \dis\sum_{i+j=k_H} v_H^2(i,j)
\; \dis\binom{m_1}{i} \dis\binom{m_2}{j} 
\l[ \dis\binom{\wm_1+i}{i} \dis\binom{\wm_2+j}{j} + 1 \r]\;.  
\label{eq.8a}
\ee
Similarly, with two operators $H$ and $G$ (with body ranks $k_H$ and $k_G$
respectively) that are independent
and both preserving $(m_1,m_2)$, $\overline{\lan H(k_H) G(k_G) H(k_H) G(k_G)
\ran^{m_1,m_2}}$ is  given by,
\be
\barr{l}
\overline{\lan H(k_H) G(k_G) H(k_H) G(k_G) \ran^{m_1,m_2}} =  \\ \\
\dis\sum_{i+j=k_H,\; t+u=k_G} \; v_H^2(i,j) \; v_G^2(t,u) \;
F(m_1,N_1,i,t) \;  F(m_2,N_2,j,u) \;.
\earr \label{eq.b13}
\ee
Also, extending the single orbit results with finite-$N$ corrections, 
we have,
\be
\barr{l}
\overline{\lan  H(k_H) G(k_G) H(k_H) G(k_G) \ran^{m_1,m_2}} 
\\ \\
=
\dis\sum_{i+j=k_H,\;t+u=k_G} \; 
\dis\sum_{\alpha_1, \beta_1, \gamma_1, \delta_1, \alpha_2, \beta_2, 
\gamma_2,\delta_2} v_H^2(i,j)\;\;{v_G^2(t,u)} 
\earr \label{eq.6a}
\ee
\be
\barr{l}
\times
\lan \alpha_1^\dg(i) \beta_1(i) \gamma_1^\dg(t) \delta_1(t)
\beta_1^\dg(i)\alpha_1(i) \delta_1^\dg(t)  \gamma_1(t) \ran^{m_1} 
\\ \\
\times
\lan \alpha_2^\dg(j) \beta_2(j) \gamma_2^\dg(u) \delta_2(u)
\beta_2^\dg(j)\alpha_2(j) \delta_2^\dg(u)  \gamma_2(u) \ran^{m_2} 
\;.\nonumber
\earr \label{eq.6a-1}
\ee
Applying Eqs. (\ref{eq.2})-(\ref{eq.5}) to the two traces in Eq.
(\ref{eq.6a}), we get 
\be
\barr{l}
\overline{\lan  H(k_H) G(k_G) H(k_H) G(k_G) \ran^{m_1,m_2}} =
\dis\sum_{i+j=k_H,\;t+u=k_G} \;v_H^2(i,j) \; v_G^2(t,u) \\ \\
\times \l[ F(m_1,N_1,i,t) F(m_2,N_2,j,u) 
+ B(m_1,N_1,i,t) B(m_2,N_2,j,u) \r.
\\ \\
+ \l. B(m_1,N_1,t,i) B(m_2,N_2,u,j) 
+ C(m_1,N_1,i,t) C(m_2,N_2,j,u) \r]\;. 
\earr \label{eq.7}
\ee
The $F(\cdots)$'s appearing in Eq. (\ref{eq.7}) are given by Eq.
(\ref{eq.3}).  Also, the $B$'s  and $C$'s are given by Eqs. (\ref{eq.4}) and
(\ref{eq.5}) respectively. Finally, in the strict dilute limit 
as $F(\cdots)$'s dominate over
$B$'s and $C$'s, we get back Eq. (\ref{eq.b13}). In all the 
applications discussed
in Chapter \ref{ch7}, we use
Eq. (\ref{eq.b13}). Now, using Eqs. (\ref{eq.b12}) and (\ref{eq.b13}), we have
\be
\barr{l}
\overline{\lan H^4(k_H) \ran^{m_1,m_2}}
= 2\; 
\l[ \dis\sum_{i+j=k_H} v_H^2(i,j)\;
T(m_1,N_1,i)\;T(m_2,N_2,j) \r]^2 \\ \\ 
+ 
\dis\sum_{i+j=k_H,\; t+u=k_H} \; v_H^2(i,j) \; 
v_H^2(t,u) \; F(m_1,N_1,i,t) \;  F(m_2,N_2,j,u) \;.
\earr \label{eq.b14}
\ee

As a simple application of Eqs. (\ref{eq.b12})  and (\ref{eq.b14}), let us
consider $\gamma_2(m,M_S)$ for EGOE(2)-$M_S$ ensemble. For this ensembles, $H$
will preserve $M_S$ and it is defined for a system of $m$ fermions carrying spin
$\cs=\spin$ degree of freedom (see also Appendix \ref{c6a1}). 
Then, we have two orbits with   $N_1 = N_2 =
\Omega$, $m_1 = m/2 + M_S$ and $m_2 = m/2 - M_S$. Here, orbit \#1 corresponds to
sp states with $m_\cs =+\spin$ and  orbit \#2 corresponds to sp states with
$m_\cs =-\spin$. Note that the fixed-$M_S$ dimension is  $D(m,M_S) =
\binom{\Omega}{m/2-M_S} \binom{\Omega}{m/2+M_S}$. By substituting $m_1 = m/2 +
M_S$ and $m_2 = m/2 - M_S$, Eqs. (\ref{eq.b12})  and (\ref{eq.b14}) will give 
$\overline{\lan H^4(2) \ran^{m,M_S}}$ and $\overline{\lan H^2(2) \ran^{m,M_S}}$,
respectively. Then, the fixed-$(m,M_S)$ excess parameter $\gamma_2(m,M_S)$ in
the dilute limit is given by,
\be
\gamma_2(m,M_S) = \dis\frac{\dis\sum_{i+j=2,\; t+u=2} \; v_H^2(i,j) \; 
v_H^2(t,u) \; F(m_1,\Omega,i,t) \;  F(m_2,\Omega,j,u)}
{\l[ \dis\sum_{i+j=2} v_H^2(i,j)\;
T(m_1,\Omega,i)\;T(m_2,\Omega,j) \r]^2} - 1 \;,
\label{eq.spn4}
\ee
with $T(\cdots)$'s and $F(\cdots)$'s given before.

\chapter{Fixed-$(m,M)$ occupation numbers}
\label{c7a3}

\renewcommand{\theequation}{I\arabic{equation}}
\setcounter{equation}{0}   

Our purpose here is to derive a simple expression for the occupation
probabilities $\lan n_{m_{z_i}}\ran^{mM}$ for $m$ fermions in $N$ sp states
labeled by $J_z$ quantum number $m_{z_i}$. Here, $M$ are the
eigenvalues of the $J_z$ operator. As $\lan n_{m_{z_i}}\ran^{mM}$ is an
expectation value, we can write  a polynomial expansion in terms of the $J_{z}$
operator \cite{Dr-77}, 
\be 
\lan n_{m_{z_i}} \ran^{mM} = \sum_{\mu}\lan n_{m_{z_i}}
P_{\mu}(\wj_z)\ran^m P_{\mu}(\whm) \;,
\label{eq.481}
\ee
where $\whm=M/\sigma_{J_z}(m)$, $\wj=J_z/\sigma_{J_z}(m)$ and 
$P_{\mu}(M)$ are orthogonal polynomials defined by the density
$\rho_{J_z}(M)$ which is close to a Gaussian. Retaining terms up to
order 2, the expansion is, 
\be 
\barr{rcl} 
\lan n_{m_{z_i}}
\ran^{mM} & = & \lan n_{m_{z_i}}\ran^m + \lan n_{m_{z_i}} \wj_z
\ran^m \whm + \;\dis\frac{\l(\lan n_{m_{z_i}} \wj_z^2 \ran^m -\lan
n_{m_{z_i}}
\ran^m \r) \l(\whm^2 - 1 \r)}{\lan J_z^4 \ran^m - 1} \\ \\
& = & \lan n_{m_{z_i}}\ran^m + \;\dis\frac{\lan n_{m_{z_i}}
J_z \ran^m M}{\sigma_{J_z}^2(m)} \\\\
& + & \;\dis\frac{1}{2} \l(\frac{\lan n_{m_{z_i}} J_z^2
\ran^m}{\sigma_{J_z}^2(m)} - \lan n_{m_{z_i}}
\ran^m\r)\l(\frac{M^2}{\sigma_{J_z}^2(m)} - 1 \r) \;.
\earr \label{eq.fj49} 
\ee 
In the above expression we used $\lan \hat{J}_z^4 \ran^m = 3$, 
the value for a Gaussian $\rho_{J_z}(M)$. Now the formulas for the 
traces in Eq. (\ref{eq.fj49}) are as follows. Firstly,
\be 
\lan n_{m_{z_i}} \ran^m = \frac{m}{N}
\lan \lan n_{m_{z_i}} \ran \ran^1 = \frac{m}{N} \;.
\label{eq.fj51}
\ee 
This implies $n_{m_{z_i}}^{\nu=0} = \hat{n}/N$. Also, 
\be 
\lan n_{m_{z_i}} J_z \ran^m = \frac{m(N-m)}{N(N-1)} \lan \lan
n_{m_{z_i}} J_z  \ran \ran^1 \;,
\label{eq.fj52}
\ee 
with $\lan \lan n_{m_{z_i}} J_z \ran \ran^1 = m_{z_i}$. Unitary decomposition
of the number operator gives,
\be 
\lan n_{m_{z_i}} J_z^2 \ran^m = \lan n_{m_{z_i}} \ran^m \lan J_z^2 \ran^m + 
\lan n_{m_{z_i}}^{\nu=1} J_z^2 \ran^m \;,
\label{eq.fj53}
\ee
\be
\barr{rcl}
 \lan n_{m_{z_i}}^{\nu=1} J_z^2 \ran^m & = &
\;\dis \frac{m(N-m)(N-2m)}{N(N-1)(N-2)}\lan \lan
 n_{m_{z_i}}^{\nu=1} J_z^2 \ran \ran^1 \;.
\earr \label{eq.fj50} 
\ee 
Now, 
\be 
\lan \lan n_{m_{z_i}}^{\nu=1}
J_z^2 \ran \ran^1 = \lan \lan (n_{m_{z_i}}-n_{m_{z_i}}^{\nu=0})
J_z^2 \ran \ran^1 = m_{z_i}^2 - \frac{1}{N}\lan \lan J_z^2\ran
\ran^1 \;,
\label{eq.fj554} 
\ee 
where we have used the result that $n_{m_{z_i}}^{\nu=0} = \hat{n}/N$ 
deduced from Eq. (\ref{eq.fj51}). Thus, 
\be 
\barr{rcl} 
\lan n_{m_{z_i}} J_z^2
\ran^m/\sigma_{J_z}^2(m) & = & \lan
n_{m_{z_i}}\ran^m+\;\dis\frac{(N-2m)(m_{z_i}^2-\lan J_z^2
\ran^1)}{(N-2)N\lan J_z^2 \ran^1} \;.
\earr \label{eq.fj555}
\ee
Substituting above traces in Eq. (\ref{eq.fj49}) we have, 
\be 
\lan n_{m_{z_i}}\ran^{m,M}=\overline{m}+\;\dis\frac{m_{z_i}M}{N\lan
J_z^2 \ran^1}-\frac{(\overline{m}-1/2)(m_{z_i}^2-\lan J_z^2
\ran^1)(M^2-\sigma_{J_z}^2(m))} {N^2(\lan J_z^2 \ran^1)^2
\overline{m}(1-\overline{m})} \;,
\label{eq.fj556}
\ee 
where $\overline{m}=m/N$. The expression for the occupation number $\lan
n_{m_{z_i}}\ran^{mM}$ is close to that obtained in \cite{Mu-00,Zel-04} where
statistical mechanics approach has been employed. Thus, we have successfully
reproduced the previously obtained results using moment method
formalism. 

\chapter{Bivariate edgeworth expansion}
\label{c7a2}

\renewcommand{\theequation}{J\arabic{equation}}
\setcounter{equation}{0}   

Given the bivariate Gaussian, in terms of the standardized variables $\wx$
and $\wy$,
\be
\eta_\cg(\wx , \wy) = \dis\frac {1}{2\pi \sqrt{(1 - \zeta^2)}} \;
exp \l\{- \dis\frac {\wx^2 - 2\zeta \wx \wy + \wy^2}
{2(1 - \zeta^2)} \r\} \;,
\label{eq.ew6}
\ee
the bivariate Edgeworth expansion for any bivariate distribution $\eta(\wx
, \wy)$ follows from,
\be
\eta(\wx , \wy) = \exp\l\{\dis\sum_{r+s \geq 3}\,(-1)^{r+s} 
\dis\frac{
k_{rs}}{r! s!}\, \dis\frac{\partial^r}{\partial \wx^r}
\dis\frac{\partial^s}{\partial \wy^s}\r\}\;\eta_\cg(\wx , \wy) \;.
\label{eq.ew7}
\ee
Assuming that the bivariate reduced cumulants $k_{r+s}$ behave as  $k_{r_s}
\propto \Upsilon^{ -(r+s-2)/2}$ where $\Upsilon$ is a system parameter,
and  collecting in the expansion of Eq. (\ref{eq.ew7}) all the terms that
behave as $\Upsilon^{-P/2}$, $P=1,2,\ldots$, we obtain the bivariate ED
expansion to order $1/P$ \cite{Ko-84,St-87},
\be
\barr{rcl}
\eta_{biv-ED}(\wx,\wy) &=&
\l\{ 1 + \l({\dis\frac{k_{30}}{ 6}} He_{30}(\wx,\wy)
+ {\dis\frac{k_{21}}{2}} He_{21}(\wx,\wy)
\r . \r . \\ \\
& + & \l.\l.  {\dis\frac{k_{12}}{2}} He_{12}(\wx,\wy) +
{\dis\frac{k_{03}}{6}} He_{03}(\wx,\wy) \r) \r. \\ \\
& + & \l. \l(\l\{ {\dis\frac{k_{40}}{24}} He_{40}
(\wx,\wy) +
{\dis\frac{k_{31}}{6}} He_{31}(\wx,\wy)
\r.\r.\r. \\ \\
& + & \l.\l.\l.  {\dis\frac{k_{22}}{4}} He_{22}(\wx,\wy)
+ {\dis\frac{k_{13}}{6}} He_{13}(\wx,\wy) +
{\dis\frac{k_{04}}{24}}
He_{04}(\wx,\wy) \r\} \r.\r.
\earr \label{eq.ew9}
\ee
\be
\barr{rcl}
& + & \l.\l. \l\{ {\dis\frac{k_{30}^2}{72}} He_{60}(\wx,
\wy)
+ {\dis\frac{k_{30} k_{21}}{12}} He_{51}(\wx,\wy)
\r.\r.\r. 
\\ \\
& + & \l.\l.\l. \l[{\dis\frac{k_{21}^2}{8}} + {\dis\frac{k_{30} k_{12}}{12}}\r]
He_{42}(\wx,\wy) \r.\r.\r. \\ \\
& + & \l.\l.\l.
\l[{\dis\frac{k_{30}k_{03}}{36}} + {\dis\frac{k_{12}k_{21}}{4}}\r]
He_{33}(\wx,\wy) \r.\r.\r. \\ \\
& + & \l.\l.\l.  \l[{\dis\frac{k_{12}^2}{8}} + {\dis\frac{k_{21}k_{03}}{12}}\r]
He_{24}(\wx,\wy) +
{\dis\frac{k_{12} k_{03}}{12}} He_{15}(\wx,\wy)
\r.\r.\r. \\ \\
& + & \l.\l.\l. {\dis\frac{k_{03}^2}{72}}
He_{06}(\wx,\wy) \r\} \r)
\r\} \eta_{\cal G}(\wx,\wy) \;.\nonumber
\earr \label{eq.ew9-1}
\ee
The bivariate  Hermite polynomials $He_{m_1m_2}(\wx,\wy)$ in Eq.
(\ref{eq.ew9}) satisfy the recursion relation,
\be
\barr{rcl}
(1 - \zeta^2)He_{m_1+1,m_2}(\wx, \wy) & = & (\wx - \zeta \wy) He_{m_1,m_2}(
\wx,\wy) \\ \\
& - & m_1 He_{m_1-1,m_2}(\wx,\wy) + m_2 \zeta He_{m_1,m_2-1}(\wx,\wy) \;.
\earr \label{ew.11}
\ee
The polynomials $He_{m_1 m_2}$ with $m_1 + m_2 \leq 2$ are
\be
\barr{rcl}
He_{00}(\wx,\wy) & = &  1 \;,\\ \\
He_{10}(\wx,\wy) & = & \l(\wx - \zeta \wy\r)/\l(1 - \zeta^2\r) \;,\\ \\
He_{20}(\wx, \wy) & = & \dis\frac{(\wx-\zeta \wy)^2}{(1-\zeta^2)^2}
-\dis\frac{1}{(1-\zeta^2)} \;,\\ \\
He_{11}(\wx,\wy) & = & \dis\frac{(\wx - \zeta \wy)(\wy - \zeta \wx)}
{(1 - \zeta^2)^2} + \frac {\zeta} {1 - \zeta^2} \;.
\earr \label{eq.ew12}
\ee
Note that $He_{m_1 m_2}(\wx, \wy) = He_{m_2 m_1}(\wy, \wx)$.

\backmatter
\begin{spacing}{0.9}

\end{spacing}
\ed